\documentclass[a4paper,fleqn,usenatbib]{mnras}

\usepackage{newtxtext,newtxmath}
\usepackage[T1]{fontenc}
\usepackage{ae,aecompl}

\usepackage{graphicx}	% Including figure files
\usepackage{amsmath}	% Advanced maths commands
\usepackage{tabularx}
\usepackage{morefloats}
\usepackage{natbib}
\usepackage{cleveref}
\usepackage{aas_macros}
\usepackage{caption}
\usepackage{subcaption}
\usepackage{epstopdf}
\usepackage{float}
\usepackage{xcolor}

\usepackage{calrsfs}
\DeclareMathAlphabet{\pazocal}{OMS}{zplm}{m}{n}

\makeatletter
\def\thickhline{%
  \noalign{\ifnum0=`}\fi\hrule \@height \thickarrayrulewidth \futurelet
   \reserved@a\@xthickhline}
\def\@xthickhline{\ifx\reserved@a\thickhline
               \vskip\doublerulesep
               \vskip-\thickarrayrulewidth
             \fi
      \ifnum0=`{\fi}}
\makeatother

\newlength{\thickarrayrulewidth}
\title[Visibility of Black Hole Shadows in Low-luminosity AGN]%{Characterizing the relation between spin and observed source geometry of accreting black holes}
{Visibility of Black Hole Shadows in Low-luminosity AGN}

\author[Bronzwaer et al.]{Thomas Bronzwaer$^{1}$, Jordy Davelaar$^{1}$, Ziri Younsi$^{2,3}$, Monika Mo{\'s}cibrodzka$^{1}$, \newauthor H{\'e}ctor Olivares$^{1}$, Yosuke Mizuno$^{3}$, Jesse Vos$^{1}$, Heino Falcke$^{1}$
\\
$^1$Department of Astrophysics/IMAPP, Radboud University Nijmegen
P.O. Box 9010, 6500 GL Nijmegen, The Netherlands
\\
$^2$ Mullard Space Science Laboratory, University College London, Holmbury St.~Mary, Dorking, Surrey, RH5 6NT, United Kingdom
\\
$^3$Institut f\"ur Theoretische Physik, Max-von-Laue-Stra{\ss}e 1, D-60438 Frankfurt am Main, Germany
}

\date{Accepted XXX. Received YYY; in original form ZZZ}

\pubyear{2020}

\begin{document}
\label{firstpage}
\pagerange{\pageref{firstpage}--\pageref{lastpage}}
\maketitle

%We employ an extensive library of general-relativistic magnetohydrodynamics (GRMHD)-based models of both the magnetically arrested disk (MAD) and the standard and normal evolution (SANE) types, and use these GRMHD models to perform general-relativistic radiative transfer (GRRT) simulations of both disk-dominated and jet-dominated radiative models. Synthetic intensity maps and profiles are computed and discussed. To investigate the role played by optical depth explicitly, we also plot a number of optical-depth maps. For each model, we also varied in black-hole spin, observer inclination, and total flux. 

%Abstract of the paper
\begin{abstract}
Accreting black holes tend to display a characteristic dark central region called the black-hole shadow, which depends only on spacetime/observer geometry and which conveys information about the black hole's mass and spin. Conversely, the observed central brightness depression, or image shadow, additionally depends on the morphology of the emission region. In this paper, we investigate the astrophysical requirements for observing a meaningful black-hole shadow in GRMHD-based models of accreting black holes. In particular, we identify two processes by which the image shadow can differ from the black-hole shadow: evacuation of the innermost region of the accretion flow, which can render the image shadow larger than the black-hole shadow, and obscuration of the black-hole shadow by optically thick regions of the accretion flow, which can render the image shadow smaller than the black-hole shadow, or eliminate it altogether. We investigate in which models the image shadows of our models match their corresponding black-hole shadows, and in which models the two deviate from each other. We find that, given a compact and optically thin emission region, our models allow for measurement of the black-hole shadow size to an accuracy of 5\%. We show that these conditions are generally met for all MAD simulations we considered, as well as some of the SANE simulations. %We conclude that observations of low-luminosity active galactic nuclei that resemble SANE models, particularly jet-dominated models, will be challenging due to the much more varied appearances such models can take, including forms that show no image shadow at all.
\end{abstract}


%%%%%%%%%%%%%%%%%%%%%%%%%%%%%%%%%%%%%%%%%%%%%%%%%%

%%%%%%%%%%%%%%%%% BODY OF PAPER %%%%%%%%%%%%%%%%%%

\section{Introduction}

The Event Horizon Telescope (EHT) is an international network of telescopes that is capable of resolving the accretion flows (jets and discs) around the central supermassive black holes of certain low-luminosity active galactic nuclei (LLAGN) \citep{EHT1}. Primary EHT targets are the LLAGN hosted by the galaxies M87  and the Milky Way, named M87* and Sagittarius A* (Sgr A*), respectively. Such sources are thought to contain advection-dominated accretion flows (ADAF's) that produce astrophysical jets (\citealt{yuan2002}; \citealt{yuan2014}). M87* has been shown to display a central brightness depression (CBD), which theoretical calculations had predicted to be a key feature of accreting (or back-lit) black holes (see, e.g., \citealt{luminet1979}, \citealt{falcke2000}, and \citealt{noble2007}). \citet{falcke2000} showed that, given an optically thin, spherical accretion flow, the observed shape of the CBD - which we shall call the image shadow (IS) - conforms to the projection of the unstable-photon region, which those authors called the black-hole shadow (BHS). The shape of the BHS depends only on the mass and spin of the black hole (\citealt{broderick2006}; \citealt{johannsen2010}; \citealt{younsi2016}), and therefore, according to the no-hair theorem, it characterises  an electrically neutral black hole completely \citep{heusler1996}. In practice, measuring the BHS to the required precision for determining the black-hole spin is extremely technically challenging, due to the intrinsic spatio-temporal variability of the source (\citealt{johannsen2010}; \citealt{broderick2014}). The mass, however, which is directly proportional to the radius of the BHS, is a more robust observable. The mass of an LLAGN such as Sgr A* may also be experimentally determined from the orbital elements of stars in its immediate environment, such as the star S2 \citep{boehle2016}, and from infrared interferometry experiments by the GRAVITY collaboration \citep{gravity2020}. In this way, combined measurements enable a strong-field test of whether the observed objects are black holes as described by Einstein's general theory of relativity (GR) as opposed to other objects described by an alternative theory (see, e.g.,~\citealt{broderick2014}; \citealt{psaltis2015bbb}; \citealt{vincent2015ccc}; \citealt{hertog2017}; \citealt{mizuno2018}; \citealt{olivares2018}; \citealt{hertog2019}). 

Unlike the BHS, the IS of an accreting black hole is determined both by the strongly curved spacetime in the vicinity of the black hole (parametrized by the black hole's mass and spin) and by the properties of the radiating plasma, such as the geometrical shape of the accretion disc and the radiating electrons' energy-distribution function (see, e.g., \citealt{broderick2006a}; \citealt{noble2007}; \citealt{moscibrodzka2009}). Although it has been demonstrated that, except in the case of an optically thick accretion disc that obscures the black hole, one expects to observe a significant CBD in GRMHD models of accreting black holes (\citealt{moscibrodzka2014}; \citealt{chan2015}), it is still true that the IS does not always match the BHS. \citet{gralla2019} showed that certain analytical models may produce circular IS's that do not match their associated BHS's. Such mismatches may also occur in physically motivated models. For example, some GRMHD-based models of accreting black holes show a clear evacuated region that scales with the ISCO (see Appendix \ref{app:evacuation}), as matter in this region will rapidly plunge into the black hole \citep{bardeen1972}. Thus, we expect to see little to no radiating matter within this region, which is wider than the horizon itself, potentially causing the IS to be dominated by the size of the evacuated region in the accretion flow. The IS will then appear to be larger than the BHS. Conversely, in the case of an optically thick accretion flow, the BHS may be completely obscured, eliminating the IS entirely. \citet{wielgus2020} develop an analytical model to represent the accretion flow around M87*, and show that the observed image features are, in general, highly dependent on the geometry of the accretion flow. Such considerations reflect the fact that an experiment is only meaningful to the extent that one understands the setup (meaning the astrophysical circumstances) of the experiment. A crucial question is just how sensitive the conclusions of the experiment are to one's knowledge of the astrophysical circumstances of the accretion flow, such as its optical depth at a given frequency and line of sight. For these reasons, the totality of astrophysical observations of a source must be considered for an astrophysical test of GR.

The question of when the IS and the BHS (mis)match generally is rendered more difficult by the vast diversity of accretion-flow geometries encountered. \citet{narayan2019} showed that, in the case of an optically thin, spherically symmetric accretion flow, the IS always matches the BHS, independently of the inner boundary of the radiating region. In this work, we look at physically motivated models of LLAGN (which are not spherical), by constructing a library of GRMHD simulations and radiative models. The GRMHD simulations are of both the magnetically arrested disc (MAD) (\citealt{bisnovatyi1974}; \citealt{narayan2003};  \citealt{tchekhovskoy2011}) and the standard and normal evolution (SANE) types \citep{narayan2012}. In the former type of simulation, strong magnetic fields arrest the accretion process close to the black hole. In SANE simulations, on the other hand, no magnetic `pile-up' occurs, causing a radical change in the observed source morphology. Whether an accretion flow is MAD or SANE depends on the initial conditions of the magnetic field and the size of the accretion disc (see, e.g., \citet{ripperda2020bbbb}. The radiative models considered in this work consist of a single-electron-temperature, disc-dominated model \citep{moscibrodzka2009}, and a two-electron-temperature, jet-dominated model (\citealt{moscibrodzka2014}; \citealt{accpart}). We also vary the black-hole spin of our models, to determine whether it has a substantial effect on our considerations; as higher prograde spins imply a smaller event horizon and innermost stable circular orbit (ISCO), the orbital velocity (and thus the effects of relativistic boosting) will be enhanced in those cases. In order to understand the relationship between the BHS and the IS for these models, we plot maps of the specific intensity at 230 GHz of all models, for a range of observer circumstances. This observing frequency is appropriate because we set the black-hole mass, as well as our distance to the source, to be those of Sgr A*, but this is done only to provide a baseline LLAGN model that is optically thin at that frequency; our aim is explicitly not to fit our models to Sgr A* itself. Instead, we wish to explore a physically plausible range of LLAGN models. For these models, we investigate how clear a BHS they show, and measure the source size and position angle. We discuss the findings of \citet{gralla2019} and \citet{narayan2019}, and explain the ways in which the IS can deviate from the BHS in the case of general (non-spherical) accretion flows.

In Section \ref{sec:theory}, we examine the BHS and the IS in detail, and discuss how they may differ. The construction of our GRMHD-based radiative models is covered in Section \ref{sec:methods}. Section \ref{sec:results} presents the resulting simulated data. Finally, our results are discussed in Section \ref{sec:discussion}.

\section{Black-hole shadow vs. image shadow}
\label{sec:theory}

The BHS \citep{falcke2000} is an optical effect - a darkening of a certain region on the sky - that occurs for any black hole, but may not always be observable. The BHS's shape matches that of the projection of the black hole's unstable-photon region on the sky, and its shape is generally (except for cases where the observer's line of sight is nearly aligned with the black-hole spin axis) informative about the black hole's properties, specifically its mass and spin \citep{younsi2016}. Since the BHS depends only on the shape of null geodesics, which are independent of radiation frequency, the BHS is achromatic, as long as the radiation wavelength is much smaller than the Schwarzschild radius, and the regime of geometrical optics applies - an assumption that is made throughout this paper.

When observing an accreting black hole, whether simulated or physical, the actually observed IS may or may not align closely with the BHS, because the IS is affected not only by the spacetime/observer geometry, but also by the observing frequency (the IS is therefore not achromatic, unlike the BHS) and by properties of the accretion flow \citep{moscibrodzka2014}. The size of the IS is a key observable of the EHT, and it has been investigated, for various accretion scenarios, using a combination of numerical simulations of the black-hole accretion flow and general-relativistic ray tracing (GRRT) algorithms (see, e.g.,~\citealt{luminet1979}; \citealt{falcke2000}; \citealt{noble2007}; \citealt{dexter2009}, \citealt{psaltis2013}; \citealt{ipole}; \citealt{raptor1}). This is necessary because the shape and size of both the BHS and the IS are a function of many variables, including the black-hole spin and the observer inclination. Expressing all lengths in terms of the black hole's gravitational radius, $R_{\rm g}=GM/c^2$, we note that the physical radius of the unstable-photon orbit region in the equatorial plane shrinks from $3 R_{\rm g}$ to $1 R_{\rm g}$ as the black-hole spin $a$ goes from 0 to 1 \citep{bardeen1972}. Similarly, the ISCO of a black hole shrinks from 6 $R_{\rm g}$ in the Schwarzschild case to 1 $R_{\rm g}$ for an extreme-Kerr black hole \citep{bardeen1972}. A consequence of these effects is that the inner region of the accretion disc will have a much higher orbital velocity compared to the low-spin case. Relativistic boosting, which causes the side of the accretion flow that approaches the observer to appear significantly brighter than the side that recedes from the observer, is strongly enhanced by this effect. Relativistic boosting is also a strong function of the observer inclination, as it becomes negligible for low inclination angles, where the observer's line-of-sight is nearly perpendicular to the flow velocity. Thus, while the apparent size of the BHS barely changes with respect to the spin and observer inclination, the same may not be true for the observed geometry of an accretion flow.

Given the uncertainties just mentioned, we investigate the relation between the BHS and the IS for the case of non-spherical accretion models of LLAGN. We identify two key processes that may cause the two to diverge:
\begin{itemize}
\item Timelike orbits within the black hole's ISCO are unstable, causing matter in this region to quickly plunge into the black hole. This produces an evacuated region of reduced density in our SANE GRMHD models (see Appendix \ref{app:evacuation}), which appears dark compared to the high-density regions, and can cause the IS to be larger than the BHS. The extent of the evacuated region, and thus the properties of the IS, depend strongly on the black-hole spin; at high spins, the evacuated region shrinks down to nearly the size of the photon ring \citep{bardeen1972}, so that the two will visually align (see Section \ref{sec:results}), while at minimal (retrograde) spins, the discrepancy reaches its maximum.

\item Besides the optical appearance of the evacuated region, GRRT calculations have shown that the shape of the IS can also be affected through obscuration by the accretion disc and/or jet \citep{noble2007, moscibrodzka2009, dexter2010}. The extent of the obscuration may range from minor to total, if the accretion disc is optically thick, and in such cases, the IS offers no clear way to determine the mass and spin of the black hole.
\end{itemize}

In case of severe or total obscuration of the BHS by the plasma, no shadow is visible at all. However, in that case, no ring-like structure is observed, either, and thus the question of whether the observation of a ring-like feature entails the observation of a BHS \citep{gralla2019} does not arise in such cases. The most interesting cases to our investigation are therefore those of partial or moderate obscuration of the BHS by the plasma.

In order to compare the BHS and IS of our LLAGN models, we employ an analytical method of describing the BHS \citep{younsi2016}, which is used to create the dashed lines indicating the BHS in the GRMHD-based images of AGN shown in this paper (see Section \ref{sec:intensity_maps}). The relationship between BHS and IS is examined in more detail, and compared to the EHT's error bars for measuring the BHS in its 2017 campaign, in Section \ref{sec:intensity_profiles} (see \citet{eht6} for a more in-depth discussion of the EHT's treatment of the BHS).

\section{Simulation Setup}
\label{sec:methods}

Simulated black-hole observations are produced in two steps: in the first step, a GRMHD simulation is employed to construct a numerical facsimile of the accreting plasma. In the second step, radiative-transfer calculations are performed to determine the appearance of the radiating plasma to a distant observer. In this section, we we discuss the specifics of our GRMHD simulations and radiative models.

\subsection{GRMHD simulations}
\label{sec:GRMHD}

We use the 3D-GRMHD simulations made for the EHT \citep{EHT5} using the GRMHD code {\tt BHAC} (\citealt{bhac}; \citealt{olivares2019}). These simulations were performed for five values for the black-hole spin: $a \in [-0.9375, -0.5, 0, 0.5, 0.9375]$. The initial conditions are those of a Fishbone-Moncrief torus \citep{fishbone1976}, which is a torus in hydrodynamic equilibrium. Both SANE and MAD versions were made of these simulations. For a 3D-GRMHD simulation, one must specify a numerical grid of three spatial dimensions ($r$, $\theta$, and $\phi$, respectively) in {\tt BHAC}; these runs were performed in spherical modified-Kerr-Schild (MKS) coordinates \citep{gammie2003}. The spatial extent of the GRMHD simulations ranges from the horizon (whose location is different for each black-hole spin) to an outer radius of $3300~M$ for the SANE runs and $2500~M$ for the MAD runs. The spatial resolution was set to $N_{r}=512$, $N_{\theta}=192$, $N_{\phi}=192$ for the SANE models, and $N_{r}=384$, $N_{\theta}=192$, $N_{\phi}=192$ for the MAD models. A lower resolution in the $r$ direction was chosen for the MAD simulations for reasons of computational efficiency; due to the smaller outer radius of the MAD simulations, the spatial resolution is similar in both cases. Geometrized units are used in {\tt BHAC}, so that $G=c=1$, so that $R_{\rm g}=M$, and $M=1$. Time is measured in units of $R_{\rm g}/c$, which is equivalent to $M$; snapshots of the instantaneous state of the plasma were created with temporal intervals of $10~M$. In each case, 200 GRMHD snapshots were selected from a range of the simulation in which the accretion rate is relatively stable (preceded by an unstable phase in which accretion starts and grows), which occurs around $t > 8000~M$ for the SANE models and $t > 10000~M$ for the MAD models. Thus, the range of simulation time in which  we perform our GRRT calculations is $2000~M$. For Sgr A*, this corresponds to approximately 11 hours, which is of the same order as an EHT observing run. Note that, in all models, the black-hole spin axis is aligned with the rotation axis of the accretion flow; for a study of the effects of misalignment of the two, see \citet{chatterjee2020} and \citet{white2020ccc}. For more details regarding the construction of the 3D-GRMHD simulations, please refer to \citealt{bhac}, \citealt{olivares2019}, \citealt{grmhd_comparison}, and \citealt{EHT5}.

%Time variability... Accretion rate as function of time...

\subsection{Radiative models}

We use the radiative-transfer code {\tt RAPTOR} \citep{raptor1}, which sets up a grid of virtual light rays (one ray per pixel in the observer's screen, thus forming a virtual camera), and then solves the equation of motion for the field of rays, whilst simultaneously solving the radiative-transfer equation along the resulting ray paths.

The dominant emission mechanism of our model is thermal synchrotron emission and self-absorption (see \citet{leung2011} for the associated emission and absorption coefficients). These quantities are computed based on the local state of the plasma. Our set of models will be based on Sgr A*; we use the black-hole mass estimate $M \simeq 4 \times 10^6 M_{\odot}$ and the distance estimate $d_{\rm obs} \simeq 8 \ {\rm kpc}$ from \citet{boehle2016}. We reiterate that it is not our intent to find a model that fits Sgr A*'s observational data, but rather to explore a physically plausible range of LLAGN models, in order to examine under which circumstances the IS matches the BHS. For this reason, we explore the effects on the LLAGN's appearance of applying different flux scalings for some models. Most of the resulting images are incompatible with observations of Sgr A*. Our findings pertain to any LLAGN with an ADAF, as long as it is observed at the right frequency. For an overview of the emission models on which ours are based, see, e.g., \citealt{moscibrodzka2009}, \citealt{moscibrodzkafalcke2013}, \citealt{moscibrodzka2014}, \citealt{moscibrodzka2016}, \citealt{accpart}, and \citealt{davelaar2019bbb}.

\begin{table}
\def\arraystretch{1.25}
\centering
%\textbf{ MAD disk, 2.5 Jy} \\
\begin{tabularx}{0.48\textwidth}{@{}p{0.078\textwidth} p{0.048\textwidth} p{0.058\textwidth} p{0.058\textwidth} p{0.058\textwidth} p{0.058\textwidth}@{}}
      \thickhline
       & $R_{\rm high}$ & $R_{\rm low}$ & N1 & N2  & N3 \\
      \hline
SANE jet & $25$ & $1$ & $512$ & $192$ & $192$\\ 
SANE disc & $1 $ & $1$ & $512$ & $192$ & $192$\\ 
MAD jet & $25 $ & $1$ & $384$& $192$ & $192$\\ 
MAD disc & $1 $ & $1$ & $384$ & $192$ & $192$\\ 
      \thickhline
    \end{tabularx} \\
\caption{Key model parameters of our four model classes. All models are computed for five black-hole spin values and four observer-inclination values.}
\label{tab:M_unit_MAuihihD_disk}
\end{table}

It is not yet known whether Sgr A* exhibits a clearly visible jet (see, e.g., \citealt{falckemanheim1993}; \citealt{falckemarkoff2013}; \citealt{issaoun2018bbb}). The most influental emission model for our work is \citet{moscibrodzka2016}, which introduced both jet-dominated and disc-dominated scenarios for accreting black holes. The two classes of models emerge by coupling the temperature ratio of the protons and the electrons to the plasma-beta parameter, $\beta= \frac{P_\mathrm{gas}}{P_\mathrm{mag}}$, which is the ratio of gas pressure to
magnetic field pressure. The coupling is done as follows:
\begin{equation}
\frac{T_{\rm p}}{T_{\rm e}} = \frac{1}{1+\beta^2} + R_{\rm high} \frac{\beta^2}{1+\beta^2},
\end{equation}
where $T_{\rm p}$ is the proton temperature, $T_{\rm e}$ the electron temperature, while $R_{\rm high}$ is a free parameter. When $\beta\ll 1$ the temperature ratio is set to unity, while for $\beta\gg 1$ the temperature ratio is set by $R_{\rm high}$.  In the case of the disc-dominated radiative model, $R_{\rm high}=1$; in the case of the jet-dominated model, $R_{\rm high}=25$ (reproducing the model shown in \citealt{accpart}), which causes the electron temperature (and thus the emitted radiation) to be suppressed in the disc regions, emphasising the jet region more.

An important assumption made in our model is that we include only thermal electron populations. For sources such as Sgr A*, emission from electrons with a non-thermal energy-distribution function is thought to be important to explain the observed time variability and the high frequency emission \citep{ball2016,mao2017,gravity2018,accpart,ripperda2020bbbb,Dexter2020,Porth2020,Petersen2020}. However, at 230 GHz, time-averaged images of this model will look similar to the thermal case when adding non-thermal electrons \citep{accpart}, although it is shown in that work that high-spin models of M87 do become more optically thin because of this addition. Thus we focus on this frequency only, and explore a wide range of spins and models.

\subsection{Time-averaging and flux calibration}

\begin{figure*}
\centering
\begin{subfigure}[b]{0.49\textwidth}
	\includegraphics[width=\textwidth]{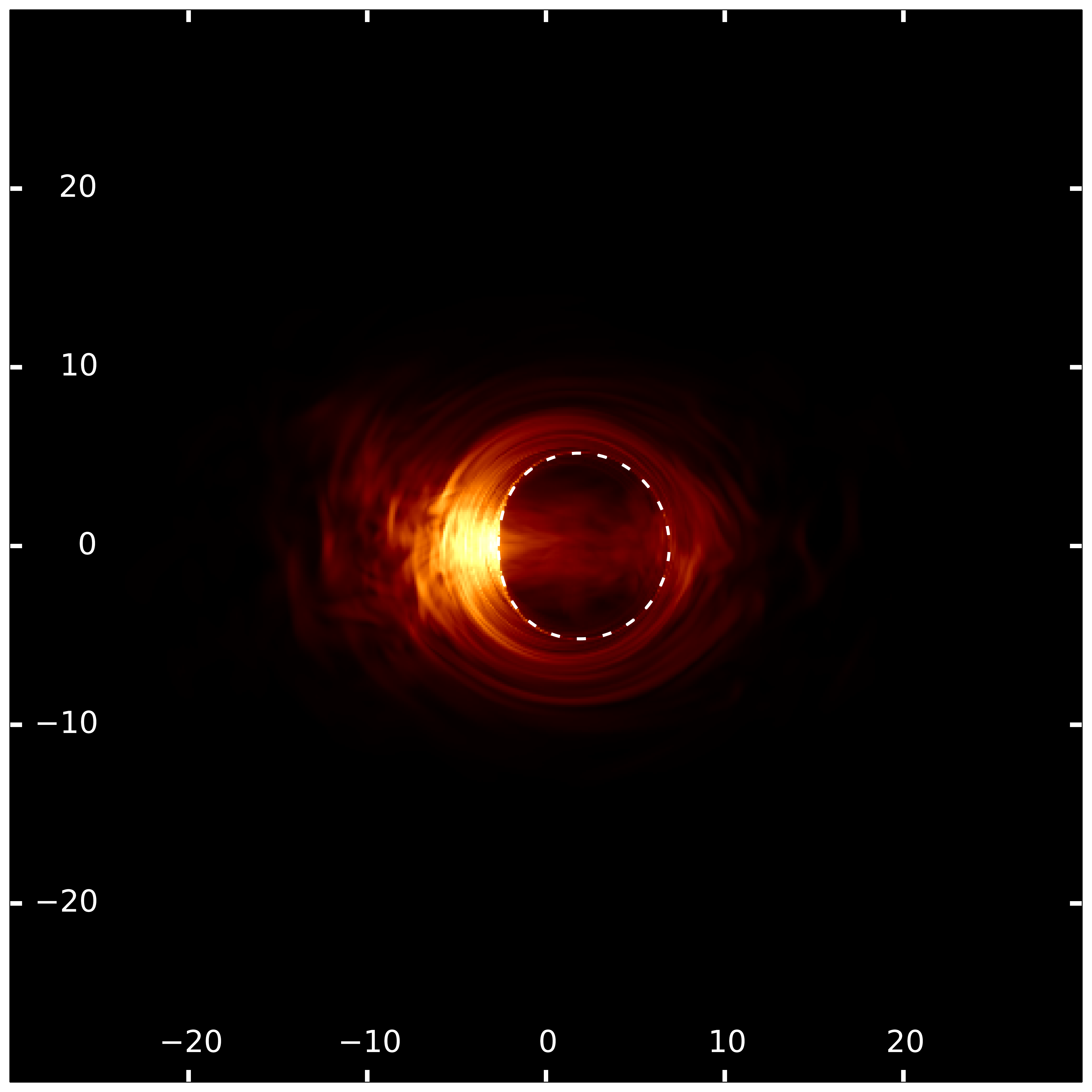}
\end{subfigure}
\begin{subfigure}[b]{0.49\textwidth}
	\includegraphics[width=\textwidth]{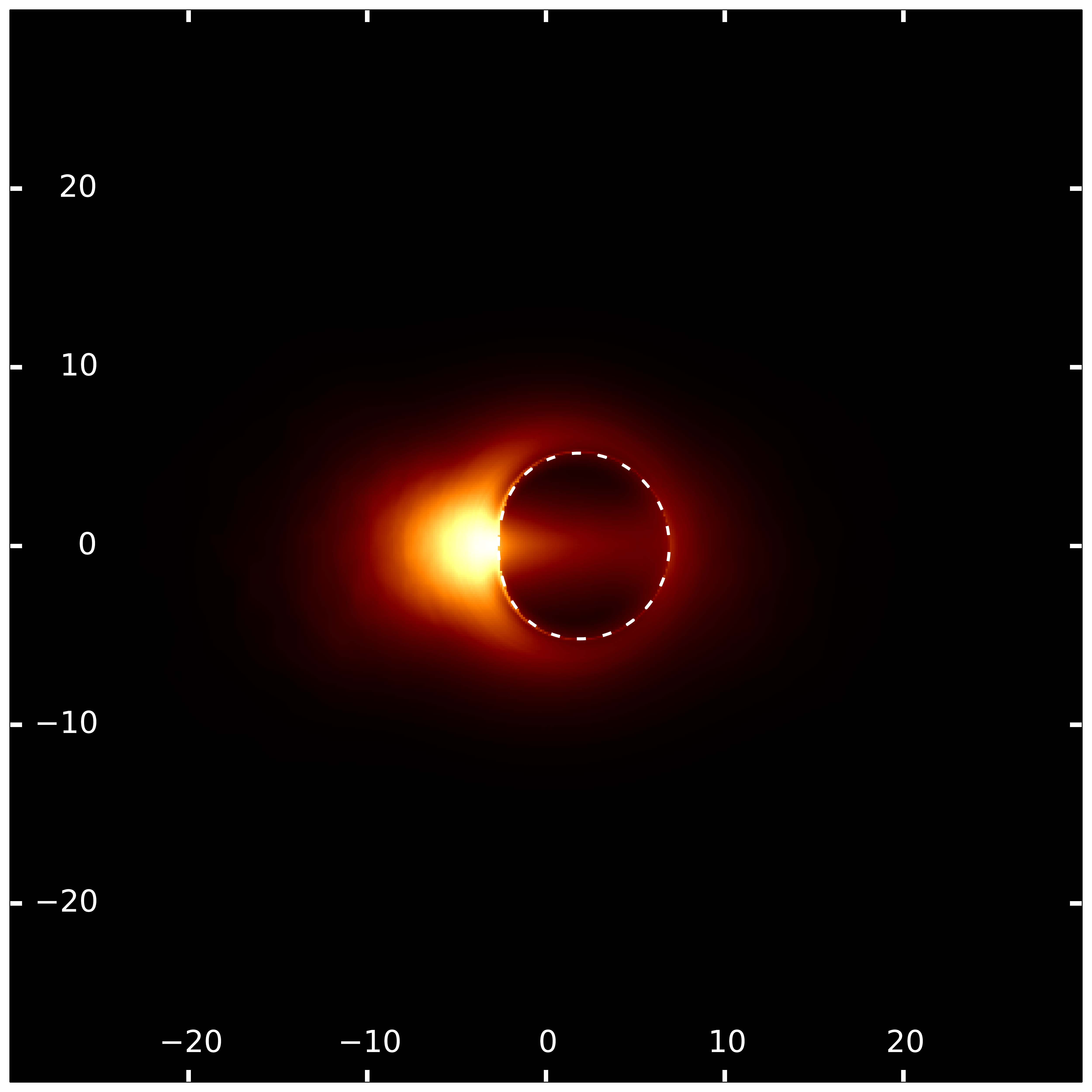}
\end{subfigure}
\caption{An instantaneous image of the accretion disc, using data from a single GRMHD slice of the SANE disc simulation (left panel), and a time-averaged image over a range of $2000~M$ (approximately 12 hours), using 200 GRMHD slices from the SANE disc simulation (right panel). In both panels, the predicted location of the BHS is indicated with a dashed line. The values for the impact parameters along the x- and y-axes are expressed in terms of $R_{\rm g}$. Both panels are of the SANE disc 3D-GRMHD simulation, for which $a=0.9375$ and $i=90^\circ$. The flux is calibrated to 2.5 Jy. Note that, since the flow moves in the azimuthal direction, time-averaging has the effect of 'smearing out' any heterogeneous features in the azimuthal direction, causing the accretion disc to resemble a 2D-GRMHD model.}\label{fig:time_averaging}
\end{figure*}

{\tt RAPTOR} is used to produce images of GRMHD snapshots. One snapshot is used for each image, which is equivalent to using the fast-light approximation, in which the speed of light is effectively treated as being infinite, and the plasma does not evolve while the light propagates through it. For a demonstration that this approximation is appropriate in the present case, see \citet{raptor1}. In order to simulate the effects of prolonged observations (which typically last hours in the case of EHT), 200 subsequent images are added together, meaning that we average over a temporal range of 2000 $M$ (see Section \ref{sec:GRMHD}), or about 12 hours (comparable with one day of EHT observations). Figure~\ref{fig:time_averaging} shows the effects of time-averaging on intensity maps of our SANE disc model with $i=90^\circ$ and $a=0.9375$. For a more in-depth look at the effects of time-averaging on the observed source size and morphology, see Appendix \ref{app:averaging}.

One of {\tt RAPTOR}'s key variables is the mass scaling factor, $\mathcal{M}$, which determines the overall accretion rate in the simulation (which would otherwise be scale-free, as GRMHD simulations are; see, e.g., \citet{bhac}). Generally, lower values of $\mathcal{M}$ entail a more compact and optically thin source size. {\tt RAPTOR} also allows calibration of $\mathcal{M}$ to achieve a desired integrated flux density. The flux densities that were used for calibration in this work take Sgr A* as an archetypical example of a LLAGN; its flux density at $230$ GHz is approximately $2.5$ Jy \citep{doeleman2008}. Since we aim to explore a range of possible LLAGN sources, we additionally explore lower fluxes of 1.25 Jy and (in the case of the SANE jet model, which is particularly variable with respect to flux scaling) 0.625 Jy. We shall see that in some models, the appearance of the IS changes drastically based on the flux scaling, while in other models, it is nearly constant. Another reason for exploring lower fluxes is that, for the M87 results, the inner few $R_{\rm g}$ accounted for roughly half of the flux observed for the unresolved core region \citep{EHT5}. 

In order to calibrate the time-averaged image to the correct integrated flux density, a binary-search algorithm was used that works in the following way:
\begin{itemize}
\item Set $\mathcal{M}_{\tt min}$ and $\mathcal{M}_{\tt max}$ (these are estimates that are set using trial and error).
\item Launch 20 {\tt RAPTOR} runs, evenly distributed over the $2000~M$ time range, using $\mathcal{M}_{\tt cur} = \left( \mathcal{M}_{\tt max} + \mathcal{M}_{\tt min} \right) / 2$.
\item Compare the average flux of the 20 runs with the target flux. If lower than target flux, $\mathcal{M}_{\tt max} = \mathcal{M}_{\tt cur}$; if higher than target flux, $\mathcal{M}_{\tt min} = \mathcal{M}_{\tt cur}$. If within range of target flux (we chose the range to be $\pm 0.01 {\rm Jy}$), stop algorithm and return $\mathcal{M}_{\tt cur}$.
\end{itemize}

Although it has been shown that the time averaging of GRMHD images creates artefacts in the context of a synthetic-data pipeline \citep{medeiros2018}, the time-averaged models offer a clearer, more simplified view of the emitting geometry (by averaging out local, random fluctiations), and illustrate a number of optical effects that are significant to this work.

\subsection{Image processing}

A recipe is needed to quantify key aspects of a given image, such as the source size. We define an image's impact parameters, $\alpha$ and $\beta$, to mean the coordinate offset of a particular ray with respect to the vertical and horizontal symmetry axes of the image, respectively. The impact parameters are expressed in terms of the gravitational radius, $R_{\rm g}$. All images in this work consist of a grid of 512 by 512 pixels. When constructing estimators for the source size, it is convenient to employ the so-called image moments; the image moment $M_{pq}$ is given by
\begin{equation}
M_{pq} = \sum_x^W \sum_y^H x^p y^q S\left(x,y\right),
\label{eqn:moment}
\end{equation}
where $x$ and $y$ are the pixel indices, $p$ and $q$ are the image moment's order in the horizontal and vertical direction, and $S\left(x,y\right)$ is the flux density at frequency $\nu$ of pixel $(x,y)$. Note that $p,q,x,y$ are nonnegative integers.

Using the image moments, various quantities can be constructed that quantify certain aspects of the image, such as an estimate of the size of the emission region. The standard deviation, which is an indicator of the source size, is given by:
\begin{equation}
\sigma_x = \sqrt{\frac{M_{20}}{M_{00}}-\left(\frac{M_{10}}{M_{00}}\right)^2}.
\label{eq:spread}
\end{equation}
The value of $\sigma_x$ depends (except for sources with circular symmetry) on the position angle of the source. Using the image moments, rotation-invariant measures of the source size - namely, the major axis, $\lambda_{\rm max}$, and the minor axis, $\lambda_{\rm min}$ - can be constructed as well. For a more complete discussion of their derivation, please refer to Appendix \ref{appA}.

\section{Results}
\label{sec:results}

In this section, we present all images and plots produced from our library of GRMHD simulations and corresponding {\tt RAPTOR} images. In Subsection \ref{sec:intensity_maps}, we plot maps of the specific intensity of the source at a frequency of 230 GHz, with the analytical prediction of the shape of the BHS overplotted in each case.  Subsection \ref{sec:intensity_profiles} contains normalised cross-cut profiles of the intensity maps, in order to get a clearer picture of the (mis)match between BHS and IS. Maps of $\tau_{\rm 230GHz}$, the optical depth at our observing frequency of 230 GHz, are presented in Subsection \ref{sec:optical_depth_maps}. Intensity maps of two of our SANE disc models, but at a higher observing frequency of 6 THz (which lies in the infrared range of the electromagnetic spectrum, a different observational window that corresponds to a different appearance of the accretion flow), are shown in Subsection \ref{sec:high_freq_maps}. Note that tables of $\lambda_{\rm max}$ and $\lambda_{\rm min}$, the major and minor axes of the source, are presented in Appendix \ref{sec:source_size_appendix}, while tables of $\mathcal{M}$, the calibration factor used for {\tt RAPTOR}, are listed in Appendix \ref{sec:flux_calibration_appendix}.

\subsection{Intensity maps for all models}
\label{sec:intensity_maps}

This subsection contains intensity maps of all GRMHD models, plotted using the square-root scale for additional clarity. The images are categorised in the next four subsections by their quasi-steady-state of the disc (MAD or SANE) and radiative model (disc or jet).

\subsubsection{SANE jet models}

\begin{figure*}
\centering
\begin{subfigure}[b]{0.49\textwidth}
	\includegraphics[width=\textwidth]{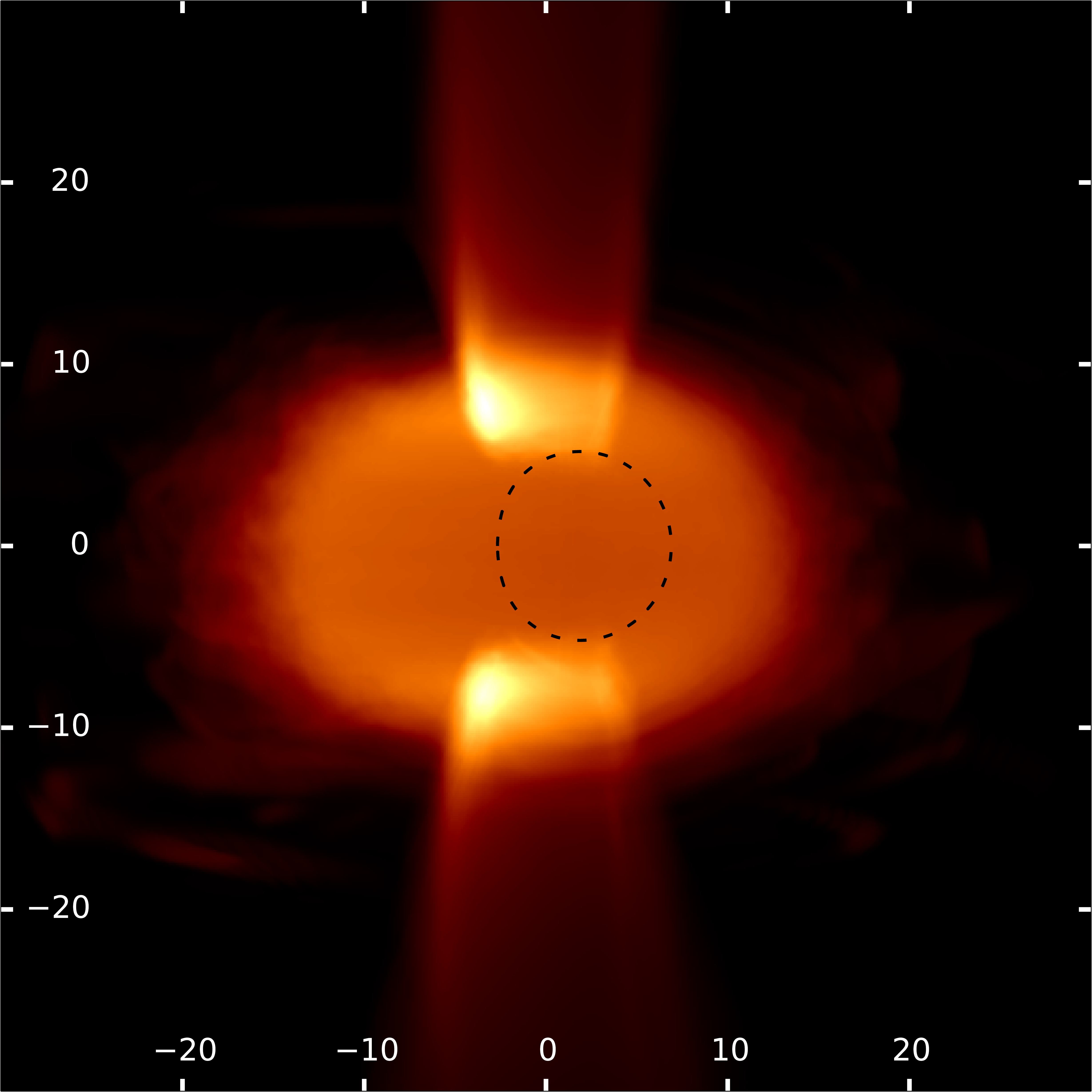}
\end{subfigure}
\begin{subfigure}[b]{0.49\textwidth}
	\includegraphics[width=\textwidth]{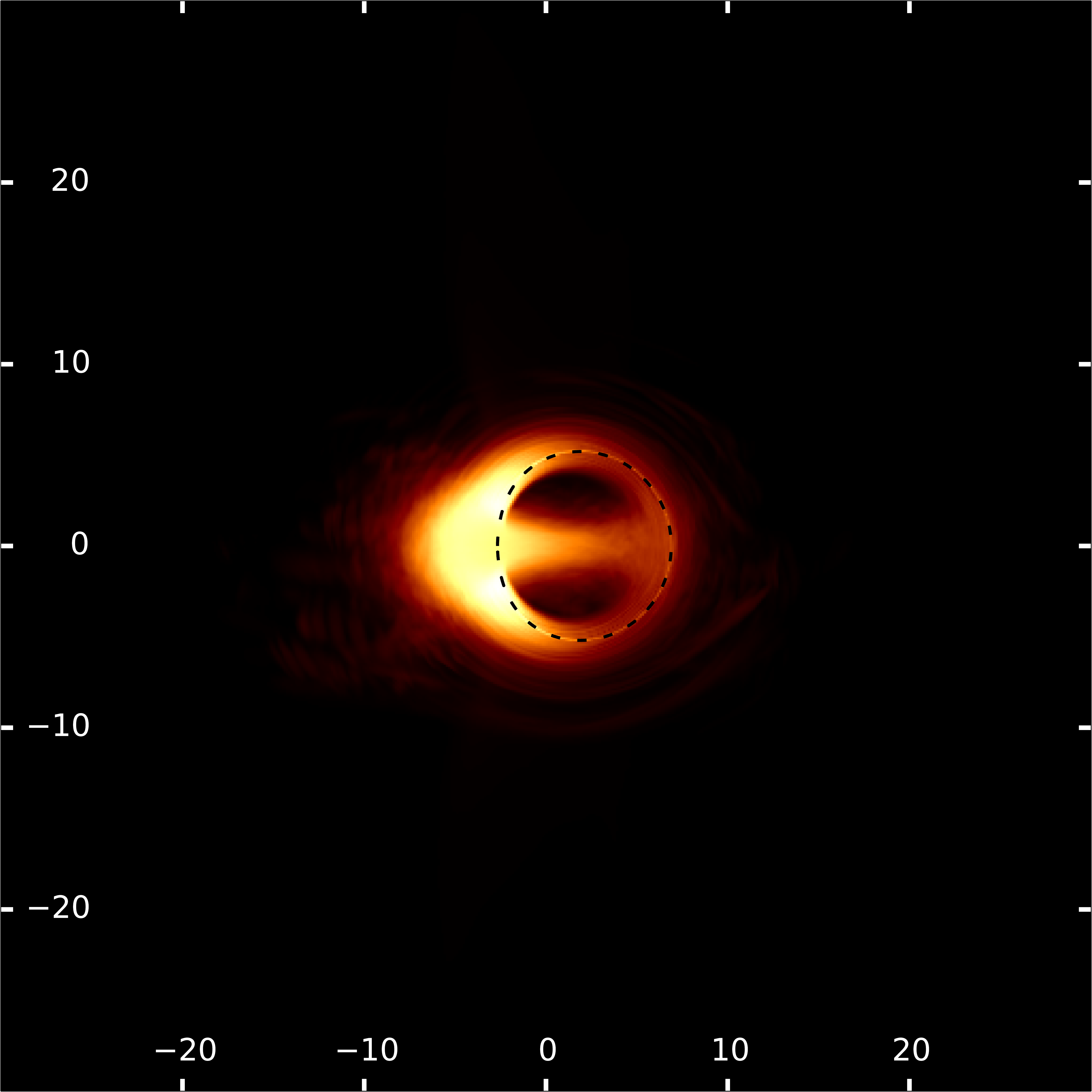}
\end{subfigure}
\caption{Time-averaged images of the SANE jet model with $a=0.9375$ and $i=90^\circ$, with an integrated flux density of 2.5 Jy (left panel) and 0.625 (right panel). The dashed line indicates the BHS. The morphology of this model is particularly sensitive to the integrated flux density. In the high-flux case, the IS vanishes, and the BHS cannot be observed. In the low-flux case, the IS is visible but partially obscured. }\label{fig:tbafig}
\end{figure*}

\begin{figure}
	\includegraphics[width=0.49\textwidth]{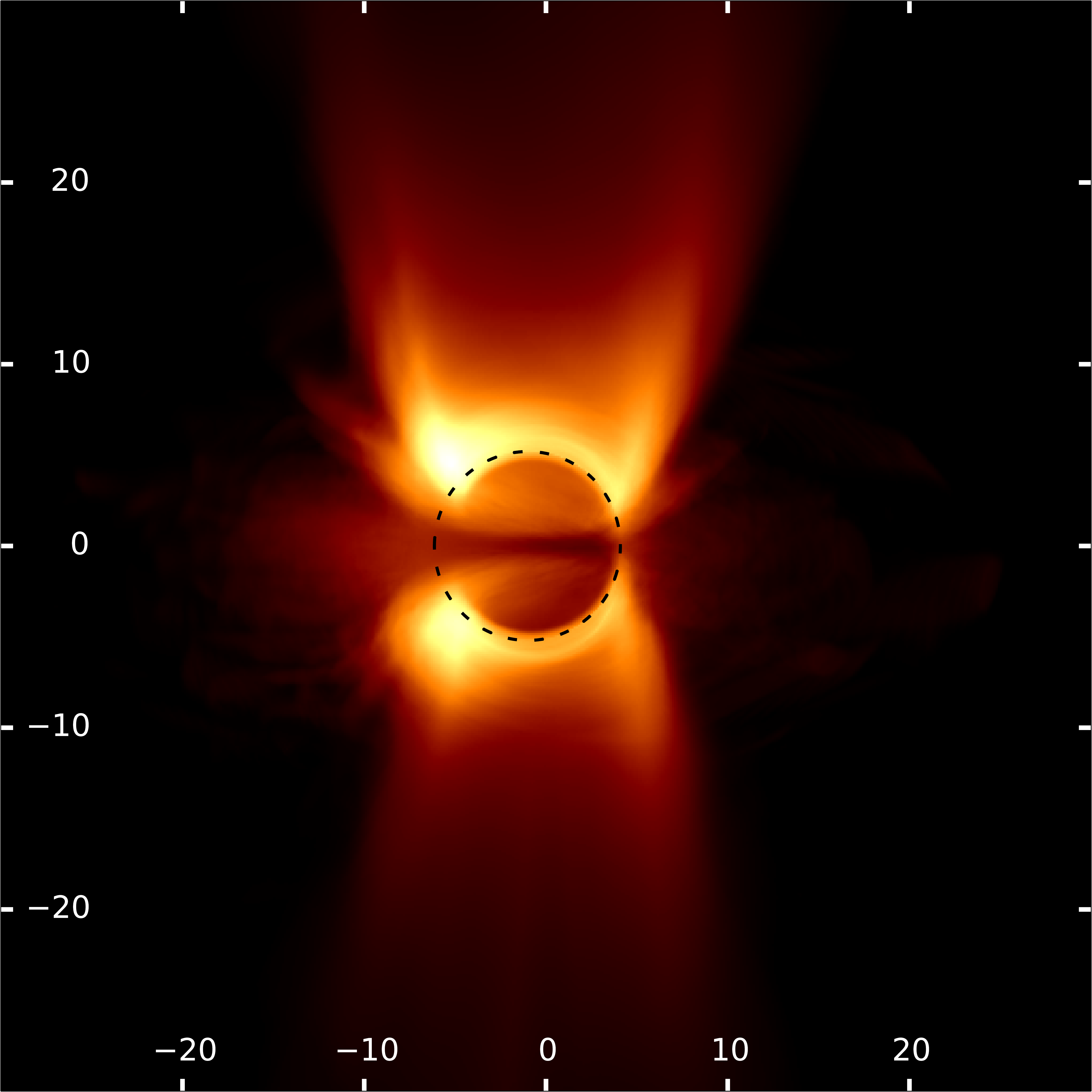}
\caption{Time-averaged image of the SANE jet model with $a=-0.5$ and $i=90^\circ$, with an integrated flux density of 2.5 Jy. The dashed line indicates the BHS. Although an obscuring disc is present, the IS still tracks the BHS.}\label{fig:tbafig2}
\end{figure}

\begin{figure}
	\includegraphics[width=0.49\textwidth]{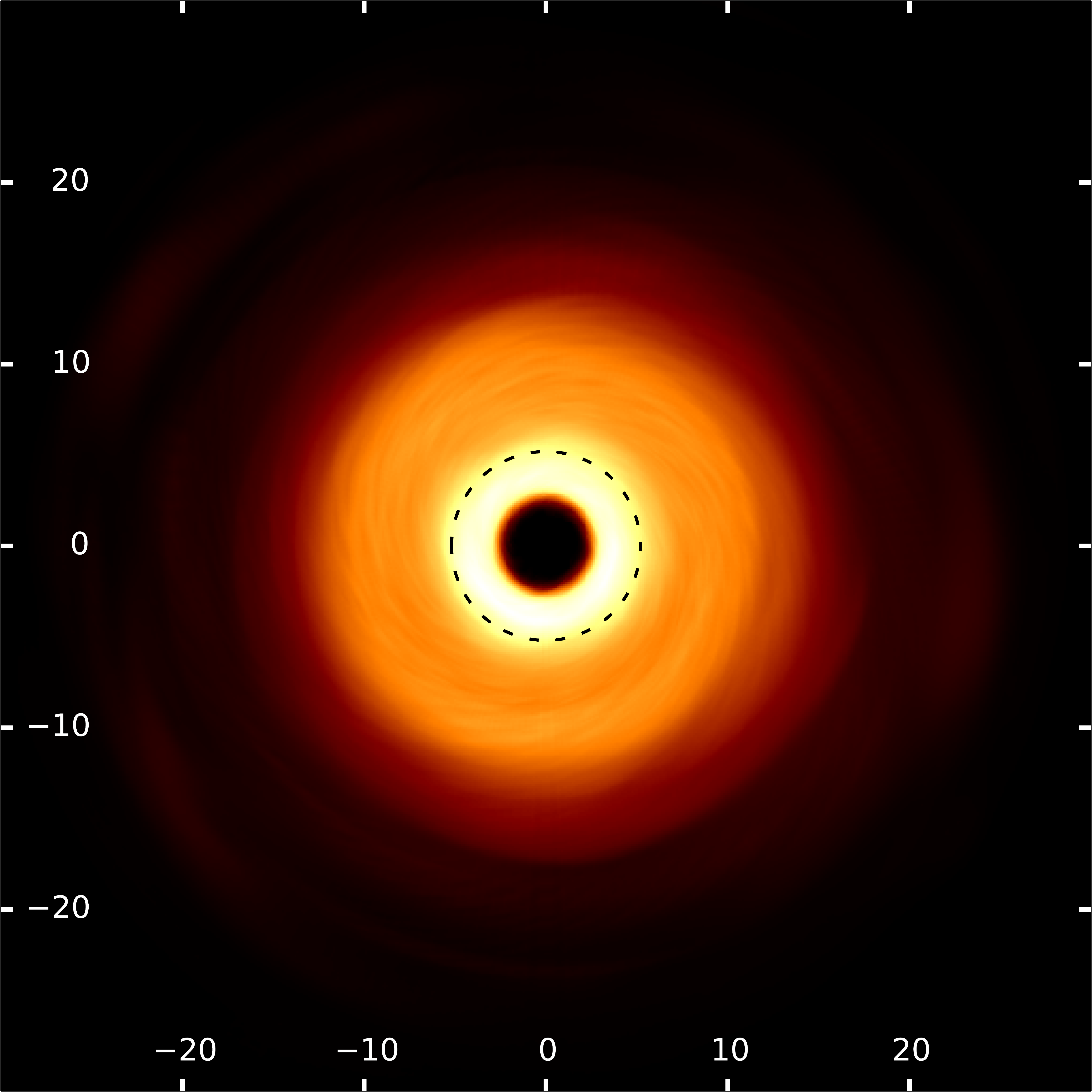}
\caption{Time-averaged images of the SANE jet model with $a=0$ and $i=1^\circ$, with an integrated flux density of 2.5 Jy. The dashed line indicates the BHS. At low inclinations, the SANE jet models show circular IS's that underestimate the BHS.}\label{fig:tbafig3}
\end{figure}

Figures \ref{fig:sane_jet_25_matrix}, \ref{fig:sane_jet_125_matrix}, and \ref{fig:sane_jet_0625_matrix} show the SANE jet model calibrated to 2.5 Jy, 1.25 Jy, and 0.625 Jy, respectively. These two-temperature SANE models are the most diverse in terms of source morphology; many, but not all, show a pronounced jet at 230 GHz. As was shown in \citet{moscibrodzka2016}, for SANE jet models of M87, the counter-jet (which faces away from the observer) can be magnified by the action of the gravitational lens, and thus appear to be larger than the observer-facing jet \citep{EHT5}. Note that the optical depth of these models varies significantly between the different fluxes (see Subsection \ref{sec:optical_depth_maps}). In particular, Fig.~\ref{fig:tbafig} shows that the $a=15/16$ SANE jet models viewed with $i=90^\circ$ range from completely optically thick at 2.5 Jy (no IS is visible at all in this case) to optically thin at 0.625 Jy, in which case the IS aligns more closely with the BHS. When viewed at low inclination angles, the jet models strongly obscure the BHS at higher fluxes; Fig.~\ref{fig:tbafig3} shows an IS that is circular and centred on the black hole, but significantly smaller than the BHS. On the other hand, in some optically thick cases, such as the model with $a=-1/2$ and  $i=90^\circ$ at 2.5 Jy (Fig.~\ref{fig:tbafig2}, the outline of the BHS is clearly visible, despite the presence of an obscuring, optically thick disc.

An interesting effect can be observed in the $a=15/16$ case for the 1.25 Jy model: this particular model is nearly isotropic in its emission. This can be seen in Table \ref{tab:M_unit_SANE_jet}: the factor $\mathcal{M}$ is constant with respect to observer inclination, with an error of a few percent. As Fig.~\ref{fig:sane_jet_125_matrix} shows, this model has the appearance of an optically thick, homogenous torus. Evidently, the effects of relativistic boosting on one side of the black-hole spin axis and de-boosting on the other side - which are significant at higher inclination angles - cancel out in this case, maintaining a more or less constant flux with respect to the inclination angle, even though the source transitions from being symmetric to asymmetric with respect to the black-hole spin axis.

%Another striking feature of these models is the fact that a diagonal cross-like structure, centered on the horizon, appears when $a \leq 0$. These structures are apparently destroyed in the case of prograde black-hole spin. 

%Note behavior of M unit and LAMBDA as function of spin for each model...

\subsubsection{SANE disc models}

\begin{figure}
	\includegraphics[width=0.49\textwidth]{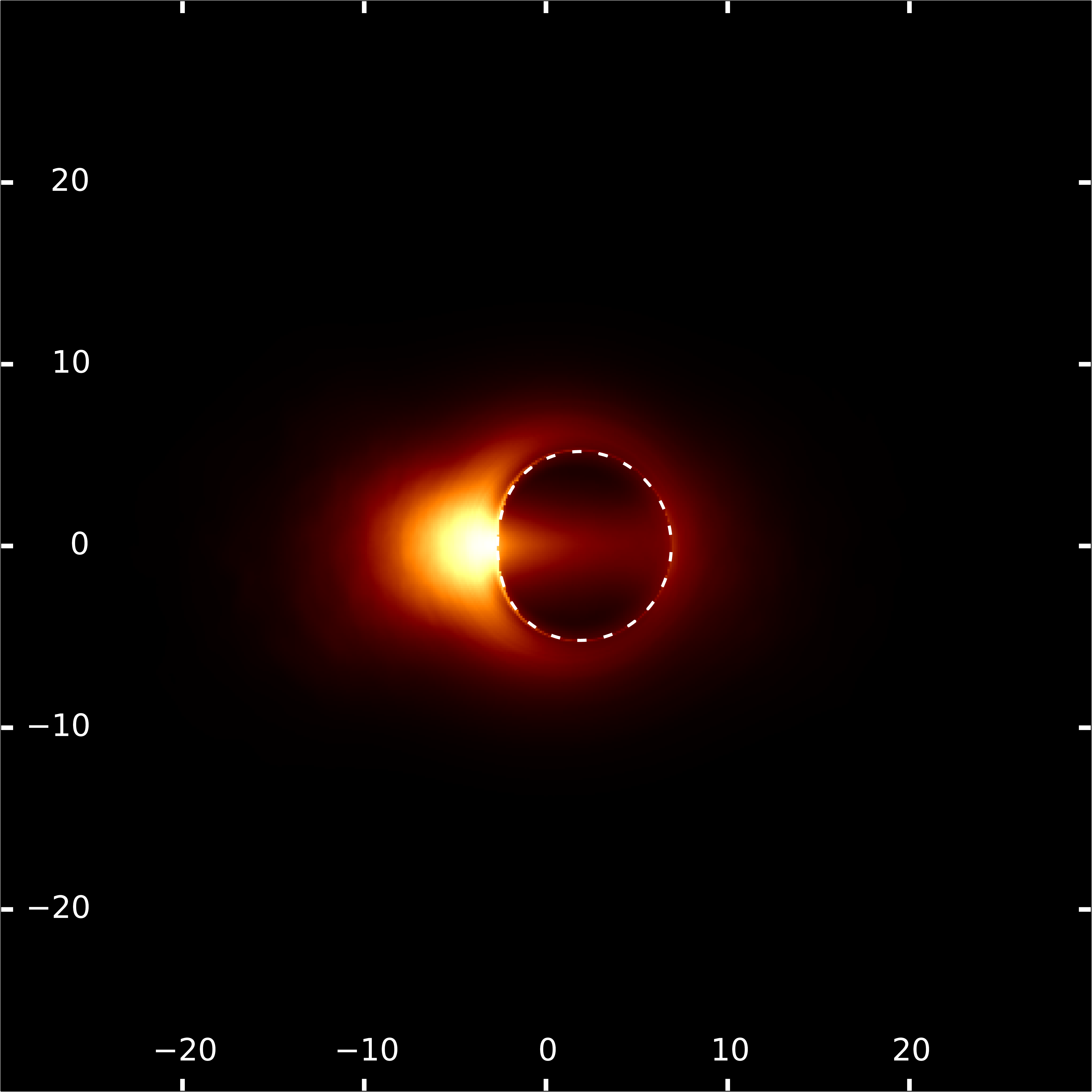}
\caption{Time-averaged images of the SANE disc model with $a=0.9375$ and $i=90^\circ$, with an integrated flux density of 2.5 Jy (note that the 1.25 Jy case looks almost identical; see Fig.~\ref{fig:sane_disk_125_matrix}). The dashed line indicates the BHS. The high black-hole spin shrinks the ISCO, increasing the accretion disc's angular momentum and thus the effect of relativistic boosting, causing one side of the disc to be much brighter than the other. This creates the optical effect of a small source size.}\label{fig:tbafig4}
\end{figure}

\begin{figure}
\centering
\begin{subfigure}[b]{0.49\textwidth}
	\includegraphics[width=\textwidth]{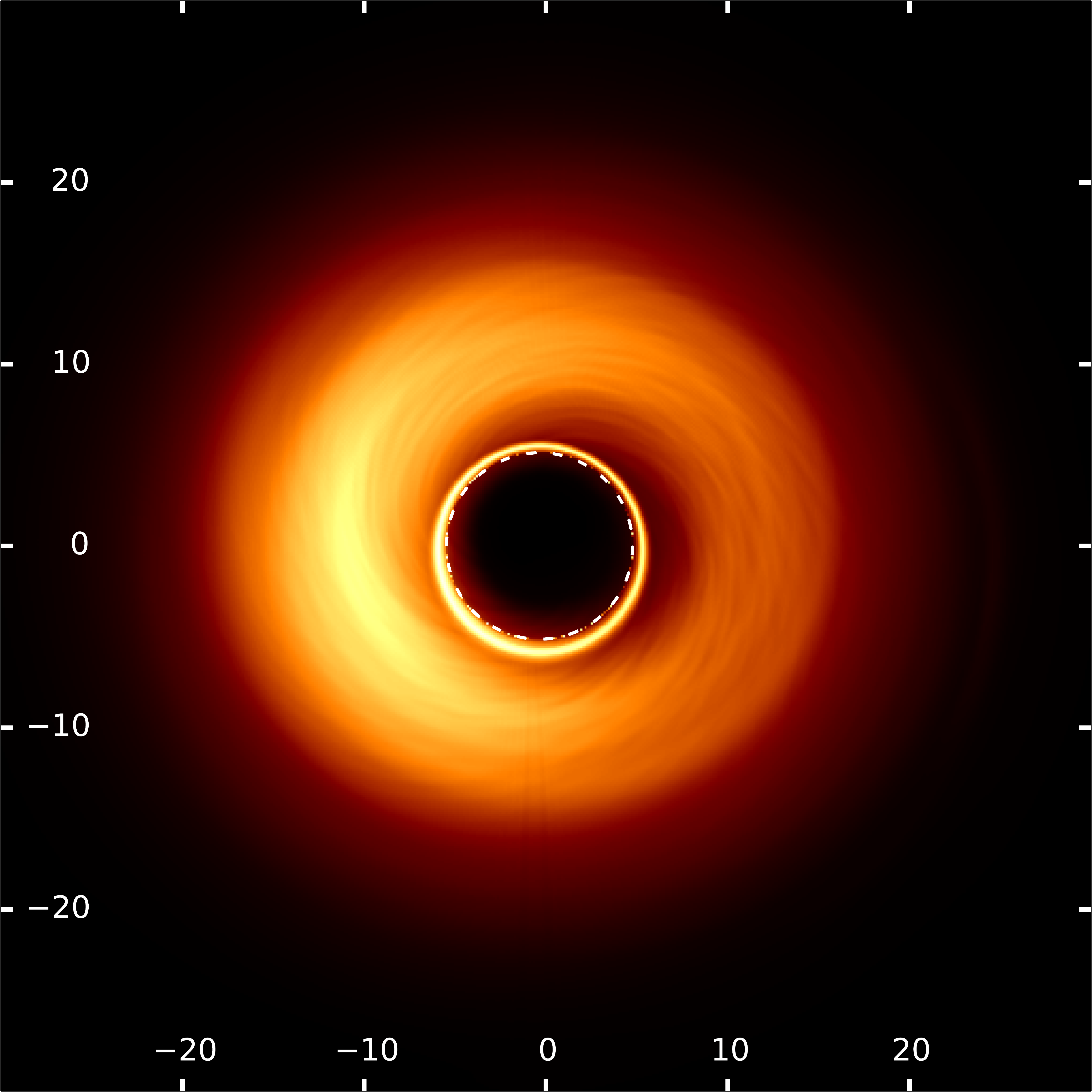}
\end{subfigure}
\begin{subfigure}[b]{0.49\textwidth}
	\includegraphics[width=\textwidth]{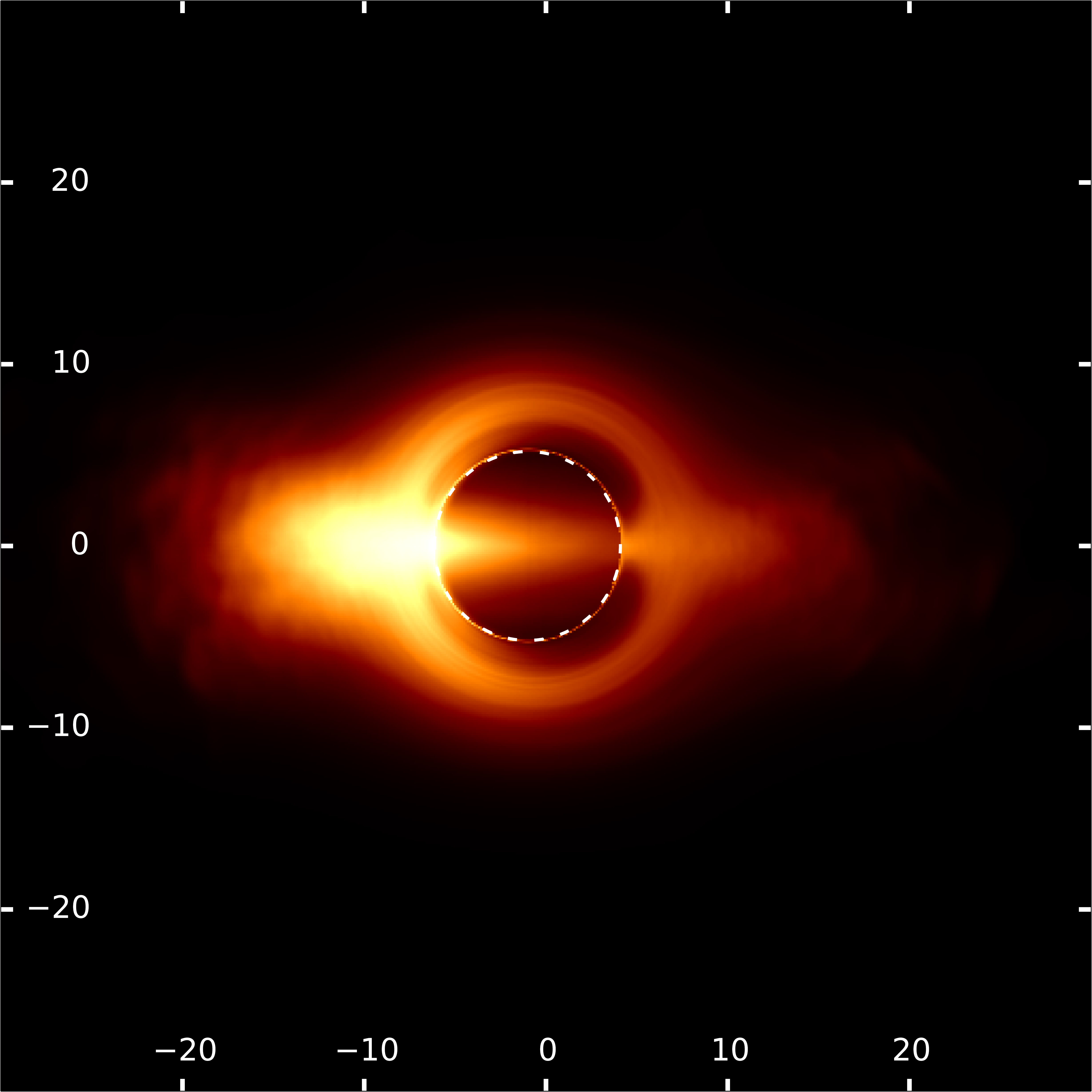}
\end{subfigure}
\caption{Time-averaged images of the SANE jet model with $a=-0.5$ and $i=20^\circ$ (left panel) and $i=90^\circ$ (right panel), with an integrated flux density of 2.5 Jy. The dashed line indicates the BHS. At lower (and especially negative) spins, the ISCO expands, causing the source size to increase, and displaying the effect of evacuation, which exaggerates the IS.}\label{fig:tbafig5}
\end{figure}

Figures \ref{fig:sane_disk_25_matrix} and \ref{fig:sane_disk_125_matrix} show the SANE disc models calibrated to 2.5 Jy and 1.25 Jy, respectively. Of all models, the disc models show the greatest variation with respect to spin, when it comes to source size (see Appendix \ref{sec:source_size_appendix}). Otherwise, their appearance is much more constant than was the case for the SANE jet models. As the SANE disc models have a lower overall accretion rate, they are less optically thick, and show less obscuration effects, than the SANE jet models. For retrograde spins, the SANE disc models, form a large torus, which (due to being geometrically extended) produces a pronounced first-order image just around the photon ring. Thus, the IS aligns with the BHS, although a secondary dark ring may be observed around it (see Subsection \ref{sec:intensity_profiles}).

Figure \ref{fig:tbafig4} shows that the high-spin SANE disc models, viewed at $i=90^\circ$, display a much smaller source size than the jet models. This occurs due to two factors; 
\begin{enumerate}
\item At high inclinations, the effects of relativistic boosting are more pronounced, exaggerating the importance of one side of the accretion flow while diminishing the other, thus decreasing the source size overall, as discussed by \citet{psaltis2015}.
\item At high spins, the ISCO shrinks to about the size of the photon sphere, meaning that the orbits are much closer to the black hole, with correspondingly higher velocities. This greatly enhances the effects of relativistic boosting, and consequentially, the de-boosted side of the accretion flow becomes negligible; we only see the boosted region of the accretion flow.
\end{enumerate}

\citet{doeleman2008} detected horizon-scale structure in Sgr A*, deriving a smaller size than the BHS by assuming a circular gaussian, but also found their data to be consistent with a larger BHS model. Such small source sizes are compatible with a high-spin source that shows strong boosting effects. Later work \citep{lu2018} refined these models, also showing evidence for emission on scales from a few Rsch to that of the BHS.

The effect of evacuation, discussed in Section \ref{sec:theory}, is pronounced in the SANE disc models with retrograde spin $(a < 0)$ (Fig.~\ref{fig:tbafig5}). These retrograde models have the largest ISCO, and thus the largest evacuated region within the accretion flow. In such cases, the IS becomes larger than the BHS (Subsection \ref{sec:intensity_profiles}). This effect is barely visible for the SANE jet models, presumably due to the presence of a bright jet base combined with a colder disc, which acts as a weak emitter near the ISCO.

\subsubsection{MAD jet models}

\begin{figure}
	\includegraphics[width=0.49\textwidth]{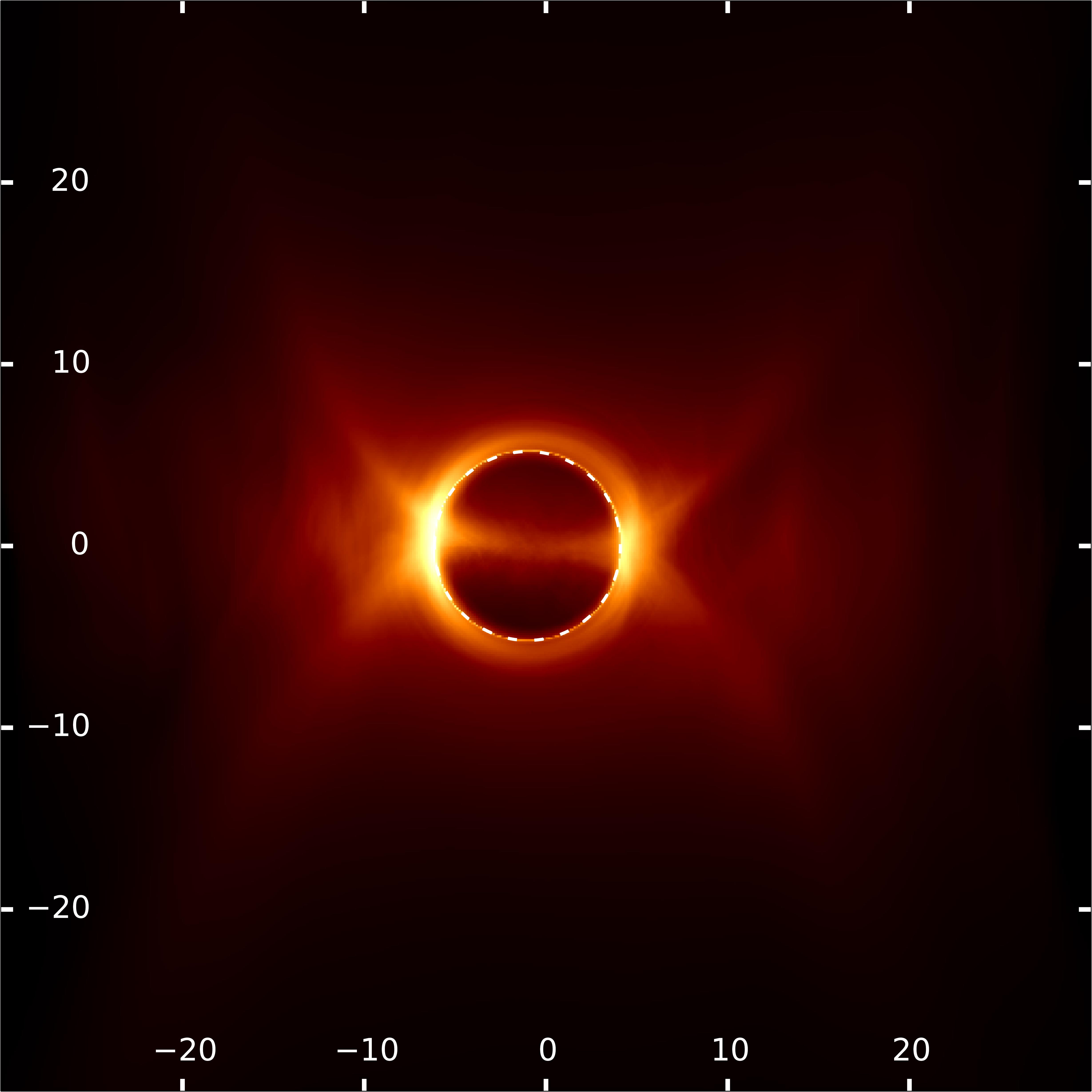}
\caption{Time-averaged images of the MAD jet model with $a=-0.5$ and $i=90^\circ$, with an integrated flux density of 2.5 Jy. The dashed line indicates the BHS. The MAD jet models are optically thin and relatively constant in appearance (Appendix \ref{appLibrary}). The model pictured here shows an apparent misalignment between the accretion flow and the black hole's rotation axis. }\label{fig:tbafig6}
\end{figure}

Figures \ref{fig:mad_jet_25_matrix} and \ref{fig:mad_jet_125_matrix} show the MAD jet model calibrated to 2.5 Jy and 1.25 Jy. One can see at a glance that the source morphologies of these models depends more weakly on the black-hole spin than the source morphologies of the SANE models do (Appendix \ref{sec:source_size_appendix}). As Table \ref{tab:M_unit_MAD_jet} shows, the same is true for $\mathcal{M}$, meaning that, much like the morphology, the luminosity of these models is a weaker function of the black-hole spin than for SANE models. A striking feature of these models is the parabolic jet base, which again appears more or less constaint for all spins. In the $a=-1/2$ models, however, this structure appears to be slightly misaligned with respect to the black-hole spin axis, as can be seen in Fig.~\ref{fig:tbafig6}. A future work might explore whether such misalignments occur regularly, and at what time-scales, etc.

Note that none of the MAD models show a clear evacuated region, even at highly retrograde spins, when the ISCO is maximally extended. This is due to the fundamentally different nature of MAD models with respect to SANE models; in MAD models, a `pile-up' of material occurs around a disruption radius, as magnetic fields disrupt the symmetrical accretion flow, causing discrete `blobs' of material to penetrate the disrupted region and accrete on to the black hole with velocities much lower than free fall \citep{bisnovatyi1974}. Since the images in this section are time-averaged, the discrete nature of these packets of accreting material is `averaged out', showing a much-simplified geometry in which no clear evacuated region occurs.

The optical depth of the MAD jet (and disc) models is substantially lower than that of the SANE models (Subsection \ref{sec:optical_depth_maps}). The combination of these factors ensures that the IS for these models represents the BHS well (see Subsection \ref{sec:intensity_profiles} for details).

\subsubsection{MAD disc models}

\begin{figure}
	\includegraphics[width=0.49\textwidth]{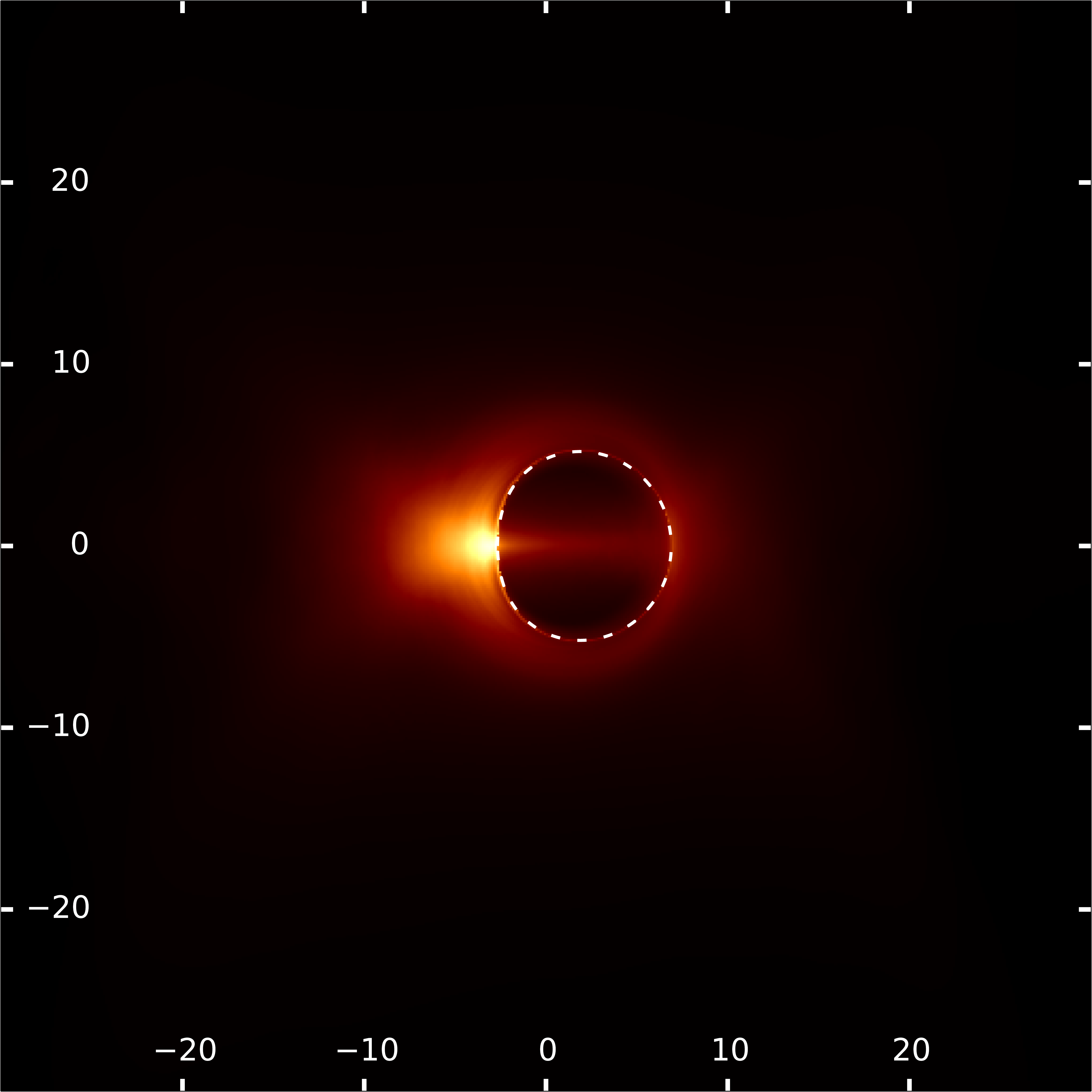}
\caption{Time-averaged images of the MAD disc model with $a=0.9375$ and $i=90^\circ$, with an integrated flux density of 2.5 Jy. The dashed line indicates the BHS. Like the MAD jet models, the MAD disc models are optically thin and relatively constant in appearance (Appendix \ref{appLibrary}).}\label{fig:vuvuzela}
\end{figure}

Figures \ref{fig:mad_disk_25_matrix} and \ref{fig:mad_disk_125_matrix} show the MAD disc model calibrated to 2.5 Jy and 1.25 Jy. As in the case of the MAD jet models, the images are optically thin, and the source sizes and flux-calibration factors are nearly identical (Appendices \ref{sec:source_size_appendix} and \ref{sec:flux_calibration_appendix}, see also Fig.~\ref{fig:vuvuzela}). Again, all IS's match the BHS, and there is no clear evacuated region within the ISCO (Subsection \ref{sec:intensity_profiles}). In other words, the single-temperature MAD disc model is visually very similar to the two-temperature MAD jet model, unlike the case of SANE, where the appearance of the two models differed drastically. This is due to the fact that for most of the material in the MAD disc, $\beta \simeq 1$, which makes the temperature ratio of the electrons and the ions insensitive to the $\beta$-parametrization \citep{EHT5}.

\subsection{Intensity profiles}
\label{sec:intensity_profiles}

In this section we examine the relationship between the BHS and the IS more closely, by plotting certain normalised cross-cut profiles of the intensity maps presented in Section \ref{sec:intensity_maps}. The cuts are made in the directions parallel to, and perpendicular to, the black-hole spin axis, at the locations where the BHS is maximally extended in those directions. We thus obtain the largest possible cross-section of the BHS. In each case, we indicate the theoretically predicted location of the BHS with red, vertical lines.

\subsubsection{Intensity profiles of SANE models}

\begin{figure}
\centering
\includegraphics[width=0.49\textwidth]{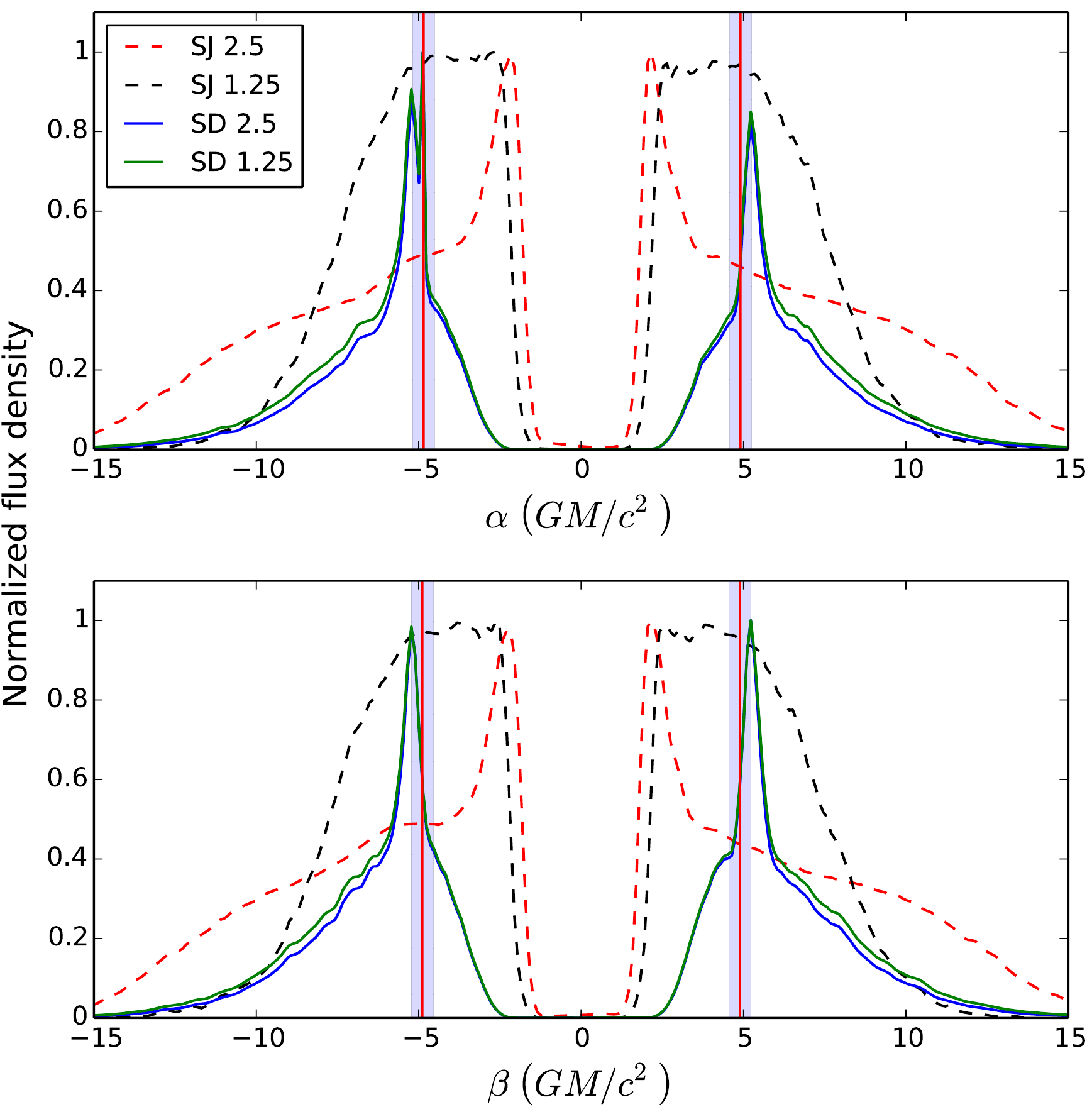}
\caption{Normalised intensity cross-cut profiles in the directions perpendicular to the BH spin axis (upper panel) and parallel to the BH spin axis (lower panel), centred on the maximal extent of the BHS in those directions, for all SANE models with $a=0.9375$ and $i=1$. The vertical red lines mark the theoretical location of the BHS, and the shaded blue regions indicate the error on the EHT measurements of M87*.}
\label{fig:SANE_a15o16_i1_profiles}
\end{figure}

Figure \ref{fig:SANE_a15o16_i1_profiles} shows the high-spin SANE models at low inclination, which is one of the most challenging models in terms of visualising the BHS, especially the jet model, for which the peak intensity does not occur within the EHT's error bars for the shadow measurement (indicated with the shaded blue regions). The disc models, on the other hand, do peak near the BHS, although they show a mildly obscured IS.

\begin{figure}
\centering
\includegraphics[width=0.49\textwidth]{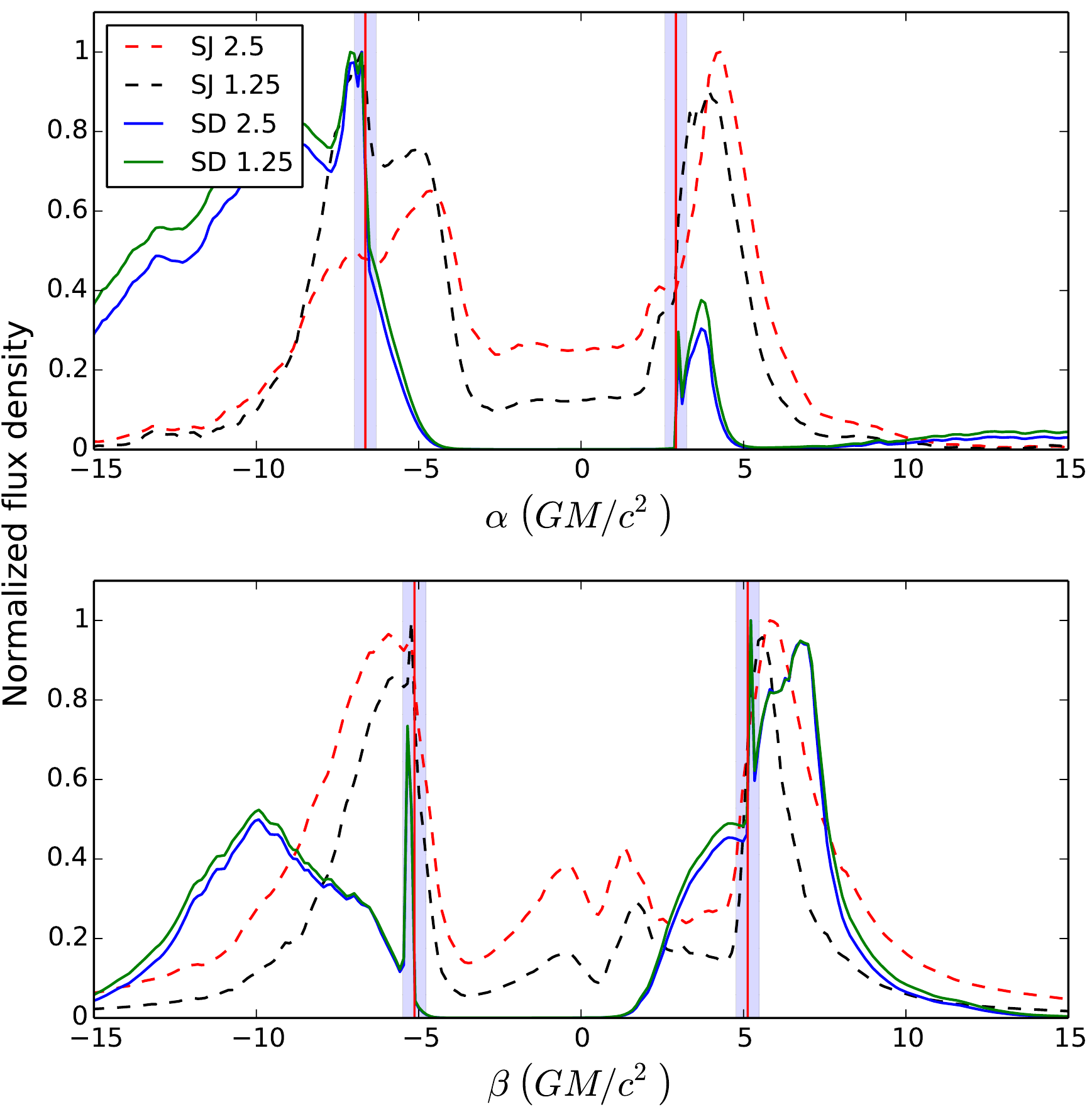}
\caption{Normalised intensity cross-cut profiles in the directions perpendicular to the BH spin axis (upper panel) and parallel to the BH spin axis (lower panel), centred on the maximal extent of the BHS in those directions, for all SANE models with $a=-0.9375$ and $i=60$. The vertical red lines mark the theoretical location of the BHS, and the shaded blue regions indicate the error on the EHT measurements of M87*. }
\label{fig:SANE_a-15o16_i60_profiles}
\end{figure}

Figure \ref{fig:SANE_a-15o16_i60_profiles} shows the highly retrograde-spin models. In this case, the flow is overall more optically thin (see Section \ref{sec:optical_depth_maps}), and the BHS can be more clearly seen. However, the effect of evacuation - i.e., the lack of radiating matter within the ISCO - can enlarge the perceived IS (see, for example, the disc models in the lower panels of those figures). %The case of a non-rotating (Schwarzschild) black hole is shown in Fig.~\ref{fig:SANE_a0_i60_profiles}; the amount of obscuration and evacuation seen here is similar to the retrograde-spin case.

%\begin{figure}
%\centering
%\includegraphics[width=0.49\textwidth]{Figures/SANE_a15o16_i90-crop.pdf}
%\caption{Normalised intensity cross-cut profiles in the directions perpendicular to the BH spin axis (upper panel) and parallel to the BH spin axis (lower panel), centred on the maximal extent of the BHS in those directions, for all SANE models with $a=0.9375$ and $i=90$. The vertical red lines mark the theoretical location of the BHS, and the shaded blue regions indicate the error on the EHT measurements of M87*.}
%\label{fig:SANE_a15o16_i90_profiles}
%\end{figure}

%\begin{figure}
%\centering
%\includegraphics[width=0.49\textwidth]{Figures/SANE_a-15o16_i90-crop.pdf}
%\caption{Normalised intensity cross-cut profiles in the directions perpendicular to the BH spin axis (upper panel) and parallel to the BH spin axis (lower panel), centred on the maximal extent of the BHS in those directions, for all SANE models with $a=-0.9375$ and $i=90$. The vertical red lines mark the theoretical location of the BHS, and the shaded blue regions indicate the error on the EHT measurements of M87*. }
%\label{fig:SANE_a-15o16_i90_profiles}
%\end{figure}

%\begin{figure}
%\centering
%\includegraphics[width=0.49\textwidth]{Figures/SANE_a0_i60-crop.pdf}
%\caption{Normalised intensity cross-cut profiles in the directions perpendicular to the BH spin axis (upper panel) and parallel to the BH spin axis (lower panel), centred on the maximal extent of the BHS in those directions, for all SANE models with $a=0$ and $i=60$. The vertical red lines mark the theoretical location of the BHS and the shaded blue regions indicate the error on the EHT measurements of M87*. }
%\label{fig:SANE_a0_i60_profiles}
%\end{figure}

\subsubsection{Intensity profiles of MAD models}

The MAD models are, overall, much less optically thick than the SANE models under the circumstances considered here (see Section \ref{sec:optical_depth_maps}). Consequently, the IS matches the BHS much better, and in most cases, the intensity profile peaks near the photon ring (i.e., within the EHT's error bars). It is also interesting to note that the lensing ring introduced by \citet{gralla2019} appears quite distinctly for these models, due to the fact that the flow is optically thin and torus-shaped.

\begin{figure}
\centering
\includegraphics[width=0.49\textwidth]{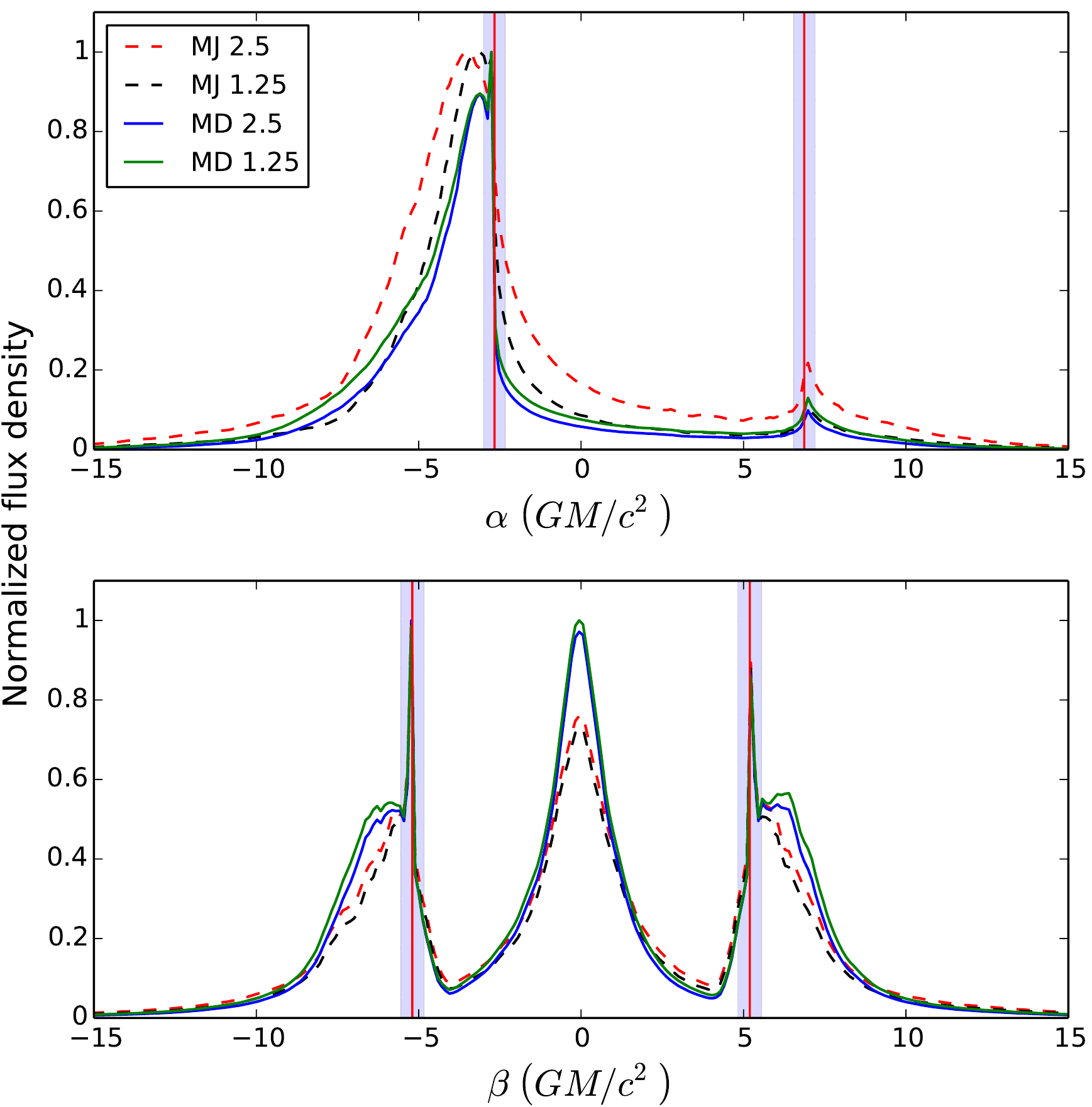}
\caption{Normalised intensity cross-cut profiles in the directions perpendicular to the BH spin axis (upper panel) and parallel to the BH spin axis (lower panel), centred on the maximal extent of the BHS in those directions, for all MAD models with $a=0.9375$ and $i=90$. The vertical red lines mark the theoretical location of the BHS, and the shaded blue regions indicate the error on the EHT measurements of M87*.}
\label{fig:MAD_a15o16_i90_profiles}
\end{figure}

\begin{figure}
\centering
\includegraphics[width=0.49\textwidth]{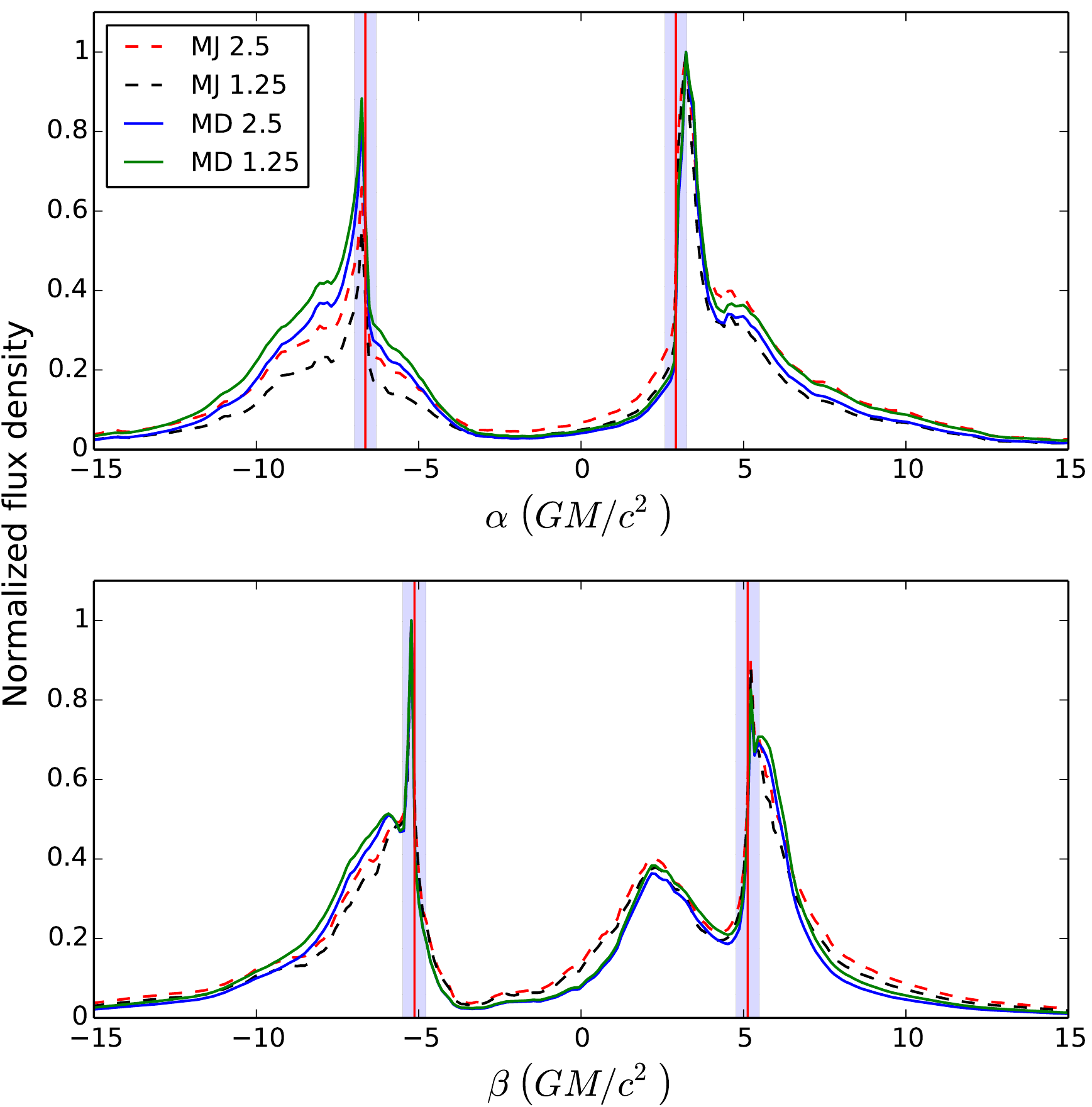}
\caption{Normalised intensity cross-cut profiles in the directions perpendicular to the BH spin axis (upper panel) and parallel to the BH spin axis (lower panel), centred on the maximal extent of the BHS in those directions, for all MAD models with $a=-0.9375$ and $i=60$. The vertical red lines mark the theoretical location of the BHS, and the shaded blue regions indicate the error on the EHT measurements of M87*. }
\label{fig:MAD_a-15o16_i60_profiles}
\end{figure}

Figure \ref{fig:MAD_a15o16_i90_profiles} shows the high-spin MAD models with $i=90^\circ$. For these models, obscuration occurs, when the accretion flow passes in front of the BHS. Figure \ref{fig:MAD_a-15o16_i60_profiles} shows the highly retrograde-spin MAD models. For these models, the intensity profile peaks near the photon ring in each case. Intriguingly, the rightmost peaks of the intensity profiles are higher than the peaks left of the BHS, even though the overall sense of rotation of the accretion flow suggests that the leftmost peaks are highly boosted (as occurs in the prograde-spin models). This suggests that the flow's rotation slows down near the horizon, possibly even reversing. This phenomenon is discussed in \citep{EHT5}.

%Figure \ref{fig:MAD_a0_i60_profiles} shows the case for $a=0$, i.e., a Schwarzschild black hole.

\subsection{Optical-depth maps}
\label{sec:optical_depth_maps}

\begin{figure*}
\centering
\begin{subfigure}[b]{0.325\textwidth}
	\includegraphics[width=\textwidth]{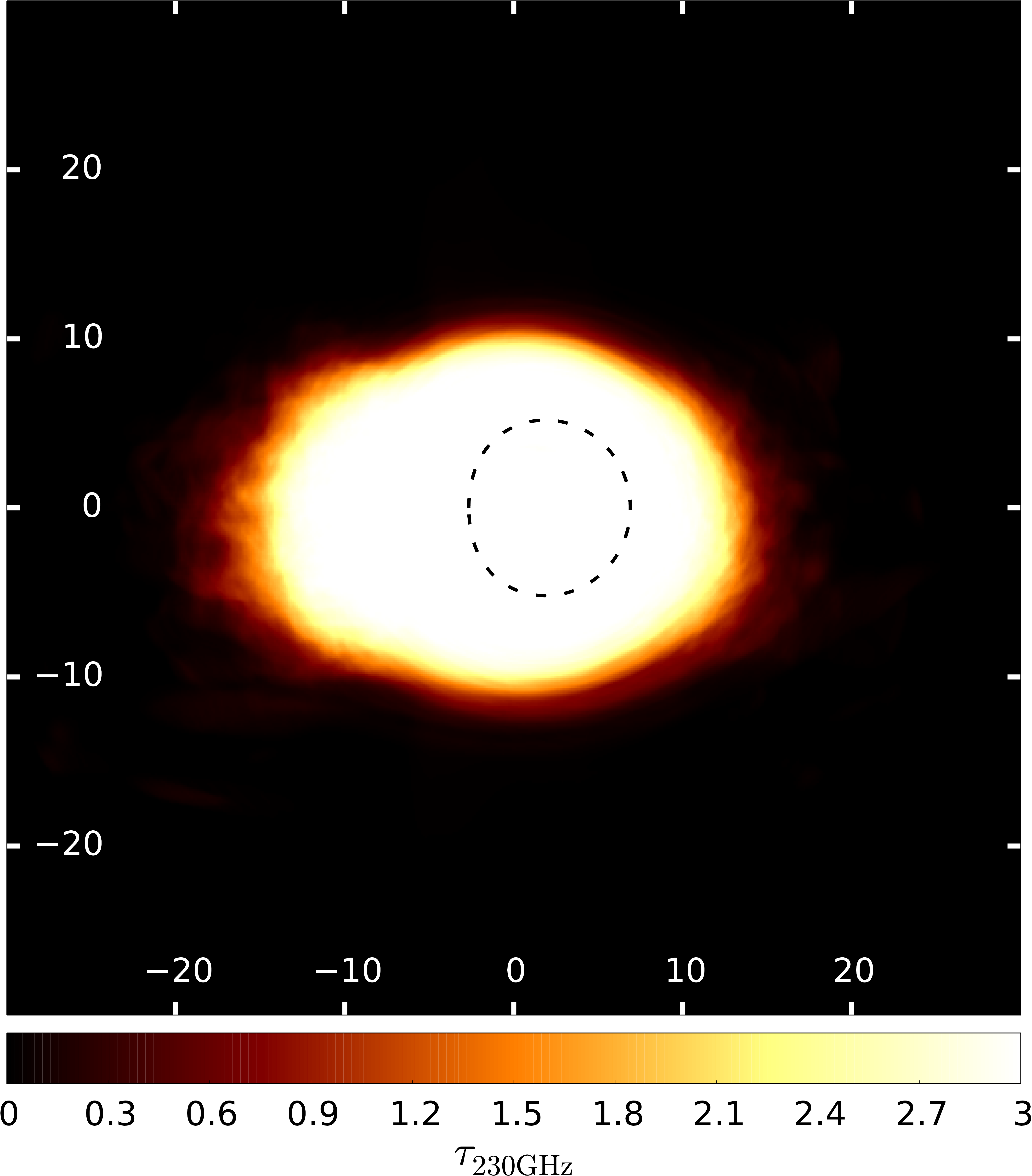}
	\caption{$a=0.9375$, $i=90^\circ$.}
\end{subfigure}
\begin{subfigure}[b]{0.325\textwidth}
	\includegraphics[width=\textwidth]{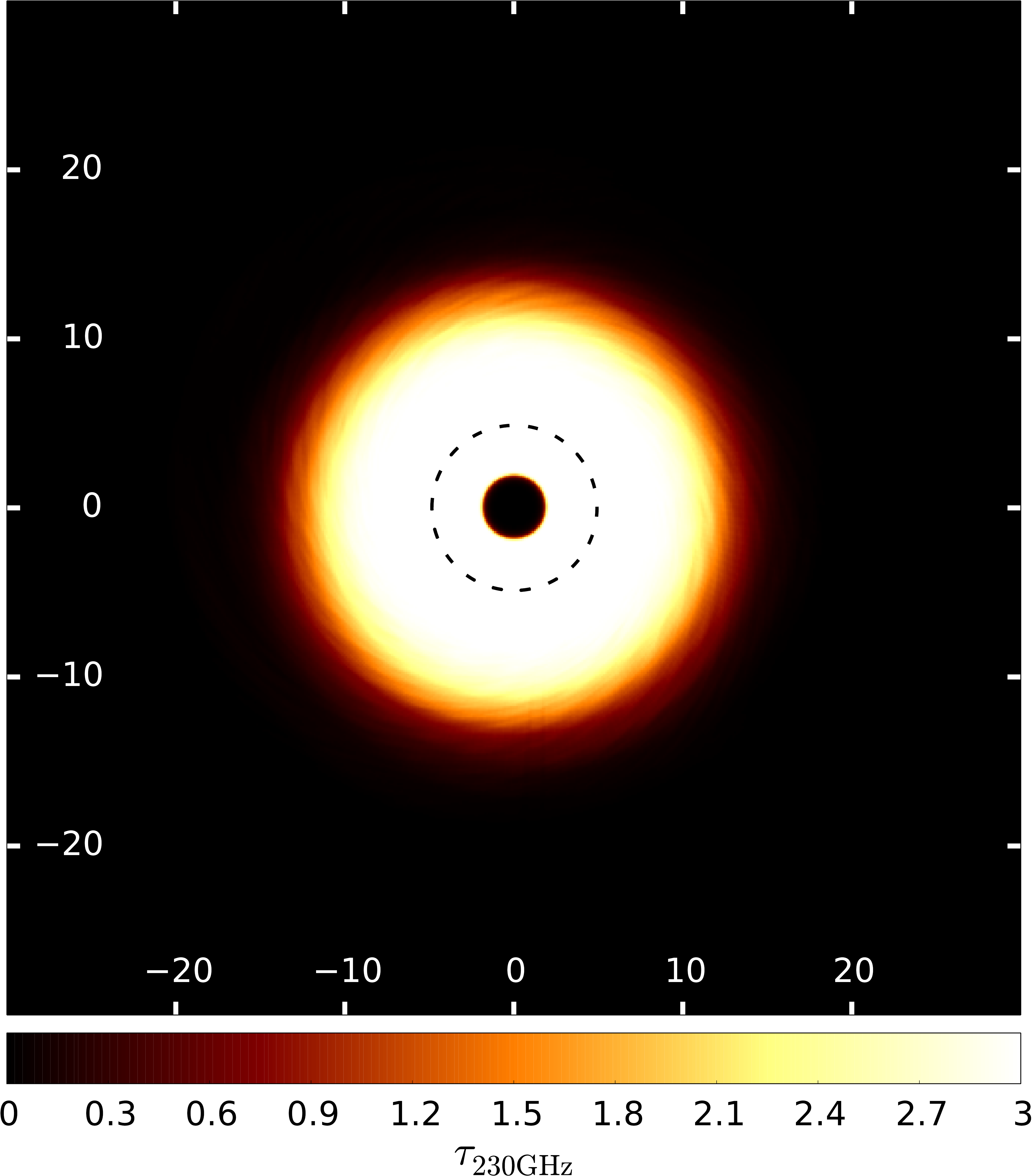}
	\caption{$a=0.9375$, $i=1^\circ$.}
\end{subfigure}
\begin{subfigure}[b]{0.325\textwidth}
	\includegraphics[width=\textwidth]{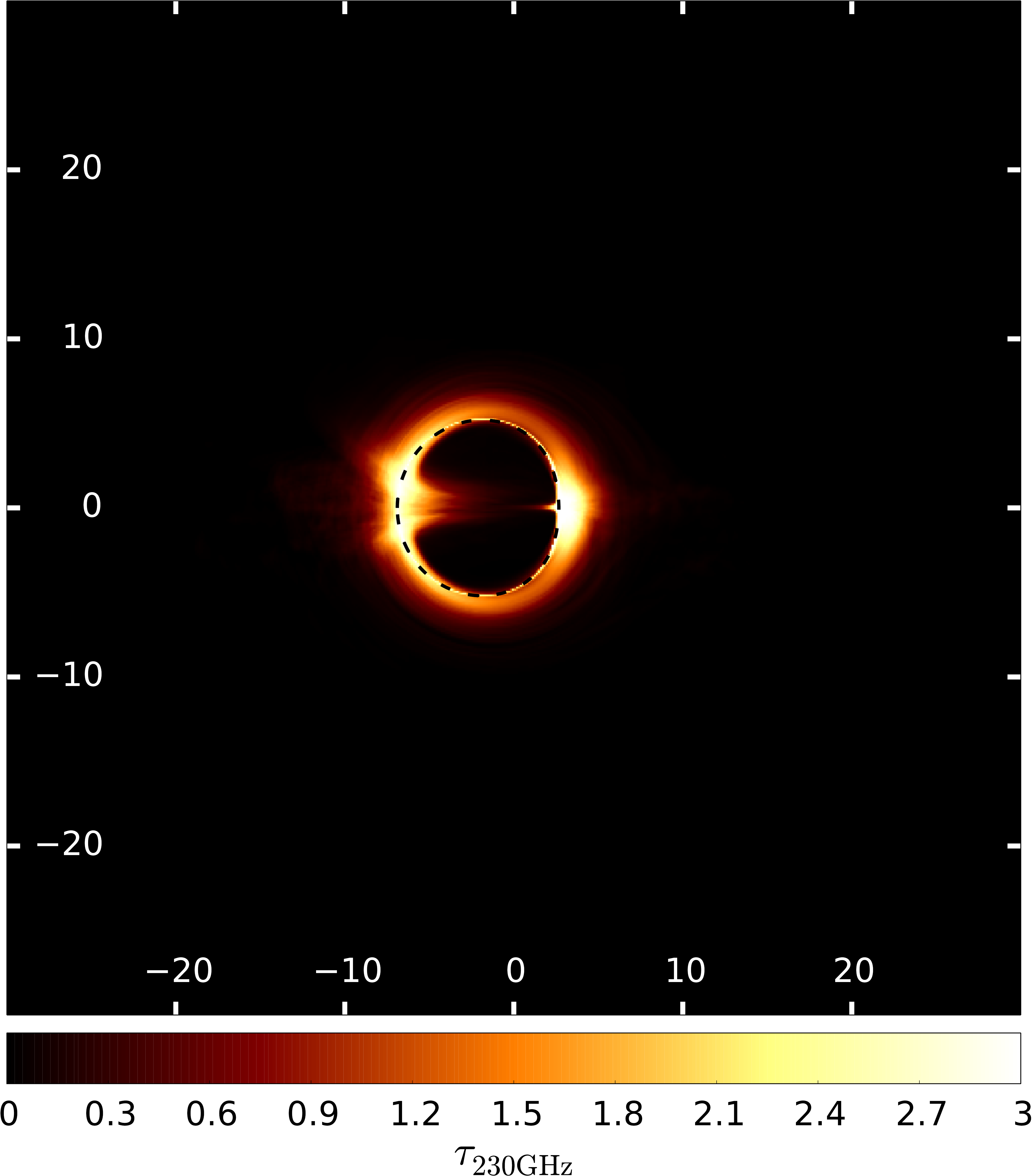}
	\caption{$a=-0.9375$, $i=90^\circ$.}
\end{subfigure}
\caption{Maps of $\tau_{\rm 230 GHz}$, the optical depth at an observing frequency of 230 GHz, for our SANE jet model with an integrated flux density of 2.5 Jy. The maximum optical depth shown is $\tau_{\rm 230 GHz} = 3$; at this optical depth, less than 5\% of radiation is transmitted, and we consider the BHS to be effectively obscured. The analytical prediction of the BHS's appearance is overplotted with a dotted line.}
\label{fig:sanejet_25_OD}
\end{figure*}

\begin{figure*}
\centering
\begin{subfigure}[b]{0.325\textwidth}
	\includegraphics[width=\textwidth]{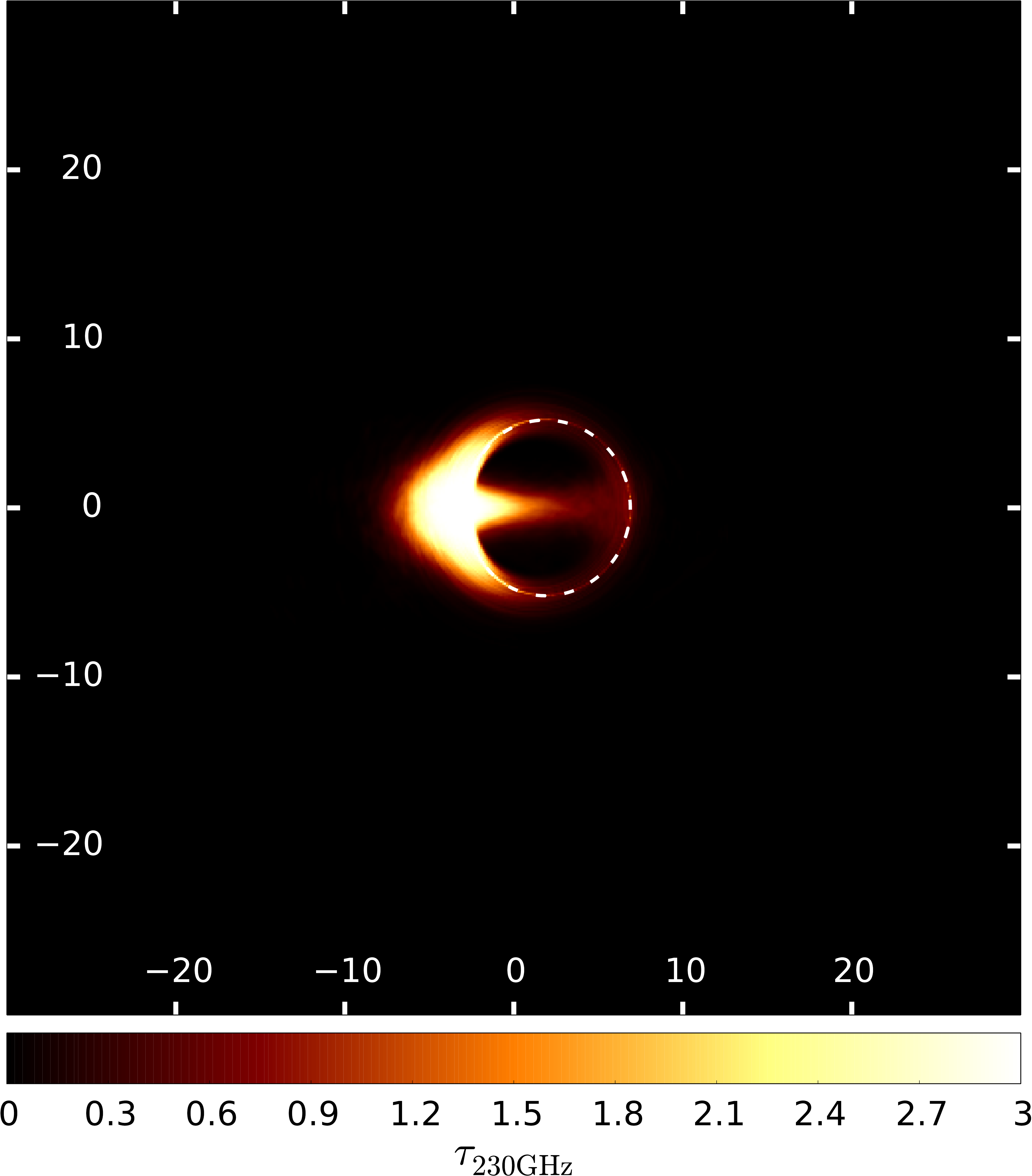}
	\caption{$a=0.9375$, $i=90^\circ$.}
\end{subfigure}
\begin{subfigure}[b]{0.325\textwidth}
	\includegraphics[width=\textwidth]{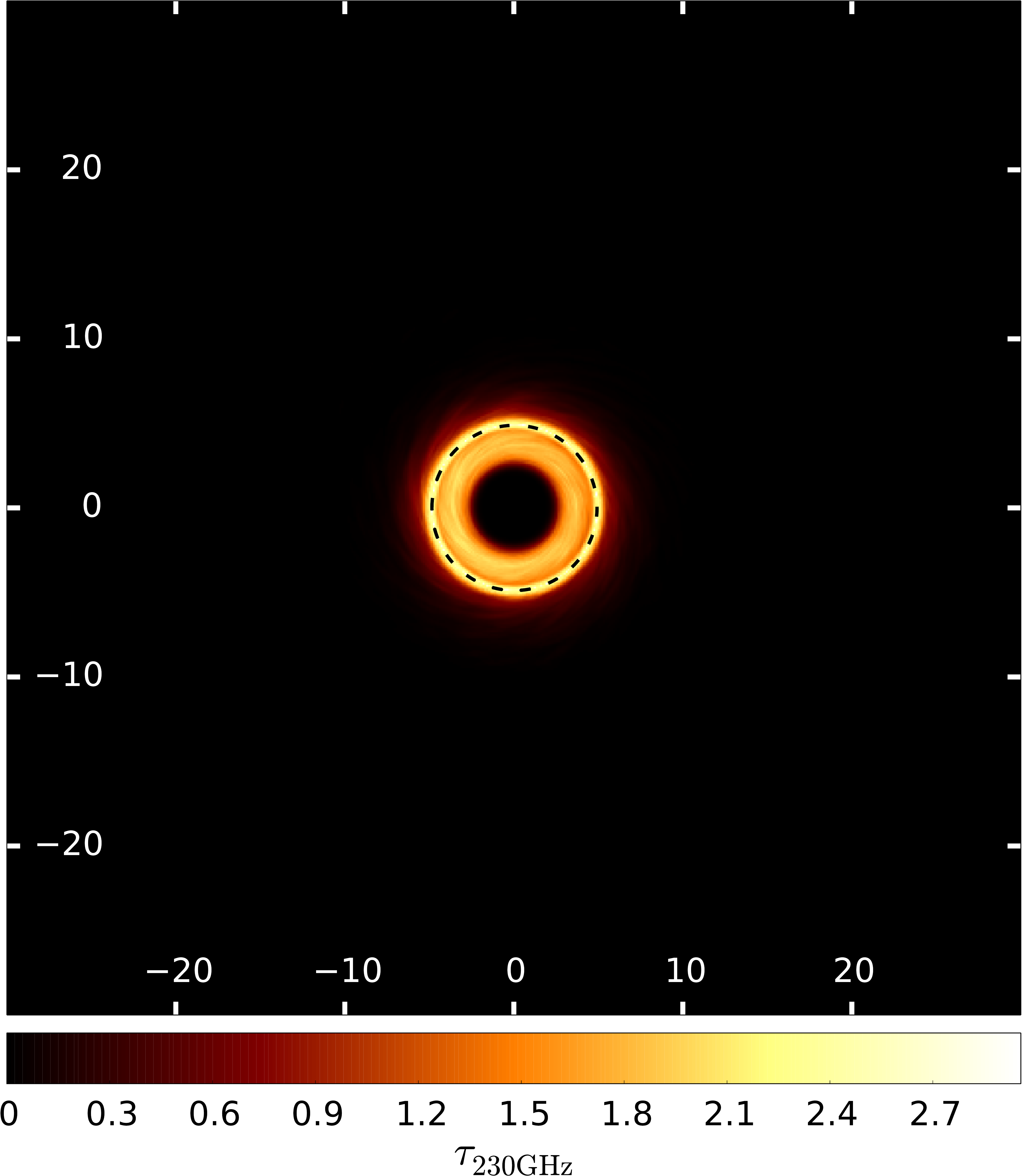}
	\caption{$a=0.9375$, $i=1^\circ$.}
\end{subfigure}
\begin{subfigure}[b]{0.325\textwidth}
	\includegraphics[width=\textwidth]{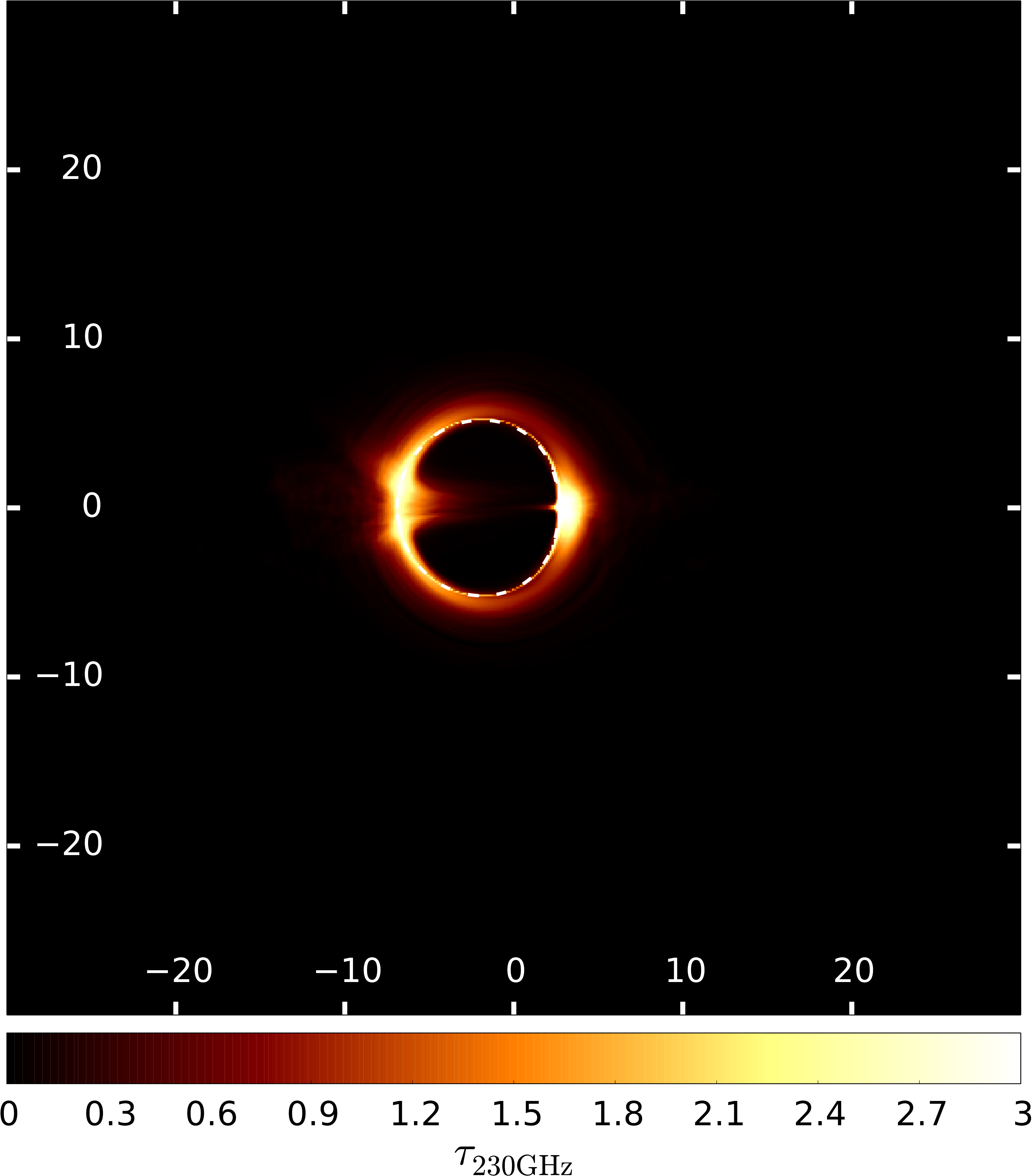}
	\caption{$a=-0.9375$, $i=90^\circ$.}
\end{subfigure}
\caption{Maps of $\tau_{\rm 230 GHz}$, the optical depth at an observing frequency of 230 GHz, for our SANE jet model with an integrated flux density of 0.625 Jy. The maximum optical depth shown is $\tau_{\rm 230 GHz} = 3$; at this optical depth, less than 5\% of radiation is transmitted, and we consider the BHS to be effectively obscured. The analytical prediction of the BHS's appearance is overplotted with a dotted line.}
\label{fig:sanejet_0625_OD}
\end{figure*}

\begin{figure*}
\centering
\begin{subfigure}[b]{0.325\textwidth}
	\includegraphics[width=\textwidth]{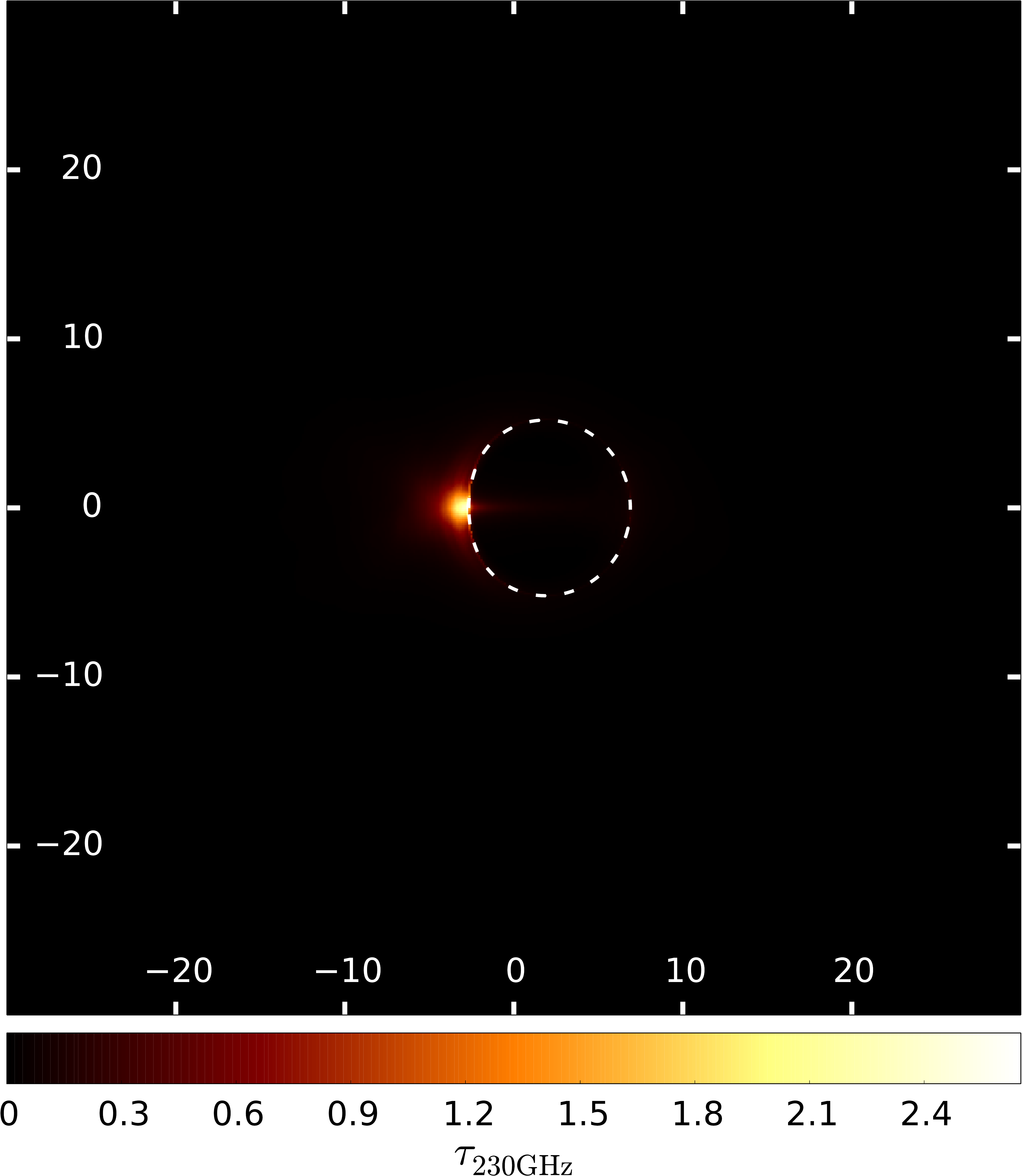}
	\caption{$a=0.9375$, $i=90^\circ$.}
\end{subfigure}
\begin{subfigure}[b]{0.325\textwidth}
	\includegraphics[width=\textwidth]{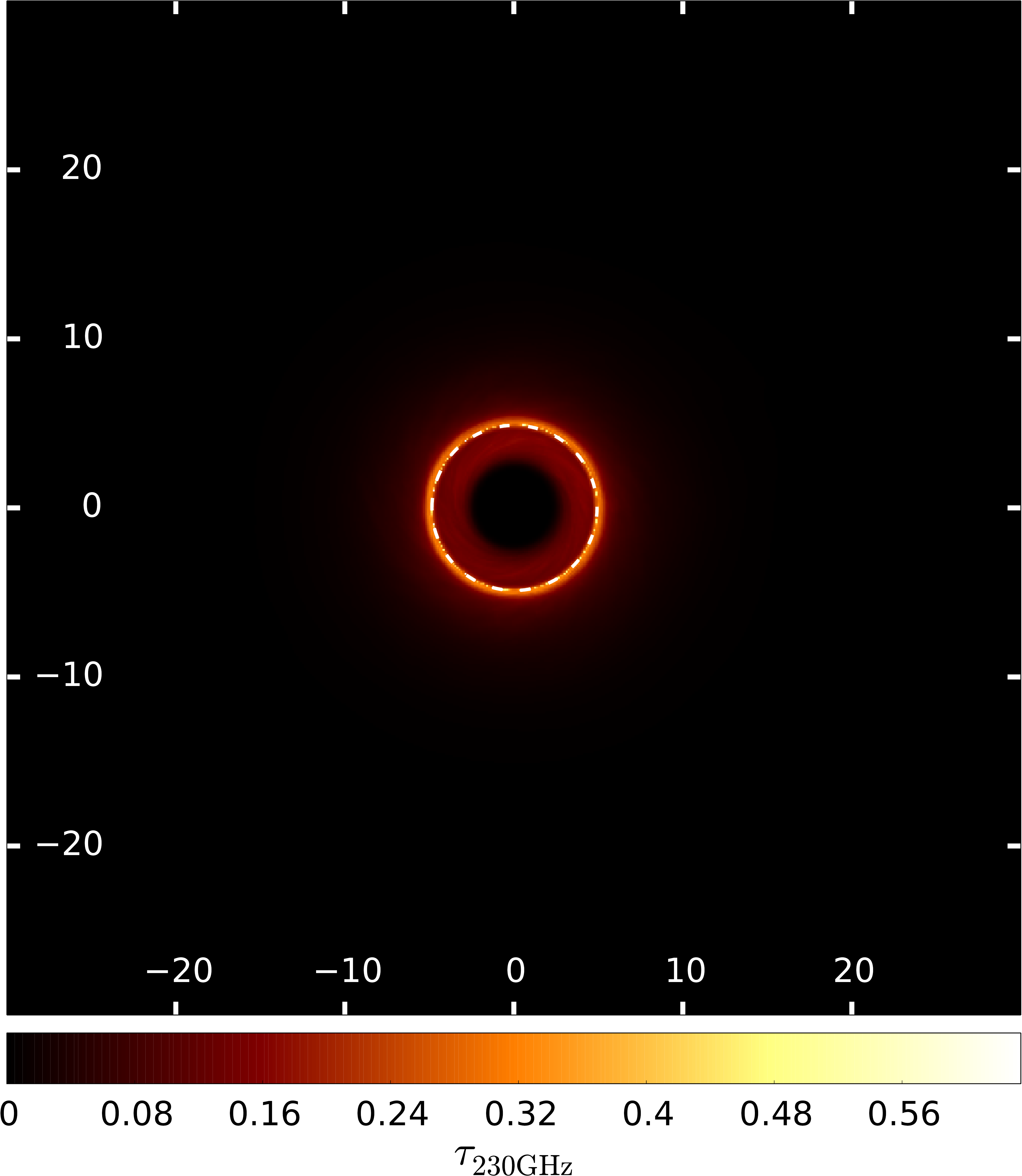}
	\caption{$a=0.9375$, $i=1^\circ$.}
\end{subfigure}
\begin{subfigure}[b]{0.325\textwidth}
	\includegraphics[width=\textwidth]{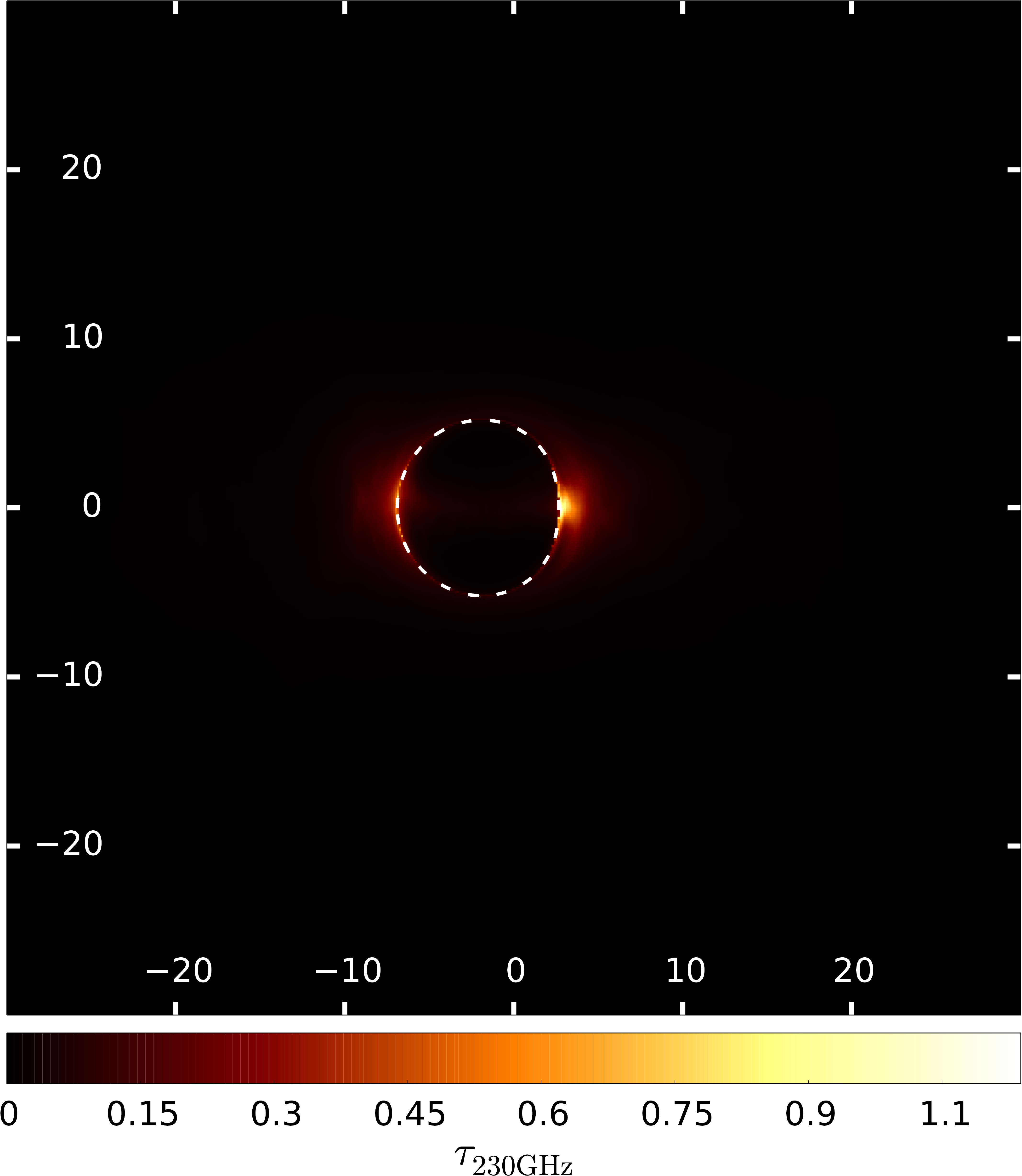}
	\caption{$a=-0.9375$, $i=90^\circ$.}
\end{subfigure}
\caption{Maps of $\tau_{\rm 230 GHz}$, the optical depth at an observing frequency of 230 GHz, for our MAD jet model with an integrated flux density of 1.25 Jy. The maximum optical depth shown is $\tau_{\rm 230 GHz} = 3$; at this optical depth, less than 5\% of radiation is transmitted, and we consider the BHS to be effectively obscured. The analytical prediction of the BHS's appearance is overplotted with a dotted line.}
\label{fig:madjet_125_OD}
\end{figure*}

In this section we present maps of $\tau_{\rm 230GHz}$, the optical depth along null geodesics, for some of our models, in order to examine the optical behavior of the accretion flow more closely. Note that the maps in this section are plotted in linear scale. Figures \ref{fig:sanejet_25_OD} and \ref{fig:sanejet_0625_OD} show optical depth maps, for various spins and inclinations, of our SANE jet models at integrated flux densities of 2.5 Jy and 0.625 Jy, respectively. The maps plot $\tau_{\rm 230GHz}$ up to a maximum of 3, at which optical depth less than 5\% of background radiation is transmitted, and we consider the BHS to be obscured. Note the optical depth of the high-prograde-spin SANE jet models at higher fluxes in particular, which almost completely disappears at the lower flux. Figure \ref{fig:madjet_125_OD} shows maps of $\tau_{\rm 230GHz}$ for our MAD jet model at 1.25 Jy. Note that this model is almost completely optically thin, never reaching the maximum optical depth of 3.

\ 

\ 

\subsection{High-frequency intensity maps}
\label{sec:high_freq_maps}

Although the BHS is achromatic, the same is not true for the IS (Section \ref{sec:theory}). Depending on the black hole's mass and accretion rate, the frequency at which the source becomes optically thin (and thus displays a clear BHS) may be different \citep{falcke2000}. However, crucially, since most AGN show a power-law regime in their spectrum, it is always possible to pick frequency in the power-law regime where the source is optically thin (see, e.g., \citet{falcke2004bbb}). For the model considered in this paper, the observing frequency we have used so far (230 GHz) lies below the SED's peak, and below its power-law region \citep{moscibrodzka2009}. Thus, we may expect the source to become more optically thin at higher frequencies, potentially obtaining a better match between the BHS and the IS in that regime.

\begin{figure*}
\centering
\begin{subfigure}[b]{0.49\textwidth}
	\includegraphics[width=\textwidth]{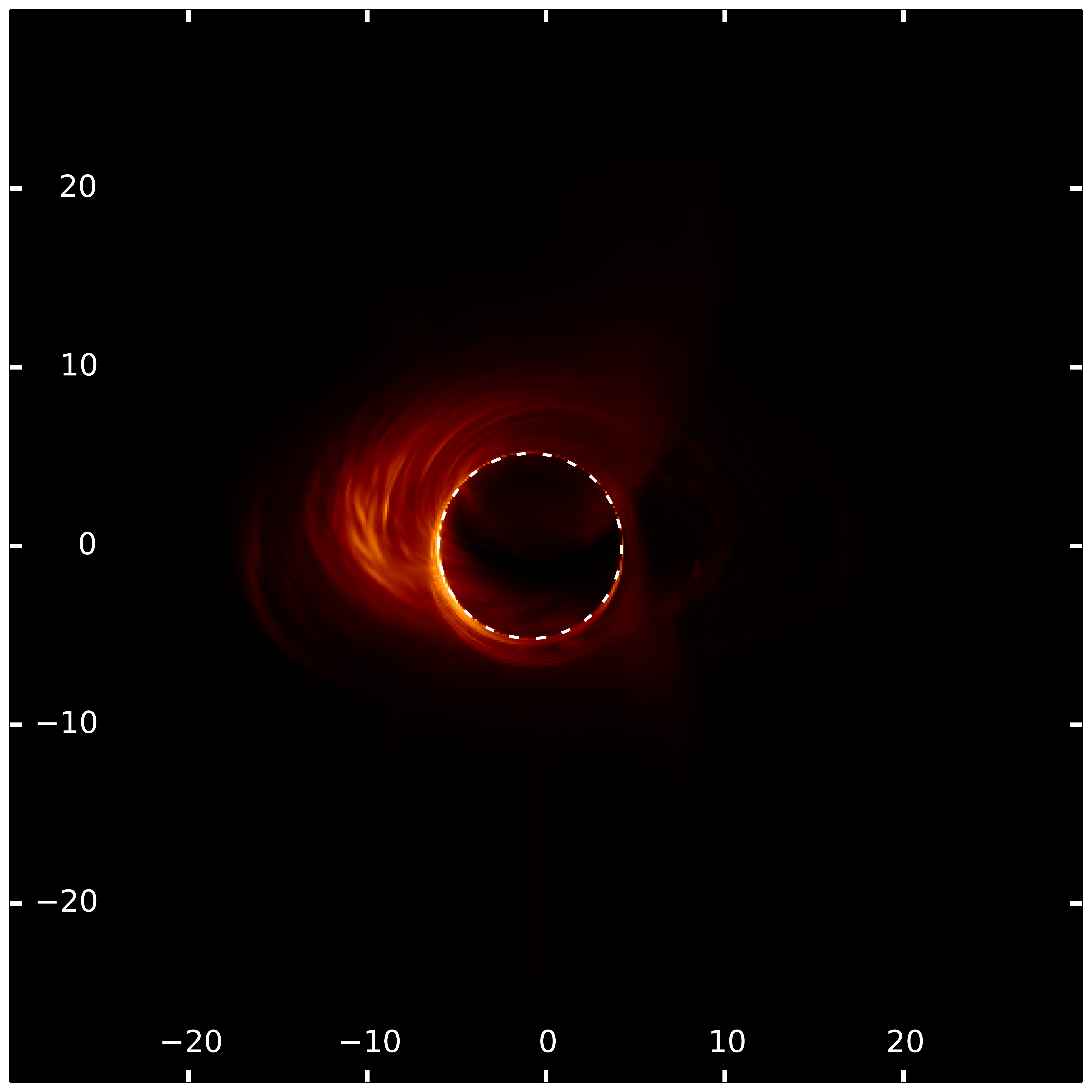}
	\caption{$a=-0.5$, $i=60^\circ$.}
\end{subfigure}
\begin{subfigure}[b]{0.49\textwidth}
	\includegraphics[width=\textwidth]{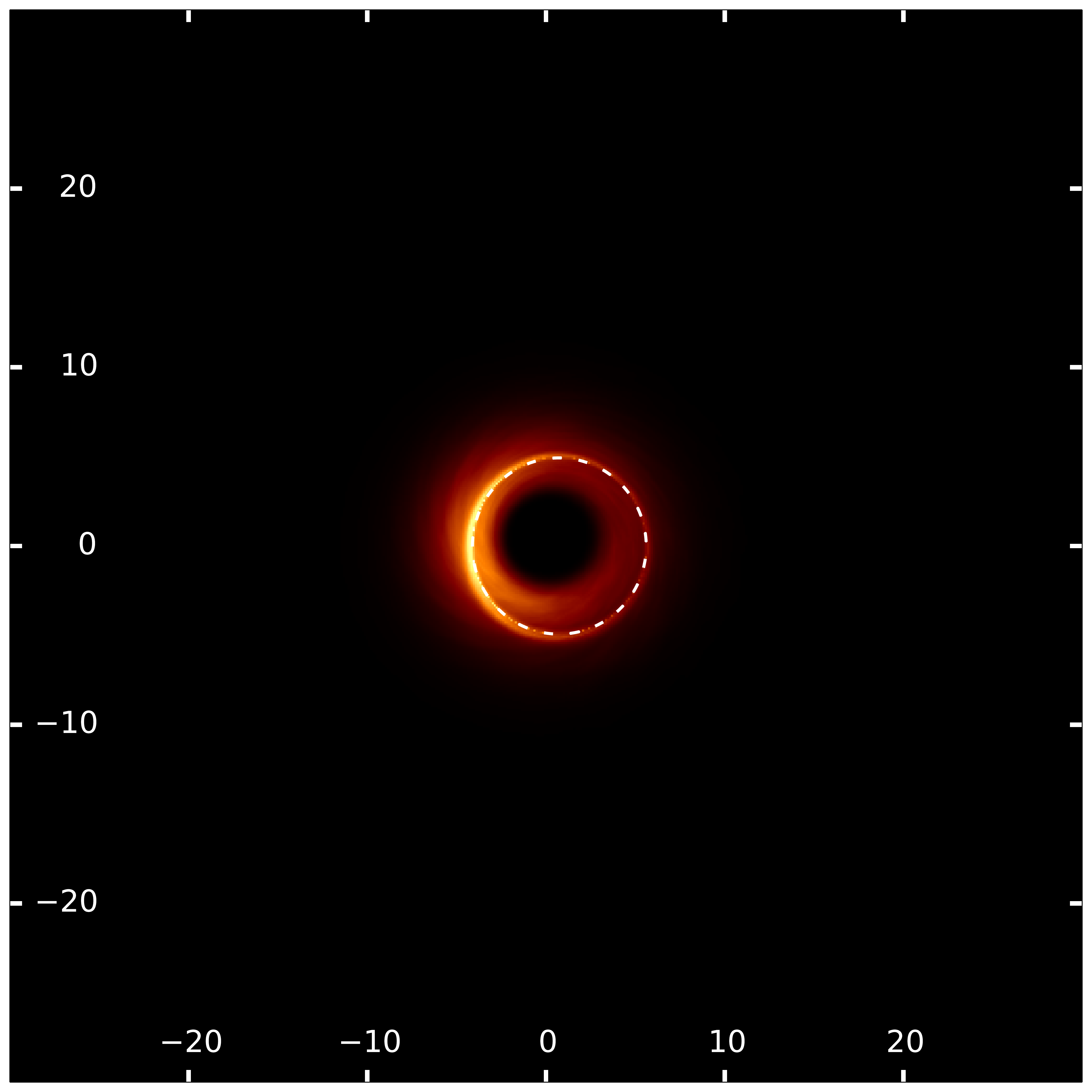}
	\caption{$a=0.9375$, $i=20^\circ$.}
\end{subfigure}
\caption{Intensity maps of our SANE disc models, calibrated to 2.5 Jy, at a frequency of 6 THz.}
\label{fig:highfreq}
\end{figure*}

\begin{figure*}
\centering
\begin{subfigure}[b]{0.49\textwidth}
	\includegraphics[width=\textwidth]{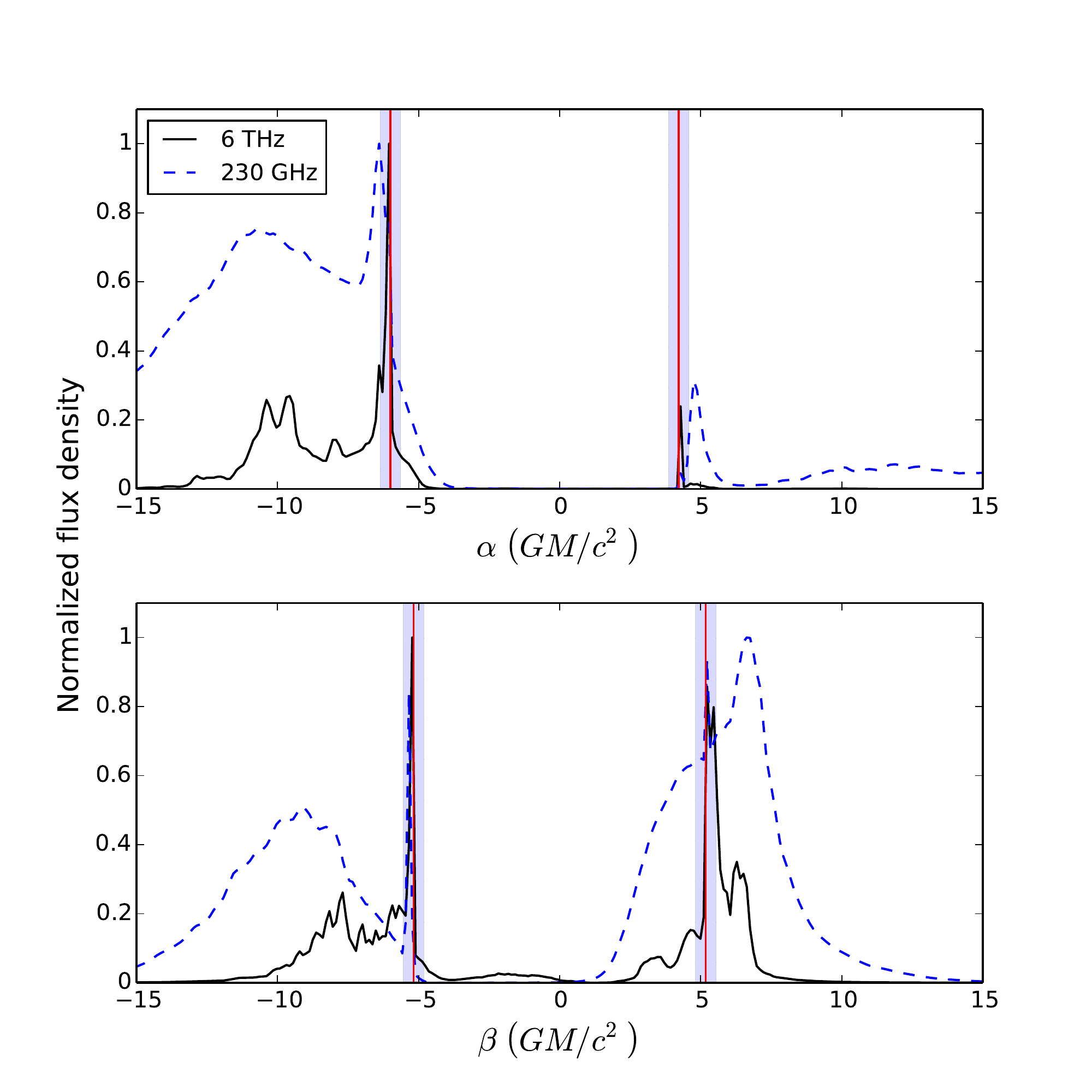}
	\caption{$a=-0.5$, $i=60^\circ$.}
\end{subfigure}
\begin{subfigure}[b]{0.49\textwidth}
	\includegraphics[width=\textwidth]{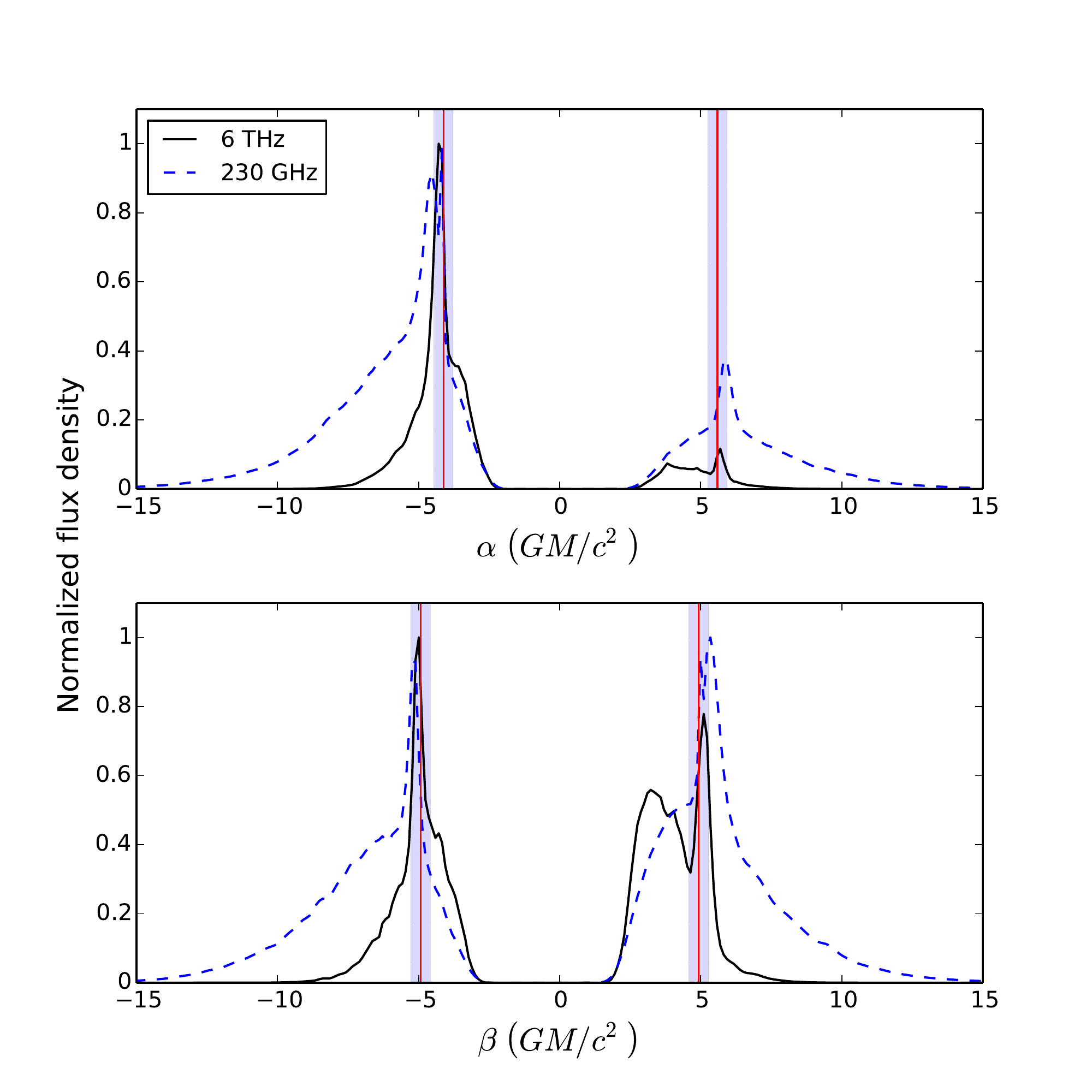}
	\caption{$a=0.9375$, $i=20^\circ$.}
\end{subfigure}
\caption{Intensity profiles of our SANE disc models, calibrated to 2.5 Jy, at observing frequencies of 230 GHz and 6 THz.}
\label{fig:highfreq_profiles}
\end{figure*}

To demonstrate the achromaticity of the BHS, as well as the decreasing optical depth at higher frequencies, we plot intensity maps of two SANE disc models at 2.5 Jy - specifically, the $a=-0.5, i=60$ and $a=0.9375, i=20$ (panels l and j in Fig.~\ref{fig:sane_disk_25_matrix}, respectively) - at an observing frequency of 6 THz in Fig.~\ref{fig:highfreq}; this frequency was chosen in order to examine the appearance of the IS in a different observational window (infrared). Intensity profiles for these models, at both 230 GHz and 6 THz, are shown in Fig.~\ref{fig:highfreq_profiles}.

Note that, at this higher (mid-IR) frequency, the overall source size is reduced, and a more pronounced lensing ring is shown. Figure \ref{fig:highfreq_profiles} shows that, at 6 THz, the peaks of the intensity profiles lie within the EHT's error bars, although some obstruction remains in the high-prograde-spin case.

\section{Discussion}
\label{sec:discussion}

In the context of accreting LLAGN, we have discussed the differences between the BHS, which depends only on the geometry of spacetime and the location of the observer, and the observable IS, which depends also on plasma conditions. We argue that, if the emission region is compact and optically thin, the IS matches the BHS to within the precision of measurements by the EHT. It was argued by \citet{gralla2019} that circumstances can none the less occur in which the IS differs significantly from the BHS, even if the IS is circular, or the source is ring-like. Their conclusions are based on analytical, spherically symmetric, optically thin accretion models that do not take the full plasma dynamics and radiation effects into account, and, in our opinion, they cannot be applied to realistic astrophysical scenarios without further refinement. In that context, we identify two specific effects that can cause the IS to deviate from the BHS in the physically motivated case presently considered (of a geometrically thick, radiatively inefficient accretion flow): obscuration of the BHS by a part of the accretion flow that passes in front of it, and enlargement of the IS due to the lack of emission coming from the evacuated innermost region of the accretion flow, which scales with the ISCO in the case of SANE models (see Appendix \ref{app:evacuation}).

\citet{narayan2019} showed that these effects do not arise in the case of spherical accretion flows; in this work, we extend that analysis to the case of non-spherical, GRMHD-based models of the accretion flow. To this end, we have constructed a library of {\tt RAPTOR} images of a set of GRMHD simulations produced using the GRMHD code {\tt BHAC} for the 2017 EHT campaign. GRMHD simulations of both the MAD and SANE types were used; different spins (from retrograde to prograde) and inclination angles were considered; two radiative models (disc- and jet-based) were considered, and finally, the analysis was repeated at two different flux calibrations, to investigate a wider regime of LLAGN models.

It was found that virtually all MAD models show clear IS's that match the BHS well. At the fluxes considered in this paper, the MAD models were highly optically thin. The emission region was compact in each case; no evacuation effect is seen, although obscuration of the BHS does occur at inclination angles near 90 degrees. The morphology of the single-temperature disc model is quite similar to that of the two-temperature jet model. Note that we have considered time-averaged images of the accretion flow in this work; instantaneous snapshots of the GRMHD data show a more complex morphology, in which the discrete packets of accreting material aren't `smeared out' in the azimuthal direction (a quantitative look at the effects of the time-averaging procedure on the appearance of the source is presented in Appendix \ref{app:averaging}). The effect of relativistic boosting are less apparent than in the SANE case. In the case of a retrograde accretion flow, the effects of relativistic boosting can be reversed; the highest intensity in those cases occurs on the side of the image where the large-scale accretion flow is receding from the observer.

For our SANE models, the relationship between the BHS and the IS is more complicated. The morphology of the two-temperature jet model, in particular, is extremely sensitive to both the black-hole spin and the integrated flux density; it can range from optically thick to optically thin. In the optically thick case, the IS can be highly distorted. In the case of high prograde spin viewed at inclination angles near 90 degrees, the shadow may disappear entirely. At lower inclination angles, the BHS tends to be partially obscured by optically thick material, rendering the IS smaller than the BHS. The SANE disc models, lacking both a bright jet base and an obscuring, colder disc, tend to show more clear shadows than the SANE jet models. However, both obscuration and evacuation (particularly for retrograde spins) can significantly affect the appearance of the IS. In all cases, the effects of evacuation disappear at high prograde spin.

We conclude that, for GRMHD-based models of LLAGN that are optically thin and compact (so that the effects of obscuration and evacuation are limited), the IS matches the BHS to within an accuracy of 5\%. Observations of LLAGN that resemble SANE models, particularly two-temperature jet models, will be challenging on account of the much more varied appearances such models can take, including forms that show no IS at all. Our examination of high-frequency (6 THz) intensity maps of the SANE disc model illustrate that the frequency at which the inner region of our models becomes optically thin is strongly model-dependent, and scales with the accretion rate and mass of the black hole \citep{falcke2004bbb}. The appropriate observing frequency for observing a clear BHS must therefore be carefully chosen.

\section*{Acknowledgements}

This work is supported by the ERC Synergy Grant "BlackHoleCam: Imaging the
Event Horizon of Black Holes" (Grant 610058).
ZY is supported by a Leverhulme Trust Early Career Fellowship. TB thanks C. Brinkerink, S. Issaoun, L. Medeiros, F. {\"O}zel, and D. Psaltis for insightful comments regarding the manuscript. The authors thank Bart Ripperda and the anonymous referee for helpful feedback. This work has made use of NASA's Astrophysics Data System (ADS). 

\section*{Data Availability}

The data underlying this article (both the raw {\tt RAPTOR} output as well as the processed plots) will be shared on reasonable request to the corresponding author (t.bronzwaer@astro.ru.nl).

\bibliography{BHC}
\bibliographystyle{apalike}

\newpage

\begin{appendix}

\section{Derivation of estimators for the source size}
\label{appA}

There are many ways to perform statistical analyses on a set of pixel intensities and locations. It is crucial to distinguish between the statistics on the set of pixel intensities, without taking into account the pixel indices (e.g., the total flux density or the average brightness of a pixel), and the statistics of features in the image plane itself (e.g., the source width). We are interested in quantitatively describing image features, which may be interpreted as lengths or distances in the image plane, weighted by the pixel intensities (i.e., a brighter pixel contributes more).

We then define a measurement $x_i$ to be the distance, along the $x$-direction, in the image plane of the $i$-th pixel to the origin of our coordinate system (which may be chosen freely without affecting the results), weighted by the intensity of the $i$-th pixel, $I_i$. The expectation value of $x^p$ is then given by
\begin{equation}
\left< x^p \right> = \frac{\sum_j I_j x_j^p}{\sum_j I_j},
\label{eq:expectationvalue}
\end{equation}
and similarly for $y^p$. In the case $p=1$, Eq.~\ref{eq:expectationvalue} is the sample mean, $\bar{x} = \left< x \right>$. The point $\left( \left< x \right>, \left< y \right> \right)$ is called the image centroid.

A measure of the spread/variance is given by the uncorrected sample standard deviation (which omits the so-called Bessel correction, as is appropriate in the case of many pixels). The standard deviation of our weighted set of measurements $\{ I_j, x_j \}$ is then given by
\begin{equation}
\sigma_x = \left[ \frac{\sum_j I_j \left( x_j - \bar{x} \right)^2}{\sum_j I_j}\right]^{\frac{1}{2}}.
\end{equation}
%The skewness of the weighted set of measurements is given by
%\begin{equation}
%\gamma = \frac{1}{\sigma^3}  \left[ \frac{ \sum_j I_j \left( x_j - \bar{x} \right)^3}{\sum_j I_j}  \right].
%\end{equation}
By expanding the bracketed terms and substituting the definition of the expectation value (Eq.~\ref{eq:expectationvalue}), we can express the spread as follows:
\begin{equation}
\sigma_x^2 = \left<x^2\right> - \left<x\right>^2,
\label{eq:spread_expectation}
\end{equation}
%and
%\begin{equation}
%\gamma = \frac{1}{\sigma^3} \left[ \left<x^3\right> - 3 \left< x \right> \left< x^2 \right> + 2 \left<x\right>^3 \right].
%\label{eq:skewness_expectation}
%\end{equation}
The definition of the image moments (Eq.~\ref{eqn:moment}) follows naturally from our interpretation of $\{ I_j x_j \}$, and we may write:
\begin{equation}
\left< x^p \right> = \frac{\sum_x^W \sum_y^H x^p I \left(x, y\right)}{\sum_x^W \sum_y^H I \left(x, y\right)},
\end{equation}
which we may rewrite, using Eq.~\ref{eqn:moment}, to obtain
\begin{equation}
\left< x^p \right> = \frac{M_{p0}}{M_{00}}
\end{equation}
(and similarly for $y^p$). Substituting these quantities into Eq.~\ref{eq:spread_expectation} then yields equation \ref{eq:spread}. Additionally, one can define the covariance matrix, $\sigma_{xy}$, a 2-by-2 matrix, as follows:
\begin{subequations}
\begin{align}
\sigma_{xx} &= \frac{M_{20}}{M_{00}}-\left(\frac{M_{10}}{M_{00}}\right)^2, \\
\sigma_{yy} &= \frac{M_{02}}{M_{00}}-\left(\frac{M_{01}}{M_{00}}\right)^2, \\
\sigma_{xy} &= \sigma_{yx} = \frac{M_{11}}{M_{00}}-\frac{M_{10} M_{01}}{M_{00}^2}.
\end{align}
\label{eq:cov_matrix}
\end{subequations}
The largest eigenvector and eigenvalue of $\sigma_{xy}$ represent the direction and magnitude of the largest spread in the data. We call $\sigma_{xy}$'s largest eigenvalue the major axis, $\lambda_{\rm max}$, and its smallest eigenvalue the minor axis, $\lambda_{\rm min}$. These quantities indicate the maximum extent of the source, independently of the source's rotation.

\clearpage

\section{Evacuated region in SANE GRMHD models}
\label{app:evacuation}

\begin{figure}
\centering
\includegraphics[width=0.49\textwidth]{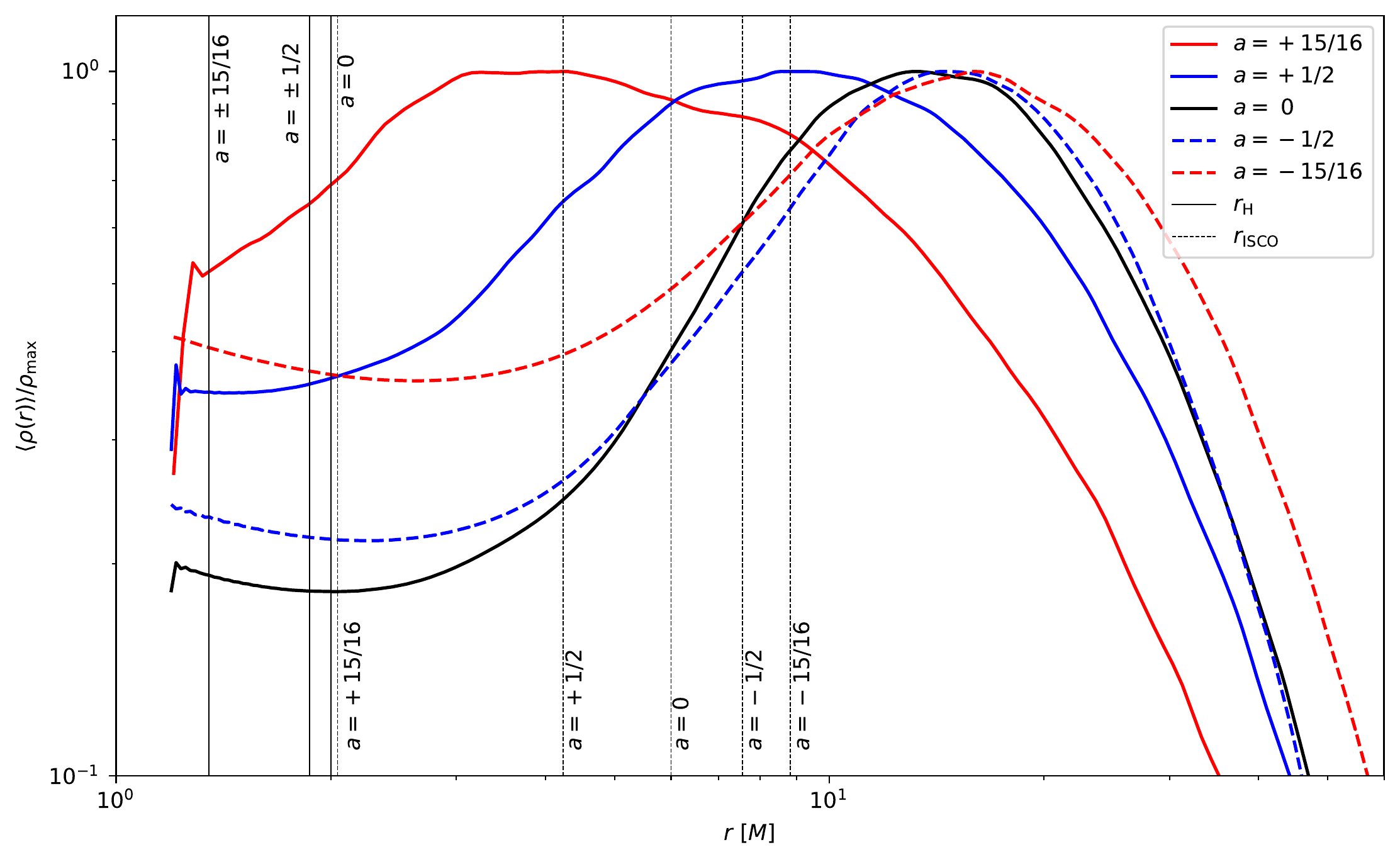}
\caption{Disc-averaged and time-averaged radial density plots of our SANE GRMHD simulations. For each spin, the locations of the event horizon and the ISCO are indicated as well.}
\label{fig:evacuation}
\end{figure}

Figure \ref{fig:evacuation} shows disc-averaged and time-averaged radial density plots for all of our SANE GRMHD simulations. The disc-averaging is done according to the following formula \citep{bhac}:
\begin{equation}
\langle \rho \left( r \right) \rangle = \frac{\int_0^{2\pi} \int_{\pi/3}^{2\pi/3} \rho \left(r, \theta, \phi \right) \sqrt{-g} \ d\theta \ d\phi}{\int_0^{2\pi} \int_{\pi/3}^{2\pi/3}  \sqrt{-g} \ d\theta \ d\phi},
\end{equation}
where $g$ is the metric determinant and the limits of integration for $\theta$ are chosen to exclude material that pertains to the atmosphere rather than the disc \citep{bhac}. Note that the evacuated region grows as the spin is reduced, so that the effect of evacuation on the IS is most apparent for the highly retrograde-spin models.

\section{Effects of time-averaging on source size and morphology}
\label{app:averaging}

\begin{figure}
\centering
\begin{subfigure}[b]{0.49\textwidth}
	\includegraphics[width=\textwidth]{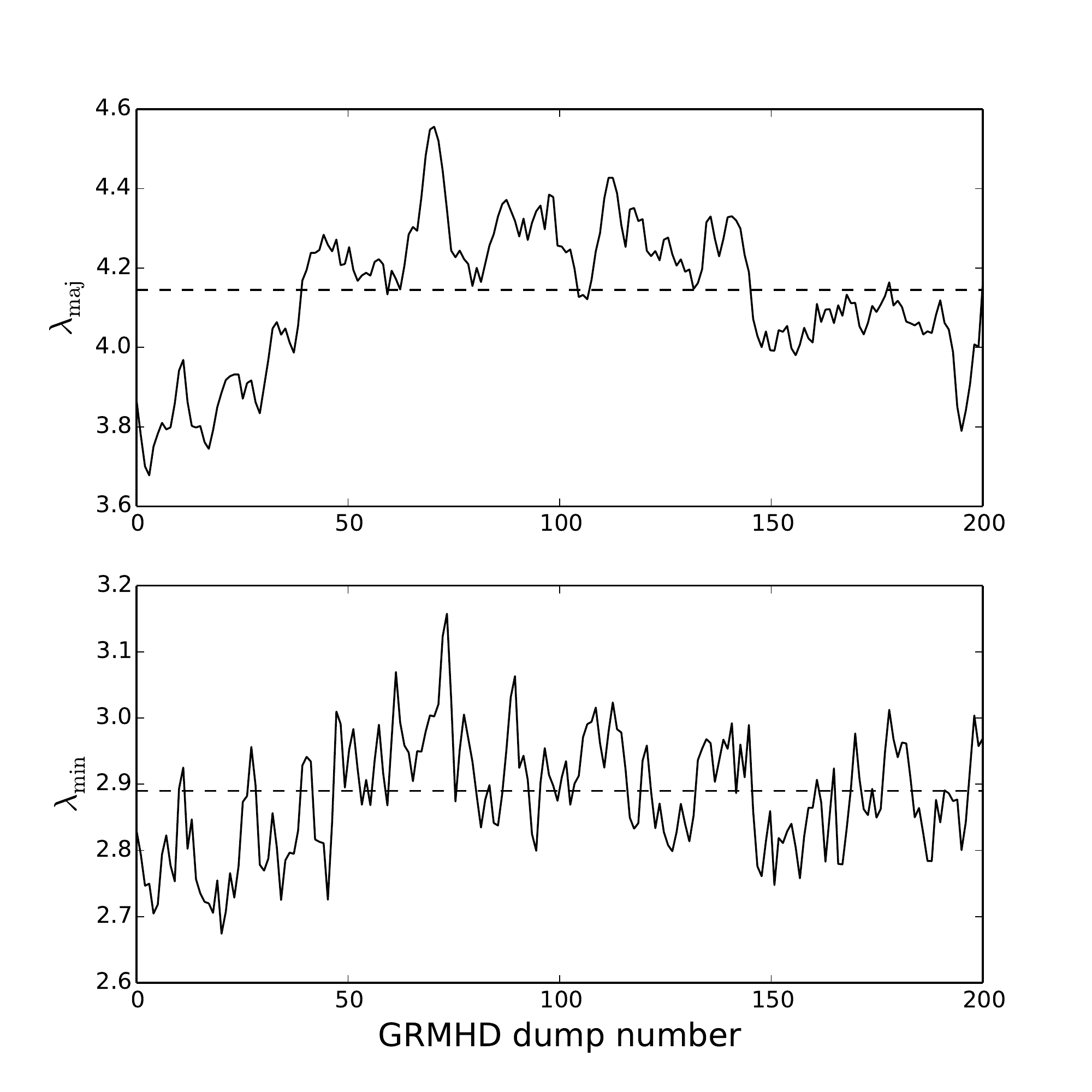}
\end{subfigure}
\begin{subfigure}[b]{0.49\textwidth}
	\includegraphics[width=\textwidth]{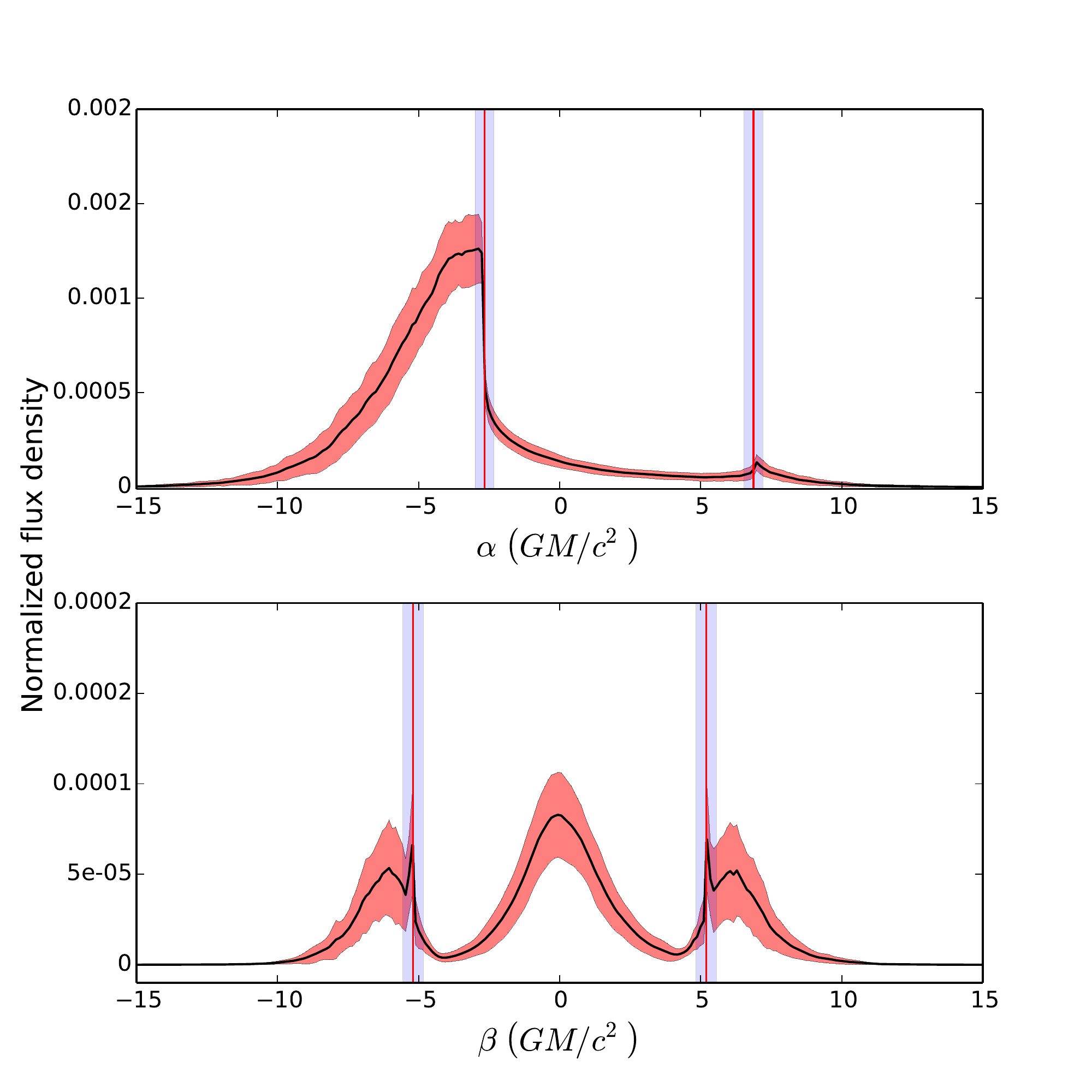}
\end{subfigure}
\caption{Top panel: major and minor axes of individual GRMHD snapshots (solid line) versus those of the time-averaged image (dashed line), for the SANE disc model with $a=0.9375$ and $i=90^\circ$ calibrated to 2.5 Jy. Bottom panel: intensity profiles, computed as in Section \ref{sec:intensity_profiles}, of the same model, plotted along with their standard deviation computed over the 200 snapshots (shaded region).}\label{fig:averaging1}
\end{figure}

\begin{figure}
\centering
\begin{subfigure}[b]{0.49\textwidth}
	\includegraphics[width=\textwidth]{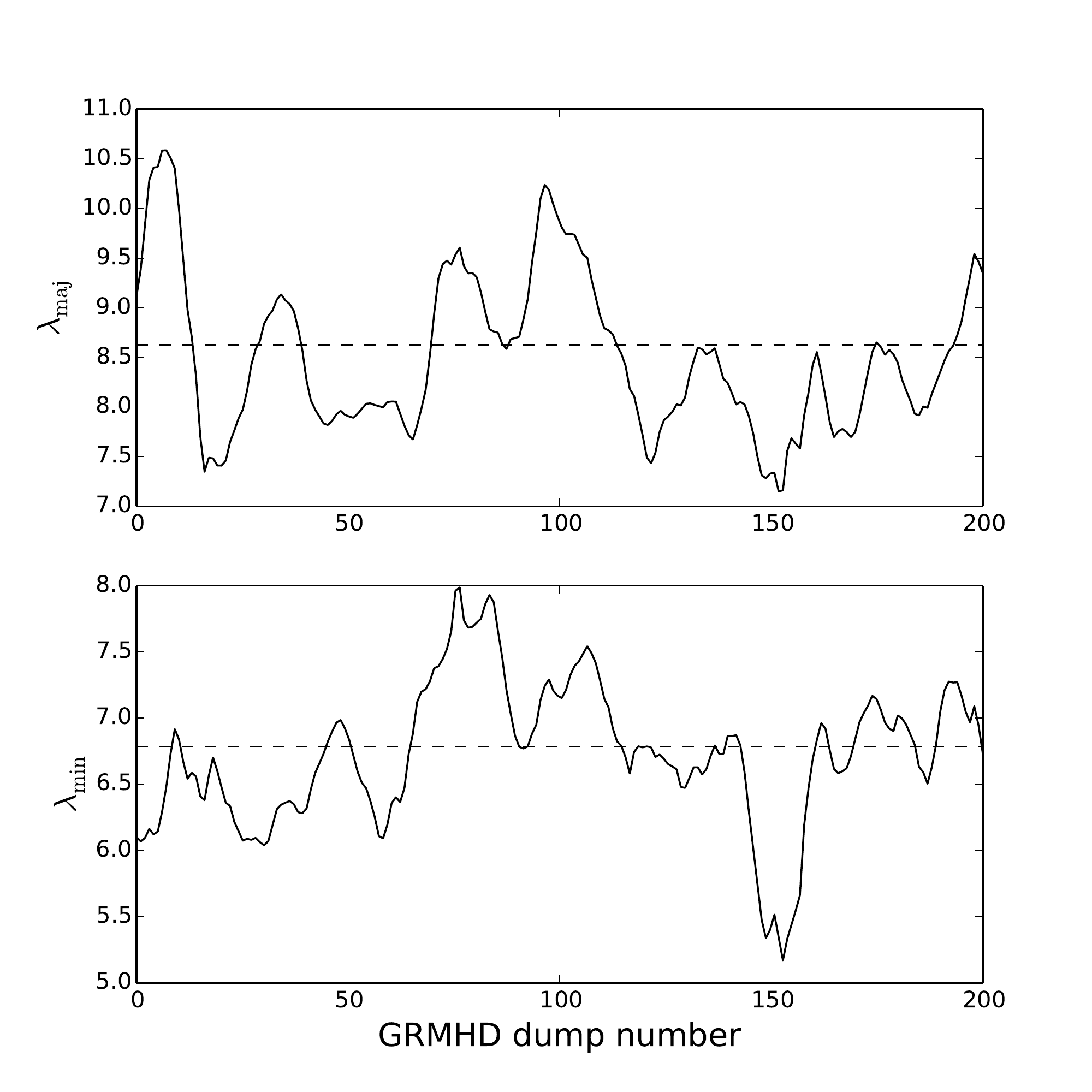}
\end{subfigure}
\begin{subfigure}[b]{0.49\textwidth}
	\includegraphics[width=\textwidth]{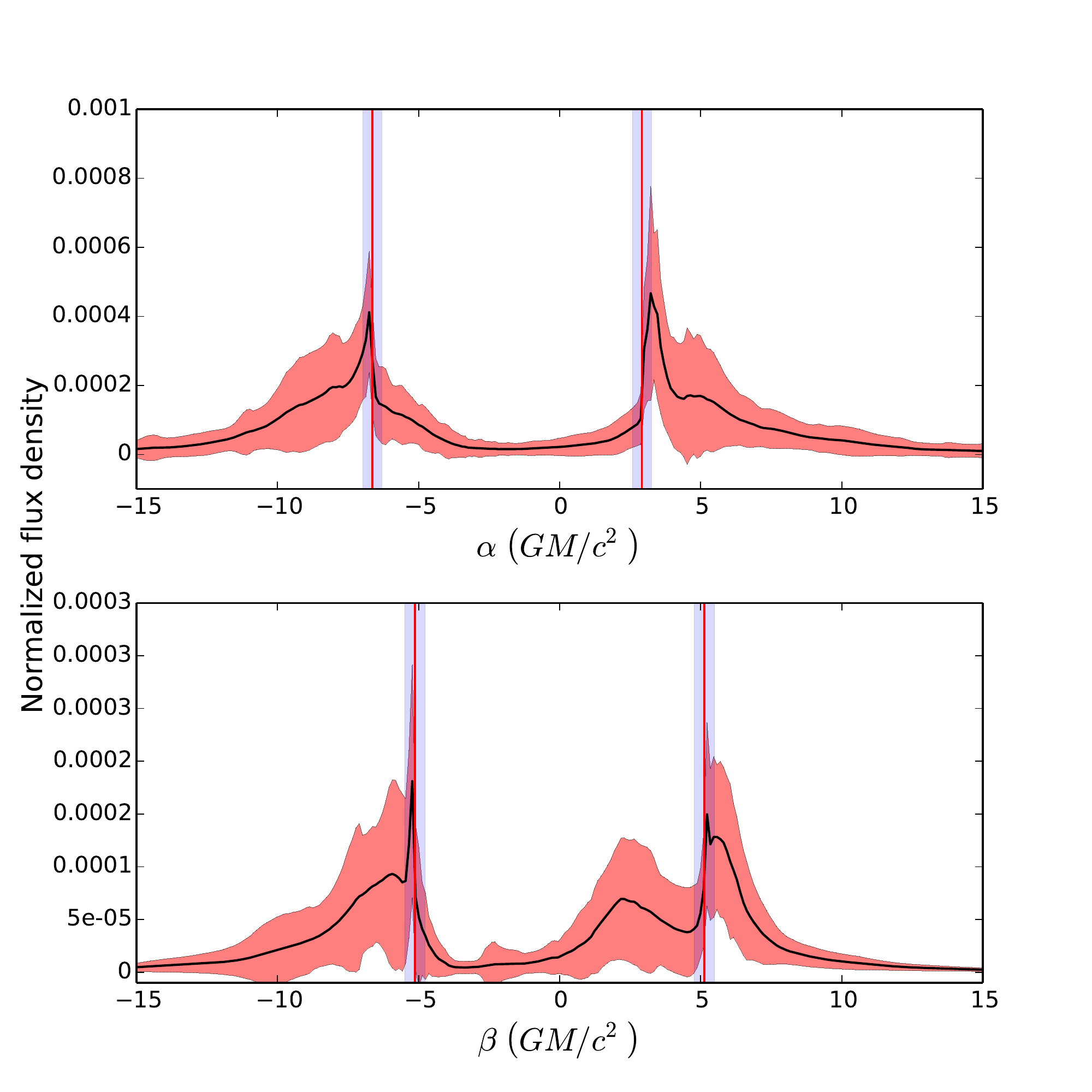}
\end{subfigure}
\caption{Top panel: major and minor axes of individual GRMHD snapshots (solid line) versus those of the time-averaged image (dashed line), for the MAD disc model with $a=-0.9375$ and $i=60^\circ$ calibrated to 2.5 Jy. Bottom panel: intensity profiles, computed as in Section \ref{sec:intensity_profiles}, of the same model, plotted along with their standard deviation computed over the 200 snapshots (shaded region).}\label{fig:averaging2}
\end{figure}

As the analyses in this paper are based on time-averaged images, we here investigate how the major and minor axes $\lambda_{\rm max}$ and  $\lambda_{\rm min}$ of single-snapshot images compare to those of the time-averaged image. Figure \ref{fig:averaging1} shows this comparison for our SANE disc model $a=0.9375$ and $i=90^\circ$ calibrated to 2.5 Jy, while Fig. \ref{fig:averaging2} shows the same for the MAD disc model with $a=-0.9375$ and $i=60^\circ$ calibrated to 2.5 Jy. In the case of the SANE disc model, the source size varies by roughly 10\%, while for the SANE disc case, it is closer to 30\%. Note that in both cases, however, the bright peaks on either side of the horizon are robust features.

\section{Time-averaged intensity maps for all models}
\label{appLibrary}

\begin{figure*}
\centering
\begin{subfigure}[b]{0.197\textwidth}
	\includegraphics[width=\textwidth]{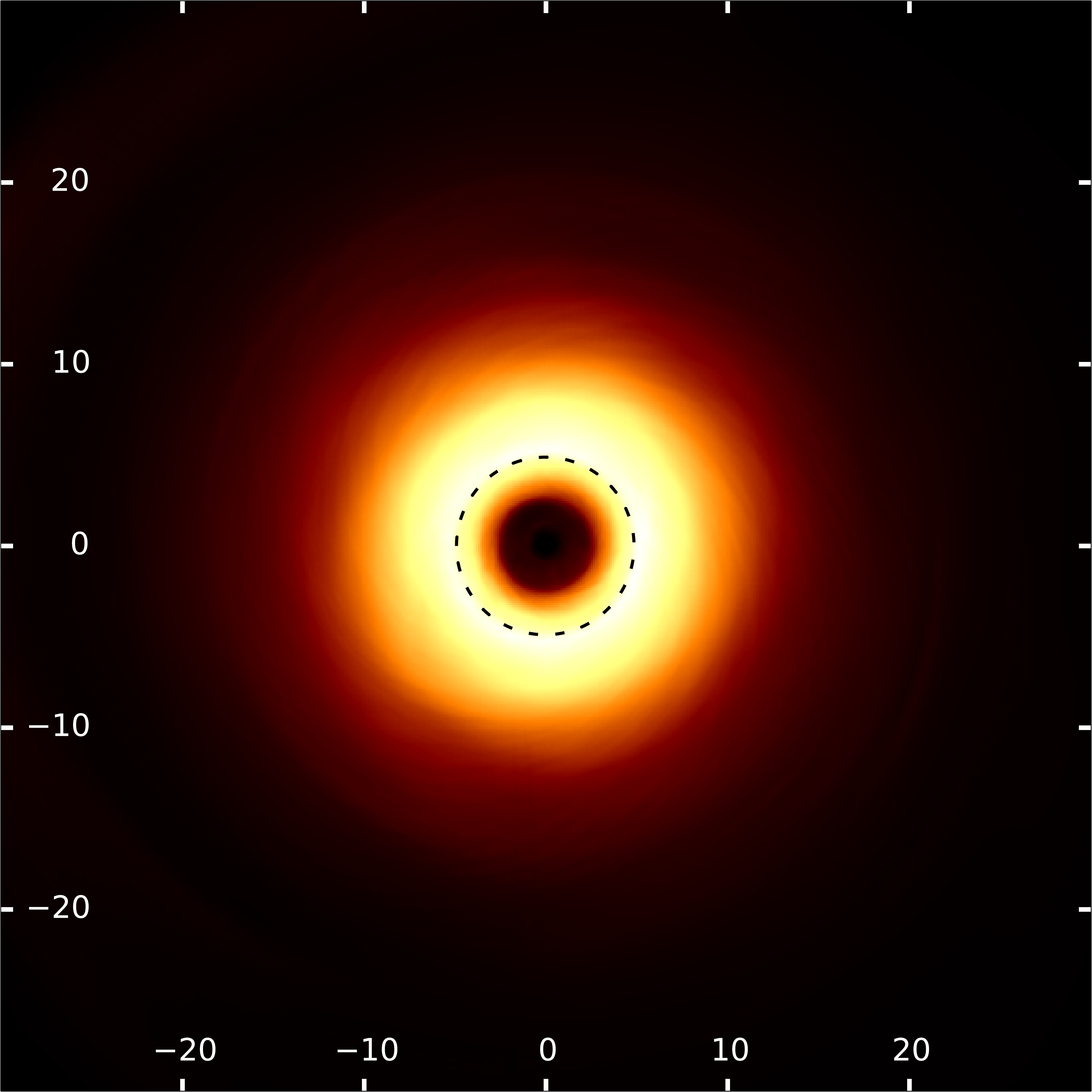}
	\caption{$a=-0.9375$, $i=1^\circ$.}
\end{subfigure}
\begin{subfigure}[b]{0.197\textwidth}
	\includegraphics[width=\textwidth]{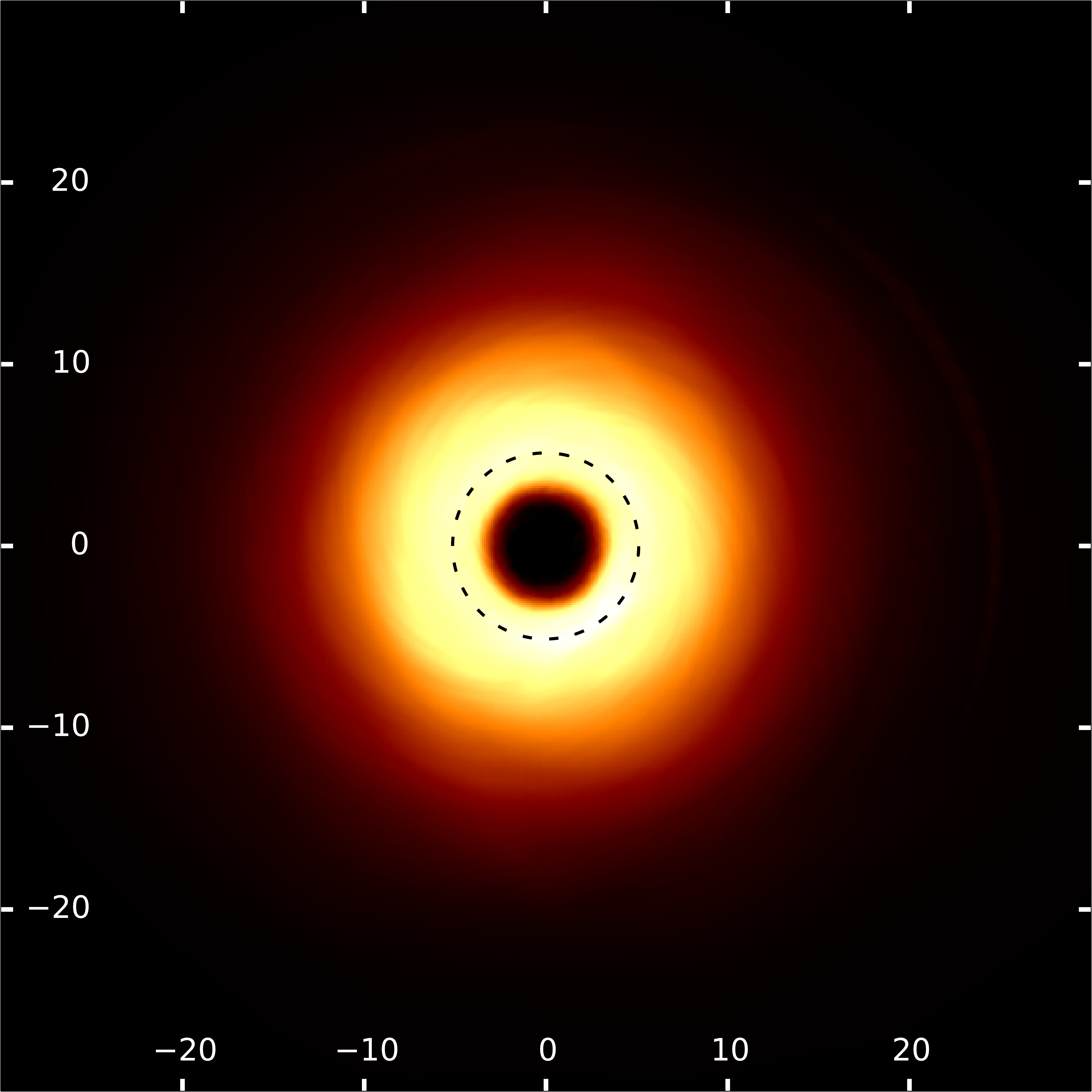}
	\caption{$a=-0.5$, $i=1^\circ$.}
\end{subfigure}
\begin{subfigure}[b]{0.197\textwidth}
	\includegraphics[width=\textwidth]{Figures/sane_jet_a0_0_25-crop}
	\caption{$a=0$, $i=1^\circ$.}
\end{subfigure}
\begin{subfigure}[b]{0.197\textwidth}
	\includegraphics[width=\textwidth]{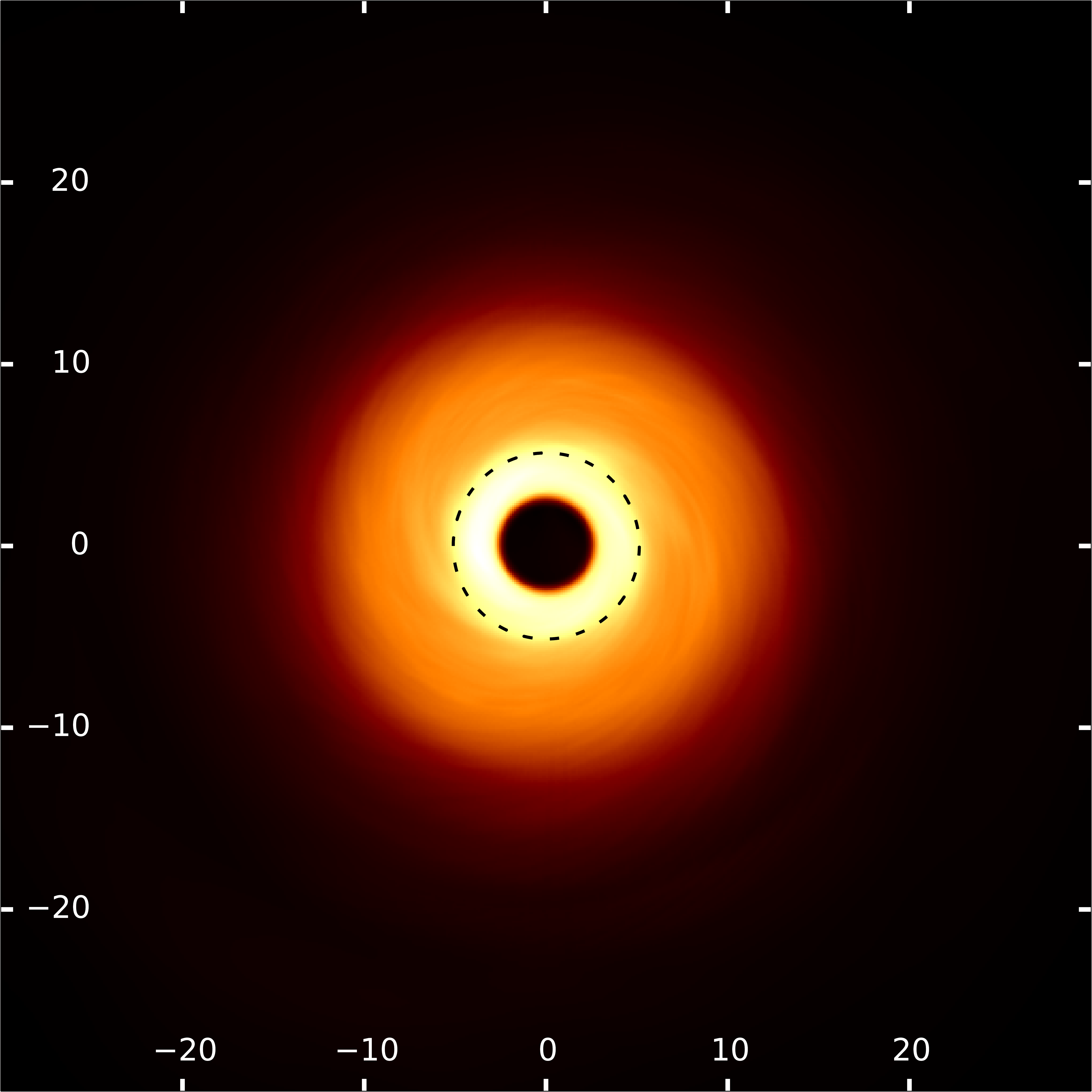}
	\caption{$a=0.5$, $i=1^\circ$.}
\end{subfigure}
\begin{subfigure}[b]{0.197\textwidth}
	\includegraphics[width=\textwidth]{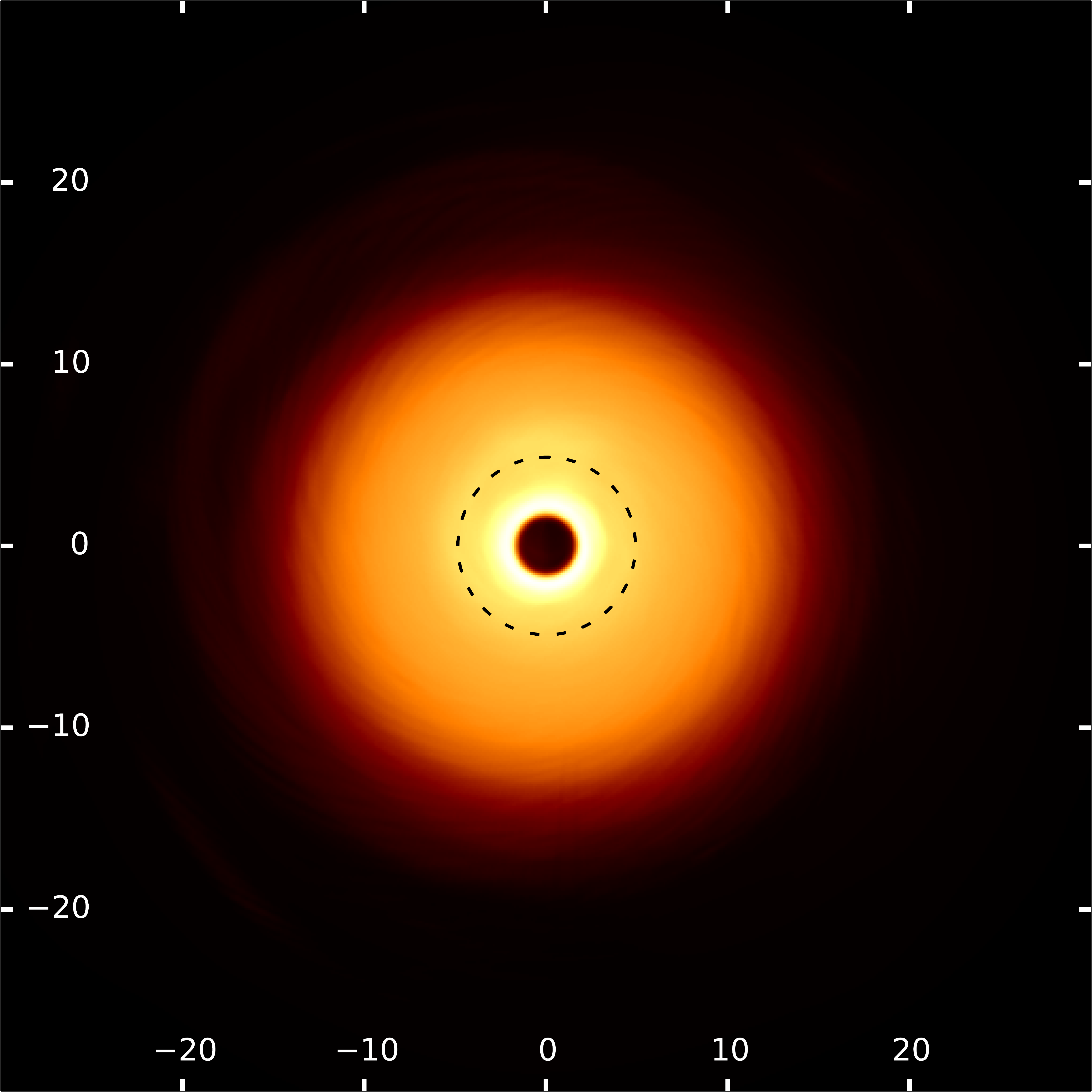}
	\caption{$a=0.9375$, $i=1^\circ$.}
\end{subfigure}
\begin{subfigure}[b]{0.197\textwidth}
	\includegraphics[width=\textwidth]{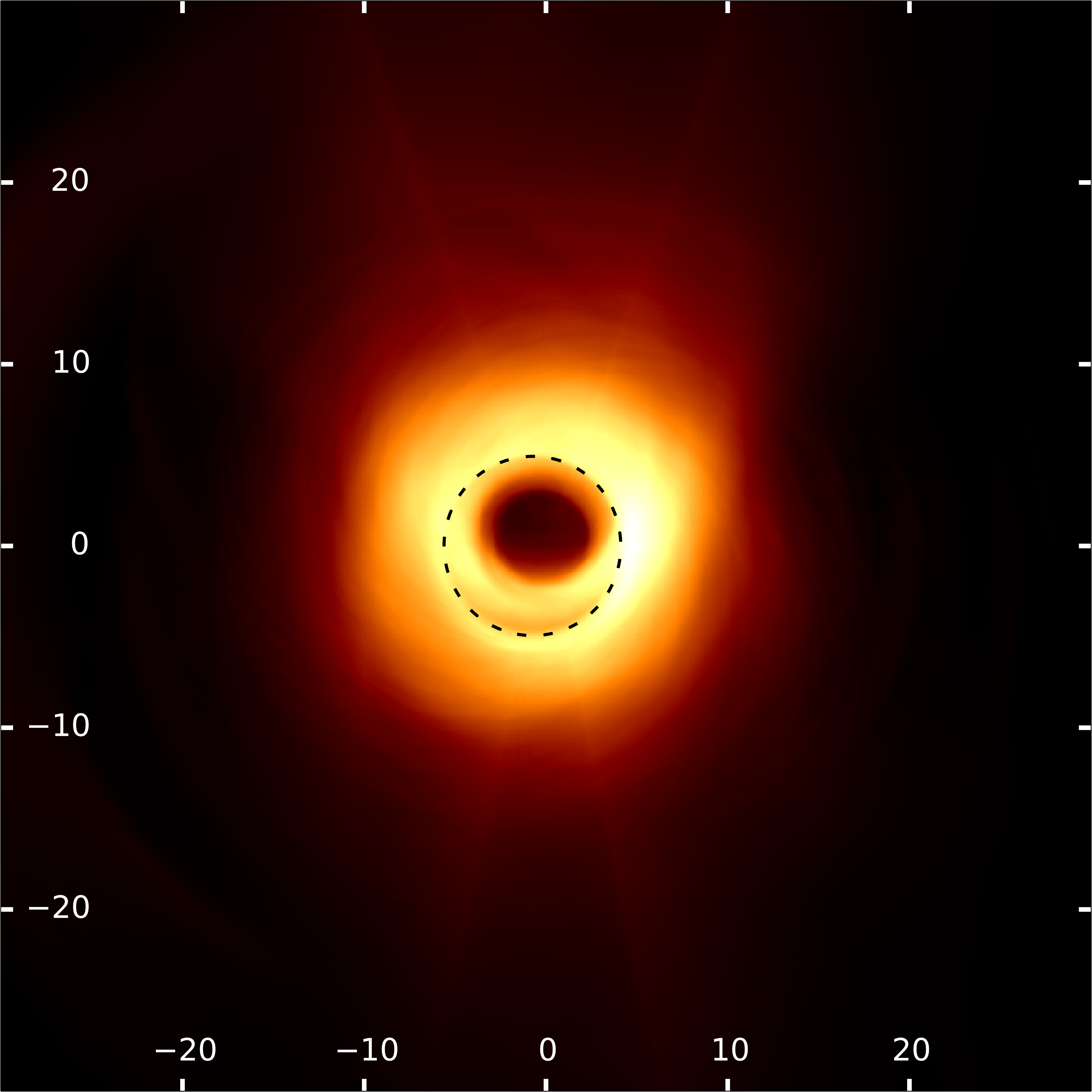}
	\caption{$a=-0.9375$, $i=20^\circ$.}
\end{subfigure}
\begin{subfigure}[b]{0.197\textwidth}
	\includegraphics[width=\textwidth]{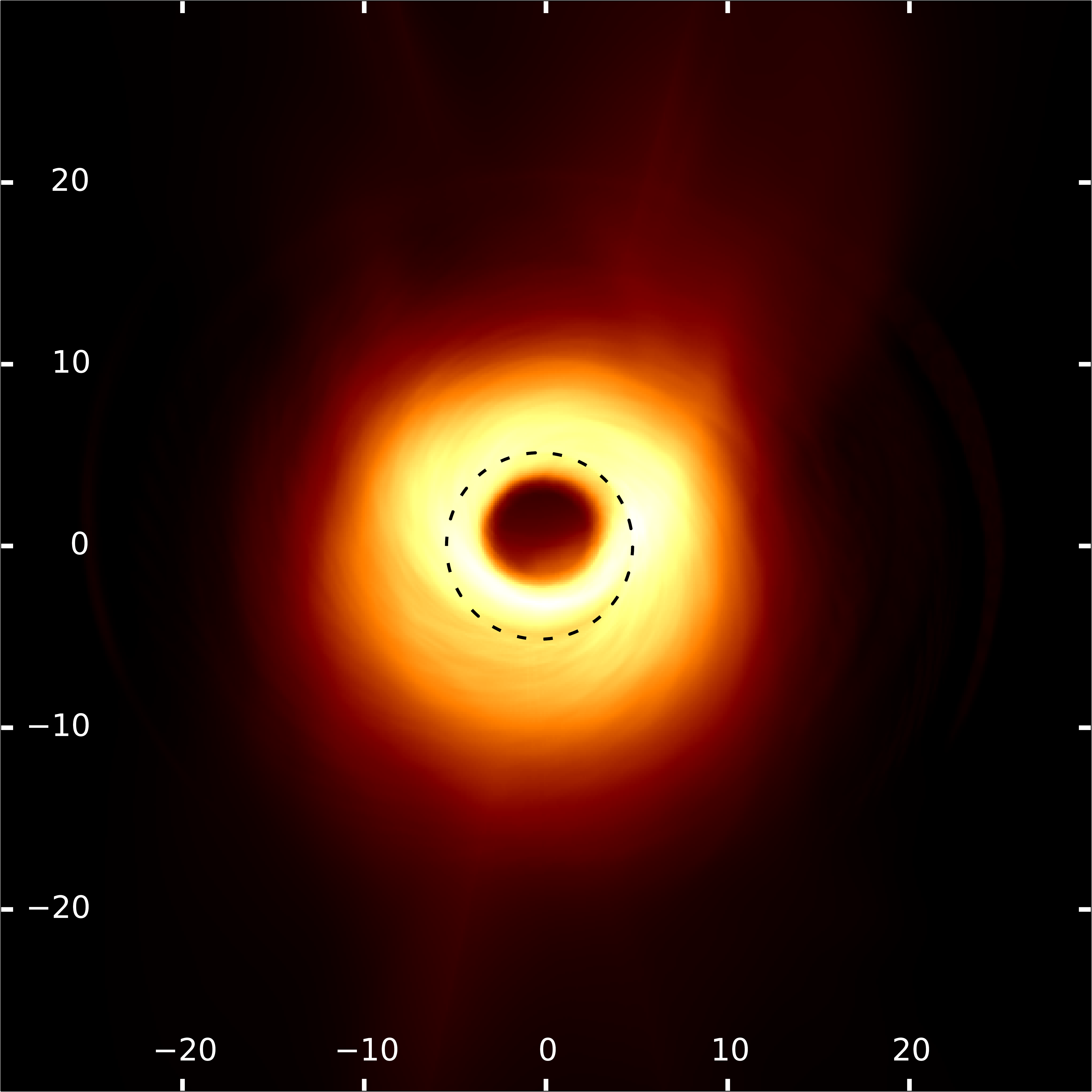}
	\caption{$a=-0.5$, $i=20^\circ$.}
\end{subfigure}
\begin{subfigure}[b]{0.197\textwidth}
	\includegraphics[width=\textwidth]{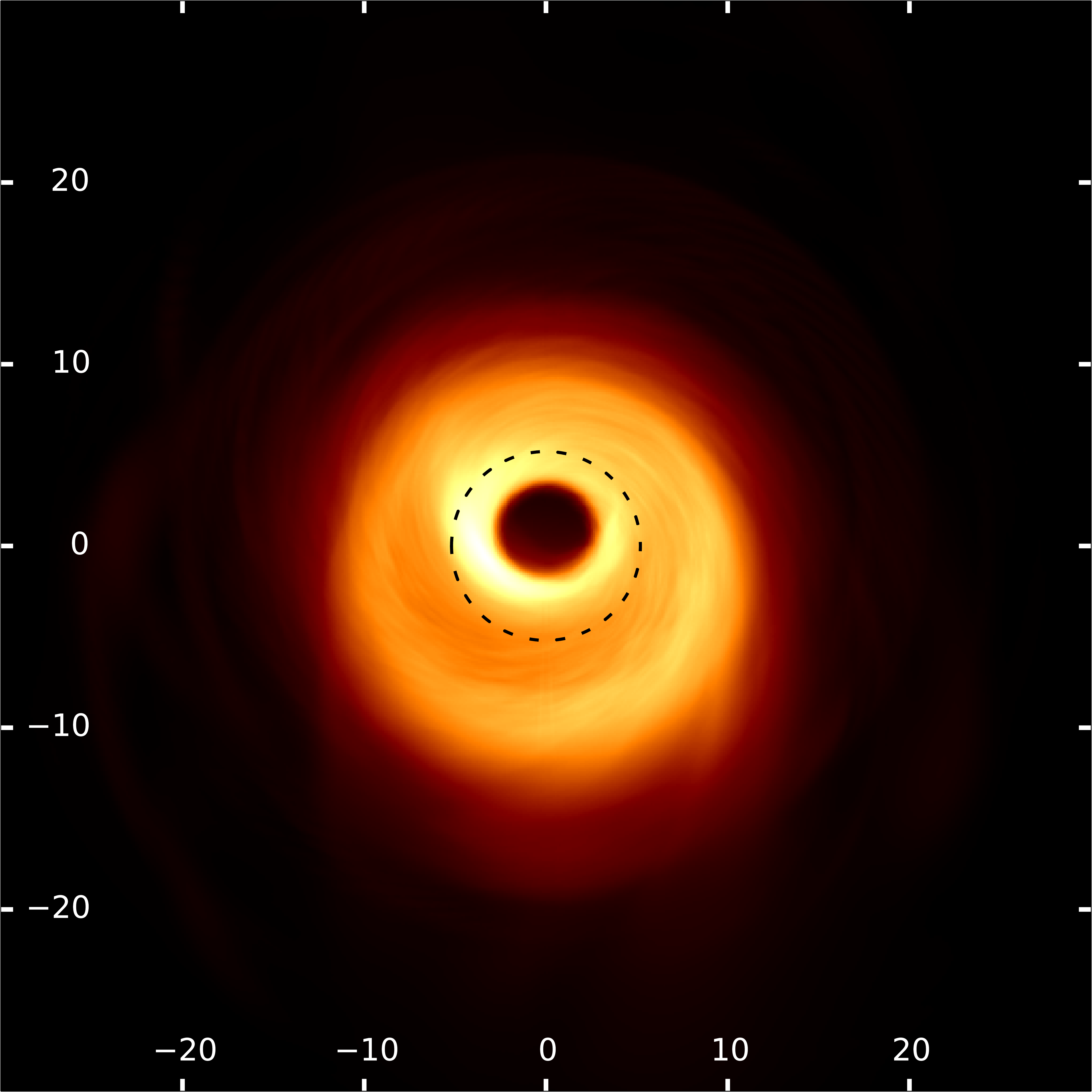}
	\caption{$a=0$, $i=20^\circ$.}
\end{subfigure}
\begin{subfigure}[b]{0.197\textwidth}
	\includegraphics[width=\textwidth]{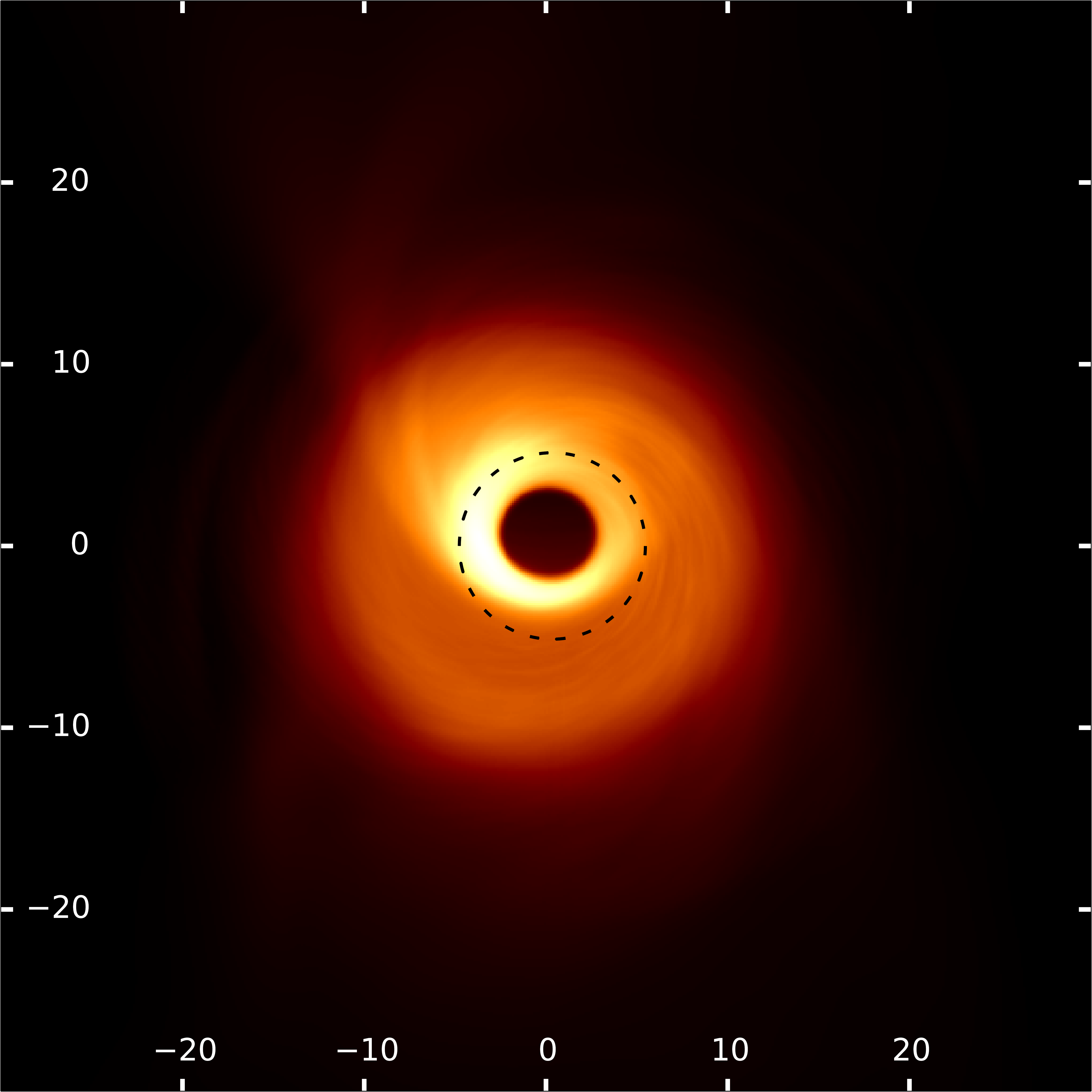}
	\caption{$a=0.5$, $i=20^\circ$.}
\end{subfigure}
\begin{subfigure}[b]{0.197\textwidth}
	\includegraphics[width=\textwidth]{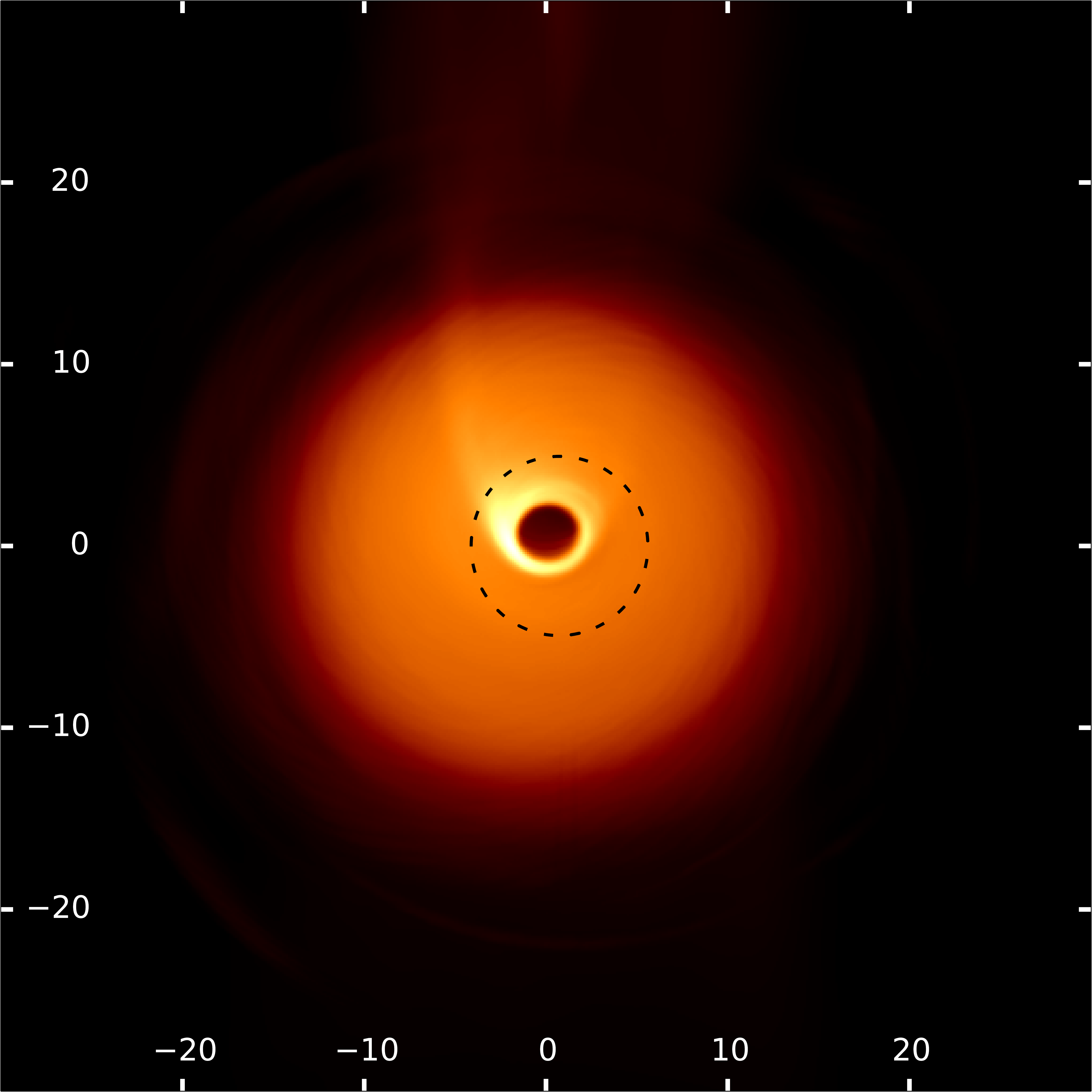}
	\caption{$a=0.9375$, $i=20^\circ$.}
\end{subfigure}
\begin{subfigure}[b]{0.197\textwidth}
	\includegraphics[width=\textwidth]{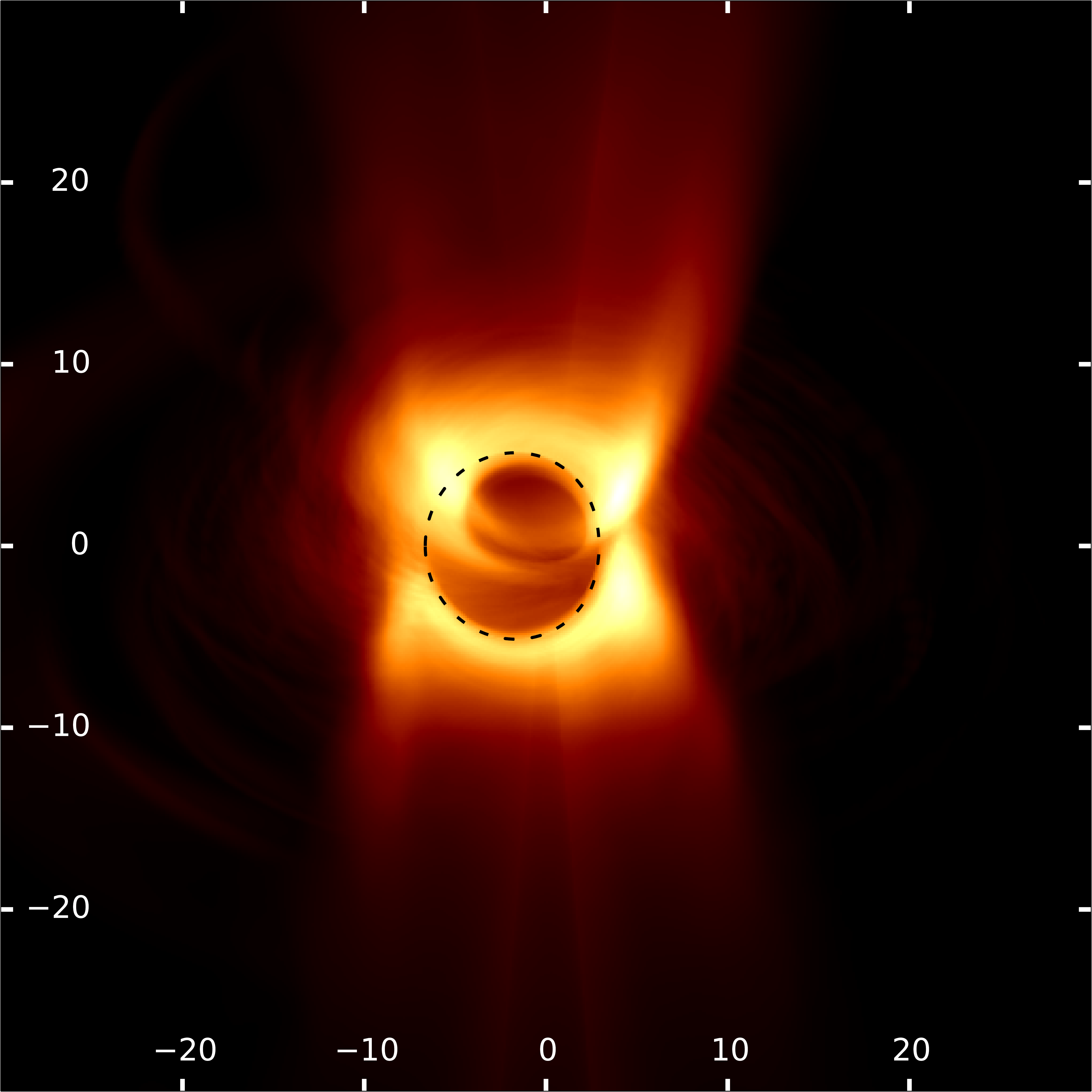}
	\caption{$a=-0.9375$, $i=60^\circ$.}
\end{subfigure}
\begin{subfigure}[b]{0.197\textwidth}
	\includegraphics[width=\textwidth]{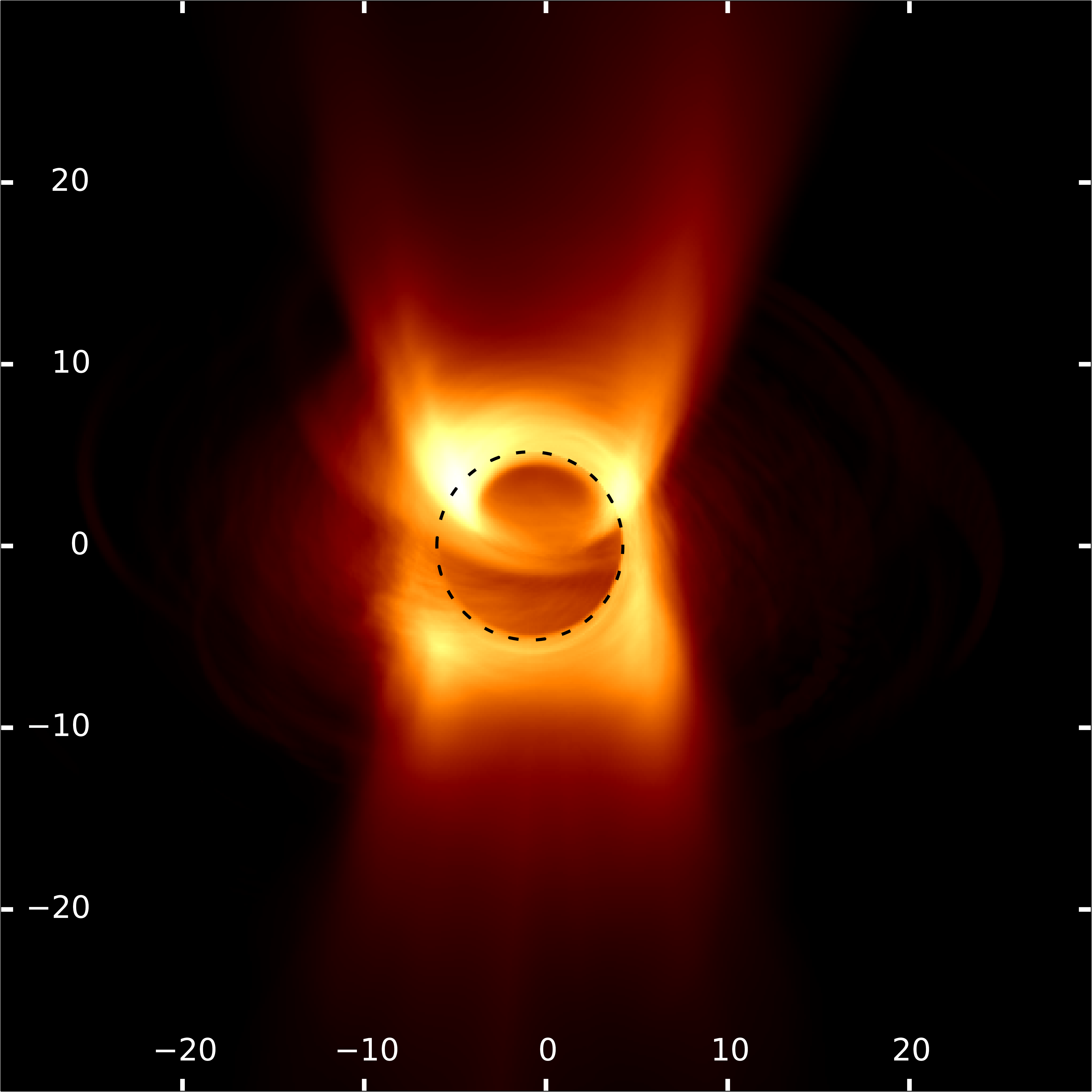}
	\caption{$a=-0.5$, $i=60^\circ$.}
\end{subfigure}
\begin{subfigure}[b]{0.197\textwidth}
	\includegraphics[width=\textwidth]{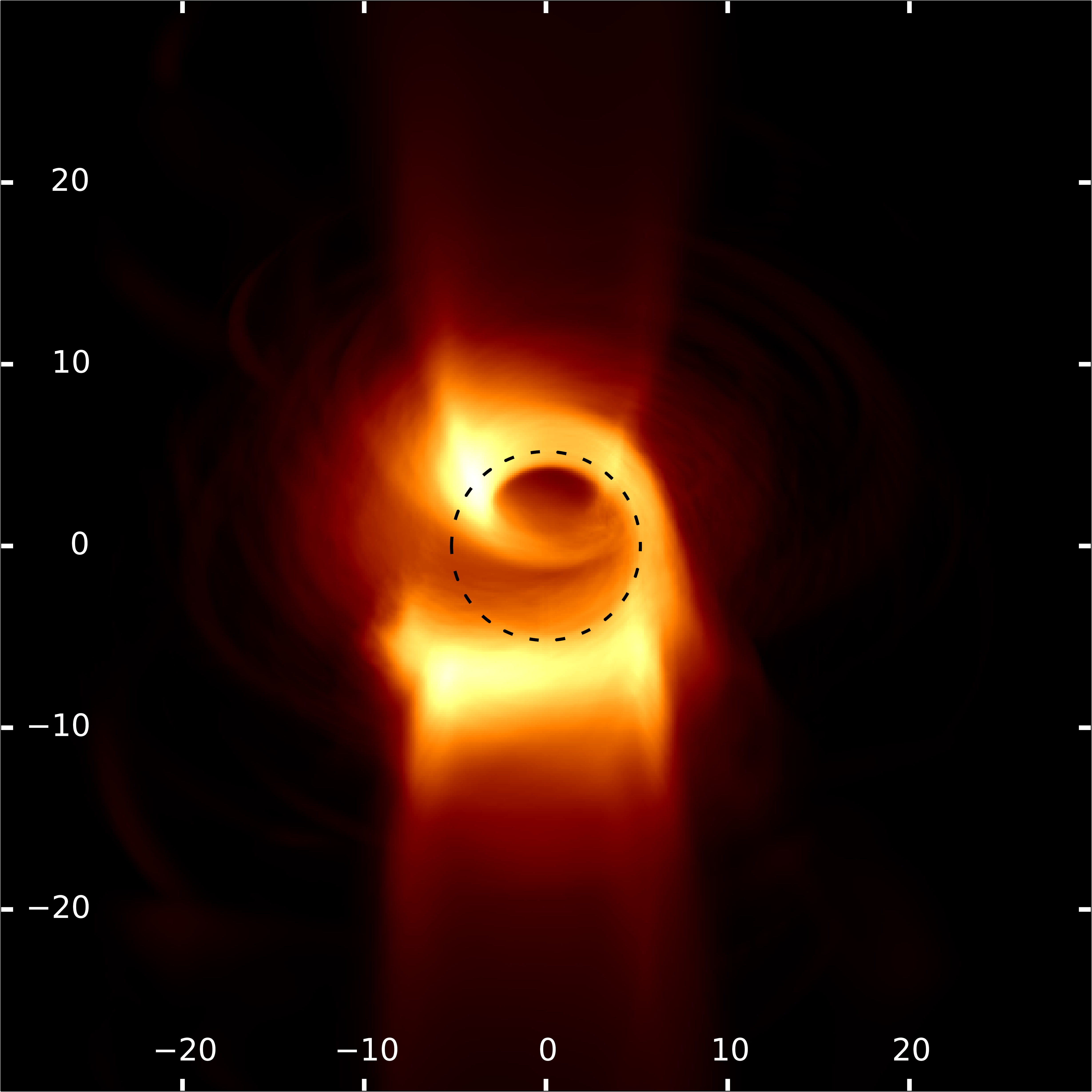}
	\caption{$a=0$, $i=60^\circ$.}
\end{subfigure}
\begin{subfigure}[b]{0.197\textwidth}
	\includegraphics[width=\textwidth]{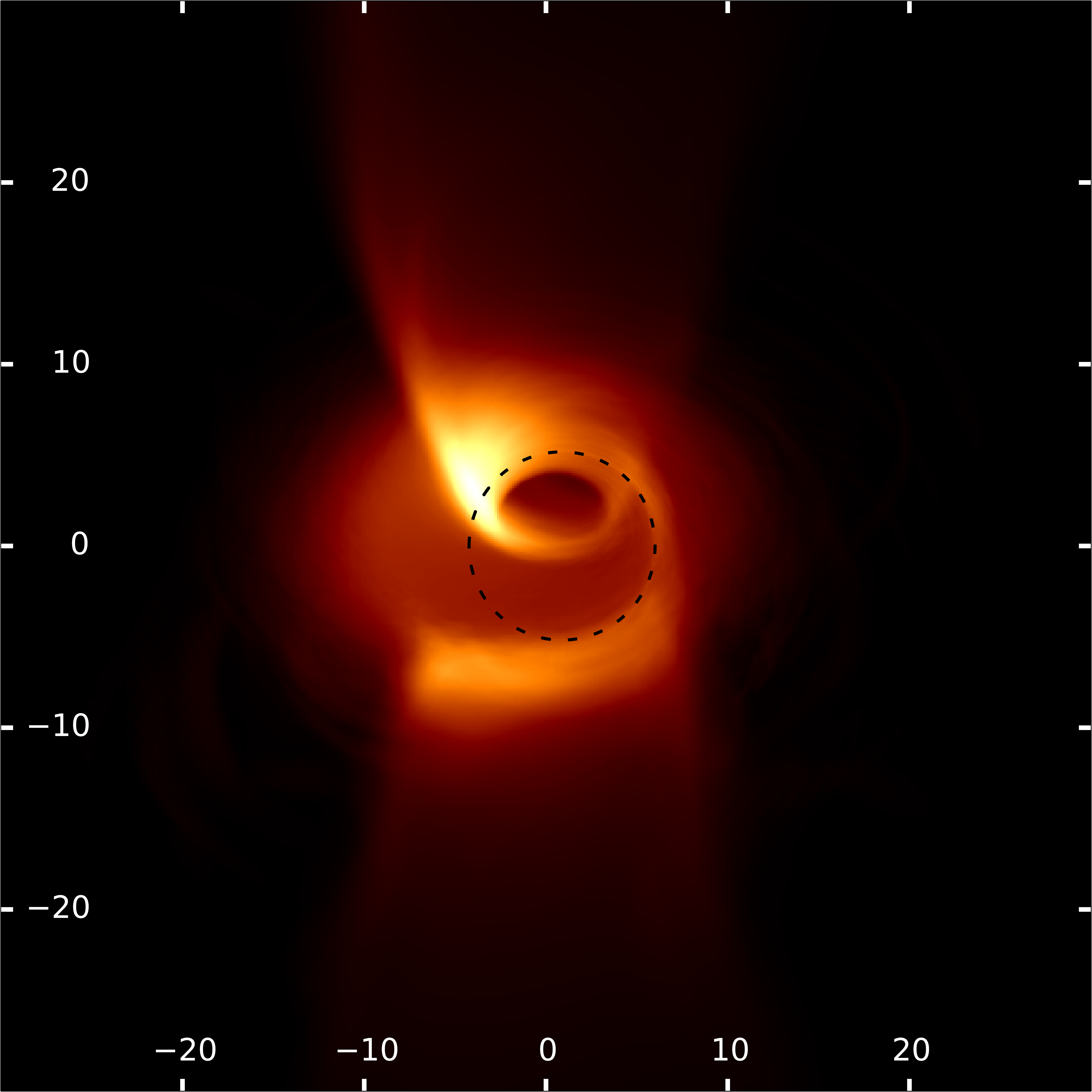}
	\caption{$a=0.5$, $i=60^\circ$.}
\end{subfigure}
\begin{subfigure}[b]{0.197\textwidth}
	\includegraphics[width=\textwidth]{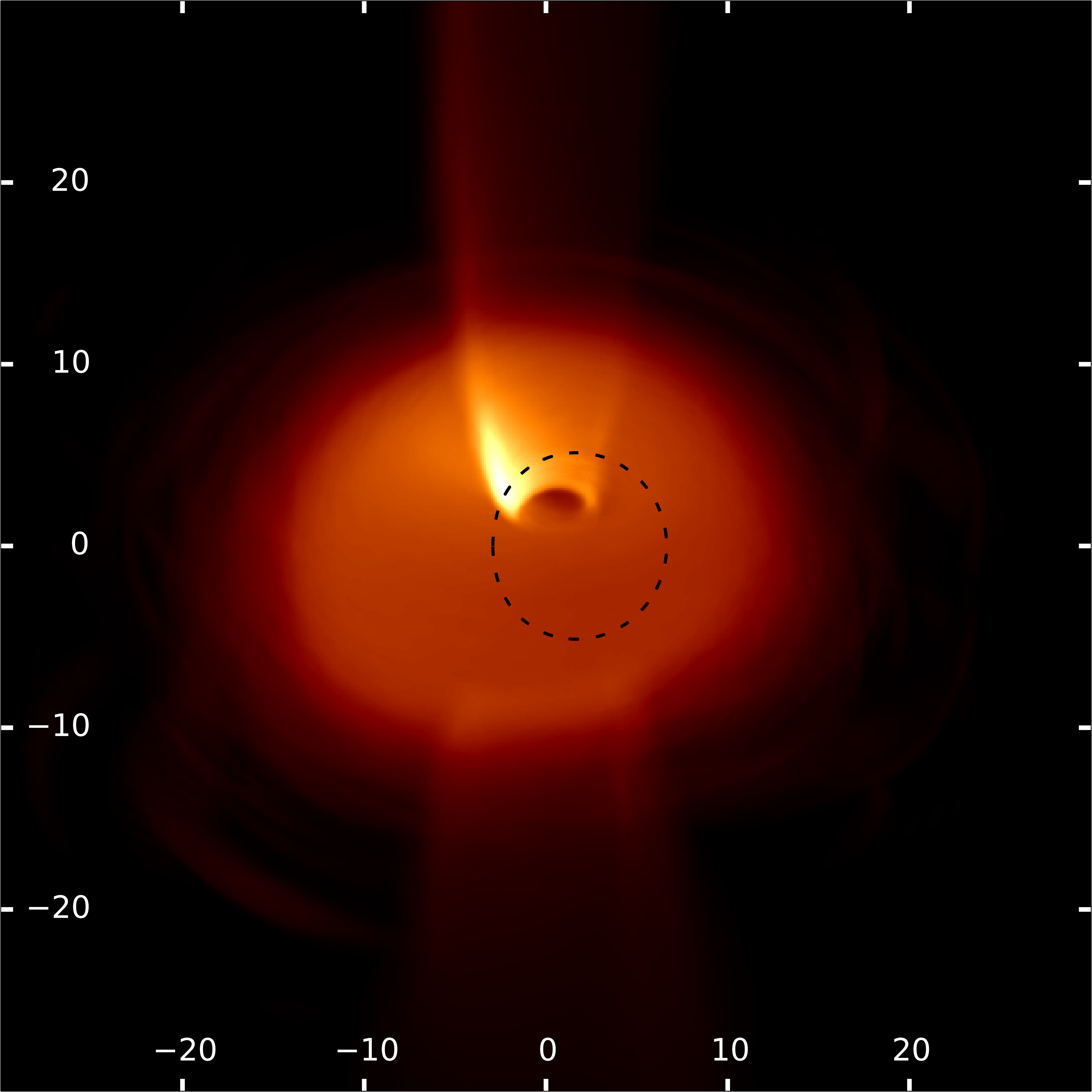}
	\caption{$a=0.9375$, $i=60^\circ$.}
\end{subfigure}
\begin{subfigure}[b]{0.197\textwidth}
	\includegraphics[width=\textwidth]{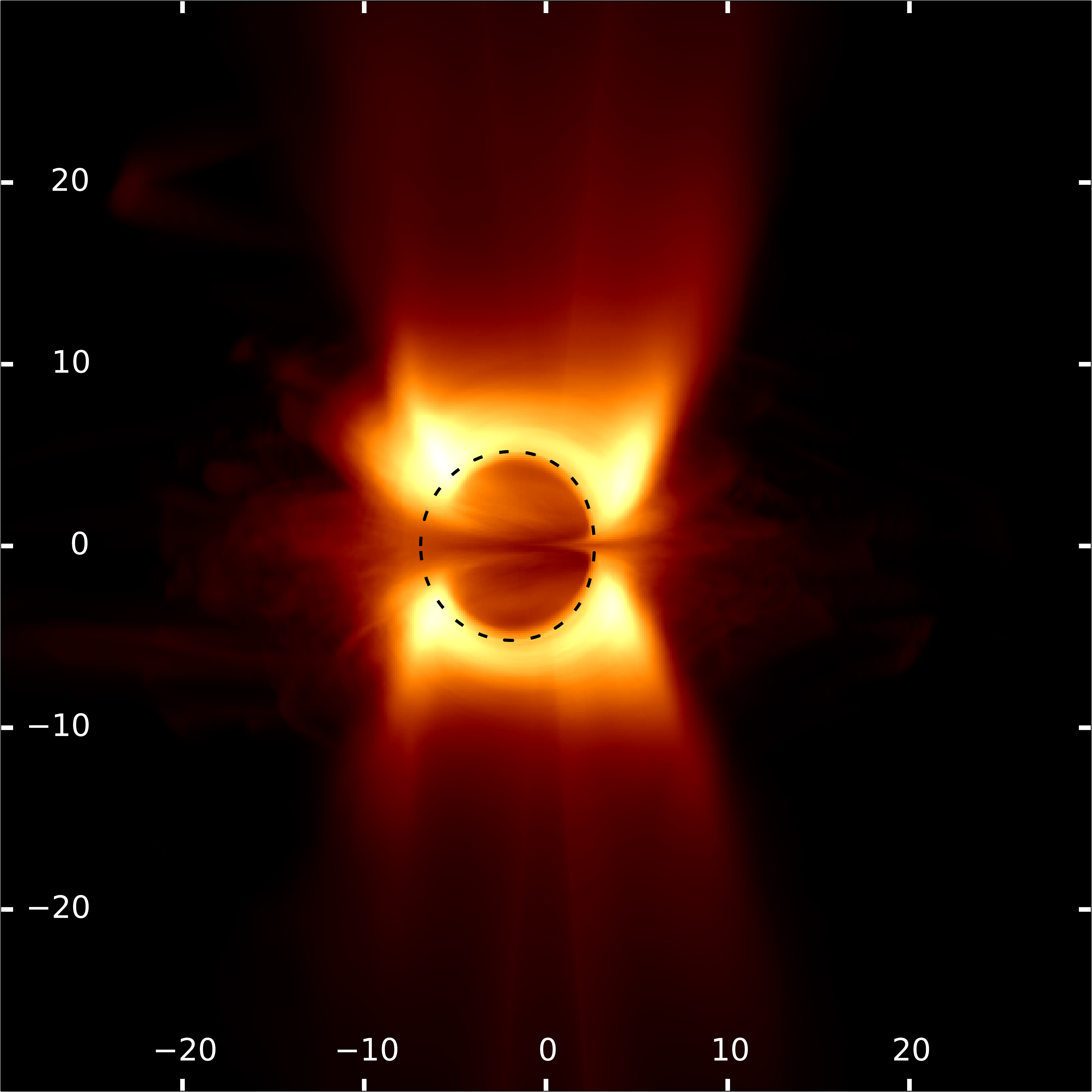}
	\caption{$a=-0.9375$, $i=90^\circ$.}
\end{subfigure}
\begin{subfigure}[b]{0.197\textwidth}
	\includegraphics[width=\textwidth]{Figures/sane_jet_a-1o2_90_25-crop}
	\caption{$a=-0.5$, $i=90^\circ$.}
\end{subfigure}
\begin{subfigure}[b]{0.197\textwidth}
	\includegraphics[width=\textwidth]{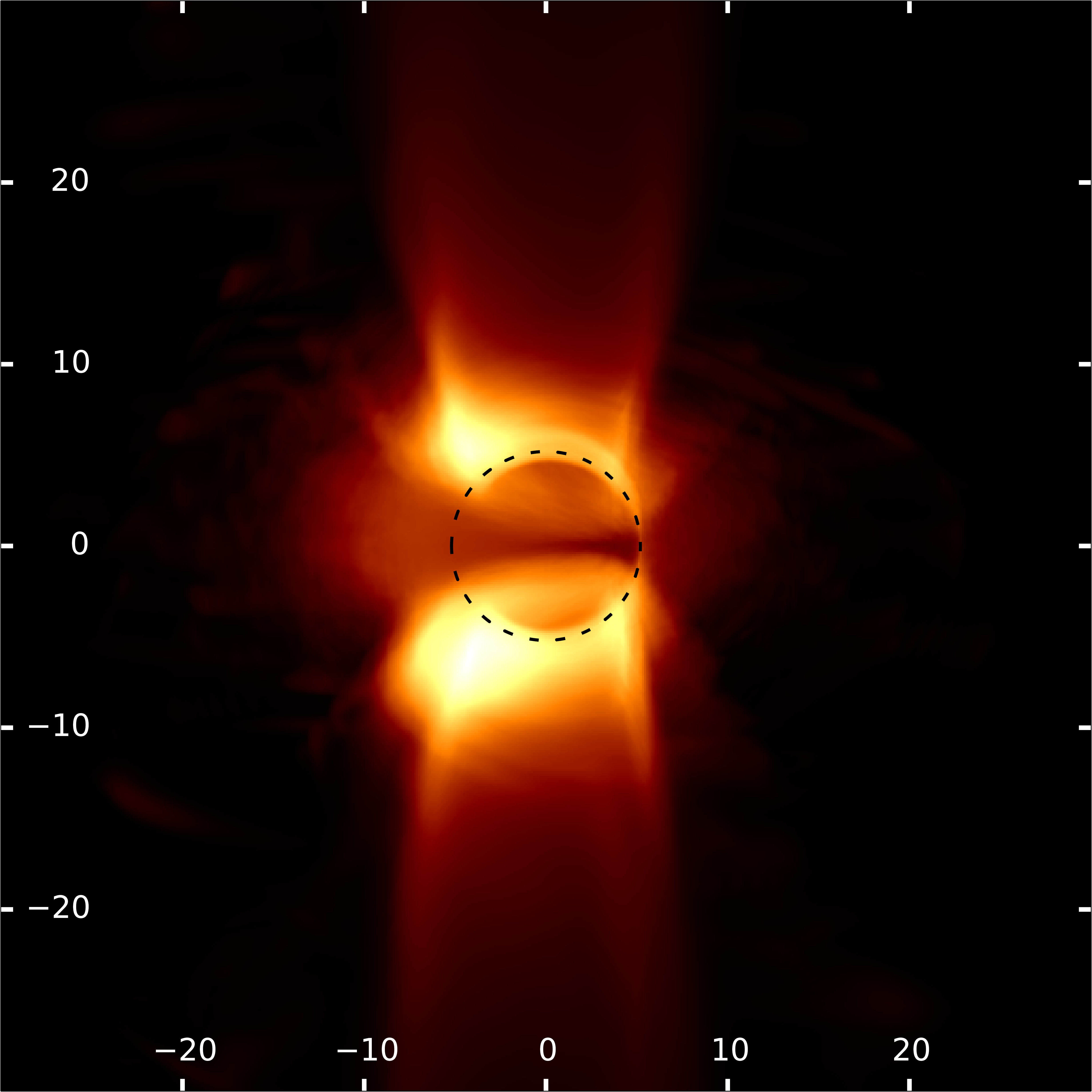}
	\caption{$a=0$, $i=90^\circ$.}
\end{subfigure}
\begin{subfigure}[b]{0.197\textwidth}
	\includegraphics[width=\textwidth]{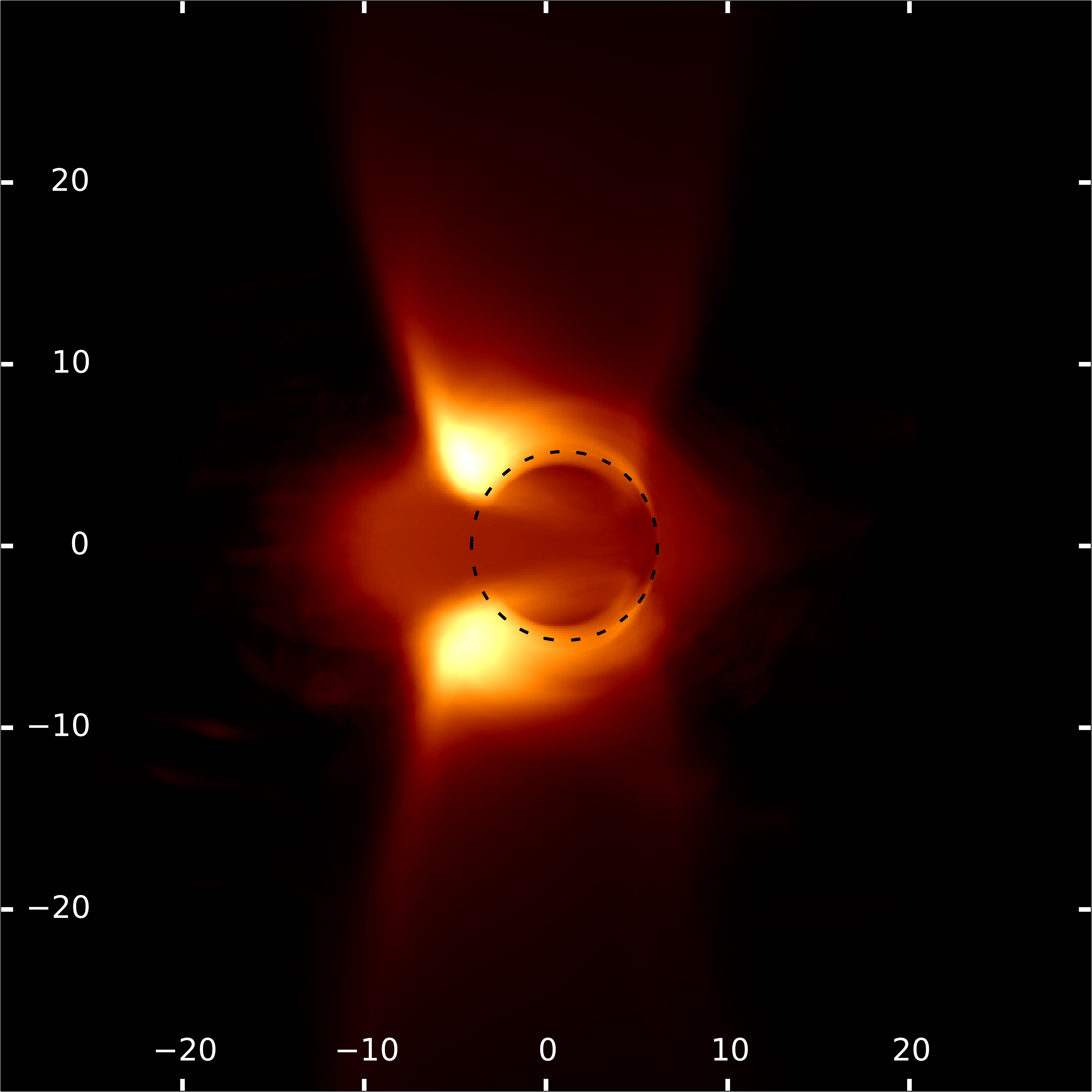}
	\caption{$a=0.5$, $i=90^\circ$.}
\end{subfigure}
\begin{subfigure}[b]{0.197\textwidth}
	\includegraphics[width=\textwidth]{Figures/sane_jet_a15o16_90_25-crop}
	\caption{$a=0.9375$, $i=90^\circ$.}
\end{subfigure}
\caption{Time-averaged, normalised intensity maps of our SANE, jet-dominated GRMHD models of Sgr A*, imaged at 230 GHz, at five different spins and four observer inclination angles, with an integrated flux density of 2.5 Jy. In each case, the photon ring, which marks the BHS, is indicated by a dashed line. The values for the impact parameters along the x- and y-axes are expressed in terms of $R_{\rm g}$. The image maps were plotted using a square-root intensity scale.}
\label{fig:sane_jet_25_matrix}
\end{figure*}

\begin{figure*}
\centering
\begin{subfigure}[b]{0.197\textwidth}
	\includegraphics[width=\textwidth]{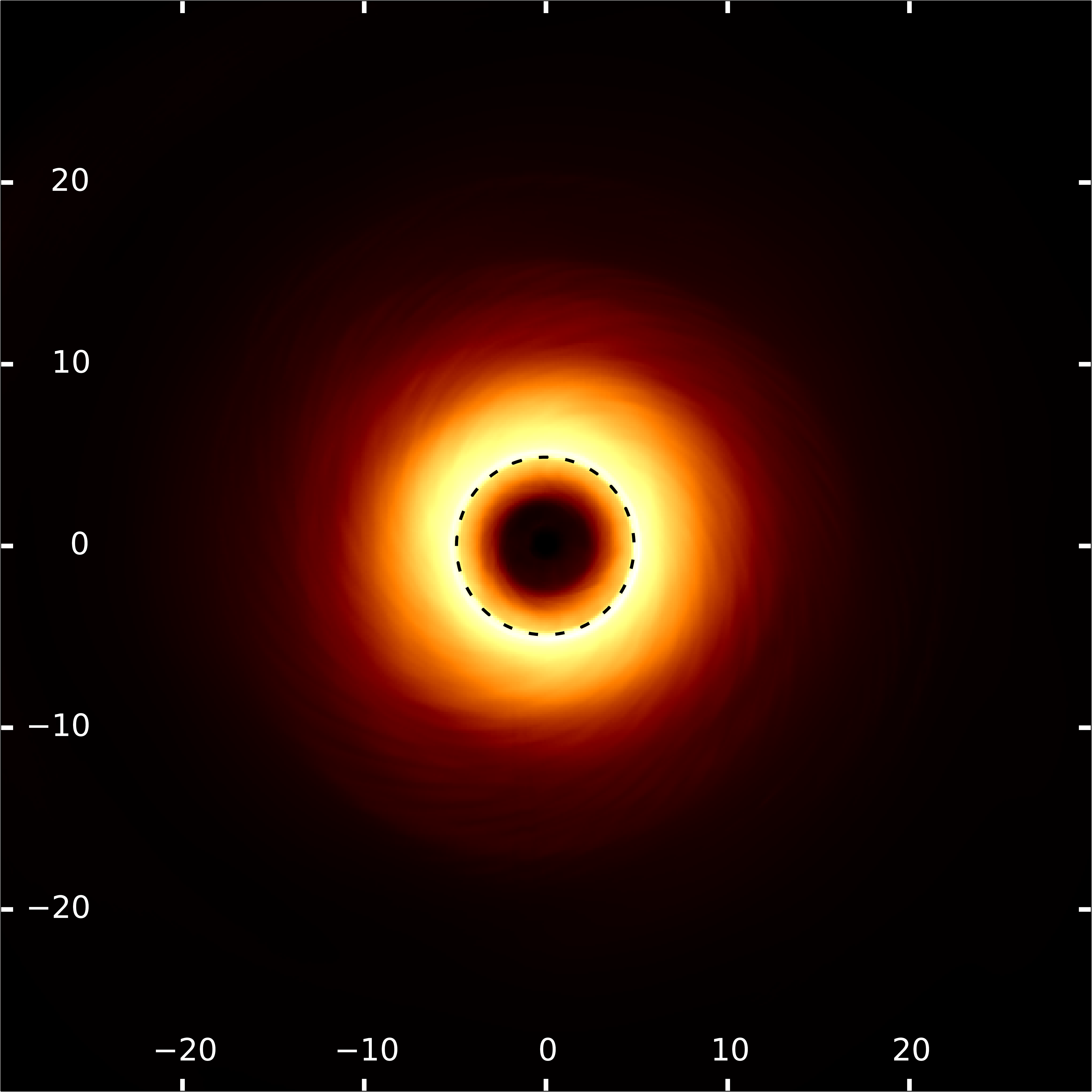}
	\caption{$a=-0.9375$, $i=1^\circ$.}
\end{subfigure}
\begin{subfigure}[b]{0.197\textwidth}
	\includegraphics[width=\textwidth]{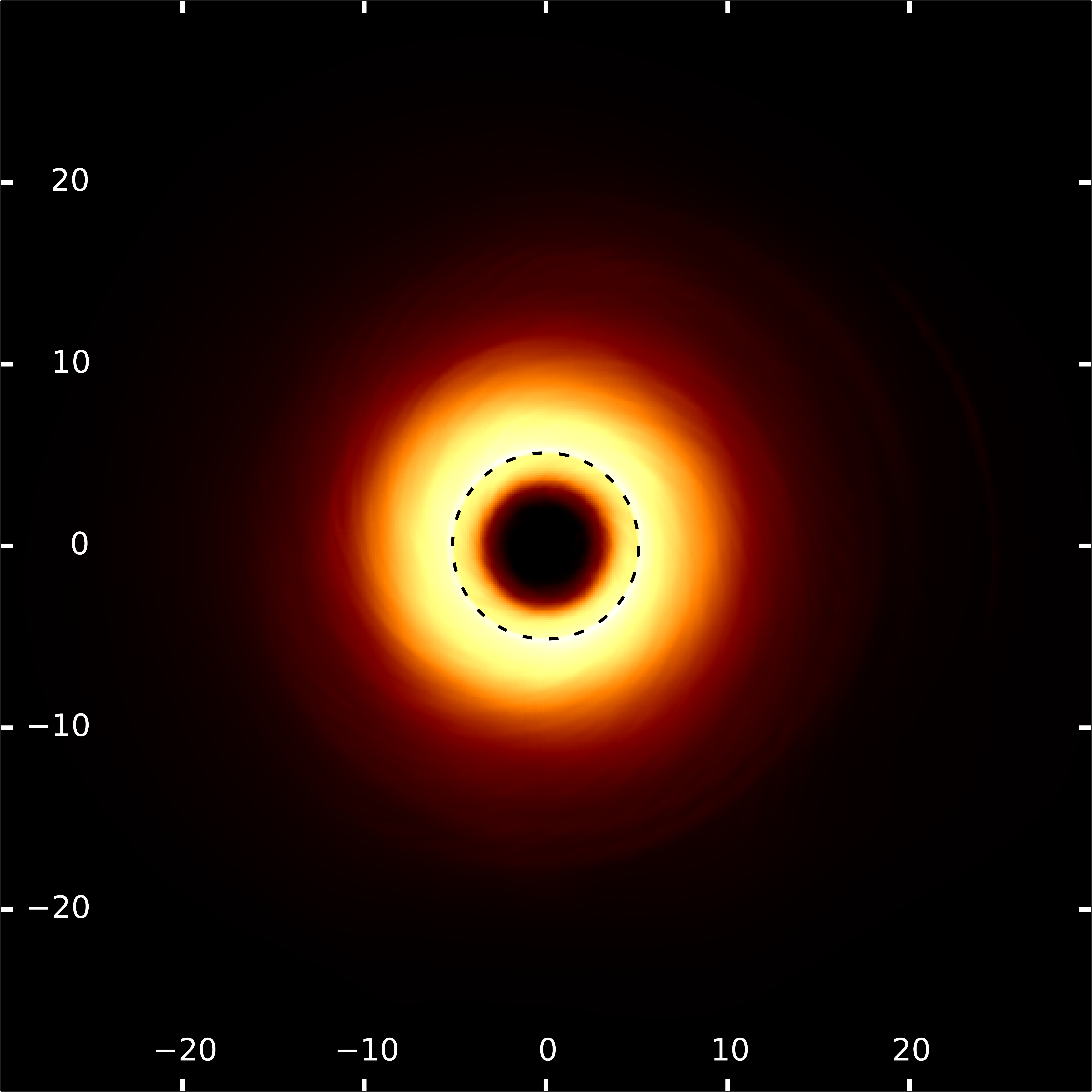}
	\caption{$a=-0.5$, $i=1^\circ$.}
\end{subfigure}
\begin{subfigure}[b]{0.197\textwidth}
	\includegraphics[width=\textwidth]{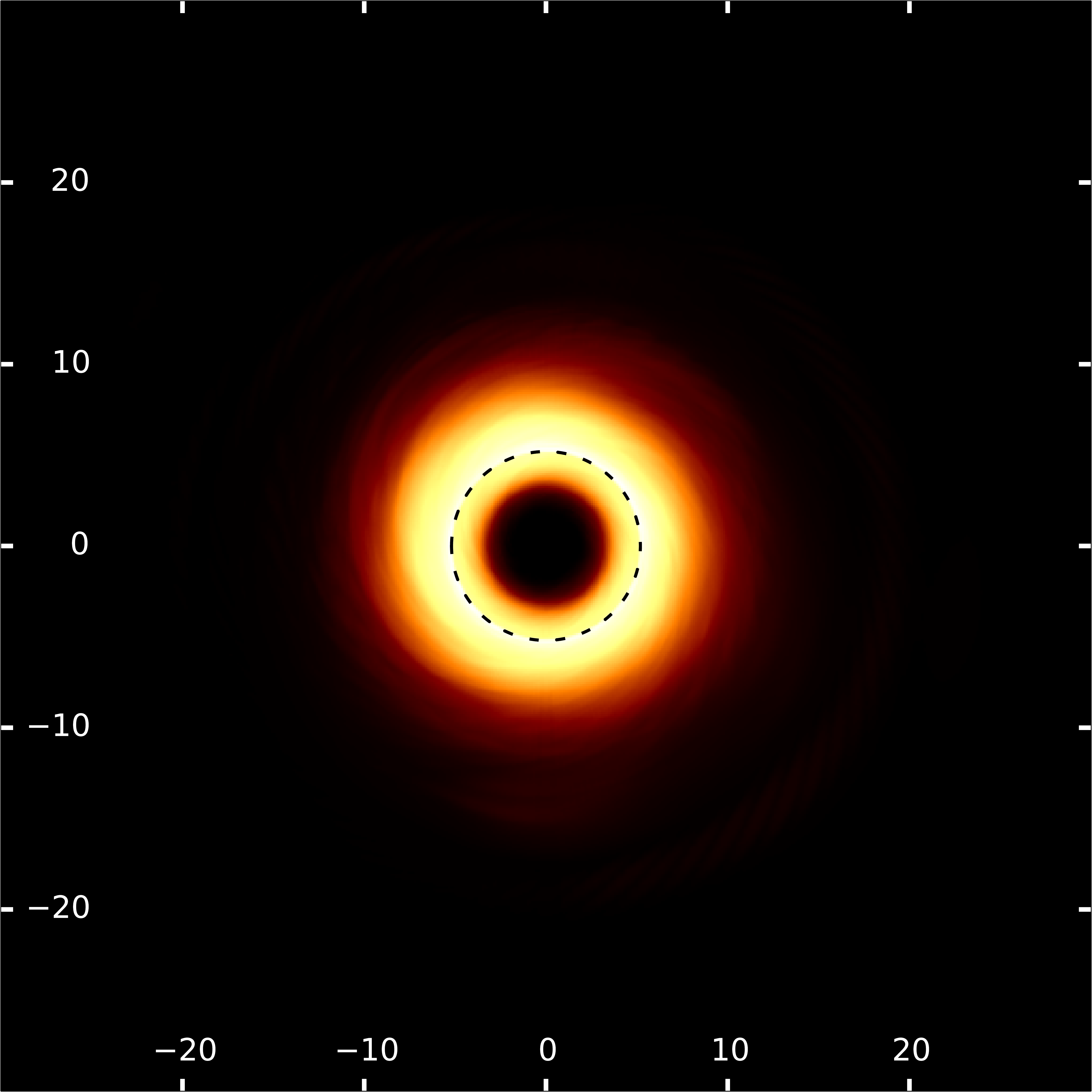}
	\caption{$a=0$, $i=1^\circ$.}
\end{subfigure}
\begin{subfigure}[b]{0.197\textwidth}
	\includegraphics[width=\textwidth]{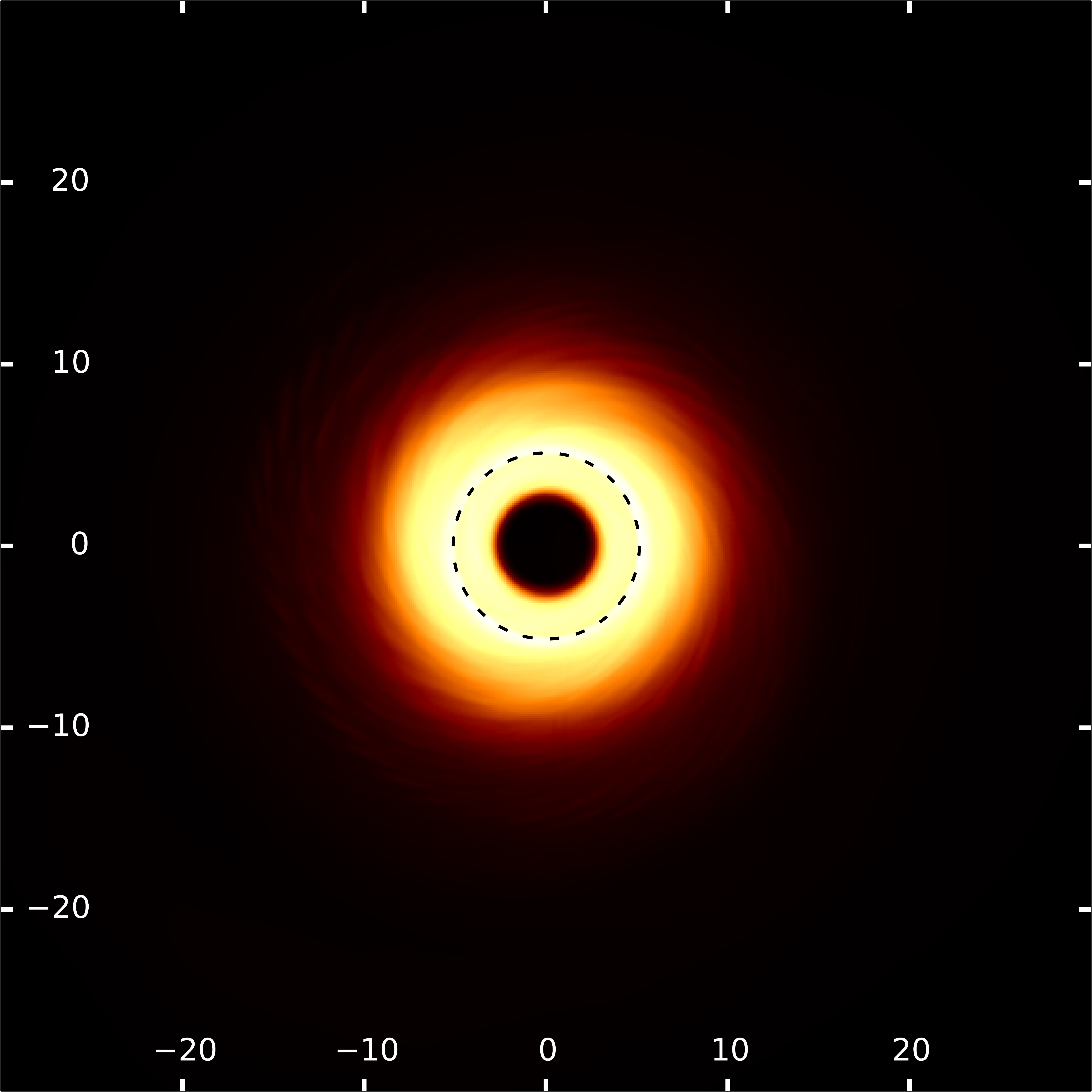}
	\caption{$a=0.5$, $i=1^\circ$.}
\end{subfigure}
\begin{subfigure}[b]{0.197\textwidth}
	\includegraphics[width=\textwidth]{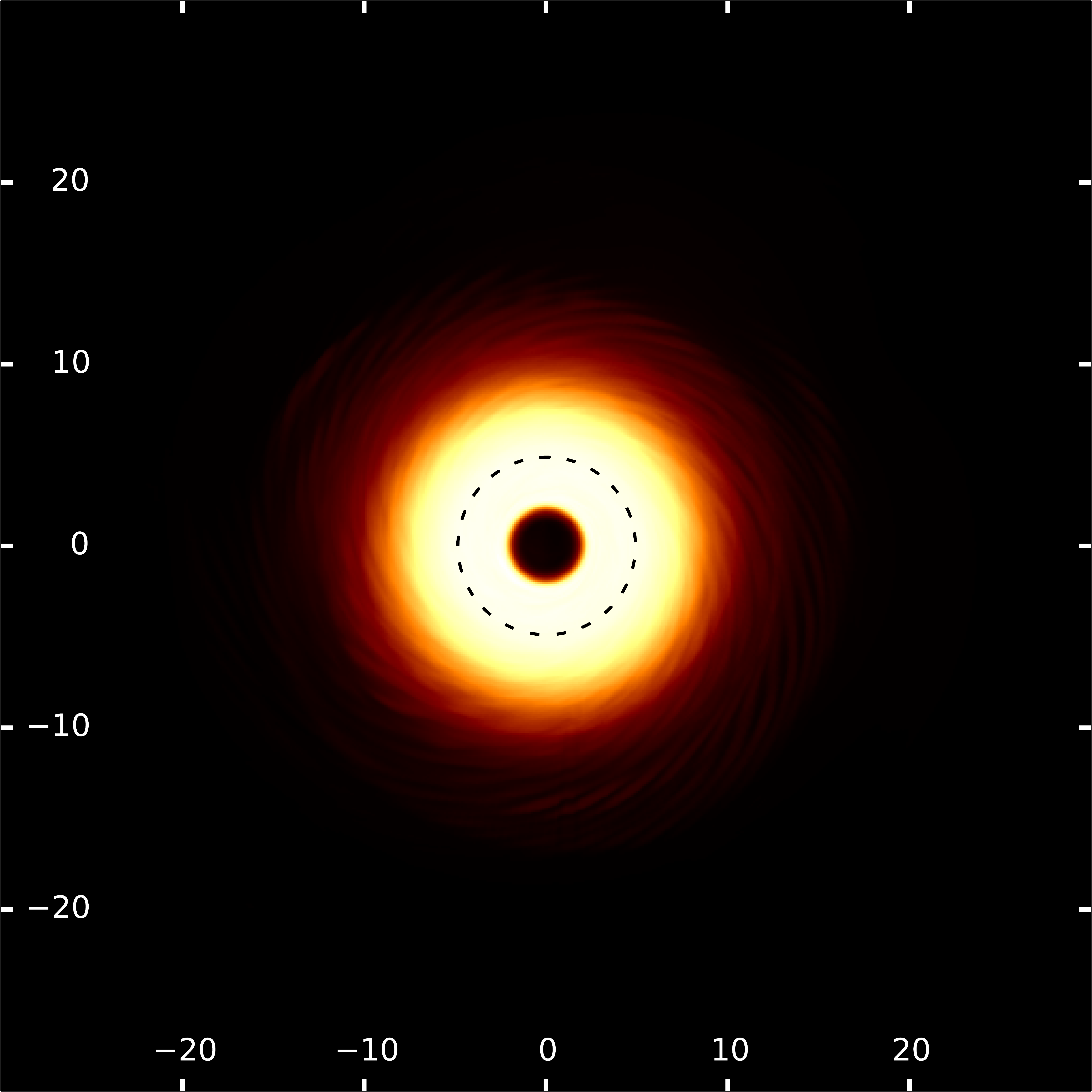}
	\caption{$a=0.9375$, $i=1^\circ$.}
\end{subfigure}
\begin{subfigure}[b]{0.197\textwidth}
	\includegraphics[width=\textwidth]{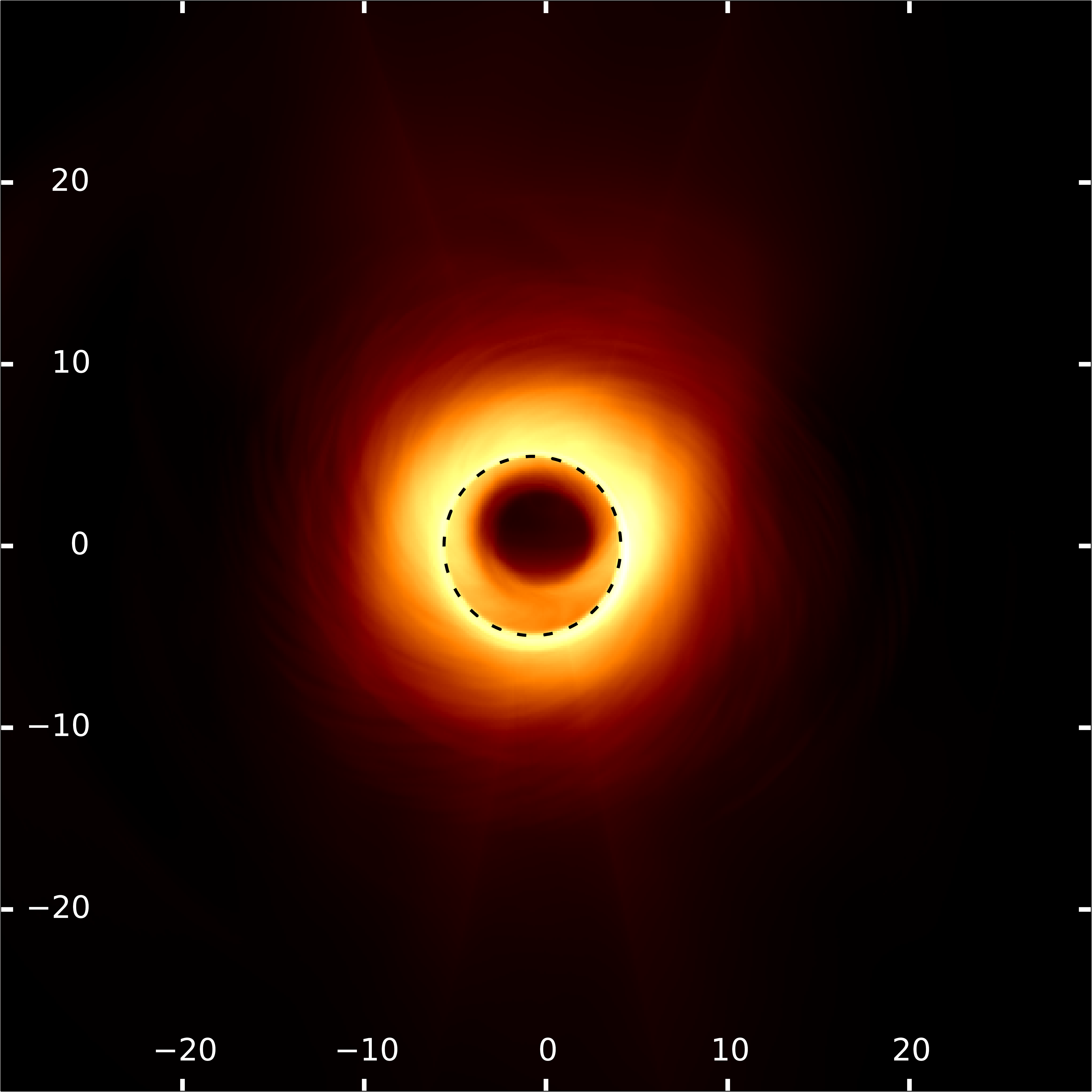}
	\caption{$a=-0.9375$, $i=20^\circ$.}
\end{subfigure}
\begin{subfigure}[b]{0.197\textwidth}
	\includegraphics[width=\textwidth]{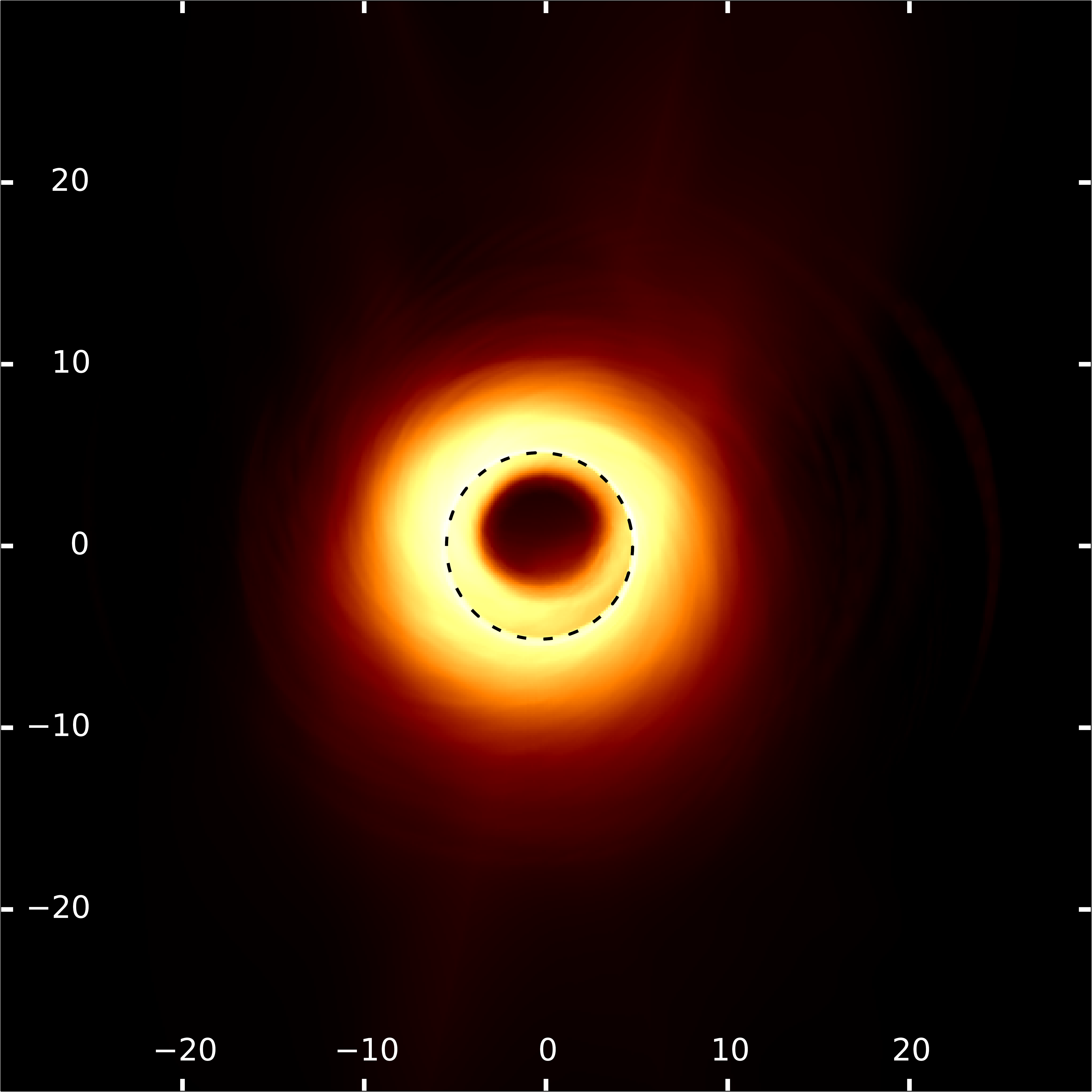}
	\caption{$a=-0.5$, $i=20^\circ$.}
\end{subfigure}
\begin{subfigure}[b]{0.197\textwidth}
	\includegraphics[width=\textwidth]{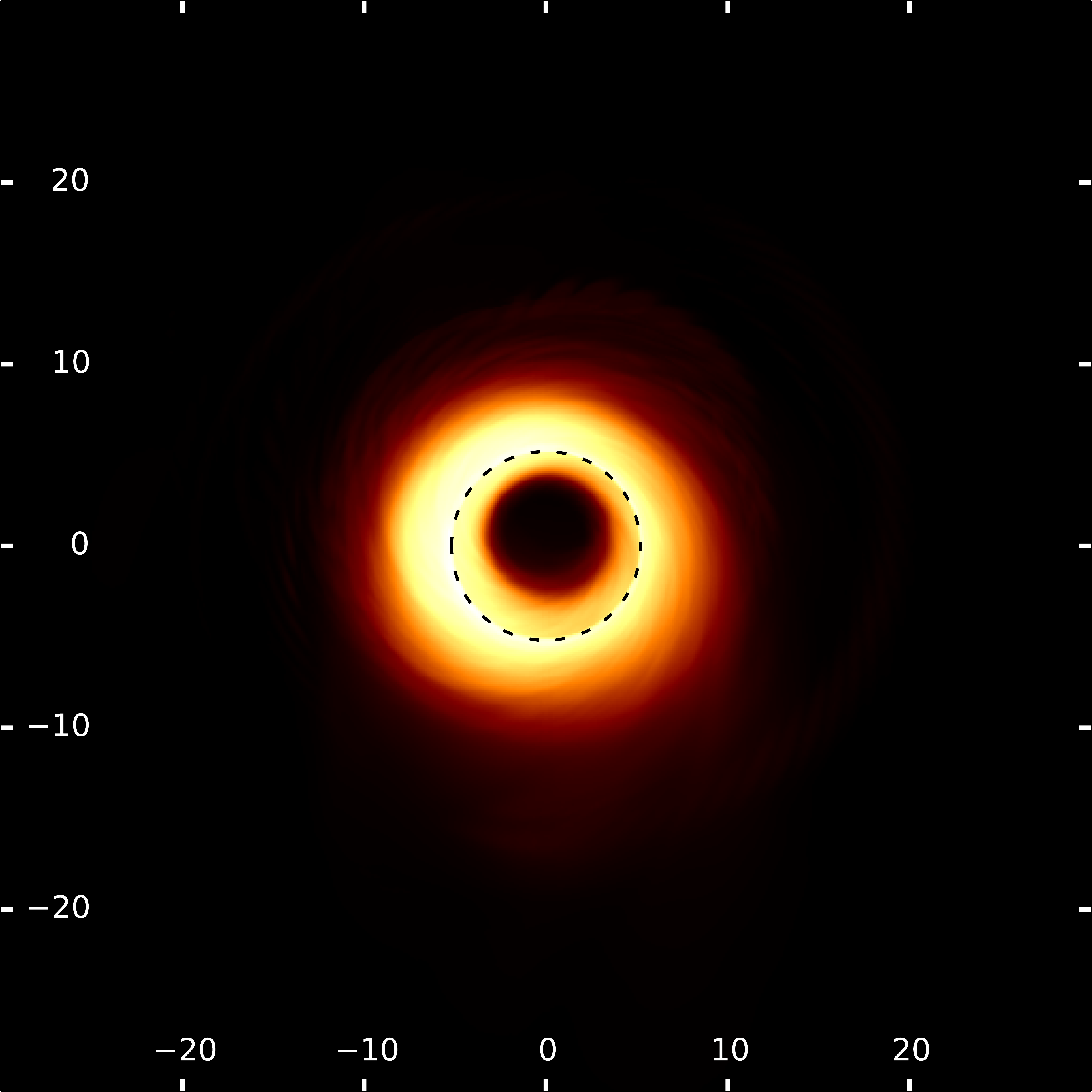}
	\caption{$a=0$, $i=20^\circ$.}
\end{subfigure}
\begin{subfigure}[b]{0.197\textwidth}
	\includegraphics[width=\textwidth]{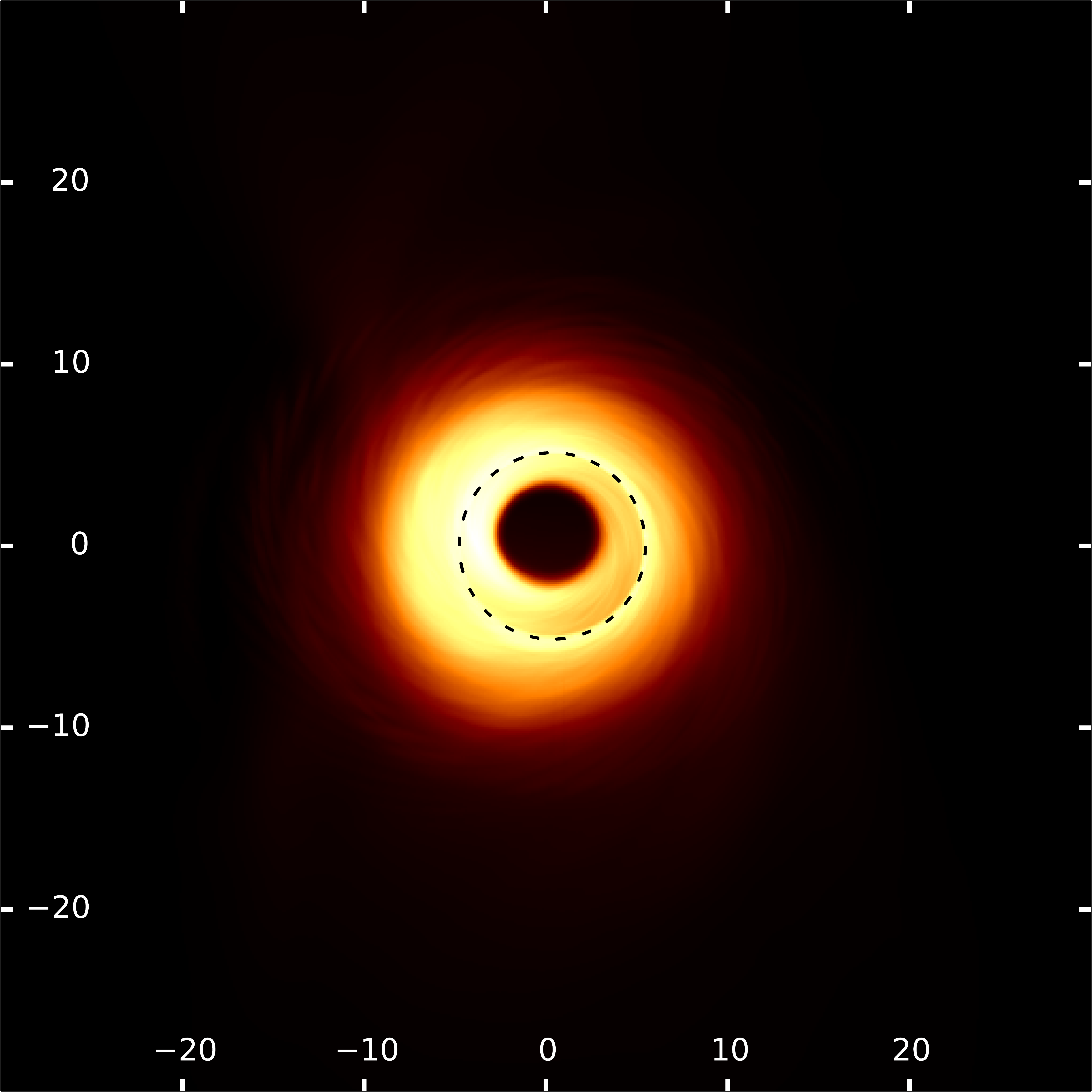}
	\caption{$a=0.5$, $i=20^\circ$.}
\end{subfigure}
\begin{subfigure}[b]{0.197\textwidth}
	\includegraphics[width=\textwidth]{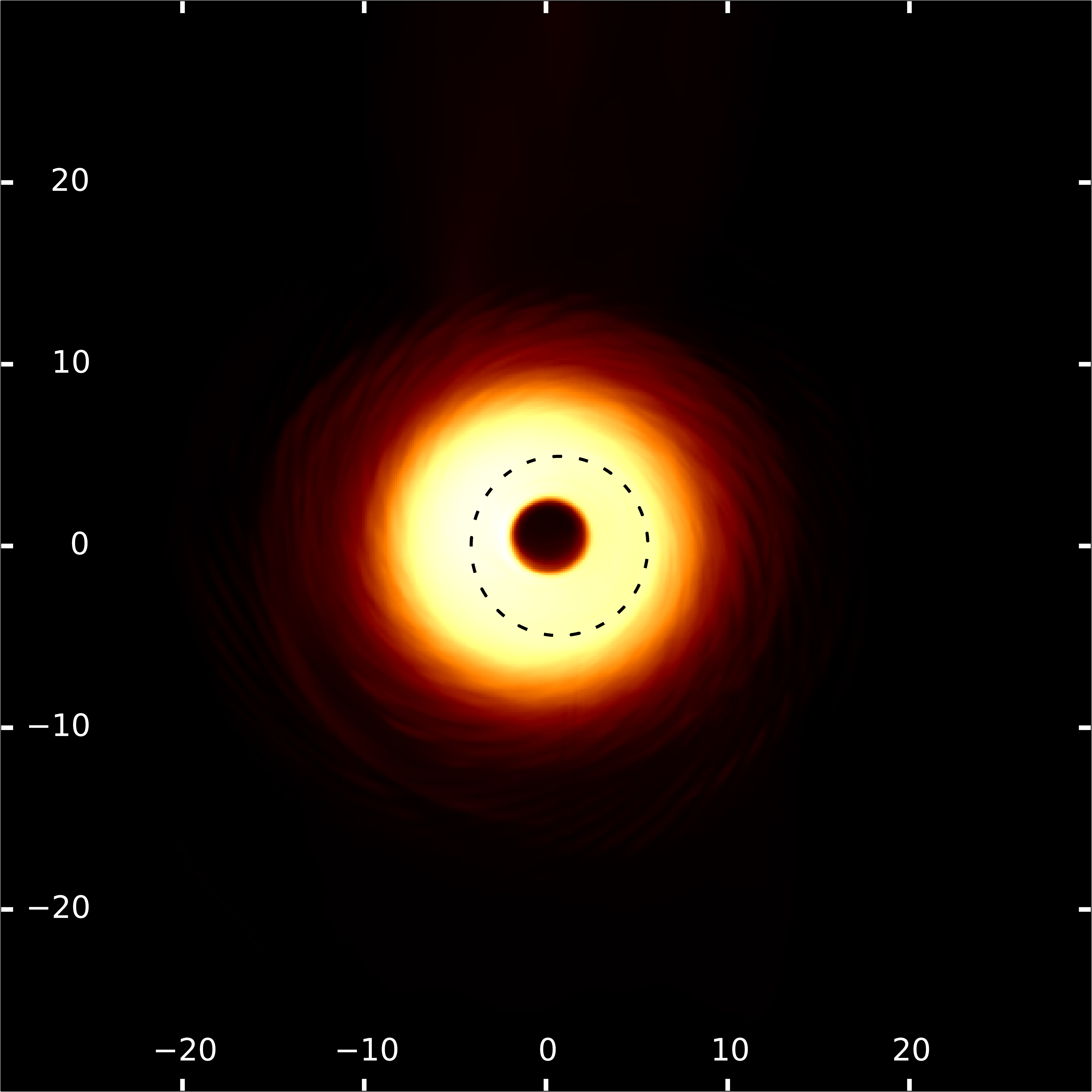}
	\caption{$a=0.9375$, $i=20^\circ$.}
\end{subfigure}
\begin{subfigure}[b]{0.197\textwidth}
	\includegraphics[width=\textwidth]{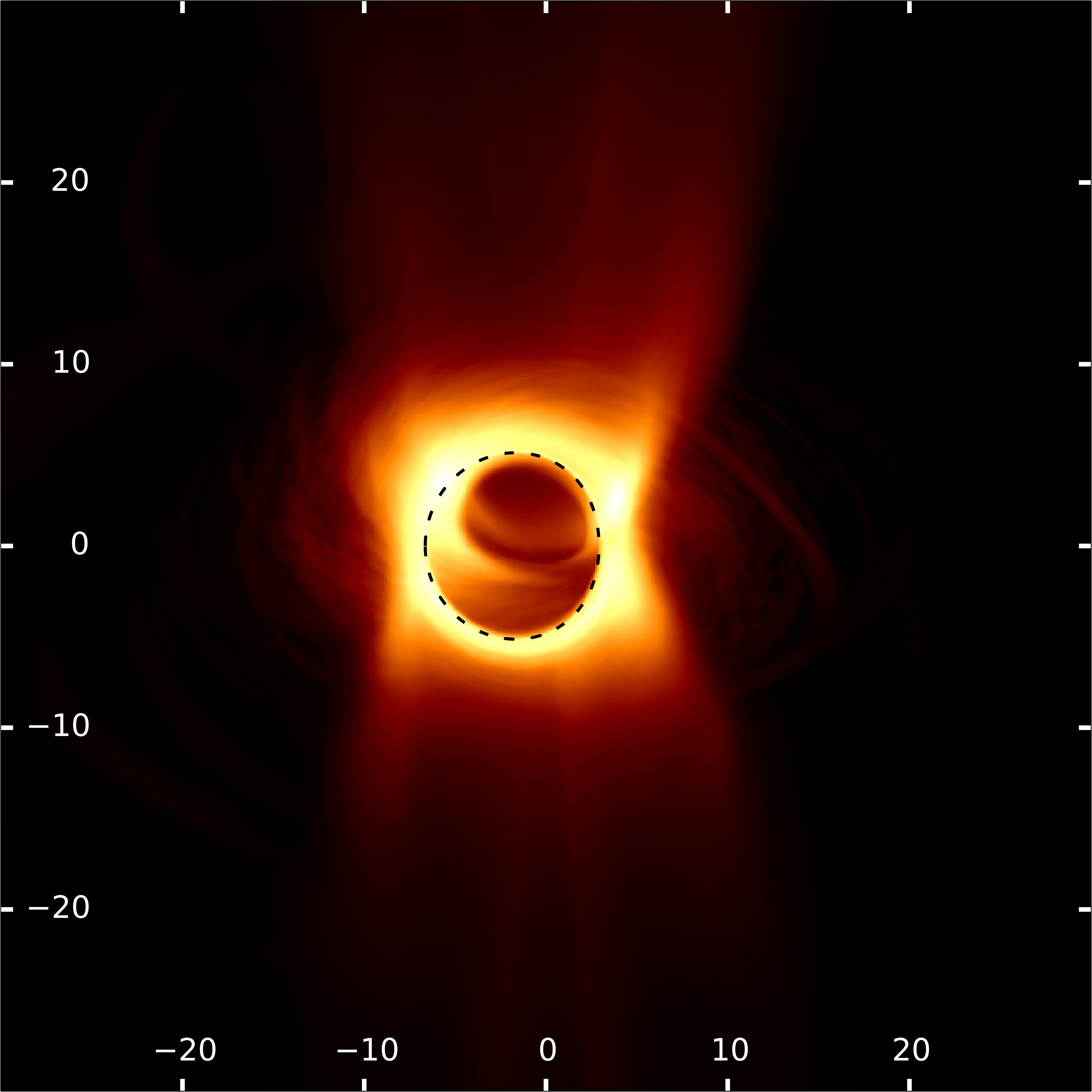}
	\caption{$a=-0.9375$, $i=60^\circ$.}
\end{subfigure}
\begin{subfigure}[b]{0.197\textwidth}
	\includegraphics[width=\textwidth]{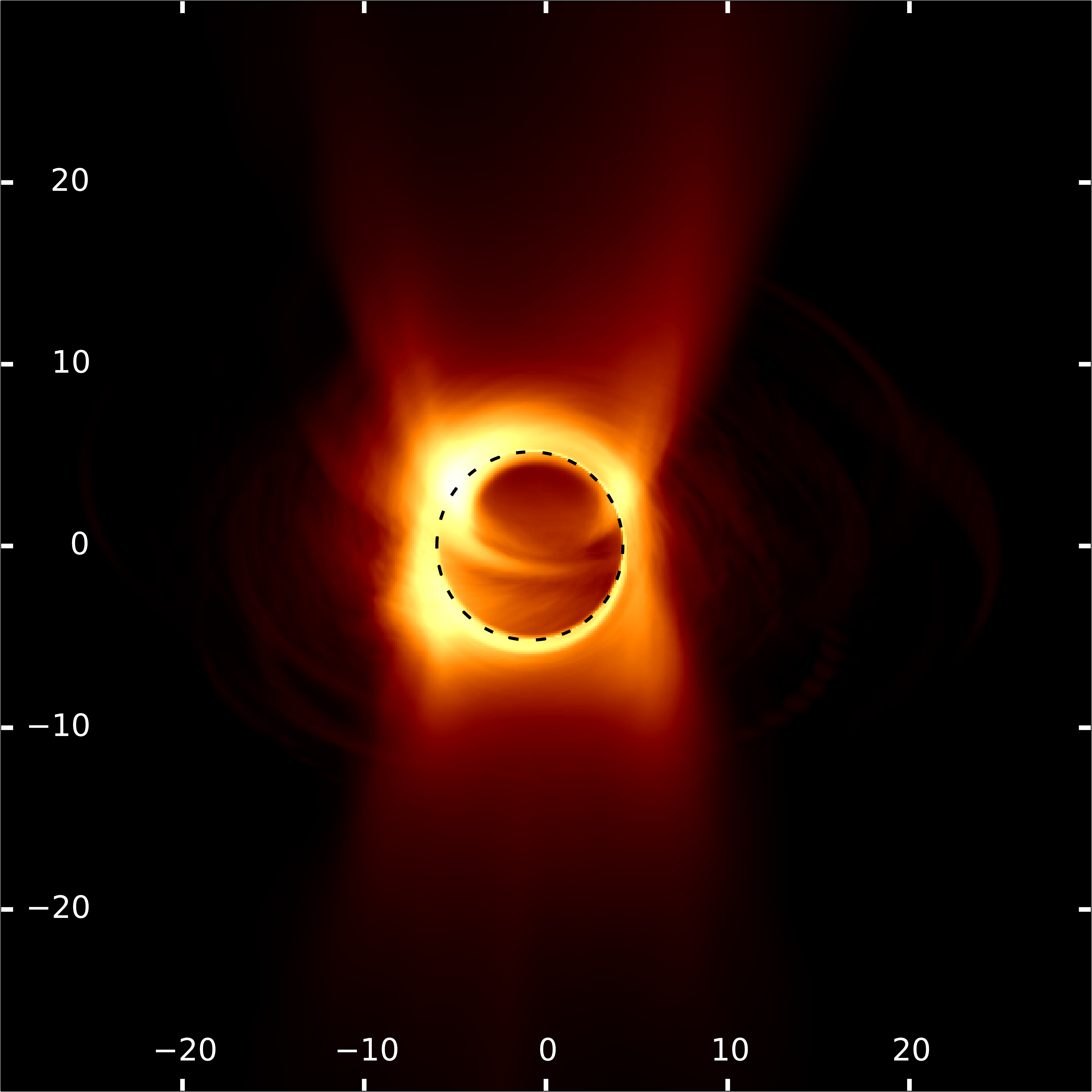}
	\caption{$a=-0.5$, $i=60^\circ$.}
\end{subfigure}
\begin{subfigure}[b]{0.197\textwidth}
	\includegraphics[width=\textwidth]{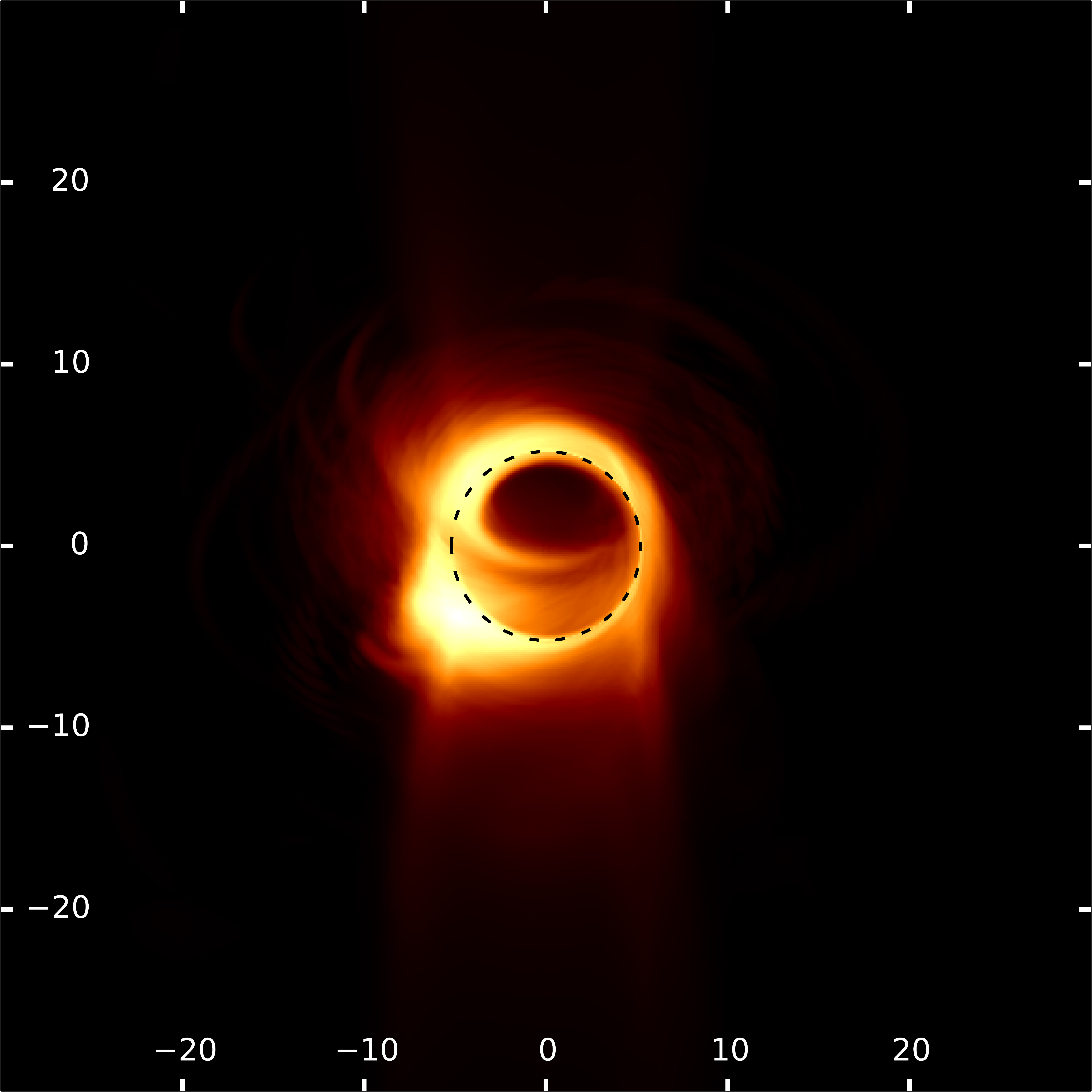}
	\caption{$a=0$, $i=60^\circ$.}
\end{subfigure}
\begin{subfigure}[b]{0.197\textwidth}
	\includegraphics[width=\textwidth]{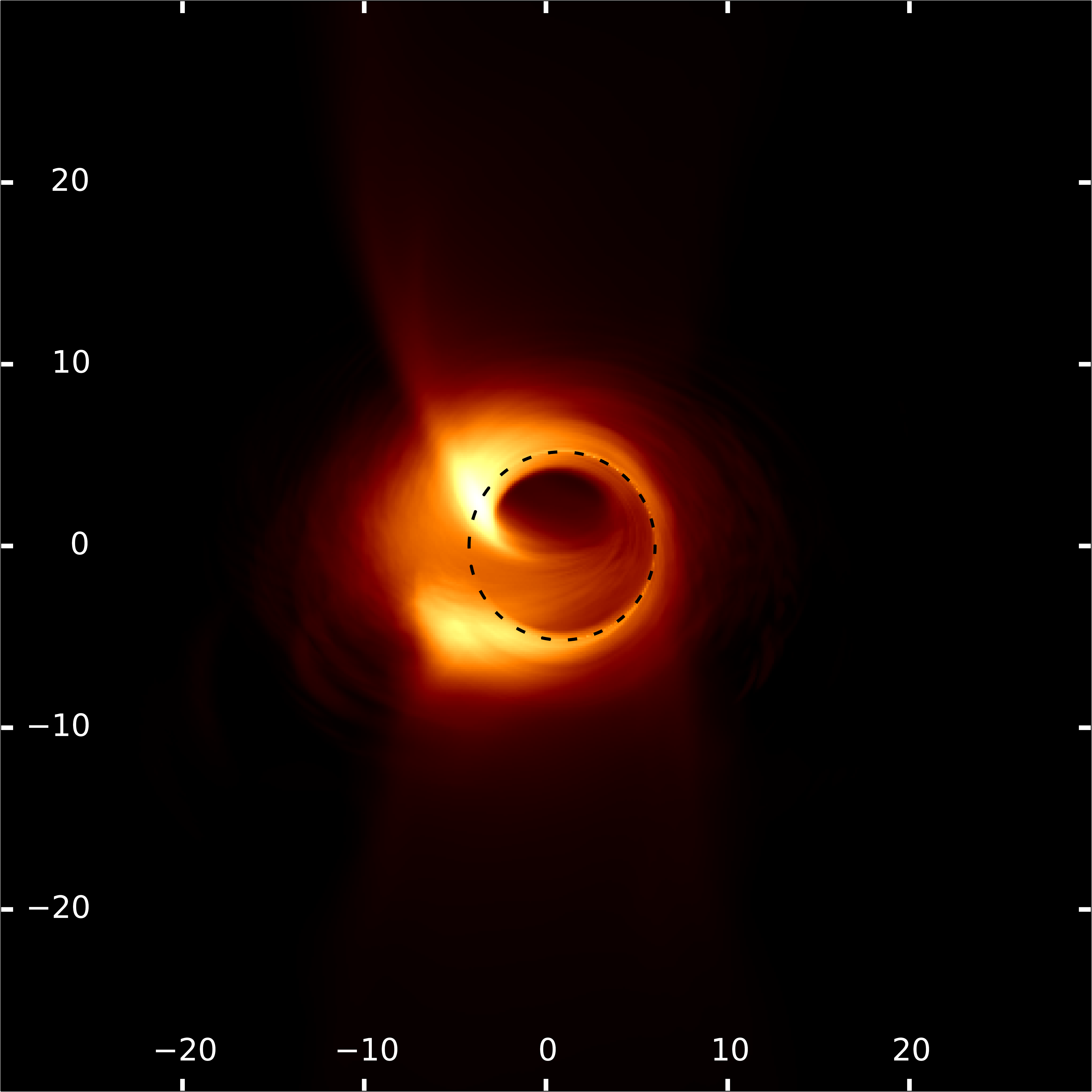}
	\caption{$a=0.5$, $i=60^\circ$.}
\end{subfigure}
\begin{subfigure}[b]{0.197\textwidth}
	\includegraphics[width=\textwidth]{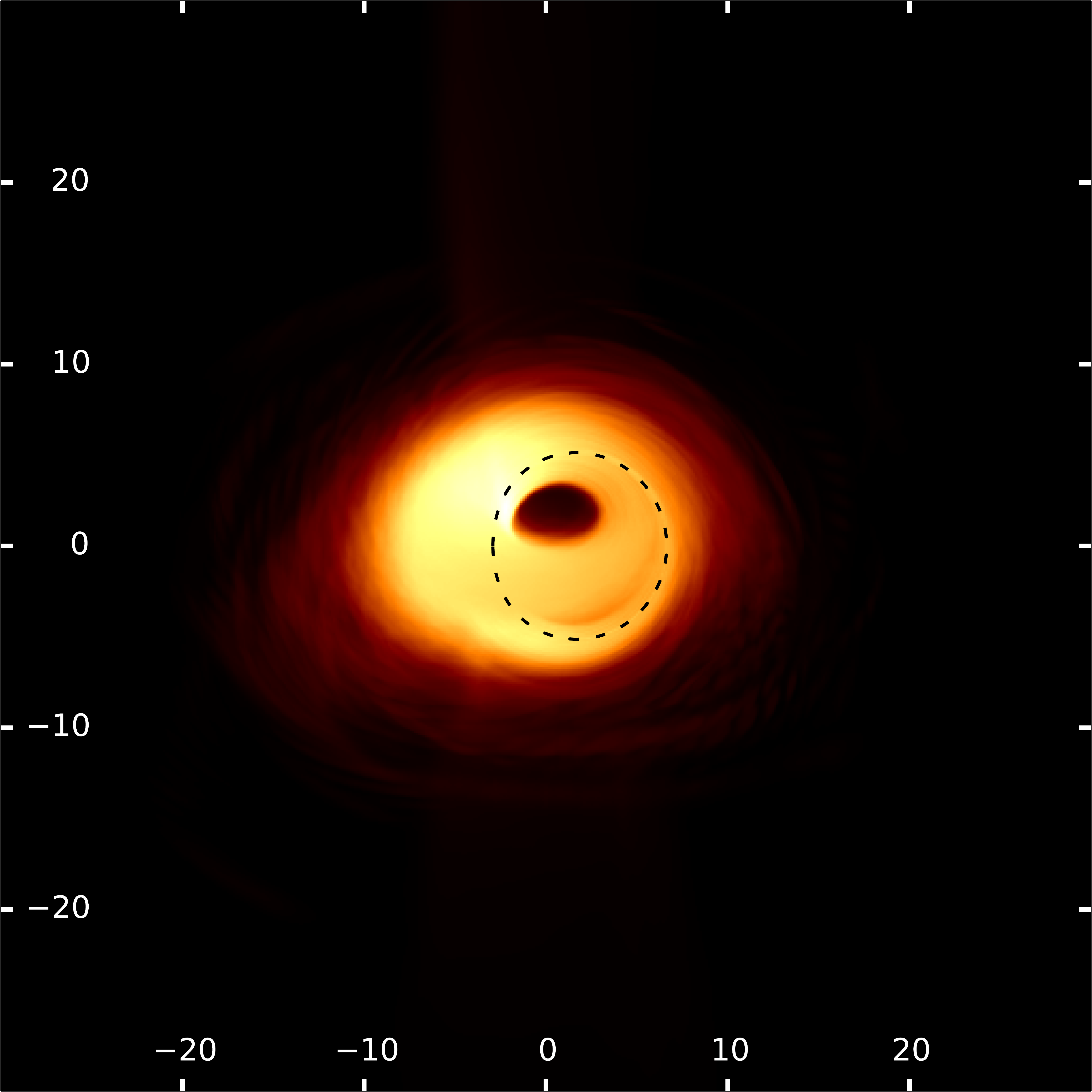}
	\caption{$a=0.9375$, $i=60^\circ$.}
\end{subfigure}
\begin{subfigure}[b]{0.197\textwidth}
	\includegraphics[width=\textwidth]{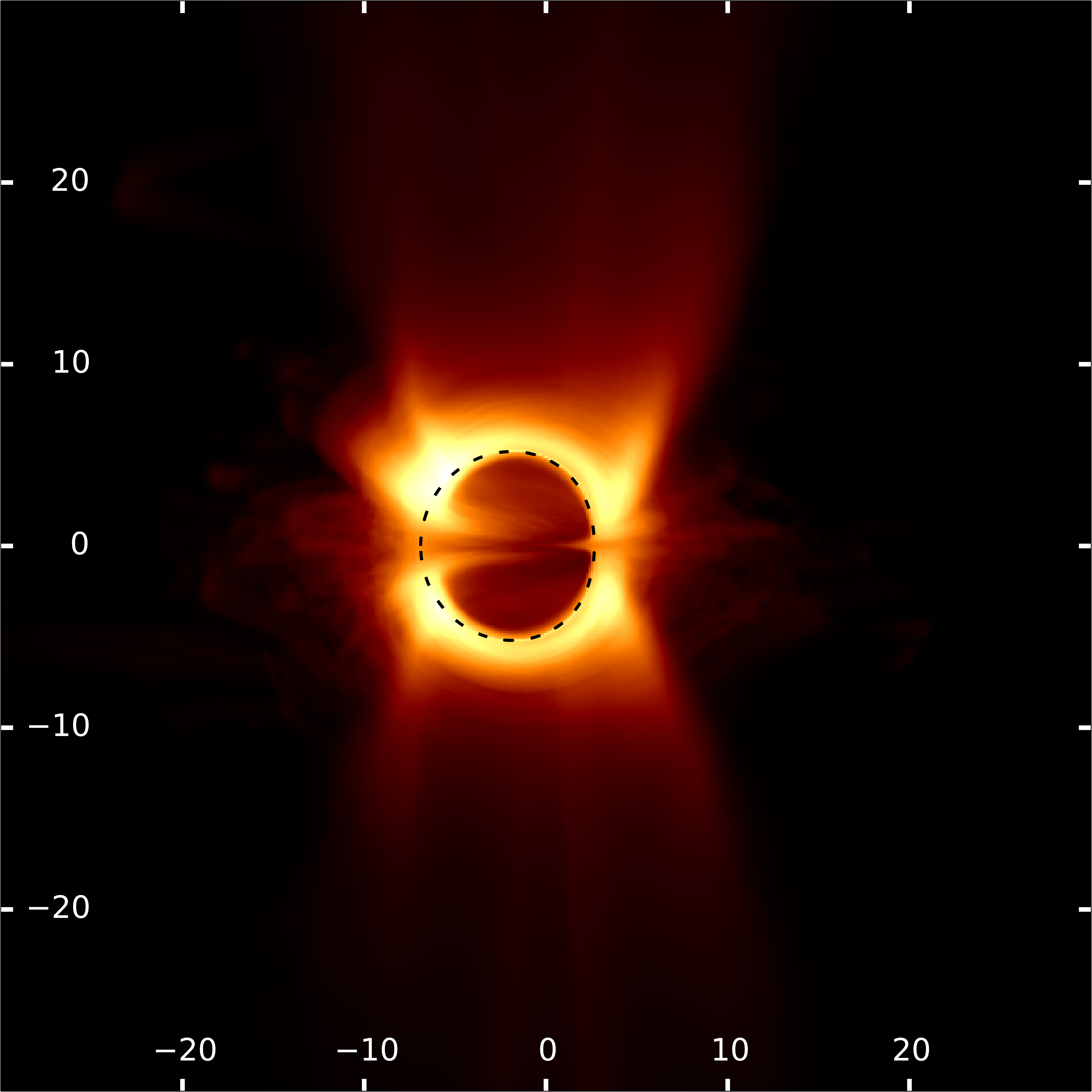}
	\caption{$a=-0.9375$, $i=90^\circ$.}
\end{subfigure}
\begin{subfigure}[b]{0.197\textwidth}
	\includegraphics[width=\textwidth]{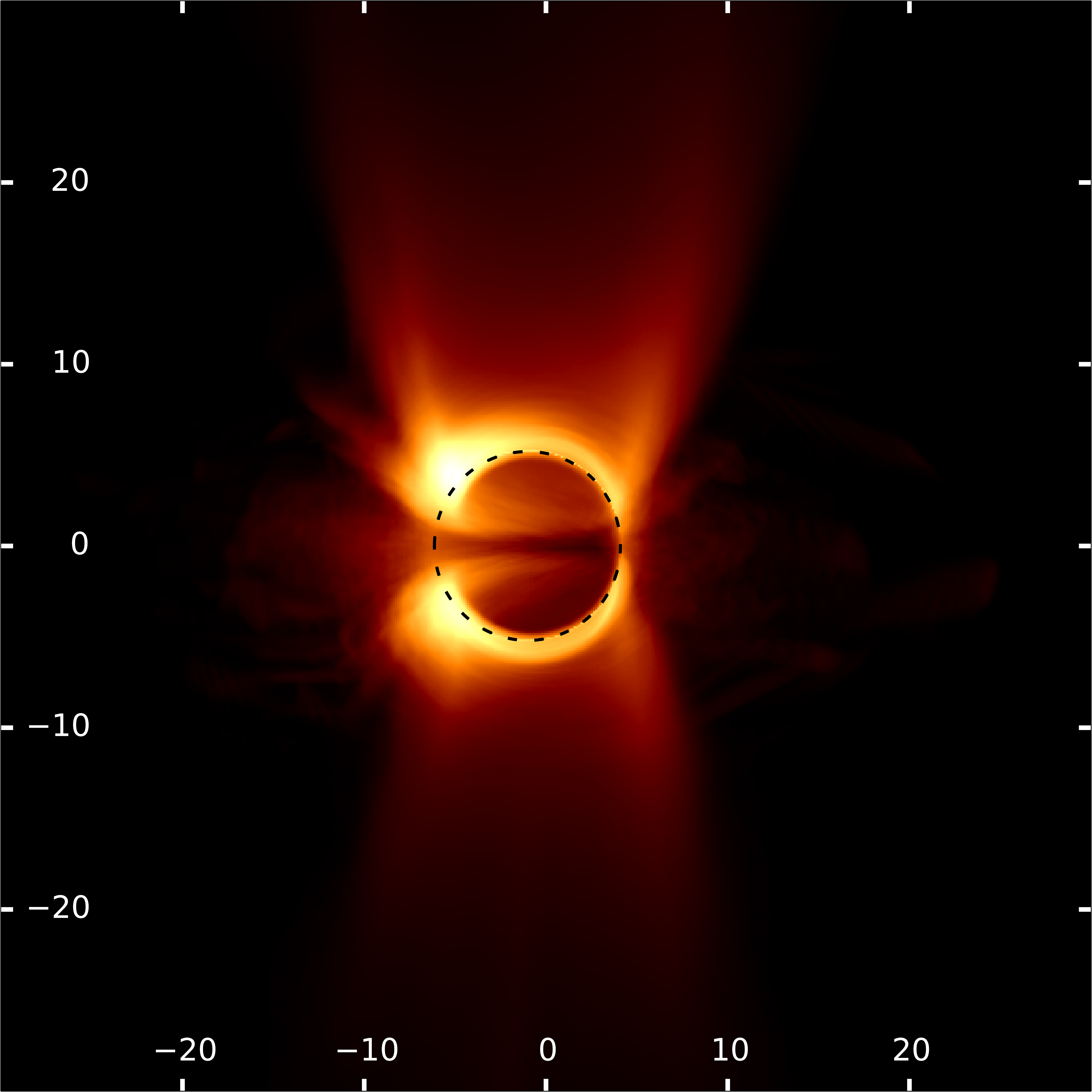}
	\caption{$a=-0.5$, $i=90^\circ$.}
\end{subfigure}
\begin{subfigure}[b]{0.197\textwidth}
	\includegraphics[width=\textwidth]{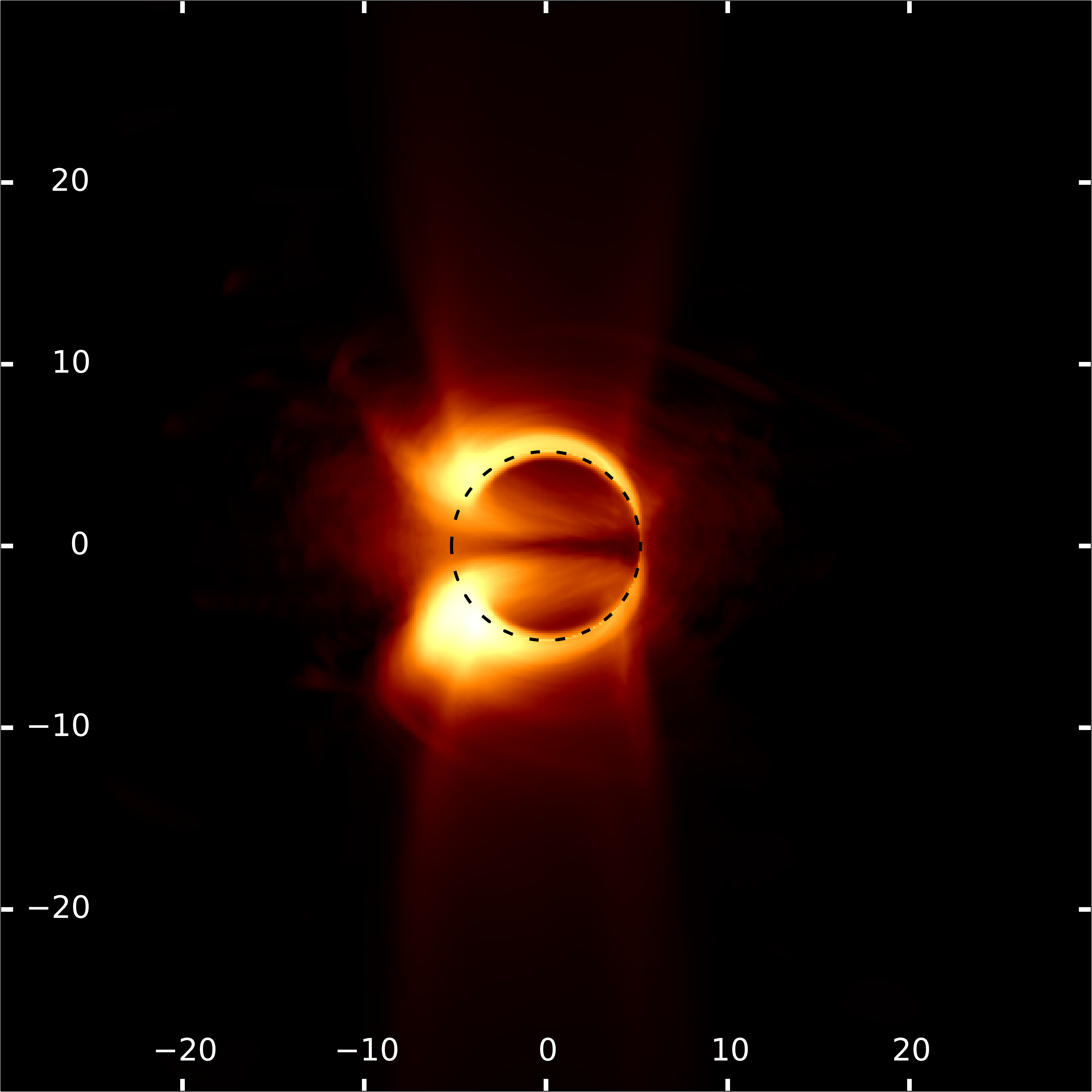}
	\caption{$a=0$, $i=90^\circ$.}
\end{subfigure}
\begin{subfigure}[b]{0.197\textwidth}
	\includegraphics[width=\textwidth]{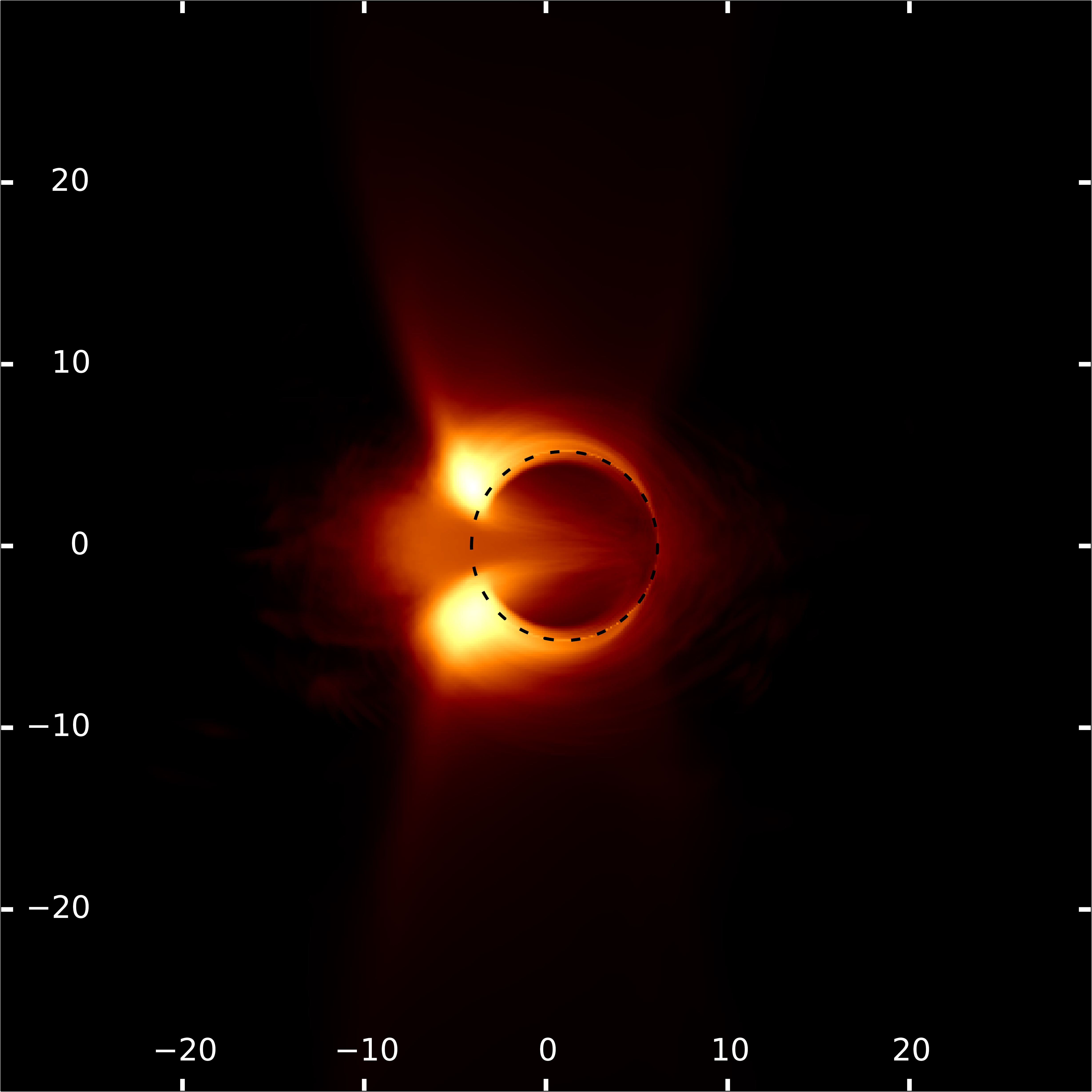}
	\caption{$a=0.5$, $i=90^\circ$.}
\end{subfigure}
\begin{subfigure}[b]{0.197\textwidth}
	\includegraphics[width=\textwidth]{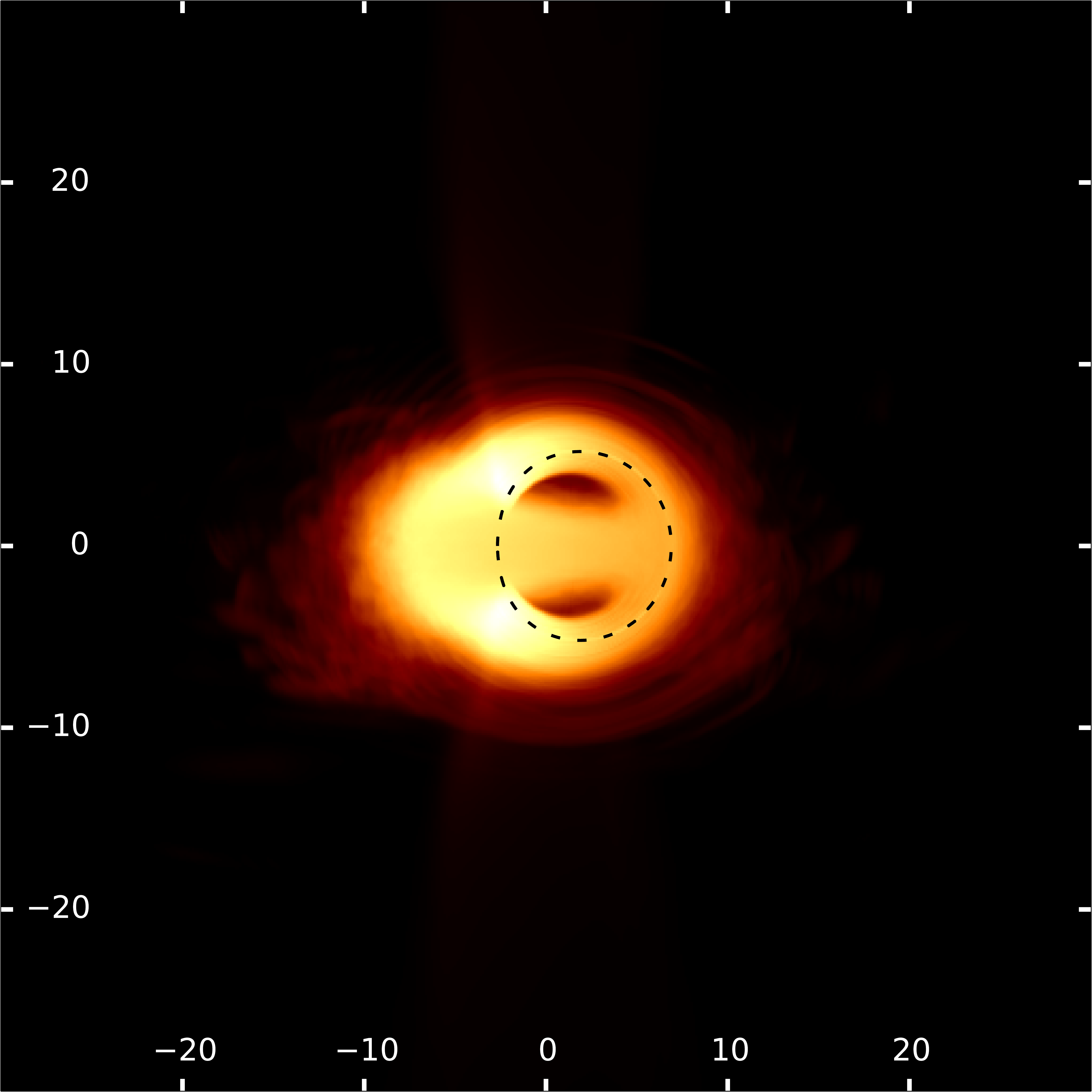}
	\caption{$a=0.9375$, $i=90^\circ$.}
\end{subfigure}
\caption{Time-averaged, normalised intensity maps of our SANE, jet-dominated GRMHD models of Sgr A*, imaged at 230 GHz, at five different spins and four observer inclination angles, with an integrated flux density of 1.25 Jy. In each case, the photon ring, which marks the BHS, is indicated by a dashed line. The values for the impact parameters along the x- and y-axes are expressed in terms of $R_{\rm g}$. The image maps were plotted using a square-root intensity scale.}
\label{fig:sane_jet_125_matrix}
\end{figure*}

\begin{figure*}
\centering
\begin{subfigure}[b]{0.197\textwidth}
	\includegraphics[width=\textwidth]{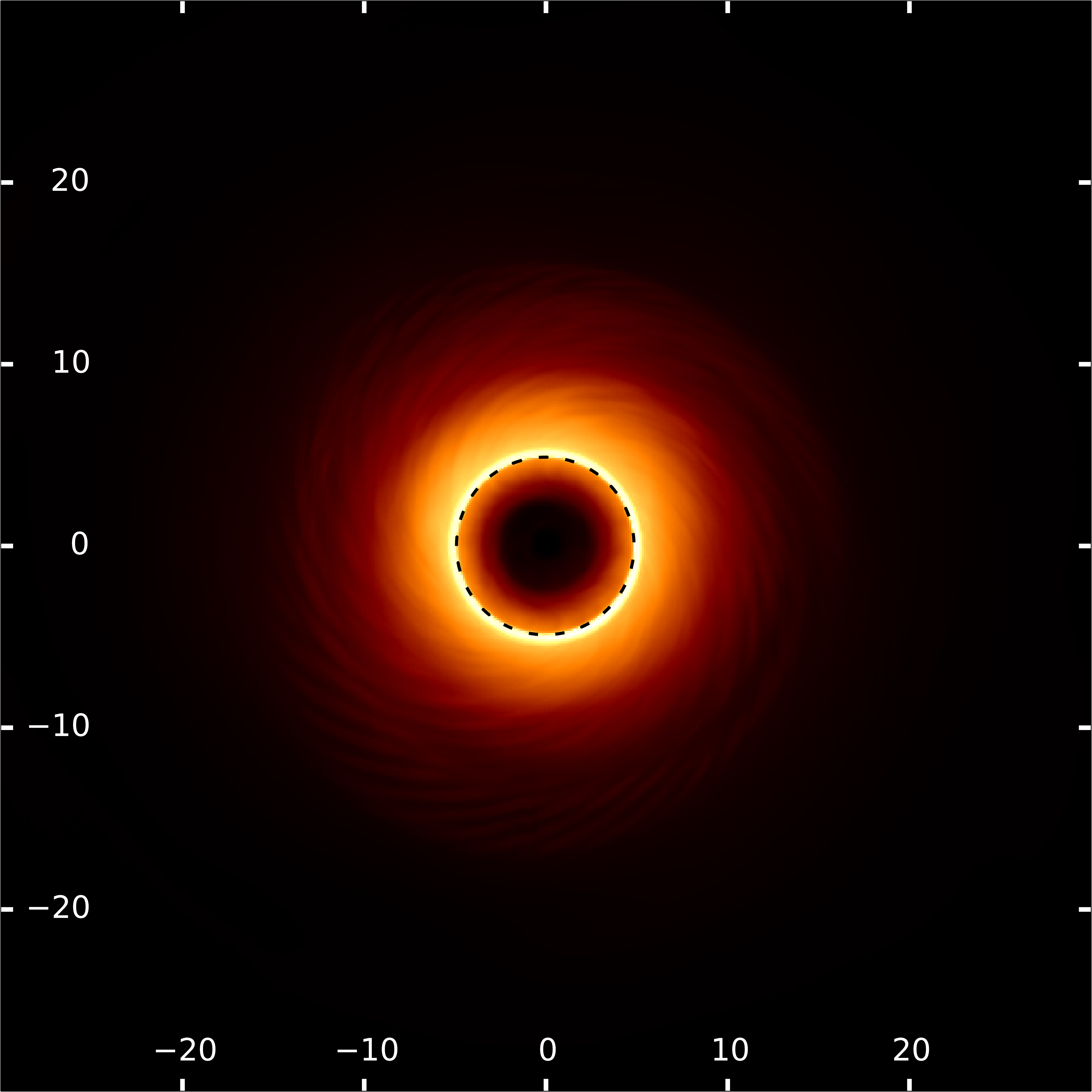}
	\caption{$a=-0.9375$, $i=1^\circ$.}
\end{subfigure}
\begin{subfigure}[b]{0.197\textwidth}
	\includegraphics[width=\textwidth]{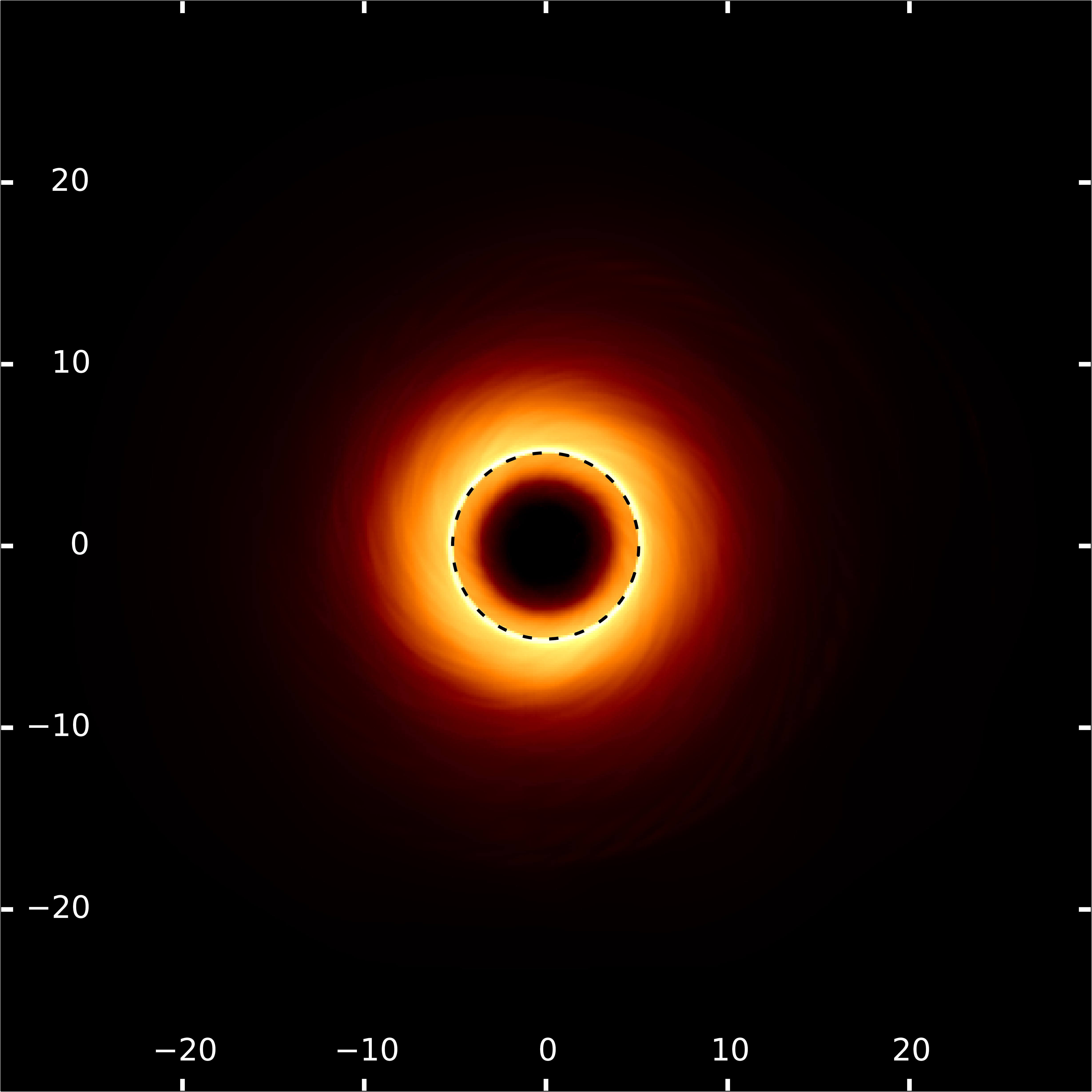}
	\caption{$a=-0.5$, $i=1^\circ$.}
\end{subfigure}
\begin{subfigure}[b]{0.197\textwidth}
	\includegraphics[width=\textwidth]{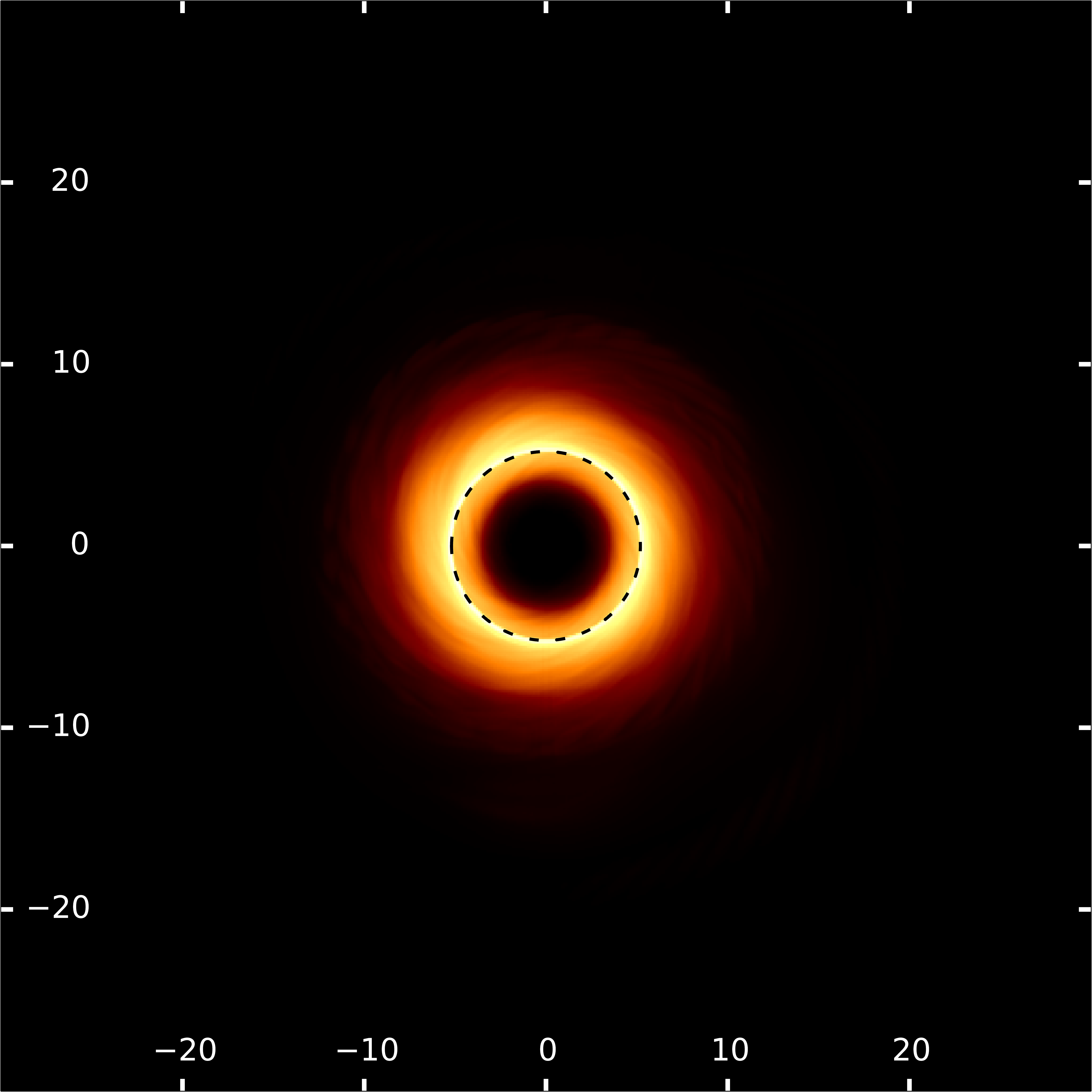}
	\caption{$a=0$, $i=1^\circ$.}
\end{subfigure}
\begin{subfigure}[b]{0.197\textwidth}
	\includegraphics[width=\textwidth]{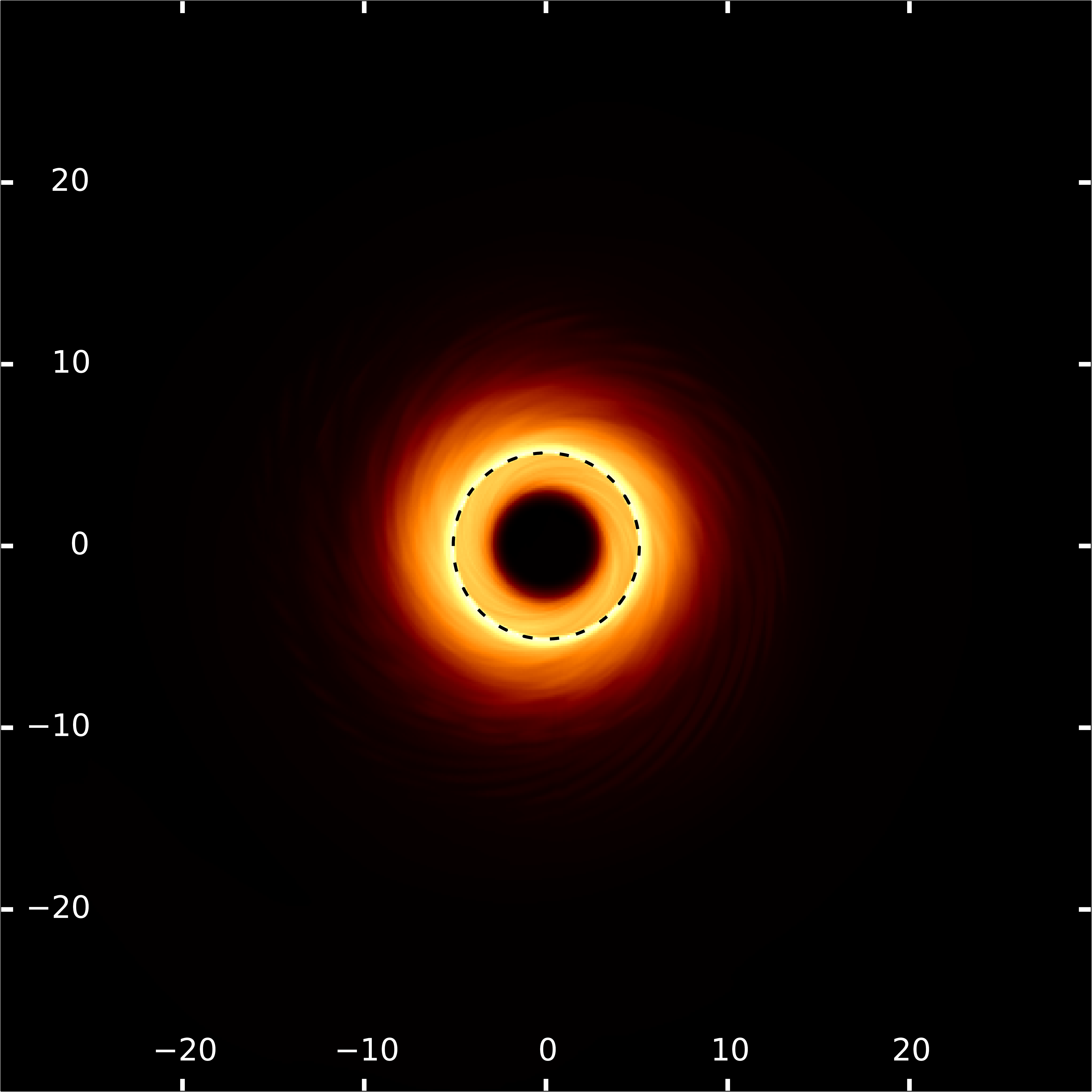}
	\caption{$a=0.5$, $i=1^\circ$.}
\end{subfigure}
\begin{subfigure}[b]{0.197\textwidth}
	\includegraphics[width=\textwidth]{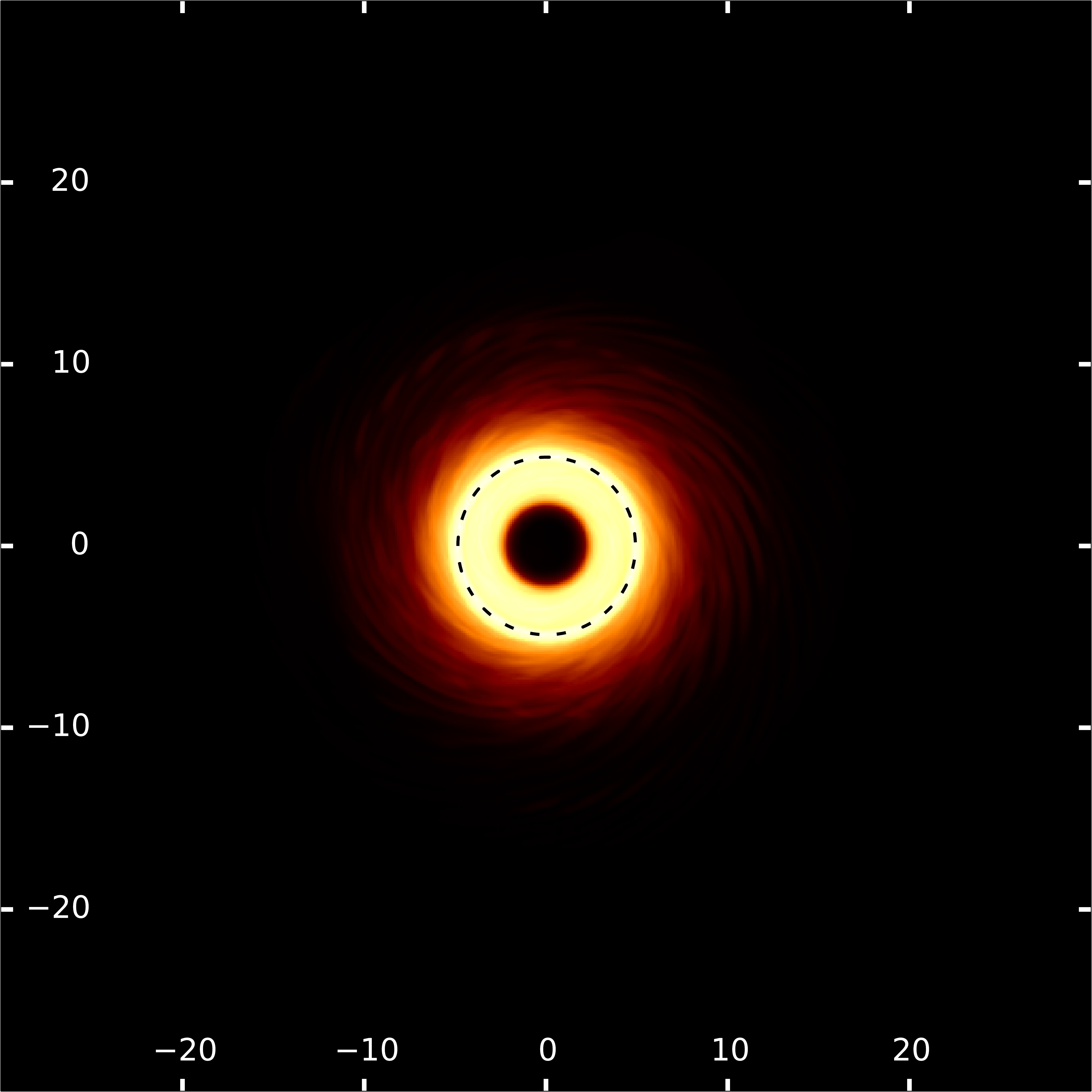}
	\caption{$a=0.9375$, $i=1^\circ$.}
\end{subfigure}
\begin{subfigure}[b]{0.197\textwidth}
	\includegraphics[width=\textwidth]{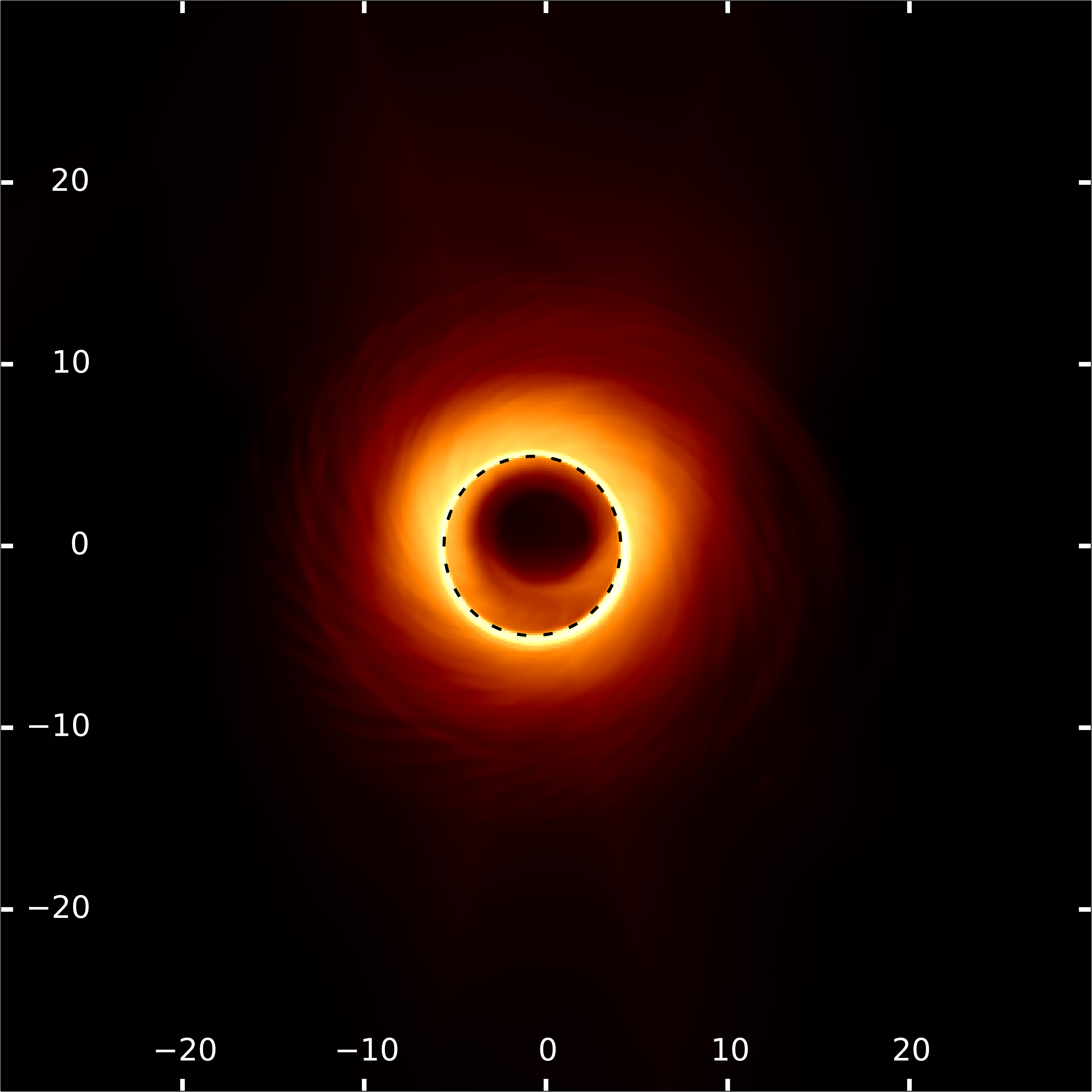}
	\caption{$a=-0.9375$, $i=20^\circ$.}
\end{subfigure}
\begin{subfigure}[b]{0.197\textwidth}
	\includegraphics[width=\textwidth]{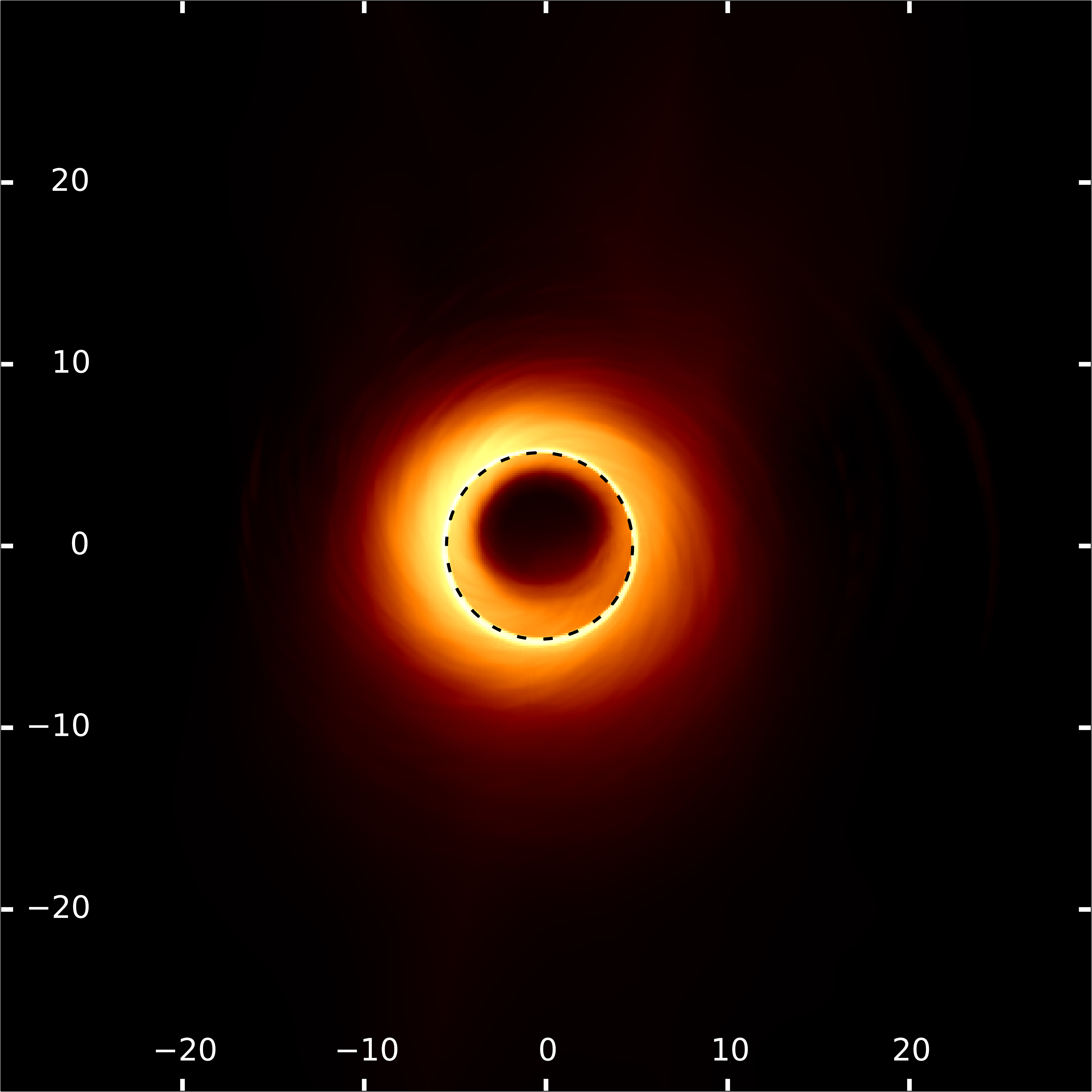}
	\caption{$a=-0.5$, $i=20^\circ$.}
\end{subfigure}
\begin{subfigure}[b]{0.197\textwidth}
	\includegraphics[width=\textwidth]{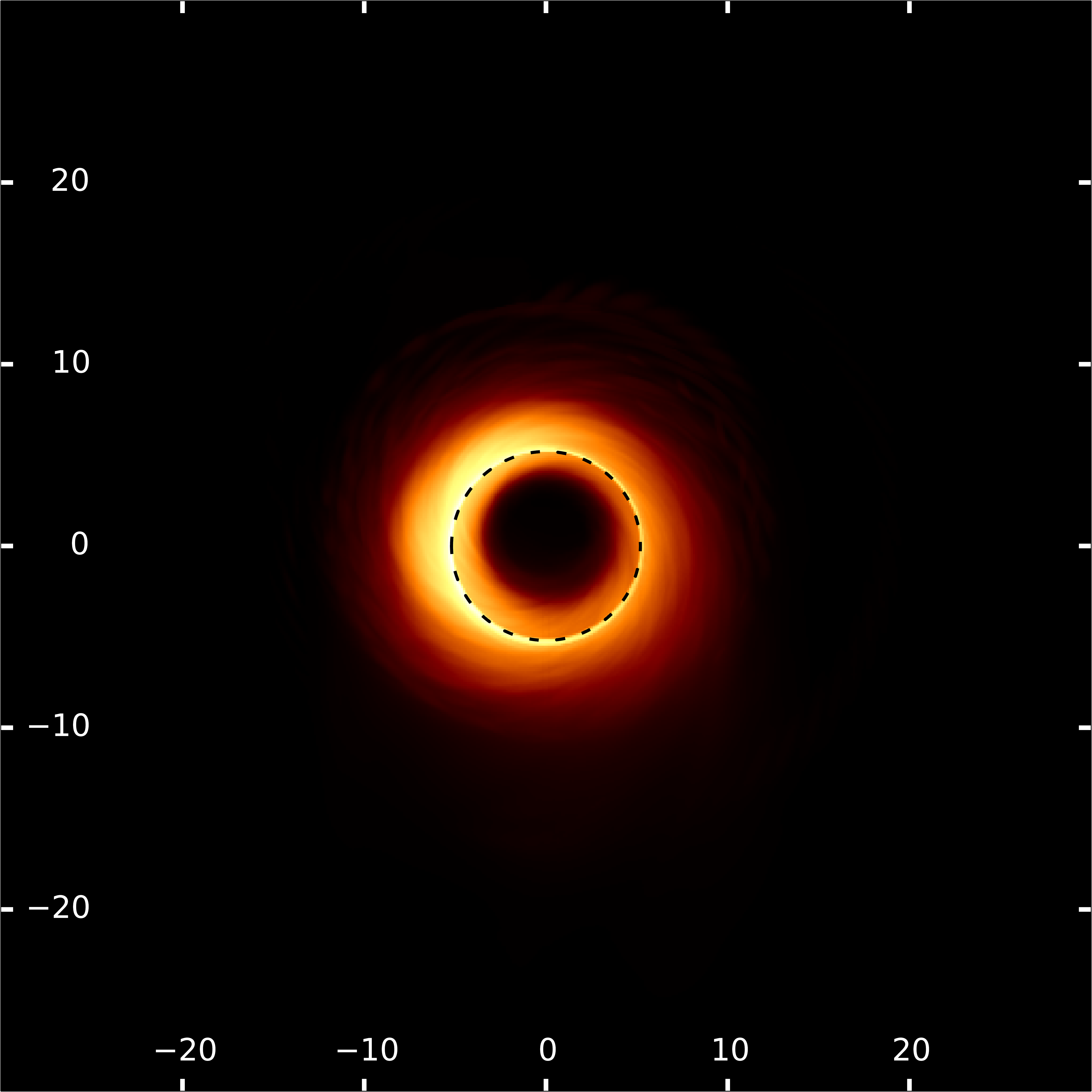}
	\caption{$a=0$, $i=20^\circ$.}
\end{subfigure}
\begin{subfigure}[b]{0.197\textwidth}
	\includegraphics[width=\textwidth]{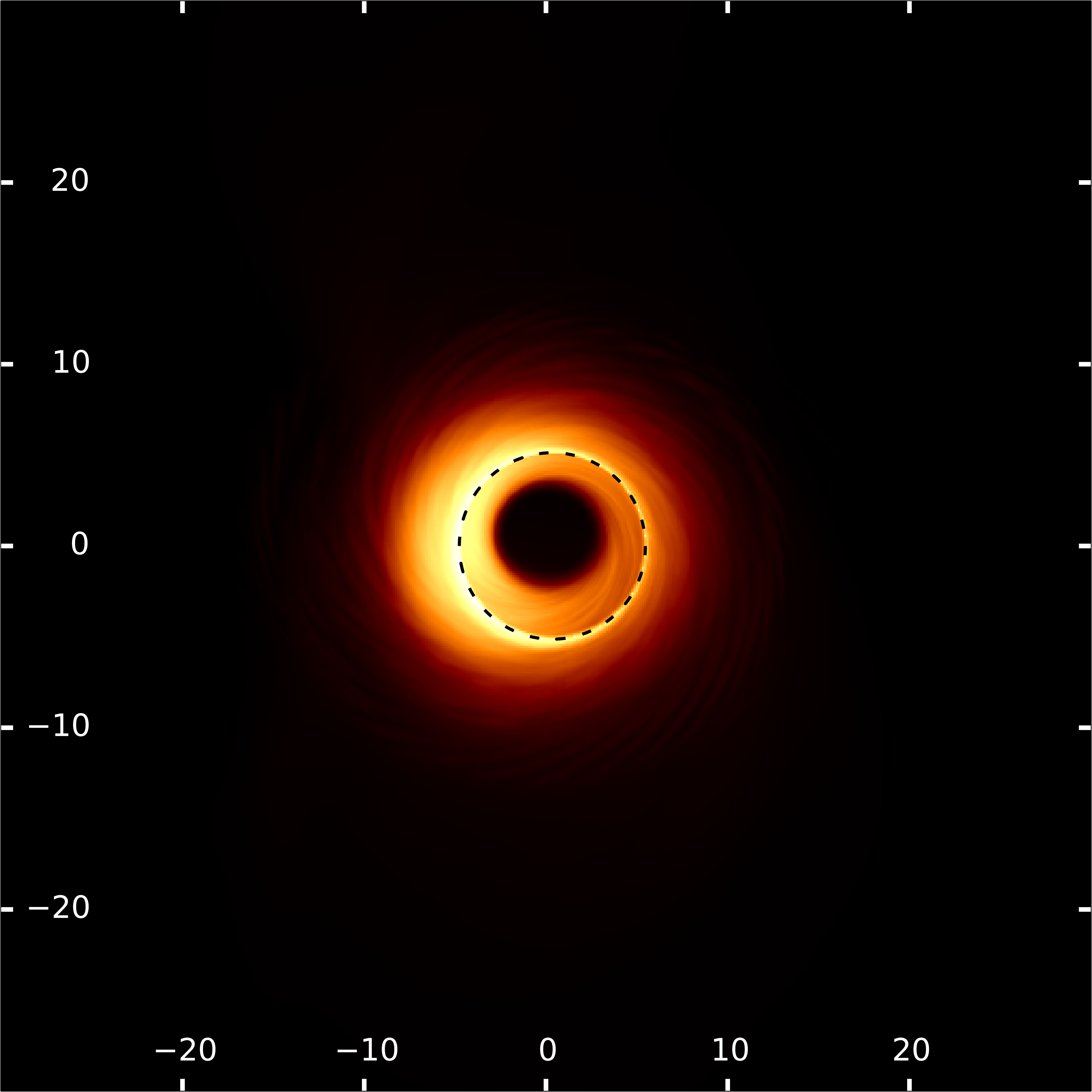}
	\caption{$a=0.5$, $i=20^\circ$.}
\end{subfigure}
\begin{subfigure}[b]{0.197\textwidth}
	\includegraphics[width=\textwidth]{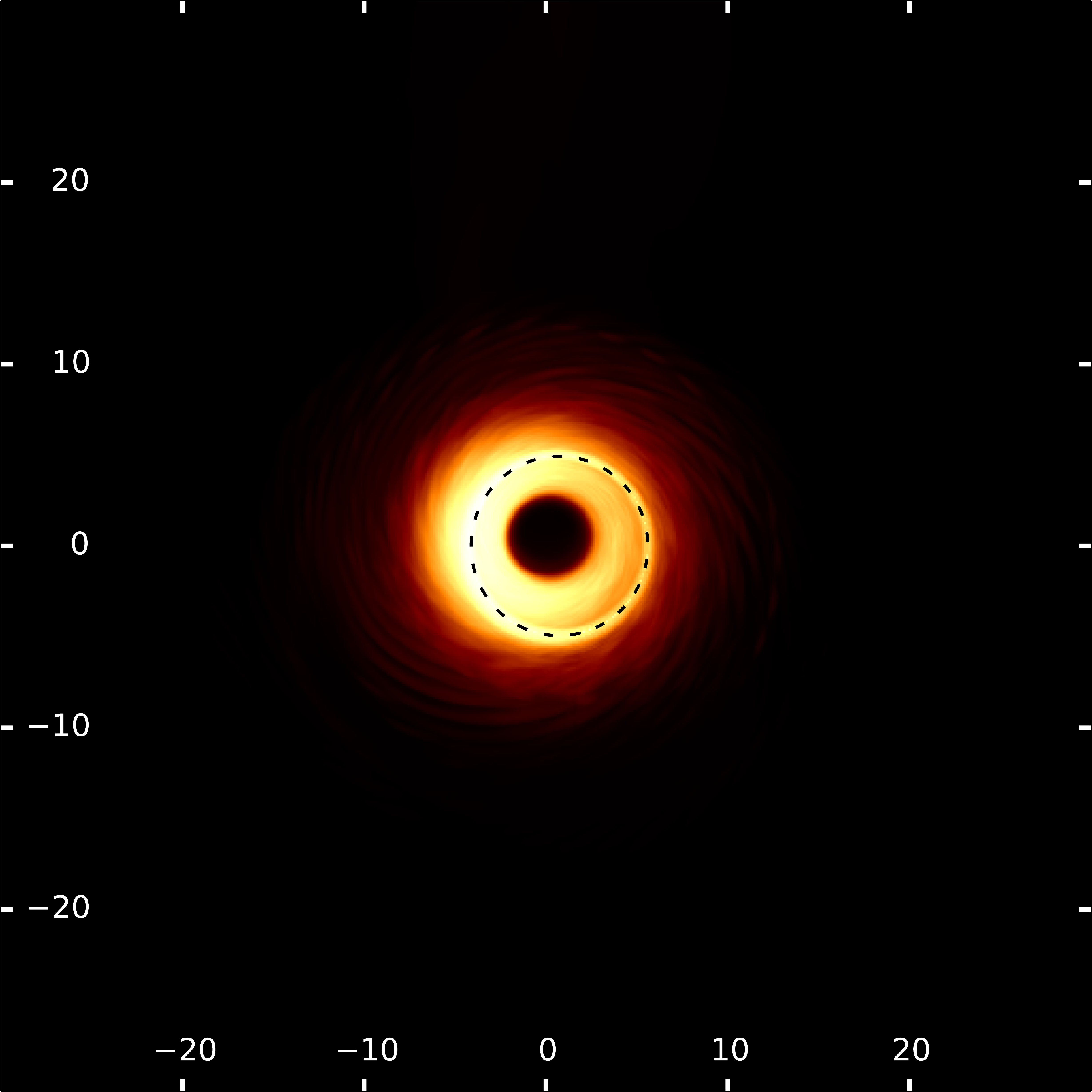}
	\caption{$a=0.9375$, $i=20^\circ$.}
\end{subfigure}
\begin{subfigure}[b]{0.197\textwidth}
	\includegraphics[width=\textwidth]{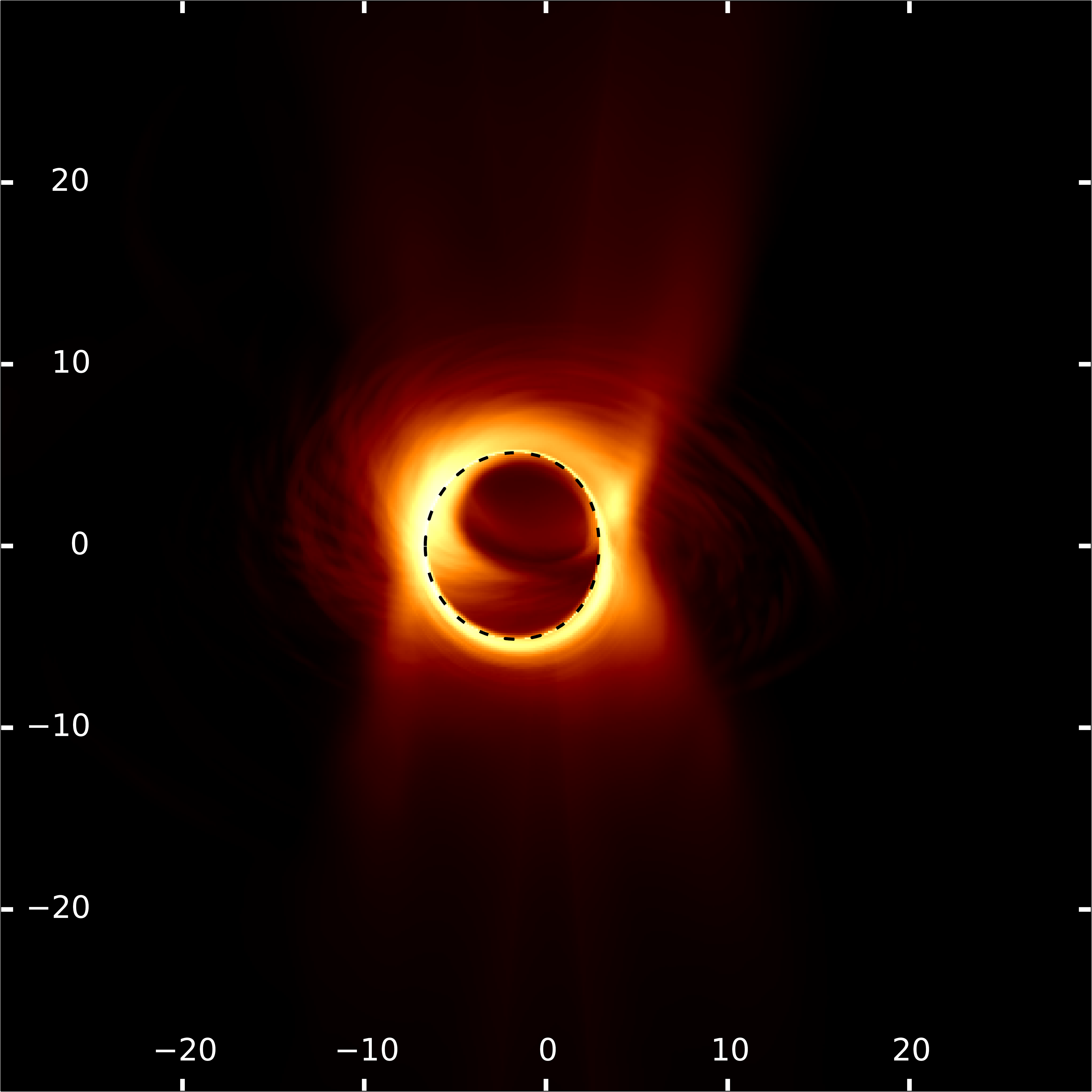}
	\caption{$a=-0.9375$, $i=60^\circ$.}
\end{subfigure}
\begin{subfigure}[b]{0.197\textwidth}
	\includegraphics[width=\textwidth]{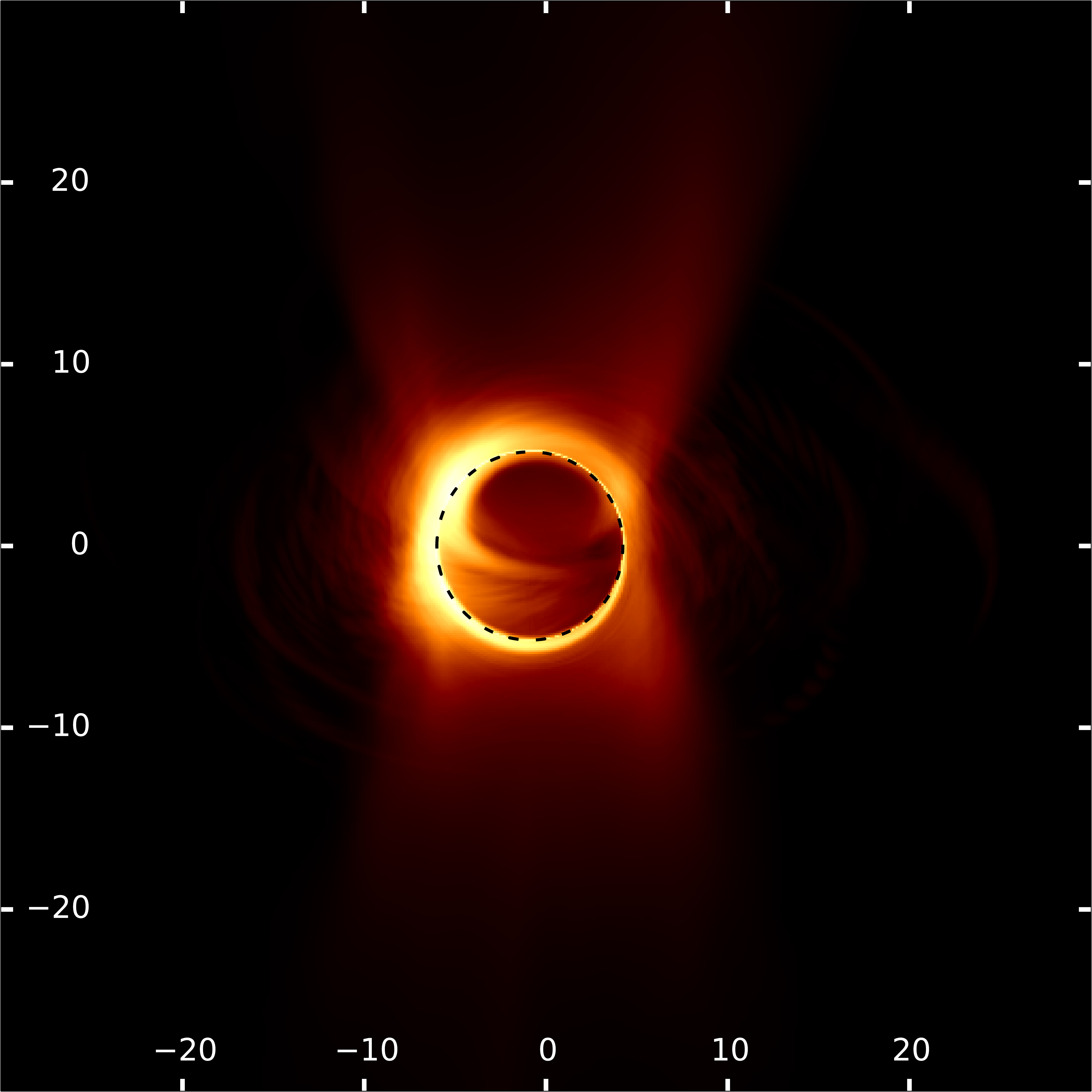}
	\caption{$a=-0.5$, $i=60^\circ$.}
\end{subfigure}
\begin{subfigure}[b]{0.197\textwidth}
	\includegraphics[width=\textwidth]{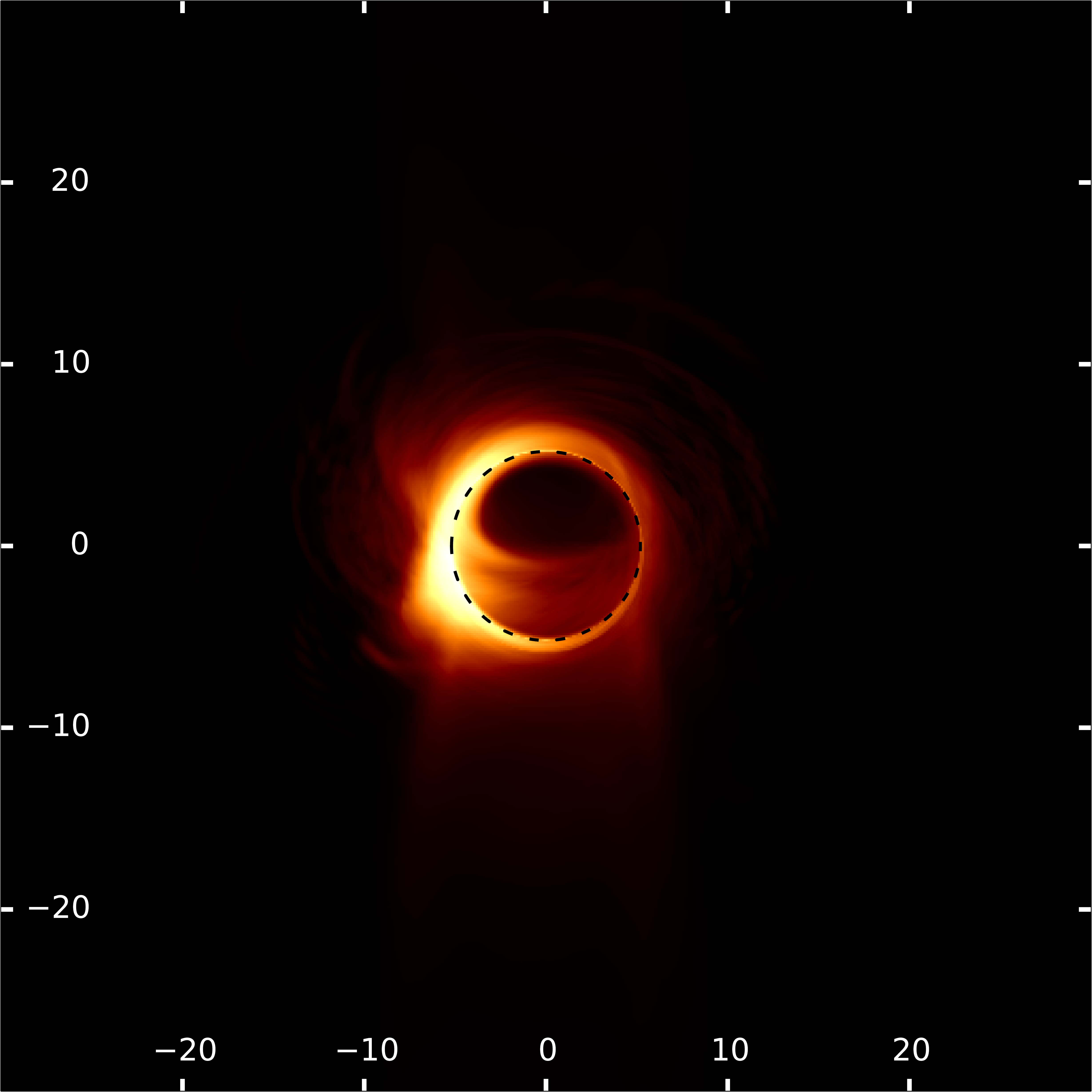}
	\caption{$a=0$, $i=60^\circ$.}
\end{subfigure}
\begin{subfigure}[b]{0.197\textwidth}
	\includegraphics[width=\textwidth]{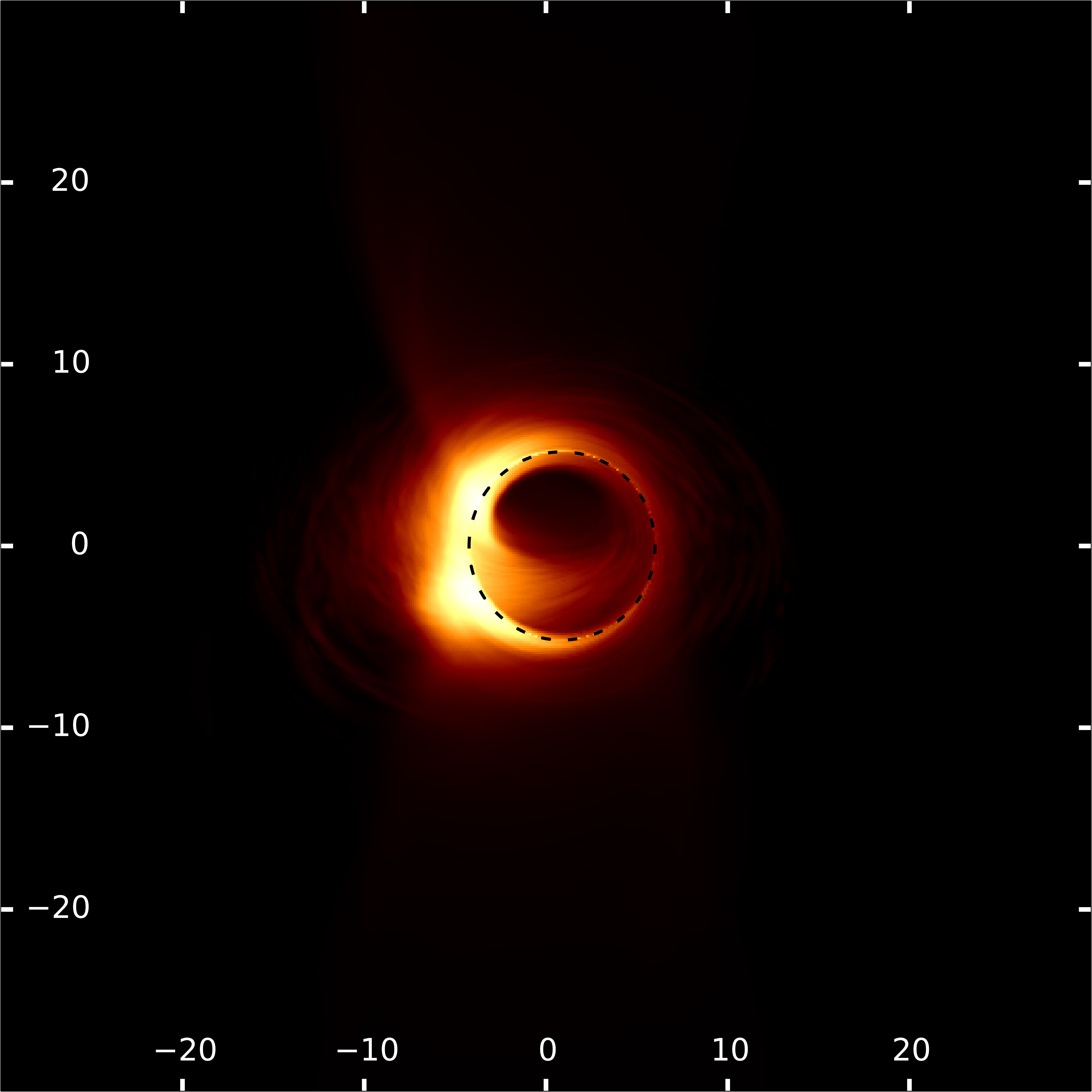}
	\caption{$a=0.5$, $i=60^\circ$.}
\end{subfigure}
\begin{subfigure}[b]{0.197\textwidth}
	\includegraphics[width=\textwidth]{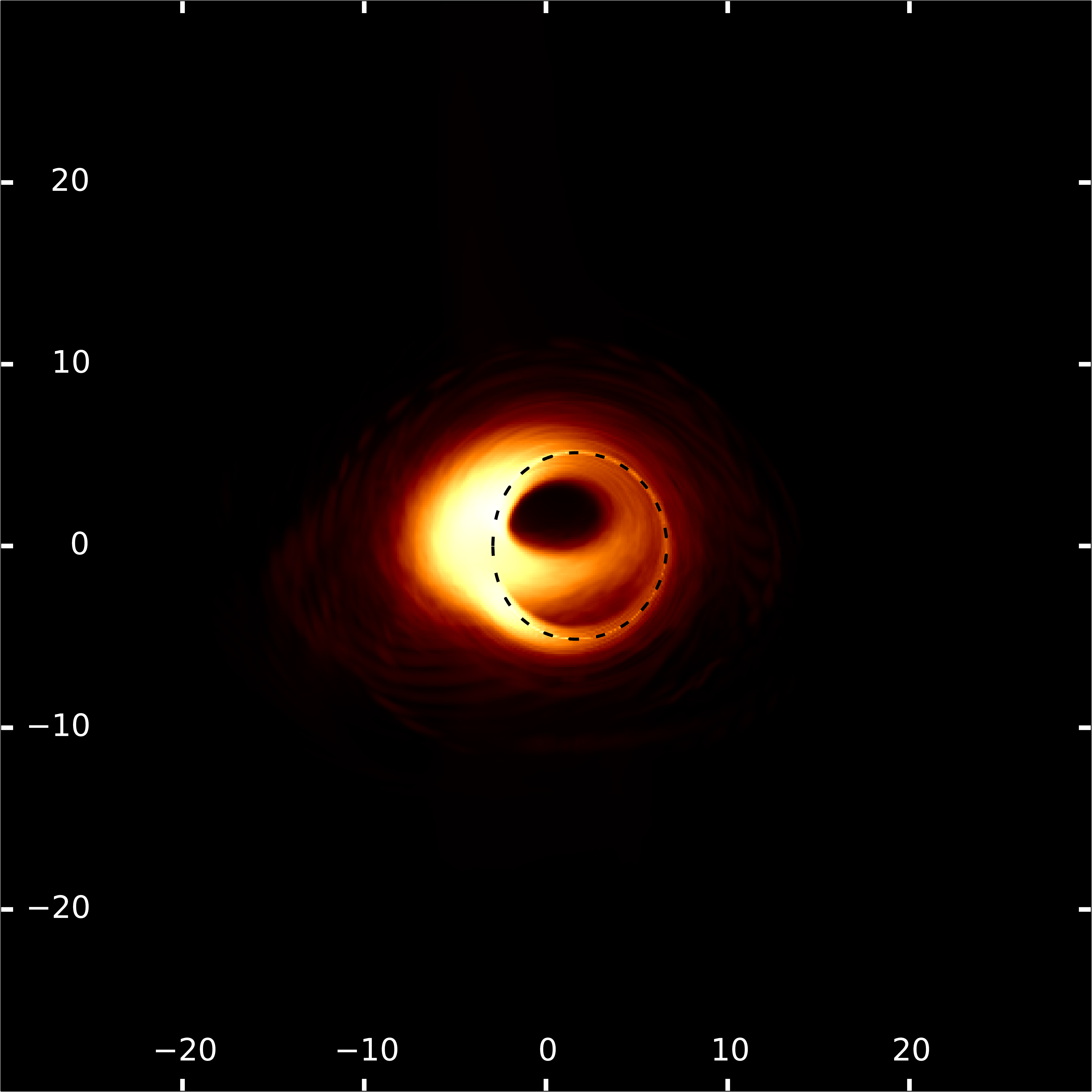}
	\caption{$a=0.9375$, $i=60^\circ$.}
\end{subfigure}
\begin{subfigure}[b]{0.197\textwidth}
	\includegraphics[width=\textwidth]{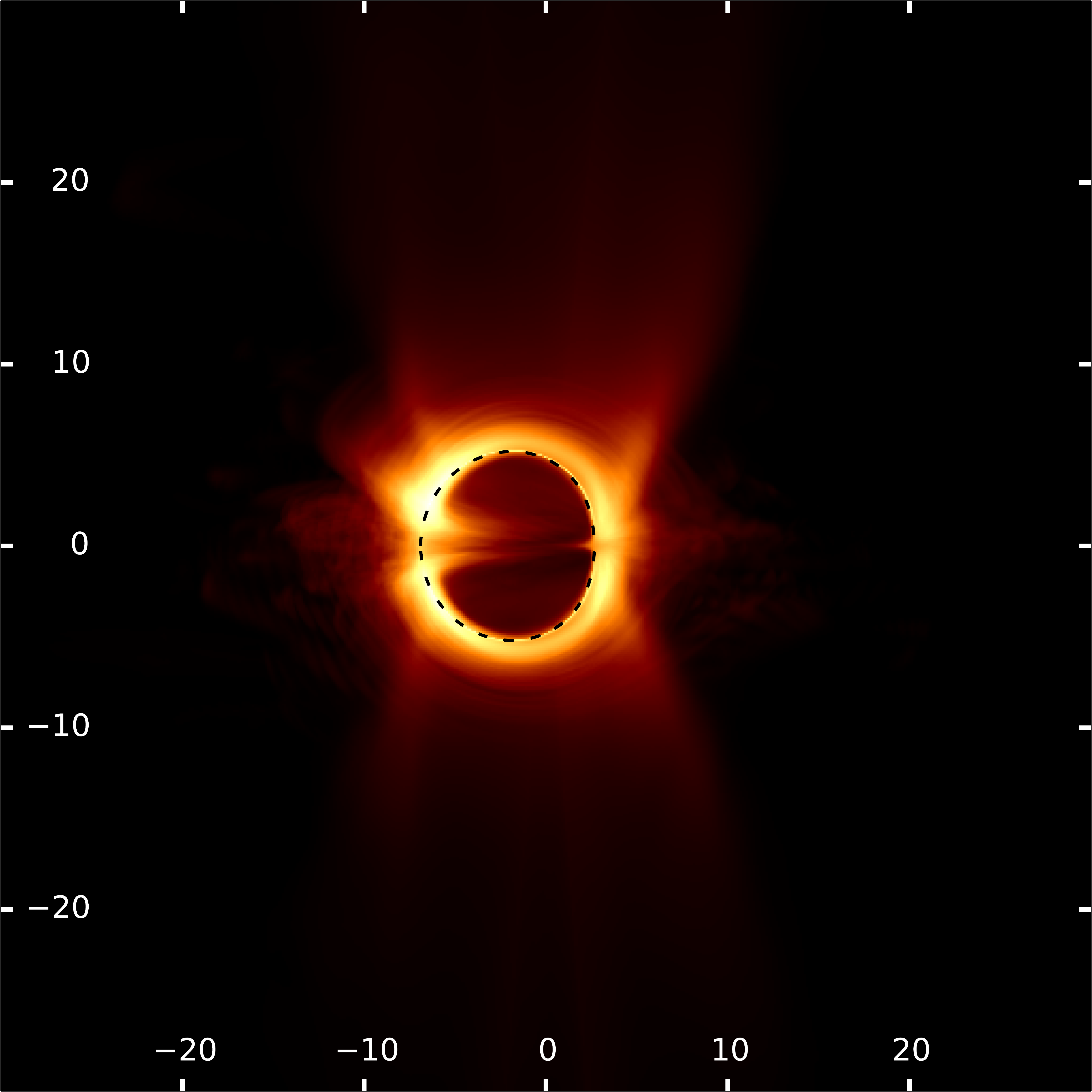}
	\caption{$a=-0.9375$, $i=90^\circ$.}
\end{subfigure}
\begin{subfigure}[b]{0.197\textwidth}
	\includegraphics[width=\textwidth]{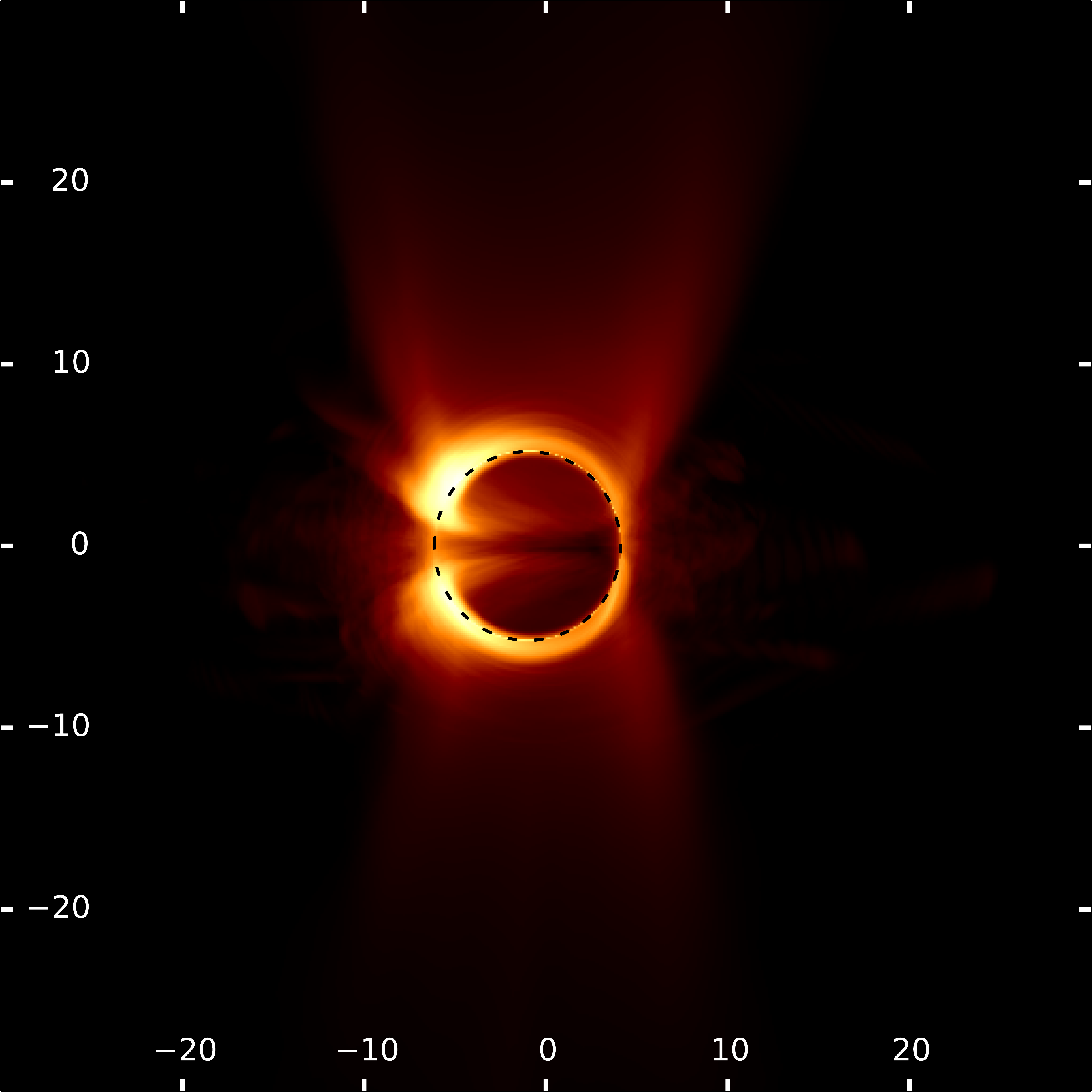}
	\caption{$a=-0.5$, $i=90^\circ$.}
\end{subfigure}
\begin{subfigure}[b]{0.197\textwidth}
	\includegraphics[width=\textwidth]{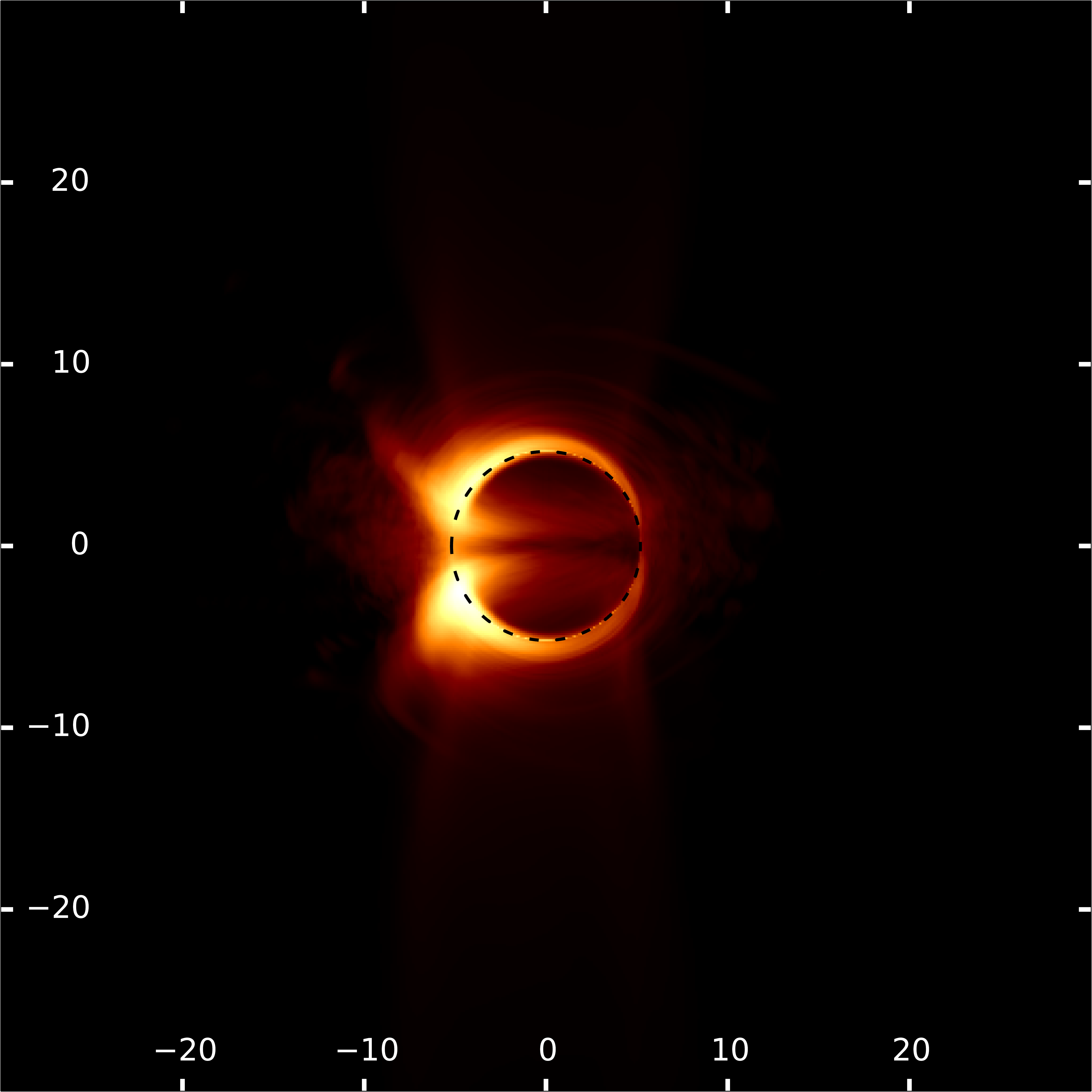}
	\caption{$a=0$, $i=90^\circ$.}
\end{subfigure}
\begin{subfigure}[b]{0.197\textwidth}
	\includegraphics[width=\textwidth]{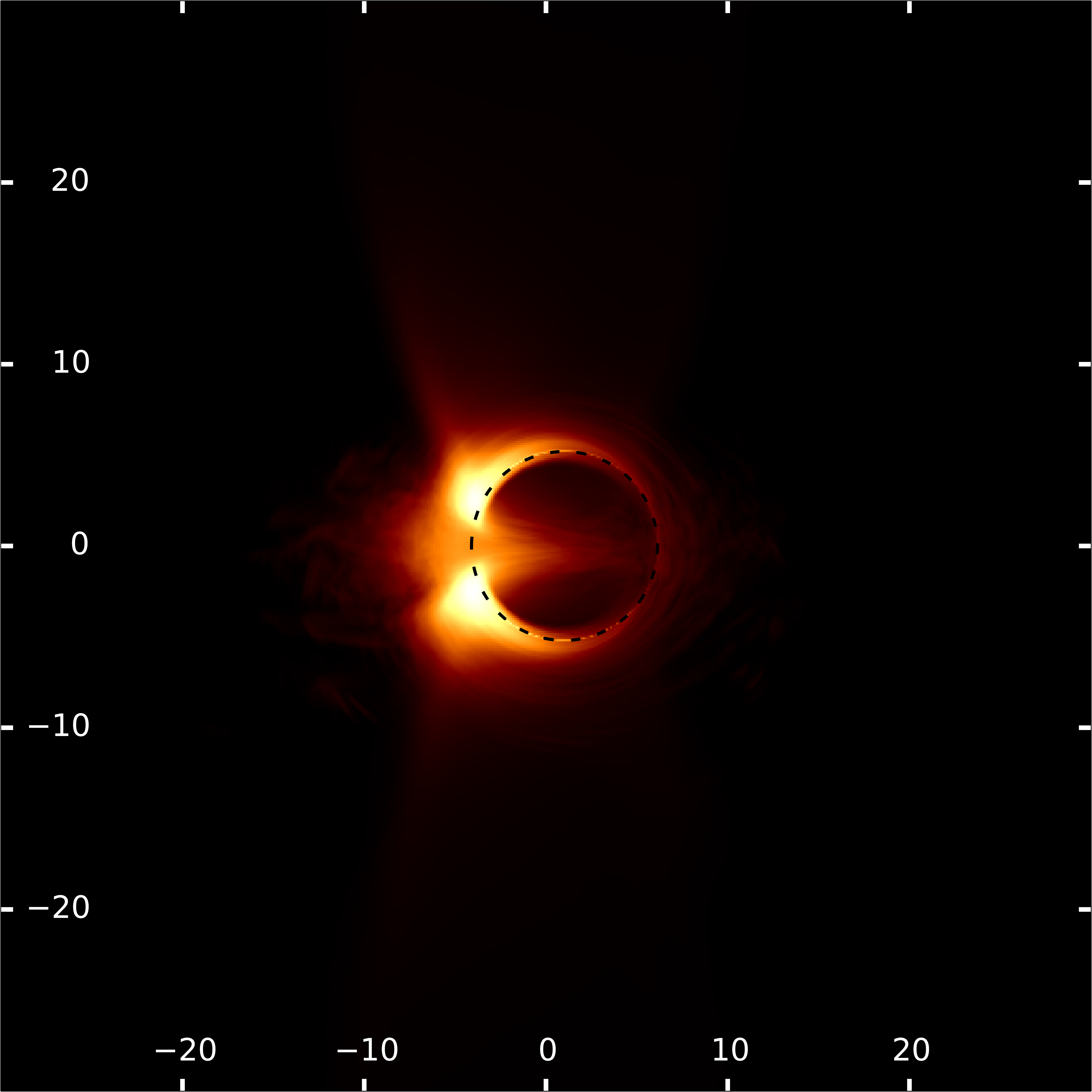}
	\caption{$a=0.5$, $i=90^\circ$.}
\end{subfigure}
\begin{subfigure}[b]{0.197\textwidth}
	\includegraphics[width=\textwidth]{Figures/sanejet_a15o16_90_0625-crop}
	\caption{$a=0.9375$, $i=90^\circ$.}
\end{subfigure}
\caption{Time-averaged, normalised intensity maps of our SANE, jet-dominated GRMHD models of Sgr A*, imaged at 230 GHz, at five different spins and four observer inclination angles, with an integrated flux density of 0.625 Jy. In each case, the photon ring, which marks the BHS, is indicated by a dashed line. The values for the impact parameters along the x- and y-axes are expressed in terms of $R_{\rm g}$. The image maps were plotted using a square-root intensity scale.}
\label{fig:sane_jet_0625_matrix}
\end{figure*}

\begin{figure*}
\centering
\begin{subfigure}[b]{0.197\textwidth}
	\includegraphics[width=\textwidth]{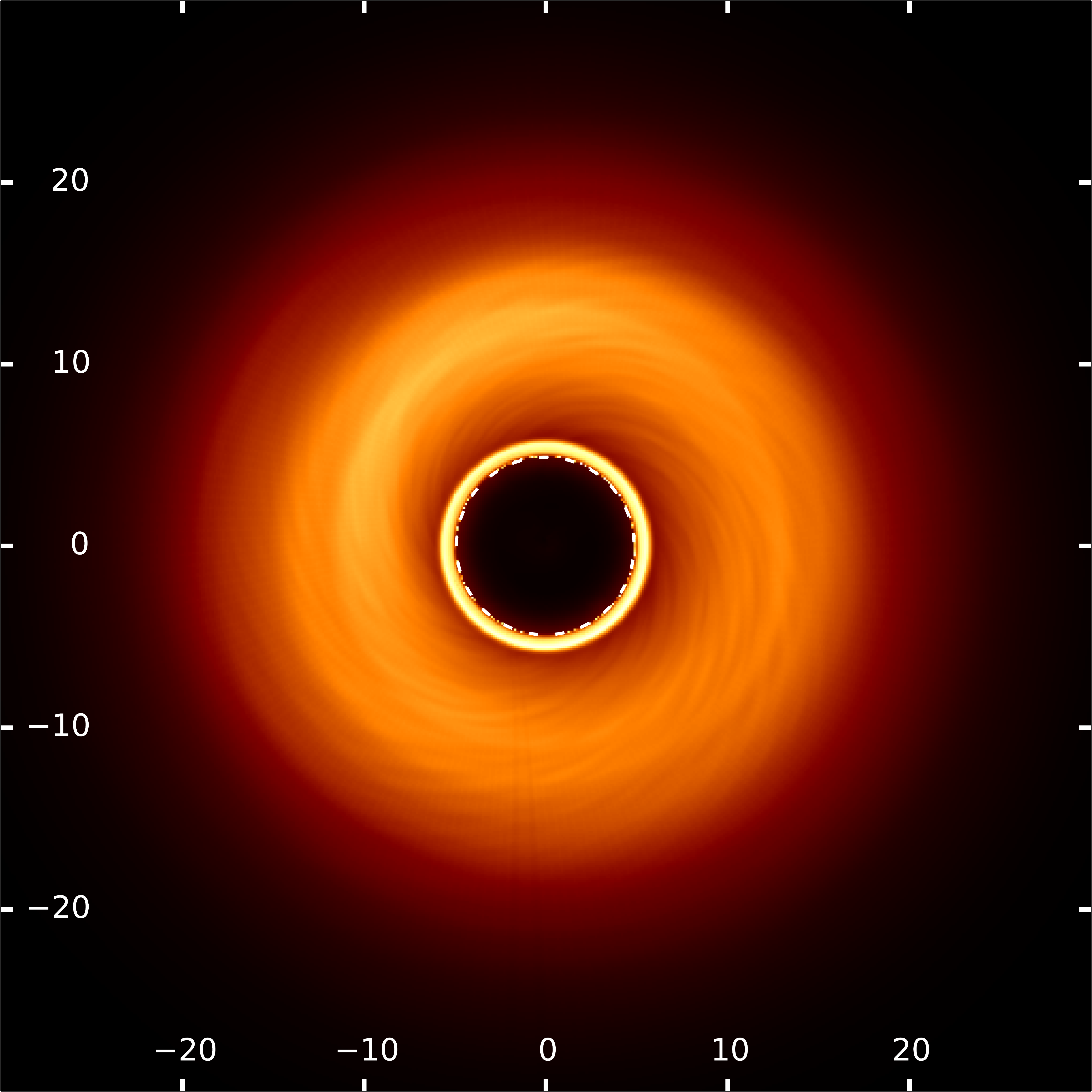}
	\caption{$a=-0.9375$, $i=1^\circ$.}
\end{subfigure}
\begin{subfigure}[b]{0.197\textwidth}
	\includegraphics[width=\textwidth]{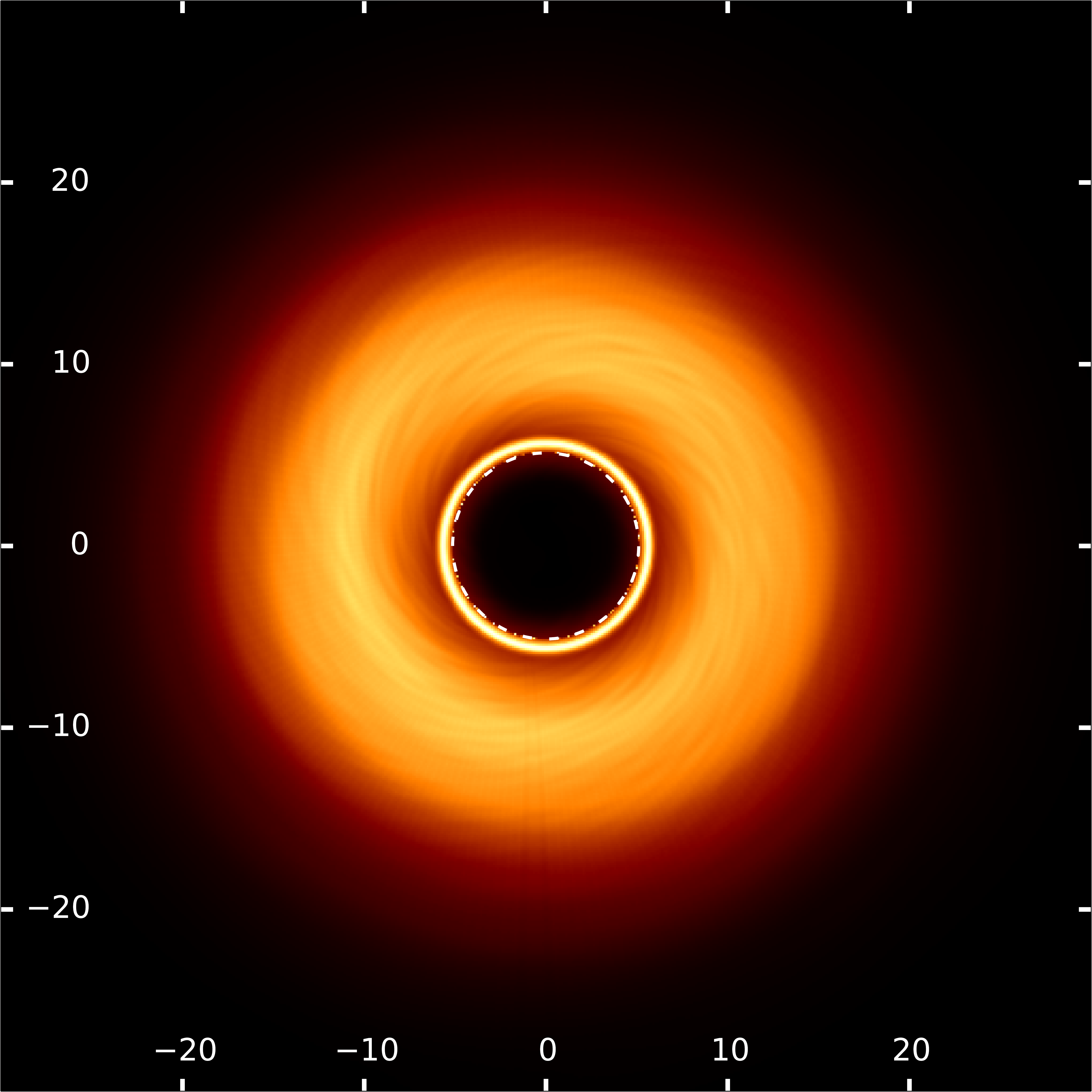}
	\caption{$a=-0.5$, $i=1^\circ$.}
\end{subfigure}
\begin{subfigure}[b]{0.197\textwidth}
	\includegraphics[width=\textwidth]{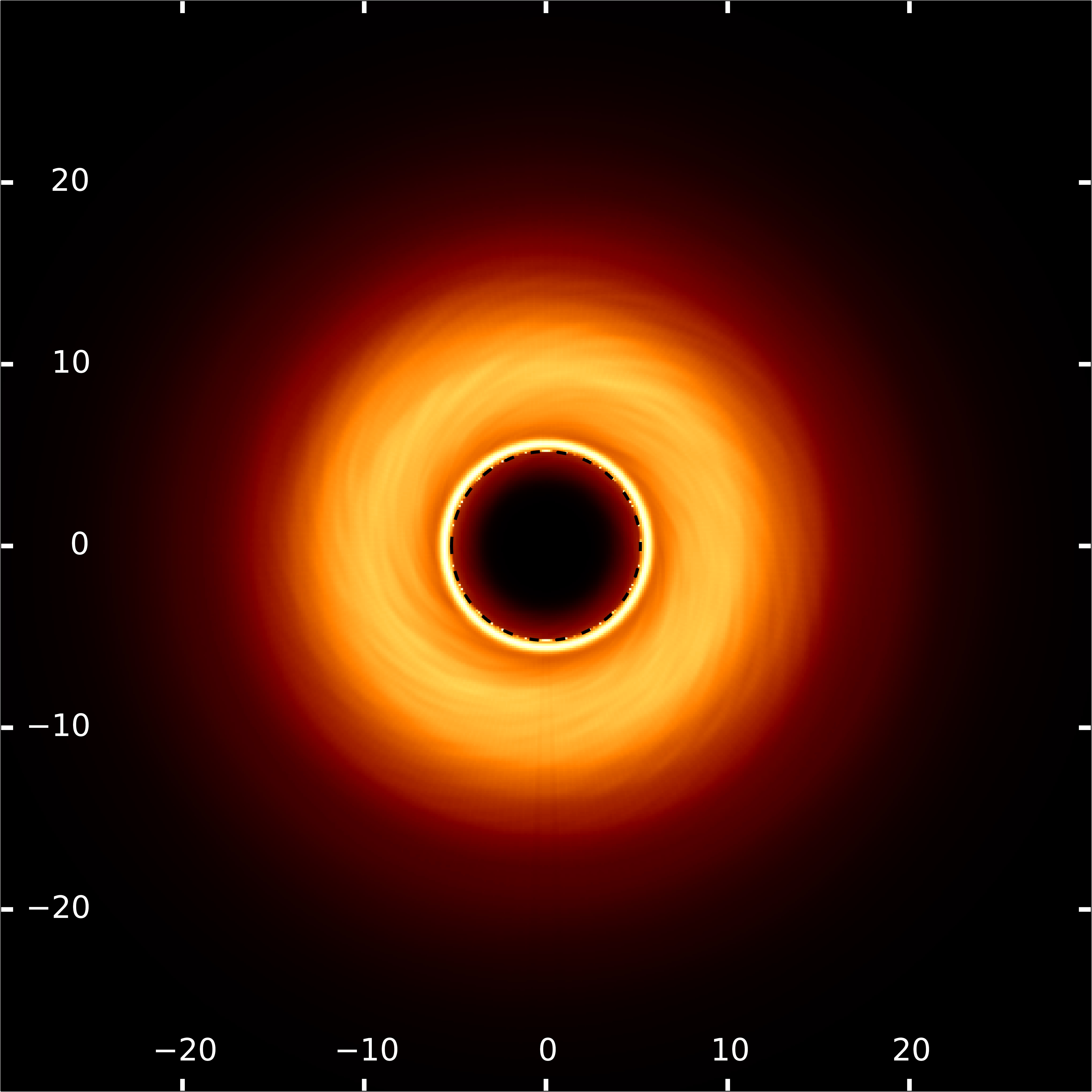}
	\caption{$a=0$, $i=1^\circ$.}
\end{subfigure}
\begin{subfigure}[b]{0.197\textwidth}
	\includegraphics[width=\textwidth]{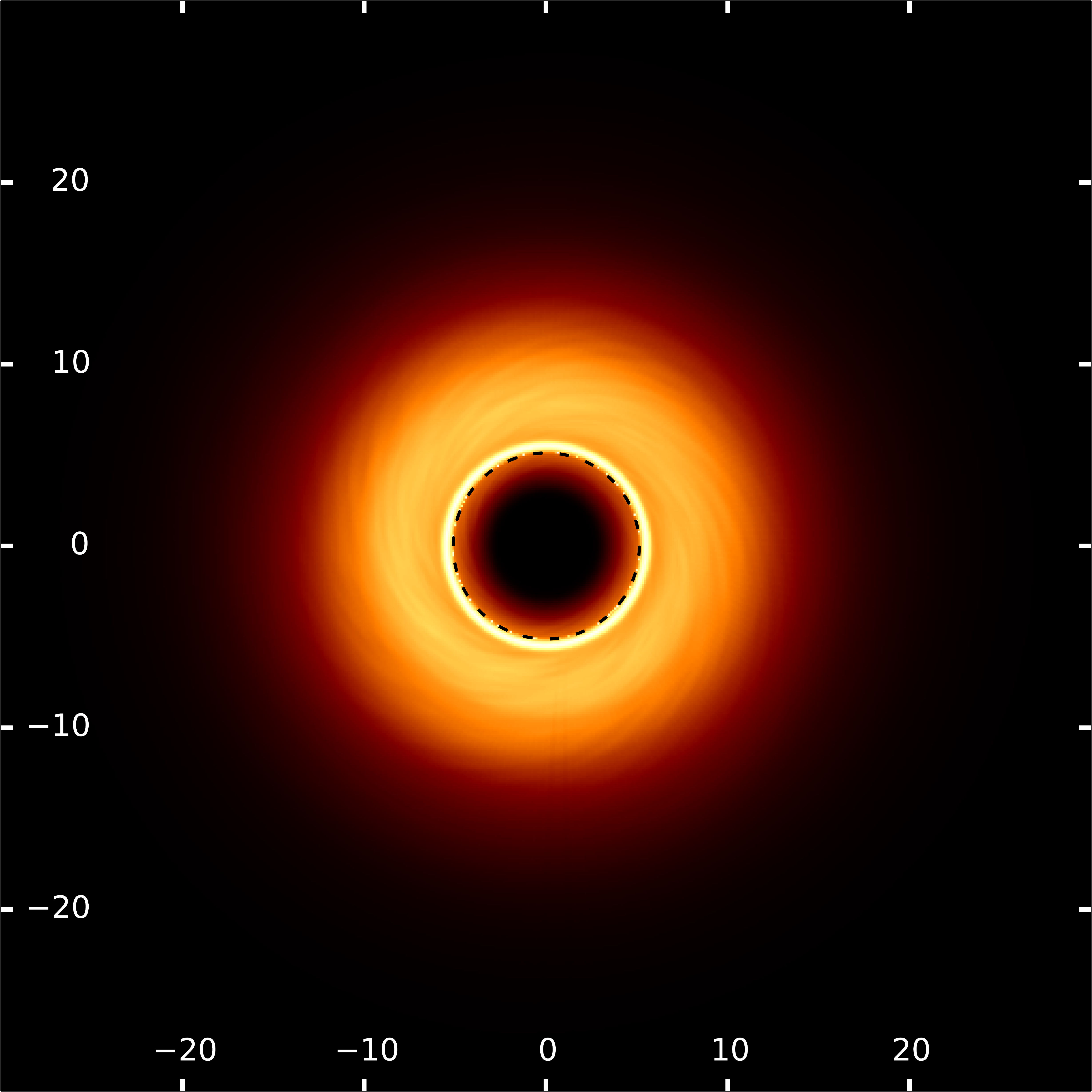}
	\caption{$a=0.5$, $i=1^\circ$.}
\end{subfigure}
\begin{subfigure}[b]{0.197\textwidth}
	\includegraphics[width=\textwidth]{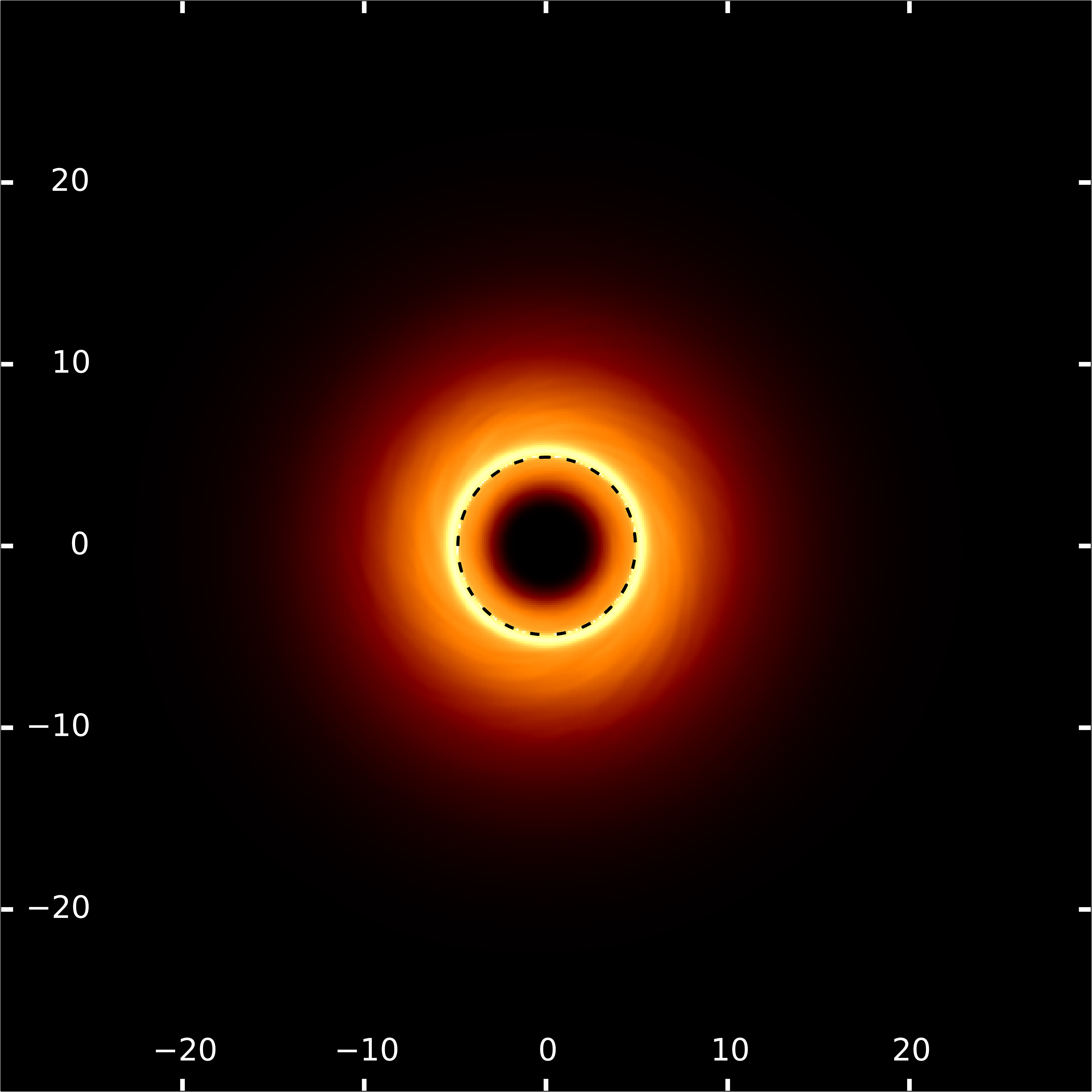}
	\caption{$a=0.9375$, $i=1^\circ$.}
\end{subfigure}
\begin{subfigure}[b]{0.197\textwidth}
	\includegraphics[width=\textwidth]{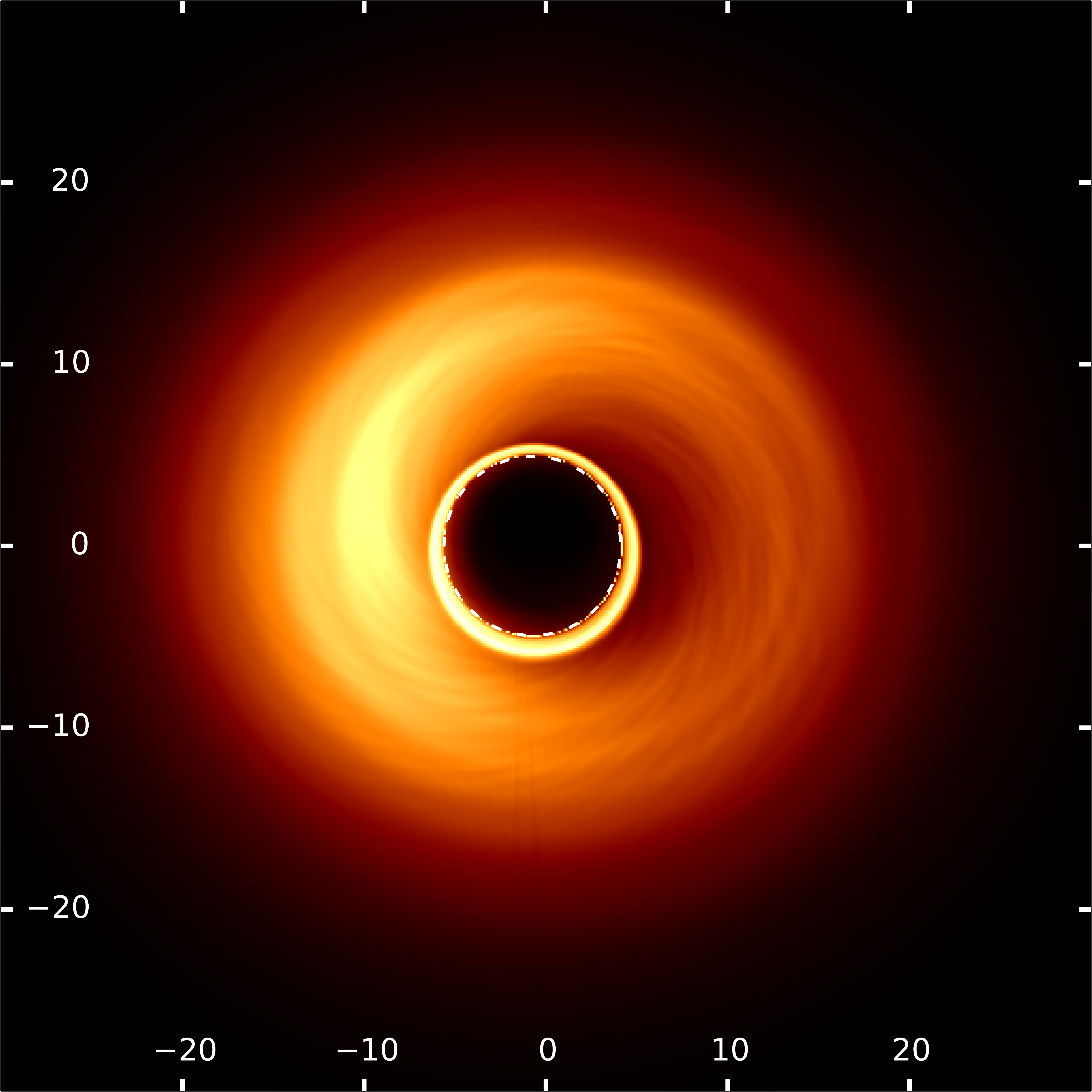}
	\caption{$a=-0.9375$, $i=20^\circ$.}
\end{subfigure}
\begin{subfigure}[b]{0.197\textwidth}
	\includegraphics[width=\textwidth]{Figures/sane_disk_a-1o2_20_25-crop}
	\caption{$a=-0.5$, $i=20^\circ$.}
\end{subfigure}
\begin{subfigure}[b]{0.197\textwidth}
	\includegraphics[width=\textwidth]{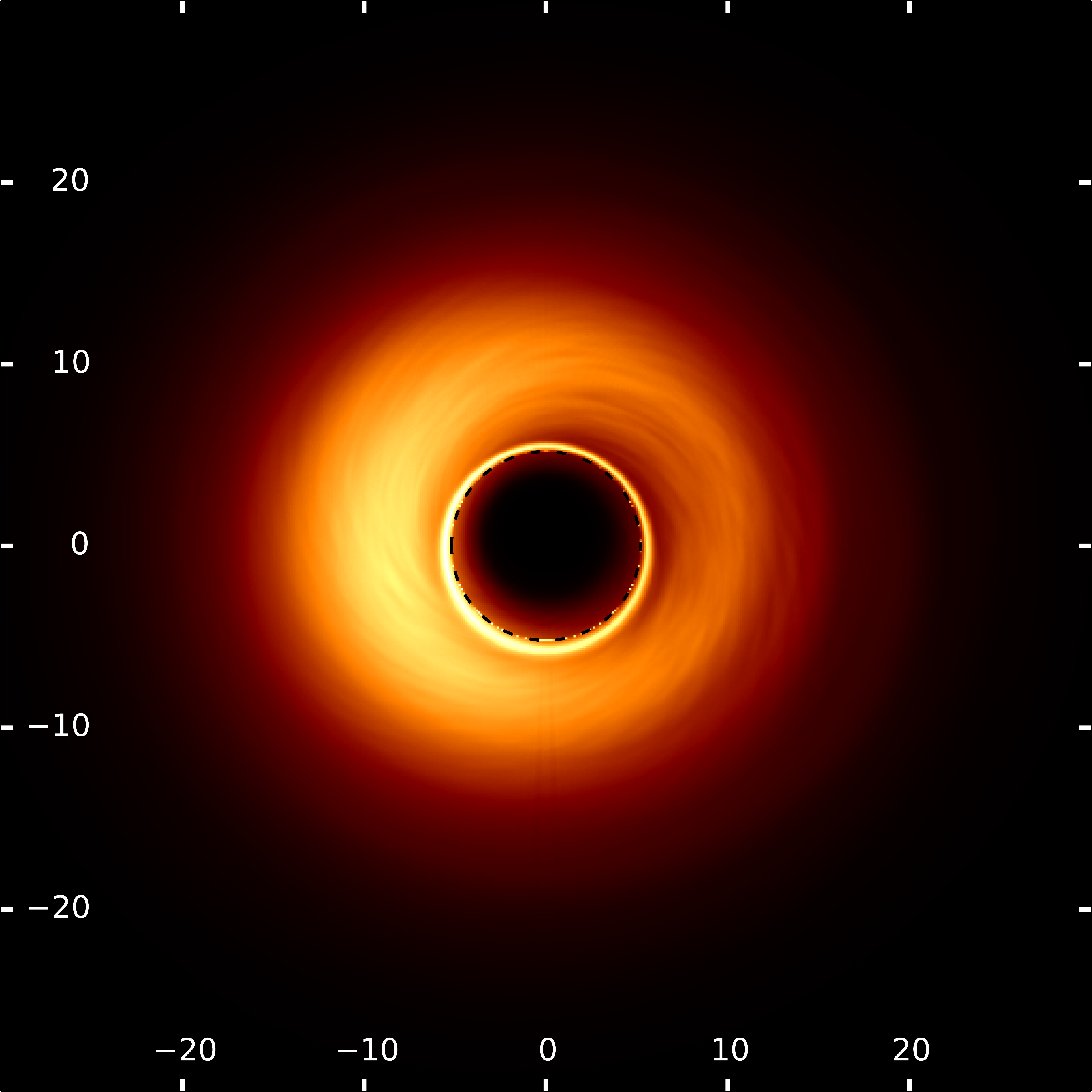}
	\caption{$a=0$, $i=20^\circ$.}
\end{subfigure}
\begin{subfigure}[b]{0.197\textwidth}
	\includegraphics[width=\textwidth]{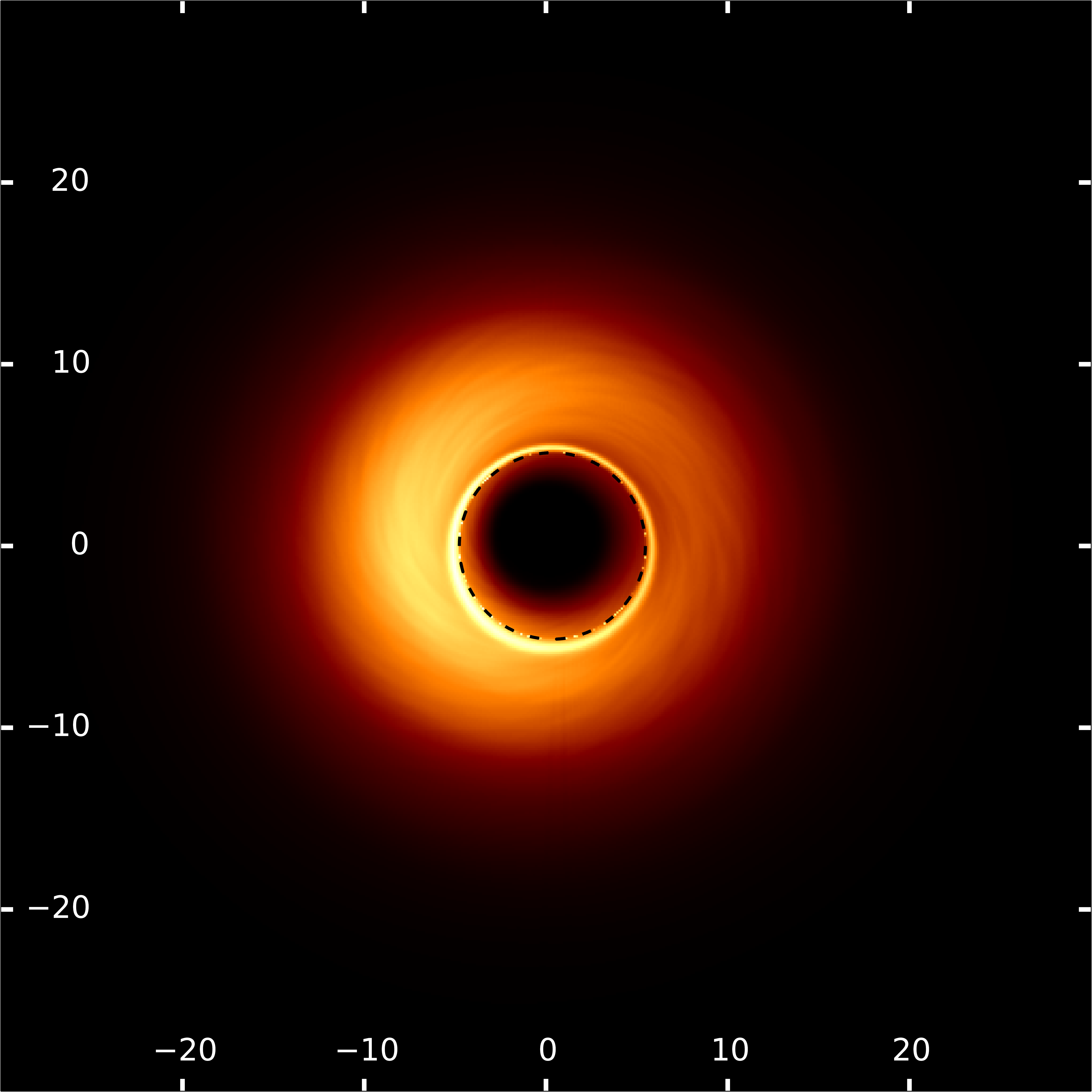}
	\caption{$a=0.5$, $i=20^\circ$.}
\end{subfigure}
\begin{subfigure}[b]{0.197\textwidth}
	\includegraphics[width=\textwidth]{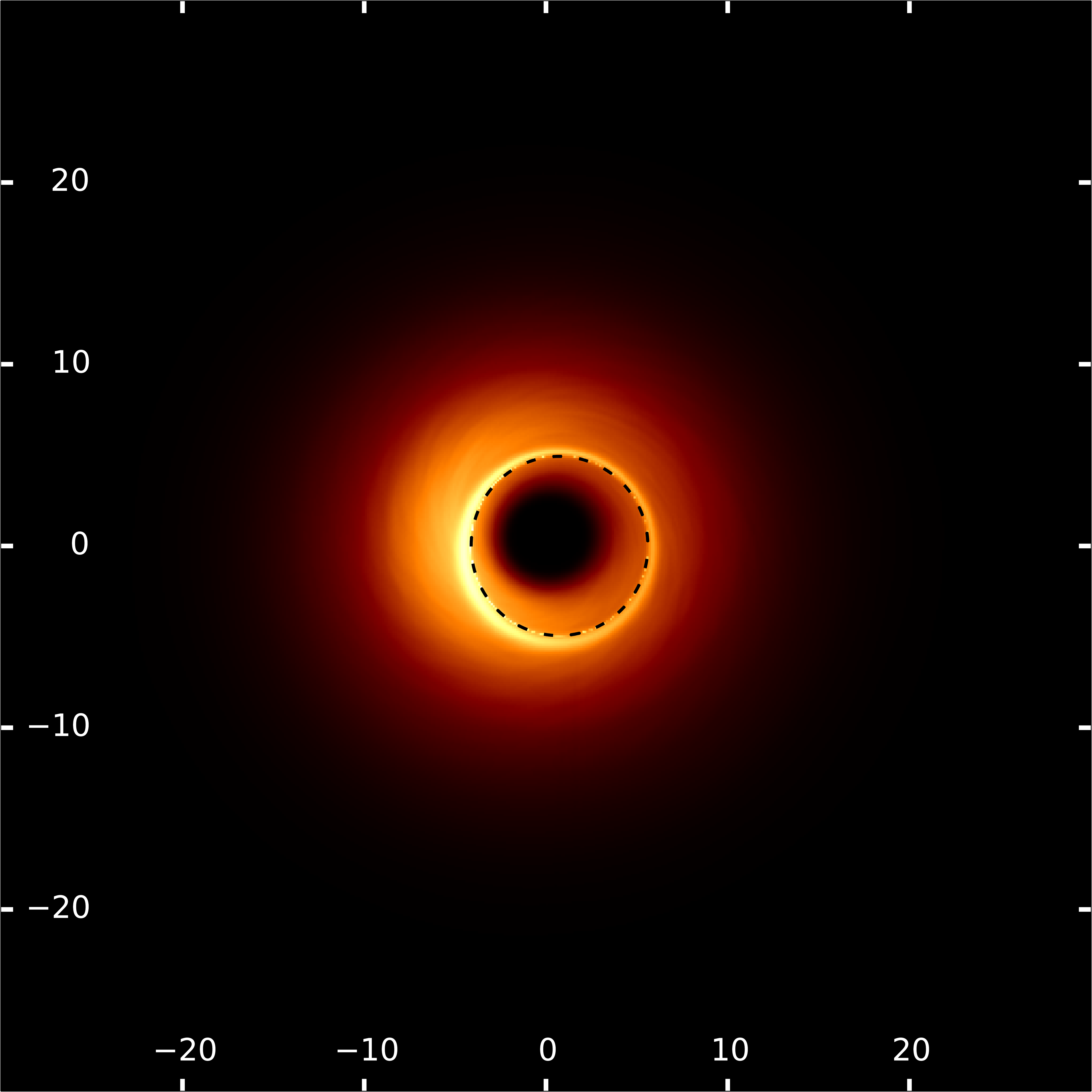}
	\caption{$a=0.9375$, $i=20^\circ$.}
\end{subfigure}
\begin{subfigure}[b]{0.197\textwidth}
	\includegraphics[width=\textwidth]{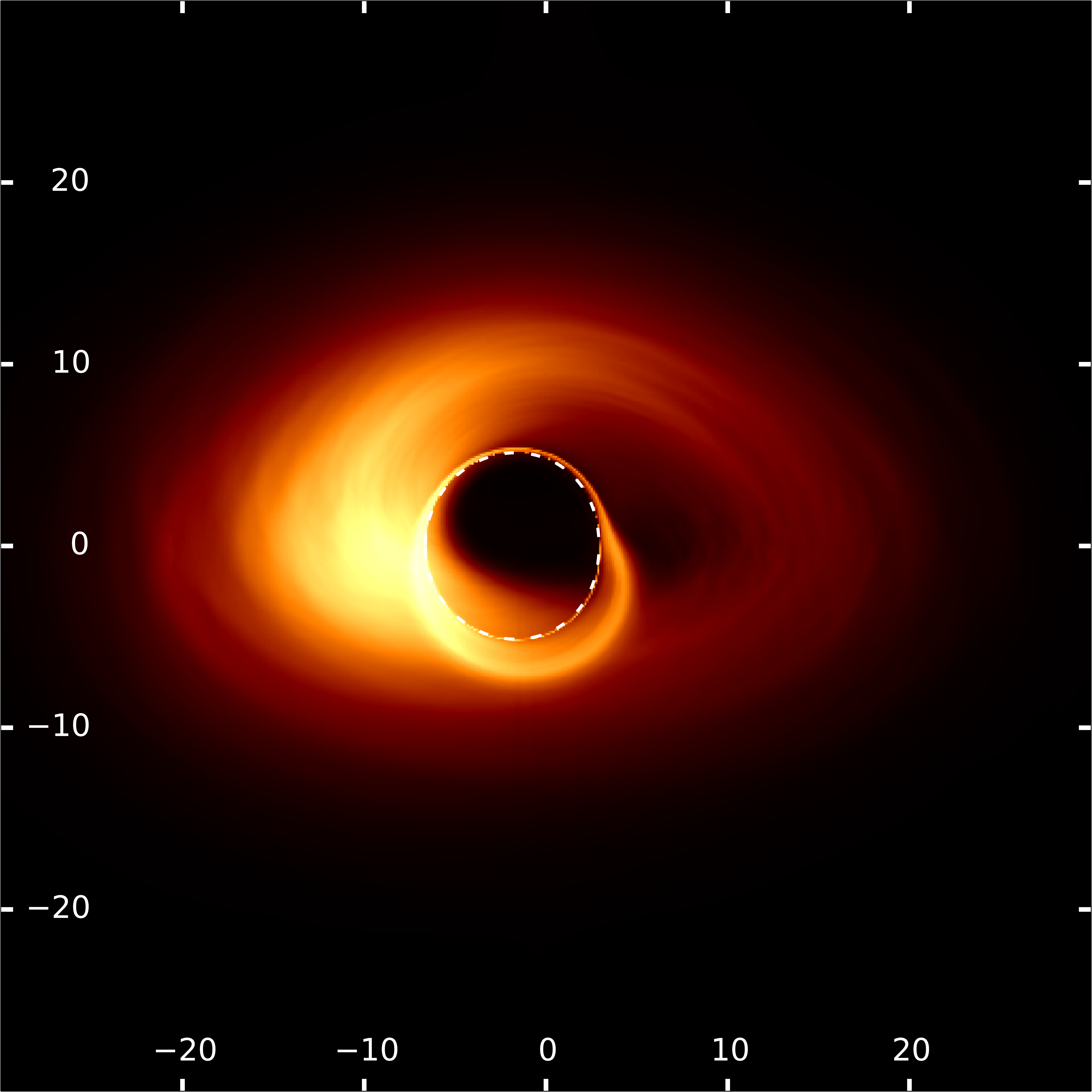}
	\caption{$a=-0.9375$, $i=60^\circ$.}
\end{subfigure}
\begin{subfigure}[b]{0.197\textwidth}
	\includegraphics[width=\textwidth]{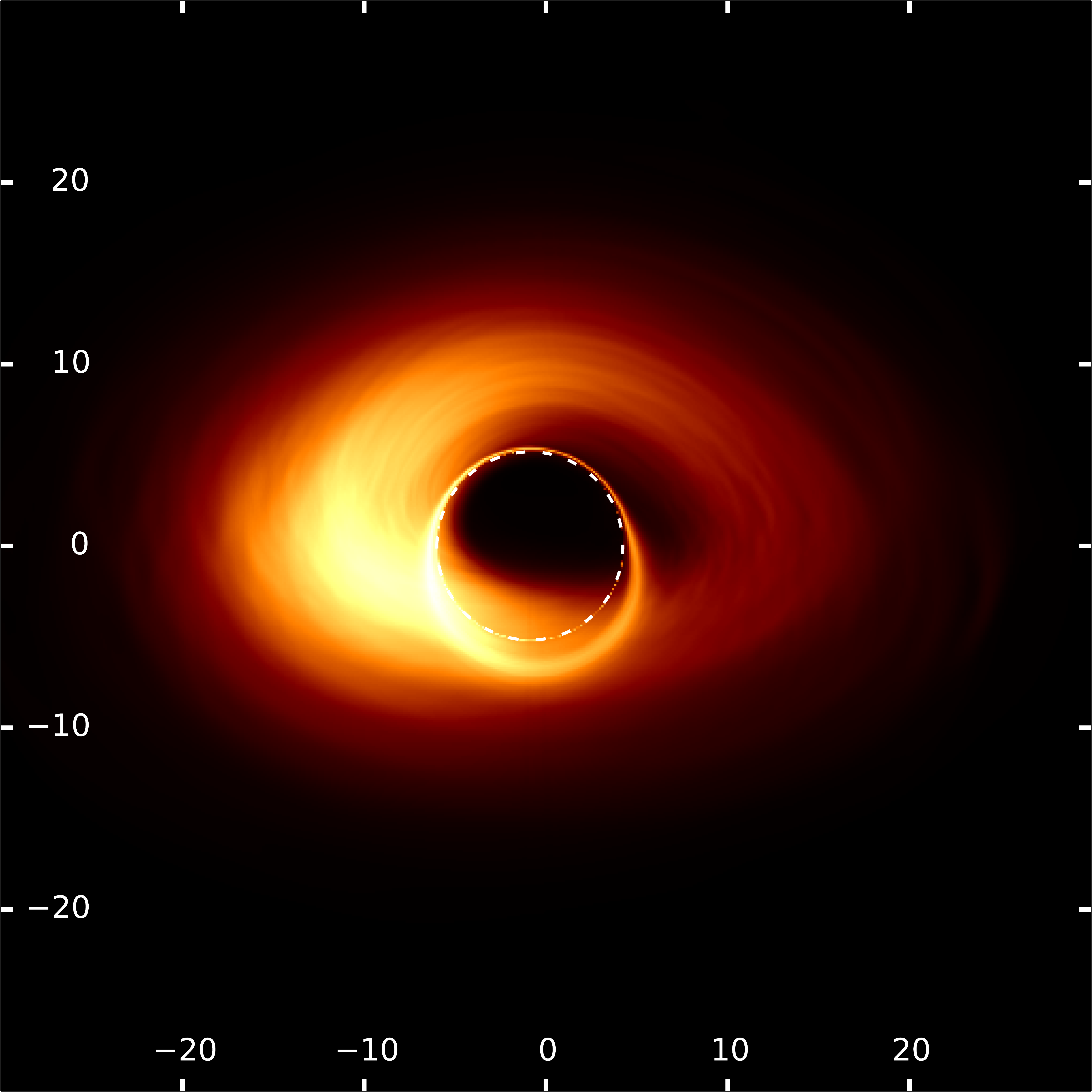}
	\caption{$a=-0.5$, $i=60^\circ$.}
\end{subfigure}
\begin{subfigure}[b]{0.197\textwidth}
	\includegraphics[width=\textwidth]{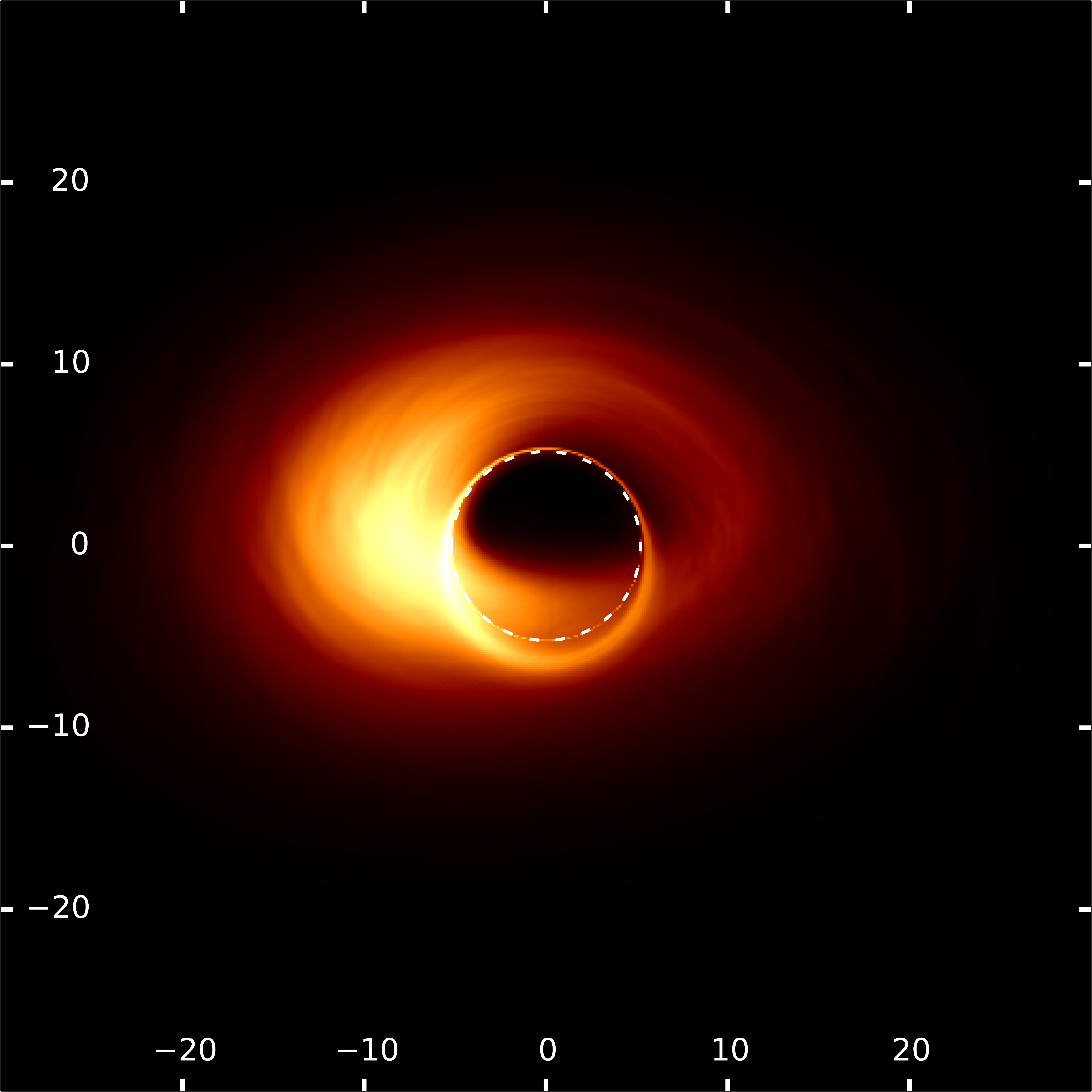}
	\caption{$a=0$, $i=60^\circ$.}
\end{subfigure}
\begin{subfigure}[b]{0.197\textwidth}
	\includegraphics[width=\textwidth]{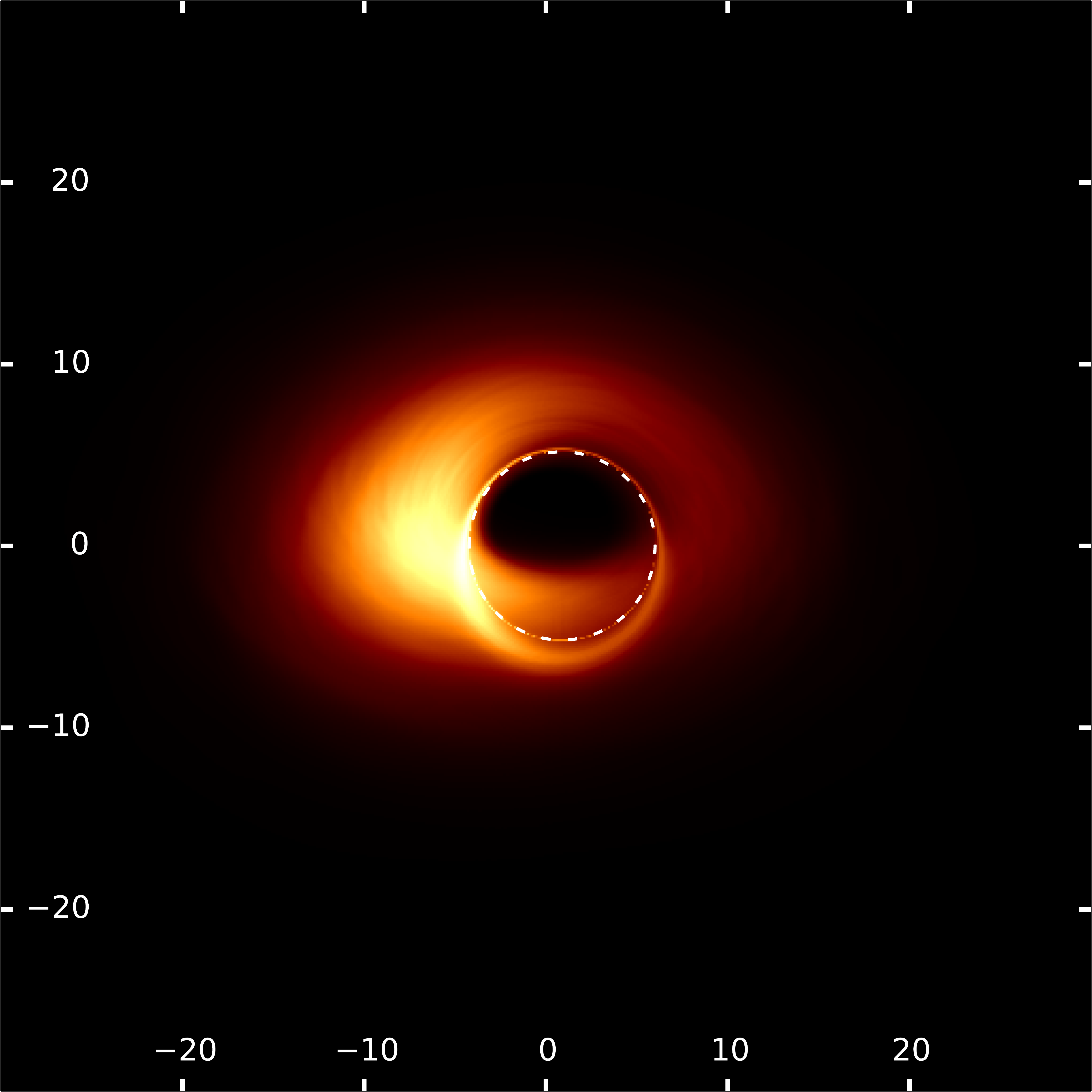}
	\caption{$a=0.5$, $i=60^\circ$.}
\end{subfigure}
\begin{subfigure}[b]{0.197\textwidth}
	\includegraphics[width=\textwidth]{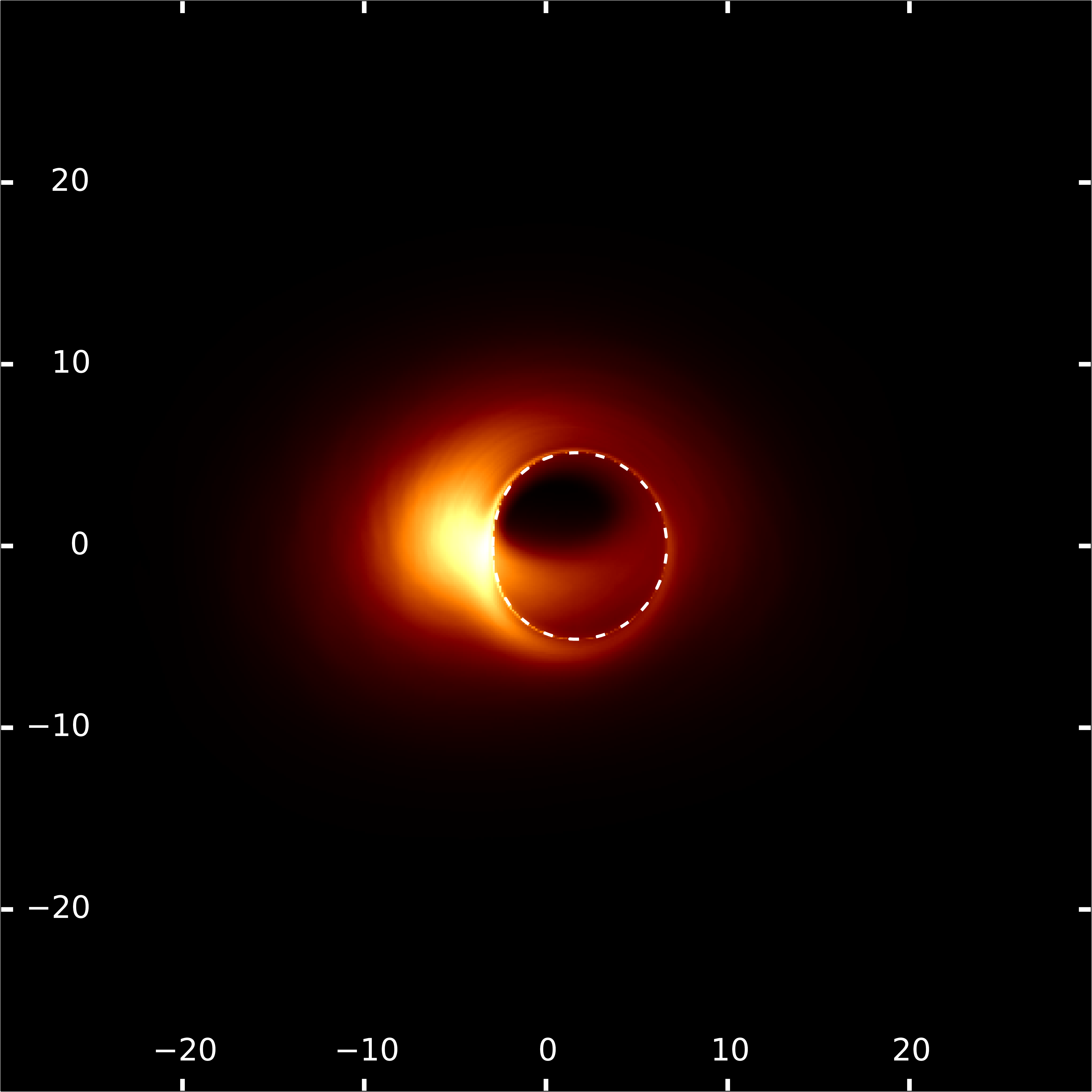}
	\caption{$a=0.9375$, $i=60^\circ$.}
\end{subfigure}
\begin{subfigure}[b]{0.197\textwidth}
	\includegraphics[width=\textwidth]{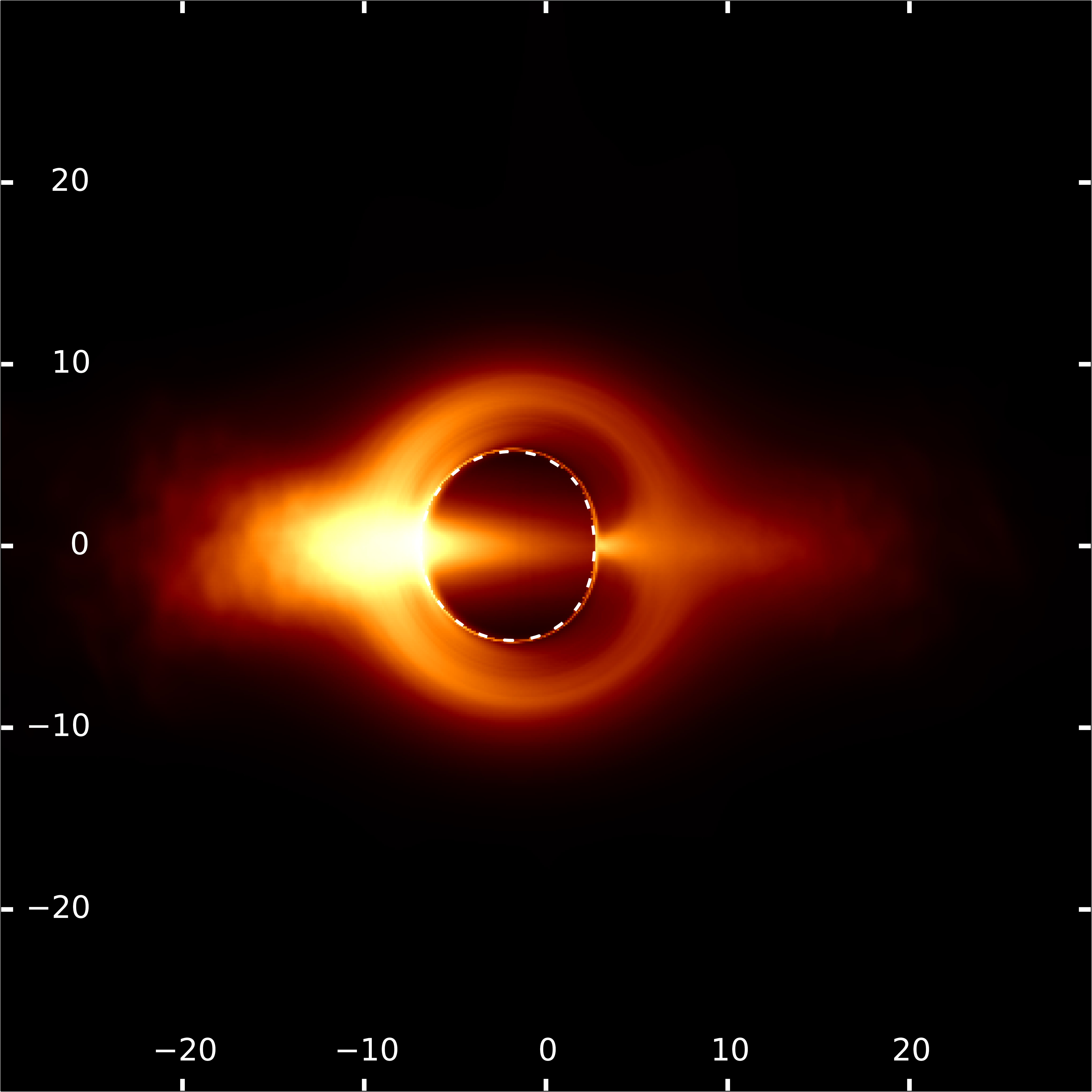}
	\caption{$a=-0.9375$, $i=90^\circ$.}
\end{subfigure}
\begin{subfigure}[b]{0.197\textwidth}
	\includegraphics[width=\textwidth]{Figures/sane_disk_a-1o2_90_25-crop}
	\caption{$a=-0.5$, $i=90^\circ$.}
\end{subfigure}
\begin{subfigure}[b]{0.197\textwidth}
	\includegraphics[width=\textwidth]{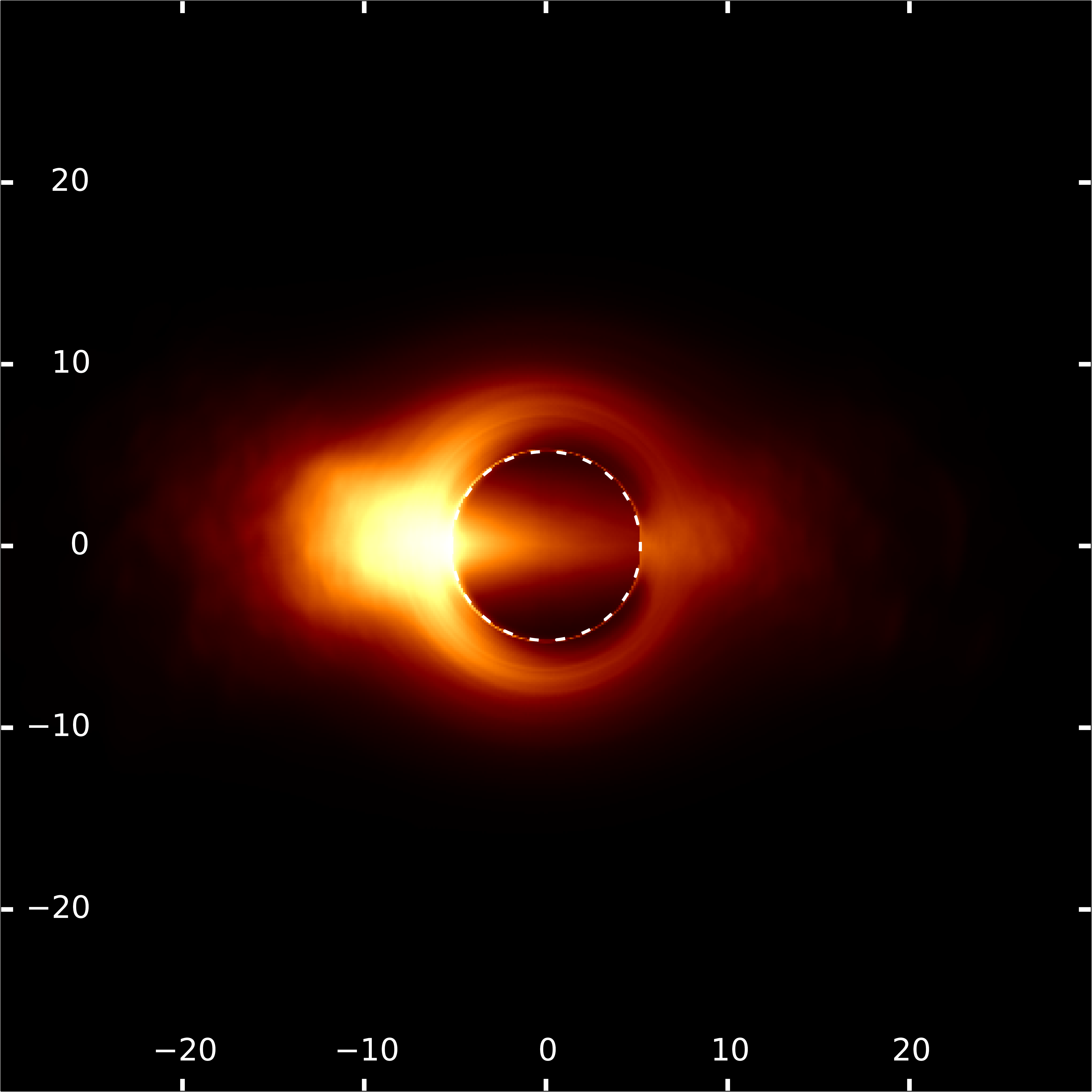}
	\caption{$a=0$, $i=90^\circ$.}
\end{subfigure}
\begin{subfigure}[b]{0.197\textwidth}
	\includegraphics[width=\textwidth]{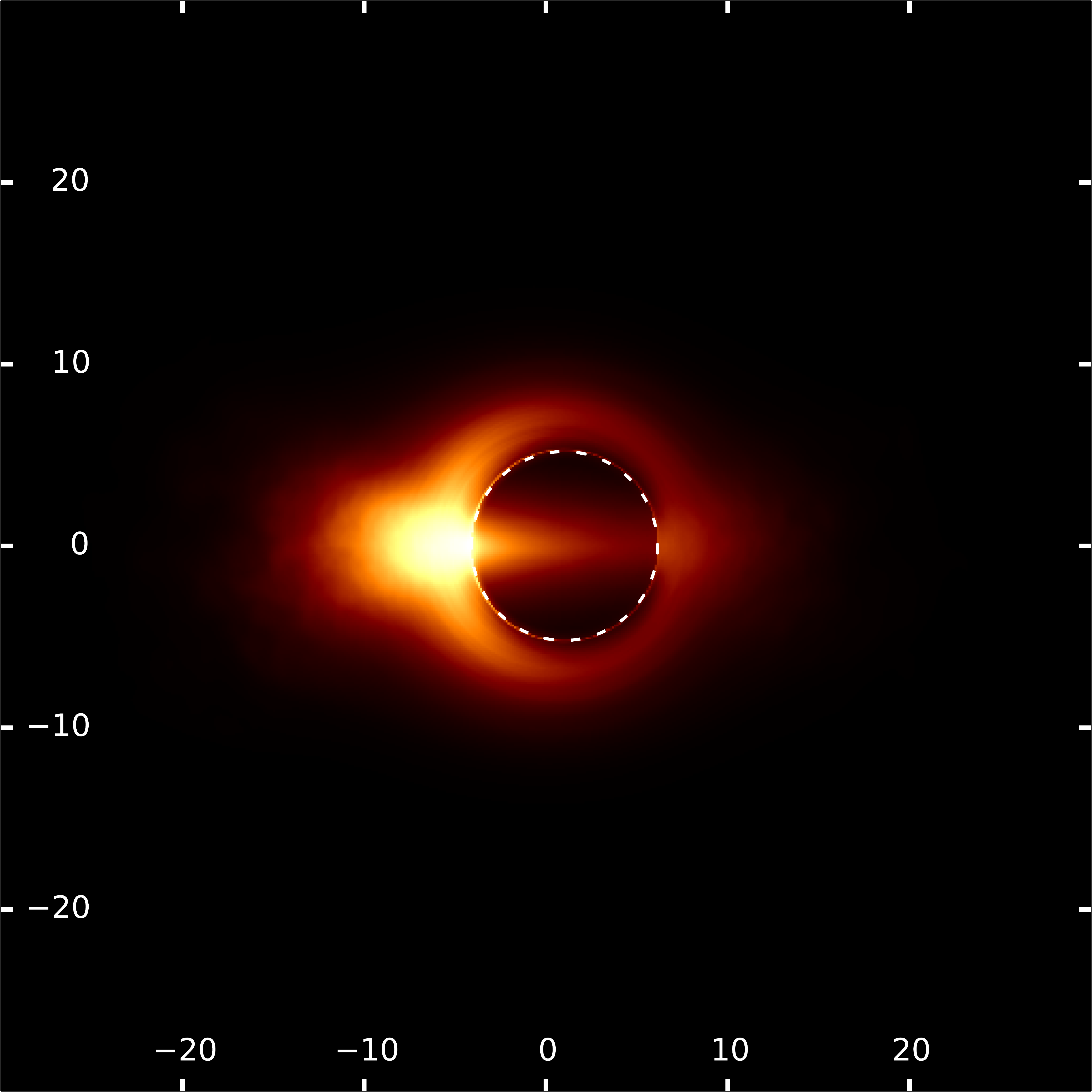}
	\caption{$a=0.5$, $i=90^\circ$.}
\end{subfigure}
\begin{subfigure}[b]{0.197\textwidth}
	\includegraphics[width=\textwidth]{Figures/sane_disk_a15o16_90_25-crop}
	\caption{$a=0.9375$, $i=90^\circ$.}
\end{subfigure}
\caption{Time-averaged, normalised intensity maps of our SANE, disc-dominated GRMHD models of Sgr A*, imaged at 230 GHz, at five different spins and four observer inclination angles, with an integrated flux density of 2.5 Jy. In each case, the photon ring, which marks the BHS, is indicated by a dashed line. The values for the impact parameters along the x- and y-axes are expressed in terms of $R_{\rm g}$. The image maps were plotted using a square-root intensity scale.}
\label{fig:sane_disk_25_matrix}
\end{figure*}

\begin{figure*}
\centering
\begin{subfigure}[b]{0.197\textwidth}
	\includegraphics[width=\textwidth]{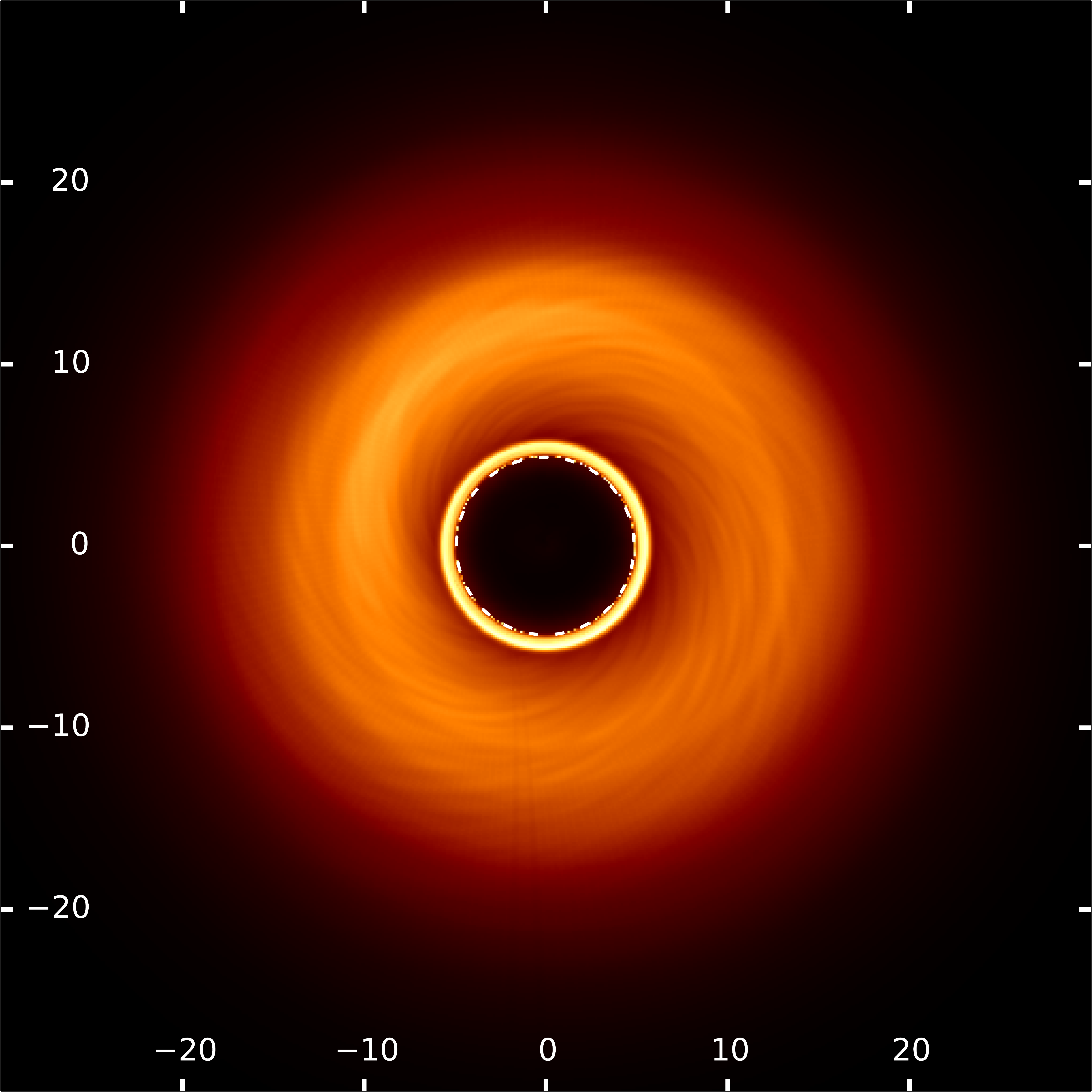}
	\caption{$a=-0.9375$, $i=1^\circ$.}
\end{subfigure}
\begin{subfigure}[b]{0.197\textwidth}
	\includegraphics[width=\textwidth]{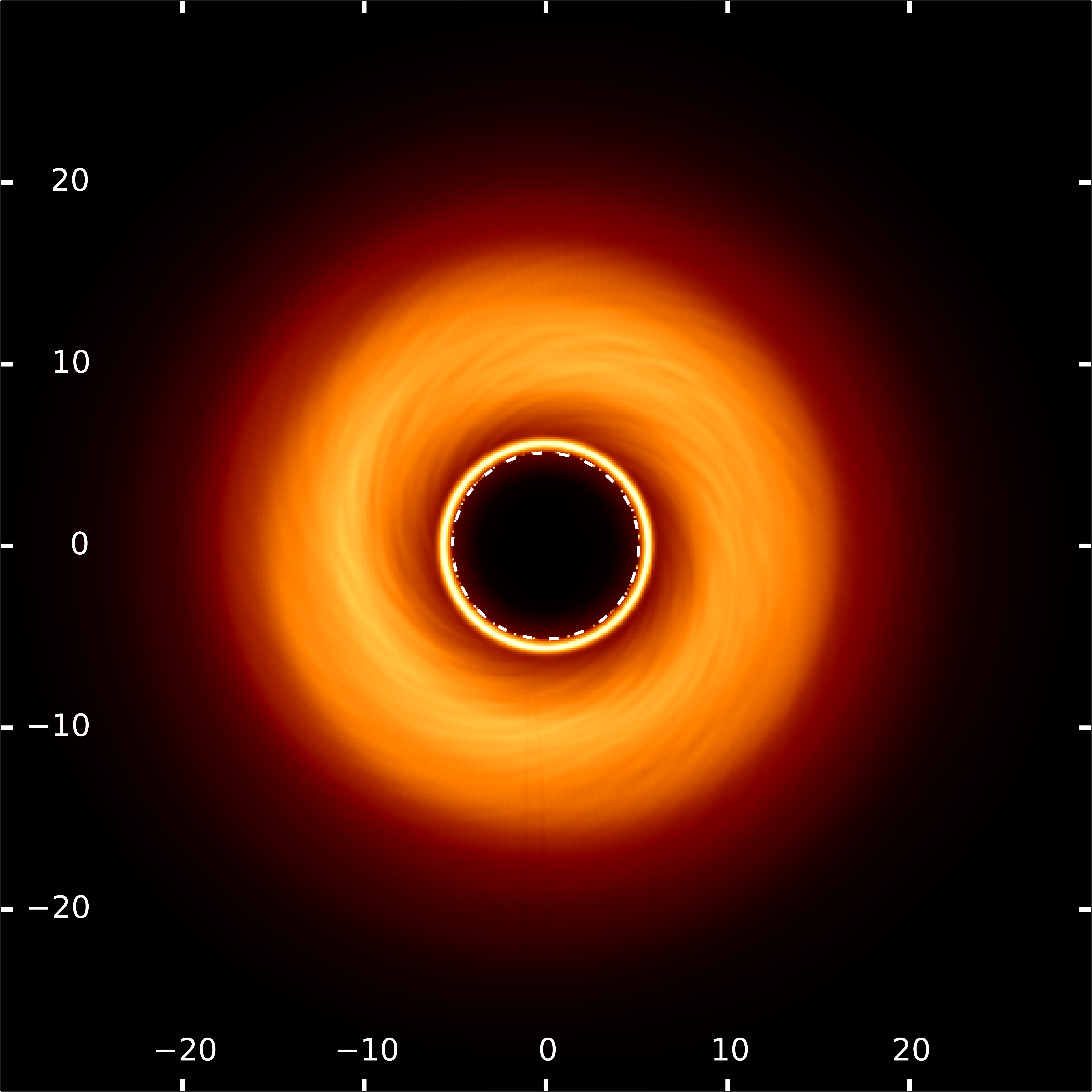}
	\caption{$a=-0.5$, $i=1^\circ$.}
\end{subfigure}
\begin{subfigure}[b]{0.197\textwidth}
	\includegraphics[width=\textwidth]{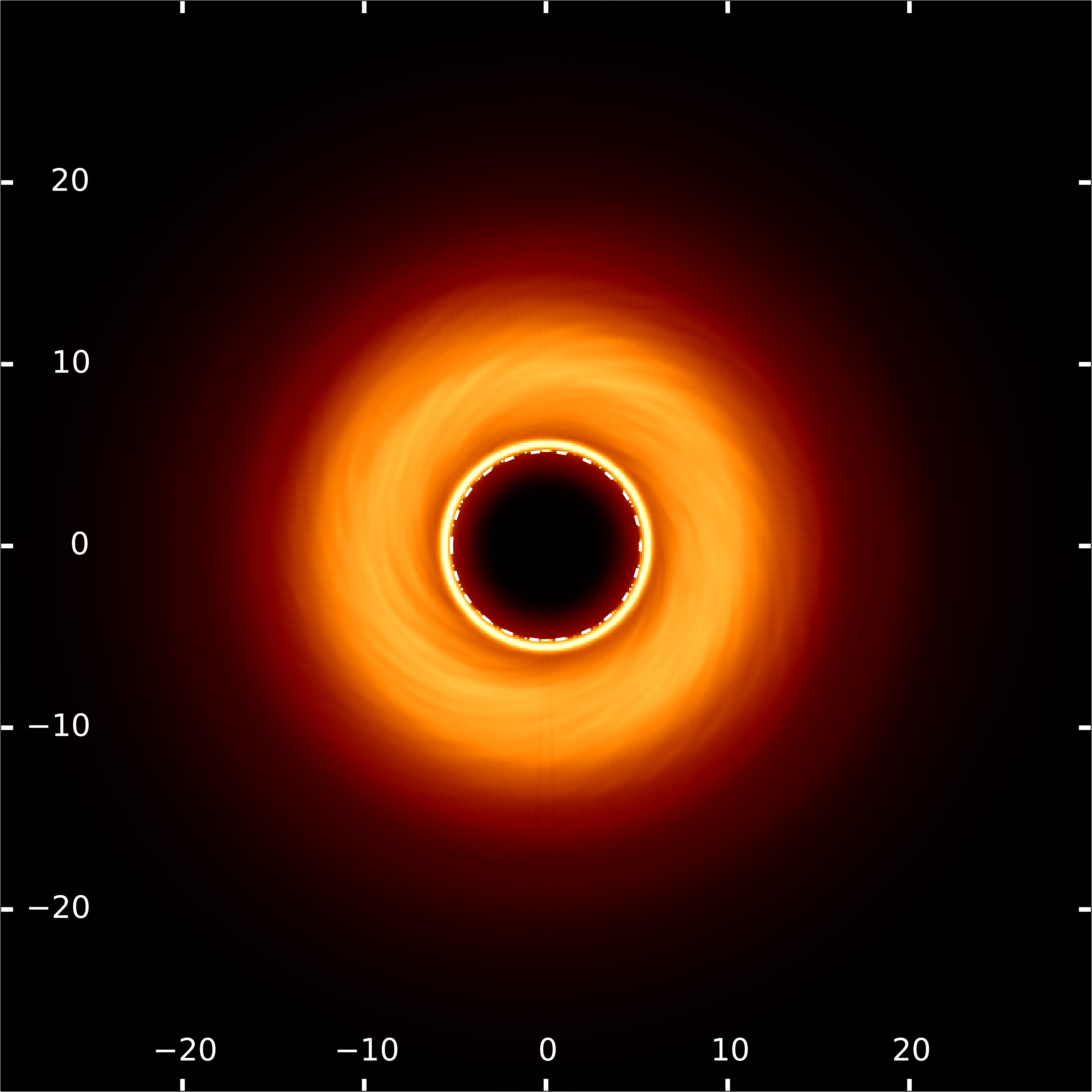}
	\caption{$a=0$, $i=1^\circ$.}
\end{subfigure}
\begin{subfigure}[b]{0.197\textwidth}
	\includegraphics[width=\textwidth]{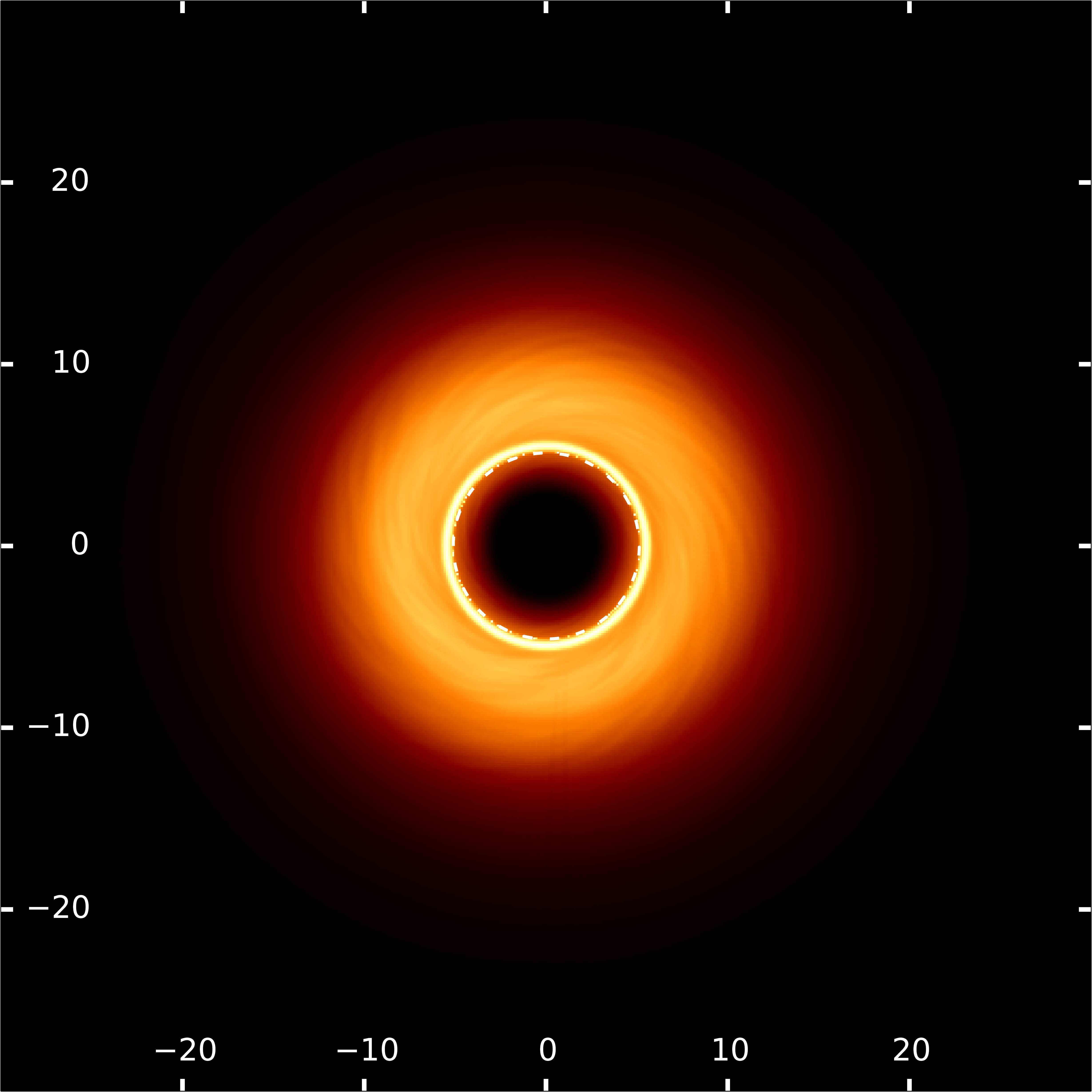}
	\caption{$a=0.5$, $i=1^\circ$.}
\end{subfigure}
\begin{subfigure}[b]{0.197\textwidth}
	\includegraphics[width=\textwidth]{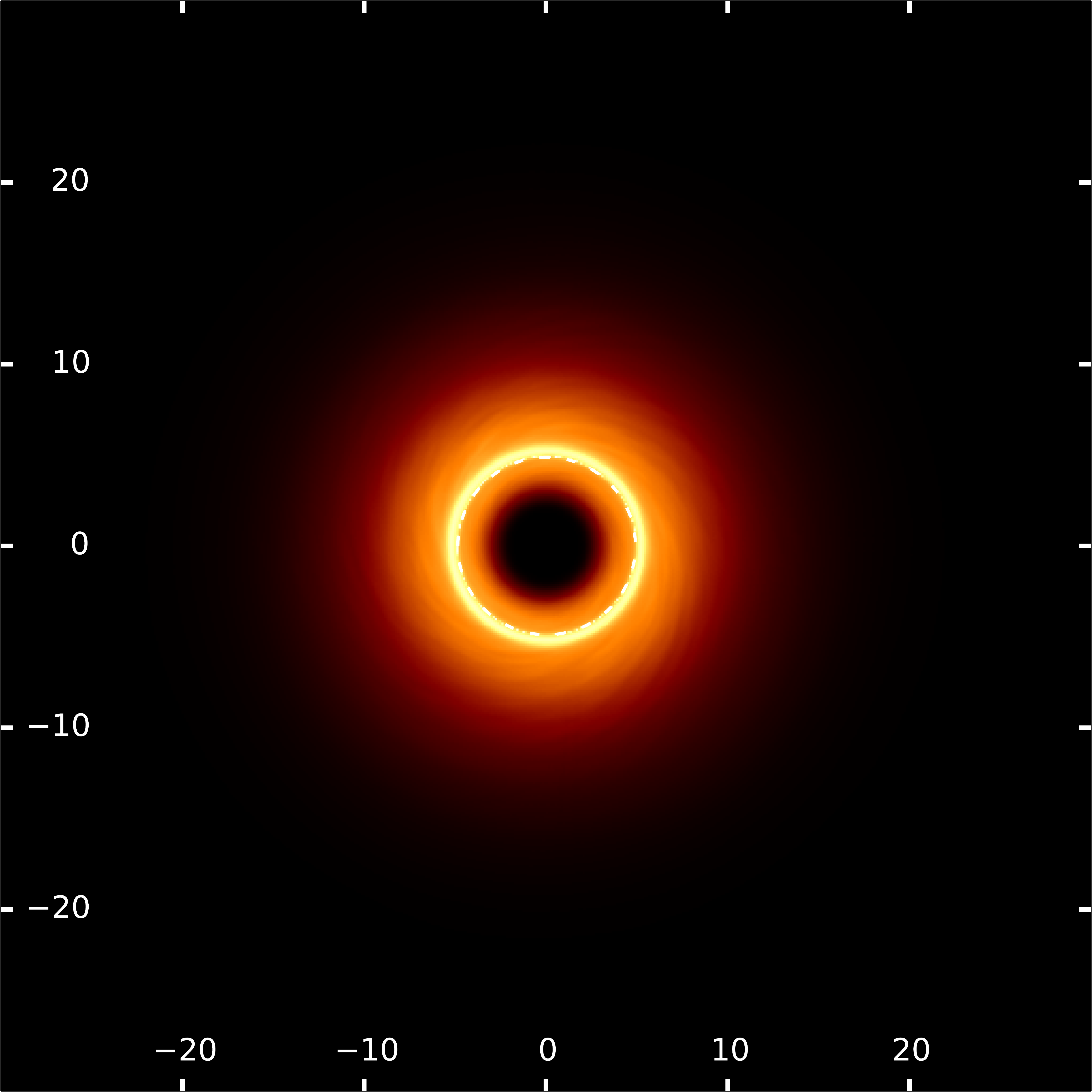}
	\caption{$a=0.9375$, $i=1^\circ$.}
\end{subfigure}
\begin{subfigure}[b]{0.197\textwidth}
	\includegraphics[width=\textwidth]{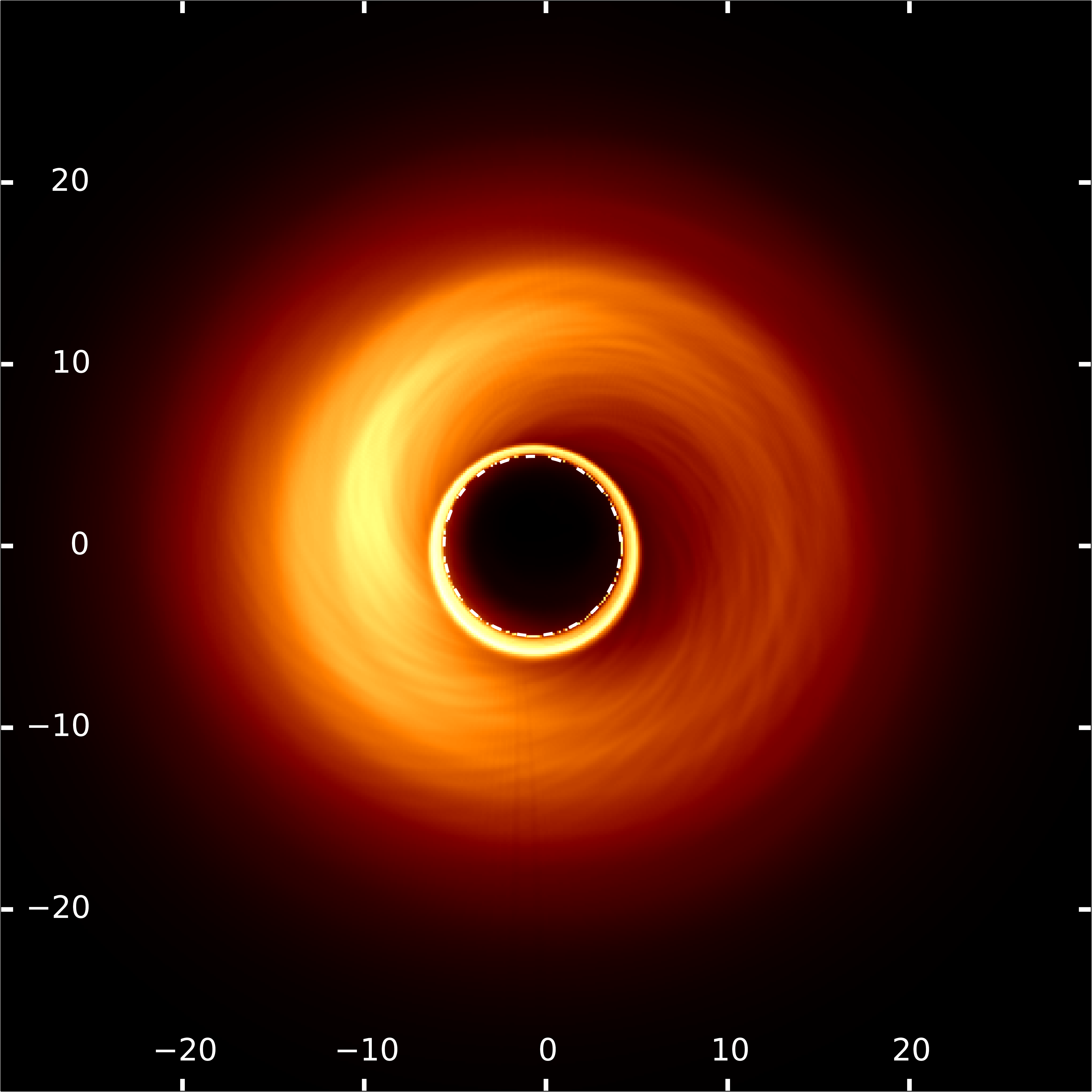}
	\caption{$a=-0.9375$, $i=20^\circ$.}
\end{subfigure}
\begin{subfigure}[b]{0.197\textwidth}
	\includegraphics[width=\textwidth]{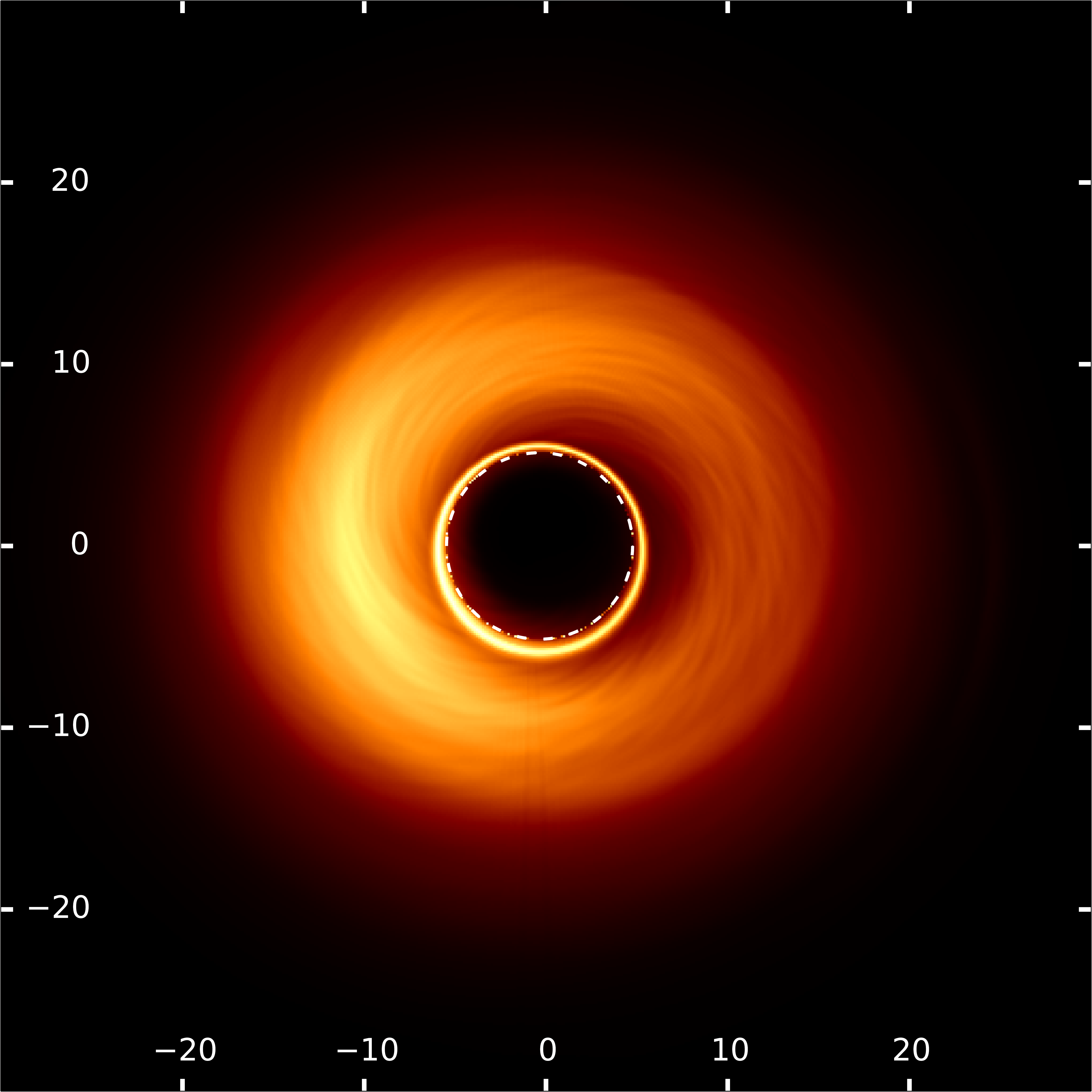}
	\caption{$a=-0.5$, $i=20^\circ$.}
\end{subfigure}
\begin{subfigure}[b]{0.197\textwidth}
	\includegraphics[width=\textwidth]{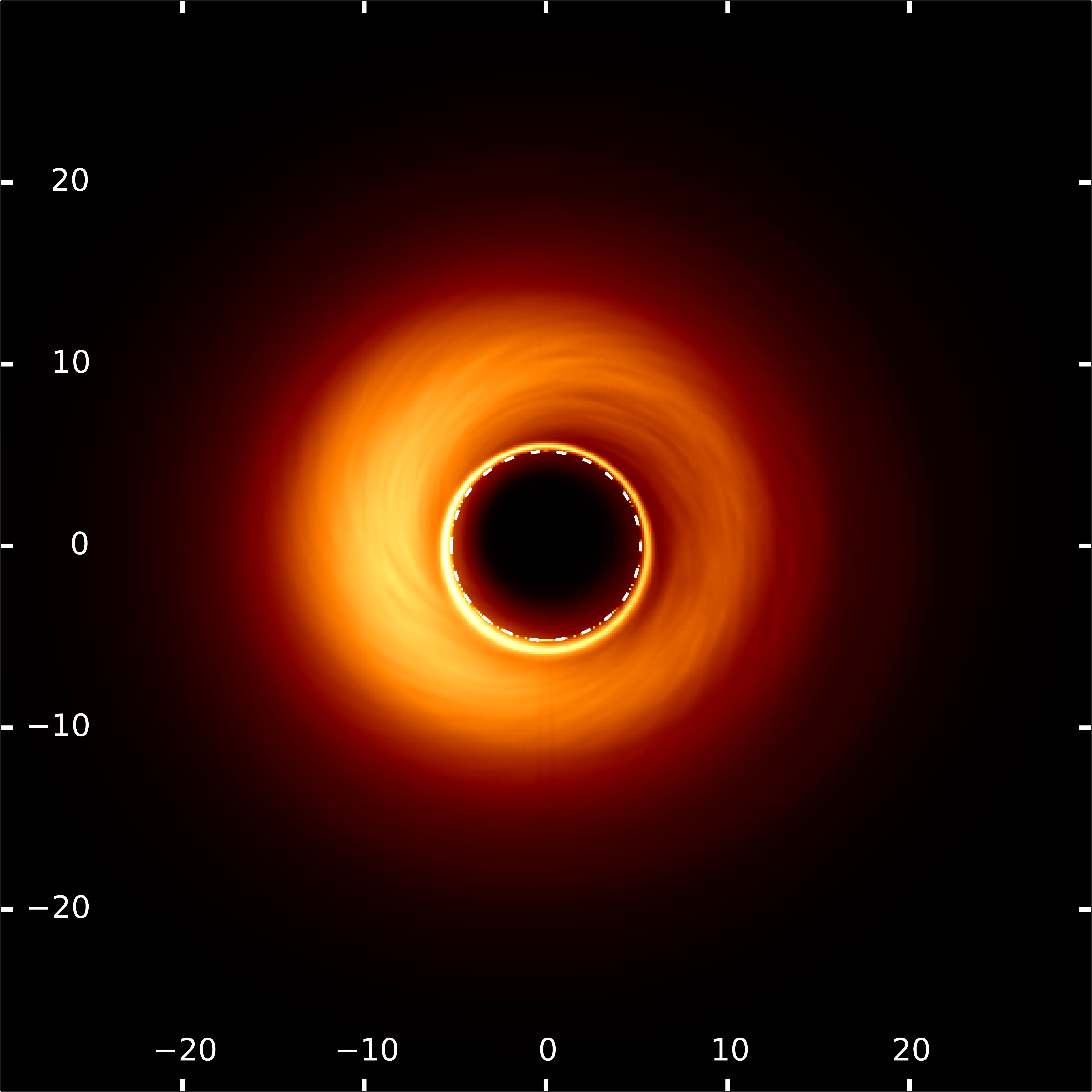}
	\caption{$a=0$, $i=20^\circ$.}
\end{subfigure}
\begin{subfigure}[b]{0.197\textwidth}
	\includegraphics[width=\textwidth]{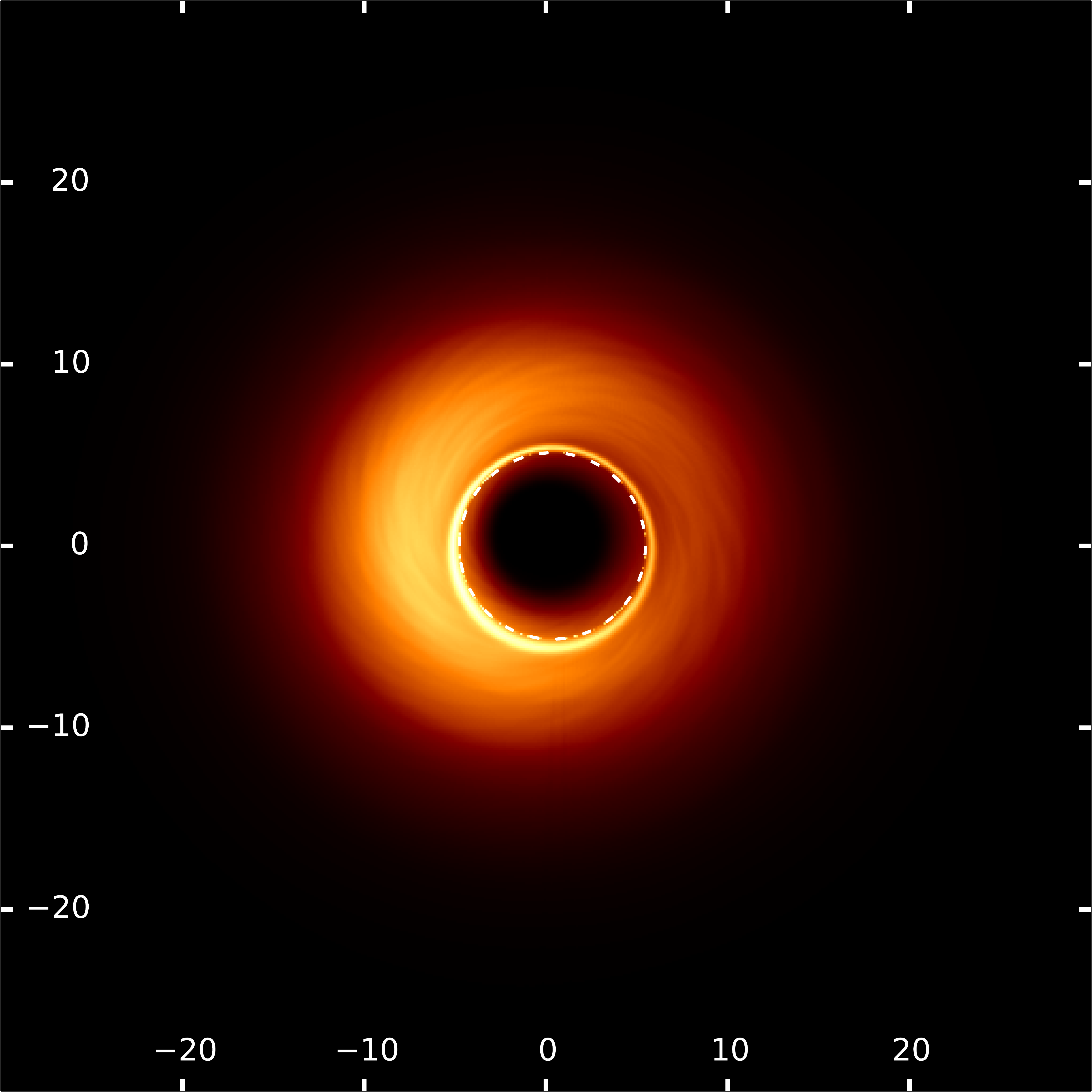}
	\caption{$a=0.5$, $i=20^\circ$.}
\end{subfigure}
\begin{subfigure}[b]{0.197\textwidth}
	\includegraphics[width=\textwidth]{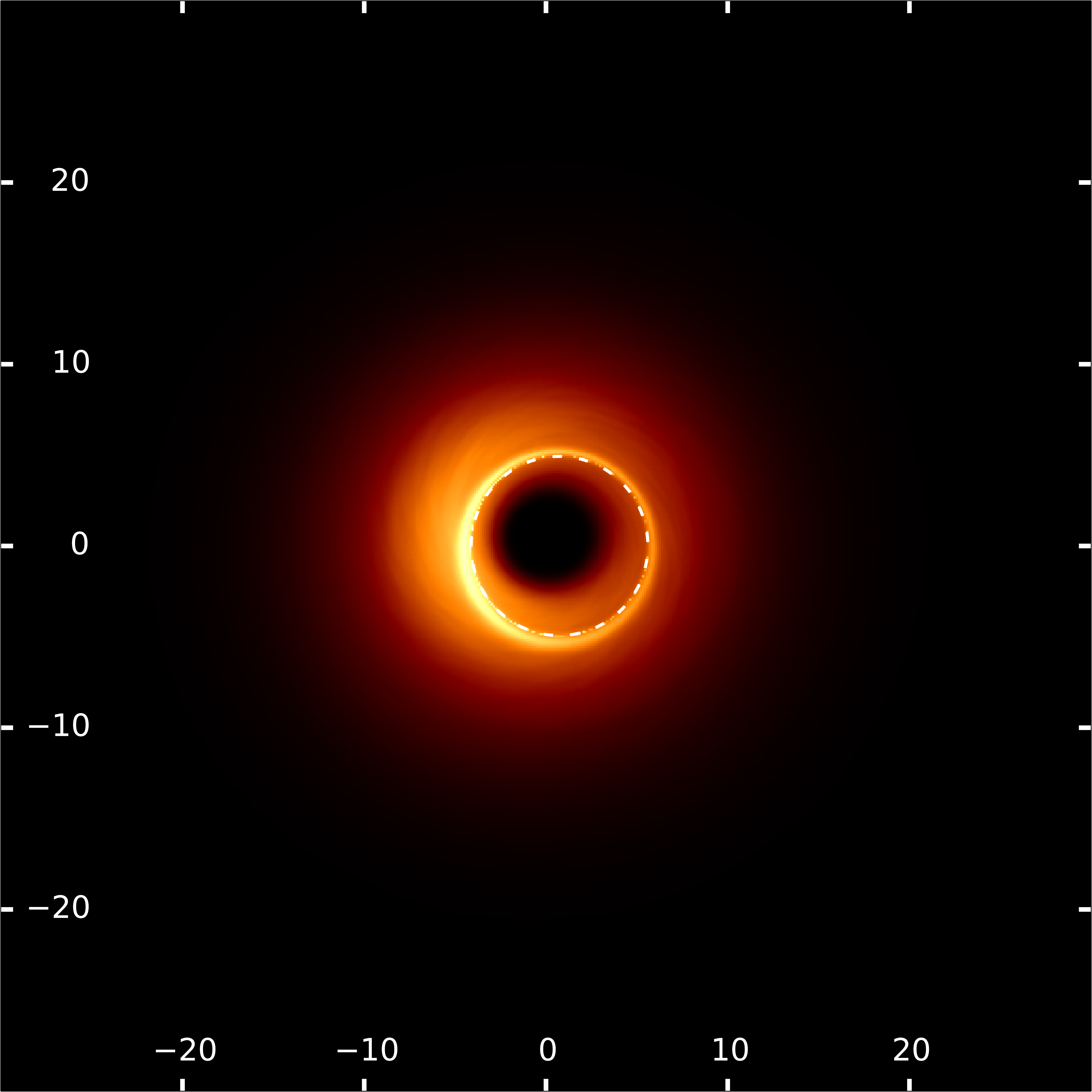}
	\caption{$a=0.9375$, $i=20^\circ$.}
\end{subfigure}
\begin{subfigure}[b]{0.197\textwidth}
	\includegraphics[width=\textwidth]{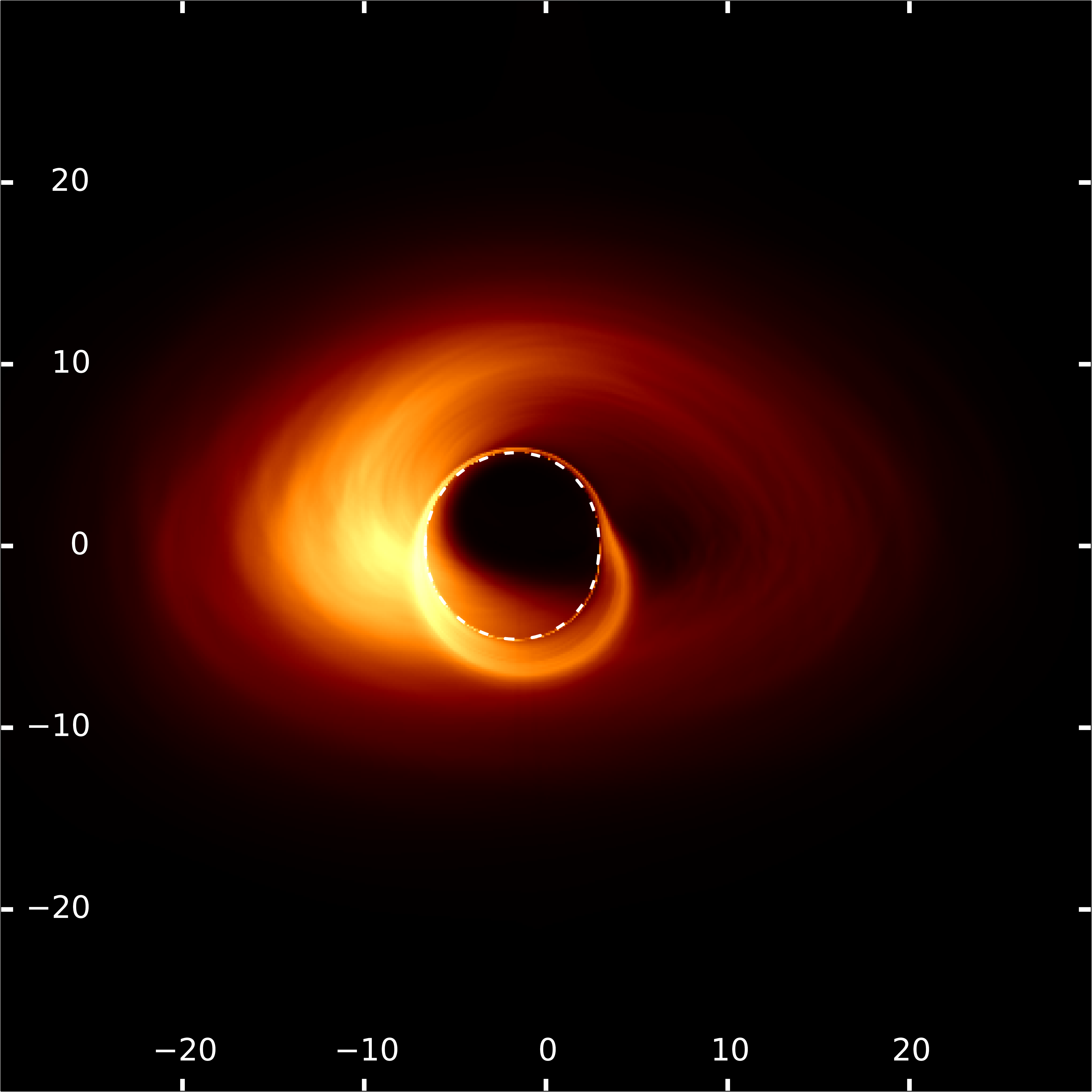}
	\caption{$a=-0.9375$, $i=60^\circ$.}
\end{subfigure}
\begin{subfigure}[b]{0.197\textwidth}
	\includegraphics[width=\textwidth]{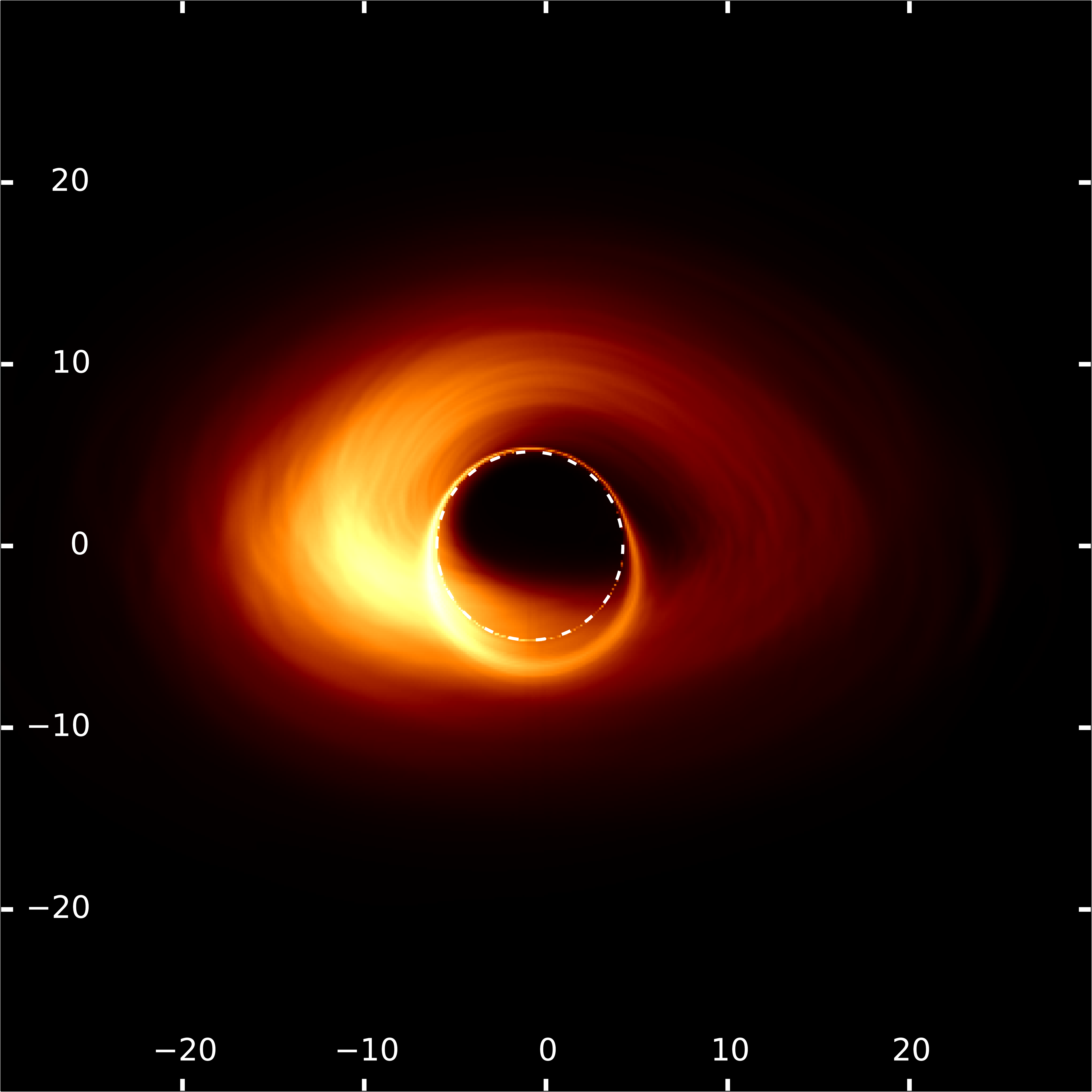}
	\caption{$a=-0.5$, $i=60^\circ$.}
\end{subfigure}
\begin{subfigure}[b]{0.197\textwidth}
	\includegraphics[width=\textwidth]{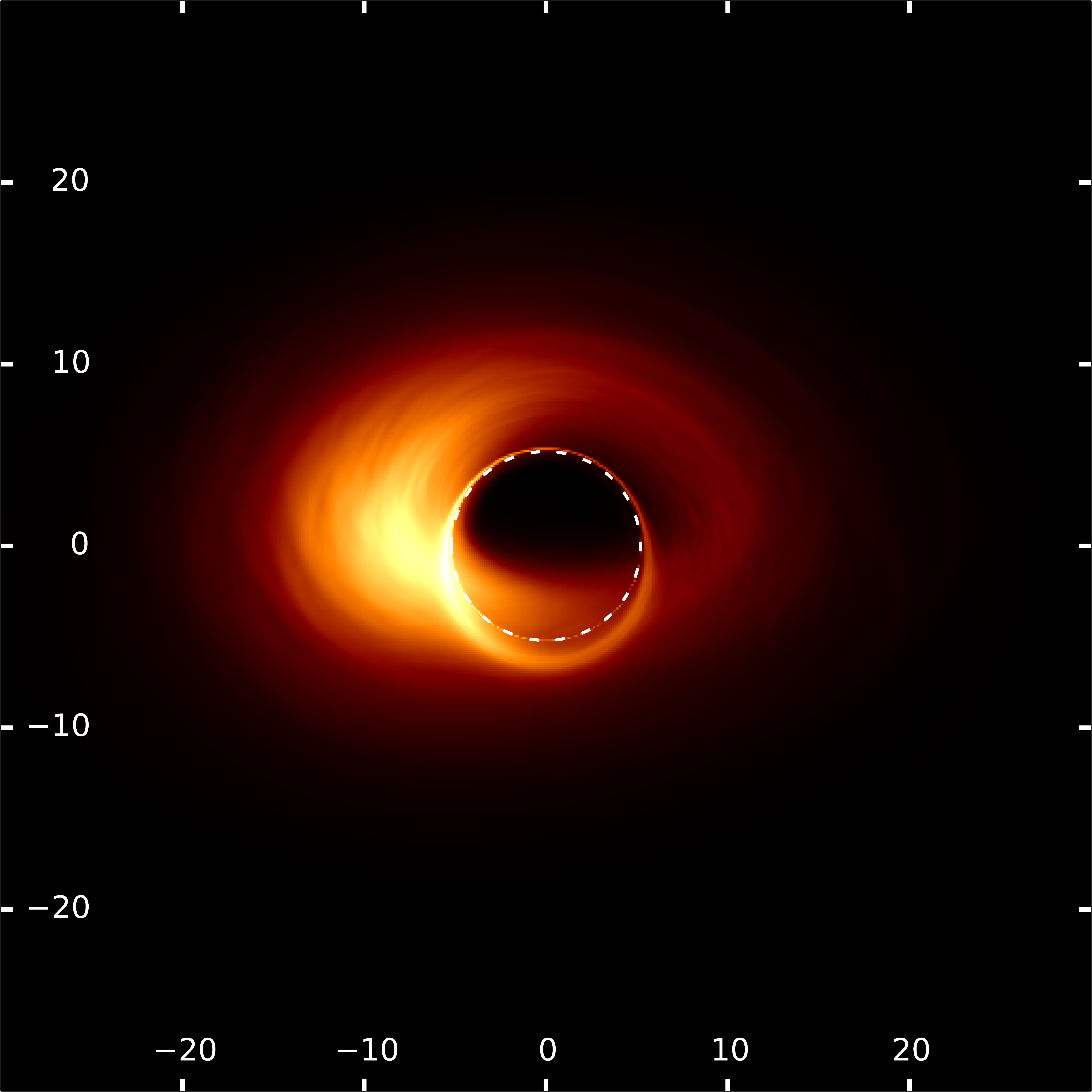}
	\caption{$a=0$, $i=60^\circ$.}
\end{subfigure}
\begin{subfigure}[b]{0.197\textwidth}
	\includegraphics[width=\textwidth]{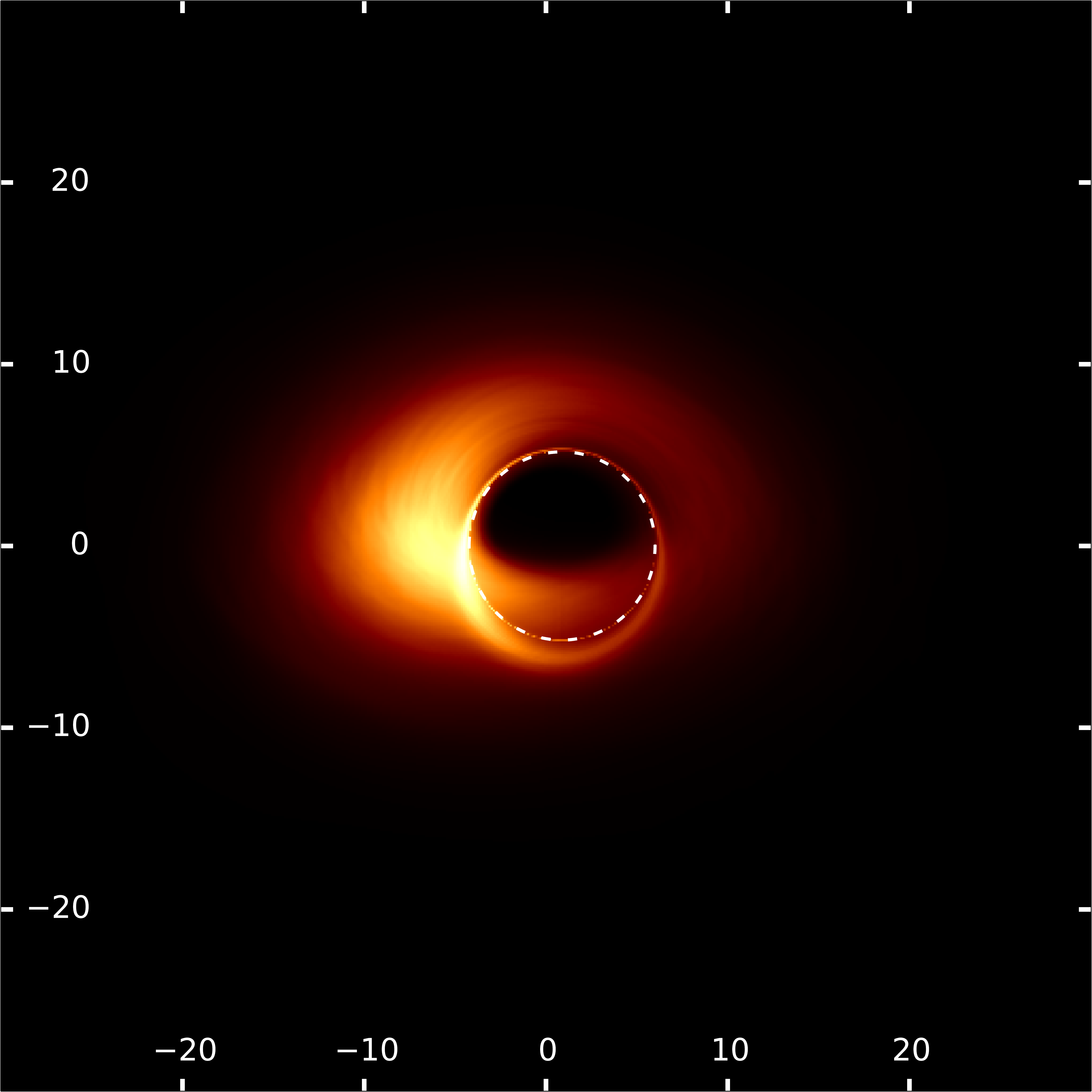}
	\caption{$a=0.5$, $i=60^\circ$.}
\end{subfigure}
\begin{subfigure}[b]{0.197\textwidth}
	\includegraphics[width=\textwidth]{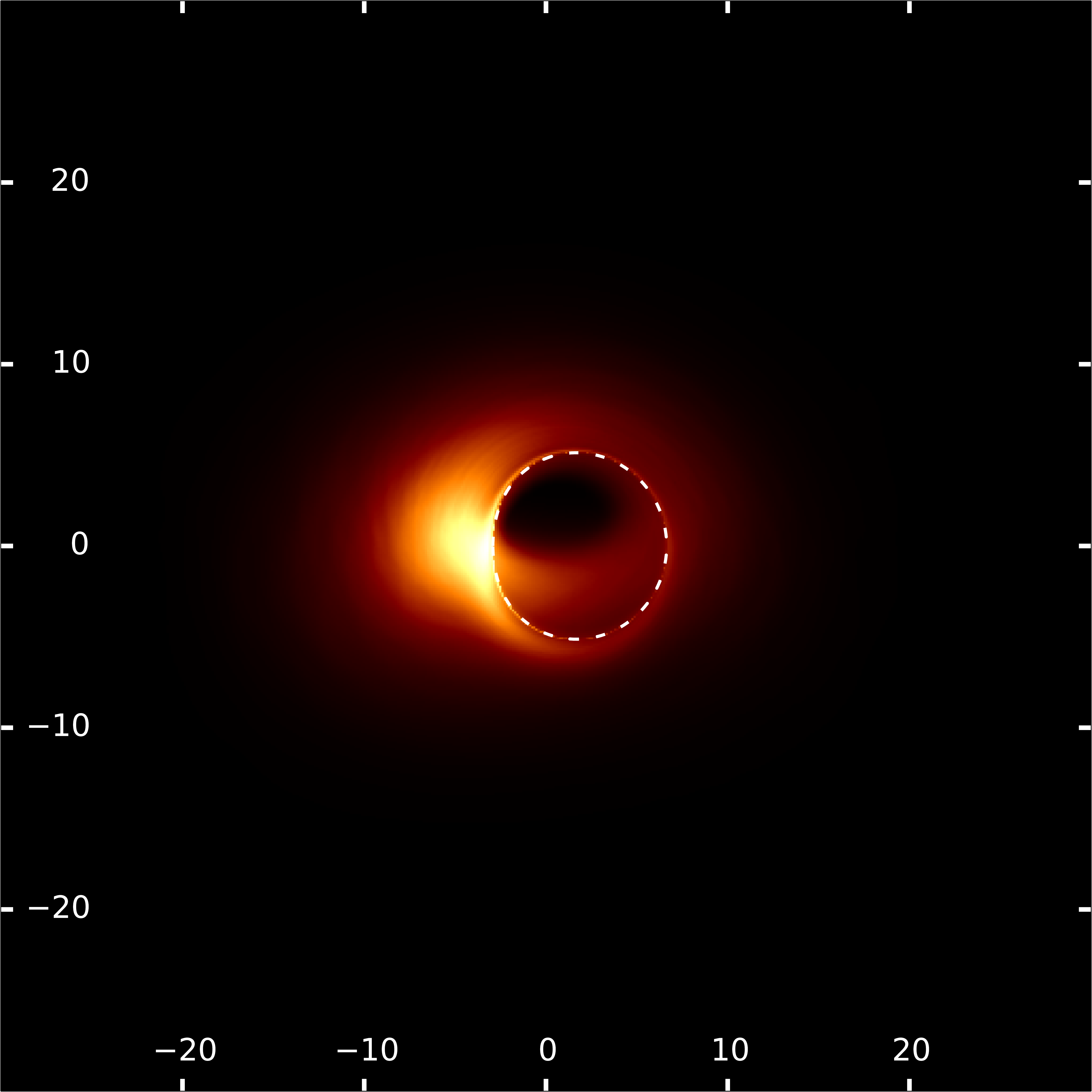}
	\caption{$a=0.9375$, $i=60^\circ$.}
\end{subfigure}
\begin{subfigure}[b]{0.197\textwidth}
	\includegraphics[width=\textwidth]{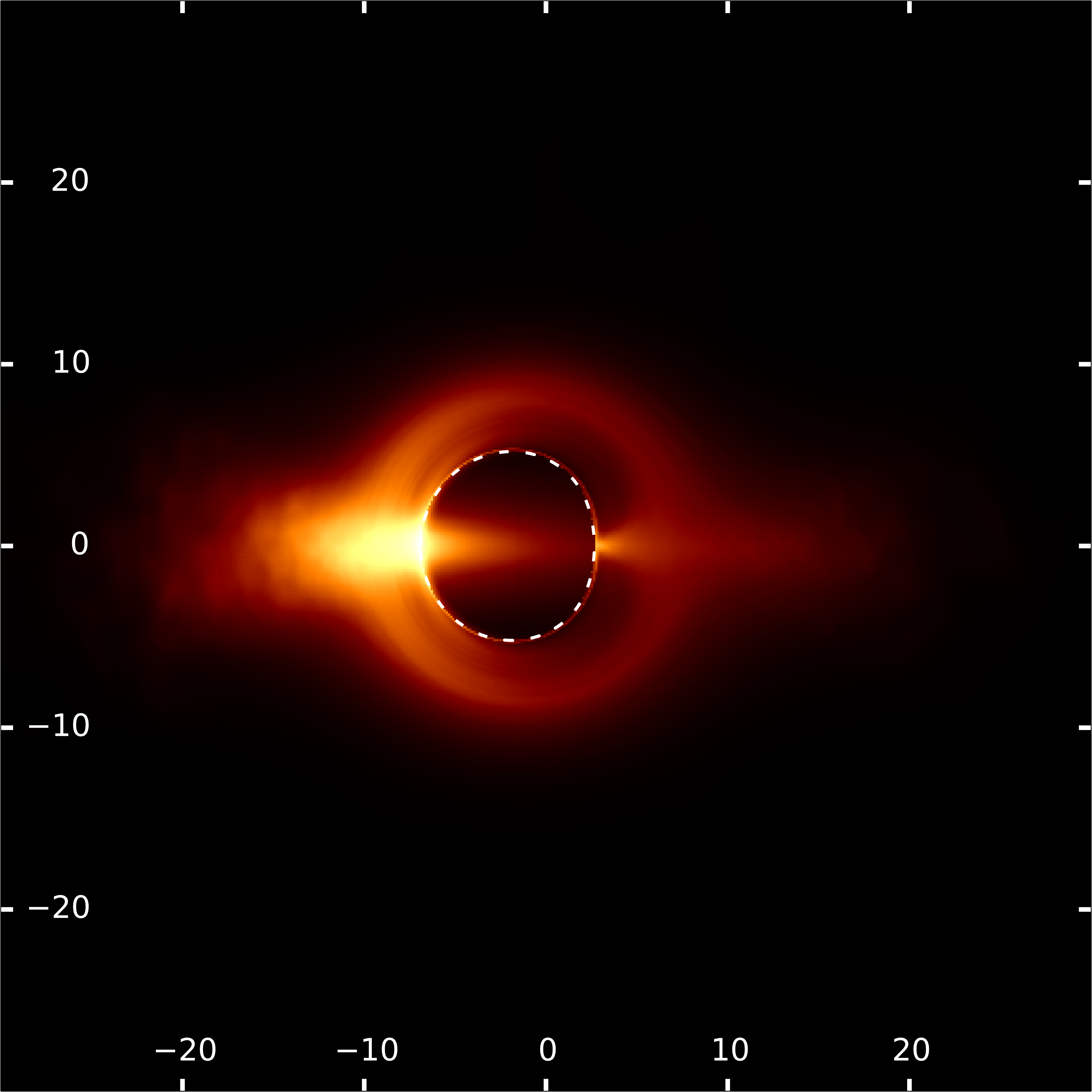}
	\caption{$a=-0.9375$, $i=90^\circ$.}
\end{subfigure}
\begin{subfigure}[b]{0.197\textwidth}
	\includegraphics[width=\textwidth]{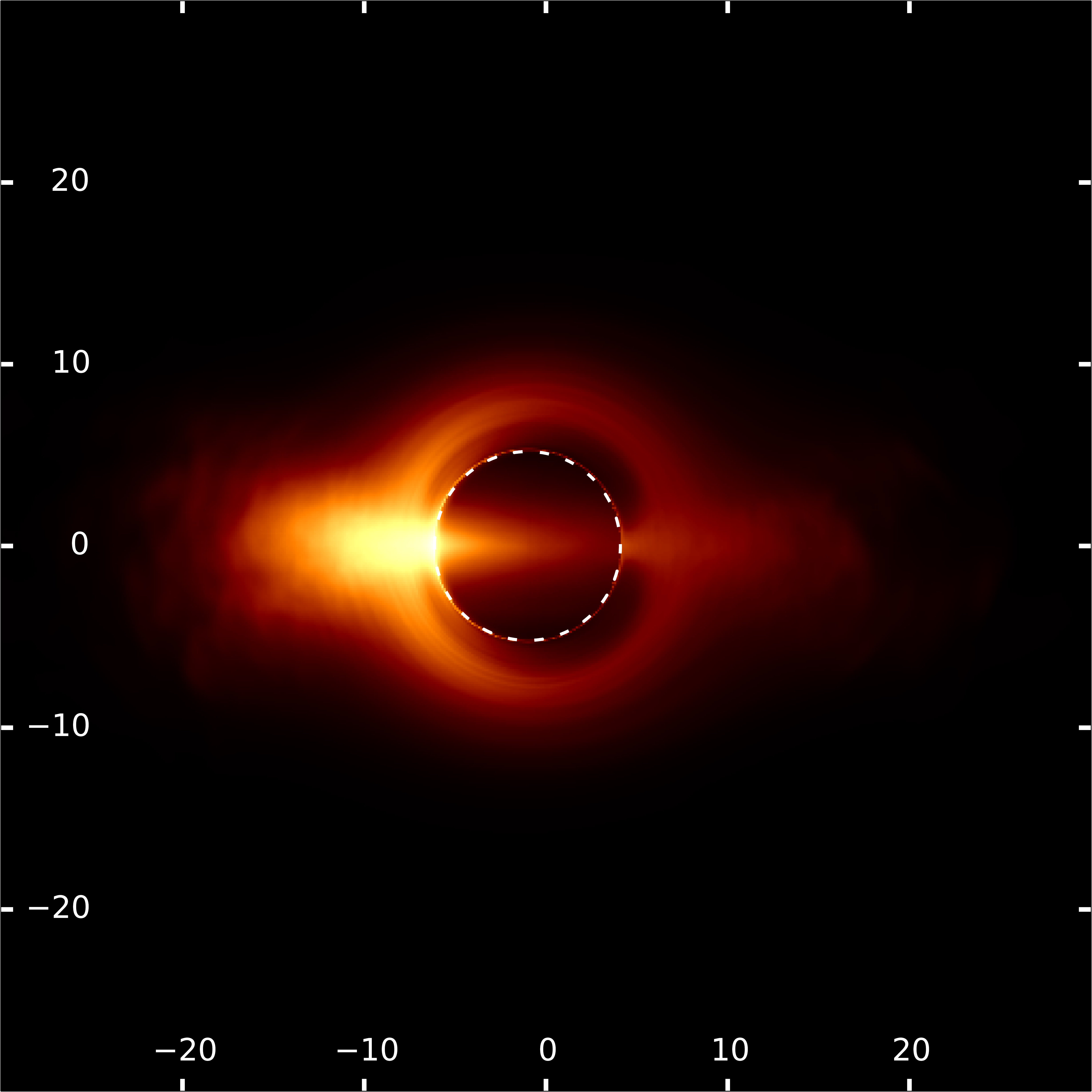}
	\caption{$a=-0.5$, $i=90^\circ$.}
\end{subfigure}
\begin{subfigure}[b]{0.197\textwidth}
	\includegraphics[width=\textwidth]{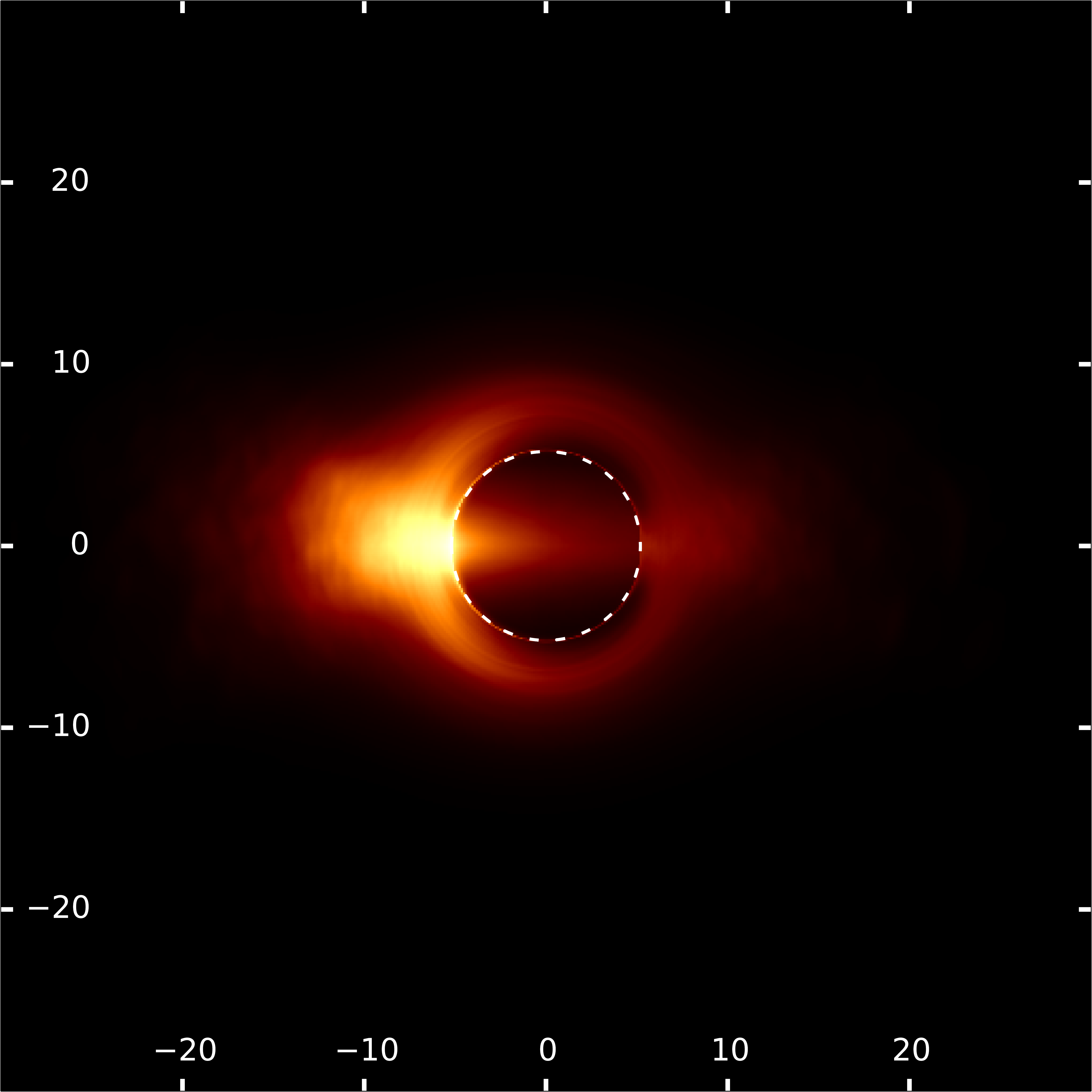}
	\caption{$a=0$, $i=90^\circ$.}
\end{subfigure}
\begin{subfigure}[b]{0.197\textwidth}
	\includegraphics[width=\textwidth]{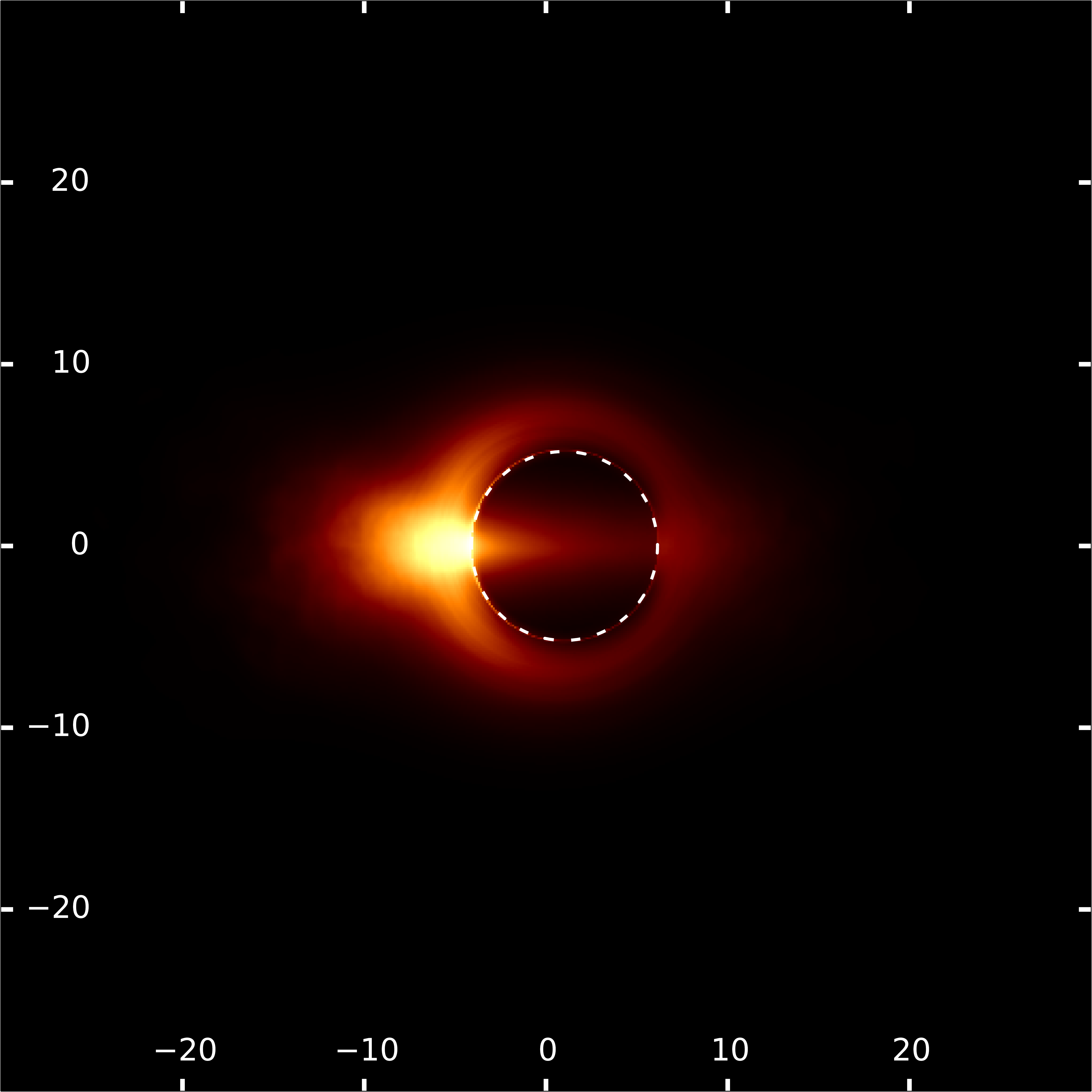}
	\caption{$a=0.5$, $i=90^\circ$.}
\end{subfigure}
\begin{subfigure}[b]{0.197\textwidth}
	\includegraphics[width=\textwidth]{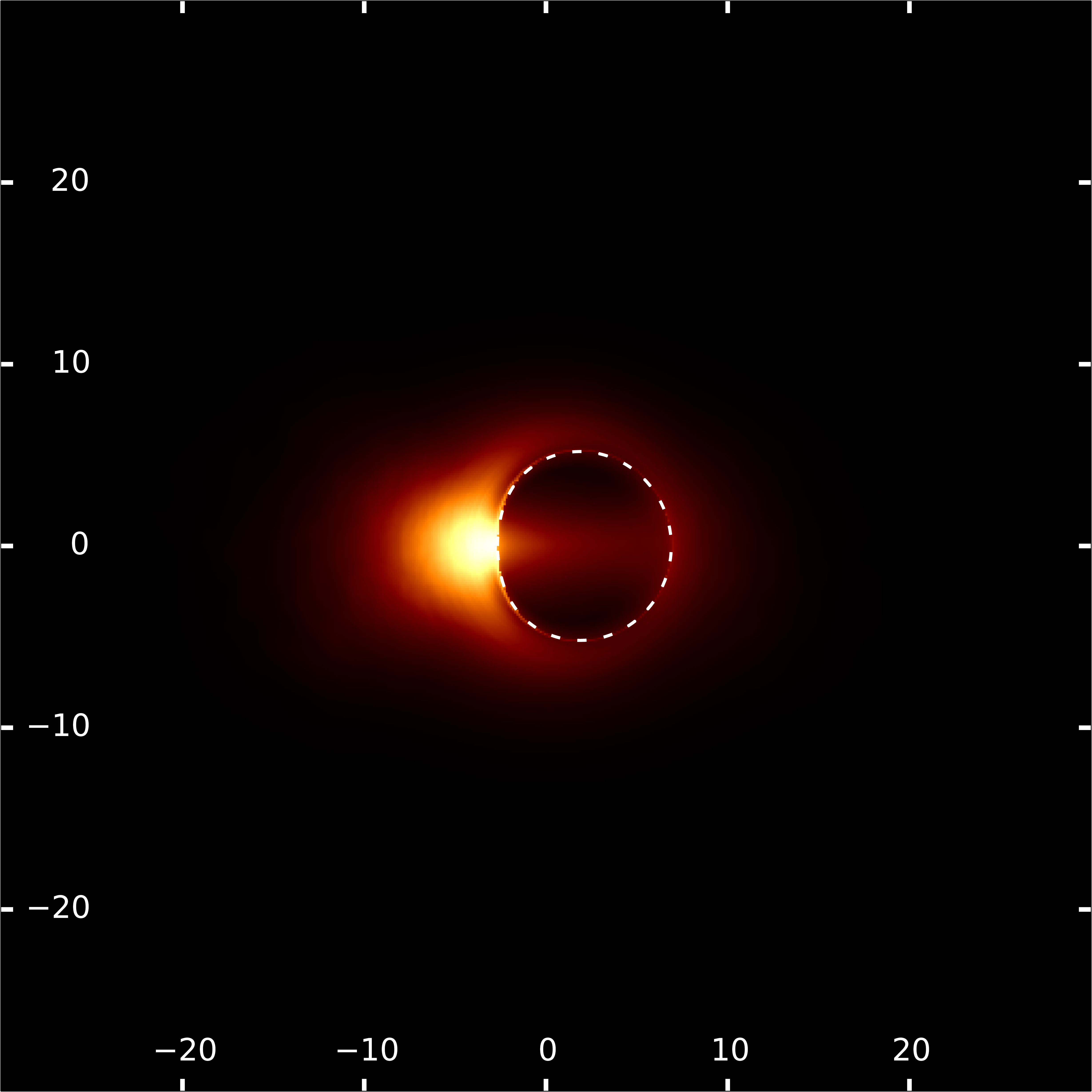}
	\caption{$a=0.9375$, $i=90^\circ$.}
\end{subfigure}
\caption{Time-averaged, normalised intensity maps of our SANE, disc-dominated GRMHD models of Sgr A*, imaged at 230 GHz, at five different spins and four observer inclination angles, with an integrated flux density of 1.25 Jy. In each case, the photon ring, which marks the BHS, is indicated by a dashed line. The values for the impact parameters along the x- and y-axes are expressed in terms of $R_{\rm g}$. The image maps were plotted using a square-root intensity scale.}
\label{fig:sane_disk_125_matrix}
\end{figure*}

\begin{figure*}
\centering
\begin{subfigure}[b]{0.197\textwidth}
	\includegraphics[width=\textwidth]{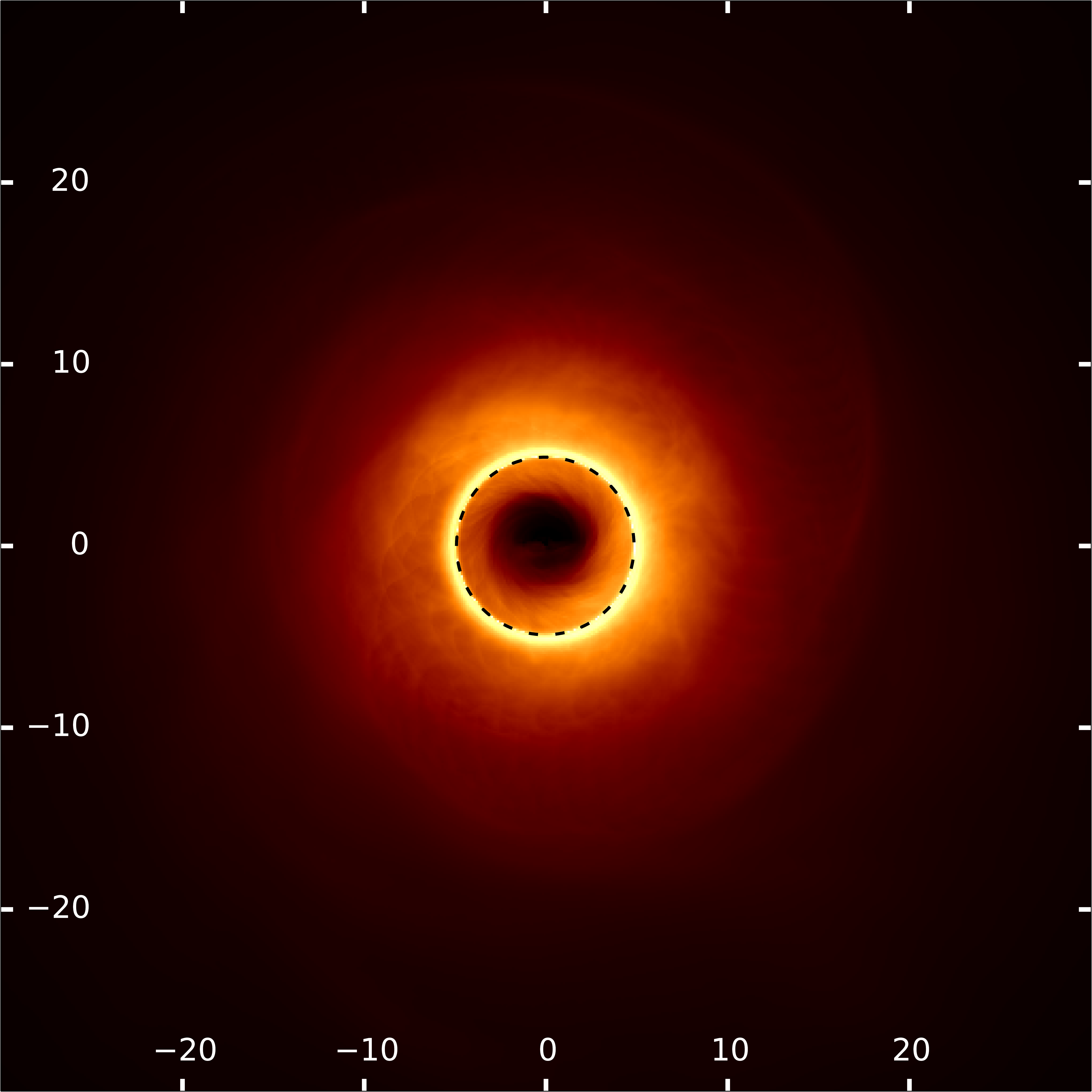}
	\caption{$a=-0.9375$, $i=1^\circ$.}
\end{subfigure}
\begin{subfigure}[b]{0.197\textwidth}
	\includegraphics[width=\textwidth]{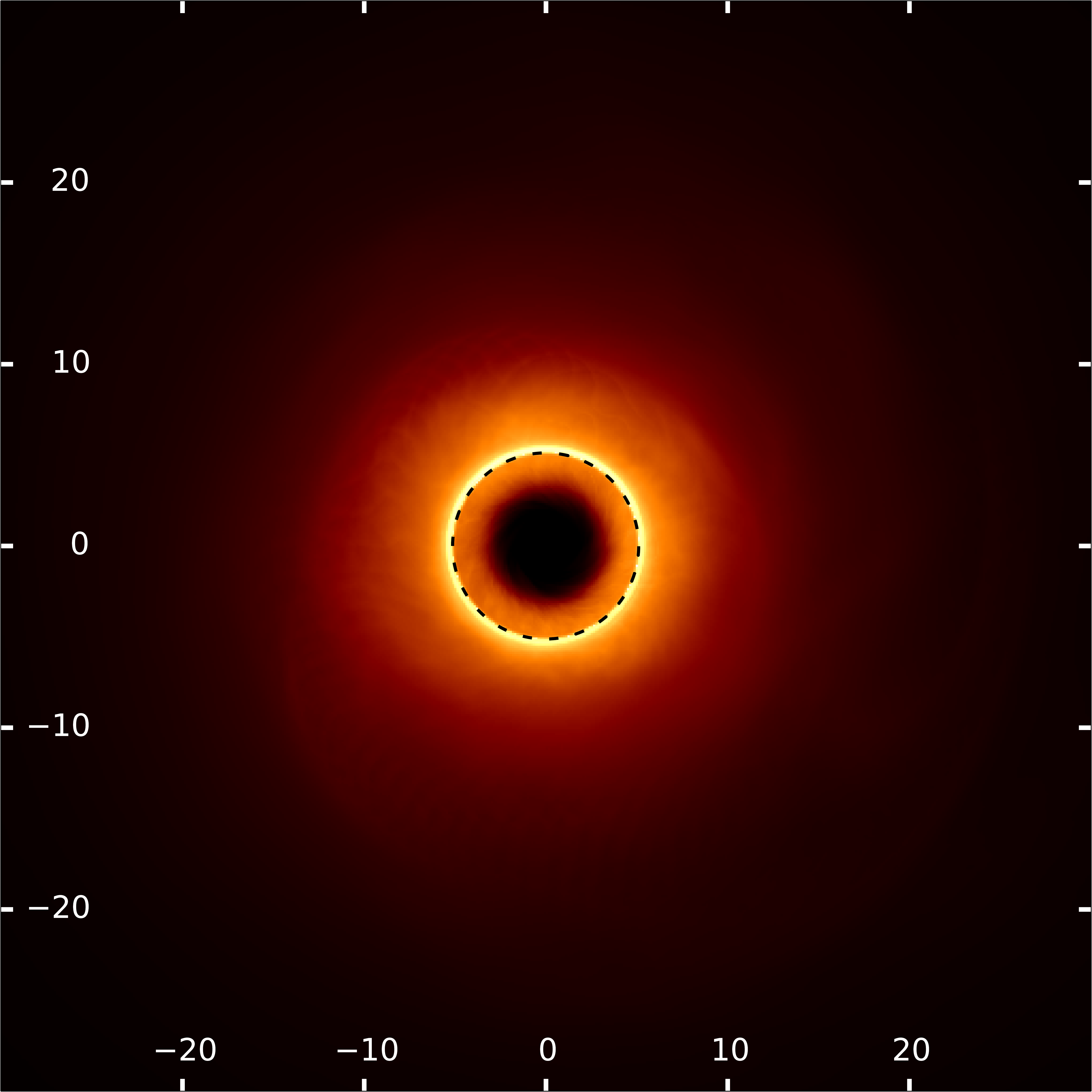}
	\caption{$a=-0.5$, $i=1^\circ$.}
\end{subfigure}
\begin{subfigure}[b]{0.197\textwidth}
	\includegraphics[width=\textwidth]{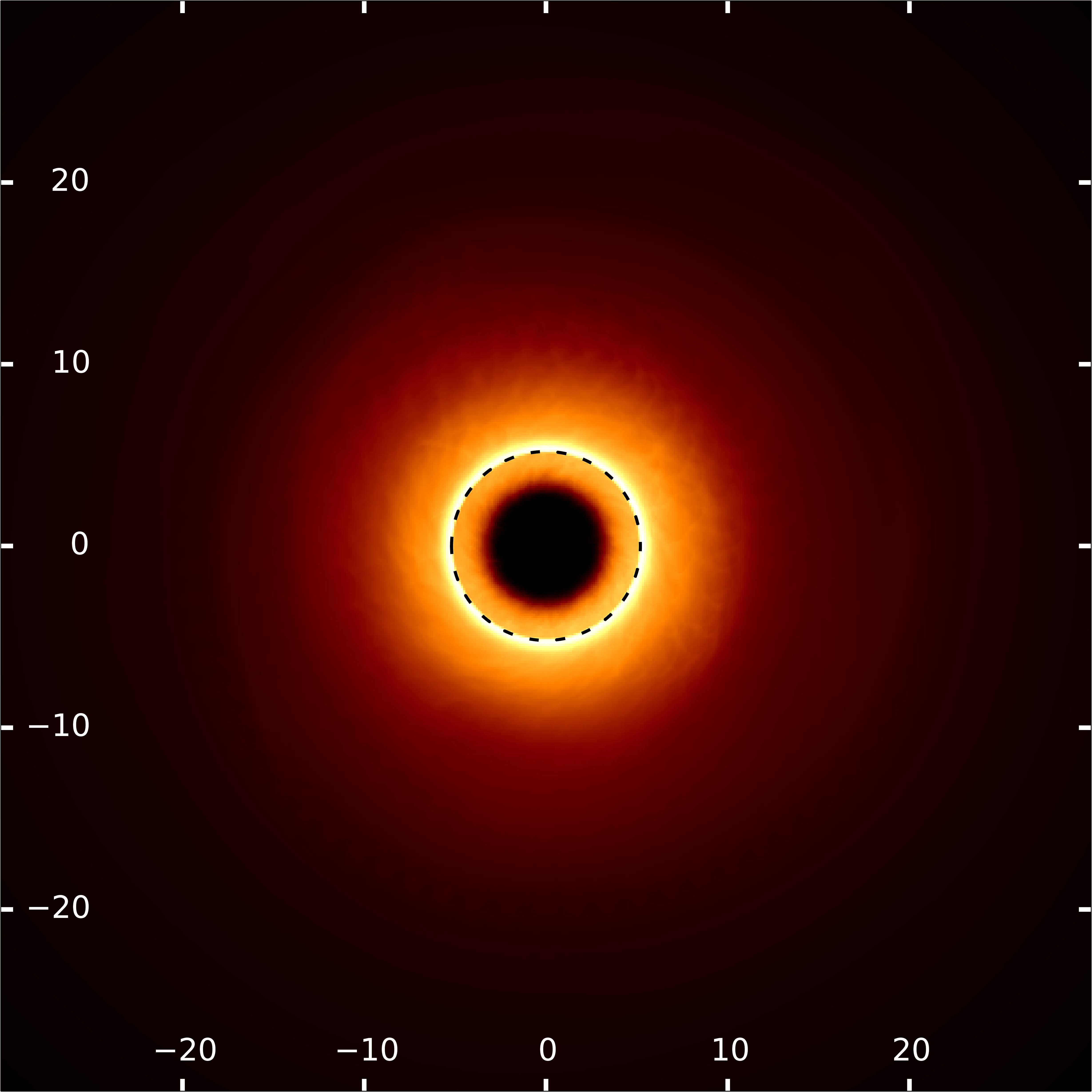}
	\caption{$a=0$, $i=1^\circ$.}
\end{subfigure}
\begin{subfigure}[b]{0.197\textwidth}
	\includegraphics[width=\textwidth]{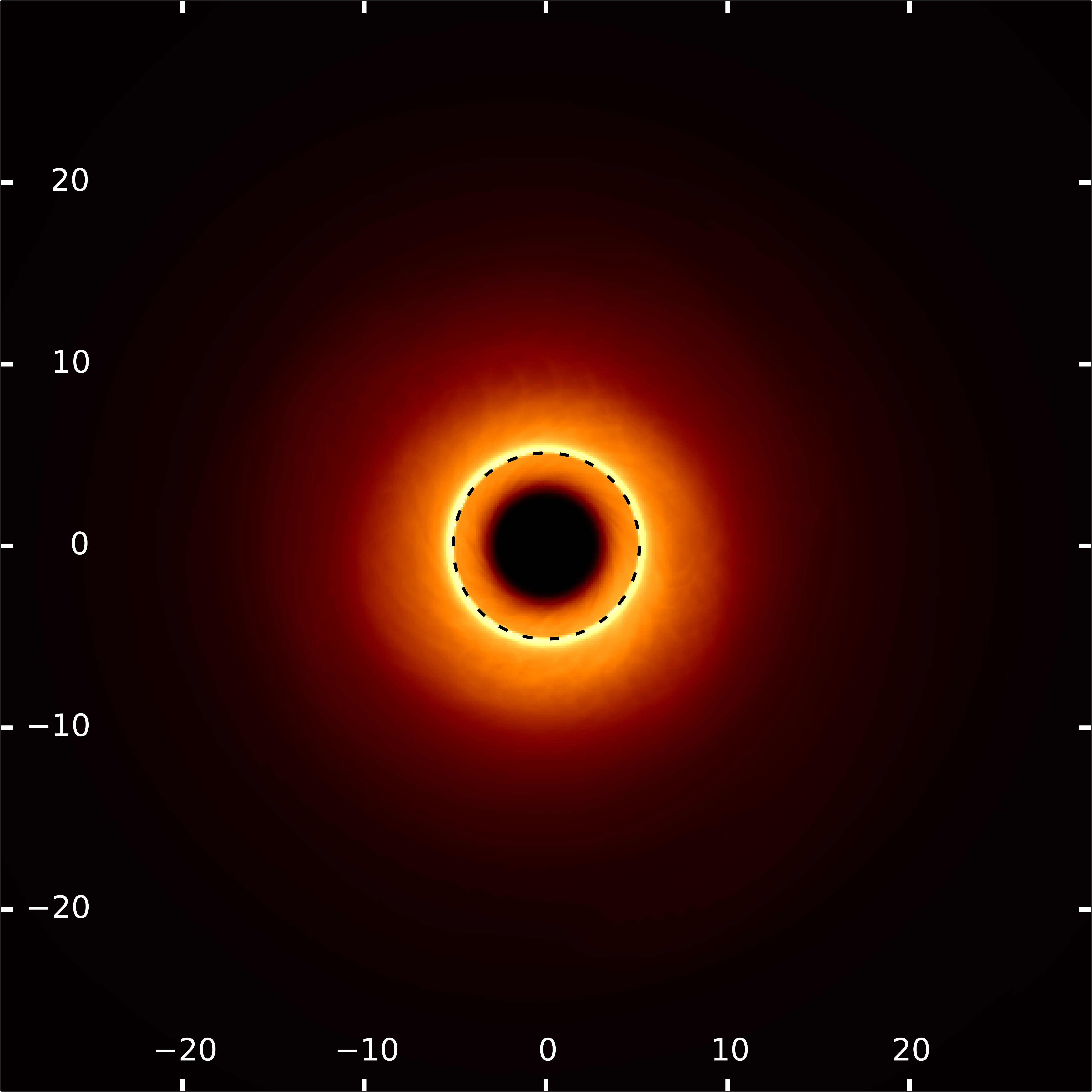}
	\caption{$a=0.5$, $i=1^\circ$.}
\end{subfigure}
\begin{subfigure}[b]{0.197\textwidth}
	\includegraphics[width=\textwidth]{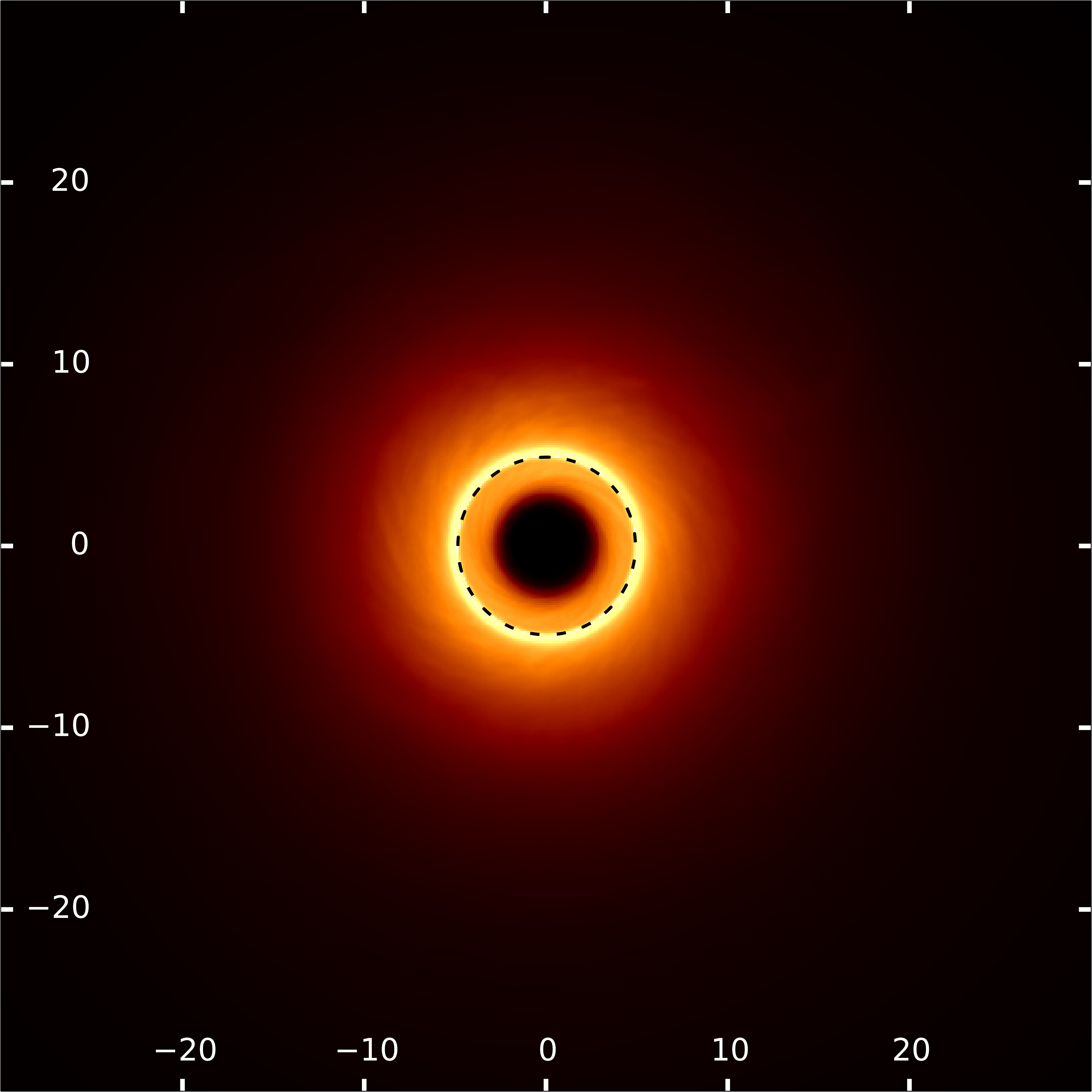}
	\caption{$a=0.9375$, $i=1^\circ$.}
\end{subfigure}
\begin{subfigure}[b]{0.197\textwidth}
	\includegraphics[width=\textwidth]{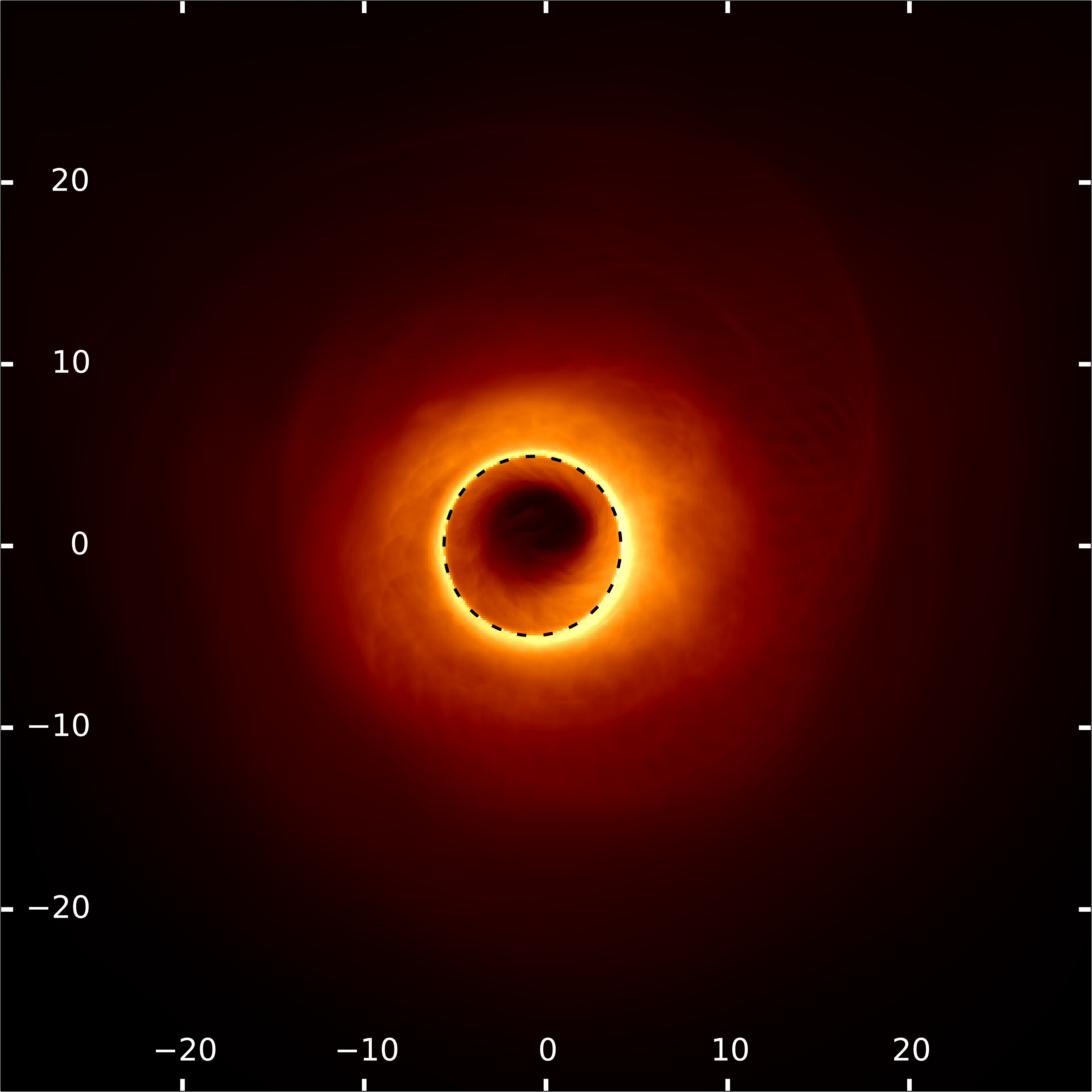}
	\caption{$a=-0.9375$, $i=20^\circ$.}
\end{subfigure}
\begin{subfigure}[b]{0.197\textwidth}
	\includegraphics[width=\textwidth]{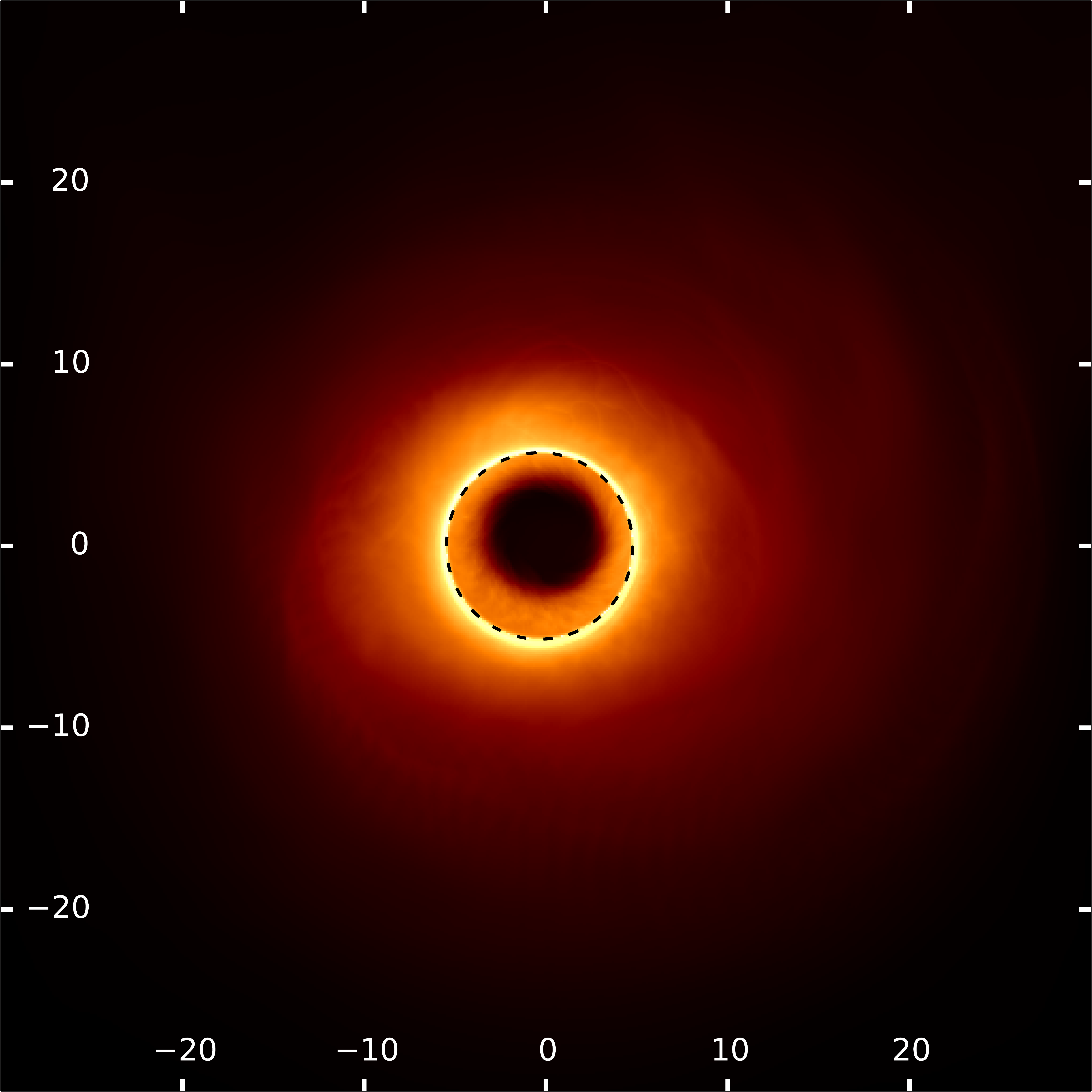}
	\caption{$a=-0.5$, $i=20^\circ$.}
\end{subfigure}
\begin{subfigure}[b]{0.197\textwidth}
	\includegraphics[width=\textwidth]{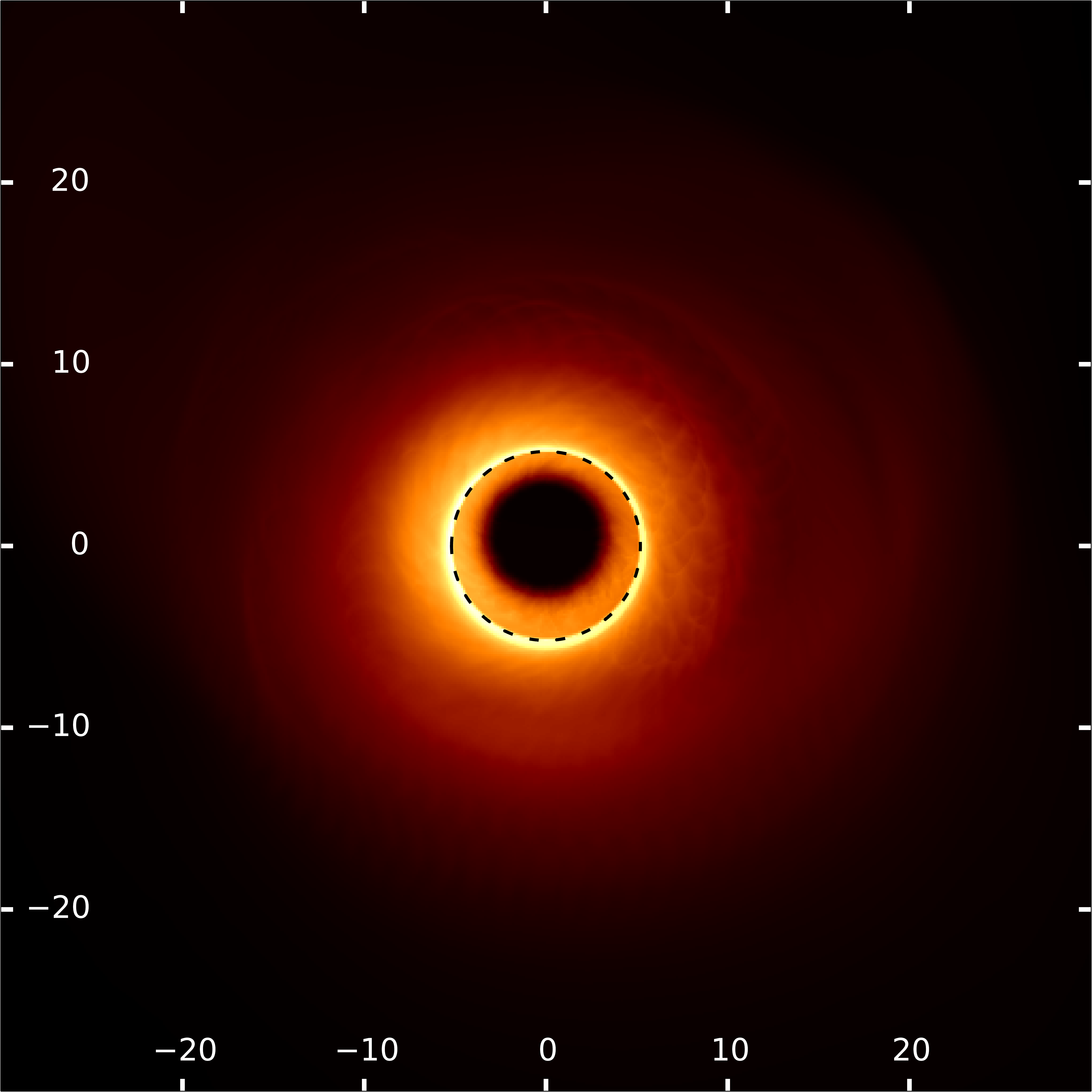}
	\caption{$a=0$, $i=20^\circ$.}
\end{subfigure}
\begin{subfigure}[b]{0.197\textwidth}
	\includegraphics[width=\textwidth]{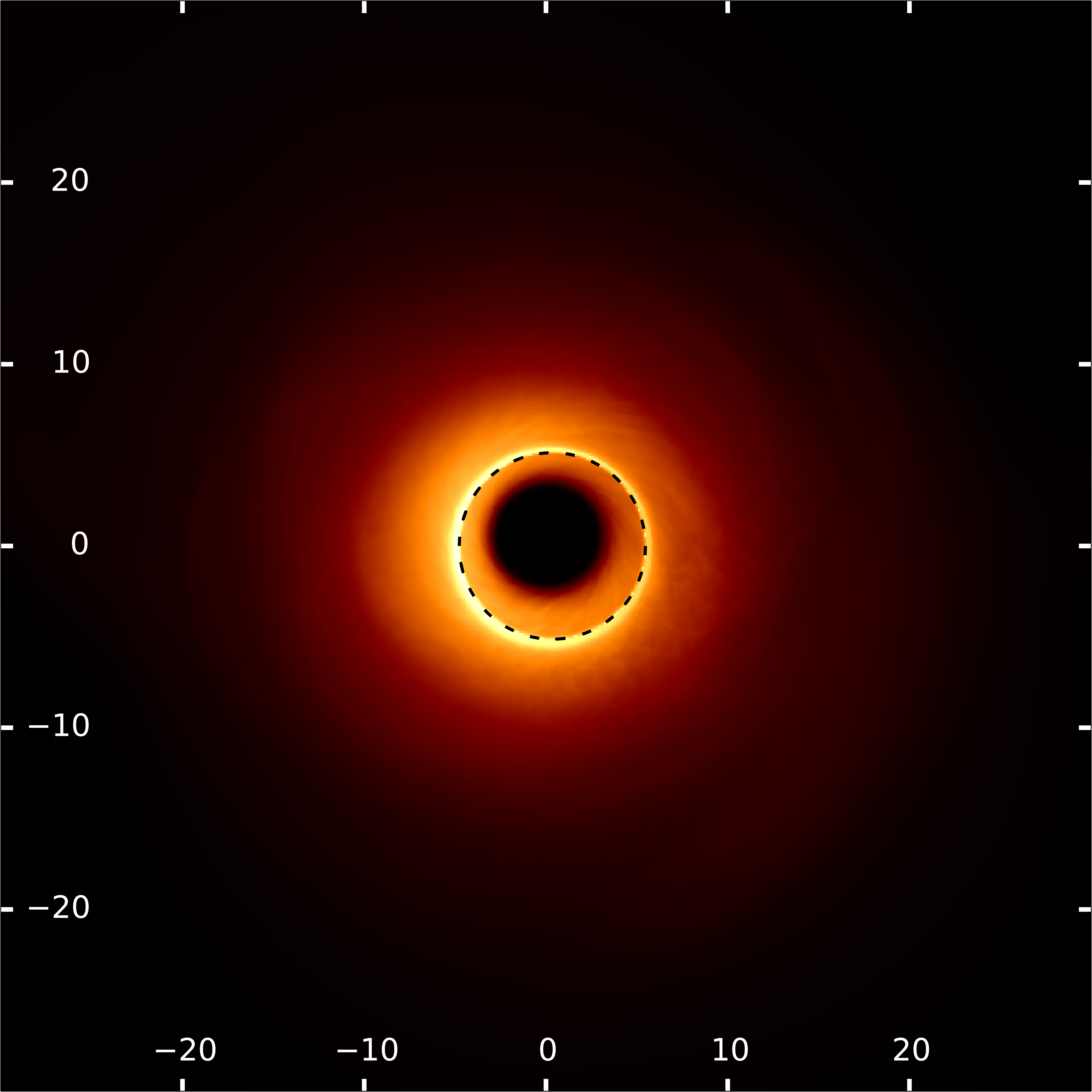}
	\caption{$a=0.5$, $i=20^\circ$.}
\end{subfigure}
\begin{subfigure}[b]{0.197\textwidth}
	\includegraphics[width=\textwidth]{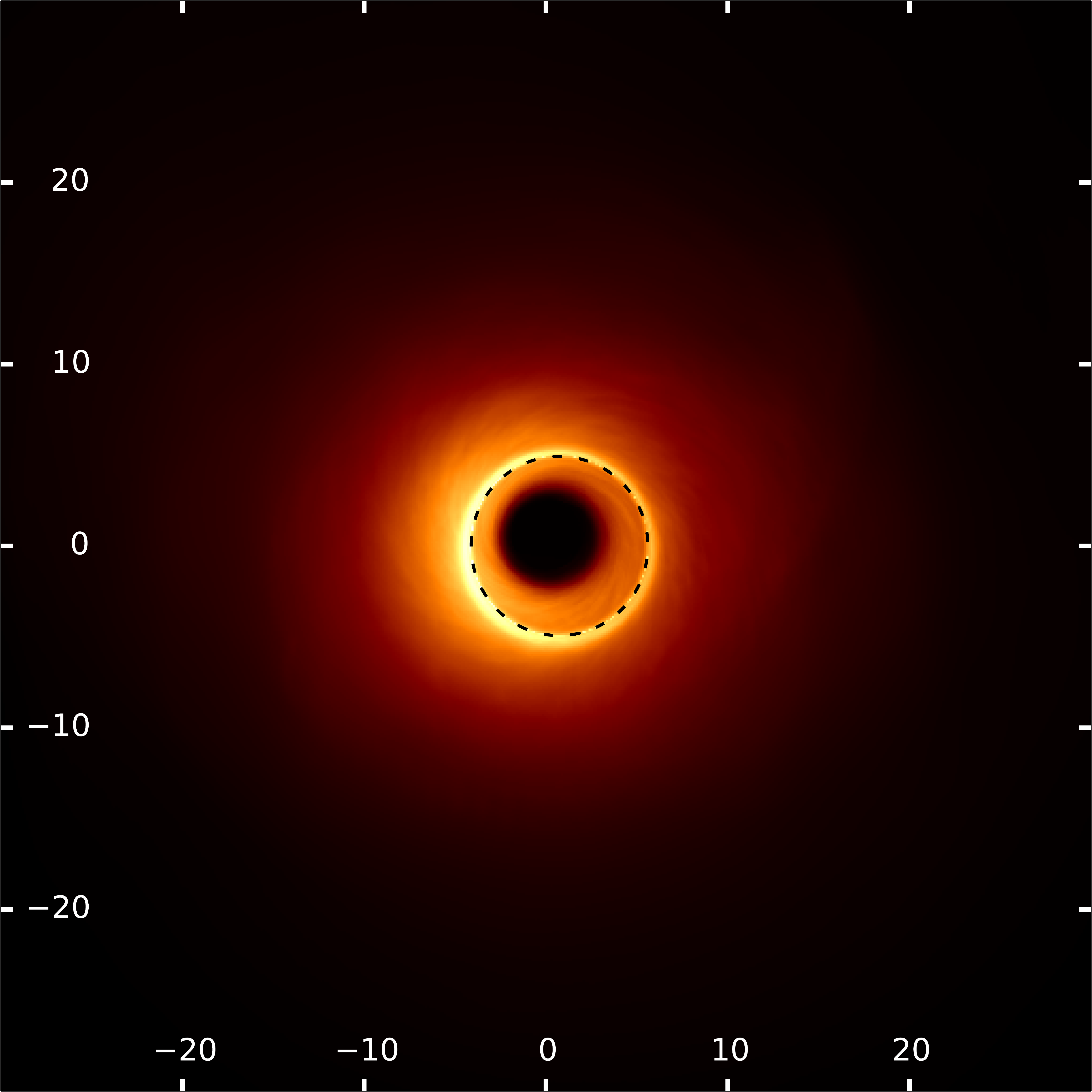}
	\caption{$a=0.9375$, $i=20^\circ$.}
\end{subfigure}
\begin{subfigure}[b]{0.197\textwidth}
	\includegraphics[width=\textwidth]{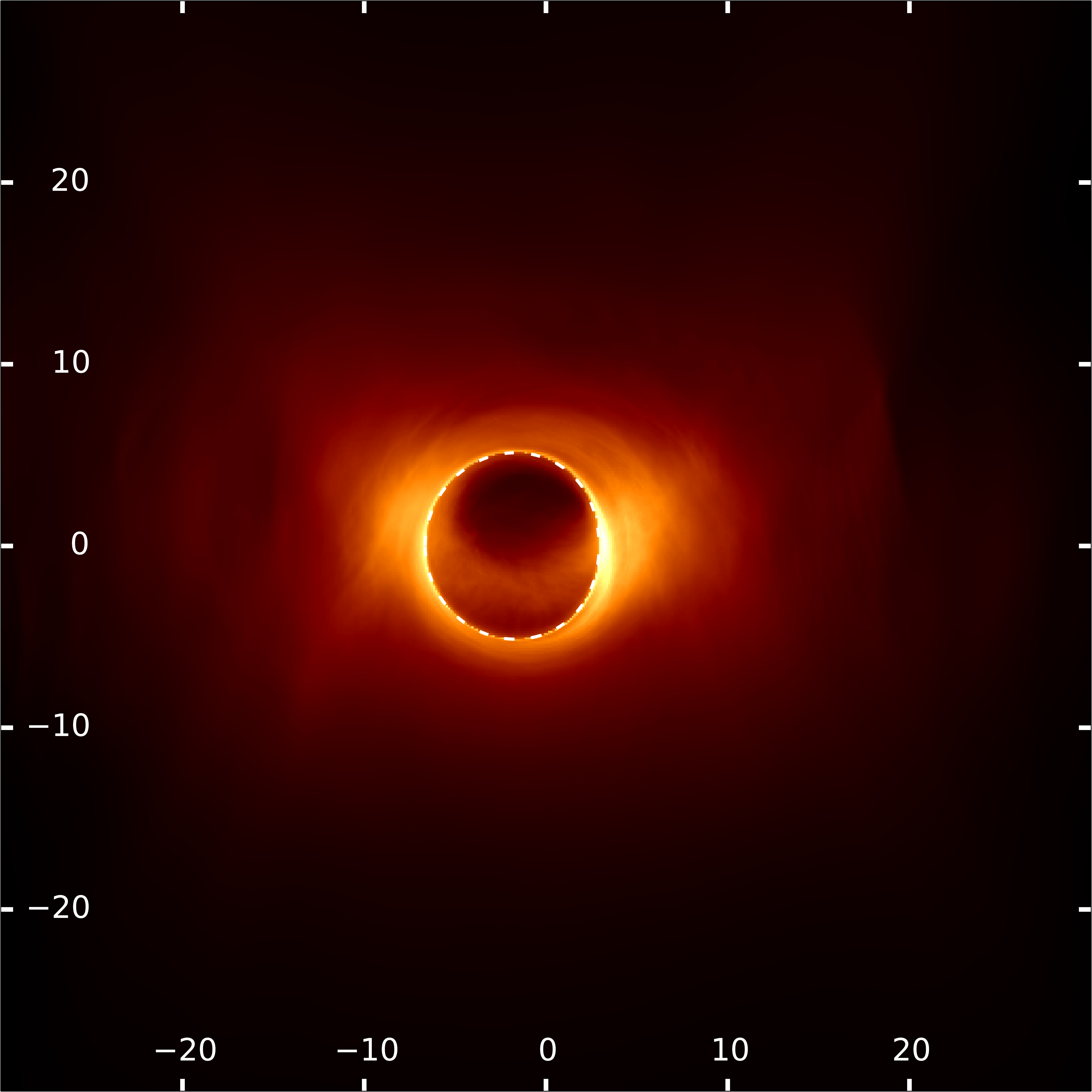}
	\caption{$a=-0.9375$, $i=60^\circ$.}
\end{subfigure}
\begin{subfigure}[b]{0.197\textwidth}
	\includegraphics[width=\textwidth]{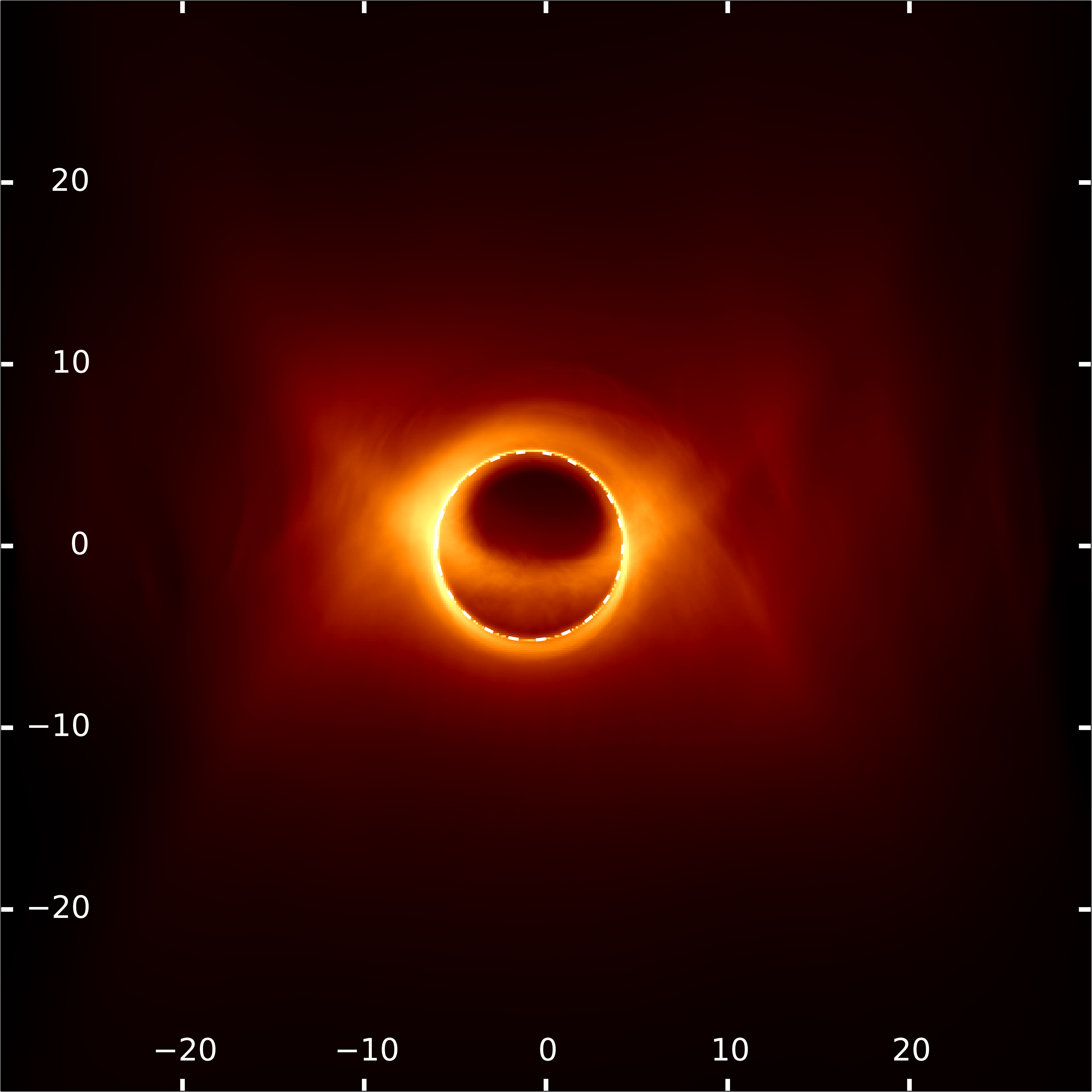}
	\caption{$a=-0.5$, $i=60^\circ$.}
\end{subfigure}
\begin{subfigure}[b]{0.197\textwidth}
	\includegraphics[width=\textwidth]{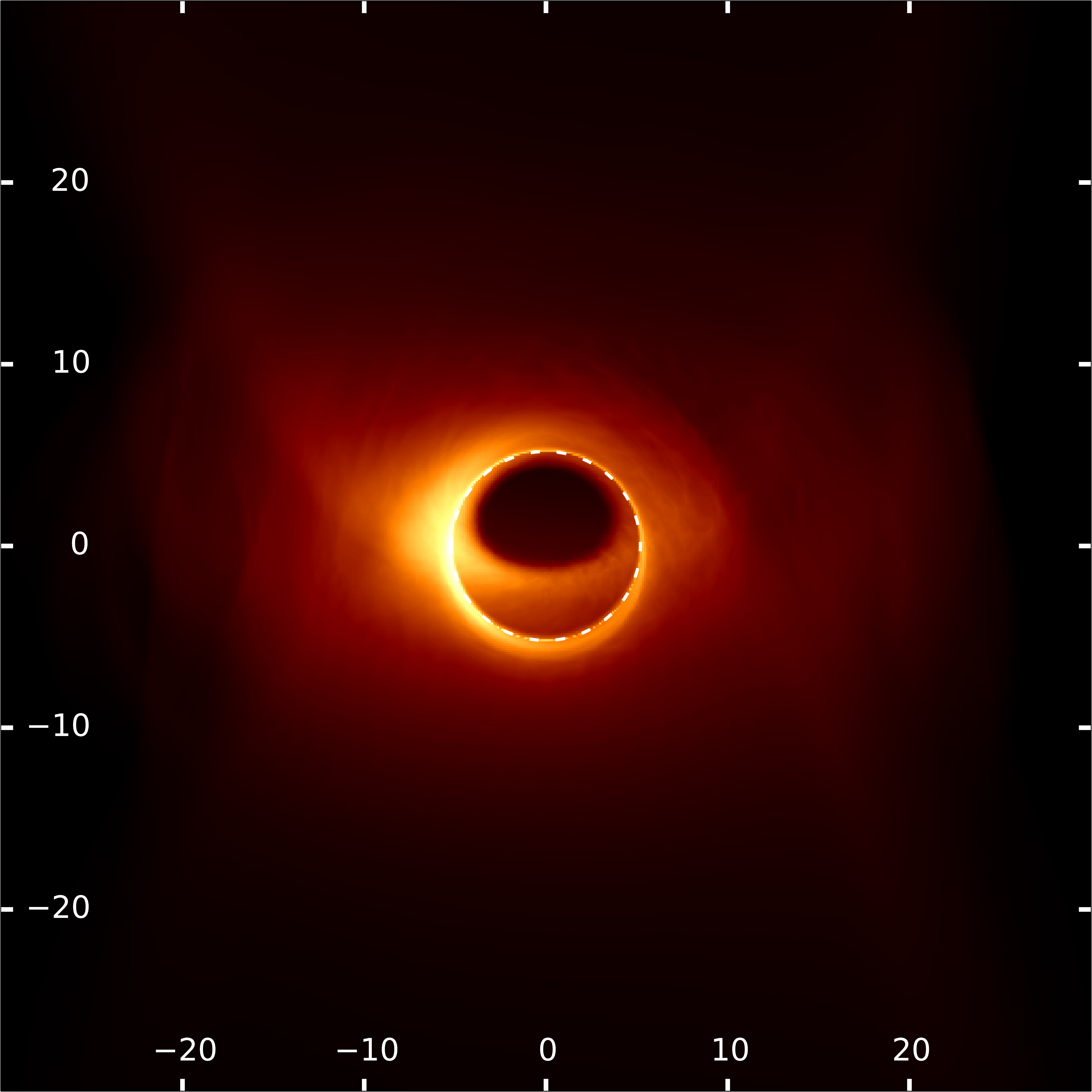}
	\caption{$a=0$, $i=60^\circ$.}
\end{subfigure}
\begin{subfigure}[b]{0.197\textwidth}
	\includegraphics[width=\textwidth]{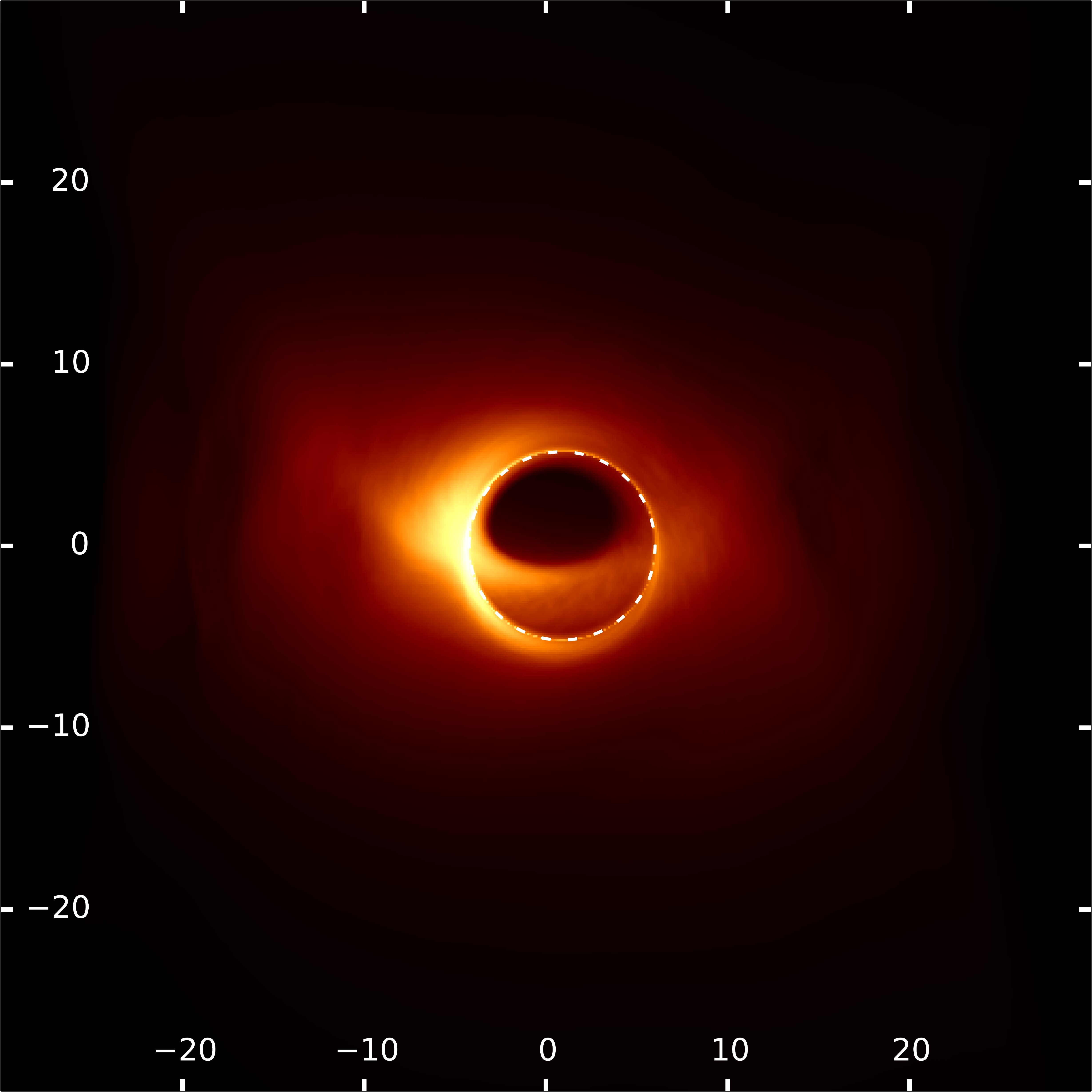}
	\caption{$a=0.5$, $i=60^\circ$.}
\end{subfigure}
\begin{subfigure}[b]{0.197\textwidth}
	\includegraphics[width=\textwidth]{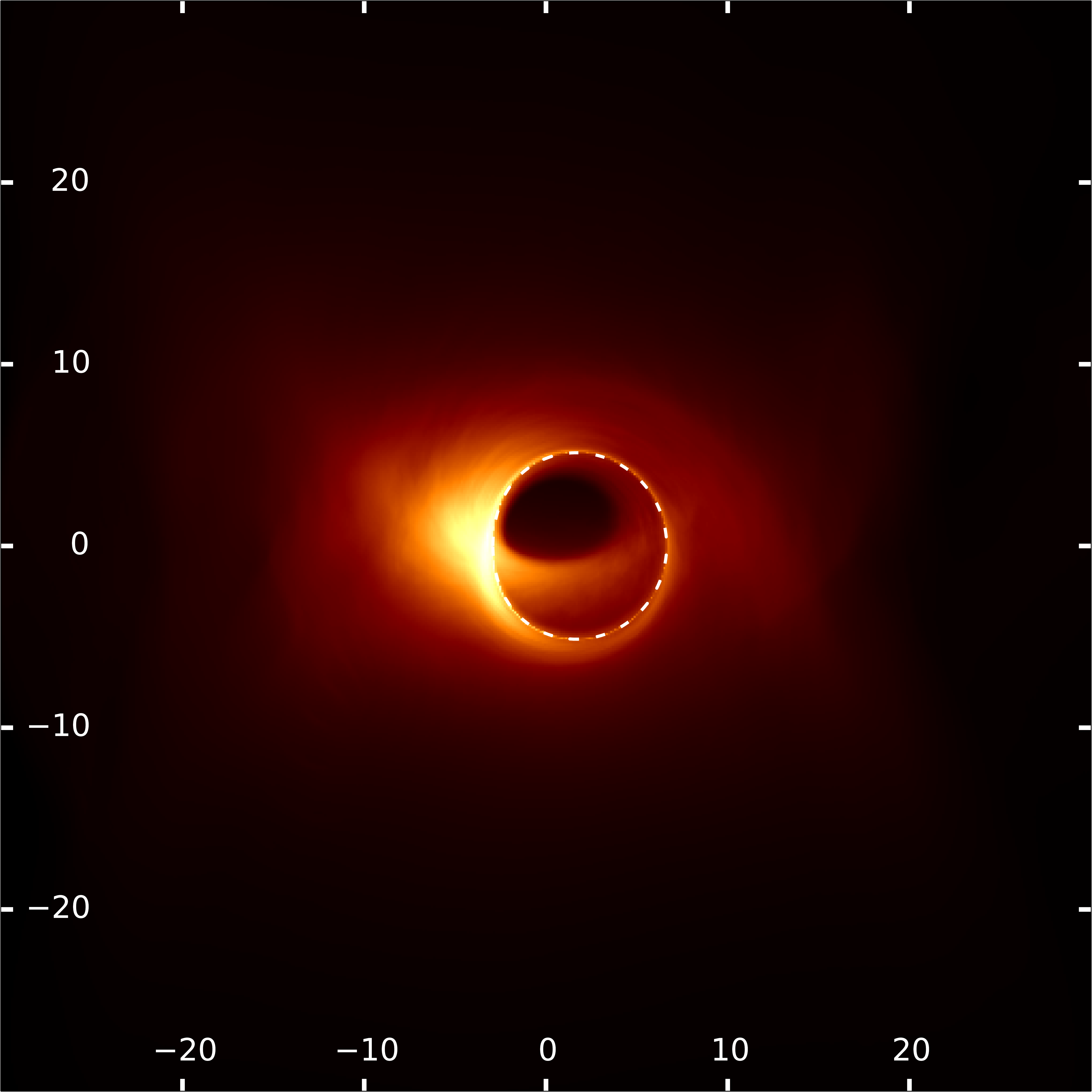}
	\caption{$a=0.9375$, $i=60^\circ$.}
\end{subfigure}
\begin{subfigure}[b]{0.197\textwidth}
	\includegraphics[width=\textwidth]{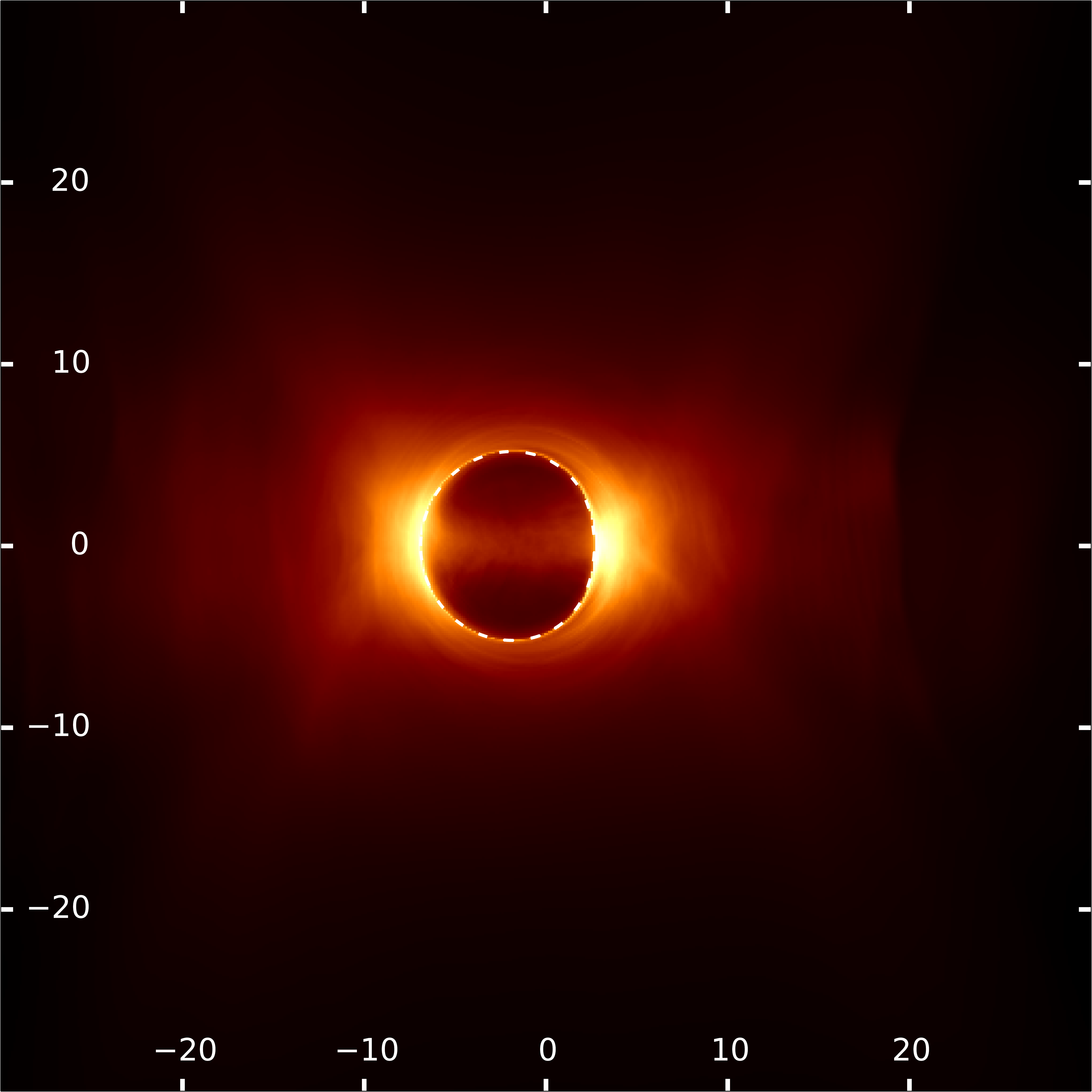}
	\caption{$a=-0.9375$, $i=90^\circ$.}
\end{subfigure}
\begin{subfigure}[b]{0.197\textwidth}
	\includegraphics[width=\textwidth]{Figures/mad_jet_a-1o2_90_25-crop}
	\caption{$a=-0.5$, $i=90^\circ$.}
\end{subfigure}
\begin{subfigure}[b]{0.197\textwidth}
	\includegraphics[width=\textwidth]{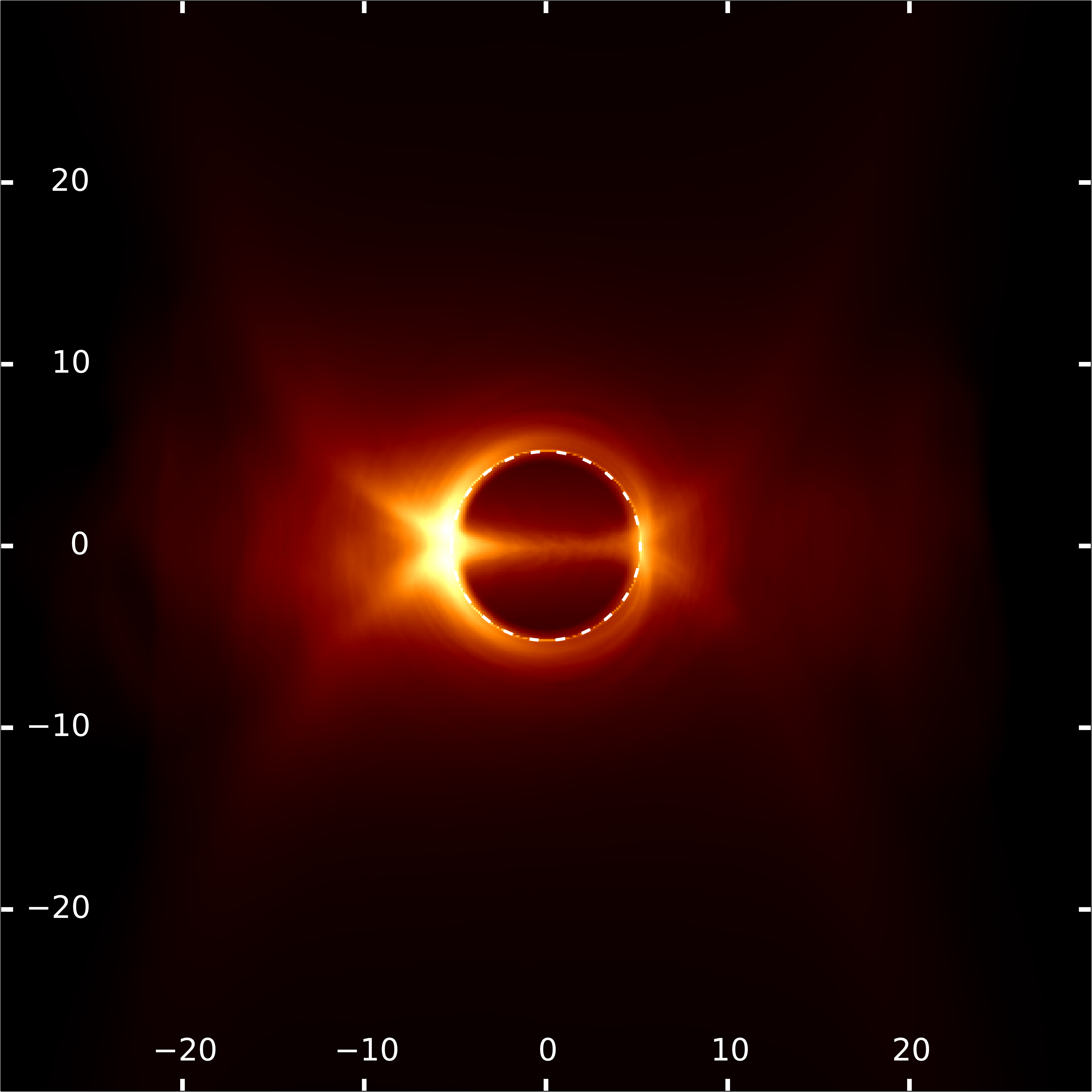}
	\caption{$a=0$, $i=90^\circ$.}
\end{subfigure}
\begin{subfigure}[b]{0.197\textwidth}
	\includegraphics[width=\textwidth]{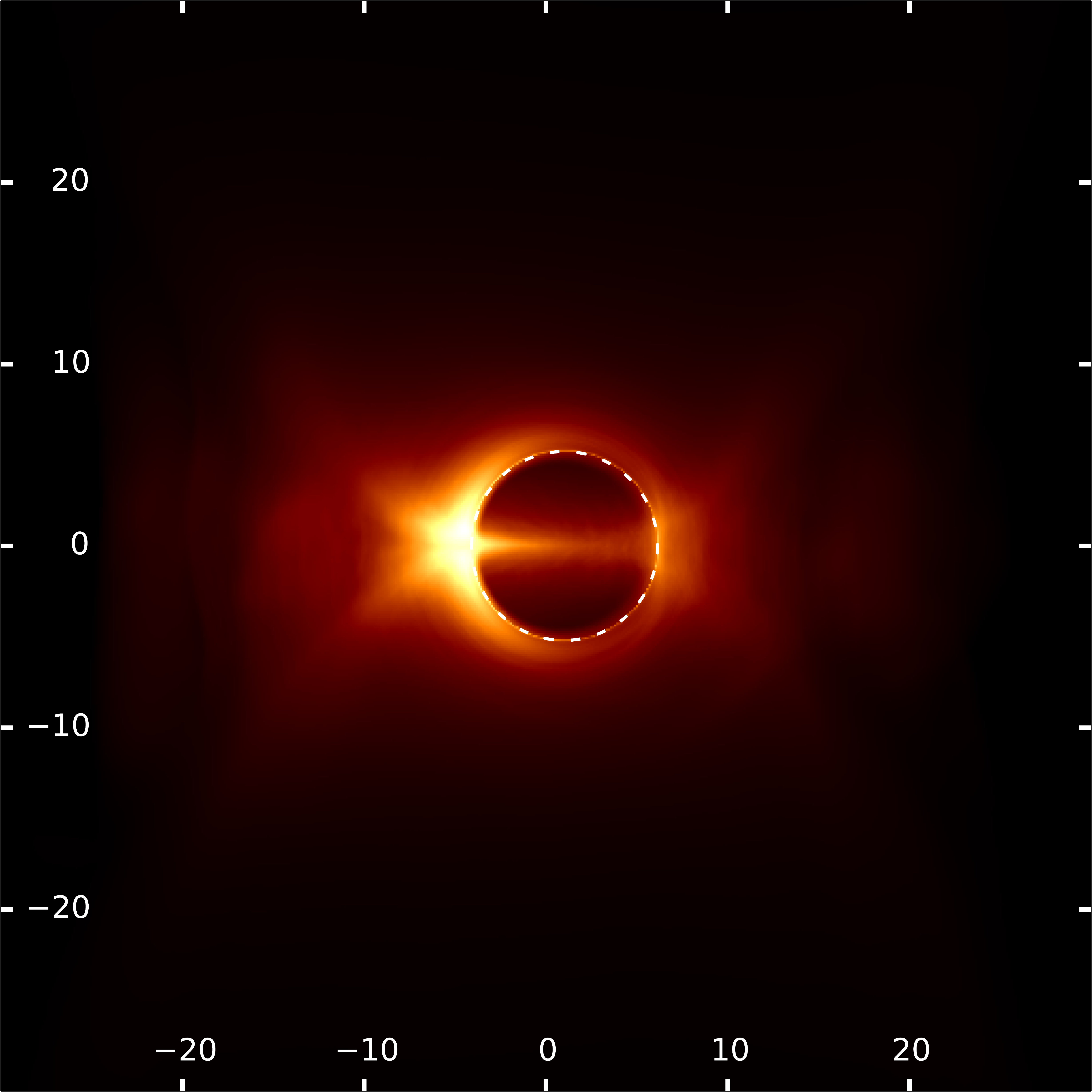}
	\caption{$a=0.5$, $i=90^\circ$.}
\end{subfigure}
\begin{subfigure}[b]{0.197\textwidth}
	\includegraphics[width=\textwidth]{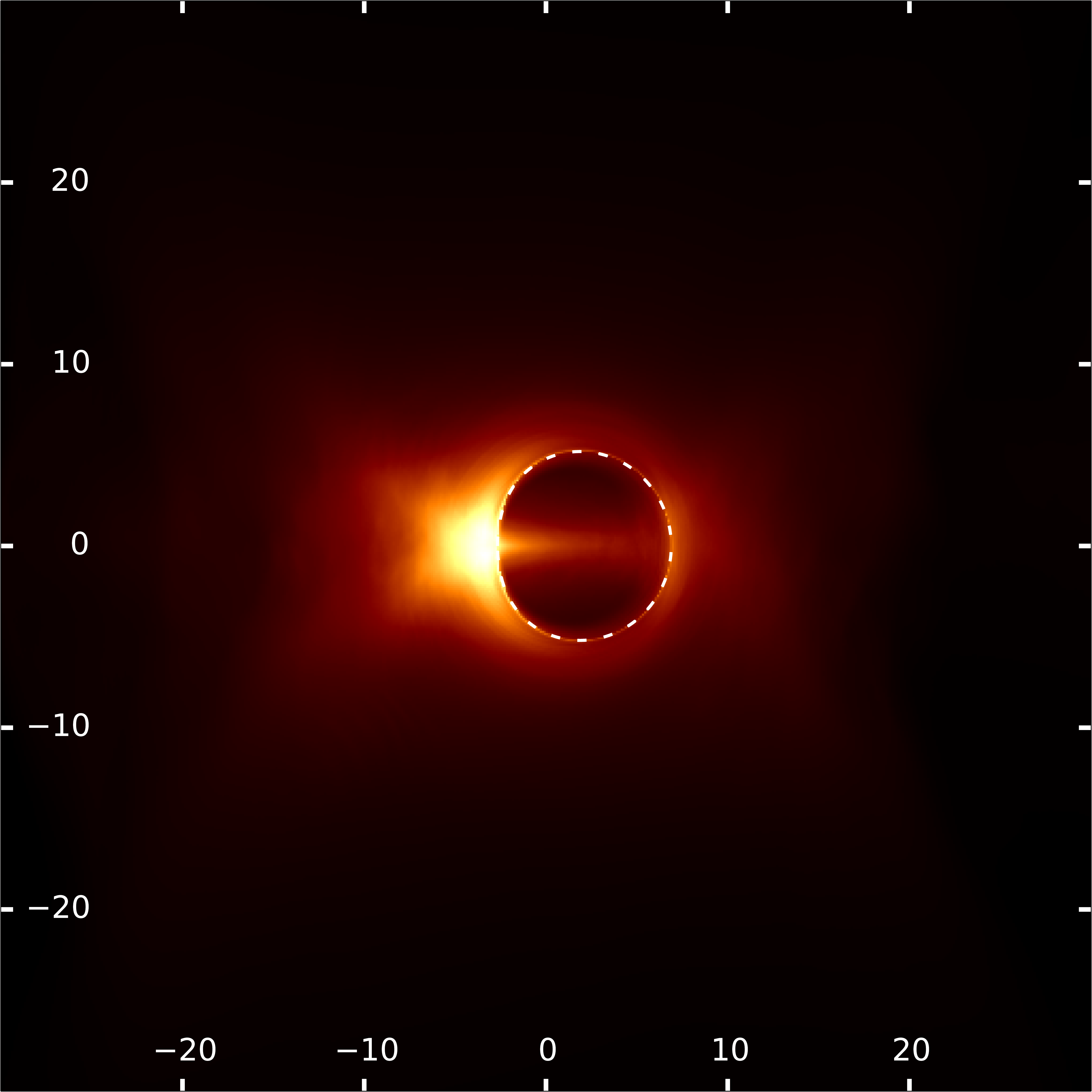}
	\caption{$a=0.9375$, $i=90^\circ$.}
\end{subfigure}
\caption{Time-averaged, normalised intensity maps of our MAD, jet-dominated GRMHD models of Sgr A*, imaged at 230 GHz, at five different spins and four observer inclination angles, with an integrated flux density of 2.5 Jy. In each case, the photon ring, which marks the BHS, is indicated by a dashed line. The values for the impact parameters along the x- and y-axes are expressed in terms of $R_{\rm g}$. The image maps were plotted using a square-root intensity scale.}
\label{fig:mad_jet_25_matrix}
\end{figure*}

\begin{figure*}
\centering
\begin{subfigure}[b]{0.197\textwidth}
	\includegraphics[width=\textwidth]{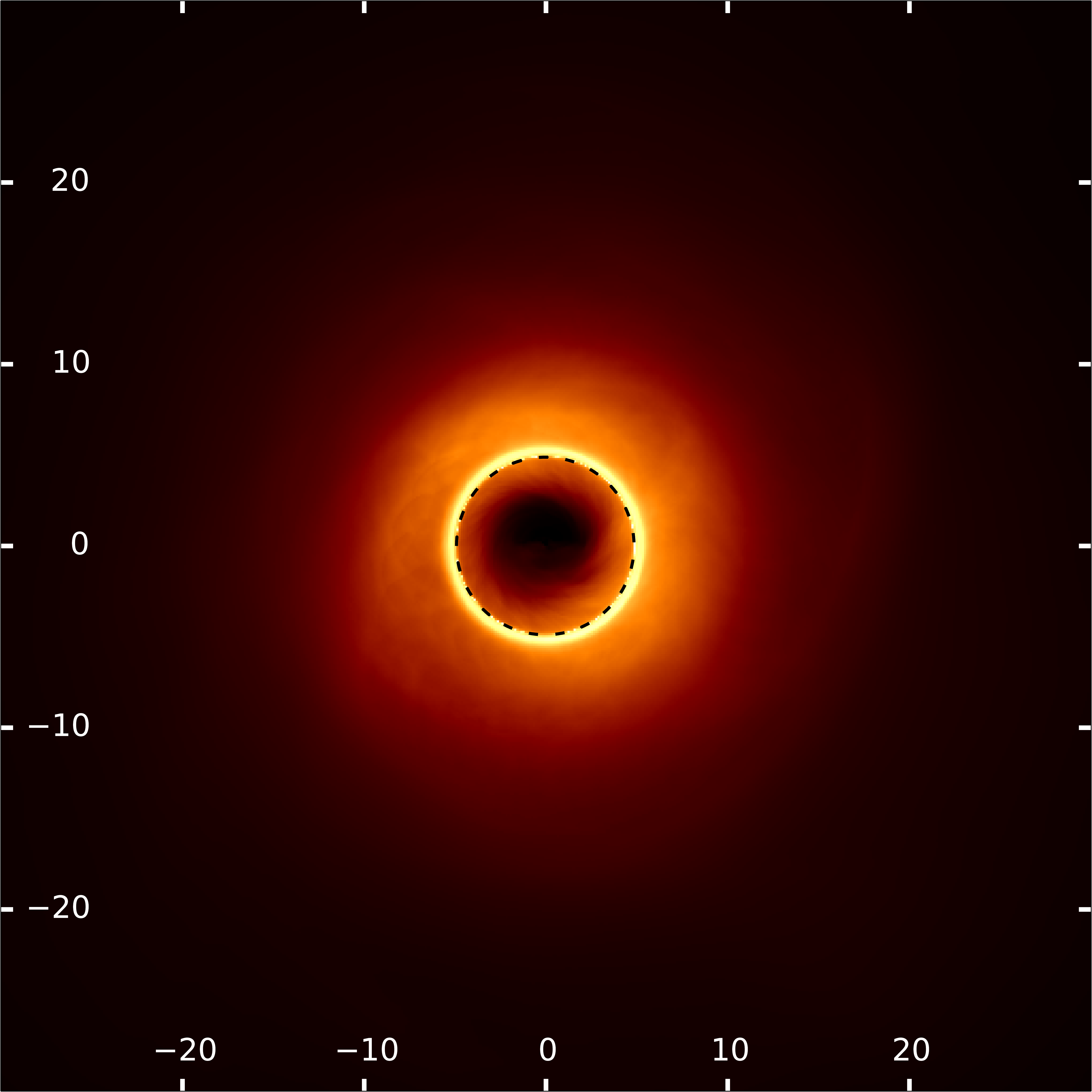}
	\caption{$a=-0.9375$, $i=1^\circ$.}
\end{subfigure}
\begin{subfigure}[b]{0.197\textwidth}
	\includegraphics[width=\textwidth]{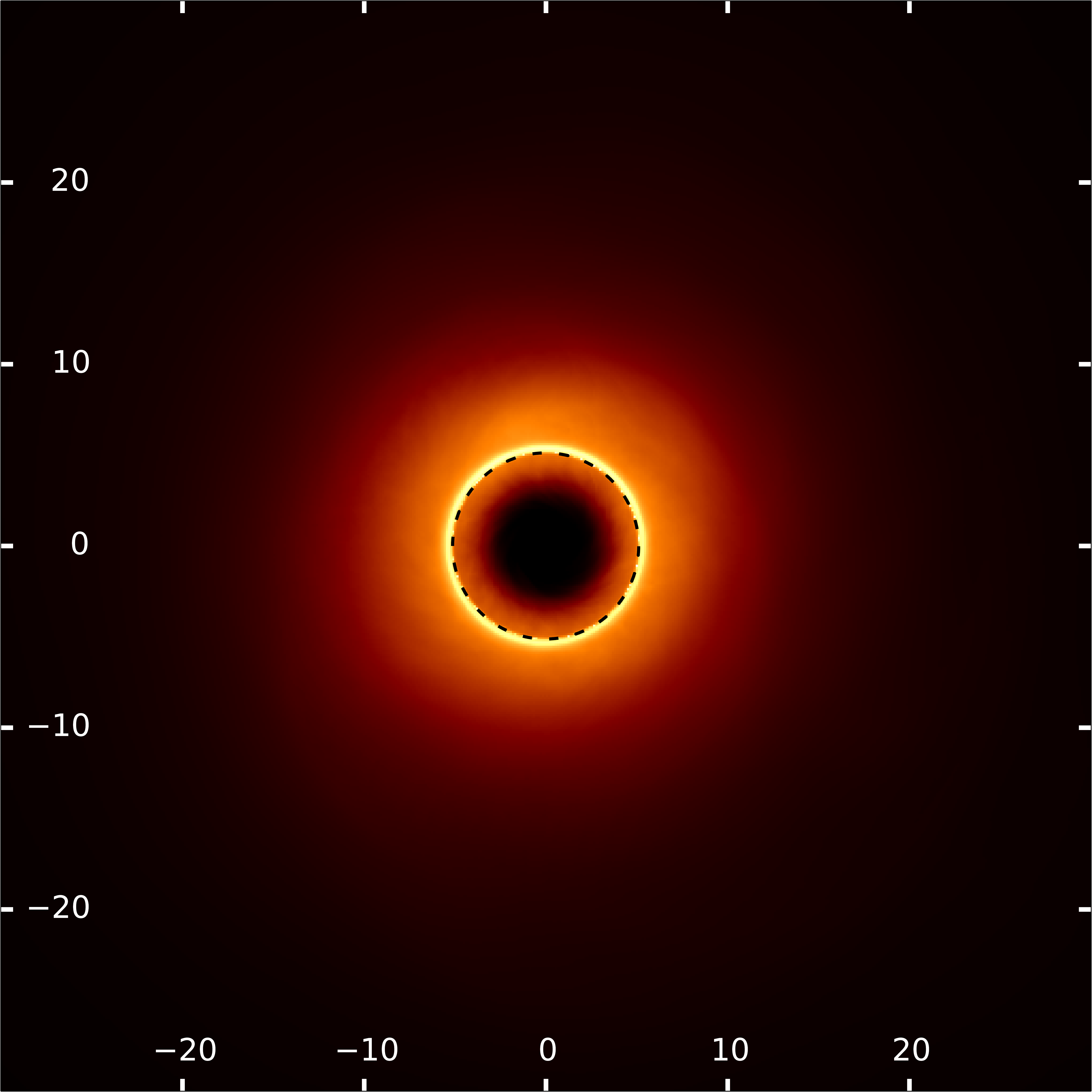}
	\caption{$a=-0.5$, $i=1^\circ$.}
\end{subfigure}
\begin{subfigure}[b]{0.197\textwidth}
	\includegraphics[width=\textwidth]{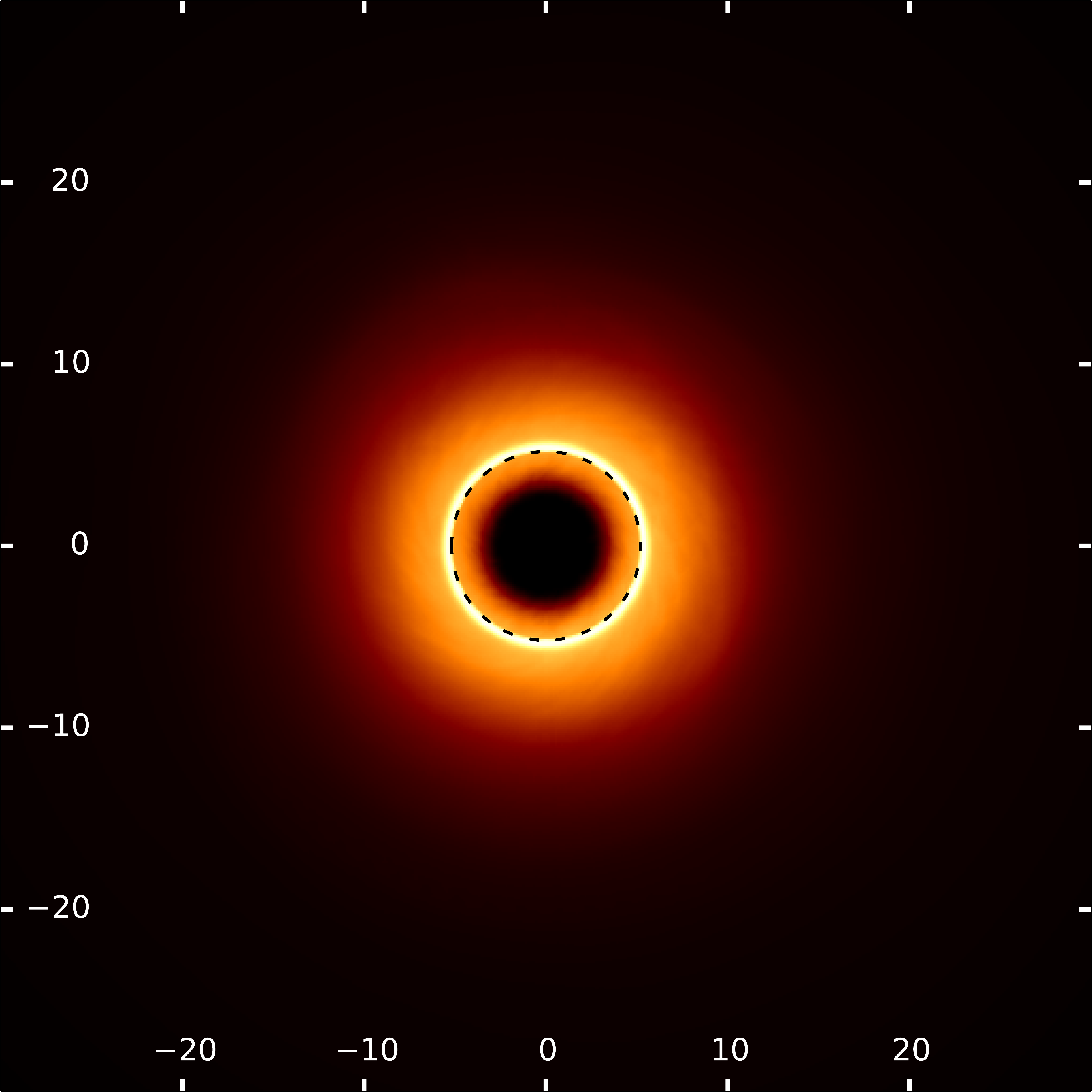}
	\caption{$a=0$, $i=1^\circ$.}
\end{subfigure}
\begin{subfigure}[b]{0.197\textwidth}
	\includegraphics[width=\textwidth]{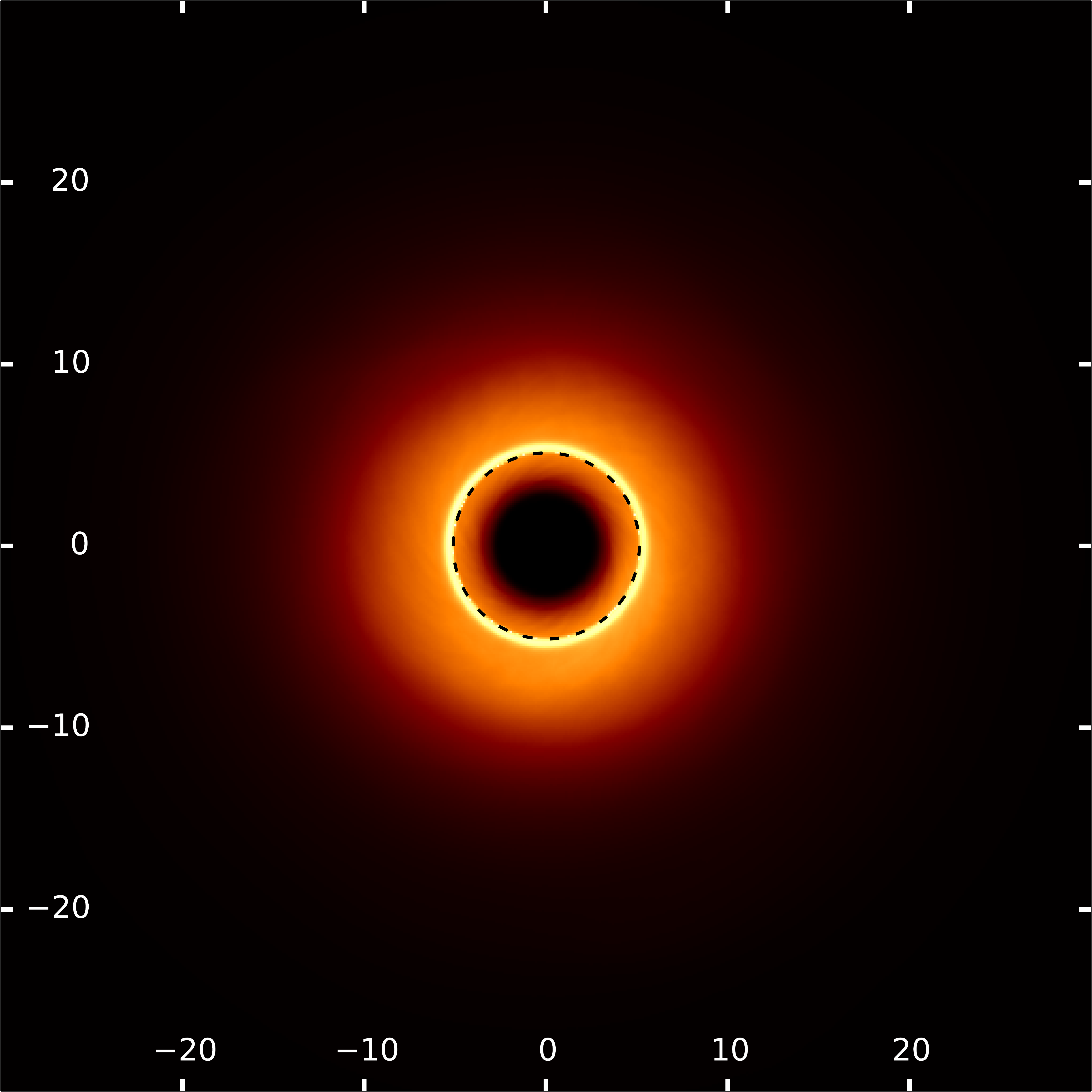}
	\caption{$a=0.5$, $i=1^\circ$.}
\end{subfigure}
\begin{subfigure}[b]{0.197\textwidth}
	\includegraphics[width=\textwidth]{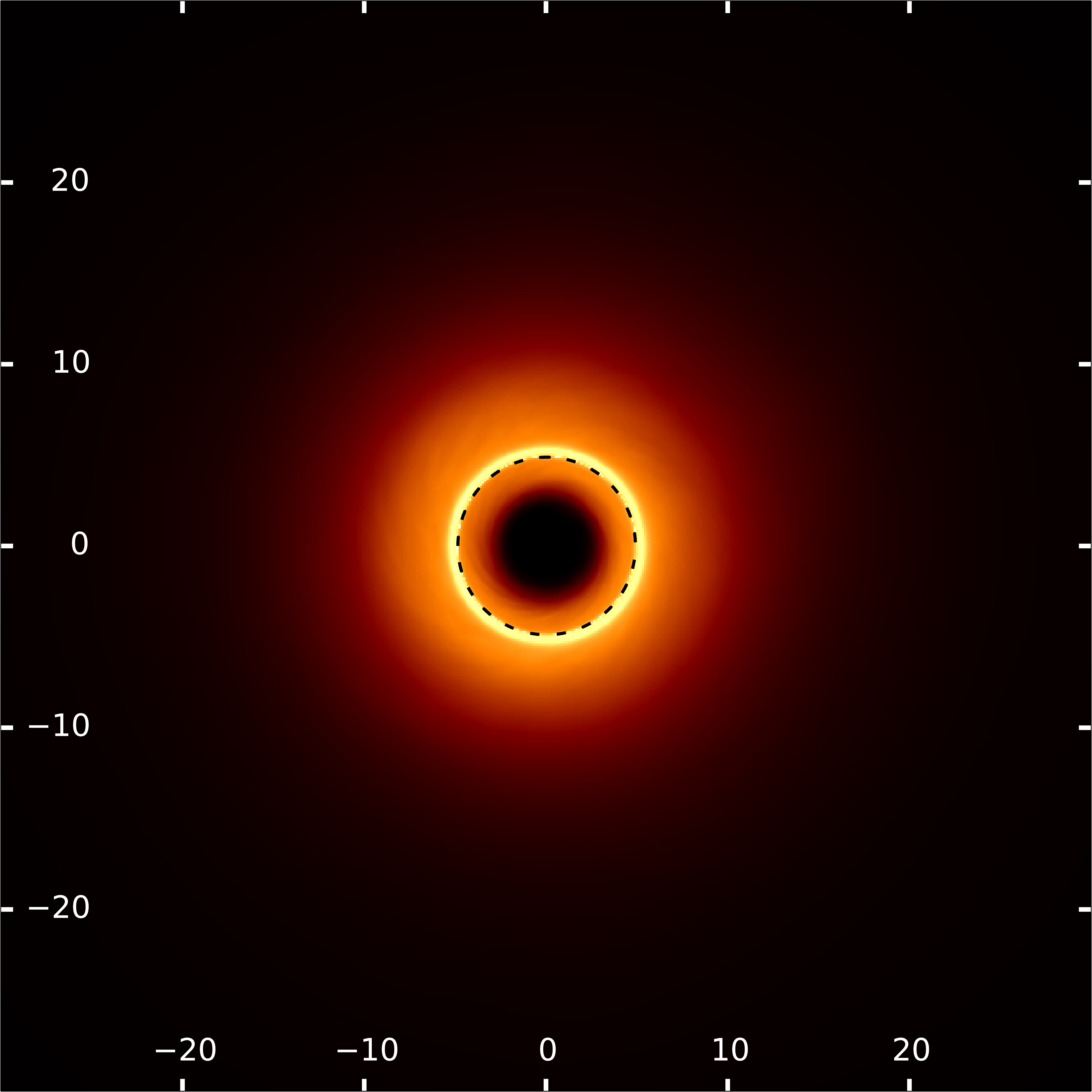}
	\caption{$a=0.9375$, $i=1^\circ$.}
\end{subfigure}
\begin{subfigure}[b]{0.197\textwidth}
	\includegraphics[width=\textwidth]{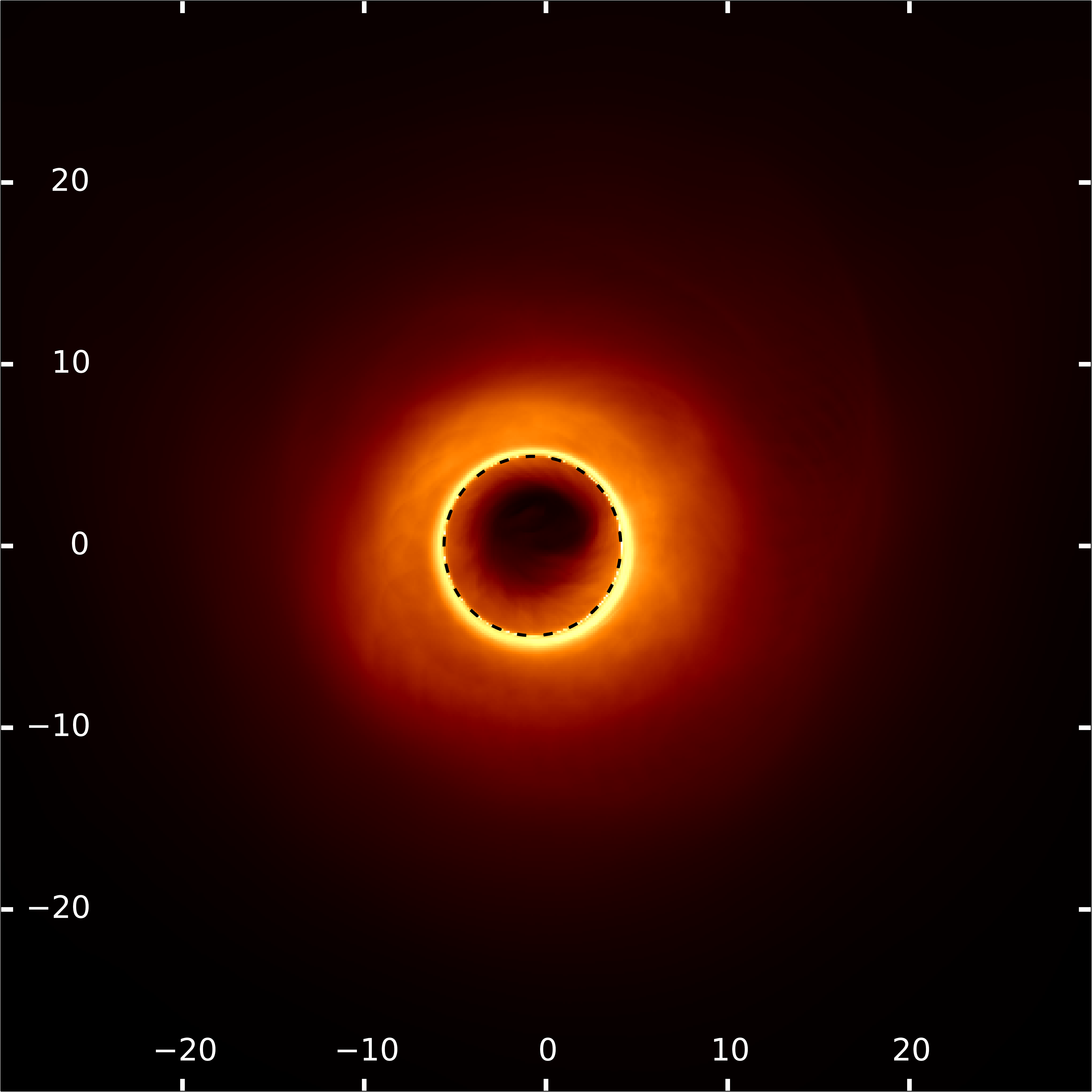}
	\caption{$a=-0.9375$, $i=20^\circ$.}
\end{subfigure}
\begin{subfigure}[b]{0.197\textwidth}
	\includegraphics[width=\textwidth]{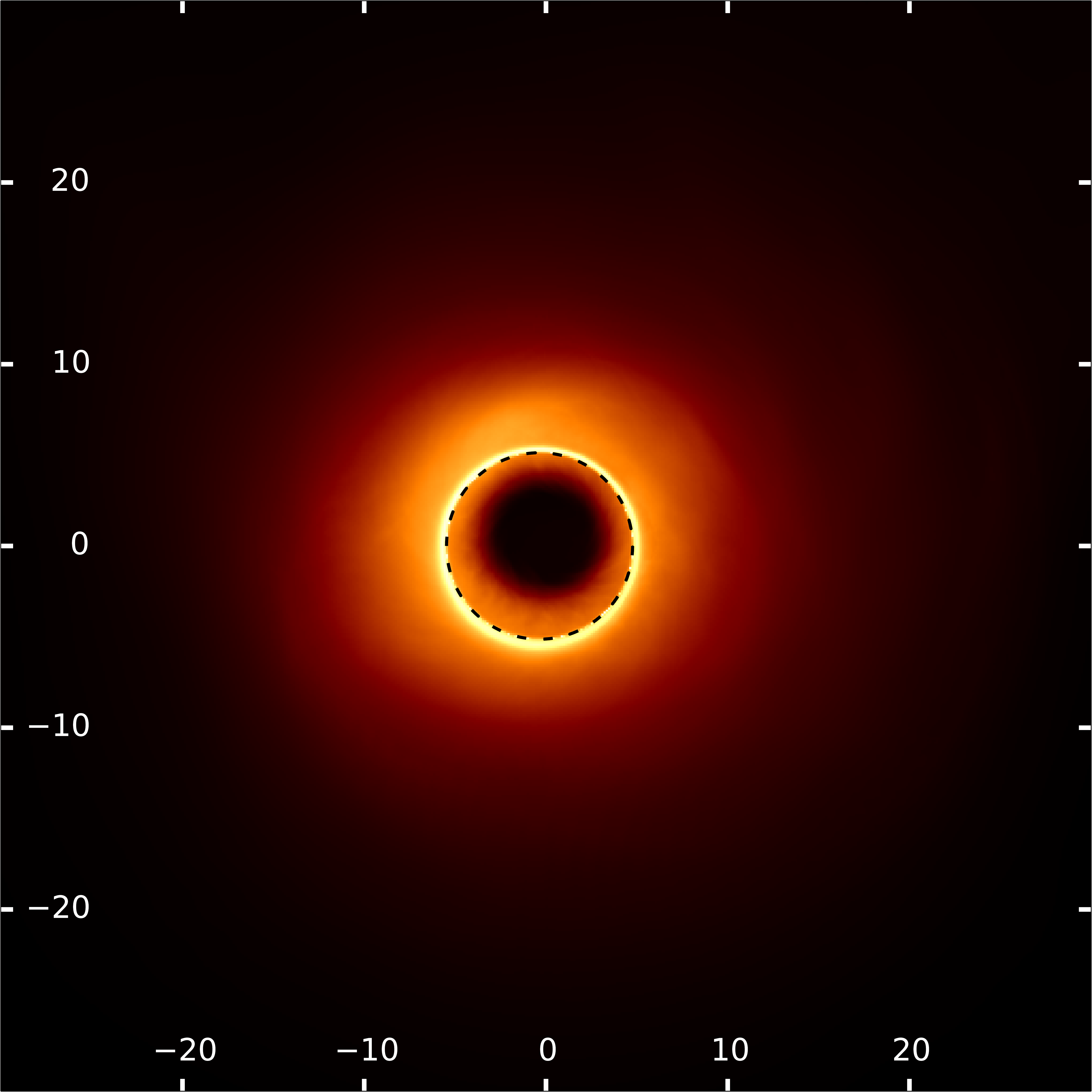}
	\caption{$a=-0.5$, $i=20^\circ$.}
\end{subfigure}
\begin{subfigure}[b]{0.197\textwidth}
	\includegraphics[width=\textwidth]{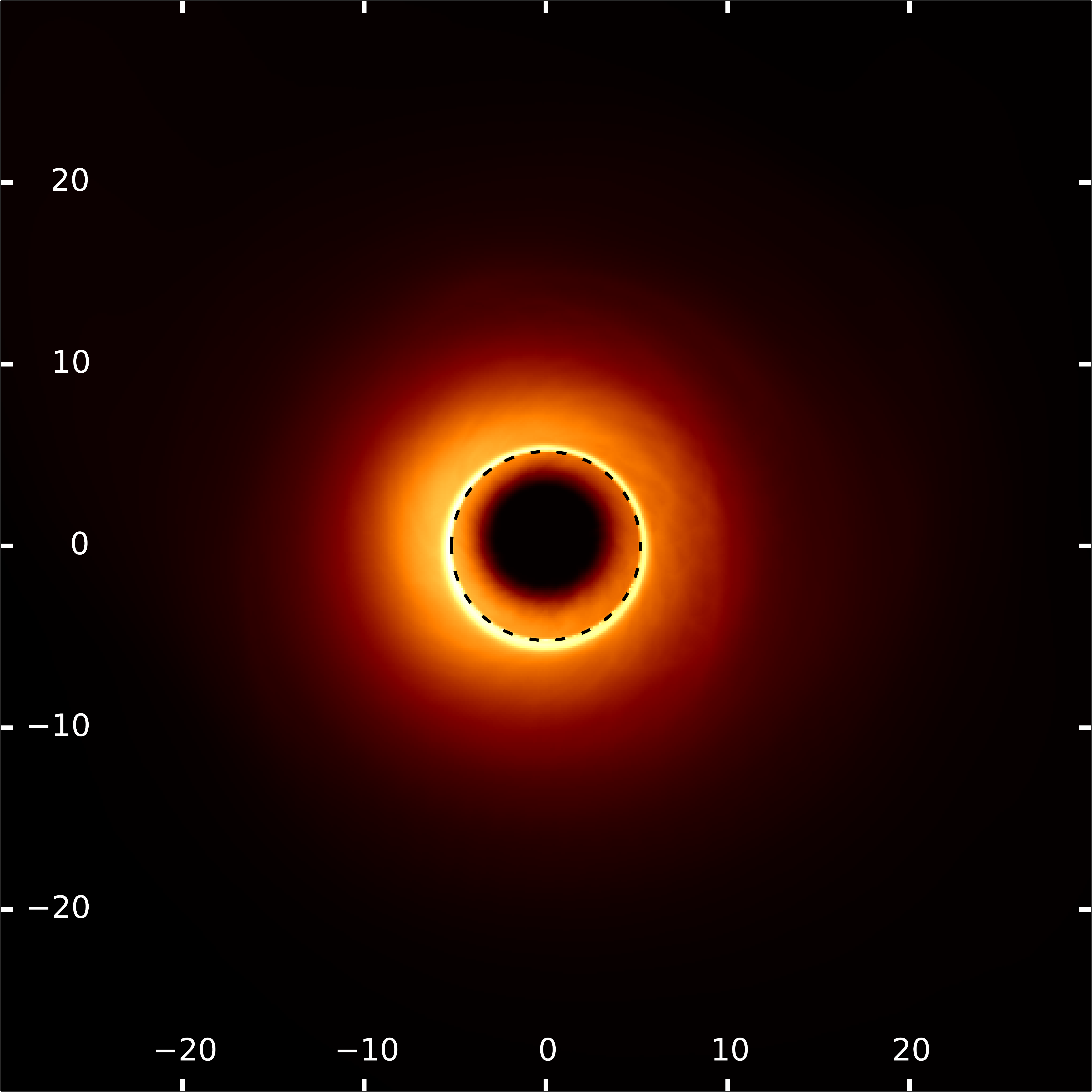}
	\caption{$a=0$, $i=20^\circ$.}
\end{subfigure}
\begin{subfigure}[b]{0.197\textwidth}
	\includegraphics[width=\textwidth]{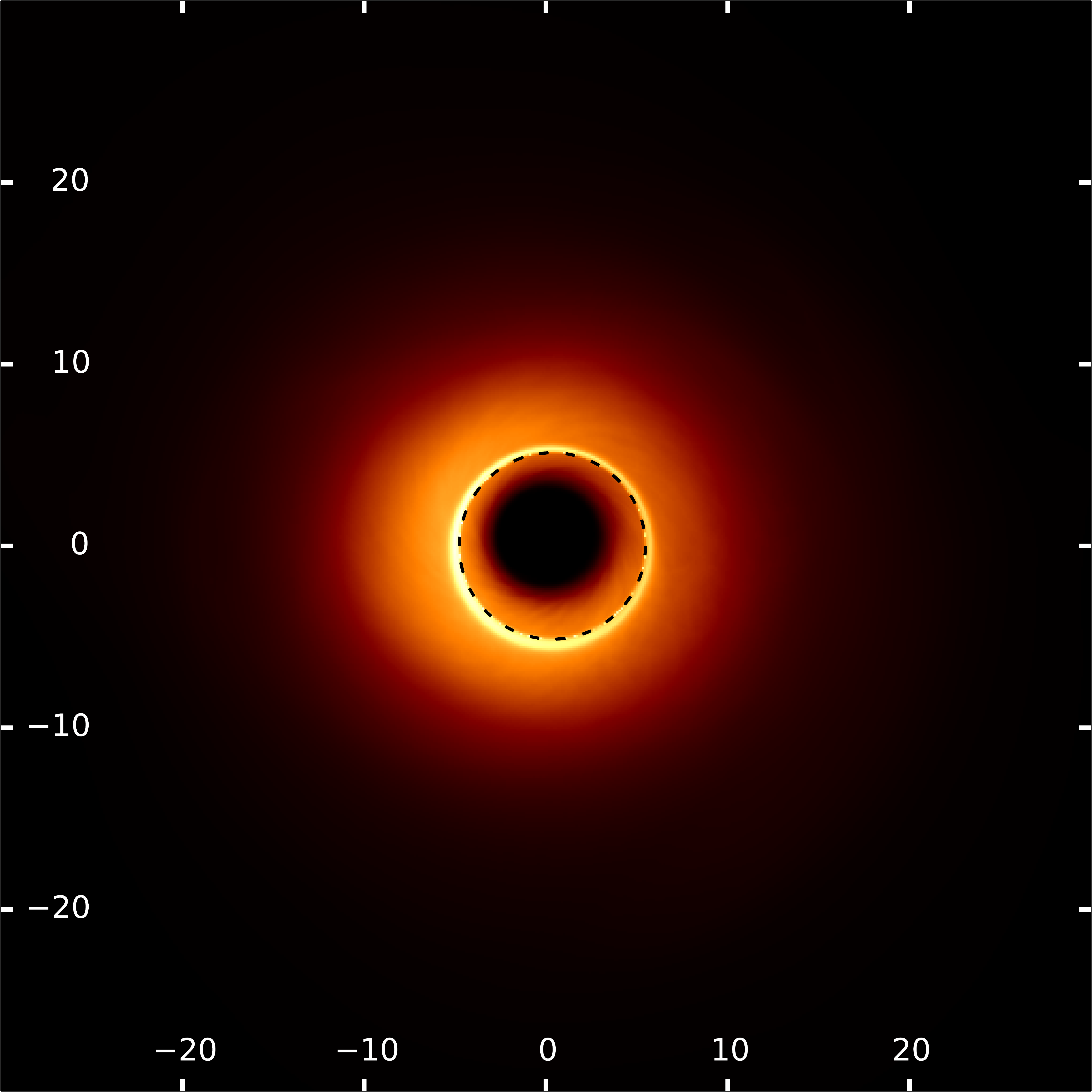}
	\caption{$a=0.5$, $i=20^\circ$.}
\end{subfigure}
\begin{subfigure}[b]{0.197\textwidth}
	\includegraphics[width=\textwidth]{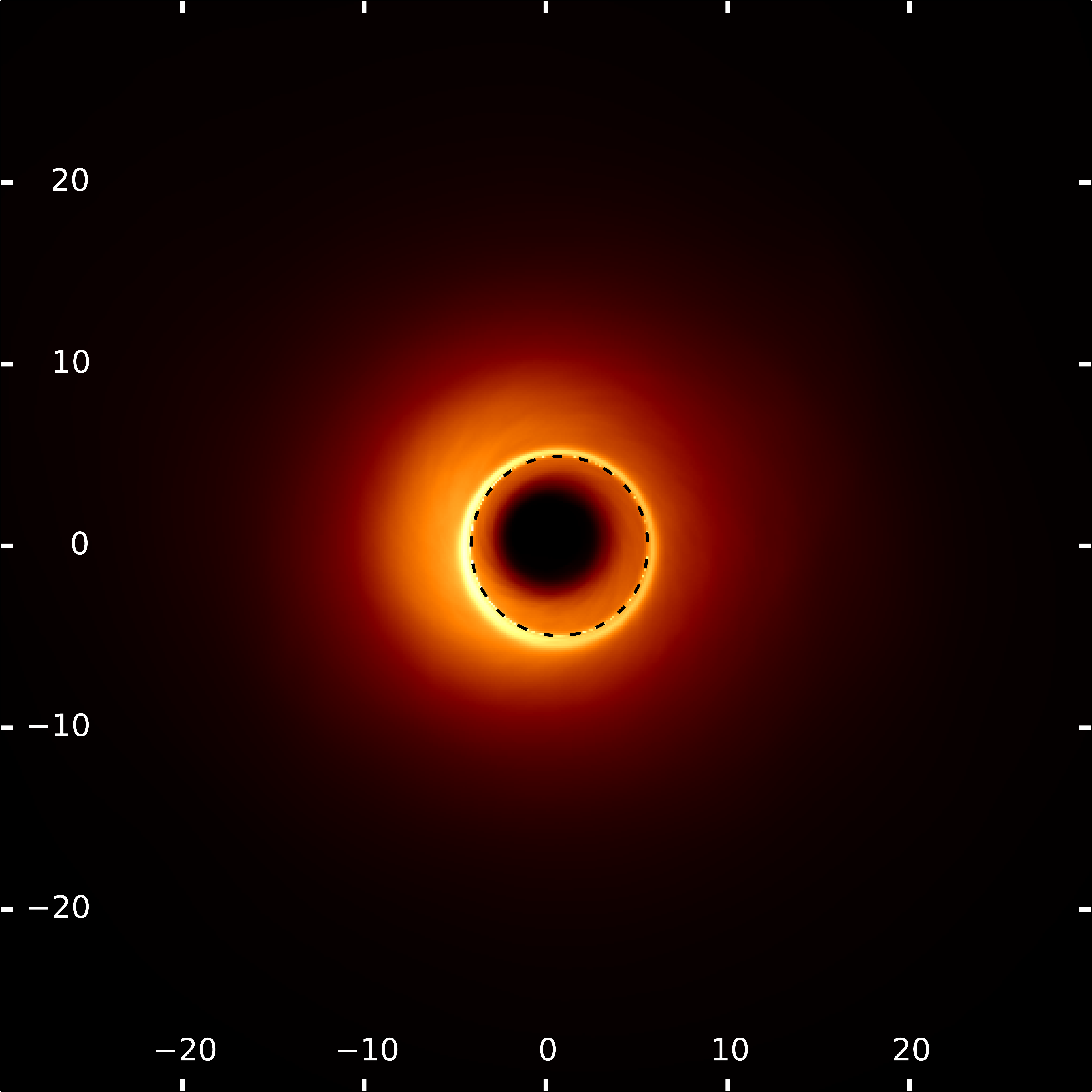}
	\caption{$a=0.9375$, $i=20^\circ$.}
\end{subfigure}
\begin{subfigure}[b]{0.197\textwidth}
	\includegraphics[width=\textwidth]{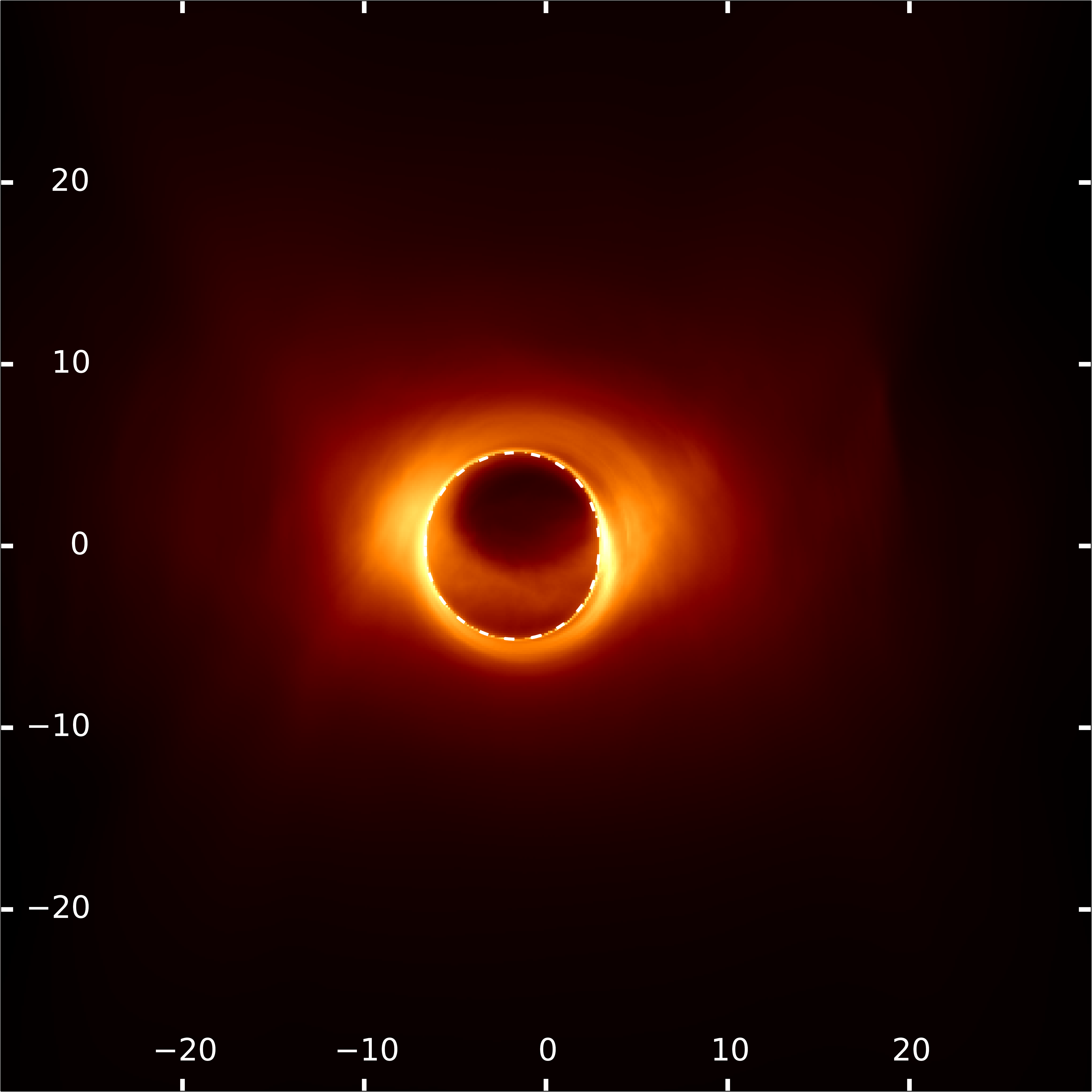}
	\caption{$a=-0.9375$, $i=60^\circ$.}
\end{subfigure}
\begin{subfigure}[b]{0.197\textwidth}
	\includegraphics[width=\textwidth]{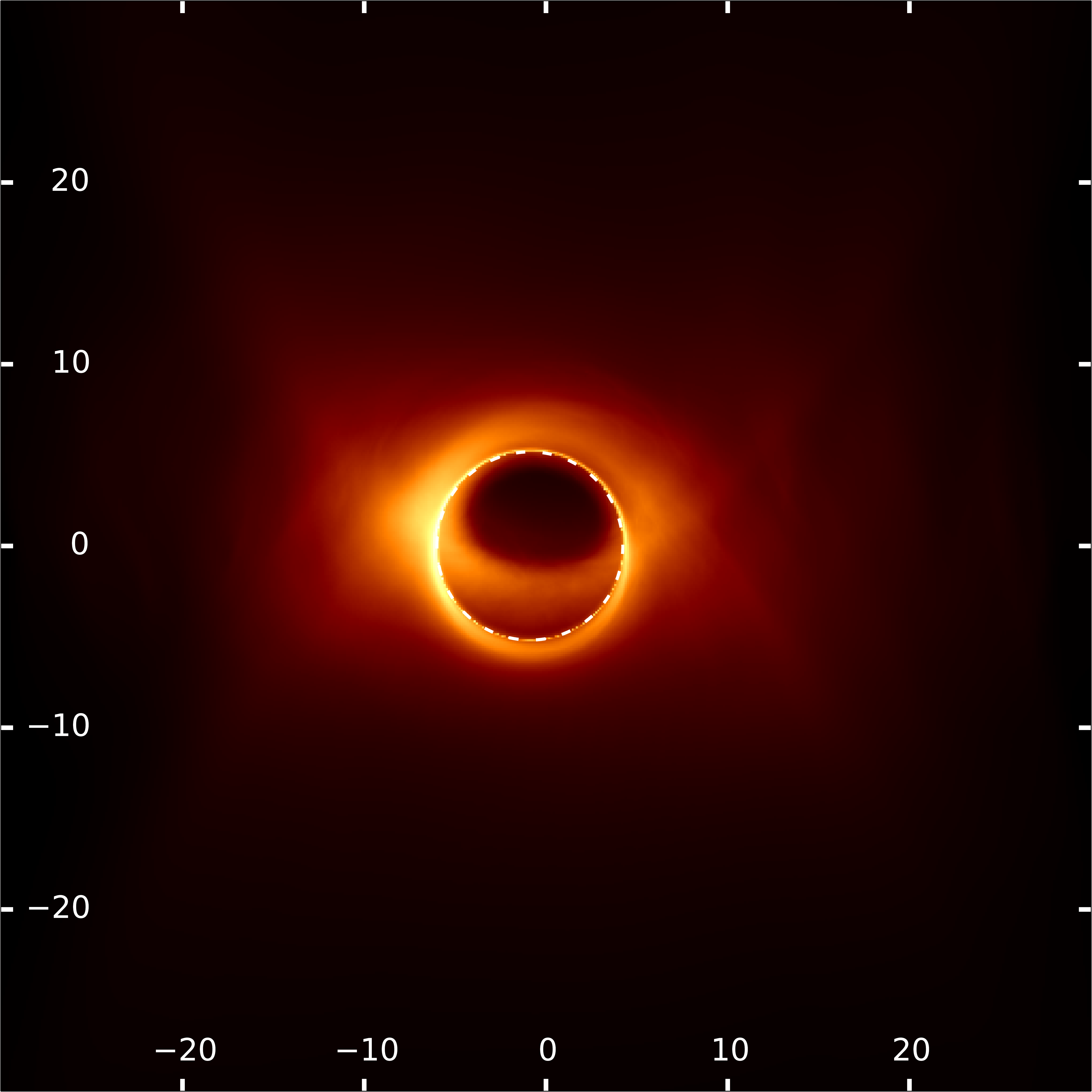}
	\caption{$a=-0.5$, $i=60^\circ$.}
\end{subfigure}
\begin{subfigure}[b]{0.197\textwidth}
	\includegraphics[width=\textwidth]{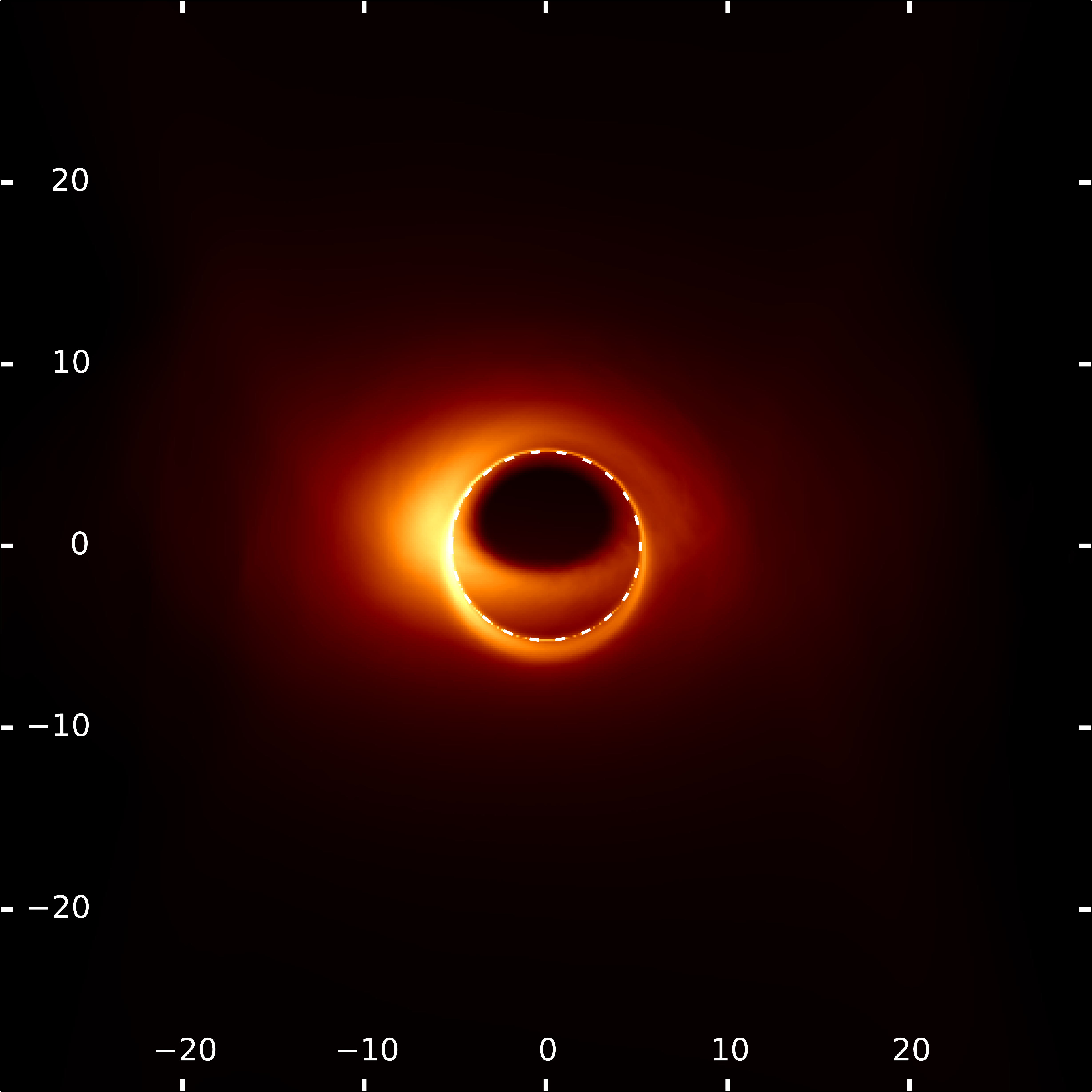}
	\caption{$a=0$, $i=60^\circ$.}
\end{subfigure}
\begin{subfigure}[b]{0.197\textwidth}
	\includegraphics[width=\textwidth]{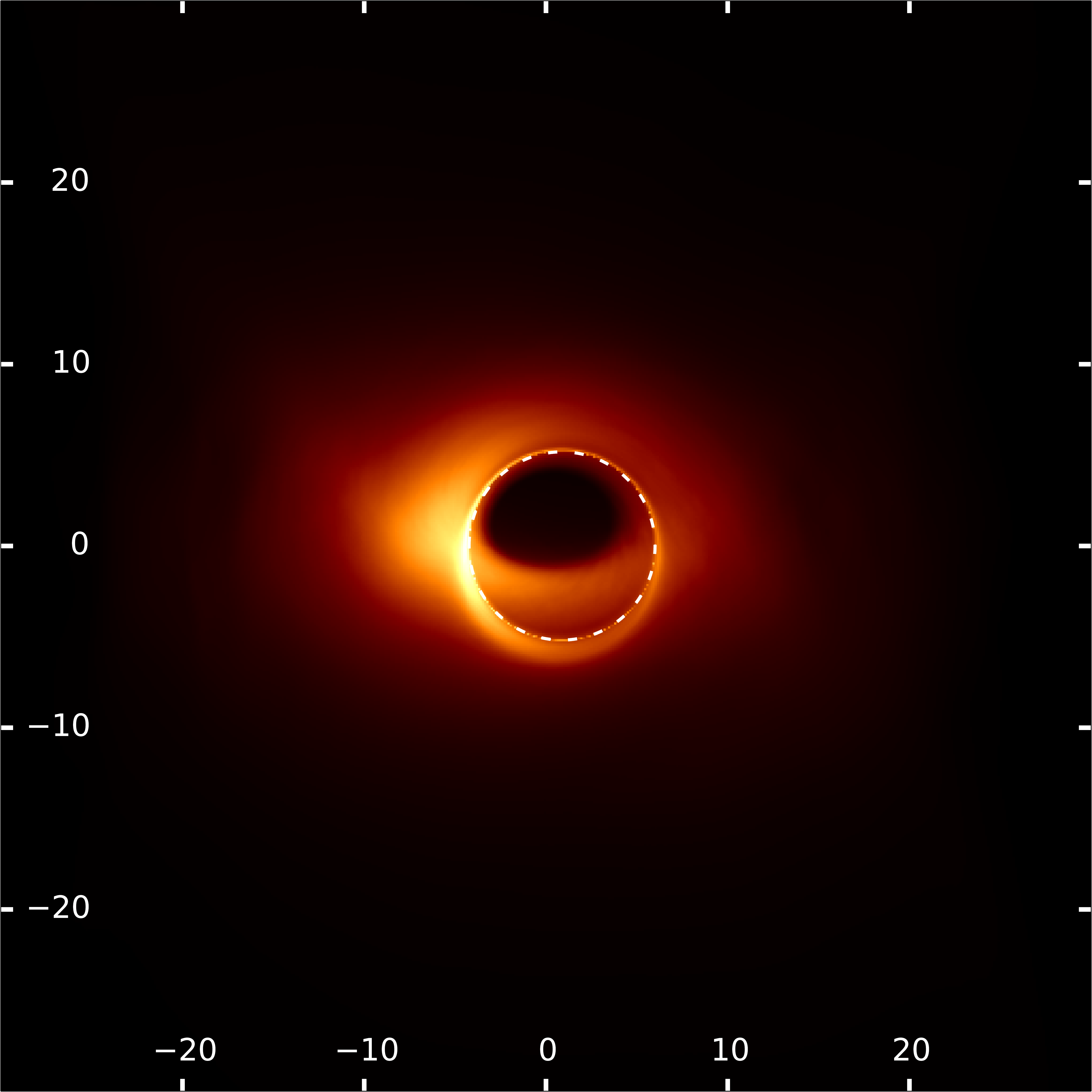}
	\caption{$a=0.5$, $i=60^\circ$.}
\end{subfigure}
\begin{subfigure}[b]{0.197\textwidth}
	\includegraphics[width=\textwidth]{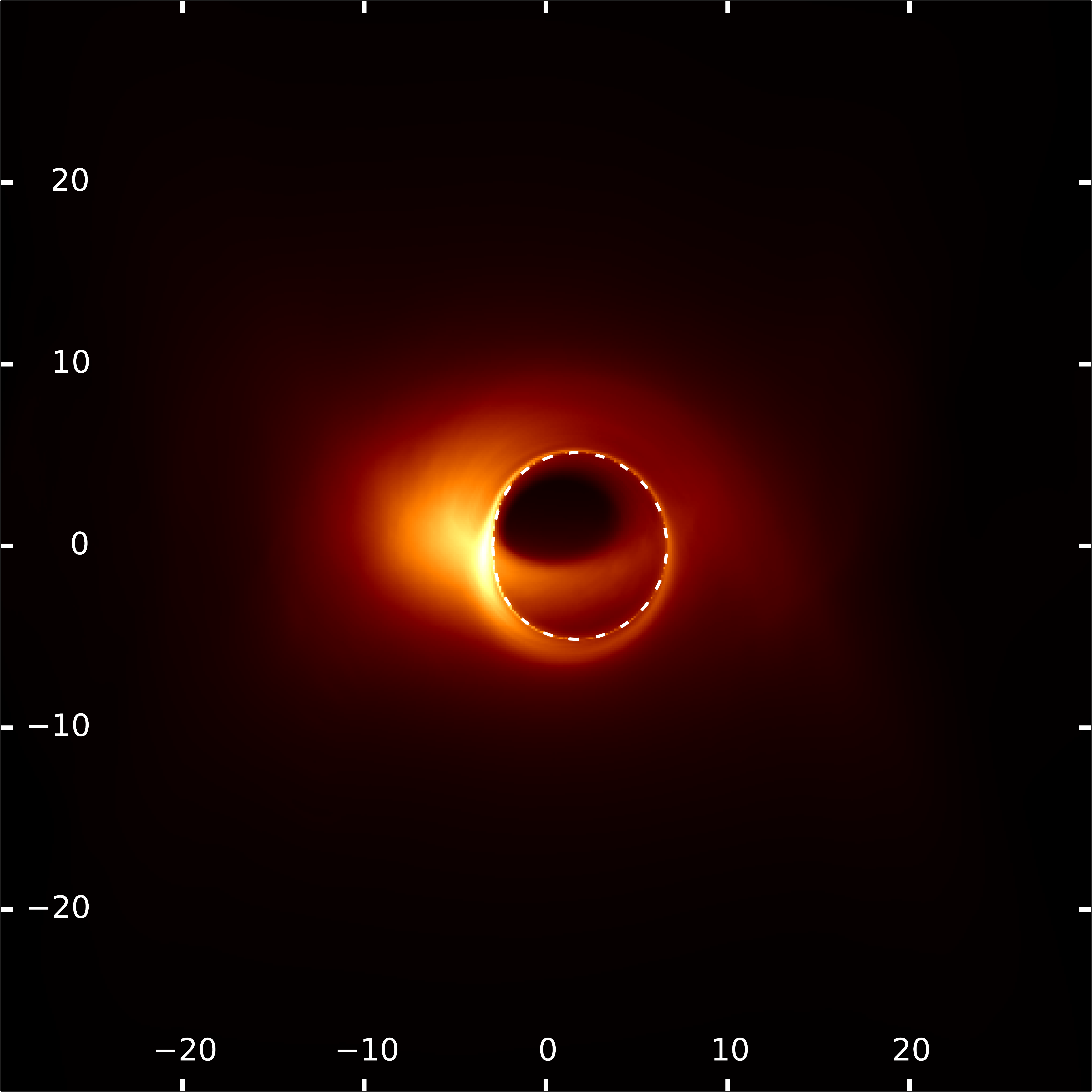}
	\caption{$a=0.9375$, $i=60^\circ$.}
\end{subfigure}
\begin{subfigure}[b]{0.197\textwidth}
	\includegraphics[width=\textwidth]{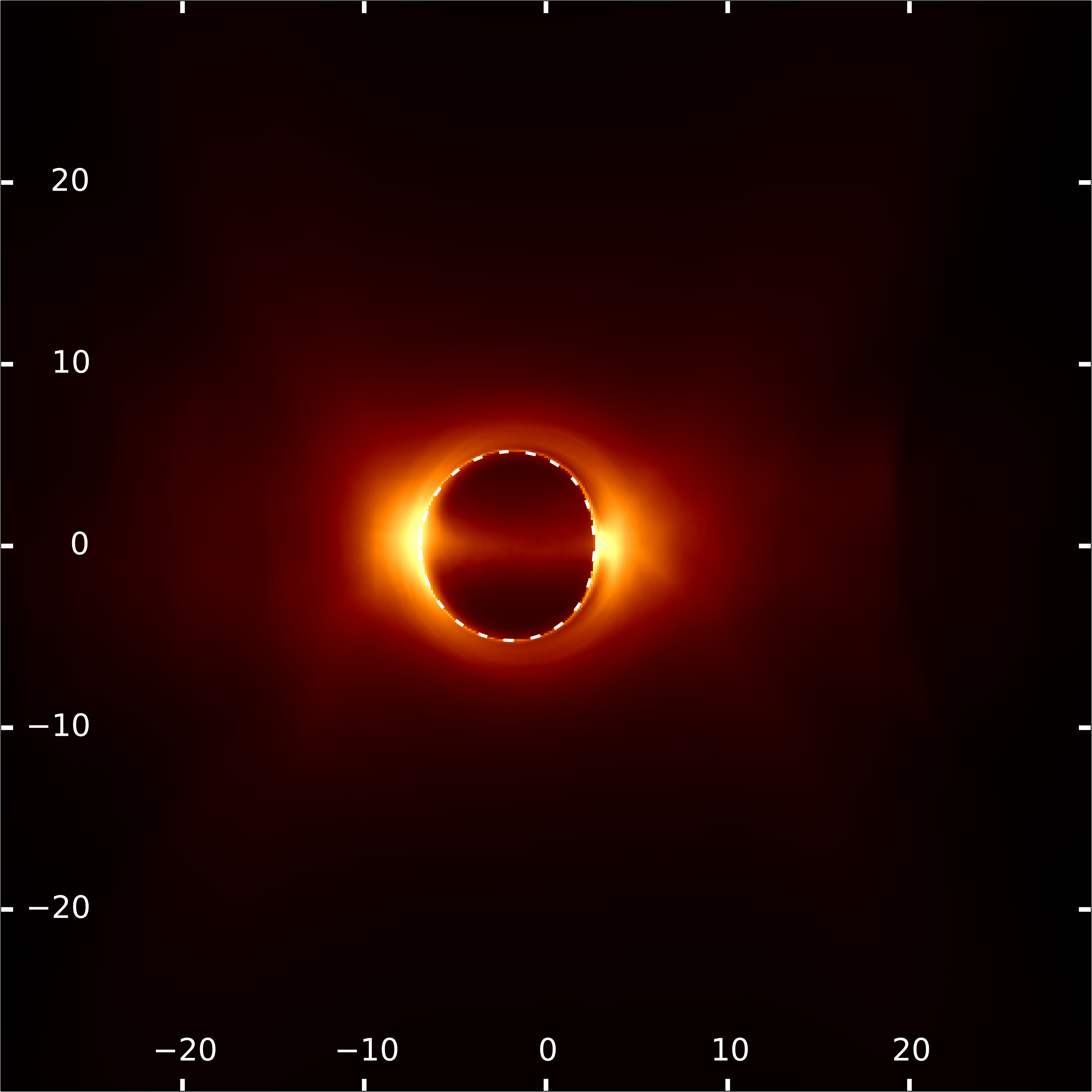}
	\caption{$a=-0.9375$, $i=90^\circ$.}
\end{subfigure}
\begin{subfigure}[b]{0.197\textwidth}
	\includegraphics[width=\textwidth]{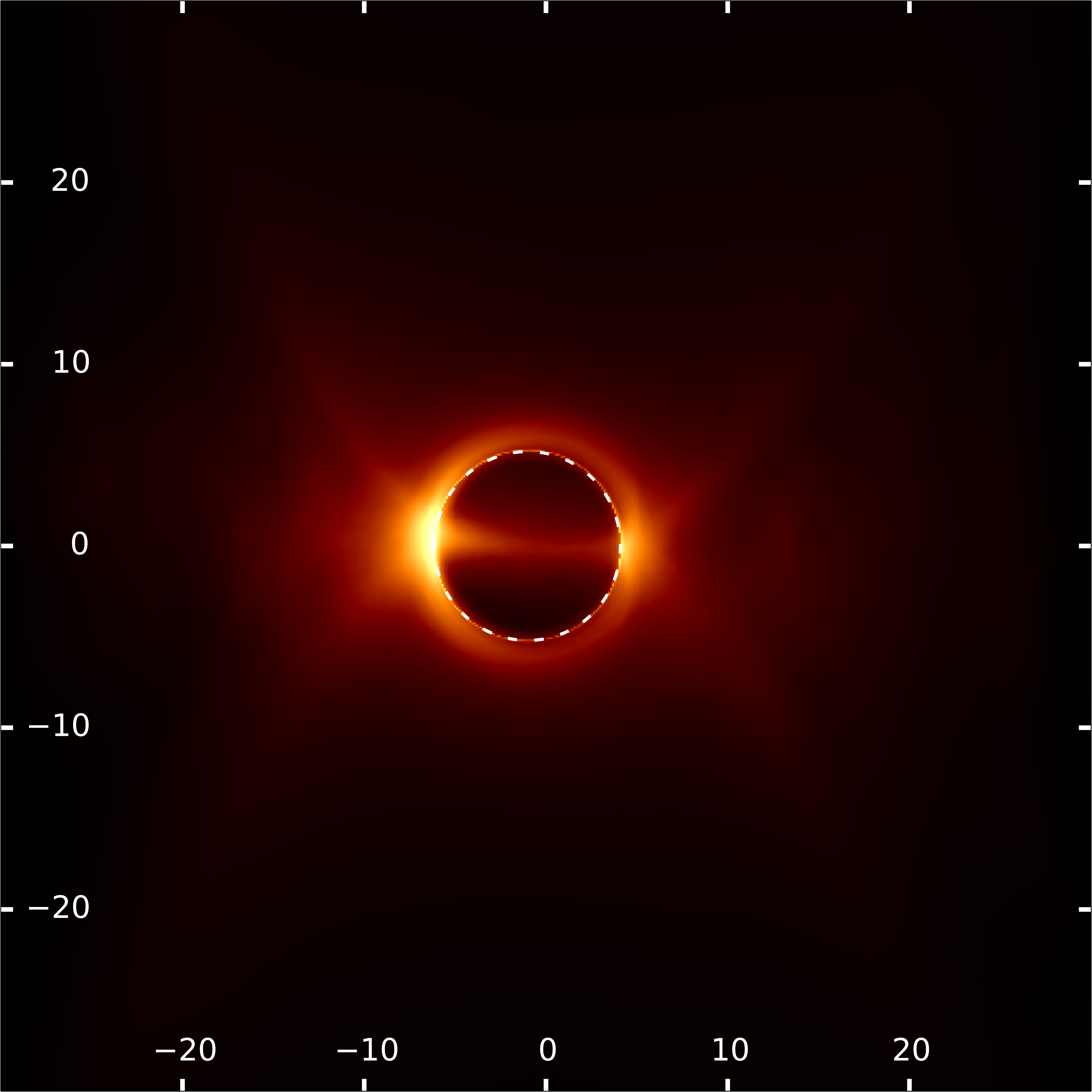}
	\caption{$a=-0.5$, $i=90^\circ$.}
\end{subfigure}
\begin{subfigure}[b]{0.197\textwidth}
	\includegraphics[width=\textwidth]{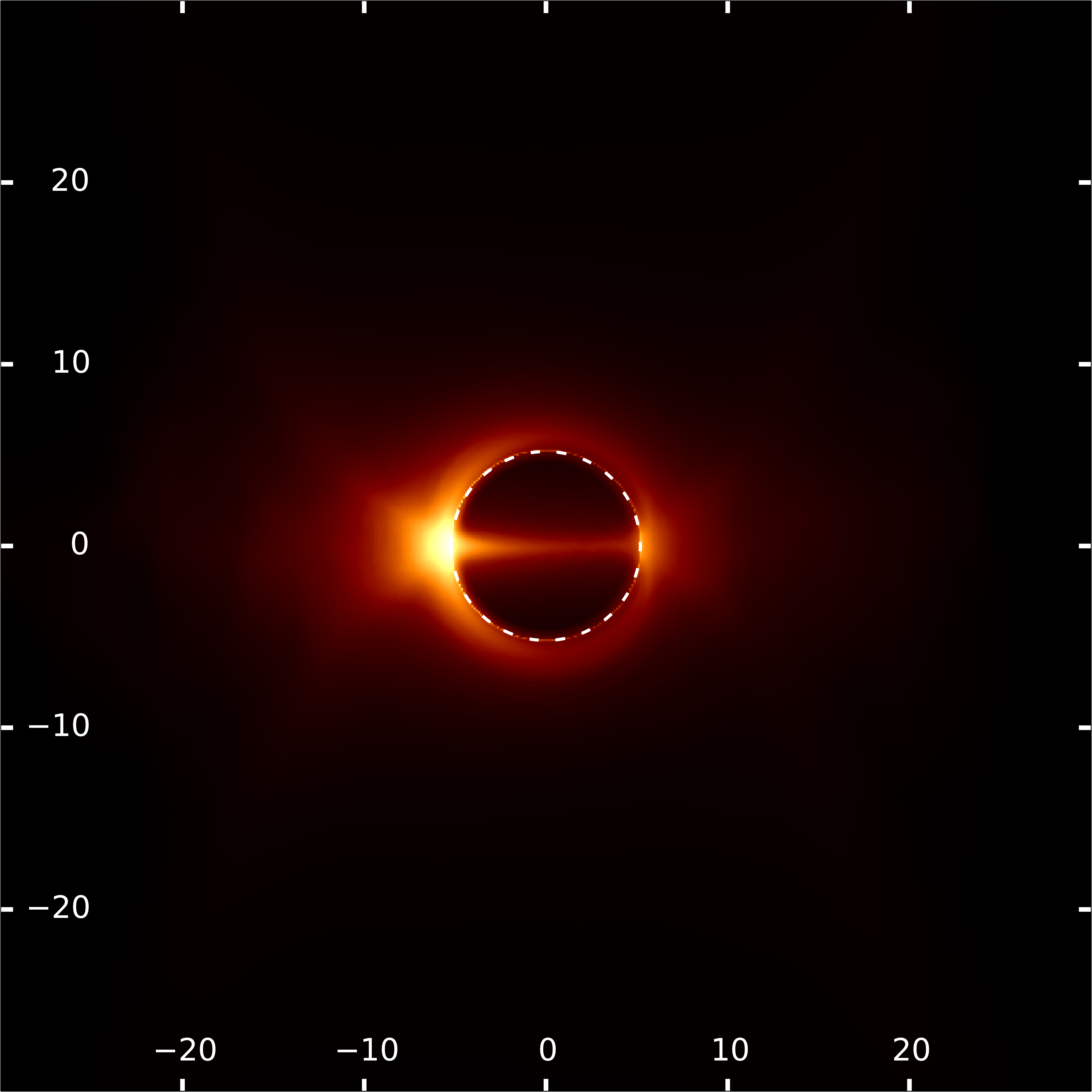}
	\caption{$a=0$, $i=90^\circ$.}
\end{subfigure}
\begin{subfigure}[b]{0.197\textwidth}
	\includegraphics[width=\textwidth]{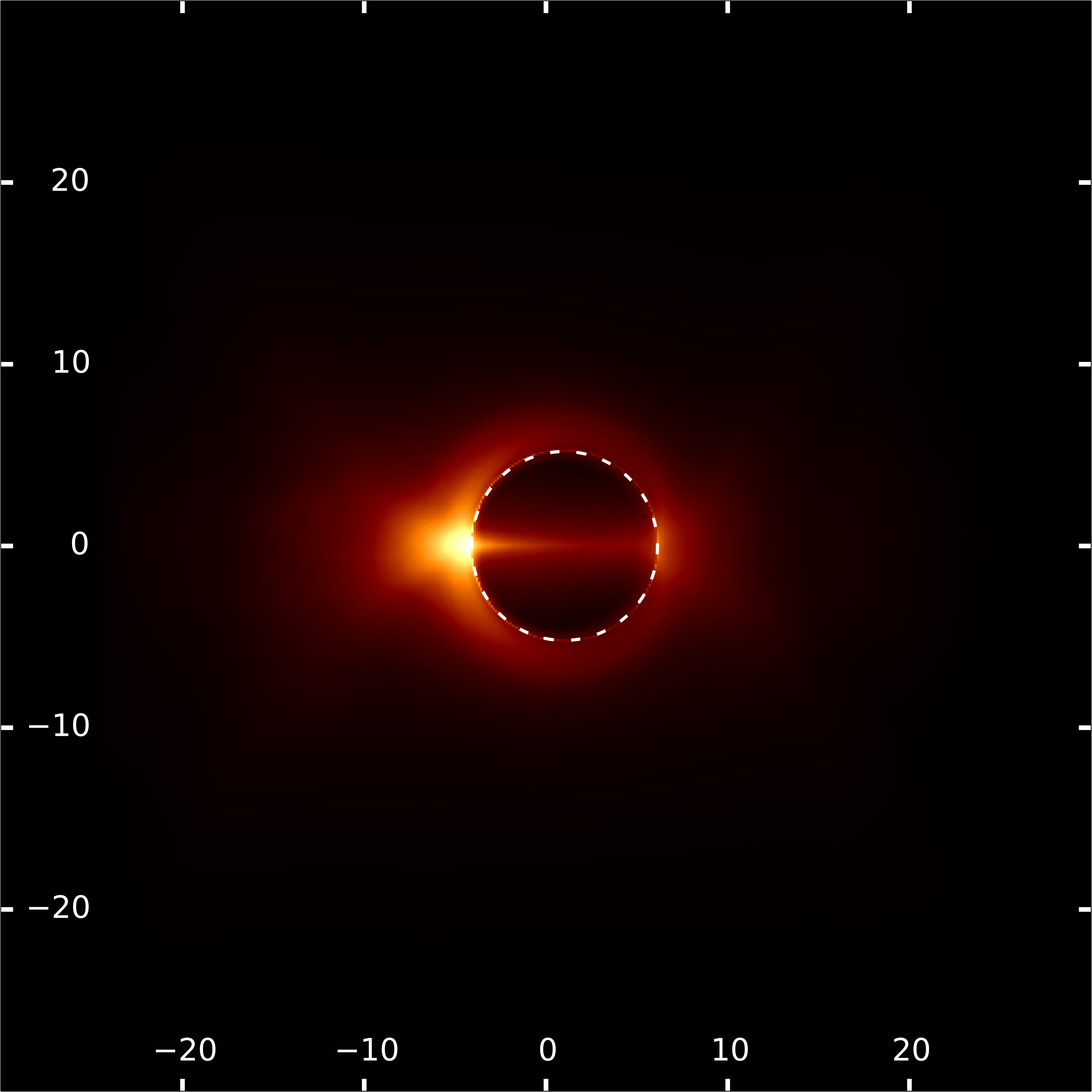}
	\caption{$a=0.5$, $i=90^\circ$.}
\end{subfigure}
\begin{subfigure}[b]{0.197\textwidth}
	\includegraphics[width=\textwidth]{Figures/mad_disk_a15o16_90_25-crop}
	\caption{$a=0.9375$, $i=90^\circ$.}
\end{subfigure}
\caption{Time-averaged, normalised intensity maps of our MAD, disc-dominated GRMHD models of Sgr A*, imaged at 230 GHz, at five different spins and four observer inclination angles, with an integrated flux density of 2.5 Jy. In each case, the photon ring, which marks the BHS, is indicated by a dashed line. The values for the impact parameters along the x- and y-axes are expressed in terms of $R_{\rm g}$. The image maps were plotted using a square-root intensity scale.}
\label{fig:mad_disk_25_matrix}
\end{figure*}

\begin{figure*}
\centering
\begin{subfigure}[b]{0.197\textwidth}
	\includegraphics[width=\textwidth]{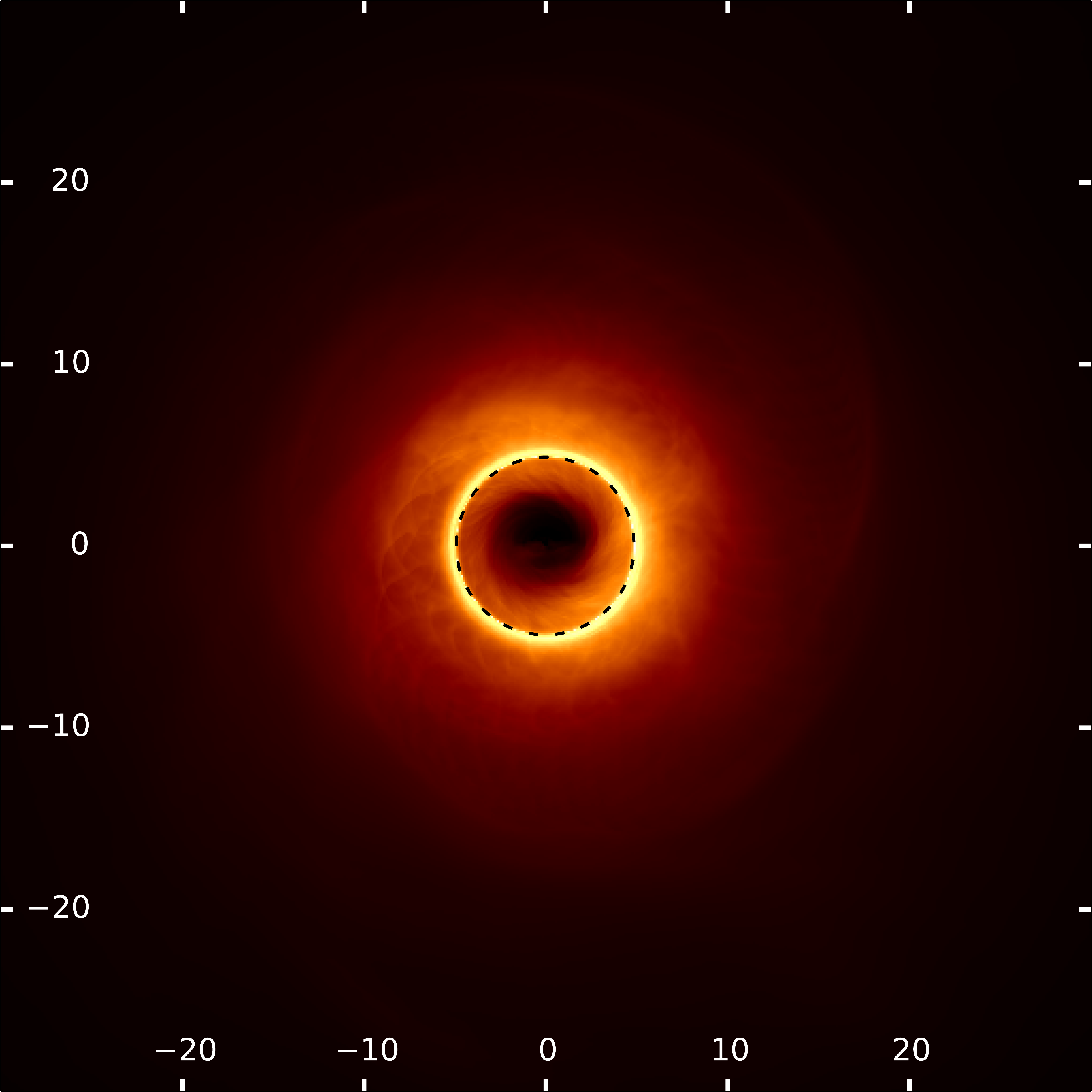}
	\caption{$a=-0.9375$, $i=1^\circ$.}
\end{subfigure}
\begin{subfigure}[b]{0.197\textwidth}
	\includegraphics[width=\textwidth]{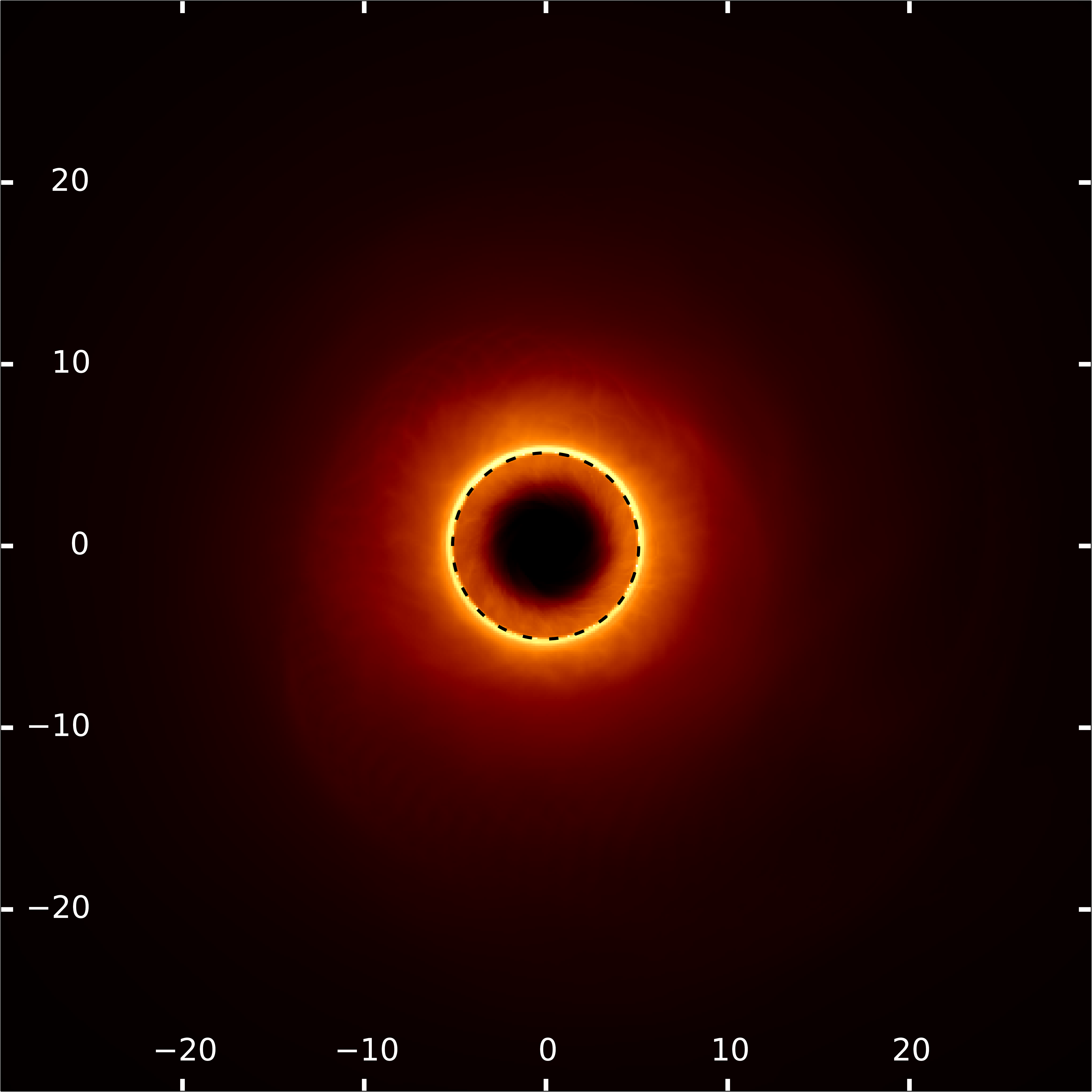}
	\caption{$a=-0.5$, $i=1^\circ$.}
\end{subfigure}
\begin{subfigure}[b]{0.197\textwidth}
	\includegraphics[width=\textwidth]{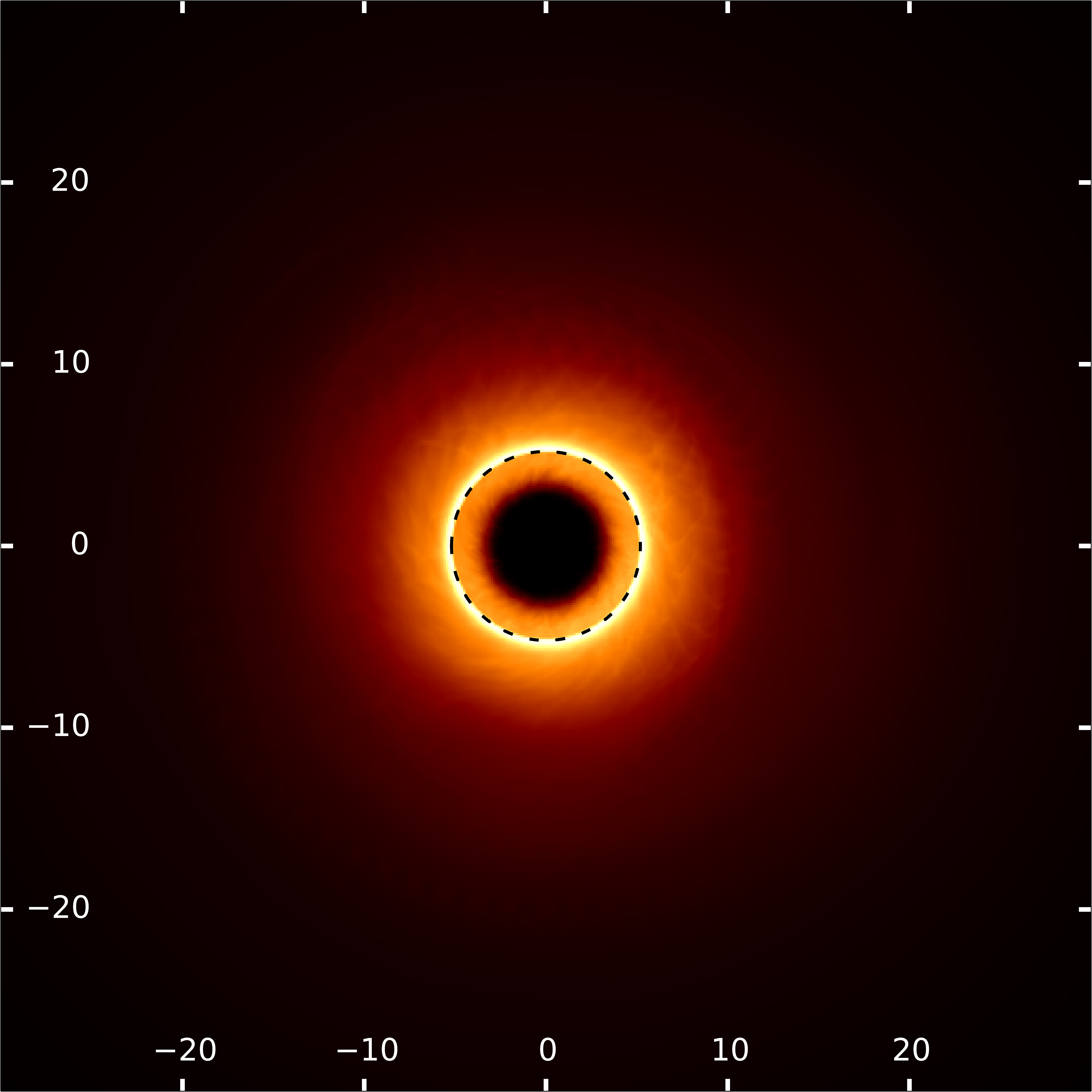}
	\caption{$a=0$, $i=1^\circ$.}
\end{subfigure}
\begin{subfigure}[b]{0.197\textwidth}
	\includegraphics[width=\textwidth]{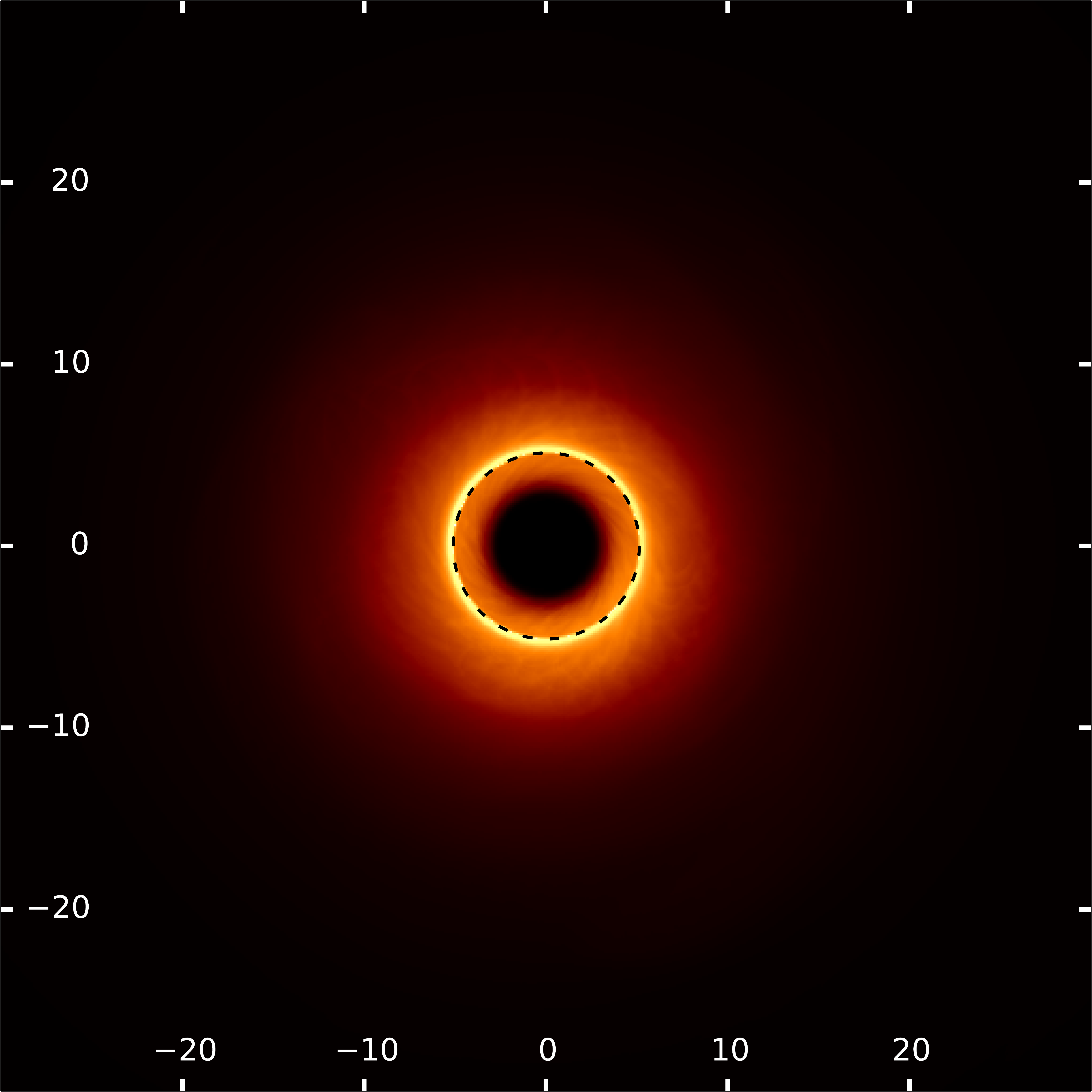}
	\caption{$a=0.5$, $i=1^\circ$.}
\end{subfigure}
\begin{subfigure}[b]{0.197\textwidth}
	\includegraphics[width=\textwidth]{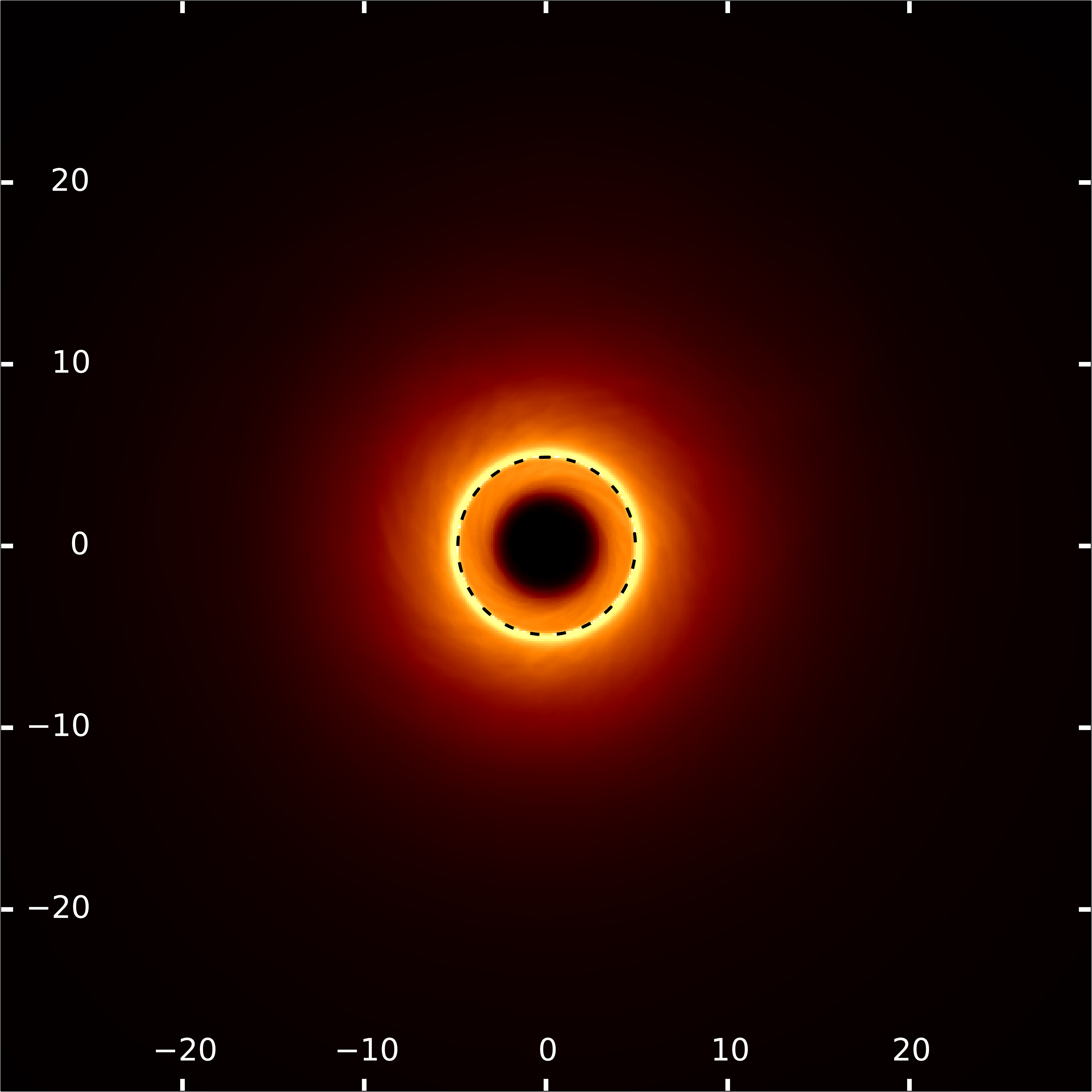}
	\caption{$a=0.9375$, $i=1^\circ$.}
\end{subfigure}
\begin{subfigure}[b]{0.197\textwidth}
	\includegraphics[width=\textwidth]{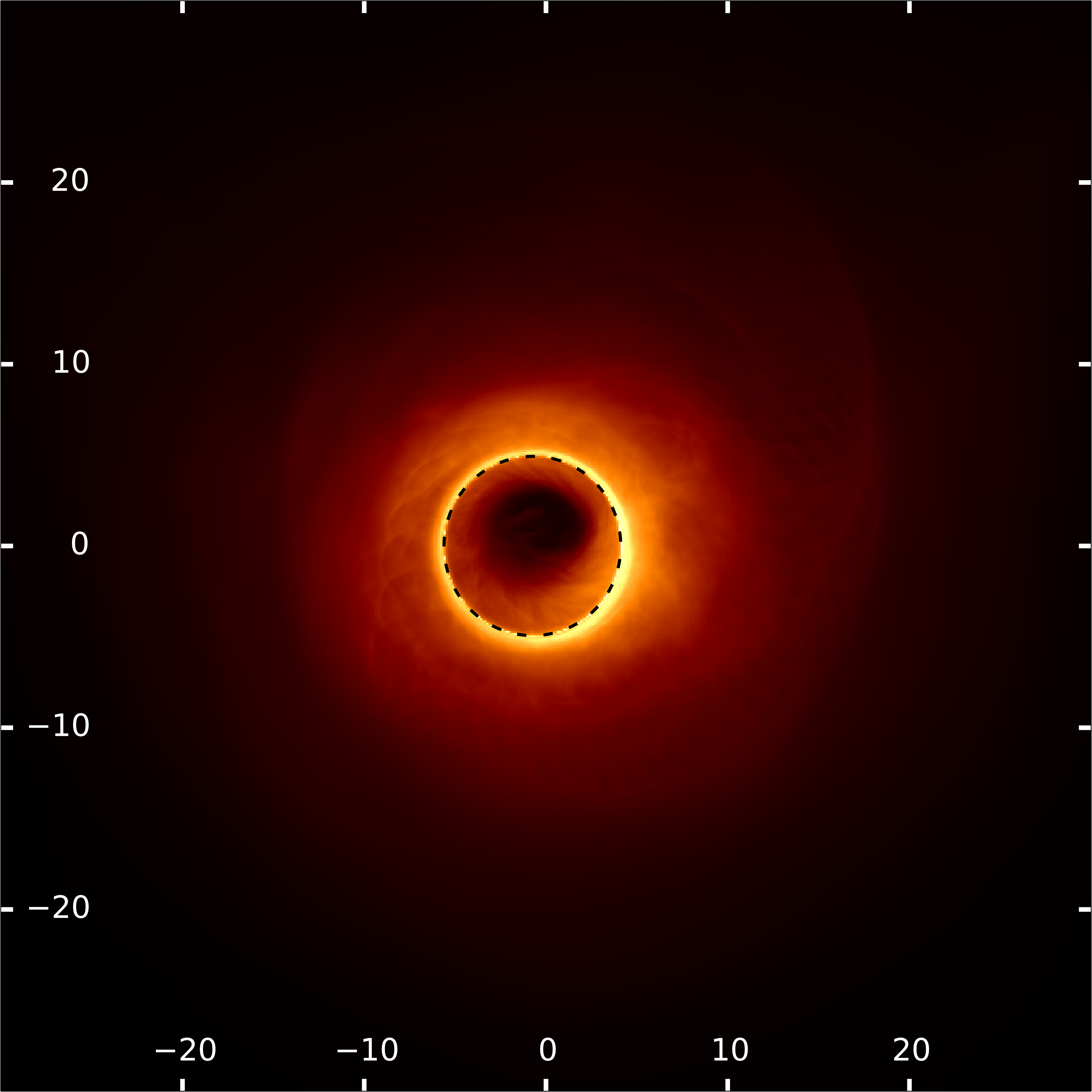}
	\caption{$a=-0.9375$, $i=20^\circ$.}
\end{subfigure}
\begin{subfigure}[b]{0.197\textwidth}
	\includegraphics[width=\textwidth]{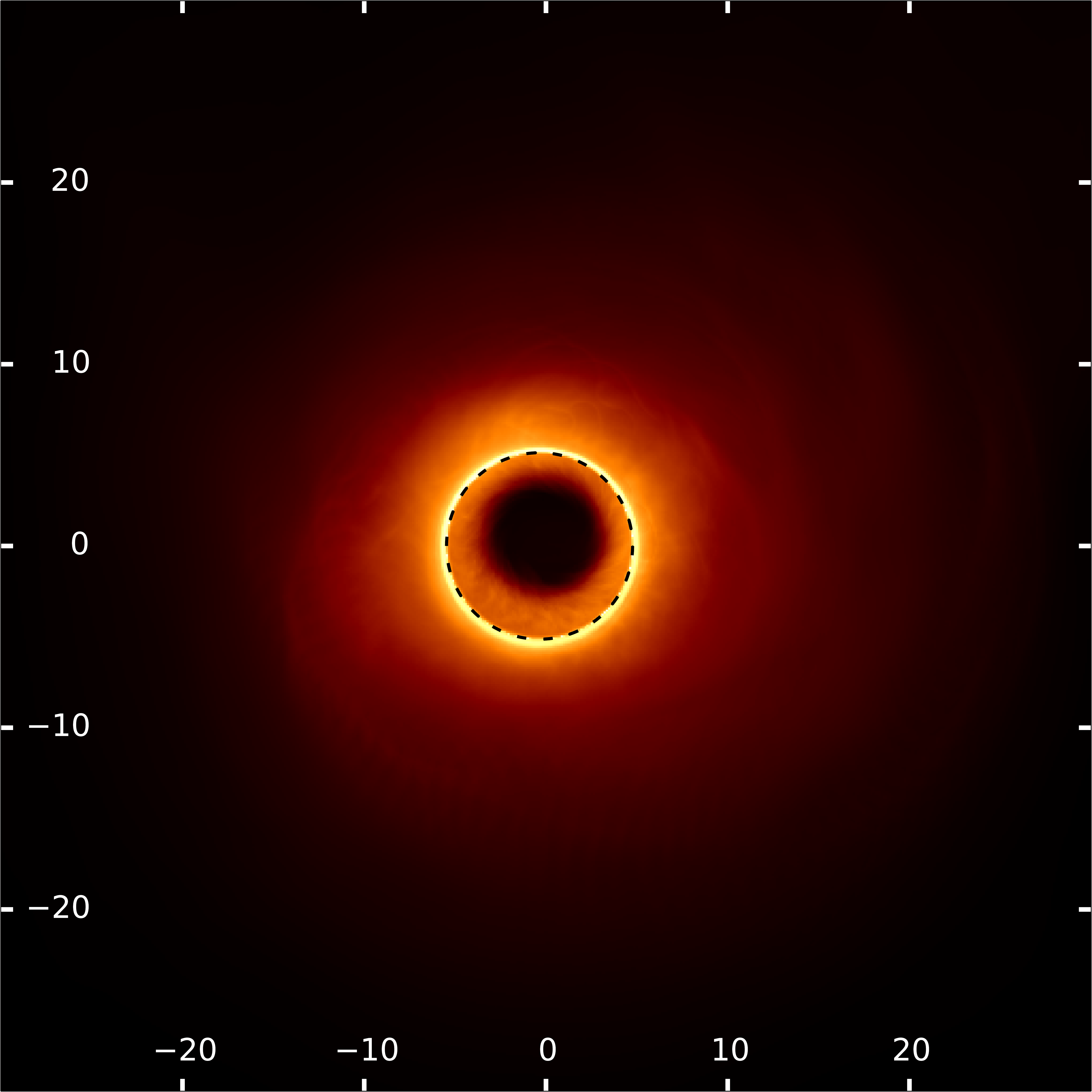}
	\caption{$a=-0.5$, $i=20^\circ$.}
\end{subfigure}
\begin{subfigure}[b]{0.197\textwidth}
	\includegraphics[width=\textwidth]{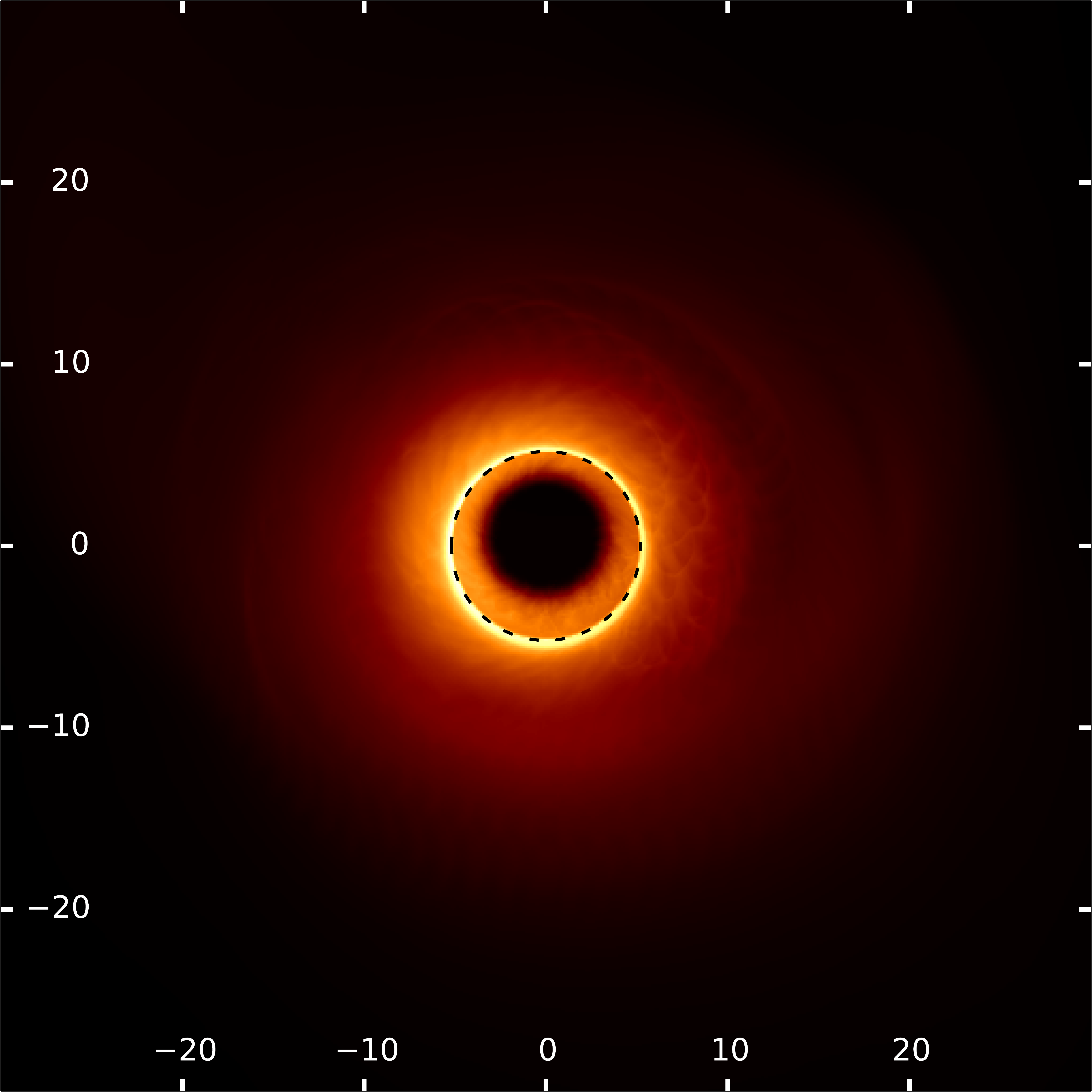}
	\caption{$a=0$, $i=20^\circ$.}
\end{subfigure}
\begin{subfigure}[b]{0.197\textwidth}
	\includegraphics[width=\textwidth]{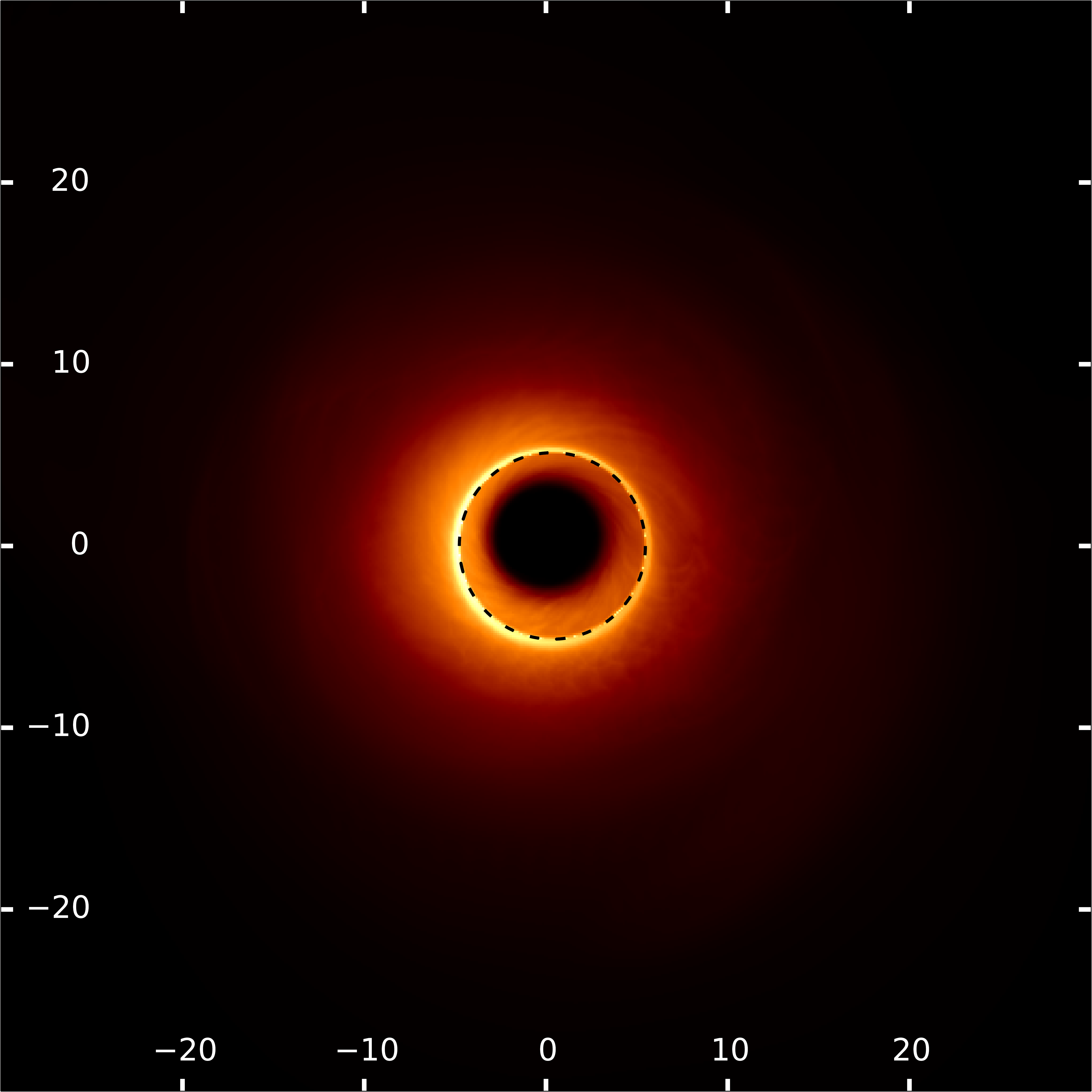}
	\caption{$a=0.5$, $i=20^\circ$.}
\end{subfigure}
\begin{subfigure}[b]{0.197\textwidth}
	\includegraphics[width=\textwidth]{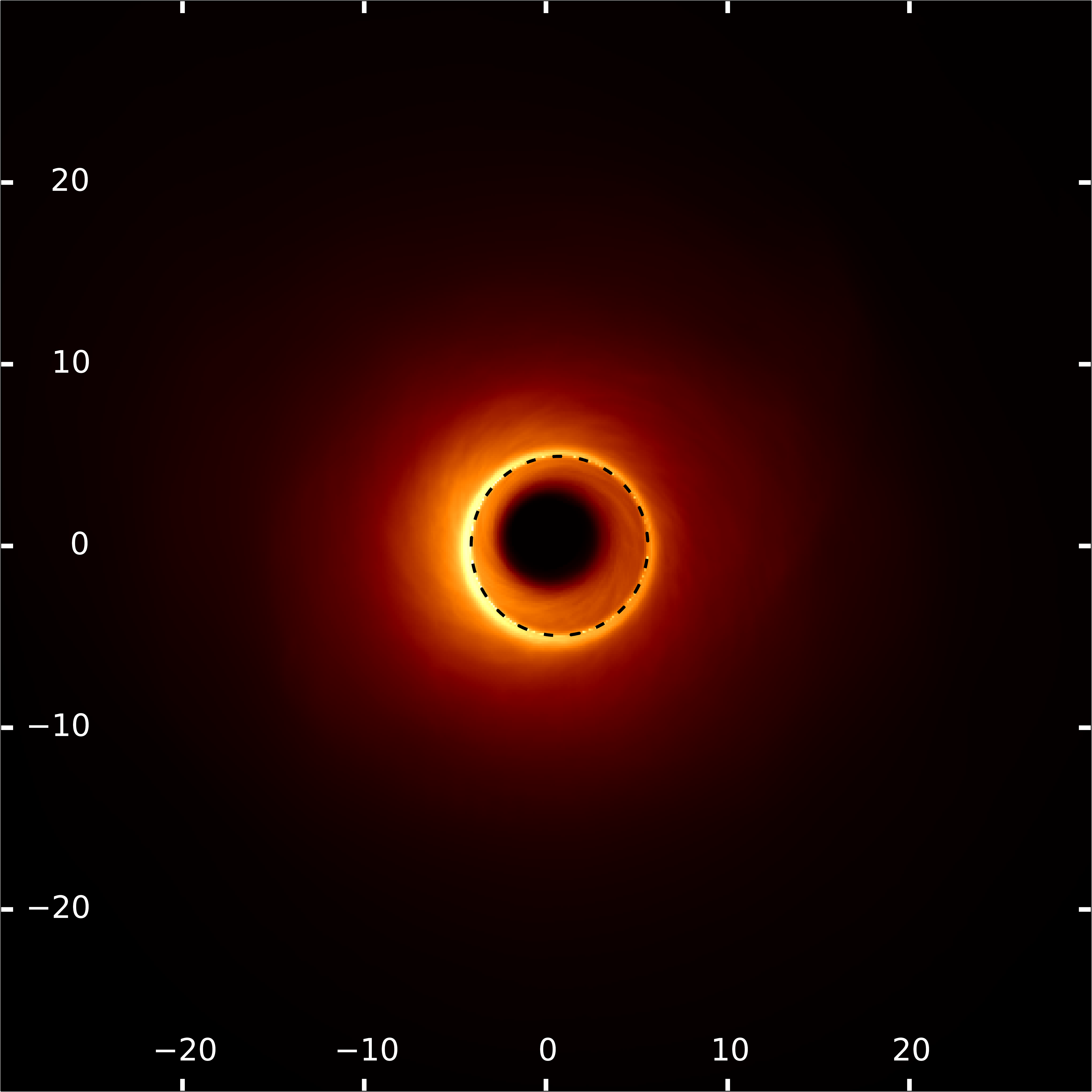}
	\caption{$a=0.9375$, $i=20^\circ$.}
\end{subfigure}
\begin{subfigure}[b]{0.197\textwidth}
	\includegraphics[width=\textwidth]{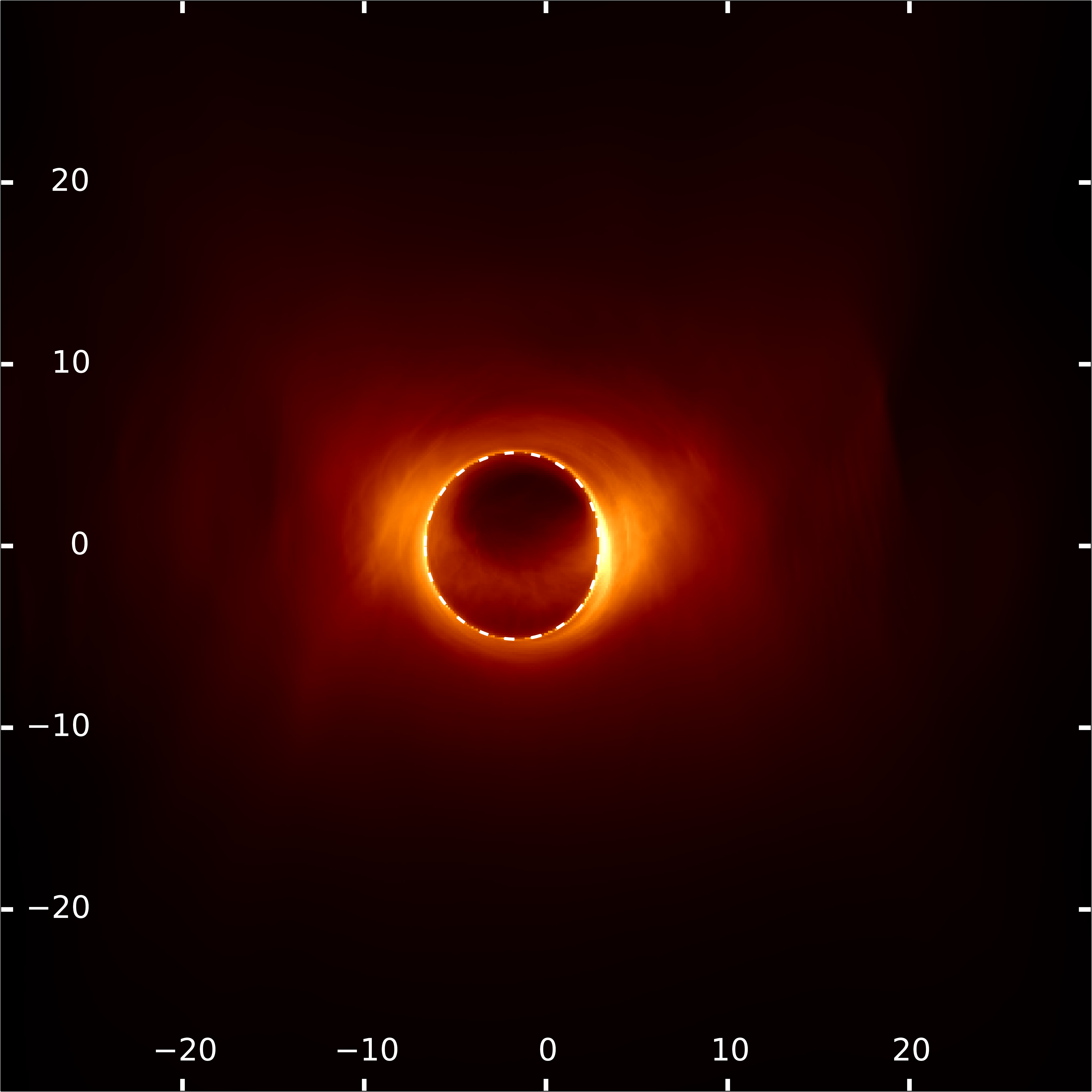}
	\caption{$a=-0.9375$, $i=60^\circ$.}
\end{subfigure}
\begin{subfigure}[b]{0.197\textwidth}
	\includegraphics[width=\textwidth]{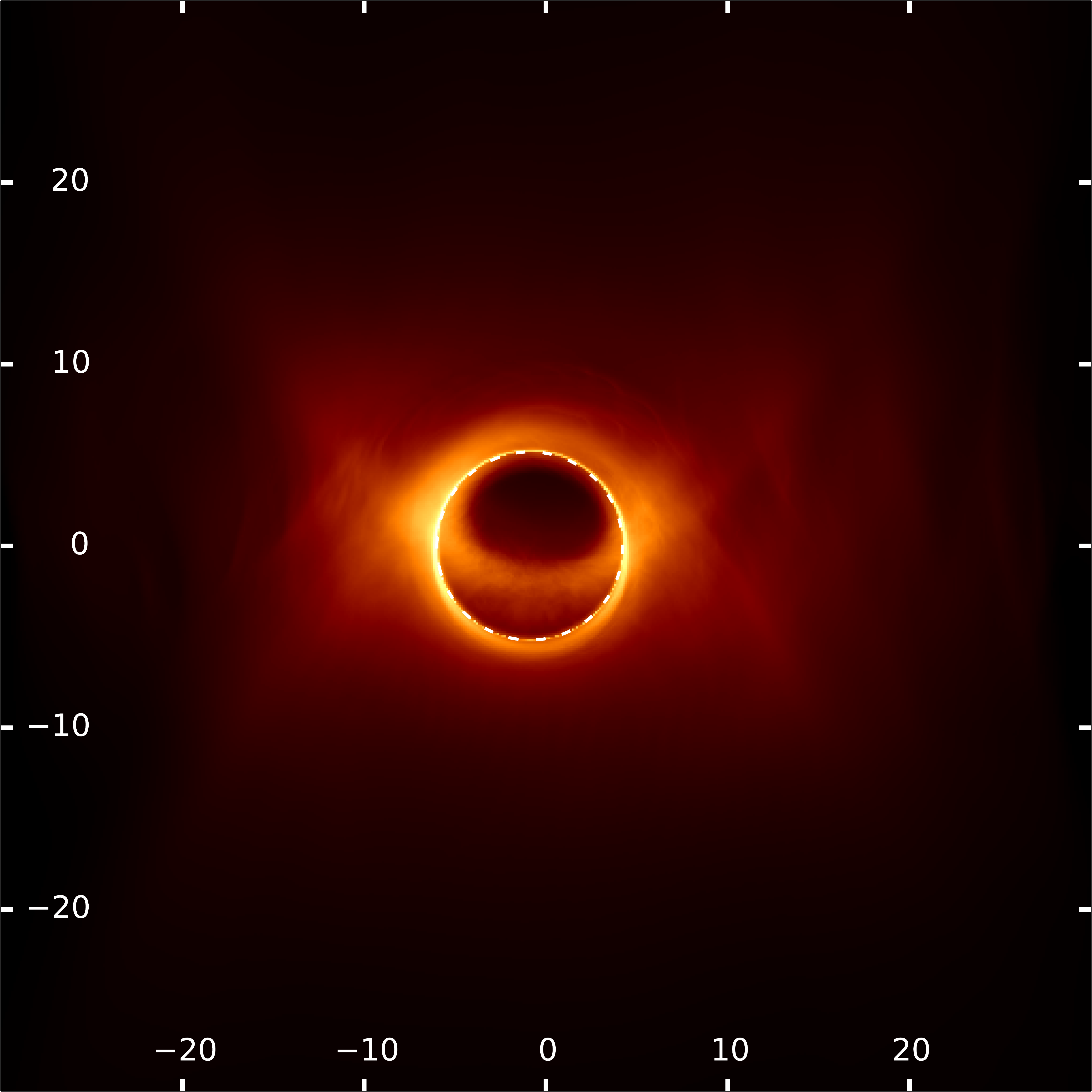}
	\caption{$a=-0.5$, $i=60^\circ$.}
\end{subfigure}
\begin{subfigure}[b]{0.197\textwidth}
	\includegraphics[width=\textwidth]{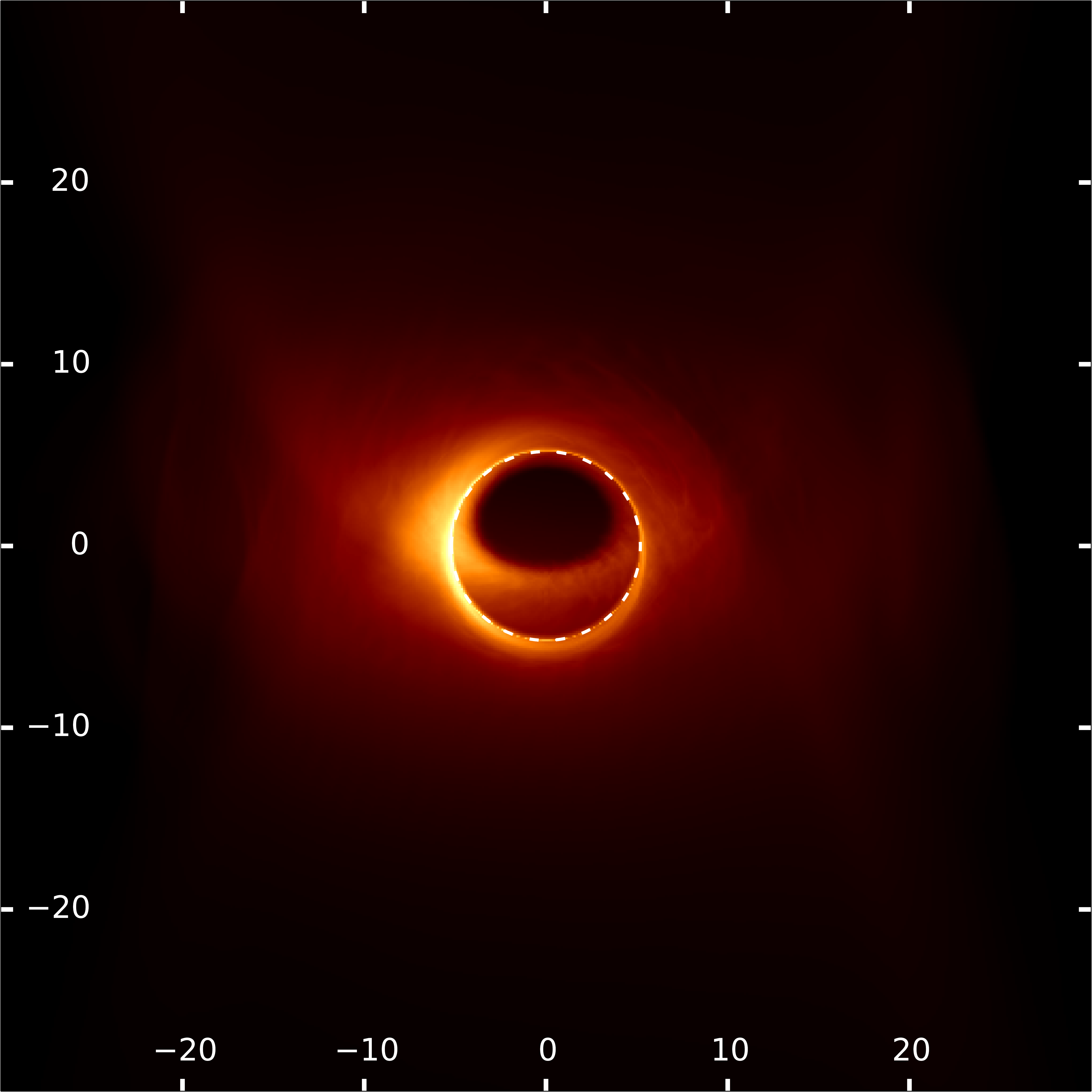}
	\caption{$a=0$, $i=60^\circ$.}
\end{subfigure}
\begin{subfigure}[b]{0.197\textwidth}
	\includegraphics[width=\textwidth]{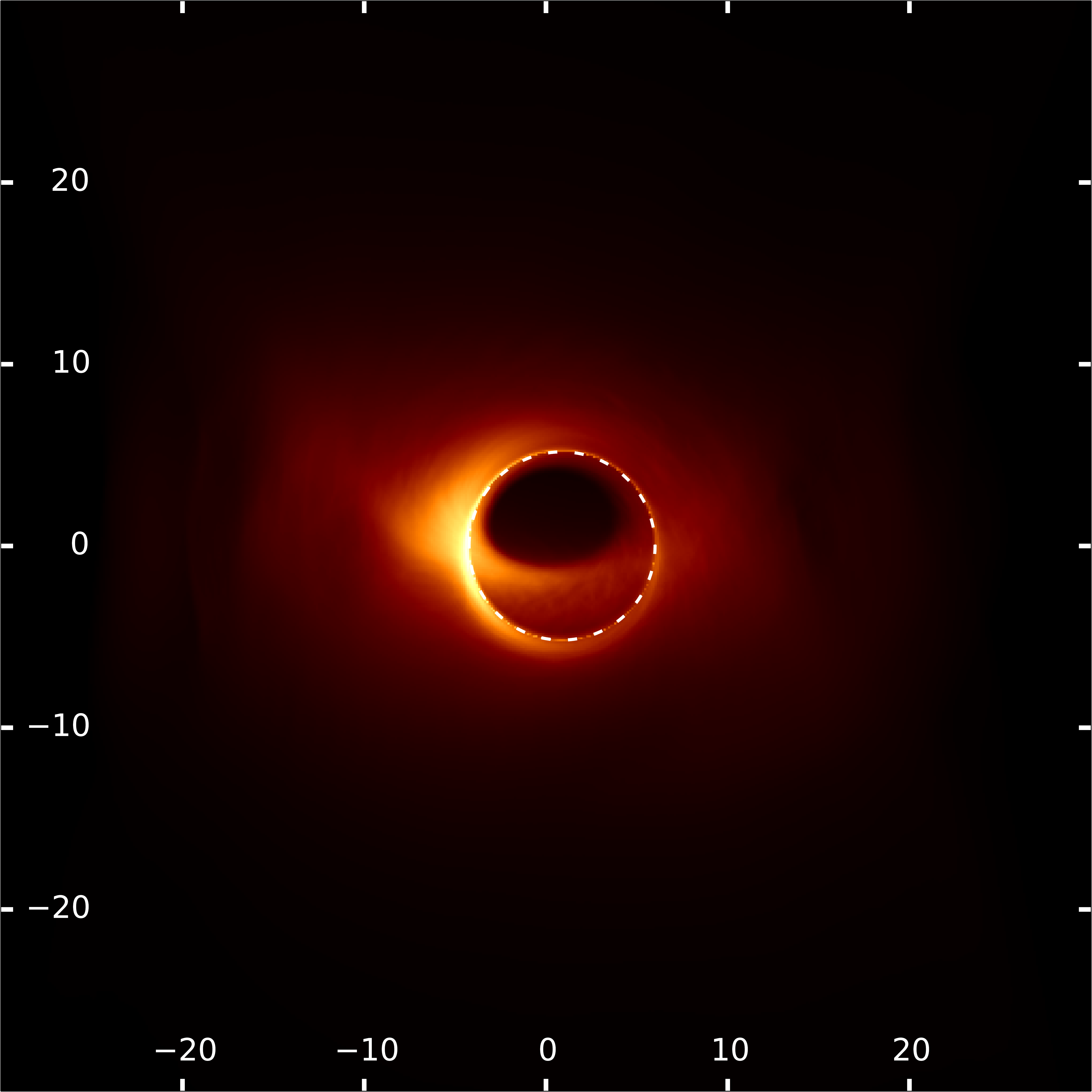}
	\caption{$a=0.5$, $i=60^\circ$.}
\end{subfigure}
\begin{subfigure}[b]{0.197\textwidth}
	\includegraphics[width=\textwidth]{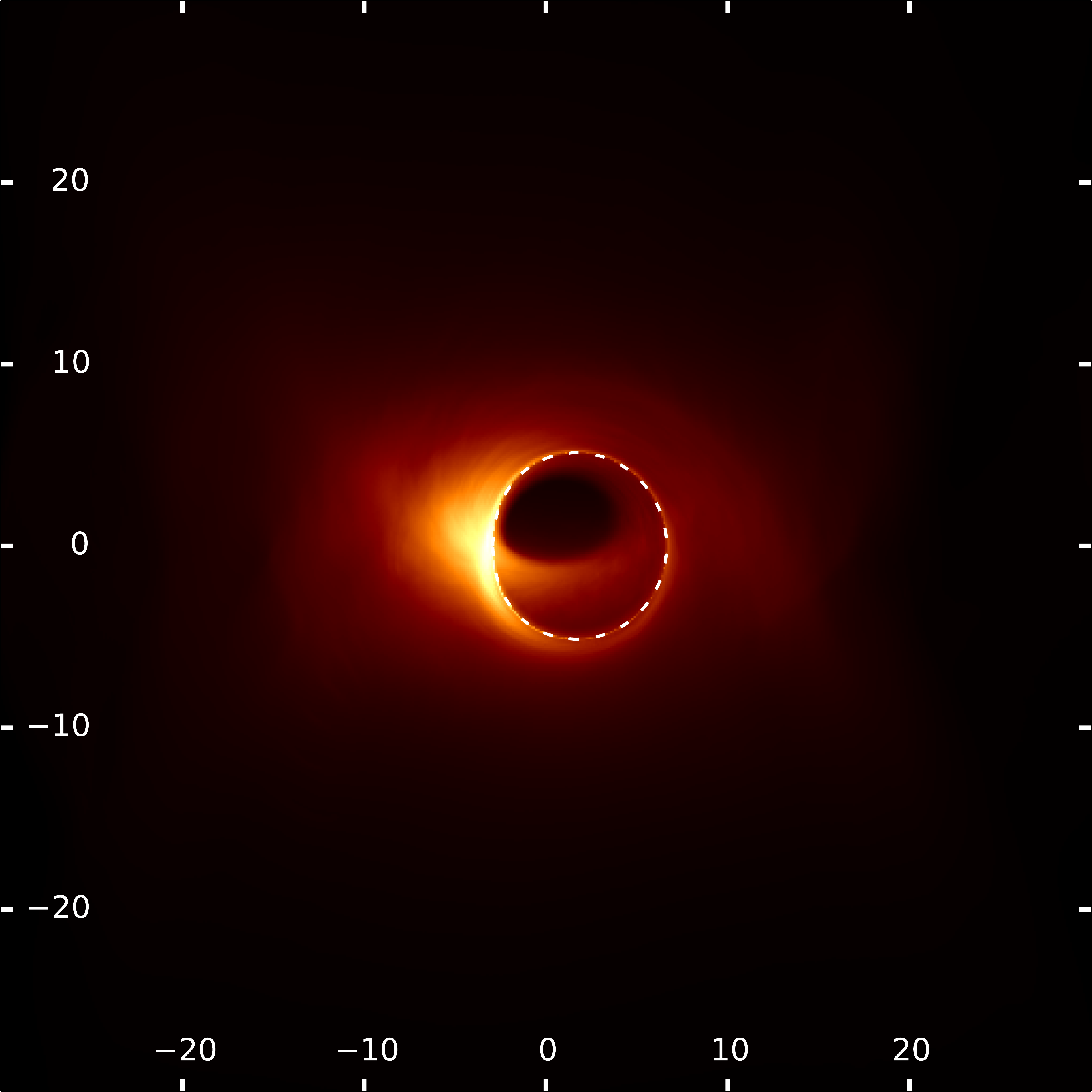}
	\caption{$a=0.9375$, $i=60^\circ$.}
\end{subfigure}
\begin{subfigure}[b]{0.197\textwidth}
	\includegraphics[width=\textwidth]{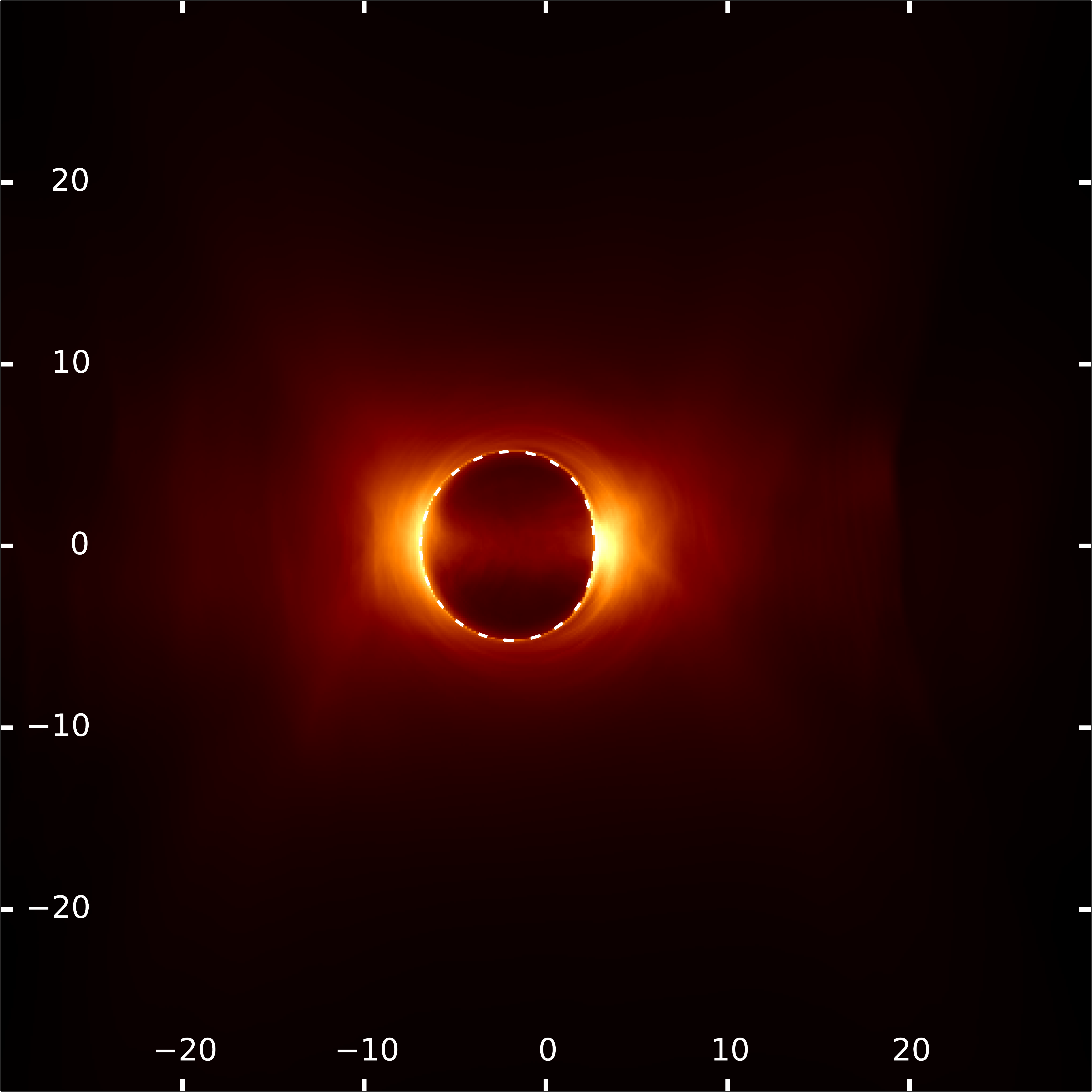}
	\caption{$a=-0.9375$, $i=90^\circ$.}
\end{subfigure}
\begin{subfigure}[b]{0.197\textwidth}
	\includegraphics[width=\textwidth]{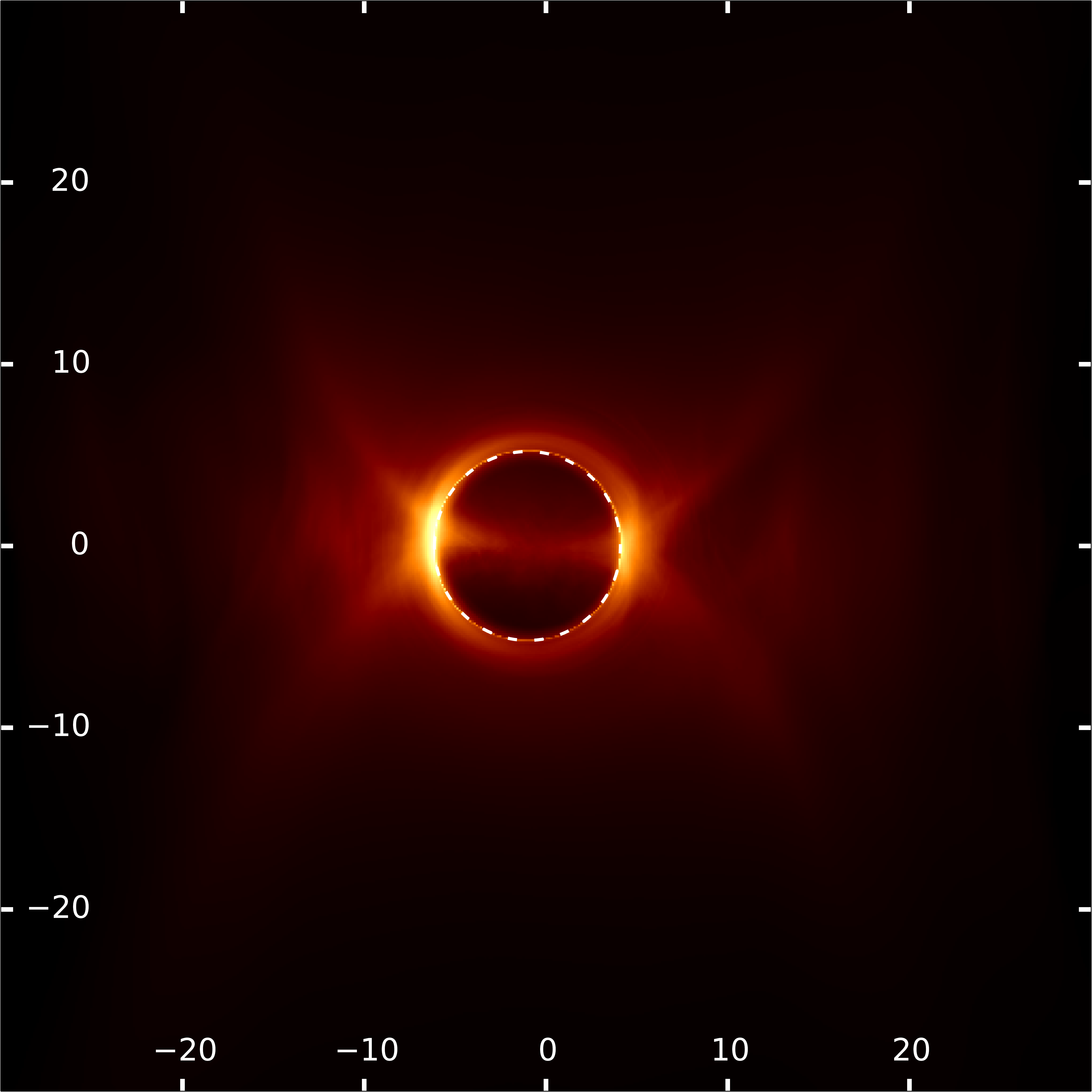}
	\caption{$a=-0.5$, $i=90^\circ$.}
\end{subfigure}
\begin{subfigure}[b]{0.197\textwidth}
	\includegraphics[width=\textwidth]{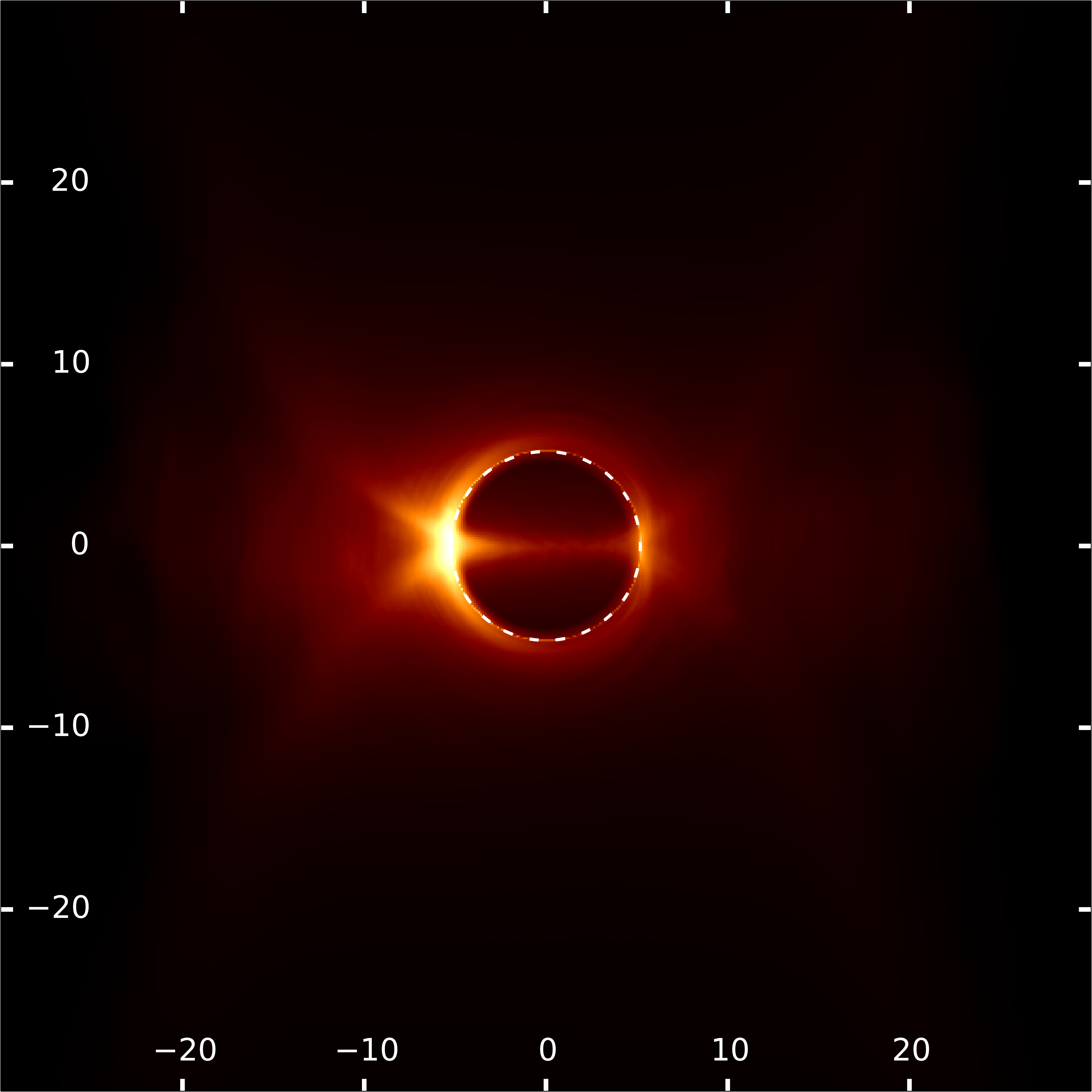}
	\caption{$a=0$, $i=90^\circ$.}
\end{subfigure}
\begin{subfigure}[b]{0.197\textwidth}
	\includegraphics[width=\textwidth]{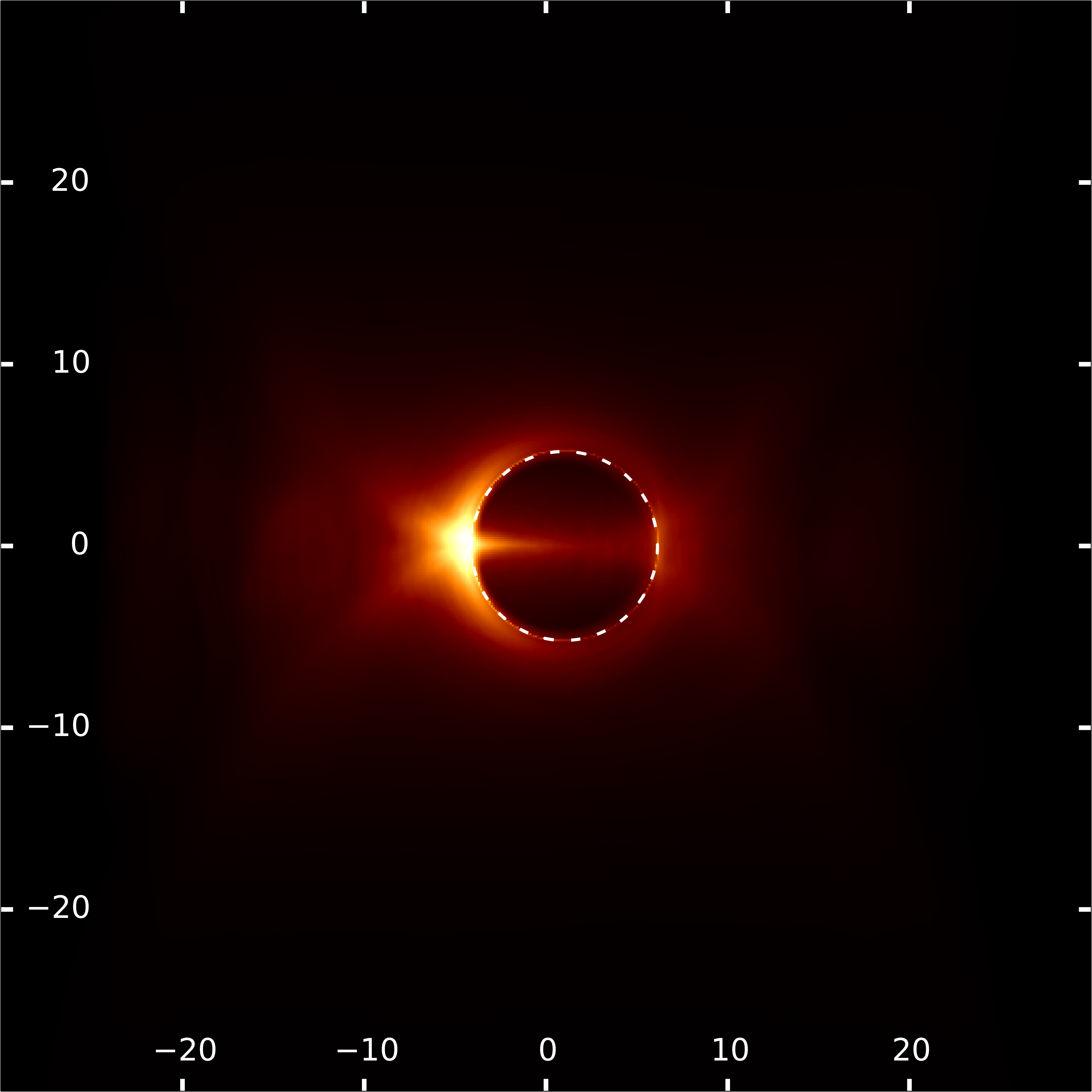}
	\caption{$a=0.5$, $i=90^\circ$.}
\end{subfigure}
\begin{subfigure}[b]{0.197\textwidth}
	\includegraphics[width=\textwidth]{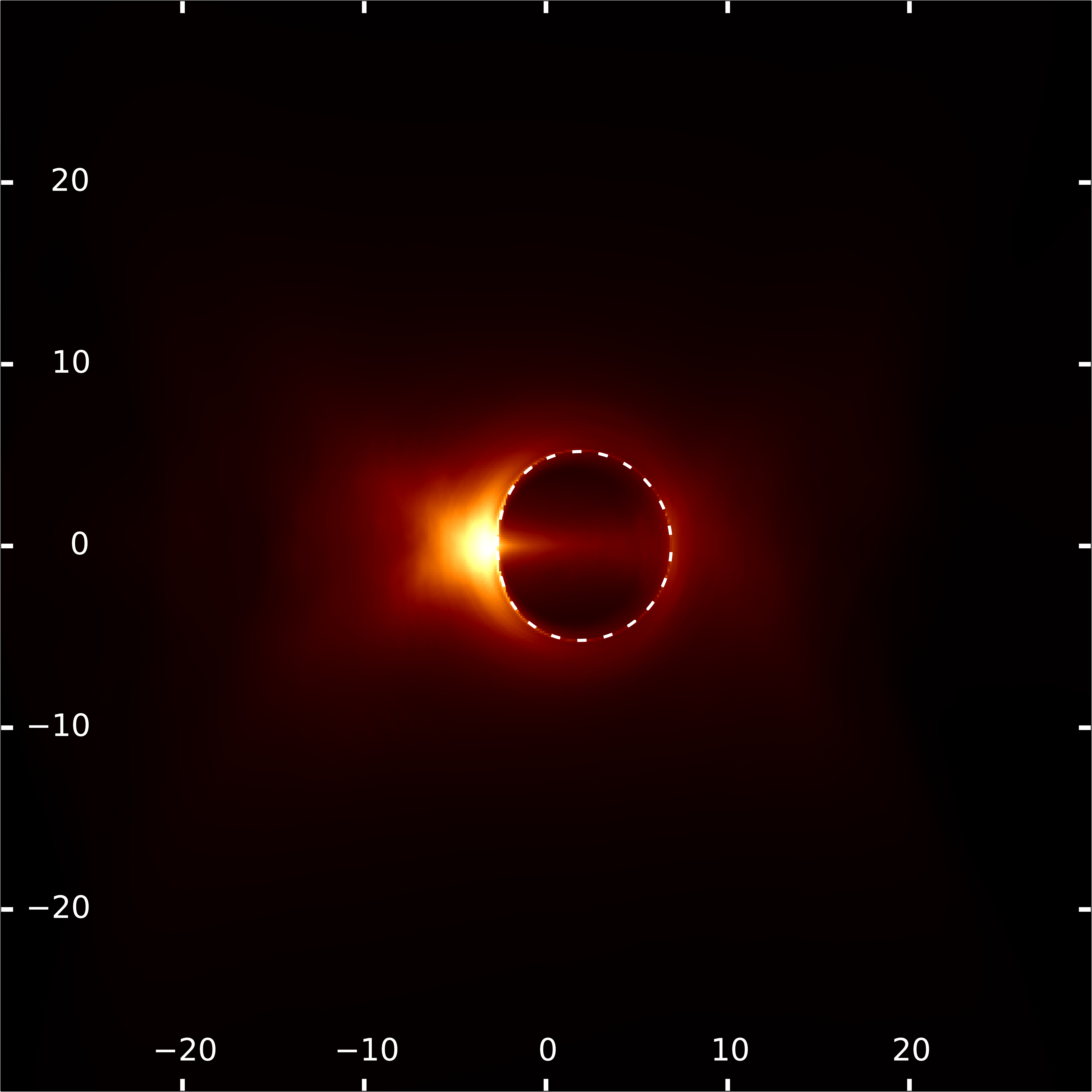}
	\caption{$a=0.9375$, $i=90^\circ$.}
\end{subfigure}
\caption{Time-averaged, normalised intensity maps of our MAD, jet-dominated GRMHD models of Sgr A*, imaged at 230 GHz, at five different spins and four observer inclination angles, with an integrated flux density of 1.25 Jy. In each case, the photon ring, which marks the BHS, is indicated by a dashed line. The values for the impact parameters along the x- and y-axes are expressed in terms of $R_{\rm g}$. The image maps were plotted using a square-root intensity scale.}
\label{fig:mad_jet_125_matrix}
\end{figure*}

\begin{figure*}
\centering
\begin{subfigure}[b]{0.197\textwidth}
	\includegraphics[width=\textwidth]{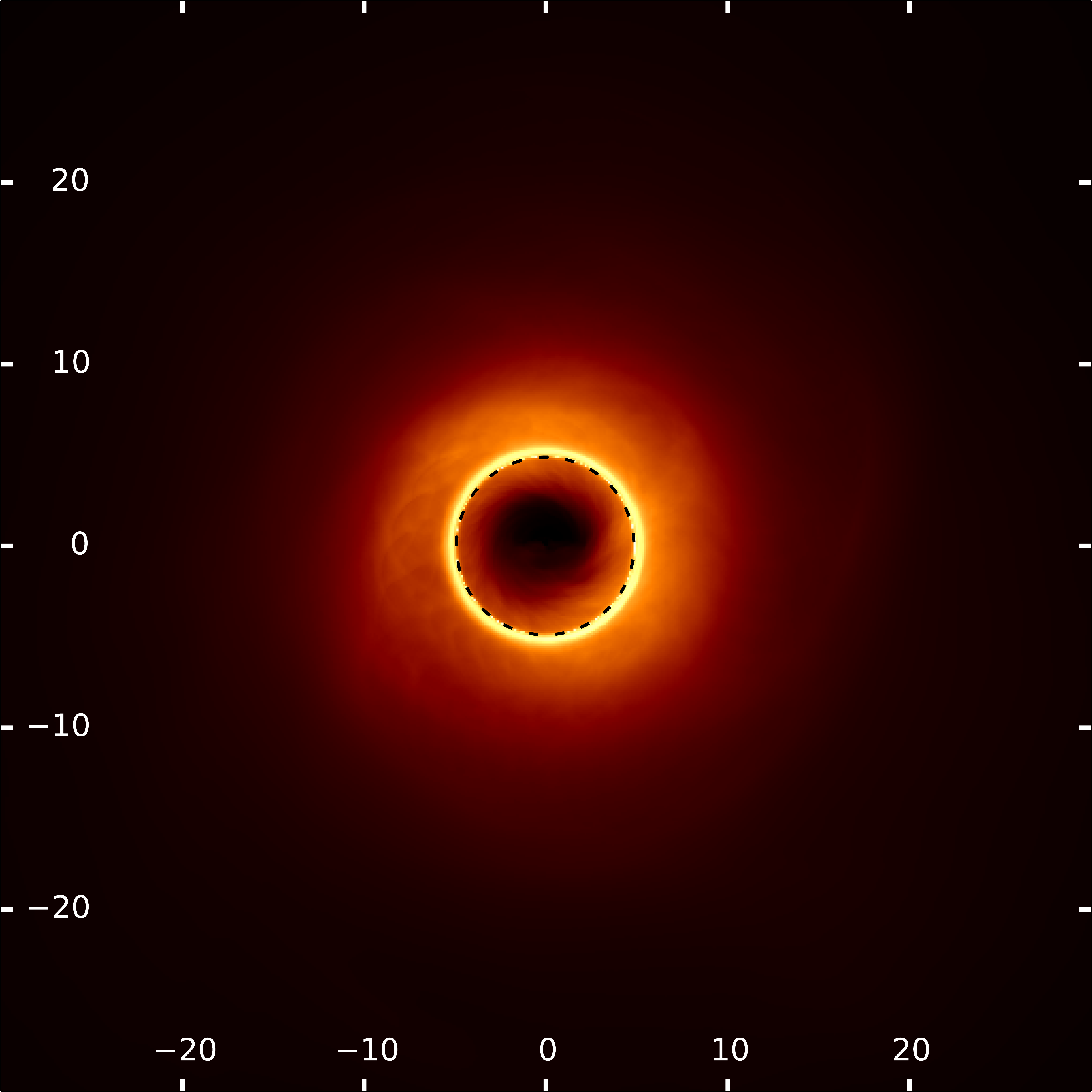}
	\caption{$a=-0.9375$, $i=1^\circ$.}
\end{subfigure}
\begin{subfigure}[b]{0.197\textwidth}
	\includegraphics[width=\textwidth]{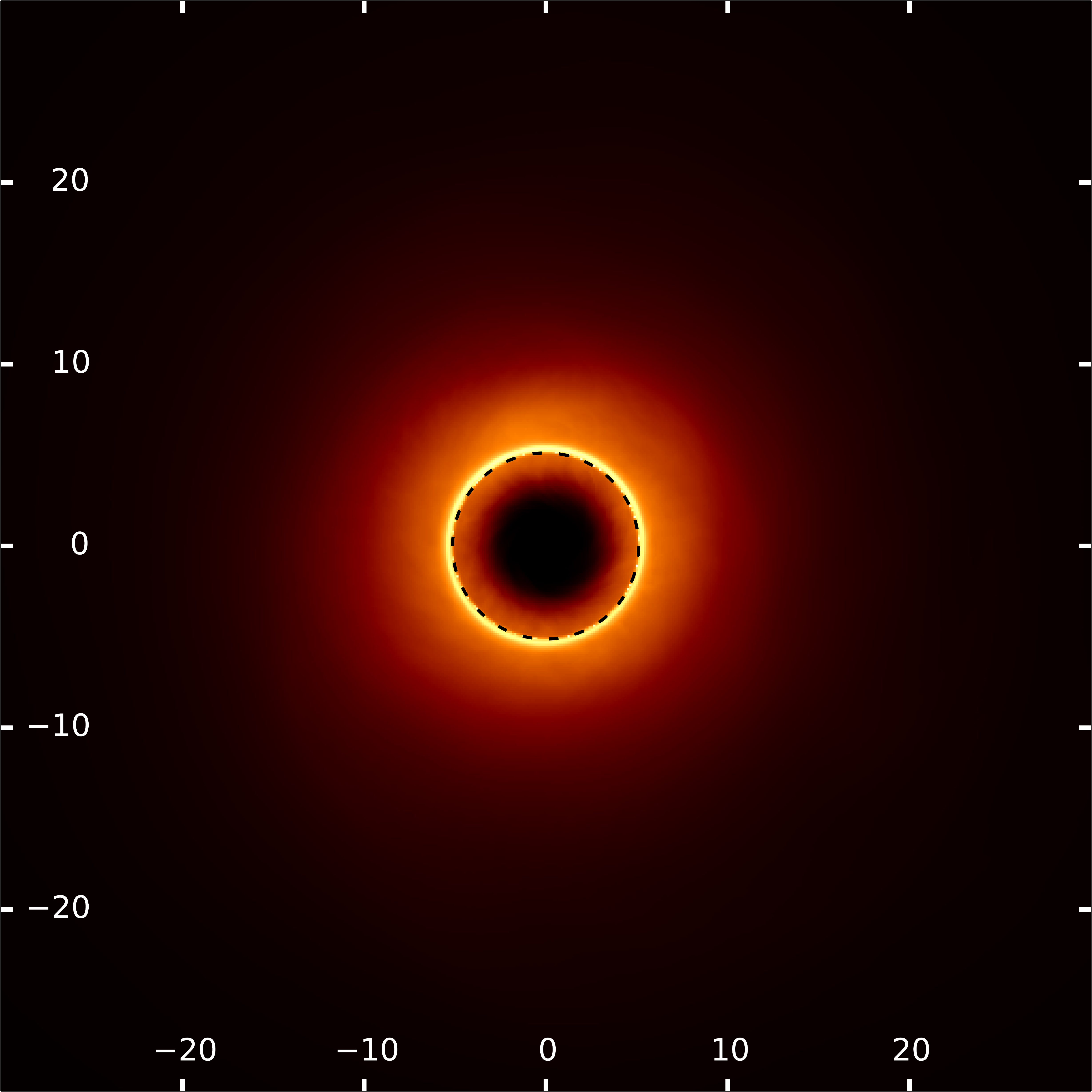}
	\caption{$a=-0.5$, $i=1^\circ$.}
\end{subfigure}
\begin{subfigure}[b]{0.197\textwidth}
	\includegraphics[width=\textwidth]{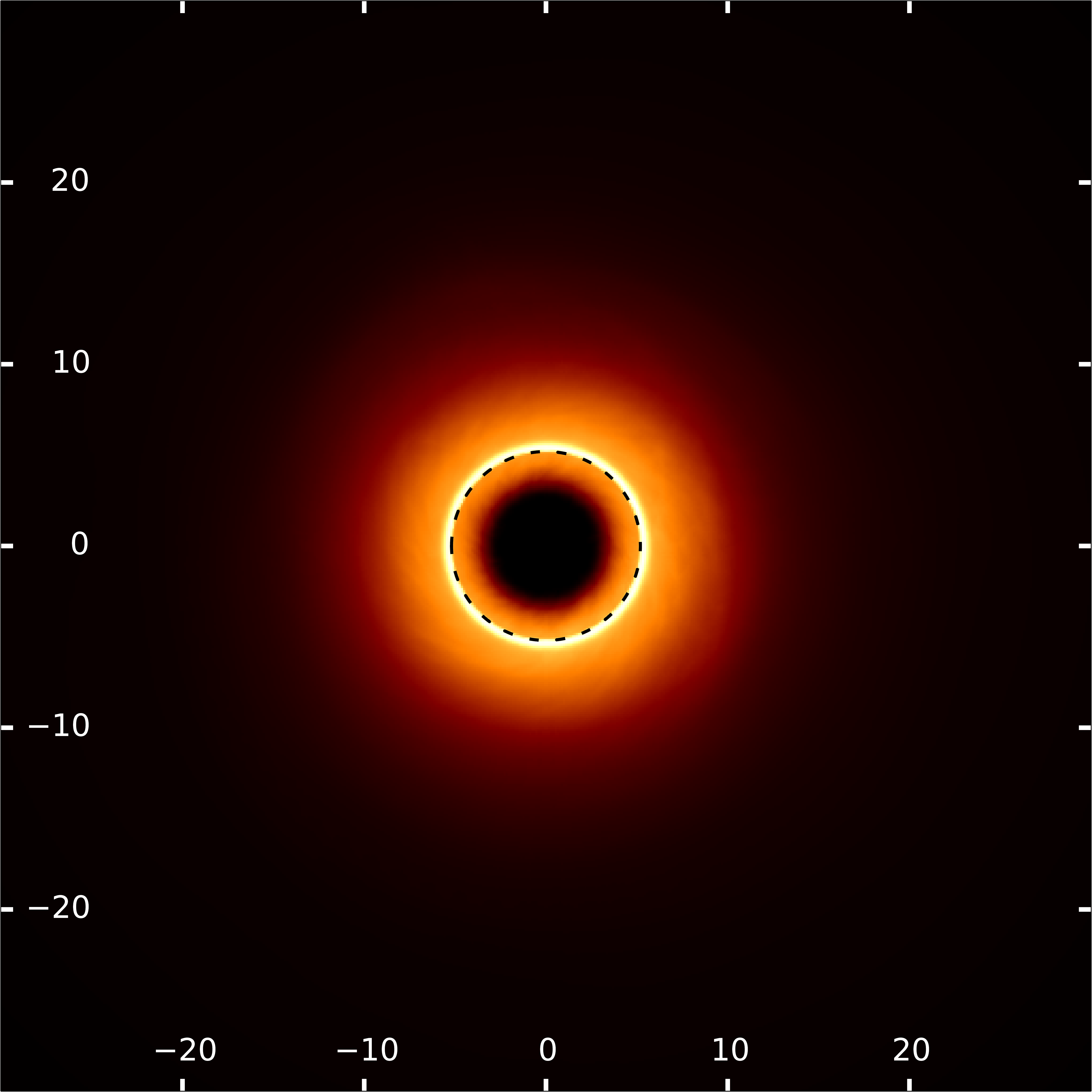}
	\caption{$a=0$, $i=1^\circ$.}
\end{subfigure}
\begin{subfigure}[b]{0.197\textwidth}
	\includegraphics[width=\textwidth]{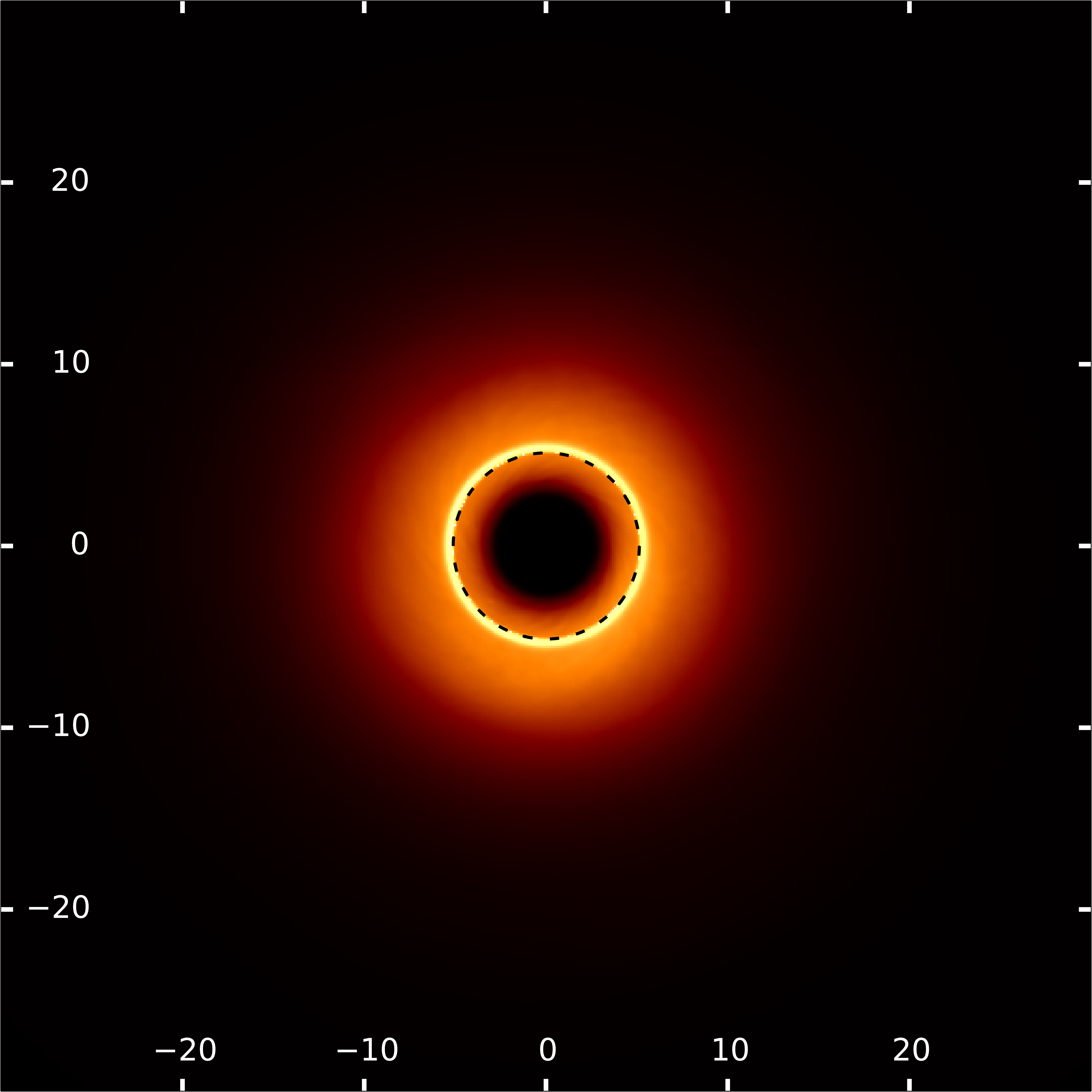}
	\caption{$a=0.5$, $i=1^\circ$.}
\end{subfigure}
\begin{subfigure}[b]{0.197\textwidth}
	\includegraphics[width=\textwidth]{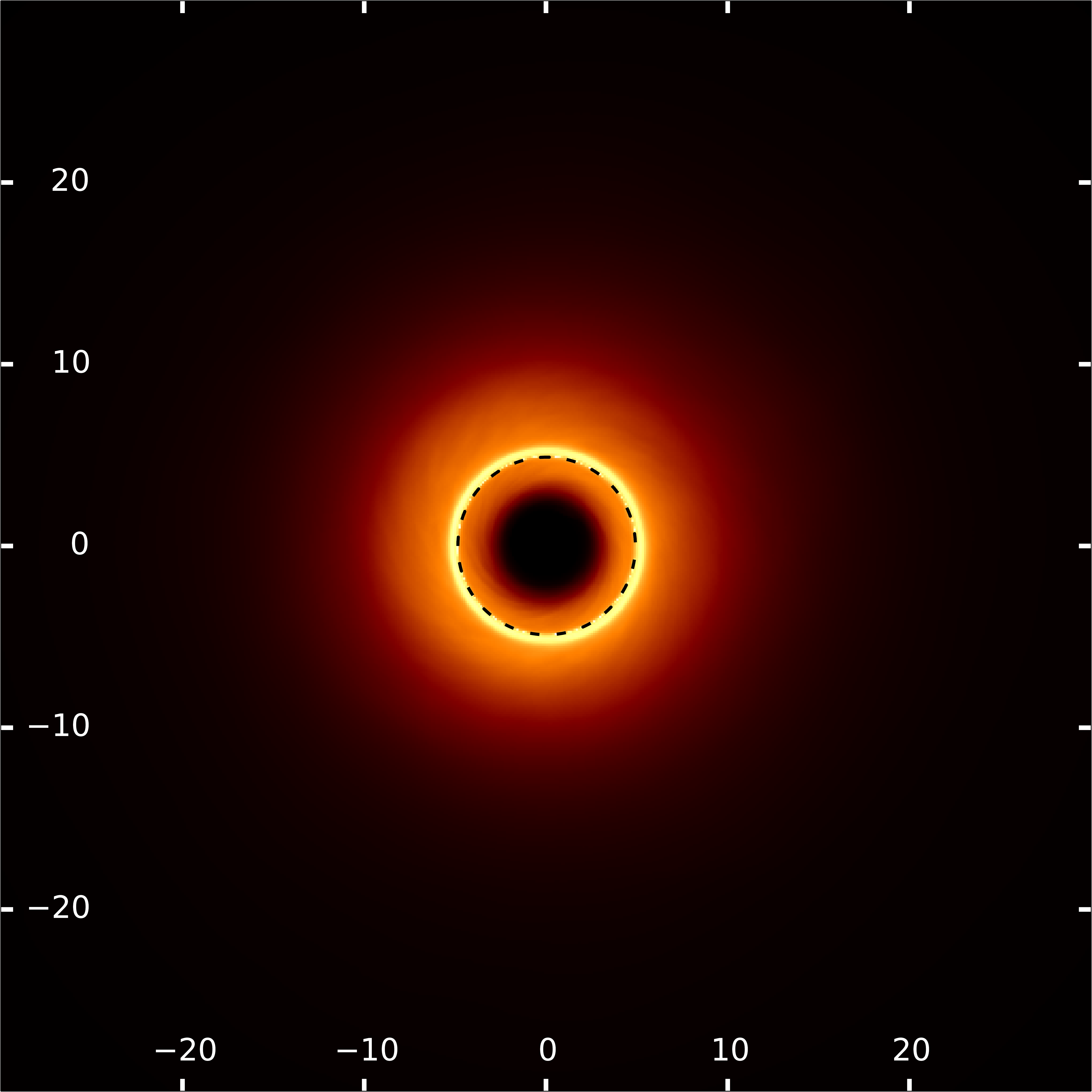}
	\caption{$a=0.9375$, $i=1^\circ$.}
\end{subfigure}
\begin{subfigure}[b]{0.197\textwidth}
	\includegraphics[width=\textwidth]{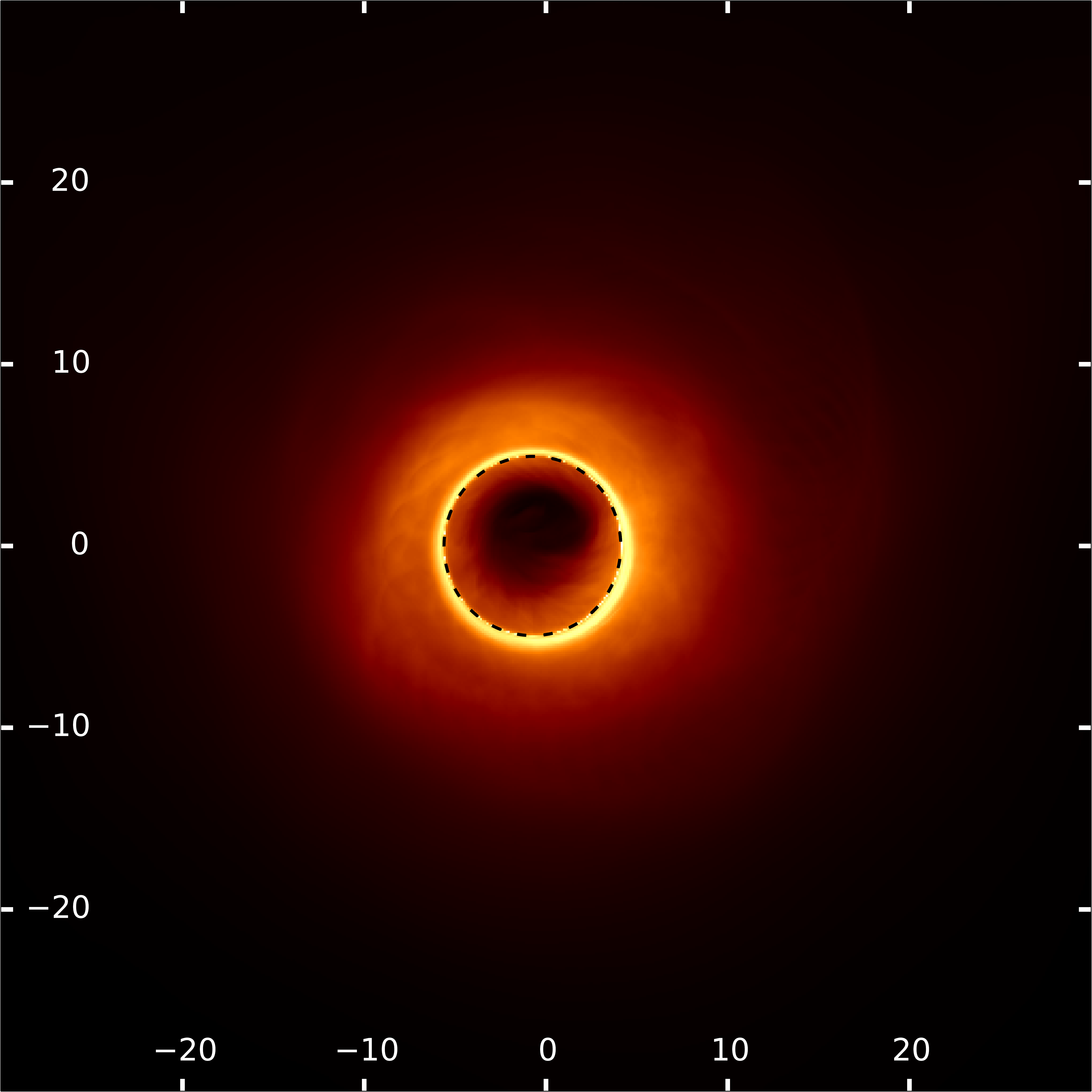}
	\caption{$a=-0.9375$, $i=20^\circ$.}
\end{subfigure}
\begin{subfigure}[b]{0.197\textwidth}
	\includegraphics[width=\textwidth]{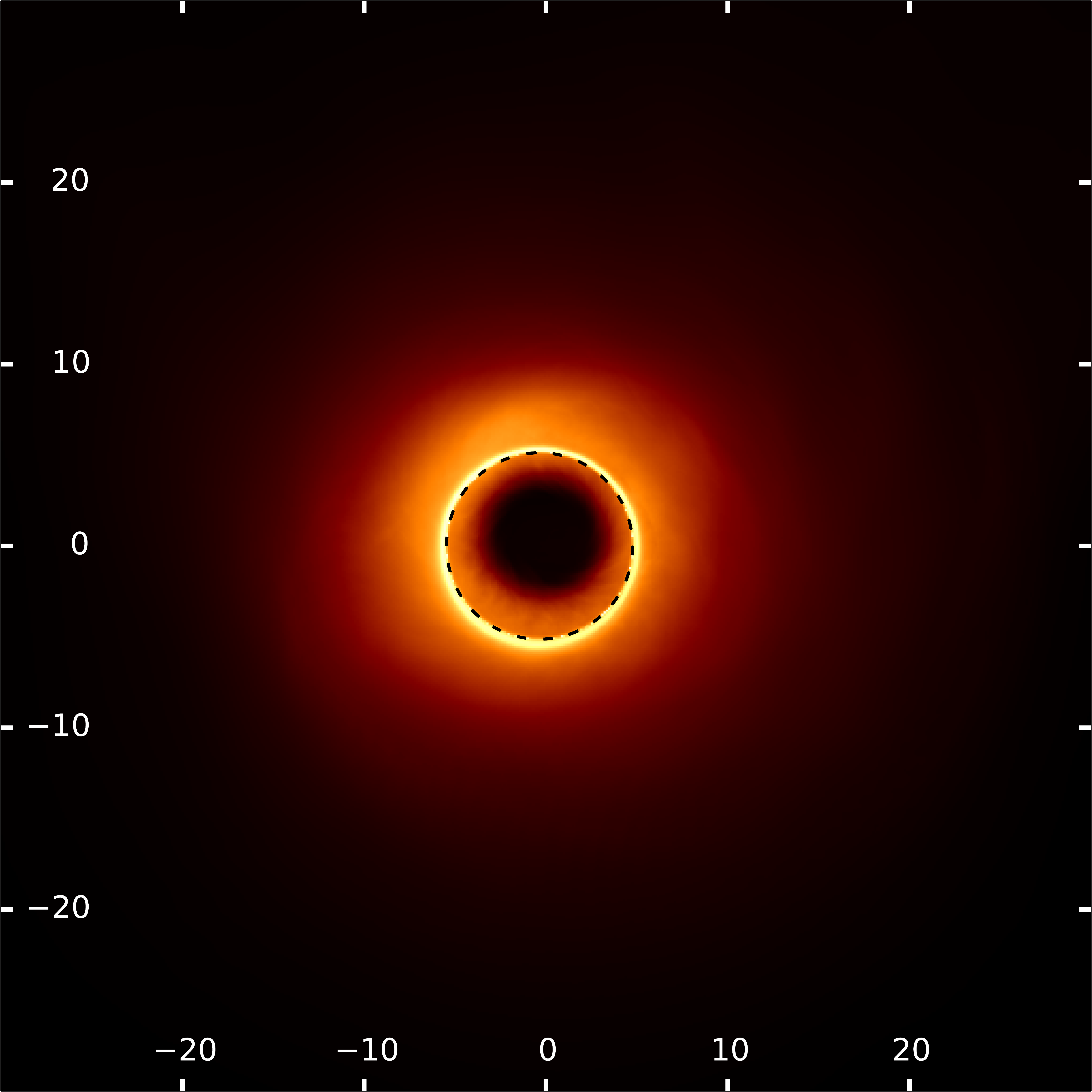}
	\caption{$a=-0.5$, $i=20^\circ$.}
\end{subfigure}
\begin{subfigure}[b]{0.197\textwidth}
	\includegraphics[width=\textwidth]{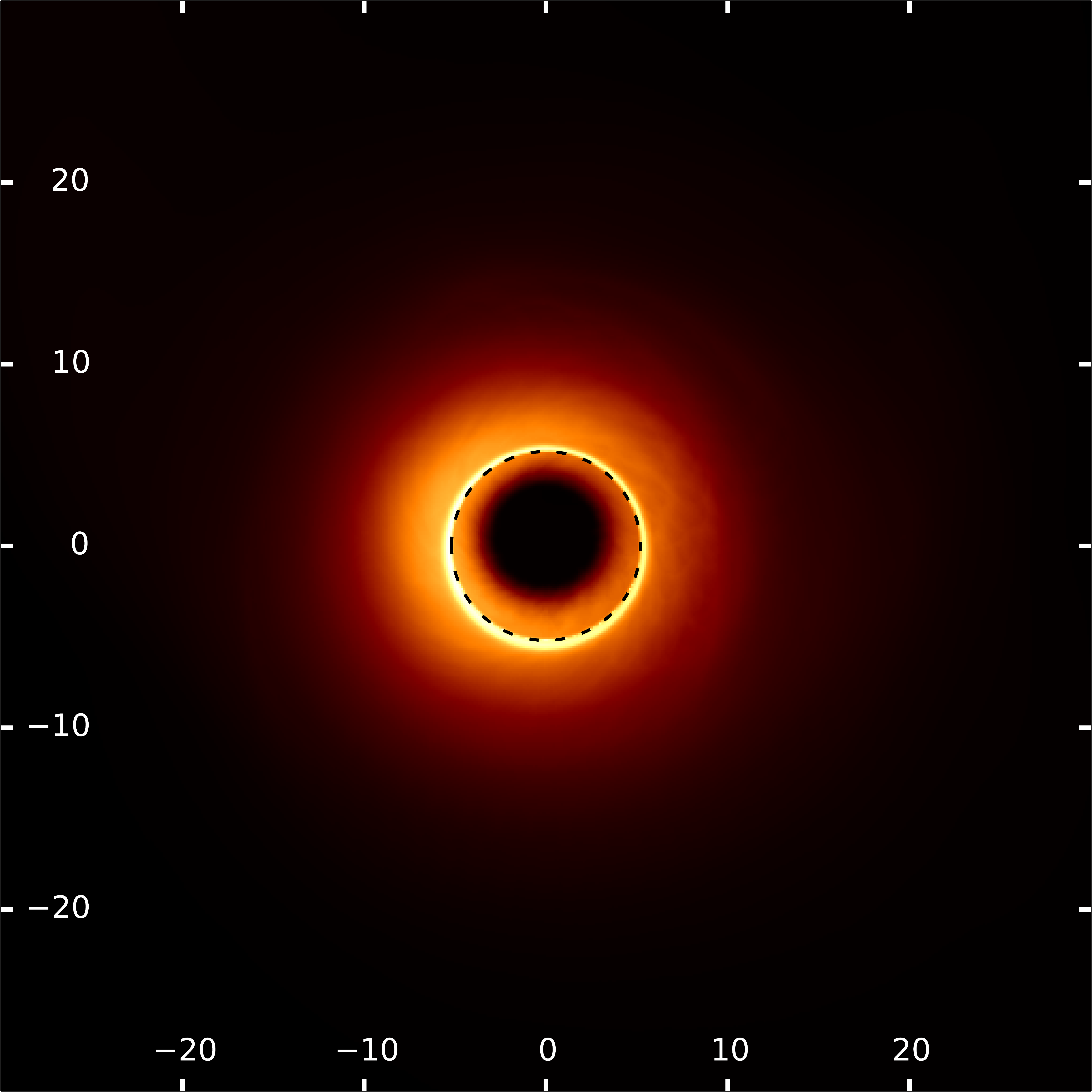}
	\caption{$a=0$, $i=20^\circ$.}
\end{subfigure}
\begin{subfigure}[b]{0.197\textwidth}
	\includegraphics[width=\textwidth]{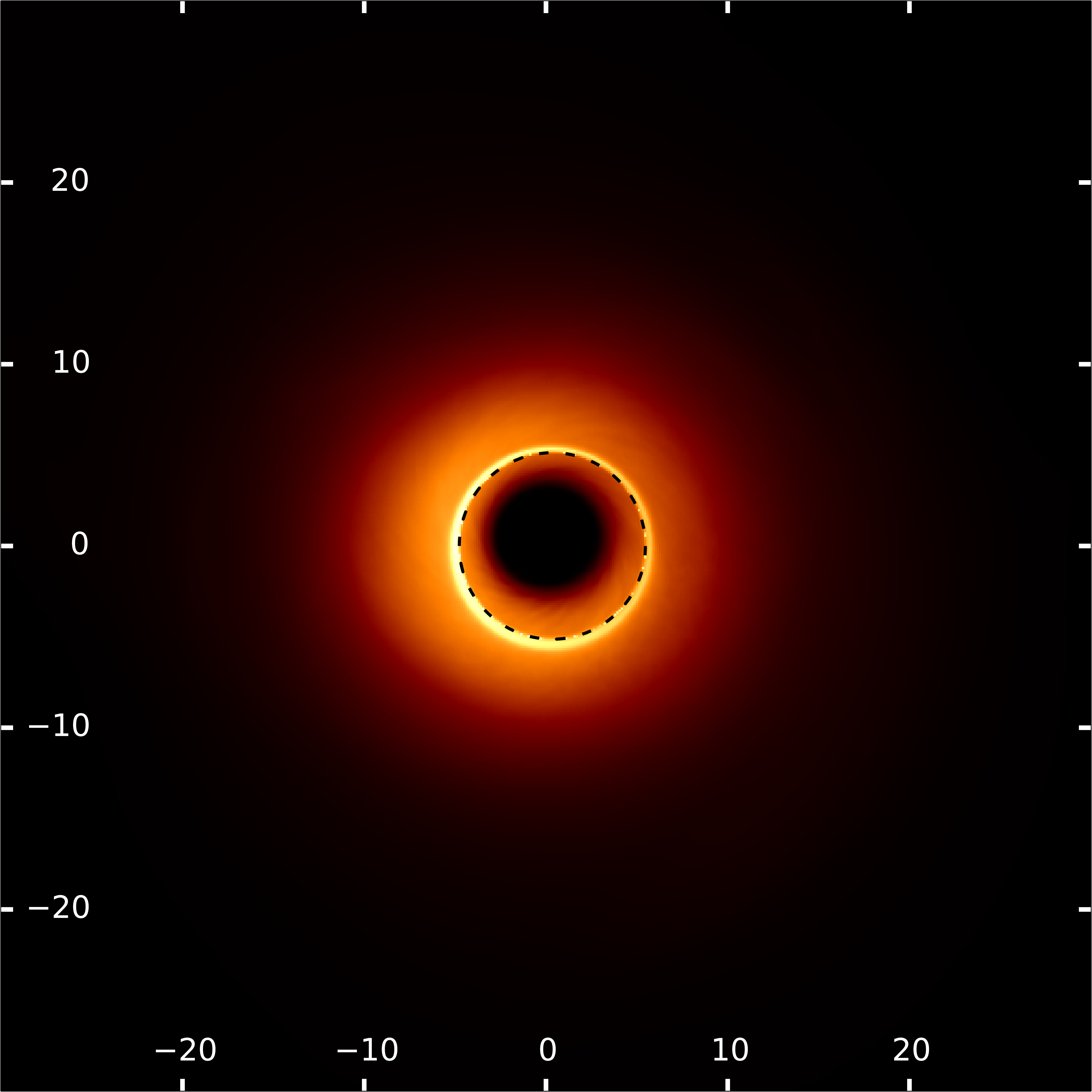}
	\caption{$a=0.5$, $i=20^\circ$.}
\end{subfigure}
\begin{subfigure}[b]{0.197\textwidth}
	\includegraphics[width=\textwidth]{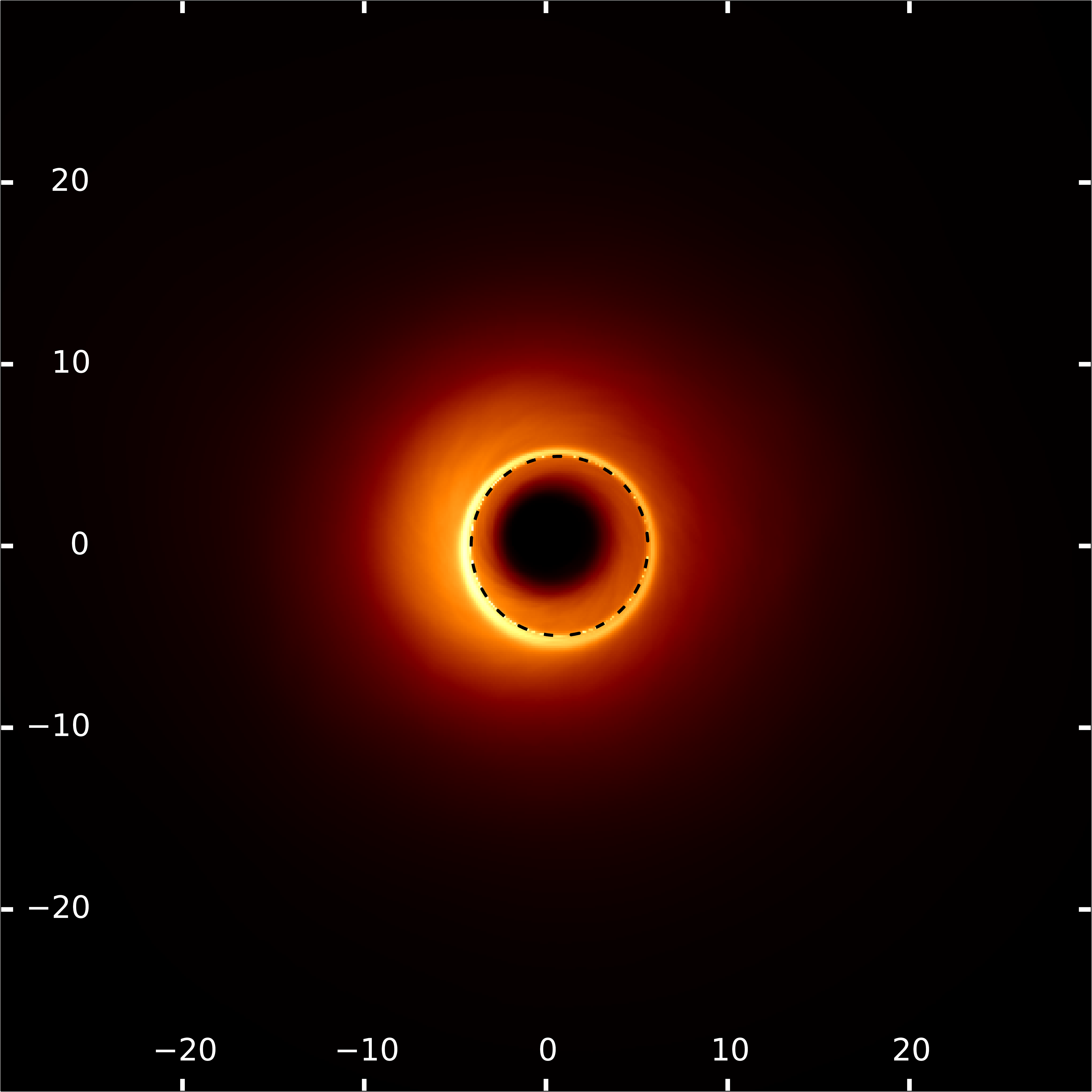}
	\caption{$a=0.9375$, $i=20^\circ$.}
\end{subfigure}
\begin{subfigure}[b]{0.197\textwidth}
	\includegraphics[width=\textwidth]{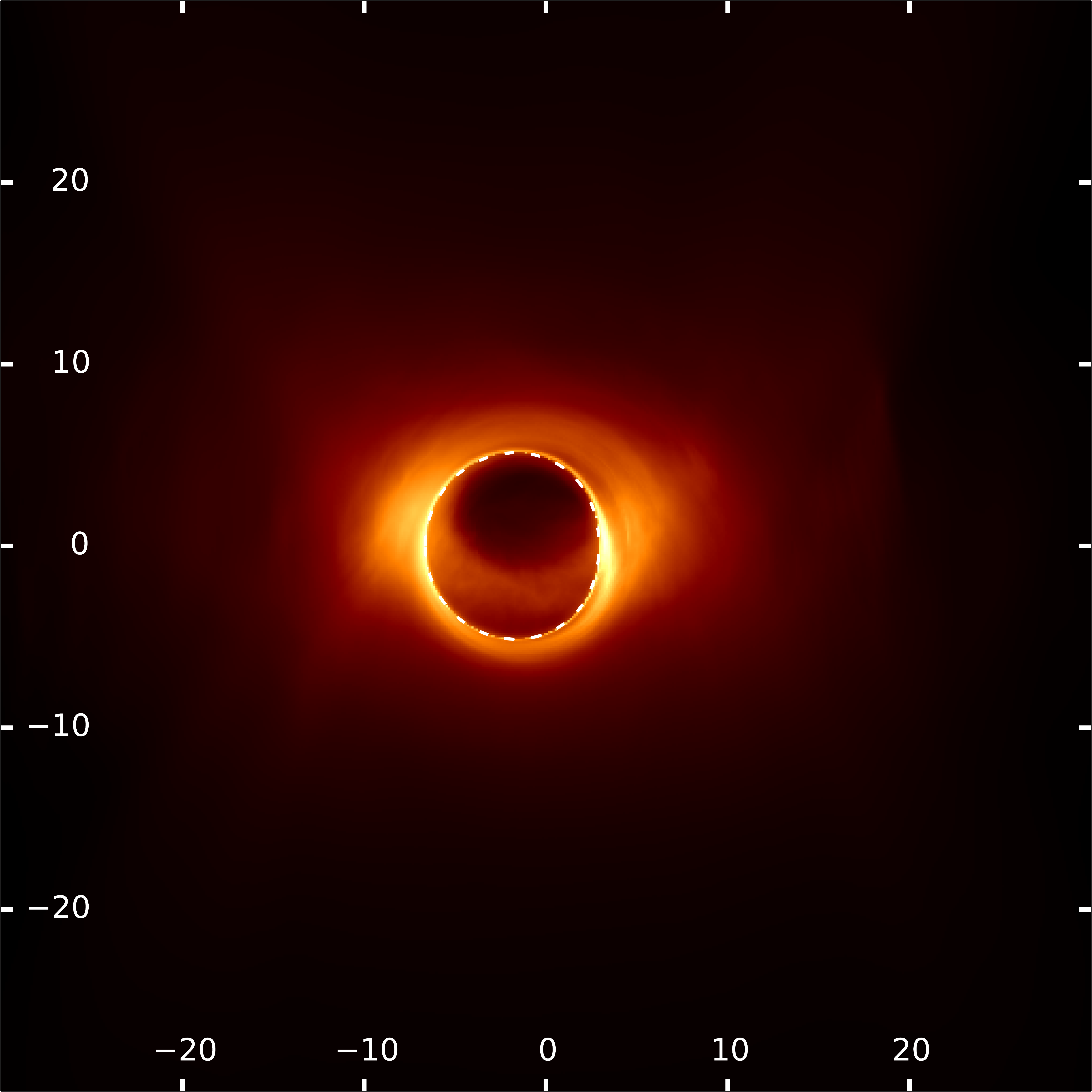}
	\caption{$a=-0.9375$, $i=60^\circ$.}
\end{subfigure}
\begin{subfigure}[b]{0.197\textwidth}
	\includegraphics[width=\textwidth]{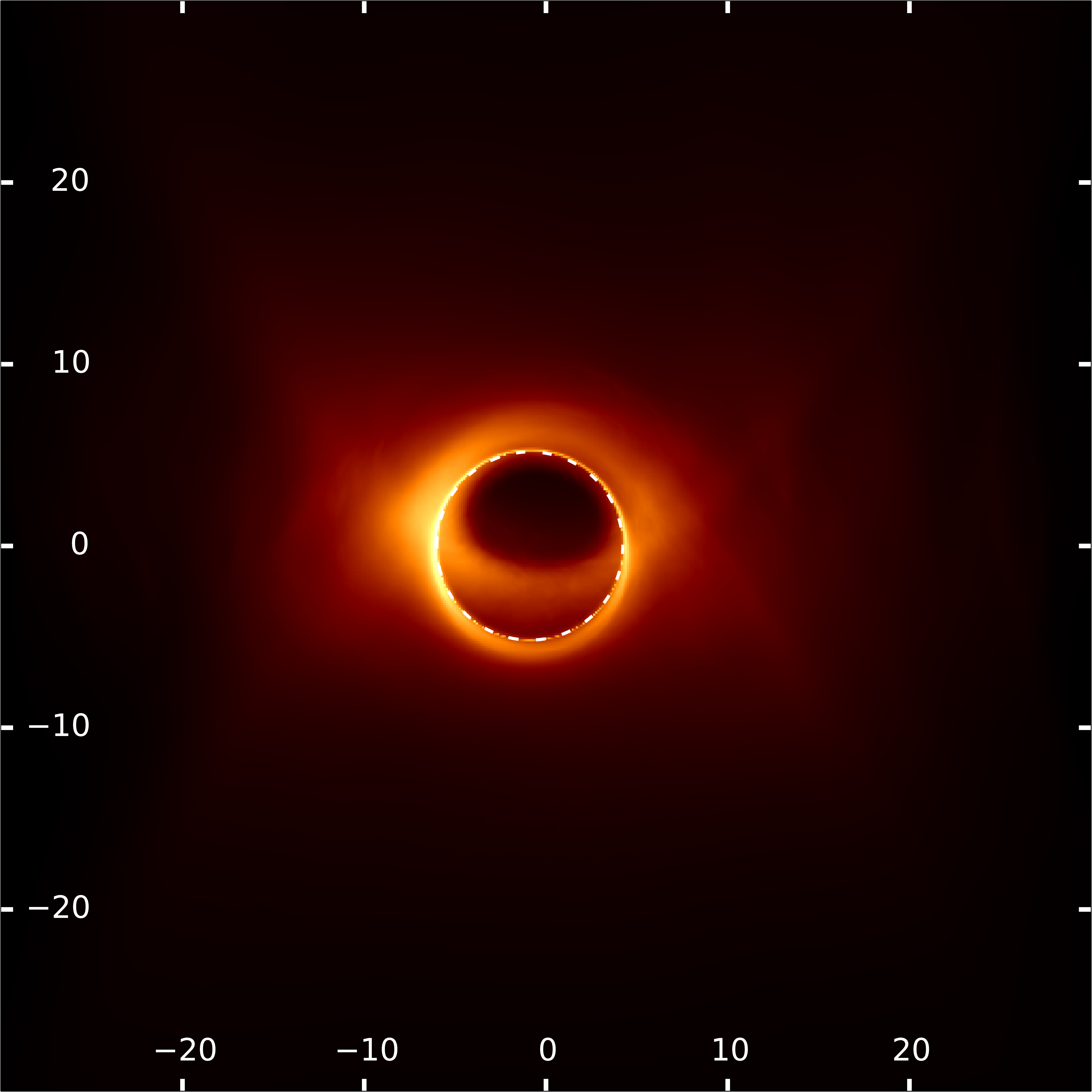}
	\caption{$a=-0.5$, $i=60^\circ$.}
\end{subfigure}
\begin{subfigure}[b]{0.197\textwidth}
	\includegraphics[width=\textwidth]{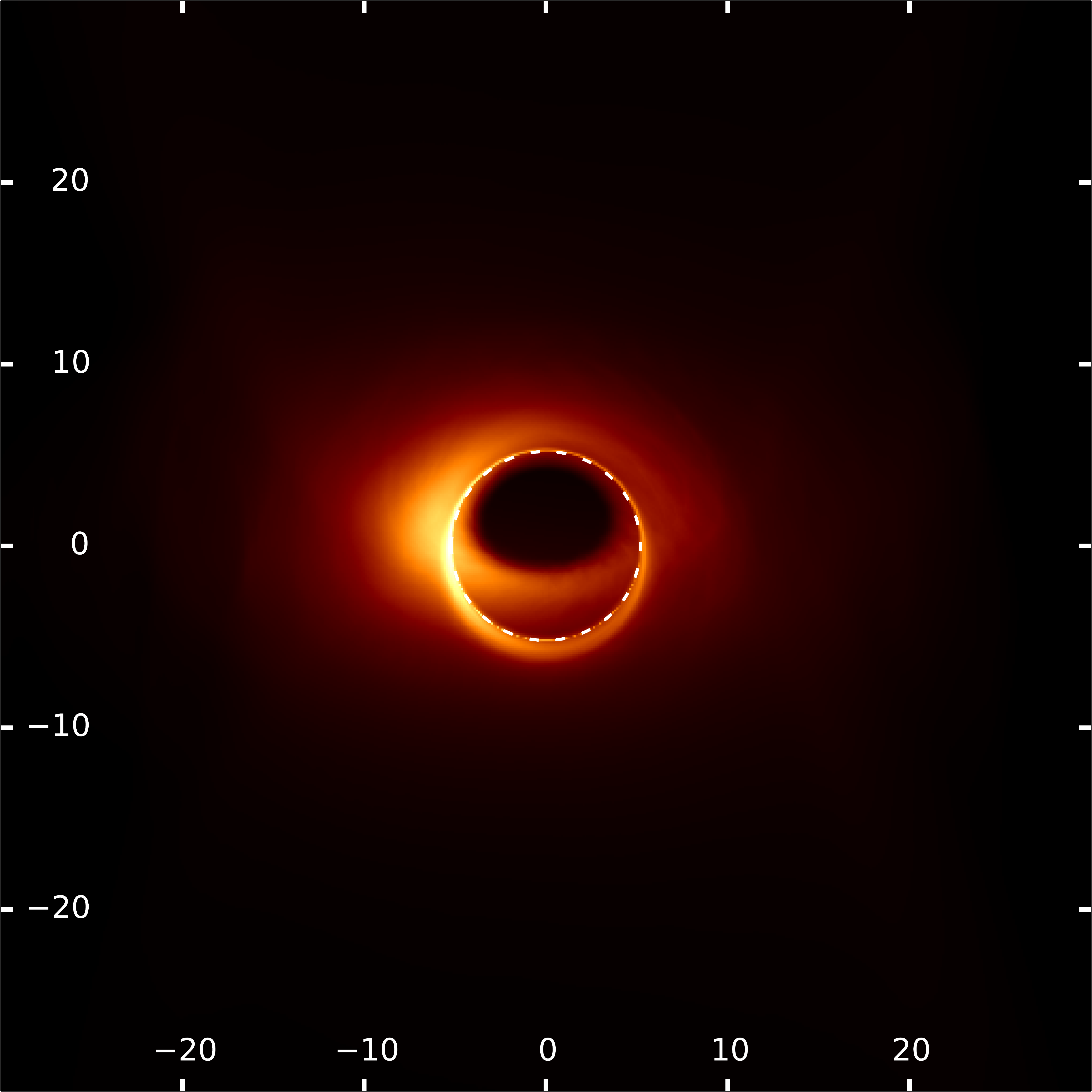}
	\caption{$a=0$, $i=60^\circ$.}
\end{subfigure}
\begin{subfigure}[b]{0.197\textwidth}
	\includegraphics[width=\textwidth]{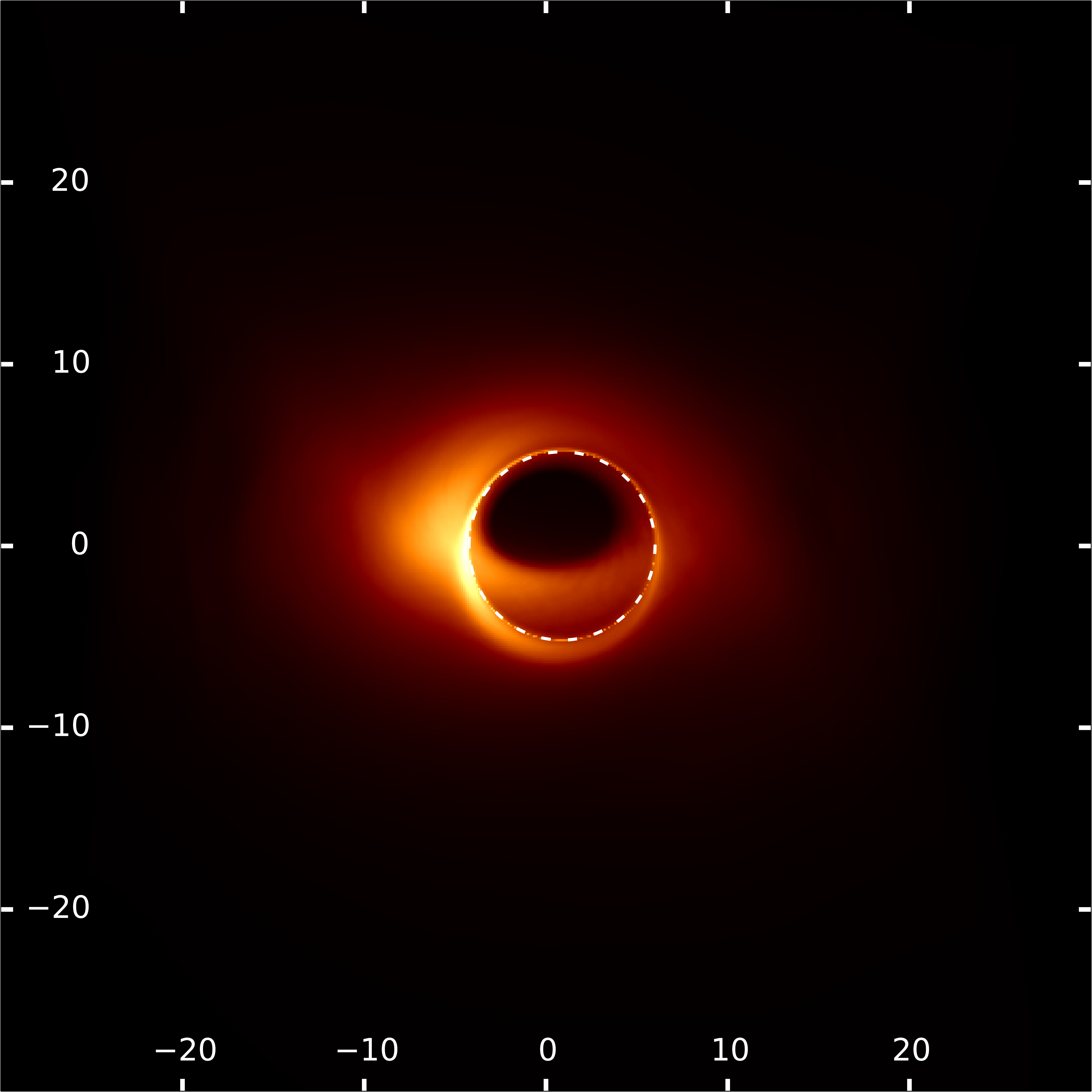}
	\caption{$a=0.5$, $i=60^\circ$.}
\end{subfigure}
\begin{subfigure}[b]{0.197\textwidth}
	\includegraphics[width=\textwidth]{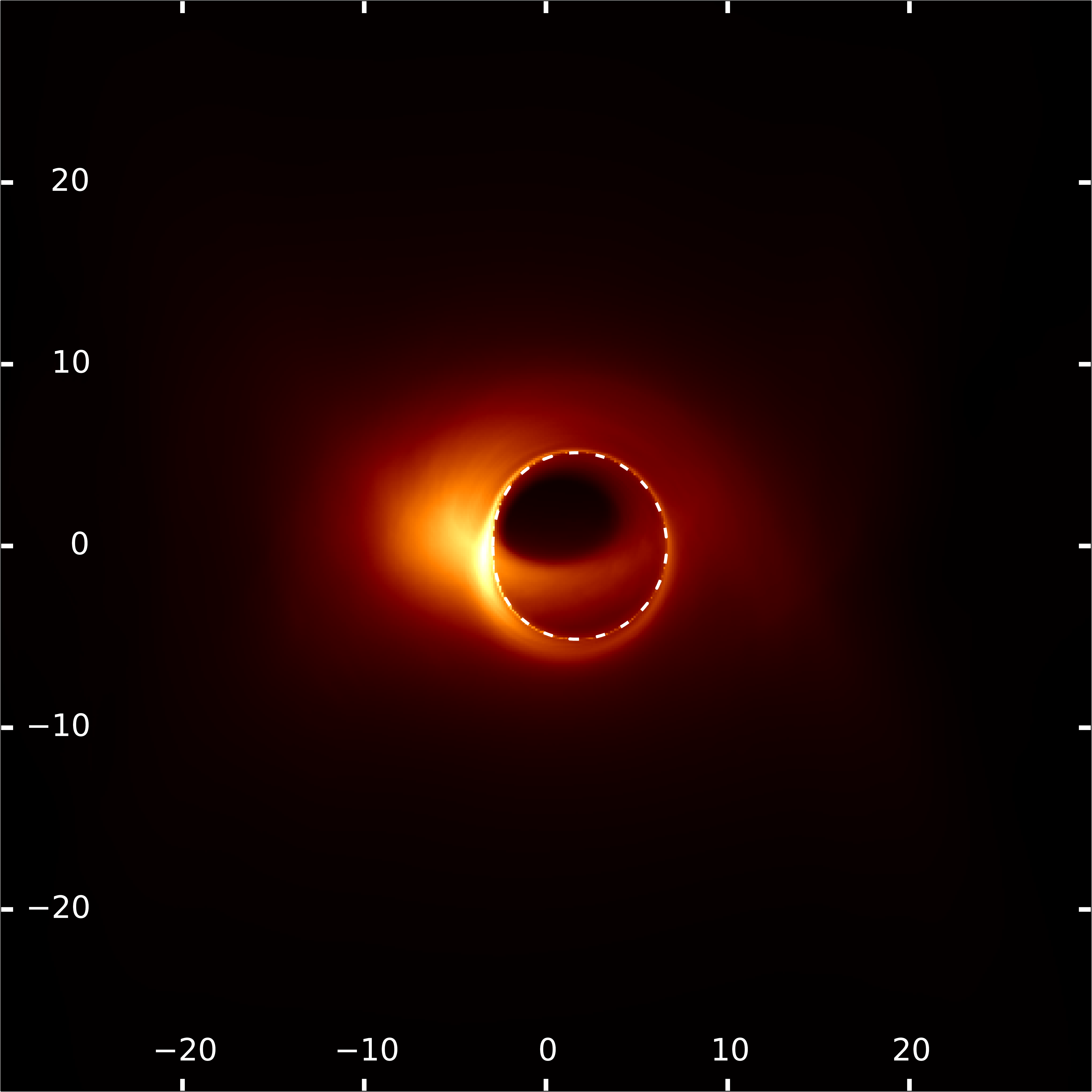}
	\caption{$a=0.9375$, $i=60^\circ$.}
\end{subfigure}
\begin{subfigure}[b]{0.197\textwidth}
	\includegraphics[width=\textwidth]{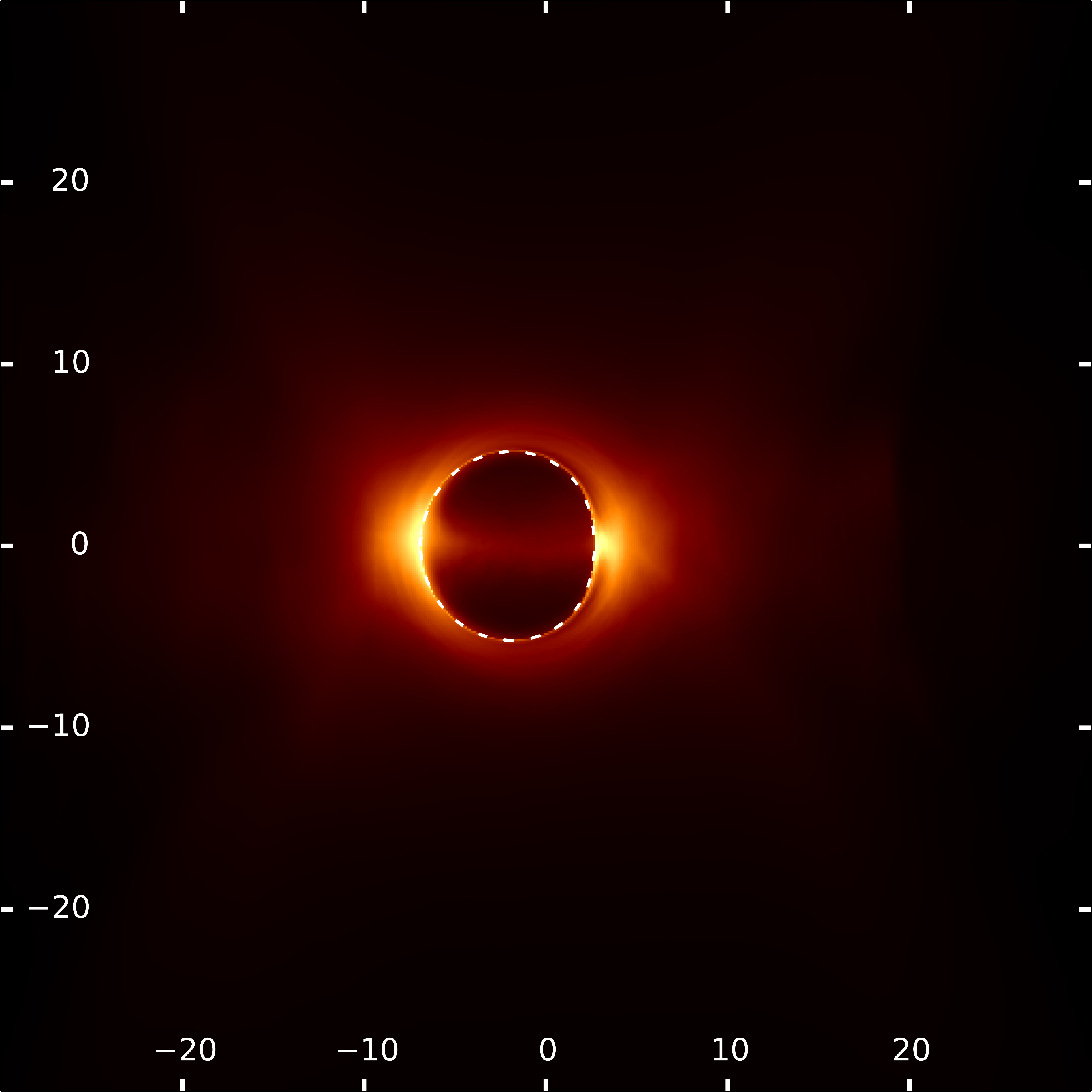}
	\caption{$a=-0.9375$, $i=90^\circ$.}
\end{subfigure}
\begin{subfigure}[b]{0.197\textwidth}
	\includegraphics[width=\textwidth]{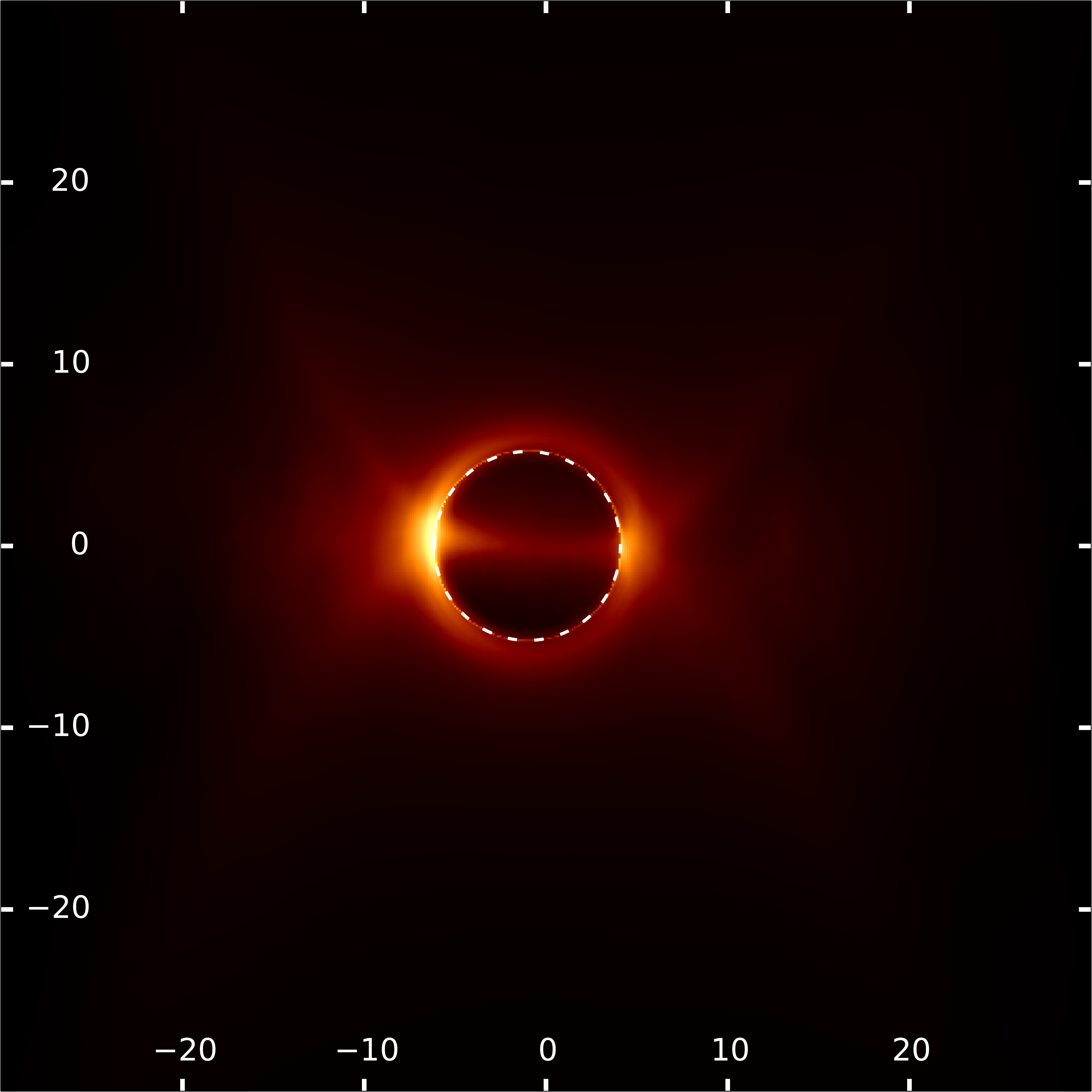}
	\caption{$a=-0.5$, $i=90^\circ$.}
\end{subfigure}
\begin{subfigure}[b]{0.197\textwidth}
	\includegraphics[width=\textwidth]{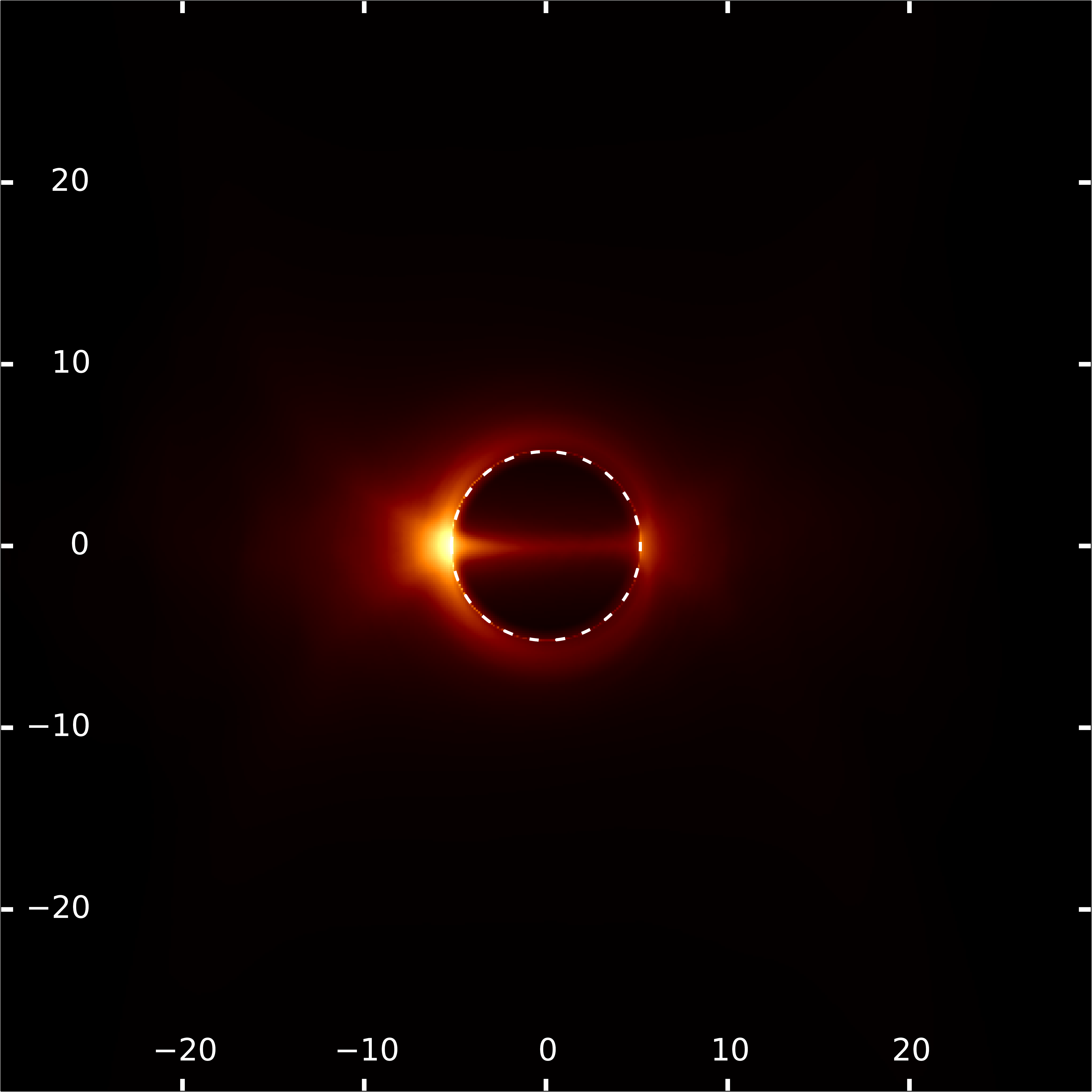}
	\caption{$a=0$, $i=90^\circ$.}
\end{subfigure}
\begin{subfigure}[b]{0.197\textwidth}
	\includegraphics[width=\textwidth]{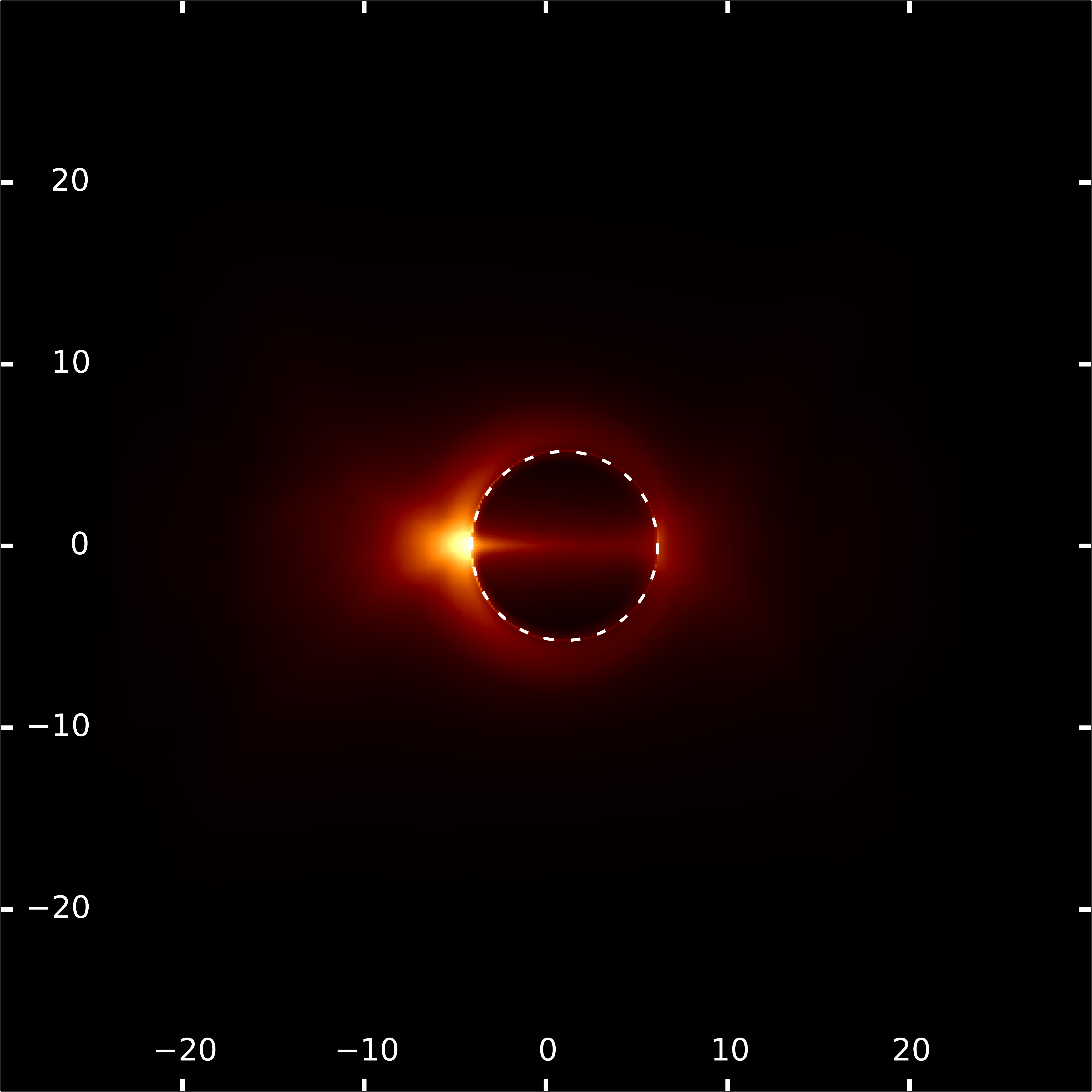}
	\caption{$a=0.5$, $i=90^\circ$.}
\end{subfigure}
\begin{subfigure}[b]{0.197\textwidth}
	\includegraphics[width=\textwidth]{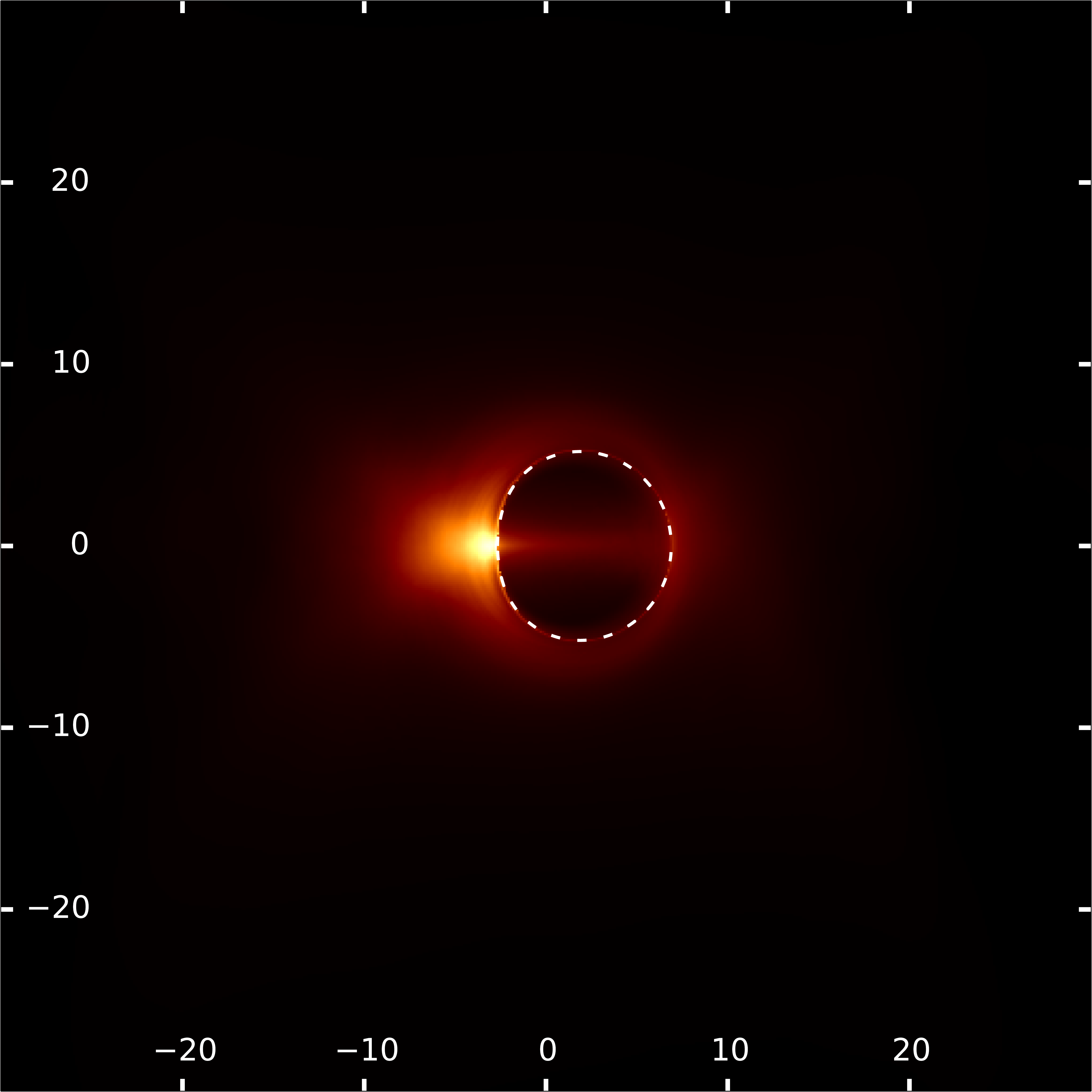}
	\caption{$a=0.9375$, $i=90^\circ$.}
\end{subfigure}
\caption{Time-averaged, normalised intensity maps of our MAD, disc-dominated GRMHD models of Sgr A*, imaged at 230 GHz, at five different spins and four observer inclination angles, with an integrated flux density of 1.25 Jy. In each case, the photon ring, which marks the BHS, is indicated by a dashed line. The values for the impact parameters along the x- and y-axes are expressed in terms of $R_{\rm g}$. The image maps were plotted using a square-root intensity scale.}
\label{fig:mad_disk_125_matrix}
\end{figure*}

Figures \ref{fig:sane_jet_25_matrix}, \ref{fig:sane_jet_125_matrix}, and \ref{fig:sane_jet_0625_matrix} show the SANE jet model calibrated to 2.5 Jy, 1.25 Jy, and 0.625 Jy, respectively. 
Figures \ref{fig:sane_disk_25_matrix} and \ref{fig:sane_disk_125_matrix} show the SANE disc models calibrated to 2.5 Jy and 1.25 Jy, respectively.
Figures \ref{fig:mad_jet_25_matrix} and \ref{fig:mad_jet_125_matrix} show the MAD jet model calibrated to 2.5 Jy and 1.25 Jy. 
Figures \ref{fig:mad_disk_25_matrix} and \ref{fig:mad_disk_125_matrix} show the MAD disc model calibrated to 2.5 Jy and 1.25 Jy.

\section{Source sizes for all models}
\label{sec:source_size_appendix}

We compute the major and minor axes, $\lambda_{\rm max}$ and  $\lambda_{\rm min}$, for the time-averaged images of all models. All major axes for the SANE models are recapitulated in Table \ref{tab:lambdamax_SANE}, while those of the MAD models are listed in Table \ref{tab:lambdamax_MAD}. Similarly, the minor axes for all SANE (MAD) models are listed in Table \ref{tab:lambdamin_SANE} (\ref{tab:lambdamin_MAD}).
\begin{table}
\bgroup
\def\arraystretch{1.25}
\centering
\textbf{ SANE jet, 2.5 Jy} \\
\begin{tabularx}{0.48\textwidth}{@{}p{0.068\textwidth} p{0.058\textwidth} p{0.058\textwidth} p{0.058\textwidth} p{0.058\textwidth} p{0.058\textwidth}@{}}
      \thickhline
      {BH spin} & \textbf{$-15/16$} & \textbf{$-1/2$} & \textbf{$0$} & \textbf{$1/2$} & \textbf{$15/16$} \\
      \hline
$i=1 ^\circ$ & $	6.29	$&$	6.45	$&$	6.73	$&$	6.35	$&$	6.54	$\\
$i=20 ^\circ$ & $	7.34	$&$	7.17	$&$	6.70	$&$	6.76	$&$	7.00	$\\
$i=60 ^\circ$ & $	8.13	$&$	8.54	$&$	7.53	$&$	7.34	$&$	7.07	$\\
$i=90 ^\circ$ & $	8.26	$&$	8.99	$&$	7.88	$&$	7.22	$&$	8.14	$\\
      \thickhline
    \end{tabularx} \\
    \ \\
    \ \\
\centering
\textbf{ SANE jet, 1.25 Jy} \\
\begin{tabularx}{0.48\textwidth}{@{}p{0.068\textwidth} p{0.058\textwidth} p{0.058\textwidth} p{0.058\textwidth} p{0.058\textwidth} p{0.058\textwidth}@{}}
      \thickhline
      {BH spin} & \textbf{$-15/16$} & \textbf{$-1/2$} & \textbf{$0$} & \textbf{$1/2$} & \textbf{$15/16$} \\
      \hline
$i=1 ^\circ$ & $	5.75	$&$	5.68	$&$	4.89	$&$	4.96	$&$	4.57	$\\
$i=20 ^\circ$ & $	6.44	$&$	6.04	$&$	4.88	$&$	4.98	$&$	4.53	$\\
$i=60 ^\circ$ & $	6.85	$&$	7.12	$&$	5.08	$&$	5.31	$&$	4.82	$\\
$i=90 ^\circ$ & $	6.90	$&$	7.56	$&$	5.52	$&$	5.22	$&$	4.82	$\\
      \thickhline
    \end{tabularx} \\
    \ \\
    \ \\
\centering
\textbf{ SANE jet, 0.625 Jy} \\
\begin{tabularx}{0.48\textwidth}{@{}p{0.068\textwidth} p{0.058\textwidth} p{0.058\textwidth} p{0.058\textwidth} p{0.058\textwidth} p{0.058\textwidth}@{}}
      \thickhline
      {BH spin} & \textbf{$-15/16$} & \textbf{$-1/2$} & \textbf{$0$} & \textbf{$1/2$} & \textbf{$15/16$} \\
      \hline
$i=1 ^\circ$ & $	5.44	$&$	5.26	$&$	4.64	$&$	4.52	$&$	3.64	$\\
$i=20 ^\circ$ & $	5.89	$&$	5.45	$&$	4.54	$&$	4.28	$&$	3.59	$\\
$i=60 ^\circ$ & $	5.94	$&$	6.12	$&$	4.12	$&$	4.06	$&$	3.36	$\\
$i=90 ^\circ$ & $	5.75	$&$	6.43	$&$	4.39	$&$	4.07	$&$	3.36	$\\
      \thickhline
    \end{tabularx}\\
    \ \\
    \ \\
    \centering
\textbf{ SANE disc, 2.5 Jy} \\
\begin{tabularx}{0.48\textwidth}{@{}p{0.068\textwidth} p{0.058\textwidth} p{0.058\textwidth} p{0.058\textwidth} p{0.058\textwidth} p{0.058\textwidth}@{}}
      \thickhline
      {BH spin} & \textbf{$-15/16$} & \textbf{$-1/2$} & \textbf{$0$} & \textbf{$1/2$} & \textbf{$15/16$} \\
      \hline
$i=1 ^\circ$ & $	9.82	$&$	9.27	$&$	8.12	$&$	6.86	$&$	5.42	$\\
$i=20 ^\circ$ & $	9.33	$&$	8.84	$&$	7.75	$&$	6.52	$&$	5.23	$\\
$i=60 ^\circ$ & $	7.64	$&$	7.40	$&$	6.36	$&$	5.25	$&$	4.45	$\\
$i=90 ^\circ$ & $	7.58	$&$	7.31	$&$	6.19	$&$	5.02	$&$	4.14	$\\
      \thickhline
    \end{tabularx} \\
    \ \\
    \ \\
\centering
\textbf{ SANE disc, 1.25 Jy} \\
\begin{tabularx}{0.48\textwidth}{@{}p{0.068\textwidth} p{0.058\textwidth} p{0.058\textwidth} p{0.058\textwidth} p{0.058\textwidth} p{0.058\textwidth}@{}}
      \thickhline
      {BH spin} & \textbf{$-15/16$} & \textbf{$-1/2$} & \textbf{$0$} & \textbf{$1/2$} & \textbf{$15/16$} \\
      \hline
$i=1 ^\circ$ & $	9.58	$&$	9.08	$&$	7.89	$&$	6.68	$&$	5.23	$\\
$i=20 ^\circ$ & $	9.05	$&$	8.59	$&$	7.47	$&$	6.31	$&$	5.02	$\\
$i=60 ^\circ$ & $	7.18	$&$	6.94	$&$	5.90	$&$	4.91	$&$	4.17	$\\
$i=90 ^\circ$ & $	6.97	$&$	6.69	$&$	5.57	$&$	4.56	$&$	3.80	$\\
      \thickhline
    \end{tabularx}
\egroup
\caption{Tables of $\lambda_{\rm max}$, the major axis length, in units of $R_{\rm g}$, for the SANE jet models.}
\label{tab:lambdamax_SANE}
\end{table}

\begin{table}
\bgroup
\def\arraystretch{1.25}
\centering
\textbf{ MAD jet, 2.5 Jy} \\
\begin{tabularx}{0.48\textwidth}{@{}p{0.068\textwidth} p{0.058\textwidth} p{0.058\textwidth} p{0.058\textwidth} p{0.058\textwidth} p{0.058\textwidth}@{}}
      \thickhline
      {BH spin} & \textbf{$-15/16$} & \textbf{$-1/2$} & \textbf{$0$} & \textbf{$1/2$} & \textbf{$15/16$} \\
      \hline
$i=1 ^\circ$ & $	7.71	$&$	7.58	$&$	7.43	$&$	6.38	$&$	6.65	$\\
$i=20 ^\circ$ & $	8.22	$&$	8.15	$&$	8.03	$&$	6.72	$&$	6.87	$\\
$i=60 ^\circ$ & $	9.45	$&$	9.30	$&$	8.50	$&$	7.35	$&$	7.35	$\\
$i=90 ^\circ$ & $	9.32	$&$	9.22	$&$	8.23	$&$	7.20	$&$	7.00	$\\
      \thickhline
    \end{tabularx} \\
    \ \\
    \ \\
\centering
\textbf{ MAD jet, 1.25 Jy} \\
\begin{tabularx}{0.48\textwidth}{@{}p{0.068\textwidth} p{0.058\textwidth} p{0.058\textwidth} p{0.058\textwidth} p{0.058\textwidth} p{0.058\textwidth}@{}}
      \thickhline
      {BH spin} & \textbf{$-15/16$} & \textbf{$-1/2$} & \textbf{$0$} & \textbf{$1/2$} & \textbf{$15/16$} \\
      \hline
$i=1 ^\circ$ & $	7.32	$&$	7.18	$&$	6.96	$&$	6.05	$&$	6.29	$\\
$i=20 ^\circ$ & $	7.81	$&$	7.76	$&$	7.59	$&$	6.37	$&$	6.48	$\\
$i=60 ^\circ$ & $	8.98	$&$	8.95	$&$	8.09	$&$	6.93	$&$	6.87	$\\
$i=90 ^\circ$ & $	8.84	$&$	8.78	$&$	7.68	$&$	6.50	$&$	6.36	$\\
      \thickhline
    \end{tabularx}
        \ \\
    \ \\
    \centering
\textbf{ MAD disc, 2.5 Jy} \\
\begin{tabularx}{0.48\textwidth}{@{}p{0.068\textwidth} p{0.058\textwidth} p{0.058\textwidth} p{0.058\textwidth} p{0.058\textwidth} p{0.058\textwidth}@{}}
      \thickhline
      {BH spin} & \textbf{$-15/16$} & \textbf{$-1/2$} & \textbf{$0$} & \textbf{$1/2$} & \textbf{$15/16$} \\
      \hline
$i=1 ^\circ$ & $	7.67	$&$	7.36	$&$	6.38	$&$	6.18	$&$	6.30	$\\
$i=20 ^\circ$ & $	7.95	$&$	7.57	$&$	6.56	$&$	6.29	$&$	6.38	$\\
$i=60 ^\circ$ & $	8.62	$&$	8.36	$&$	6.83	$&$	6.53	$&$	6.63	$\\
$i=90 ^\circ$ & $	8.39	$&$	8.14	$&$	6.62	$&$	6.18	$&$	6.17	$\\
      \thickhline
    \end{tabularx} \\
    \ \\
    \ \\
\centering
\textbf{ MAD disc, 1.25 Jy} \\
\begin{tabularx}{0.48\textwidth}{@{}p{0.068\textwidth} p{0.058\textwidth} p{0.058\textwidth} p{0.058\textwidth} p{0.058\textwidth} p{0.058\textwidth}@{}}
      \thickhline
      {BH spin} & \textbf{$-15/16$} & \textbf{$-1/2$} & \textbf{$0$} & \textbf{$1/2$} & \textbf{$15/16$} \\
      \hline
$i=1 ^\circ$ & $	7.38	$&$	7.04	$&$	6.08	$&$	5.93	$&$	6.02	$\\
$i=20 ^\circ$ & $	7.65	$&$	7.26	$&$	6.25	$&$	6.02	$&$	6.09	$\\
$i=60 ^\circ$ & $	8.25	$&$	8.04	$&$	6.48	$&$	6.21	$&$	6.29	$\\
$i=90 ^\circ$ & $	7.99	$&$	7.76	$&$	6.14	$&$	5.78	$&$	5.77	$\\
      \thickhline
    \end{tabularx}
\egroup
\caption{Tables of $\lambda_{\rm max}$, the major axis length, in units of $R_{\rm g}$, for the MAD models.}
\label{tab:lambdamax_MAD}
\end{table}

\begin{table}
\bgroup
\def\arraystretch{1.25}
\centering
\textbf{ SANE jet, 2.5 Jy} \\
\begin{tabularx}{0.48\textwidth}{@{}p{0.068\textwidth} p{0.058\textwidth} p{0.058\textwidth} p{0.058\textwidth} p{0.058\textwidth} p{0.058\textwidth}@{}}
      \thickhline
      {BH spin} & \textbf{$-15/16$} & \textbf{$-1/2$} & \textbf{$0$} & \textbf{$1/2$} & \textbf{$15/16$} \\
      \hline
$i=1 ^\circ$ & $	6.23	$&$	6.33	$&$	6.72	$&$	6.21	$&$	6.47	$\\
$i=20 ^\circ$ & $	6.04	$&$	6.04	$&$	6.01	$&$	5.92	$&$	6.53	$\\
$i=60 ^\circ$ & $	5.54	$&$	5.41	$&$	4.81	$&$	4.79	$&$	6.80	$\\
$i=90 ^\circ$ & $	5.37	$&$	5.21	$&$	4.38	$&$	4.24	$&$	7.13	$\\
      \thickhline
    \end{tabularx} \\
    \ \\
    \ \\
\centering
\textbf{ SANE jet, 1.25 Jy} \\
\begin{tabularx}{0.48\textwidth}{@{}p{0.068\textwidth} p{0.058\textwidth} p{0.058\textwidth} p{0.058\textwidth} p{0.058\textwidth} p{0.058\textwidth}@{}}
      \thickhline
      {BH spin} & \textbf{$-15/16$} & \textbf{$-1/2$} & \textbf{$0$} & \textbf{$1/2$} & \textbf{$15/16$} \\
      \hline
$i=1 ^\circ$ & $	5.69	$&$	5.55	$&$	4.77	$&$	4.90	$&$	4.53	$\\
$i=20 ^\circ$ & $	5.54	$&$	5.42	$&$	4.70	$&$	4.85	$&$	4.49	$\\
$i=60 ^\circ$ & $	5.20	$&$	4.99	$&$	4.15	$&$	4.07	$&$	4.10	$\\
$i=90 ^\circ$ & $	5.09	$&$	4.80	$&$	3.77	$&$	3.50	$&$	3.97	$\\
      \thickhline
    \end{tabularx} \\
    \ \\
    \ \\
\centering
\textbf{ SANE jet, 0.625 Jy} \\
\begin{tabularx}{0.48\textwidth}{@{}p{0.068\textwidth} p{0.058\textwidth} p{0.058\textwidth} p{0.058\textwidth} p{0.058\textwidth} p{0.058\textwidth}@{}}
      \thickhline
      {BH spin} & \textbf{$-15/16$} & \textbf{$-1/2$} & \textbf{$0$} & \textbf{$1/2$} & \textbf{$15/16$} \\
      \hline
$i=1 ^\circ$ & $	5.34	$&$	5.11	$&$	4.47	$&$	4.38	$&$	3.64	$\\
$i=20 ^\circ$ & $	5.24	$&$	5.02	$&$	4.35	$&$	4.24	$&$	3.56	$\\
$i=60 ^\circ$ & $	4.94	$&$	4.59	$&$	3.43	$&$	3.09	$&$	3.09	$\\
$i=90 ^\circ$ & $	4.85	$&$	4.40	$&$	3.23	$&$	2.74	$&$	2.90	$\\
      \thickhline
    \end{tabularx} \\
        \ \\
    \ \\
    \centering
\textbf{ SANE disc, 2.5 Jy} \\
\begin{tabularx}{0.48\textwidth}{@{}p{0.068\textwidth} p{0.058\textwidth} p{0.058\textwidth} p{0.058\textwidth} p{0.058\textwidth} p{0.058\textwidth}@{}}
      \thickhline
      {BH spin} & \textbf{$-15/16$} & \textbf{$-1/2$} & \textbf{$0$} & \textbf{$1/2$} & \textbf{$15/16$} \\
      \hline
$i=1 ^\circ$ & $	9.78	$&$	9.17	$&$	8.06	$&$	6.81	$&$	5.39	$\\
$i=20 ^\circ$ & $	9.09	$&$	8.63	$&$	7.61	$&$	6.44	$&$	5.15	$\\
$i=60 ^\circ$ & $	5.89	$&$	5.75	$&$	5.08	$&$	4.38	$&$	3.61	$\\
$i=90 ^\circ$ & $	4.07	$&$	4.03	$&$	3.77	$&$	3.31	$&$	2.89	$\\
      \thickhline
    \end{tabularx} \\
    \ \\
    \ \\
\centering
\textbf{ SANE disc, 1.25 Jy} \\
\begin{tabularx}{0.48\textwidth}{@{}p{0.068\textwidth} p{0.058\textwidth} p{0.058\textwidth} p{0.058\textwidth} p{0.058\textwidth} p{0.058\textwidth}@{}}
      \thickhline
      {BH spin} & \textbf{$-15/16$} & \textbf{$-1/2$} & \textbf{$0$} & \textbf{$1/2$} & \textbf{$15/16$} \\
      \hline
$i=1 ^\circ$ & $	9.56	$&$	8.97	$&$	7.83	$&$	6.64	$&$	5.20	$\\
$i=20 ^\circ$ & $	8.83	$&$	8.40	$&$	7.35	$&$	6.26	$&$	4.96	$\\
$i=60 ^\circ$ & $	5.65	$&$	5.48	$&$	4.80	$&$	4.19	$&$	3.42	$\\
$i=90 ^\circ$ & $	3.78	$&$	3.72	$&$	3.46	$&$	3.08	$&$	2.71	$\\
      \thickhline
    \end{tabularx}
\egroup
\caption{Tables of $\lambda_{\rm min}$, the minor axis length, in units of $R_{\rm g}$, for the SANE  models.}
\label{tab:lambdamin_SANE}
\end{table}

\begin{table}
\bgroup
\def\arraystretch{1.25}
\centering
\textbf{ MAD jet, 2.5 Jy} \\
\begin{tabularx}{0.48\textwidth}{@{}p{0.068\textwidth} p{0.058\textwidth} p{0.058\textwidth} p{0.058\textwidth} p{0.058\textwidth} p{0.058\textwidth}@{}}
      \thickhline
      {BH spin} & \textbf{$-15/16$} & \textbf{$-1/2$} & \textbf{$0$} & \textbf{$1/2$} & \textbf{$15/16$} \\
      \hline
$i=1 ^\circ$ & $	7.66	$&$	7.46	$&$	7.39	$&$	6.30	$&$	6.59	$\\
$i=20 ^\circ$ & $	7.52	$&$	7.30	$&$	7.18	$&$	6.18	$&$	6.41	$\\
$i=60 ^\circ$ & $	7.28	$&$	7.21	$&$	6.67	$&$	5.38	$&$	5.69	$\\
$i=90 ^\circ$ & $	6.59	$&$	6.69	$&$	5.88	$&$	4.82	$&$	5.11	$\\
      \thickhline
    \end{tabularx} \\
    \ \\
    \ \\
\centering
\textbf{ MAD jet, 1.25 Jy} \\
\begin{tabularx}{0.48\textwidth}{@{}p{0.068\textwidth} p{0.058\textwidth} p{0.058\textwidth} p{0.058\textwidth} p{0.058\textwidth} p{0.058\textwidth}@{}}
      \thickhline
      {BH spin} & \textbf{$-15/16$} & \textbf{$-1/2$} & \textbf{$0$} & \textbf{$1/2$} & \textbf{$15/16$} \\
      \hline
$i=1 ^\circ$ & $	7.27	$&$	7.07	$&$	6.93	$&$	5.97	$&$	6.23	$\\
$i=20 ^\circ$ & $	7.12	$&$	6.96	$&$	6.81	$&$	5.87	$&$	6.07	$\\
$i=60 ^\circ$ & $	6.94	$&$	6.92	$&$	6.33	$&$	5.06	$&$	5.30	$\\
$i=90 ^\circ$ & $	6.26	$&$	6.32	$&$	5.43	$&$	4.29	$&$	4.59	$\\
      \thickhline
    \end{tabularx} \\
        \ \\
    \ \\
    \centering
\textbf{ MAD disc, 2.5 Jy} \\
\begin{tabularx}{0.48\textwidth}{@{}p{0.068\textwidth} p{0.058\textwidth} p{0.058\textwidth} p{0.058\textwidth} p{0.058\textwidth} p{0.058\textwidth}@{}}
      \thickhline
      {BH spin} & \textbf{$-15/16$} & \textbf{$-1/2$} & \textbf{$0$} & \textbf{$1/2$} & \textbf{$15/16$} \\
      \hline
$i=1 ^\circ$ & $	7.61	$&$	7.22	$&$	6.36	$&$	6.14	$&$	6.25	$\\
$i=20 ^\circ$ & $	7.27	$&$	7.02	$&$	6.16	$&$	5.96	$&$	6.08	$\\
$i=60 ^\circ$ & $	6.78	$&$	6.58	$&$	5.19	$&$	4.69	$&$	5.05	$\\
$i=90 ^\circ$ & $	6.02	$&$	5.93	$&$	4.24	$&$	3.77	$&$	4.26	$\\
      \thickhline
    \end{tabularx} \\
    \ \\
    \ \\
\centering
\textbf{ MAD disc, 1.25 Jy} \\
\begin{tabularx}{0.48\textwidth}{@{}p{0.068\textwidth} p{0.058\textwidth} p{0.058\textwidth} p{0.058\textwidth} p{0.058\textwidth} p{0.058\textwidth}@{}}
      \thickhline
      {BH spin} & \textbf{$-15/16$} & \textbf{$-1/2$} & \textbf{$0$} & \textbf{$1/2$} & \textbf{$15/16$} \\
      \hline
$i=1 ^\circ$ & $	7.32	$&$	6.91	$&$	6.07	$&$	5.89	$&$	5.97	$\\
$i=20 ^\circ$ & $	6.97	$&$	6.73	$&$	5.88	$&$	5.71	$&$	5.82	$\\
$i=60 ^\circ$ & $	6.56	$&$	6.36	$&$	4.95	$&$	4.48	$&$	4.79	$\\
$i=90 ^\circ$ & $	5.79	$&$	5.65	$&$	3.95	$&$	3.52	$&$	3.97	$\\
      \thickhline
    \end{tabularx}
\egroup
\caption{Tables of $\lambda_{\rm min}$, the minor axis length, in units of $R_{\rm g}$, for the MAD models.}
\label{tab:lambdamin_MAD}
\end{table}

\section{Flux calibration for all models}
\label{sec:flux_calibration_appendix}

This appendix lists the values for $\mathcal{M}$, the mass-unit calibration factor used in {\tt RAPTOR}, which determines the overall accretion rate of the LLAGN. The unit of measurement is the gram. The values used in this paper are recapitulated in Tables \ref{tab:M_unit_SANE_jet} through \ref{tab:M_unit_MAD_disk}.

\begin{table}
\bgroup
\def\arraystretch{1.25}
\centering
\textbf{ SANE jet, 2.5 Jy} \\
\begin{tabularx}{0.48\textwidth}{@{}p{0.068\textwidth} p{0.058\textwidth} p{0.058\textwidth} p{0.058\textwidth} p{0.058\textwidth} p{0.058\textwidth}@{}}
      \thickhline
      {BH spin} & \textbf{$-15/16$} & \textbf{$-1/2$} & \textbf{$0$} & \textbf{$1/2$} & \textbf{$15/16$} \\
      \hline
$i=1 ^\circ$ & $9.159\mathrm{e}{21} $ & $1.134\mathrm{e}{22}$ & $7.641\mathrm{e}{21}$ & $2.83\mathrm{e}{21}$ & $1.929\mathrm{e}{21}$\\ 
$i=20 ^\circ$ & $9.112\mathrm{e}{21} $ & $1.055\mathrm{e}{22}$ & $5.233\mathrm{e}{21}$ & $2.763\mathrm{e}{21}$ & $1.981\mathrm{e}{21}$\\ 
$i=60 ^\circ$ & $8.173\mathrm{e}{21} $ & $7.91\mathrm{e}{21}$ & $3.587\mathrm{e}{21}$ & $2.227\mathrm{e}{21}$ & $2.447\mathrm{e}{21}$\\ 
$i=90 ^\circ$ & $7.924\mathrm{e}{21} $ & $7.23\mathrm{e}{21}$ & $3.46\mathrm{e}{21}$ & $1.877\mathrm{e}{21}$ & $3.162\mathrm{e}{21}$\\ 
      \thickhline
    \end{tabularx} \\
    \ \\
    \ \\
\centering
\textbf{ SANE jet, 1.25 Jy} \\
\begin{tabularx}{0.48\textwidth}{@{}p{0.068\textwidth} p{0.058\textwidth} p{0.058\textwidth} p{0.058\textwidth} p{0.058\textwidth} p{0.058\textwidth}@{}}
      \thickhline
      {BH spin} & \textbf{$-15/16$} & \textbf{$-1/2$} & \textbf{$0$} & \textbf{$1/2$} & \textbf{$15/16$} \\
      \hline
$i=1 ^\circ$ & $5.21\mathrm{e}{21} $ & $5.544\mathrm{e}{22}$ & $1.144\mathrm{e}{21}$ & $8.879\mathrm{e}{20}$ & $3.646\mathrm{e}{20}$\\ 
$i=20 ^\circ$ & $5.1\mathrm{e}{21} $ & $5.265\mathrm{e}{22}$ & $1.113\mathrm{e}{21}$ & $9.065\mathrm{e}{20}$ & $3.498\mathrm{e}{20}$\\ 
$i=60 ^\circ$ & $4.461\mathrm{e}{21} $ & $4.236\mathrm{e}{21}$ & $1.144\mathrm{e}{21}$ & $8.879\mathrm{e}{20}$ & $3.646\mathrm{e}{20}$\\ 
$i=90 ^\circ$ & $4.28\mathrm{e}{21} $ & $3.9\mathrm{e}{21}$ & $1.291\mathrm{e}{21}$ & $7.761\mathrm{e}{20}$ & $3.498\mathrm{e}{20}$\\ 
      \thickhline
    \end{tabularx} \\
    \ \\
    \ \\
\centering
\textbf{ SANE jet, 0.625 Jy} \\
\begin{tabularx}{0.48\textwidth}{@{}p{0.068\textwidth} p{0.058\textwidth} p{0.058\textwidth} p{0.058\textwidth} p{0.058\textwidth} p{0.058\textwidth}@{}}
      \thickhline
      {BH spin} & \textbf{$-15/16$} & \textbf{$-1/2$} & \textbf{$0$} & \textbf{$1/2$} & \textbf{$15/16$} \\
      \hline
$i=1 ^\circ$ & $3.106\mathrm{e}{21} $ & $3.171\mathrm{e}{22}$ & $6.182\mathrm{e}{20}$ & $4.532\mathrm{e}{20}$ & $1.205\mathrm{e}{20}$\\ 
$i=20 ^\circ$ & $3.043\mathrm{e}{21} $ & $2.980\mathrm{e}{22}$ & $5.932\mathrm{e}{20}$ & $4.348\mathrm{e}{20}$ & $1.156\mathrm{e}{20}$\\ 
$i=60 ^\circ$ & $2.526\mathrm{e}{21} $ & $2.373\mathrm{e}{21}$ & $4.923\mathrm{e}{20}$ & $3.762\mathrm{e}{20}$ & $9.402\mathrm{e}{19}$\\ 
$i=90 ^\circ$ & $2.277\mathrm{e}{21} $ & $2.14\mathrm{e}{21}$ & $5.46\mathrm{e}{20}$ & $3.536\mathrm{e}{20}$ & $8.655\mathrm{e}{19}$\\ 
      \thickhline
    \end{tabularx}
\egroup
\caption{Tables of $\mathcal{M}$, the flux-calibration factor, for the SANE jet models, in grams.}
\label{tab:M_unit_SANE_jet}
\end{table}

\begin{table}
\bgroup
\def\arraystretch{1.25}
\centering
\textbf{ SANE disc, 2.5 Jy} \\
\begin{tabularx}{0.48\textwidth}{@{}p{0.068\textwidth} p{0.058\textwidth} p{0.058\textwidth} p{0.058\textwidth} p{0.058\textwidth} p{0.058\textwidth}@{}}
      \thickhline
      {BH spin} & \textbf{$-15/16$} & \textbf{$-1/2$} & \textbf{$0$} & \textbf{$1/2$} & \textbf{$15/16$} \\
      \hline
$i=1 ^\circ$ & $1.948\mathrm{e}{20} $ & $9.496\mathrm{e}{19}$ & $4.059\mathrm{e}{19}$ & $1.894\mathrm{e}{19}$ & $6.264\mathrm{e}{18}$\\ 
$i=20 ^\circ$ & $1.877\mathrm{e}{20} $ & $9.243\mathrm{e}{19}$ & $3.937\mathrm{e}{19}$ & $1.835\mathrm{e}{19}$ & $6.042\mathrm{e}{18}$\\ 
$i=60 ^\circ$ & $1.644\mathrm{e}{20} $ & $8.242\mathrm{e}{19}$ & $3.326\mathrm{e}{19}$ & $1.568\mathrm{e}{19}$ & $4.687\mathrm{e}{18}$\\ 
$i=90 ^\circ$ & $1.543\mathrm{e}{20} $ & $7.678\mathrm{e}{19}$ & $3.068\mathrm{e}{19}$ & $1.427\mathrm{e}{19}$ & $4.013\mathrm{e}{18}$\\ 
      \thickhline
    \end{tabularx} \\
    \ \\
    \ \\
\centering
\textbf{ SANE disc, 1.25 Jy} \\
\begin{tabularx}{0.48\textwidth}{@{}p{0.068\textwidth} p{0.058\textwidth} p{0.058\textwidth} p{0.058\textwidth} p{0.058\textwidth} p{0.058\textwidth}@{}}
      \thickhline
      {BH spin} & \textbf{$-15/16$} & \textbf{$-1/2$} & \textbf{$0$} & \textbf{$1/2$} & \textbf{$15/16$} \\
      \hline
$i=1 ^\circ$ & $1.474\mathrm{e}{20} $ & $7.153\mathrm{e}{19}$ & $3.016\mathrm{e}{19}$ & $1.37\mathrm{e}{19}$ & $4.352\mathrm{e}{18}$\\ 
$i=20 ^\circ$ & $1.414\mathrm{e}{20} $ & $6.916\mathrm{e}{19}$ & $2.897\mathrm{e}{19}$ & $1.325\mathrm{e}{19}$ & $4.179\mathrm{e}{18}$\\ 
$i=60 ^\circ$ & $1.211\mathrm{e}{20} $ & $5.962\mathrm{e}{19}$ & $2.366\mathrm{e}{19}$ & $1.097\mathrm{e}{19}$ & $3.169\mathrm{e}{18}$\\ 
$i=90 ^\circ$ & $1.092\mathrm{e}{20} $ & $5.388\mathrm{e}{19}$ & $2.124\mathrm{e}{19}$ & $9.712\mathrm{e}{19}$ & $2.642\mathrm{e}{18}$\\ 
      \thickhline
    \end{tabularx}
\egroup
\caption{Tables of $\mathcal{M}$, the flux-calibration factor, for the SANE disc models, in grams.}
\label{tab:M_unit_SANE_disk}
\end{table}

\begin{table}
\bgroup
\def\arraystretch{1.25}
\centering
\textbf{ MAD jet, 2.5 Jy} \\
\begin{tabularx}{0.48\textwidth}{@{}p{0.068\textwidth} p{0.058\textwidth} p{0.058\textwidth} p{0.058\textwidth} p{0.058\textwidth} p{0.058\textwidth}@{}}
      \thickhline
      {BH spin} & \textbf{$-15/16$} & \textbf{$-1/2$} & \textbf{$0$} & \textbf{$1/2$} & \textbf{$15/16$} \\
      \hline
$i=1 ^\circ$ & $3.396\mathrm{e}{18} $ & $3.246\mathrm{e}{18}$ & $3.021\mathrm{e}{18}$ & $1.722\mathrm{e}{18}$ & $1.108\mathrm{e}{18}$\\ 
$i=20 ^\circ$ & $3.076\mathrm{e}{18} $ & $2.941\mathrm{e}{18}$ & $2.762\mathrm{e}{18}$ & $1.632\mathrm{e}{18}$ & $1.05\mathrm{e}{18}$\\ 
$i=60 ^\circ$ & $2.034\mathrm{e}{18} $ & $1.998\mathrm{e}{18}$ & $1.937\mathrm{e}{18}$ & $1.229\mathrm{e}{18}$ & $7.91\mathrm{e}{17}$\\ 
$i=90 ^\circ$ & $1.684\mathrm{e}{18} $ & $1.699\mathrm{e}{18}$ & $1.639\mathrm{e}{18}$ & $1.123\mathrm{e}{18}$ & $6.668\mathrm{e}{17}$\\ 
      \thickhline
    \end{tabularx} \\
    \ \\
    \ \\
\centering
\textbf{ MAD jet, 1.25 Jy} \\
\begin{tabularx}{0.48\textwidth}{@{}p{0.068\textwidth} p{0.058\textwidth} p{0.058\textwidth} p{0.058\textwidth} p{0.058\textwidth} p{0.058\textwidth}@{}}
      \thickhline
      {BH spin} & \textbf{$-15/16$} & \textbf{$-1/2$} & \textbf{$0$} & \textbf{$1/2$} & \textbf{$15/16$} \\
      \hline
$i=1 ^\circ$ & $2.225\mathrm{e}{18} $ & $2.132\mathrm{e}{18}$ & $1.98\mathrm{e}{18}$ & $1.123\mathrm{e}{18}$ & $7.23\mathrm{e}{17}$\\ 
$i=20 ^\circ$ & $2.037\mathrm{e}{18} $ & $1.946\mathrm{e}{18}$ & $1.826\mathrm{e}{18}$ & $1.064\mathrm{e}{18}$ & $6.85\mathrm{e}{17}$\\ 
$i=60 ^\circ$ & $1.351\mathrm{e}{18} $ & $1.337\mathrm{e}{18}$ & $1.1286\mathrm{e}{18}$ & $7.982\mathrm{e}{17}$ & $5.091\mathrm{e}{17}$\\ 
$i=90 ^\circ$ & $1.114\mathrm{e}{18} $ & $1.115\mathrm{e}{18}$ & $1.055\mathrm{e}{18}$ & $6.374\mathrm{e}{17}$ & $4.14\mathrm{e}{17}$\\ 
      \thickhline
    \end{tabularx}
\egroup
\caption{Tables of $\mathcal{M}$, the flux-calibration factor, for the MAD jet models, in grams.}
\label{tab:M_unit_MAD_jet}
\end{table}

\begin{table}
\bgroup
\def\arraystretch{1.25}
\centering
\textbf{ MAD disc, 2.5 Jy} \\
\begin{tabularx}{0.48\textwidth}{@{}p{0.068\textwidth} p{0.058\textwidth} p{0.058\textwidth} p{0.058\textwidth} p{0.058\textwidth} p{0.058\textwidth}@{}}
      \thickhline
      {BH spin} & \textbf{$-15/16$} & \textbf{$-1/2$} & \textbf{$0$} & \textbf{$1/2$} & \textbf{$15/16$} \\
      \hline
$i=1 ^\circ$ & $1.963\mathrm{e}{18} $ & $1.692\mathrm{e}{18}$ & $1.316\mathrm{e}{18}$ & $7.879\mathrm{e}{17}$ & $5.962\mathrm{e}{17}$\\ 
$i=20 ^\circ$ & $1.835\mathrm{e}{18} $ & $1.596\mathrm{e}{18}$ & $1.257\mathrm{e}{18}$ & $7.635\mathrm{e}{17}$ & $5.725\mathrm{e}{17}$\\ 
$i=60 ^\circ$ & $1.34\mathrm{e}{18} $ & $1.208\mathrm{e}{18}$ & $9.885\mathrm{e}{17}$ & $6.149\mathrm{e}{17}$ & $4.511\mathrm{e}{17}$\\ 
$i=90 ^\circ$ & $1.109\mathrm{e}{18} $ & $1.023\mathrm{e}{18}$ & $8.131\mathrm{e}{17}$ & $5.048\mathrm{e}{17}$ & $3.652\mathrm{e}{17}$\\ 
      \thickhline
    \end{tabularx} \\
    \ \\
    \ \\
\centering
\textbf{ MAD disc, 1.25 Jy} \\
\begin{tabularx}{0.48\textwidth}{@{}p{0.068\textwidth} p{0.058\textwidth} p{0.058\textwidth} p{0.058\textwidth} p{0.058\textwidth} p{0.058\textwidth}@{}}
      \thickhline
      {BH spin} & \textbf{$-15/16$} & \textbf{$-1/2$} & \textbf{$0$} & \textbf{$1/2$} & \textbf{$15/16$} \\
      \hline
$i=1 ^\circ$ & $1.321\mathrm{e}{18} $ & $1.144\mathrm{e}{18}$ & $8.891\mathrm{e}{17}$ & $5.277\mathrm{e}{17}$ & $3.922\mathrm{e}{17}$\\ 
$i=20 ^\circ$ & $1.24\mathrm{e}{18} $ & $1.084\mathrm{e}{18}$ & $8.5\mathrm{e}{17}$ & $5.091\mathrm{e}{17}$ & $3.783\mathrm{e}{17}$\\ 
$i=60 ^\circ$ & $8.971\mathrm{e}{17} $ & $8.127\mathrm{e}{17}$ & $6.608\mathrm{e}{17}$ & $4.066\mathrm{e}{17}$ & $2.941\mathrm{e}{17}$\\ 
$i=90 ^\circ$ & $7.361\mathrm{e}{17} $ & $6.789\mathrm{e}{17}$ & $5.277\mathrm{e}{17}$ & $3.247\mathrm{e}{17}$ & $2.328\mathrm{e}{17}$\\ 
      \thickhline
    \end{tabularx}
\egroup
\caption{Tables of $\mathcal{M}$, the flux-calibration factor, for the MAD disc models, in grams.}
\label{tab:M_unit_MAD_disk}
\end{table}

\end{appendix}
\label{lastpage}

\end{document}